\documentclass[useAMS,usenatbib]{mn2e}
\usepackage{graphicx}
\usepackage{natbib}
\usepackage{rotating}
\usepackage{amsmath}
\usepackage{amssymb}
\usepackage{array}
\usepackage{captcont}
\usepackage{txfonts}
\newcounter{one}   
\newcounter{two}   
\newcounter{three} 
\newcounter{four}  
\newcounter{five}  
\newcounter{six}   

\title{Parametric Strong Gravitational Lensing Analysis of Abell 1689}
\author[A. Halkola, S. Seitz and M. Pannella]
       {A. Halkola$^1$\thanks{E-mail:halkola@usm.uni-muenchen.de},
	 S. Seitz$^{1}$ and 
	 M. Pannella$^{2}$\\
	 $^{1}$Universit\"ats-Sternwarte M\"unchen, Scheinerstra\ss e 1, D-81679 M\"unchen, Germany\\
	 $^{2}$Max-Planck-Institut f\"ur extraterrestrische Physik, Giessenbachstra\ss e, Postfach 1312, D-85741 Garching, Germany}

\begin{document}
	 
  \date{Accepted ????. Received ????; in original form ????}
  \pagerange{\pageref{firstpage}--\pageref{lastpage}} \pubyear{????}

  \maketitle

  \label{firstpage}

  \begin{abstract}

    We measure the mass distribution of galaxy cluster Abell 1689
    within $0.3$ Mpc/h$_{70}$ of the cluster centre using its strong
    lensing effect on 32 background galaxies, which are mapped in
    altogether 107 multiple images. The multiple images are based on
    those of \citet{broadhurst:05} with modifications to both include
    new and exclude some of the original image systems. The cluster
    profile is explored further out to $\sim$~2.5~Mpc/h$_{70}$ with
    weak lensing shear measurements from \citet{broadhurst:05b}.

    The masses of $\sim$200 cluster galaxies are measured with
    Fundamental Plane in order to accurately model the small scale
    mass structure in the cluster. The cluster galaxies are modelled
    as elliptical truncated isothermal spheres. The scaling of the
    truncation radii with the velocity dispersions of galaxies are
    assumed to match those of i) field galaxies \citep{hoekstra:04}
    and ii) theoretical expectations for galaxies in dense
    environments \citep{merritt:83}. The dark matter component of the
    cluster is described by either non-singular isothermal ellipsoids
    (NSIE) or elliptical versions of the universal dark matter profile
    (ENFW). To account for substructure in the dark matter we allow
    for two dark matter haloes.

    The fitting of a single isothermal sphere to the {\it smooth} DM
    component results in a velocity dispersion of
    1450$^{+39}_{-31}$~km/s and a core radius of 77$^{+10}_{-8}$~kpc/h
    while an NFW profile has a a virial radius of
    2.86$\pm$0.16~Mpc/h$_{70}$ and a concentration of
    4.7$^{+0.6}_{-0.5}$.

    The {\it total} mass profile is well described by either an NSIS
    profile with $\sigma$=1514$_{-17}^{+18}$~km/s and core radius of
    $r_c$=71$\pm$5~kpc/h$_{70}$, or an NFW profile with C=6.0$\pm$0.5
    and r$_{200}$=2.82$\pm$0.11~Mpc/h$_{70}$.  The errors are assumed
    to be due to the error in assigning masses to the individual
    galaxies in the galaxy component. Their small size is due to the
    very strong constraints imposed by multiple images and the ability
    of the smooth dark matter component to adjust to uncertainties in
    the galaxy masses. The derived total mass is in good agreement
    with the mass profile of \citet{broadhurst:05} despite the
    considerable differences in the methodology used.

    Using also weak lensing shear measurements from
    \citet{broadhurst:05b} we can constrain the profile further out to
    r$\sim$2.5~Mpc/h$_{70}$. The best fit parameters change to
    $\sigma$=1499$\pm$15~km/s and $r_c$=66$\pm$5~kpc/h$_{70}$ for the
    NSIS profile and C=7.6$\pm$0.5 and
    r$_{200}$=2.55$\pm$0.07~Mpc/h$_{70}$ for the NFW profile.

    Using the same image configuration as \citet{broadhurst:05} we
    obtain a strong lensing model that is superior to that of
    \citet{broadhurst:05} (rms of 2.7'' compared to 3.2''). This is
    very surprising considering the larger freedom in the surface mass
    profile in their grid modelling.
  \end{abstract}


%

\section{Introduction}

Abell 1689 at a redshift of 0.18 is one of the richest clusters of
galaxies on the sky. Its closeness and richness should allow a
straightforward mass determination using the gravitational lens effect
on background galaxies, the dynamics of cluster members and the
\mbox{X-ray} emission of the intra-cluster gas. Nevertheless, these
methods have come up with strikingly different results in the past.

Observations with the Chandra \citep{xue:02} and XMM-Newton
\citep{andersson:04} satellites yield masses roughly a factor 2 lower
than strong lensing measurements \citep[e.g.][]{dye:01}.  The first
line-of-sight velocity measurements of cluster members
\citep{teague:90} had resulted in a velocity dispersion of $\sigma
\approx 2355$~km/s, compared to a value of only $\sigma\approx
1028$~km/s for an isothermal fit to weak lensing measurements by, for
example, \citet{king:02b}. Therefore the isothermal sphere velocity
dispersion estimates of the cluster from strong lensing, \mbox{X-ray}
and weak lensing analysis originally implied a mass estimate different
at a factor of up to 5.

The apparently incompatible weak and strong lensing results for an
isothermal sphere are most puzzling since both methods measure the
(same) line-of-sight projected two dimensional surface mass density of
the cluster. If parameters obtained with these two methods on
different angular scales do not agree for a given mass profile, it
implies that i) the assumed mass profile does not describe the true
mass distribution at all, or ii) that one analysis (more likely the
weak lensing analysis) suffers from underestimated systematic errors.
\citet{broadhurst:05} have shown that this is the case for A1689,
i.e. that in previous analyses the contamination of the 'background
galaxies' with cluster members could have biased the lensing signal of
the background galaxies. Their background galaxies show (compared to
\citet{clowe:01,king:02b}) a factor of roughly two higher lensing
signal on large scales and make the order of magnitude mass estimate
in the weak and strong lensing analysis agree.

The discrepant results from the cluster dynamics and the \mbox{X-ray}
data relative to the strong lensing analysis can potentially be
explained, if some assumptions in the interpretation of galaxies'
dynamics and the gas \mbox{X-ray} emissivity, i.e. having {\it one}
spherically symmetric isothermal structure in dynamical equilibrium,
are not valid. Indeed, \citet{girardi:97} identified three
substructures (using spectroscopic data from \citet{teague:90}) in
Abell 1689 which are well separated in velocity but overlap along the
line of sight. This reduced the previous value of the cluster's
velocity dispersion from $2355$~km/s by \citet{teague:90} to
$1429$~km/s, a value in good agreement with strong lensing
results. Evidence for substructure and merging was also found in
velocity differences of \mbox{X-ray} emission lines and in
\mbox{X-ray} temperature maps by \citet{andersson:04}. They pointed
out that the (by a factor of two low) \mbox{X-ray} mass estimate would
double if two equal mass structures along the line-of-sight are
responsible for the \mbox{X-ray} emission in stead of just one
structure.  The \mbox{X-ray} surface brightness map, however, and the
weak lensing data of A1689 indicate an almost circular (2D projected)
mass distribution centred on the cD galaxy. This is not necessarily
contradicting the substructure results summarised above, as long as
the two major contributions in mass are on the same line of sight, and
each of them is a fairly relaxed structure. The issue of
relax/unrelaxed systems and the cluster \mbox{X-ray} temperature -
mass relation is studied in 10 \mbox{X-ray} luminous galaxies by
\citet{smith:05} using Chandra \mbox{X-ray} and weak and strong
lensing. They find that a large fraction of their clusters are
experiencing, or recovering from, a cluster-cluster merger and that
the scatter in the cluster \mbox{X-ray} temperature - mass relation is
significantly larger than expected from theory.

The current status of the strong and weak lensing, the dynamical and
the \mbox{X-ray} mass estimates for A1689 are defined by the works of
\citet{broadhurst:05}, \citet{broadhurst:05b}, \citet{girardi:97} and
\citet{andersson:04} respectively. The mass estimates agree well with
the exception of the still low \mbox{X-ray} mass estimate. The
\mbox{X-ray} mass can be brought in line with the masses from other
methods if two equal mass substructures along the LOS are responsible
for the \mbox{X-ray} emission \citep{andersson:04}.

\citet{broadhurst:05b} also carried out a combined strong and weak
lensing analysis.  They rule out a softened isothermal profile at a
10-$\sigma$ level. According to their work, a universal dark matter
profile (NFW) with a concentration of $C=13.7^{+1.4}_{-1.1}$ and a
virial radius of $r_{vir}=2.04\pm0.07~Mpc/h_{100}$ fits the shear and
magnification based on weak and strong lensing data well. They point
out that the surprisingly large concentration for A1689 together with
results from other clusters could point to an unknown mechanism for
the formation of galaxy clusters. The strong rejection of an
isothermal sphere type profile has also been reported in galaxy
cluster Cl 0024+1654 by \citet{kneib:03} who probed the cluster
profile to very large clustercentric radius (5~Mpc).

\citet{oguri:05} however have demonstrated that halo triaxiality can
lead to large apparent central concentrations, if these haloes are
analysed assuming spherical symmetry. This is because a highly
elongated structure along the LOS imitates a high central density if
investigated in projection only. Allowing for triaxiality the central
concentration is less well constrained and not in disagreement with
results from numerical simulations of cluster mass profiles.\\

In our work we want to address the following points:
\begin{itemize}
\item Does one obtain the same mass profile with an analysis of the
strong lensing effect using a different method? We concentrate on
differences between grid method used by \citet{broadhurst:05} and our
parametric method.

\item Are mass profiles from weak (WL) and strong (SL) lensing at all
compatible with each other, or does the combination of both
observations already rule out an NFW or an isothermal profile?

\item How much does the large concentration and the level of
compatibility of the WL and SL results depend on the values of two
outermost WL shear data points?

\item How good is the relative performance of an isothermal sphere
vs. an NFW profile, based on both the strong and weak lensing
analyses?
\end{itemize}

We aim to investigate all these points with parametric cluster
models. Our basic assumptions are that substructure follows galaxies
and the cluster mass can be described by mass associated with the
galaxies (both luminous and dark) plus a smooth component. The
multiple image configurations determine if there are one or more of
these smooth components necessary, e.g. two haloes of similar mass
like in A370 (see e.g. \citet{kneib:93,abdelsalam:98}) , or a massive
halo plus less massive, group-like components.

We describe the smooth component by 2 parametric halo
profiles. Deviations from symmetry are accounted for by elliptical
deflection angles (ENFW) or potentials (NSIE) depending on the
model. The first halo profile is the so called universal dark matter
profile (hereafter NFW profile), an outcome of numerical simulations
of cold dark matter cosmologies \citep{navarro:96}. The second is a
non-singular isothermal ellipsoid (NSIE) which naturally reproduces
the observed flat rotation curves of both late and early-type
galaxies.

The profiles of cluster galaxies are described with an elliptical
truncated isothermal sphere profile \citep{blandford:87} whose
velocity dispersions are determined using both Fundamental-Plane and
Faber-Jackson relations.\\

Provided a range of plausible radial mass profiles are tested with
parametric halo profiles significantly better performance of grid
methods, like the one used by \citet{broadhurst:05} (also
\citet{diego:05a,diego:05b}), can give clues to the existence of dark
matter substructure not traced by galaxies (dark mini haloes) if these
dark haloes are numerous/massive enough to influence the lensing
observables on a relevant level. The difference can also result from
the details of a particular modelling, e.g. the treatment of the
cluster galaxy component and in the dark matter profiles used in
modelling the smooth DM of the cluster.\\

In section \ref{sec:data} we give a brief summary of the data and data
analysis used in this paper. Section \ref{sec:models} describes our
method to obtain the best fitting lensing models, results are given
and discussed in section \ref{sec:result}. We draw conclusions in
section \ref{sec:conclusion}.\\

The cosmology used throughout this paper is $\Omega_{m}$=0.30,
$\Omega_{\Lambda}$=0.70 and H$_{0}$=70~km/s/Mpc, unless otherwise
stated.


\section{Data and Data Analysis}
\label{sec:data}
We have used archived optical HST (WFPC2 and ACS) data in filters
F555W and F814W (WFPC2) and F475W, F625W, F775W and F850LP (ACS). A
summary of the data can be seen in Table~\ref{tab:data}. The
relatively wide field-of-view (202''x202'') and high resolution
(pixelscale 0.05'') of ACS allow us to probe A1689 over the area where
most of the multiple images are formed. WFPC2 data are used to further
constrain the photometric redshifts in the central region of A1689
covered by the observations.


\subsection{Data Reduction}


\subsubsection{HST - WFPC2}
The WFPC2 data in filters F555W and F814W come from HST proposal 6004
by Tyson. We use pipeline flatfielded images which were combined using
iraf\footnote{IRAF is distributed by the National Optical Astronomy
Observatories, which are operated by the Association of Universities
for Research in Astronomy, Inc., under cooperative agreement with the
National Science Foundation.} tasks in combination with psf fitting
cosmic ray rejection algorithms developed in house
\citep{goessl:02}. The steps of data reduction were as follows. First
all features with FWHM less than 1 pixel and a high signal to noise
were marked as cosmic rays and not used in any further analysis. In
the second step the four chips of each WFPC2 exposure were transformed
to a single coordinate system. In this step both the geometrical
distortions of the WFPC2 chips as well as translation and rotation
between the different CCDs and exposures were taken care off. The
description of \citet{holtzman:95} was used to remove the geometrical
distortions. The different chips have slightly different photometric
zeropoints and before the images were stacked all the images were
normalised to the zeropoint of the planetary camera. The stacking was
done by taking a kappa-sigma clipped mean of each pixel. It was found
during the reduction process that the two stage cosmic ray rejection
was necessary in order to remove all the cosmic rays efficiently. Most
of the cosmic rays were removed by psf- fitting in the first stage and
larger pointlike cosmic rays were removed in the stacking stage by the
kappa-sigma clipping.

\begin{table}
  \centering
  \caption[]{\label{tab:data}A summary of the data used in this study. }
  \begin{tabular}{lrcl}
    \hline
    \bf{Filter}&\bf{t (ks)}&\bf{\# of exposures}&\bf{psf (\arcsec)}\\
    \hline
    \hline
    F555W (WFPC2)&44.2&17&0.17\\
    F814W (WFPC2)&5.0&5&0.20\\
    \hline
    F475W  (ACS)&9.5&8&0.11\\
    F625W  (ACS)&9.5&8&0.10\\
    F775W  (ACS)&9.5&8&0.10\\
    F850LP (ACS)&16.6&14&0.11\\
    \hline
  \end{tabular}
\end{table}

\subsubsection{HST - ACS}
The advanced Camera for Surveys has a larger field of view than WFPC2
at a similar resolution. The data in filters F475W, F625W, F775W and
F850LP come from HST proposal 9289 by Ford. The "on the fly
re-processing" (OTFR) provides flatfielded and calibrated data. The
individual exposures were transformed to a common coordinate system
using PyDrizzle in pyraf. The cosmic rays were again removed in 2
stages as with the WFPC2 data.


\subsection{Object Catalogues}
The object catalogue was obtained with SExtractor
\citep{bertin:96}. To optimise the extraction parameters of Sextractor
- detection threshold and number of contiguous pixels - a procedure
similar to that in \citet{heidt:03} was followed. Sources were
detected from a signal-to-noise weighted sum of the four ACS filters. The
photometry for the detected sources was then done in all
filters. Output of SExtractor used later in the lensing model include
total and aperture photometry as well as source ellipticities and
their position angles.


\subsection{Spectroscopic redshifts}
The redshift data are collected from several published studies of
Abell 1689 (\citealt{girardi:97}, \citealt{balogh:02},
\citealt{duc:02}, \citealt{golse:phd}\footnote{Available at
http://tel.ccsd.cnrs.fr/documents/archives0/00/00/22/79/}).  In total
84 spectroscopic redshifts were available in the ACS field of
A1689. Except \citeauthor{golse:phd}, these studies concentrate on the
line-of-sight velocities of the cluster members in order to obtain the
velocity dispersion and hence an estimate of the dynamical mass of the
cluster. The redshift information is used to get secure cluster
members for the lensing analysis, to exclude galaxies which are not
part of the cluster and to compare photometric redshifts with
spectroscopic ones.


\subsection{Photometric redshifts}

\begin{figure}
  \centering
  \includegraphics[height=0.9\columnwidth]{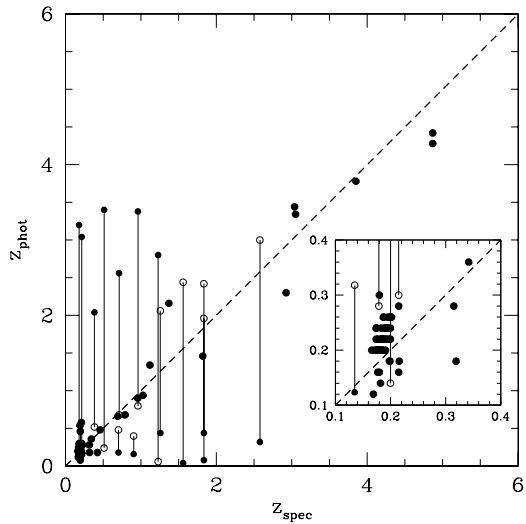}
  \caption{Comparison between spectroscopic and photometric
  redshifts. Notice how the reddest cluster ellipticals have been
  pushed to higher redshifts to compensate for the relative blueness
  of the model elliptical SED in the SED library used to compute the
  photometric redshifts. Filled symbols show the most likely redshift,
  open symbols the second most likely for the objects where it is
  closer to the true redshift.}
  \label{fig:z_spec_phot}
\end{figure}

\begin{figure}
  \centering
  \includegraphics[height=0.9\columnwidth]{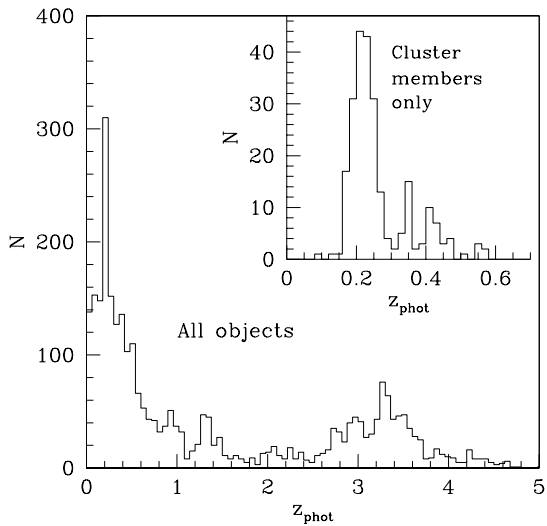}
  \caption{Photometric redshift distribution of objects in the field
  of A1689. The cluster is clearly visible as a peak at
  z$\sim$0.2. The inlaid figure shows objects considered as cluster
  galaxies. In a number of cases the 4000~$\AA$ break of cluster
  galaxies was misinterpreted as Lyman break and hence a redshift
  z$\sim$3 was assigned (see text for details).}
  \label{fig:z_spec_phot2}
\end{figure}

We have calculated photometric redshifts using the method described in
\citet{bender:01}. For the central regions we were able to use all 6
filters, otherwise only the 4 available ACS filters were
used. Comparison between photometric redshifts and available
spectroscopic ones is shown in Figure
\ref{fig:z_spec_phot}. Unfortunately spectroscopic redshifts are not
numerous and those that exist are mainly for cluster members. The
model elliptical galaxy in the SED library used to calculate the
photometric redshifts is bluer than the reddest cluster members and
hence the cluster galaxies are pushed to redshifts slightly higher
than that of the cluster. The slope of the cluster redsequence causes
the large spread in photometric redshifts of the cluster members. The
photometric redshift distribution of all objects is shown in Figure
\ref{fig:z_spec_phot2}. The cluster appears as a narrow peak at
z$\sim$0.2 on the redshift histogram. Objects at z$\sim$3-4 are
either gravitationally lensed background galaxies or cluster galaxies
whose 4000~$\AA$ break was misidentified as Lyman break. This places
them to redshifts $\sim$3. To clearly discriminate between the two
breaks we would need both redder and especially bluer filters than the
ones available. This confusion is not important in our case since
photometric redshifts are used to constrain redshifts of the multiple
images. For these the low redshift peak can be excluded because
gravitational lensing is inefficient if the background source is close
to the lens.

\section{Lensing Models}
\label{sec:models}

In this section we describe how the lensing models of A1689 were
constructed; the different mass components of the cluster, the
multiple image systems and the optimisation of model parameters. The
lensing profiles are described in detail in appendix
\ref{app:lensing}.


\subsection{Cluster Galaxy Component}
\label{sec:galaxies}

\begin{figure}
  \centering
  \includegraphics[height=\columnwidth]{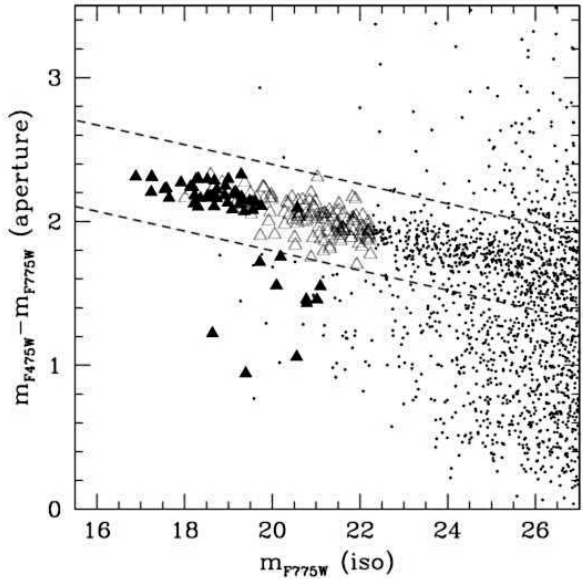}
  \caption{Cluster redsequence in ACS F475W/F775W colour-magnitude
    diagramme. The solid triangles represent cluster galaxies with
    spectroscopic redshifts, open triangles all other objects which
    were considered as cluster members in the lensing analysis. The
    remaining objects are represented by dots. A clear cluster
    redsequence can be seen. See section \ref{sec:galaxies} for
    details on the selection criterion for cluster membership.}
  \label{fig:redseq}
\end{figure}

The cluster galaxies were selected using the redsequence method
supported by spectroscopic redshifts where available. The
spectroscopic data are taken from \citet{girardi:97} (using data from
\citet{teague:90}), \citet{balogh:02} and \citet{duc:02}. We have
excluded 6 objects from the cluster catalogue obtained with the
redsequence method based on fore/background objects listed in
\citet{balogh:02} and \citet{duc:02}. Fig. \ref{fig:redseq} shows a
colour-magnitude diagramme of the cluster. We chose to use filters
F475W and F775W from the ACS observations as the cluster redsequence
is seen particularly clearly in these two filters. Galaxies included
in the lensing analysis are marked by triangles in Figure
\ref{fig:redseq}. Solid triangles show the galaxies which have been
spectroscopically confirmed to be cluster members in one or more of
the referred papers. To find more members a redsequence in the CM
diagramme was determined by fitting a line to the bright end of the
redsequence. Those galaxies whose F475W-F775W colour deviated by less
than 0.3 magnitudes from the fitted sequence were included as cluster
members (region between the two inclined dashed lines in Figure
\ref{fig:redseq}).

\begin{figure}
  \centering
  \includegraphics[height=\columnwidth, bb=60 180 580 700, clip=]{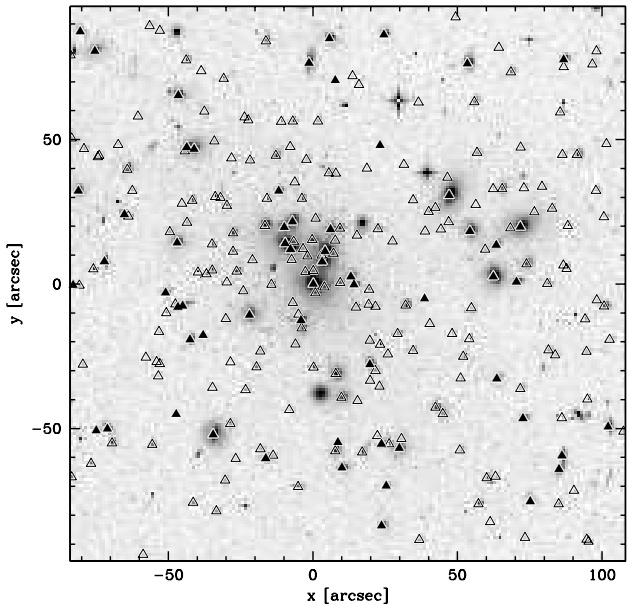}
  \caption{Positions of cluster galaxies in the field of Abell
  1689. The symbols correspond to those in
  Fig.~\ref{fig:redseq}. Origin of the coordinate system is on the
  central cD galaxy. The box width is 200'' (corresponding to
  $\sim$625kpc at z=0.18).}
  \label{fig:A1689}
\end{figure}

Figure \ref{fig:A1689} shows the positions of cluster members in the
field of A1689 (using the same symbols as in Figure
\ref{fig:redseq}).\\

The cluster galaxies were modelled with an elliptical BBS profile (see
appendix \ref{app:lensing} for further details). We have treated the
two parameter profile as a one parameter profile by assuming that
$s=s(\sigma)$. We find the fit to multiple images with a galaxy
component described by the values for $s_{*}$ (=185~$h^{-1}$~kpc) and
$\sigma_{*}$ (=136~km/s) and the scaling
($s=s_*\times(\sigma/\sigma_*)^2$) found for galaxies in the
Red-Sequence Cluster Survey \citep{hoekstra:04} is very poor. This
means that the galaxy haloes in A1689 must be significantly
stripped. The stripping of galaxies in cluster environment has been
reported earlier by e.g. \citet{natarajan:98,natarajan:02} who have
used galaxy--galaxy lensing in clusters to study the propertias of
galaxy haloes in 6 clusters at redshifts Z=0.17-0.58. They found
strong evidence for tidally truncated haloes around the galaxies
compared to galaxies in the field.

We base our models for the tidal stripping of galaxies on
observational work by \citet{hoekstra:04} for galaxies in the field
($s \propto \sigma^2$) and theoretical expectations for galaxies in
cluster environment ($s \propto \sigma$) \citep{merritt:83}. We only
take the scaling of the truncation radii with the velocity dispersions
of the galaxies from the aforementioned works and find the
normalisation of the truncation radius, $s_*$, to fit the multiple
images. The two scaling laws adopted in the paper are then 1)
$s=s_*\times(\sigma_{gal}/136km/s)^2$ and 2)
$s=s_*\times(\sigma_{gal}/136km/s)$, where for both scaling laws and
all Models we have found the $s_*$ that best reproduces the observed
multiple images. The scaling law for the truncation of galaxies in
cluster environment will be treated in more detail in a forth coming
publication (Halkola et al., 2006 in preparation).

The positions, ellipticities (of surface brightness) and position
angles were taken from SExtractor output parameters. The velocity
dispersions of cluster galaxies were determined mostly using the
Fundamental Plane. For a small number of galaxies also the
Faber-Jackson relation was used.


\subsubsection{Central Velocity Dispersions of Cluster Galaxies \& Halo
Velocity Dispersions}

The Fundamental Plane (hereafter FP) links together, in a tight way,
kinematic (velocity dispersion), photometric (effective surface
brightness) and morphological (half light radius) galaxy properties
\citep{dressler:87,djorgovski:87,bender:92}. We assume that the
central velocity dispersions of a galaxy, as derived from the FP, is
equal to the halo velocity dispersion, and that mass in disk can be
neglected.

The FP relation allows us to estimate the velocity dispersion of
galaxies more accurately than the standard Faber-Jackson relation
approach \citep{faber:76}. We model the 2--dimensional light profiles
of cluster galaxies with PSF--convolved Sersic \citep{sersic:68}
profiles using two packages, GALFIT \citep{peng:02} and GIM2D
\citep{simard:99}, to have a better handle on the systematics. The
analysis was performed on the F775W ACS image. 176 objects with AB
magnitudes brighter than 22 were fitted. The point spread function
used to convolve the models was derived by stacking stars identified
in the field. The results coming out from the two completely different
softwares agree very well.

In order to be able to use a FP determination for cluster galaxies at
redshift $\sim 0.2$ in restframe Gunn~r filter
\citep{joergensen:96,ziegler:01,fritz:05}, all the observed
F775W$_{AB}$ surface brightnesses (extinction corrected) were
converted to restframe Gunn~r$_{GT}$ ones and corrected for the
cosmological dimming.  Since the observed F775W passband is close to
restframe Gunn~r at the redshift of A1689, the conversion factor
between observed F775W and restframe Gunn~r is small.\\

The mean observed surface brightness within $r_e$ is:
\begin{equation}
  \langle\mu_e\rangle_ {F775W}= F775W_{observed} +2.5 \log (2\pi)+5
  \log(r_e) - 10 \log(1+z),
\end{equation}
where the last term corrects for the dimming due to the expansion of
the Universe. It is then converted to restframe Gunn~r$_{GT}$ by:

\begin{equation}
  \label{sbrest}
  \mathrm{\langle \mu_e\rangle}_r  = \mathrm{\langle \mu_e\rangle}_{F775W}
  - A_{F775W} + K(r,F775W,z) + GT_{corr},
\end{equation}

The Galactic extinction correction $A_{F775W}$ is calculated from the
list of A/E(B-V) in Table 6 of \citet{schlegel:98},
along with their estimate of E(B-V) calculated from COBE and IRAS maps
as well as the Leiden-Dwingeloo maps of HI emission. We adopted for
$A_{F775W}$ a value of 0.06.

The "k-correction colour", K(r,F775W,z), is the difference between rest
frame Gunn~r and observed F775W magnitude and includes also the $2.5
\log (1+z)$ term. It was obtained by using an elliptical template from
CWW \citep{coleman:80} and synthetic SEDs obtained for old stellar
populations (10 Gyr, i.e. z$_f$ = 5 observed at z=0.2) with the BC2003
Bruzual and Charlot models \citep{bruzual:03}. All models give a
conversion factor of approximately 0.174.  The correction needed to
pass from the AB photometric system to the Gunn\&Thuan system is
GT$_{\textrm{corr}} \approx 0.17$ .

We used the FP coefficients from \citeauthor{fritz:05}.  For the
Gunn~r band then
\begin{equation}
  1.048*\log R_e=1.24*\log \sigma-0.82* \langle I\rangle_e +ZP_{FPr},
\end{equation}
where the $\langle I\rangle_e$ term, i.e. the mean surface brightness
in units of L$_\odot$/pc$^2$, is given for the Gunn~r band by the
equation:
\begin{equation}
  \log \langle I\rangle_e = -0.4 (\langle\mu_r\rangle_e - 26.4).
\end{equation}

The zero--point of the FP $ZP_{FPr}$ is a quantity changing with both
the cluster peculiarity and, mainly, with the cluster redshift. We
used for $ZP_{FPr}$ the value published in \citet{fritz:05}. Their
study was focused on A2218 and A2390, two massive clusters at almost
the same redshift as A1689. They applied a bootstrap bisector method
in estimating the $ZP_{FPr}$ and relative uncertainties, finding a
value of 0.055$\pm$0.022.

Finally, we inserted the values derived from our morphological fitting
procedures into the FP relation. The uncertainties on the derived
velocity dispersions were estimated by taking into account the errors
on the morphological parameters, the propagated photometric
uncertainties, the error on the $ZP_{FPr}$ value and the intrinsic
scatter of the FP relation, which gives the main contribution.  We
found that an estimate of 0.1 in log($\sigma$) is a good value for the
total uncertainty in velocity dispersion for objects having a velocity
dispersion greater than 70~km/s. For lower velocity dispersions down
to 24~km/s, we assumed an overall uncertainty of 0.2 dex. The fitted
parameters for the 80 most massive galaxies are tabulated in Table
\ref{tab:app:FP}.

\begin{figure}
  \centering
  \includegraphics[height=\columnwidth]{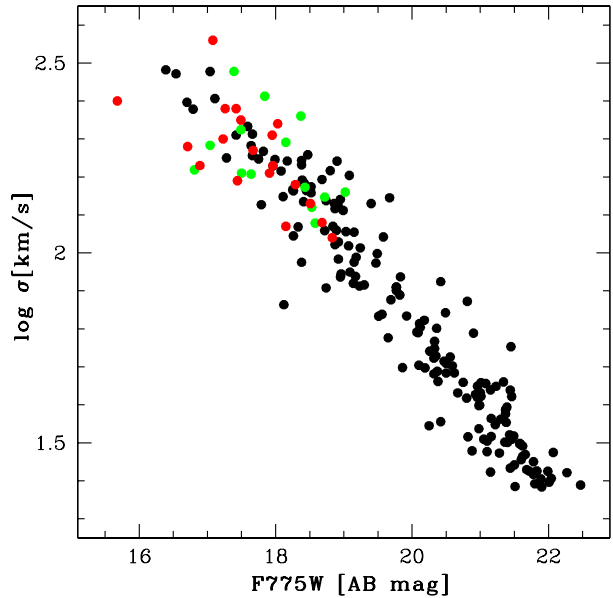}
  \caption{Total observed magnitude - $\sigma$ relation for three
    z$\sim$0.2 clusters. Red points are measured values taken from
    \citet{ziegler:01} for A2218 (z=0.18), green points are values
    taken from \citet{fritz:05} for A2390 (z=0.23) and full (empty)
    black points refer to the velocity dispersion estimates obtained
    in this paper for A1689 using the GIM2D (GALFIT) morphology. The
    literature values have been transformed to F775W$_{AB}$ magnitudes
    by applying relatively small colour terms (0.04,-0.4) and the AB
    correction (0.4).}
  \label{fig:FJ}
\end{figure}

Additionally, we have obtained the velocity dispersions of 26 galaxies
using the Faber-Jackson relations derived using the 176 galaxies for
which we have obtained the velocity dispersions via FP. These are all
faint galaxies with $\sigma<60km/S$.

The Einstein-radius of an isothermal sphere can then be written as
$\theta_E=1.4''(\sigma/220 km/s)^2$ D, where D is a geometrical factor
of order unity depending on redshifts of the objects and cosmology
(0.78 $<$ D $<$ 0.92 for 1 $<z_{s}<$ 6 and our adopted
cosmology). Cluster members with $\sigma>24~\mathrm{km/s}$ were
included in the galaxy component of the cluster. This limit is
somewhat arbitrary and below the luminosity limit where FP and FJ are
determined. An Einstein radius smaller than the pixel size of the ACS
ensures that all galaxies which could significantly affect image
morphologies locally are included when external shear and convergence
from the other cluster galaxies and the cluster halo are present.


\subsubsection{Ellipticities of Cluster Galaxies and Their Haloes}

\citet{blandford:87}, \citet{kormann:94} and others have noted that
for elliptical potentials the accompanying surface mass density can
have negative values. We have used elliptical potential for our NSIE
profile since it is straightforward to implement (all parameters of
interest can be calculated from the analytic derivatives of the
potential). An alternative approach is to have an elliptical mass
distribution as demonstrated by \citet{kormann:94} but the expressions
for $\alpha\textrm{, }\kappa\textrm{ and }\gamma$ are considerably
more complicated.

\begin{figure}
  \centering \includegraphics[height=\columnwidth]{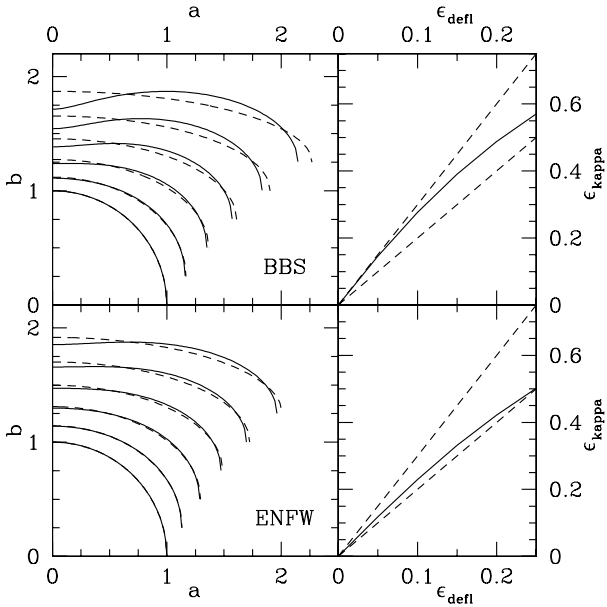}
  \caption{ Ellipticity of surface mass density vs. profile
    ellipticity (in deflection angle) for the two profiles (BBS is on
    the top two panels and ENFW on the bottom two). The ellipticity of
    $\kappa$ is estimated by fitting an ellipse to lines of constant
    surface mass density. Right panels show $\kappa$=0.2 isodensity
    contours in solid and fitted ellipses in dashed lines. Profile
    ellipticities are {0.0, 0.05, 0.10, 0.15, 0.20, 0.25} and increase
    from bottom to top. The curves have an offset of 0.25 in b for
    clarity. Left panels show $\epsilon_{kappa}$ as a function of
    profile ellipticity $\epsilon_{defl}$. The dashed lines are
    $\epsilon_{kappa}$=3~$\epsilon_{defl}$ and
    $\epsilon_{kappa}$=2~$\epsilon_{defl}$ lines.  }
  \label{fig:ellipticity}
\end{figure}

\begin{figure}
  \centering \includegraphics[height=\columnwidth]{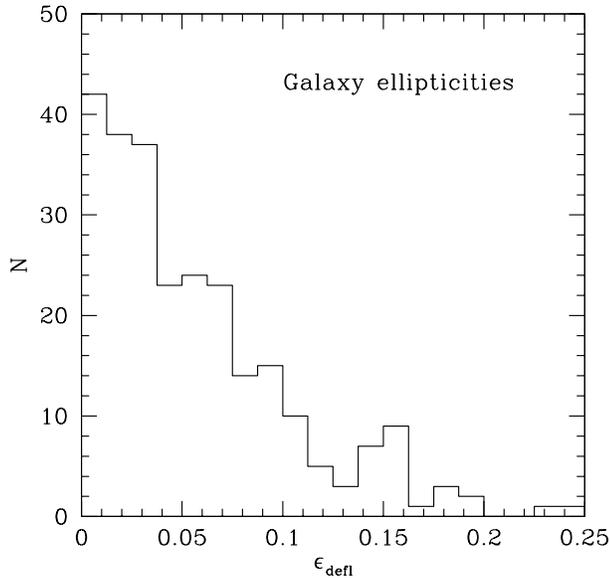}
  \caption{ A histogram of the ellipticities, $\epsilon_{defl}$, for
  the BBS profiles used to model the cluster galaxies. Most of the
  galaxies have an ellipticity well below 0.15 where the isodensity
  contours of surface mass density start to appear peanut shaped.}
  \label{fig:gal_ell}
\end{figure}

For the ENFW and BBS profiles we have introduced the ellipticity to
the deflection angle, see appendix \ref{app:lensing} for details and
references. The effect of using an elliptical deflection angle instead
of an elliptical mass distribution is shown in Figure
\ref{fig:ellipticity}. The ellipticities of the mass distributions
were estimated by fitting an ellipse to $\kappa$=0.2 isodensity
contours for both BBS (top) and ENFW (bottom) haloes with 6 different
ellipticities (0, 0.05, 0.1, 0.15, 0.2 and 0.25). On the right panel a
quarter of the $\kappa$=0.2 isodensity contour for different profile
ellipticities are drawn in solid. Dashed lines show the best fit
ellipses. For the BBS model the isodensity contours start to deviate
from an ellipse at $\epsilon_{defl}>$0.15, where the contours first
appear boxy before turning peanut shaped. For the ENFW profile the
contours are only slightly peanut shaped at
$\epsilon_{pot}$=0.25. Left panels of Figure \ref{fig:ellipticity}
show how the ellipticity of mass deviates from
$\epsilon_{kappa}$=2$\epsilon_{defl}$ and
$\epsilon_{kappa}$=3$\epsilon_{defl}$ lines shown dashed. We have
assumed that light and mass have the same ellipticity and used the
relation in Figure \ref{fig:ellipticity} to convert the measured
galaxy ellipticities ($\epsilon_{kappa}$) to BBS model ellipticities
($\epsilon_{defl}$). A histogram of the ellipticities of the included
cluster galaxies are shown in Figure \ref{fig:gal_ell}.\\

\subsection{Dark Matter Not Associated with Galaxies}
\label{sec:DM}
Different studies have shown that Abell 1689 is not a simple structure
with only one component. In an early strong lensing analysis of A1689
\citet{miralda:95} assumed two halo components based on the
distribution of galaxies. \citet{girardi:97}, on the other hand,
spectroscopically identified three distinct groups in Abell 1689. More
recently \citet{andersson:04} have also found evidence for
substructure and possible merger in \mbox{X-ray} data. This prompted
us to model the cluster dark matter in A1689 with two dark matter
haloes. The use of more haloes in the modelling is not desired since
this increases problems related to a large number-of free parameters;
larger parameter space to explore, increased difficulty of finding
global minimum and degeneracies between the free parameters.  Both
haloes have 6 free parameters: position (x,y), ellipticity, position
angle and in the case of NSIE profile velocity dispersion $\sigma$ and
core radius $r_c$ and for the NFW profile virial radius $r_{200}$ and
concentration parameter C.

For all the following modelling we have constrained the first halo to
reside within 50'' in x and y from the cD galaxy in order to reduce
the volume of the parameter space and to reduce the degeneracy between
the parameters of the two haloes. We do not want to tightly connect
the halo with any of the galaxies but use the position of the cD as a
first guess for the position of the cluster centre. This is supported
by \mbox{X-ray} maps of Abell 1689 \citep{xue:02,andersson:04} as well
as weak lensing studies \citep{king:02b} which place the centre of the
mass distribution very near the cD galaxy. The position of the second
halo was initially set to coincide with the visually identified
substructure to the north-east of the cluster centre but was left
unconstrained in the optimisation.


\subsection{Multiple Images \& Arcs}

It is evidently of great importance for the modelling to find as many
multiple images as possible. The colour and surface brightness of an
object are unaffected by gravitational lensing and we have hence used
the colour, surface brightness and the morphology of the images to
identify multiple image systems. We have first identified arcs and
obvious multiple images which were then used to find an initial set of
halo parameters. Initial constraints include images from image systems
1, 3, 4, 5, 6, 12 and arcs that contain images systems 8, 14, 20 and
32. A model based on these images could now be used to search for more
images for the existing image systems as well as new image systems
which could be included in the model and constrain the model
parameters further. New images were searched for by looking for image
positions whose source lies within a specified distance, e.g. 5'',
from the source of an existing image. This method is basically the
same as described in \citet{schramm:87} and \citet{kayser:88}.

The images we have found are with a few exceptions also identified in
the pioneering work of \citet{broadhurst:05} and essentially form a
subset of their images. For our analysis we have merged the two image
catalogues to obtain a catalogue of 107 multiple images in 32 image
systems one of which is an long arc. In the merging we have split the
image system 12 from \citet{broadhurst:05} to two separate systems (12
and 13) with 2 additional images from our catalogue. The splitting
includes separating two images with the same spectroscopic redshift
into two different image systems. We have done this based on the
morphology of the images and our lensing models and we believe these
to originate from 2 different sources. Also \citet{seitz:98} in their
analysis of cluster MS-1512 reported two sources at the same
redshift. In the case of MS-1512 \citet{teplitz:04} used near-infrared
spectroscopy to confirm that the sources were indeed separate with a
difference of only 400~km/s in velocity (0.0013 in redshift). To
positively identify a set of images to originate from a single source
is very difficult without extremely accurate spectroscopy or obviously
the same (complex) morphology. The field of A1689 has a vast number of
images that can potentially be erroneously assigned to any multiple
image system. In our work we have rather excluded an image than
include it in an image system. For this reason we have also
excluded image system 20 from \citet{broadhurst:05} from our
analysis. Image systems not used by \citet{broadhurst:05} are systems
31 and 32, a system with 2 images and a long arc respectively.

Of the images systems used 8 have an even number of images. The
missing images in these cases are always demagnified and based on the
lensing models mostly expected to lie near a galaxy making their
identification very difficult.

\begin{figure*}
  \centering
  \includegraphics[width=2.0\columnwidth, bb=60 60 580 580, clip=]{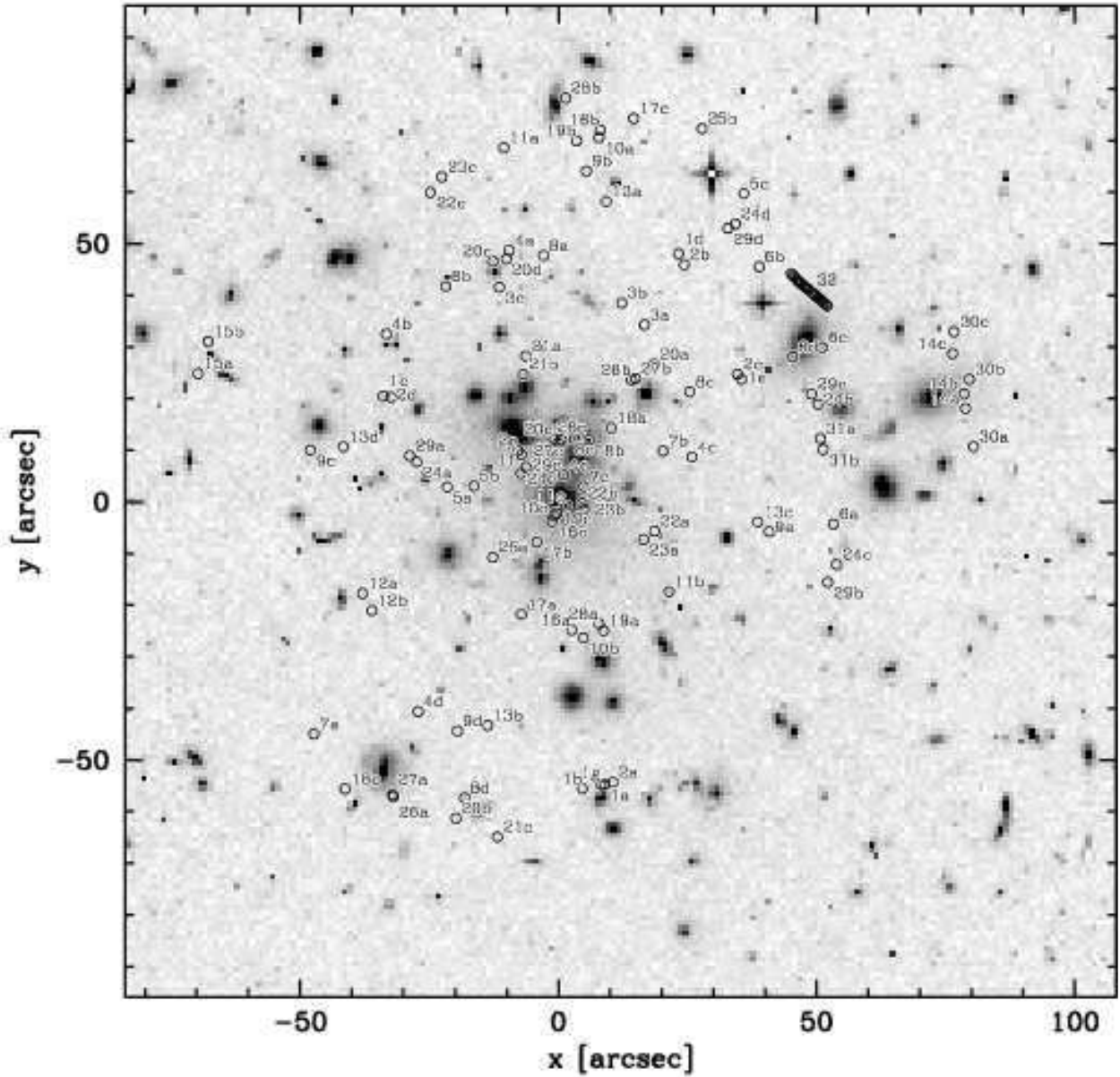}
  \caption{Positions of multiple images used in constraining the model
  parameters. The images from each multiple image system have the same
  number, the images within an image system are coded with
  letters. The box side is $\sim$600~kpc with our cosmology.}
  \label{fig:A1689_images}
\end{figure*}

In Figure \ref{fig:A1689_images} we show all the multiple images used
in this study. More details, such as positions and redshifts, of the
image systems, arcs and images can be found in Appendix
\ref{app:images} Tables \ref{tab:app:summary} and
\ref{tab:app:details} as well as \citet{broadhurst:05}.

\begin{figure}
  \centering
  \includegraphics[width=\columnwidth]{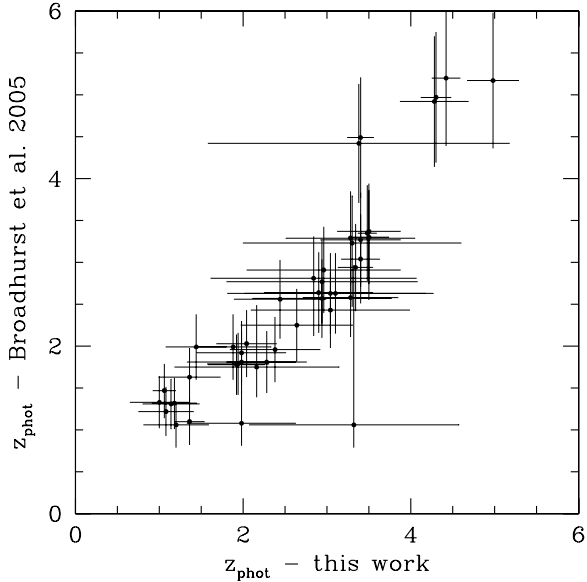}\\
  \caption{A comparison of photometric redshifts of multiple images
  from this work to those from \citet{broadhurst:05}. The
  correspondence is very good and for all objects within the
  errors. The one object with a z$_{phot}\sim$1 from
  \citet{broadhurst:05} and z$_{phot}\sim$3.4 from this work has a
  spectroscopic redshift of z$_{spec}=$3.04.}
  \label{fig:zphot_comp}
\end{figure}

A comparison between photometric redshifts from this work and those of
\citet{broadhurst:05} is shown in Figure \ref{fig:zphot_comp}. The
overall agreement is very good. The one object with a z$\sim$1 from
\citet{broadhurst:05} and z$\sim$3.4 from this work belongs to image
system 1 and is one of the few objects with a spectroscopic redshift
(z$_{spec}$=3.0).

\begin{figure}
  \centering
  \includegraphics[width=\columnwidth,bb=18 440 592 718,clip=]{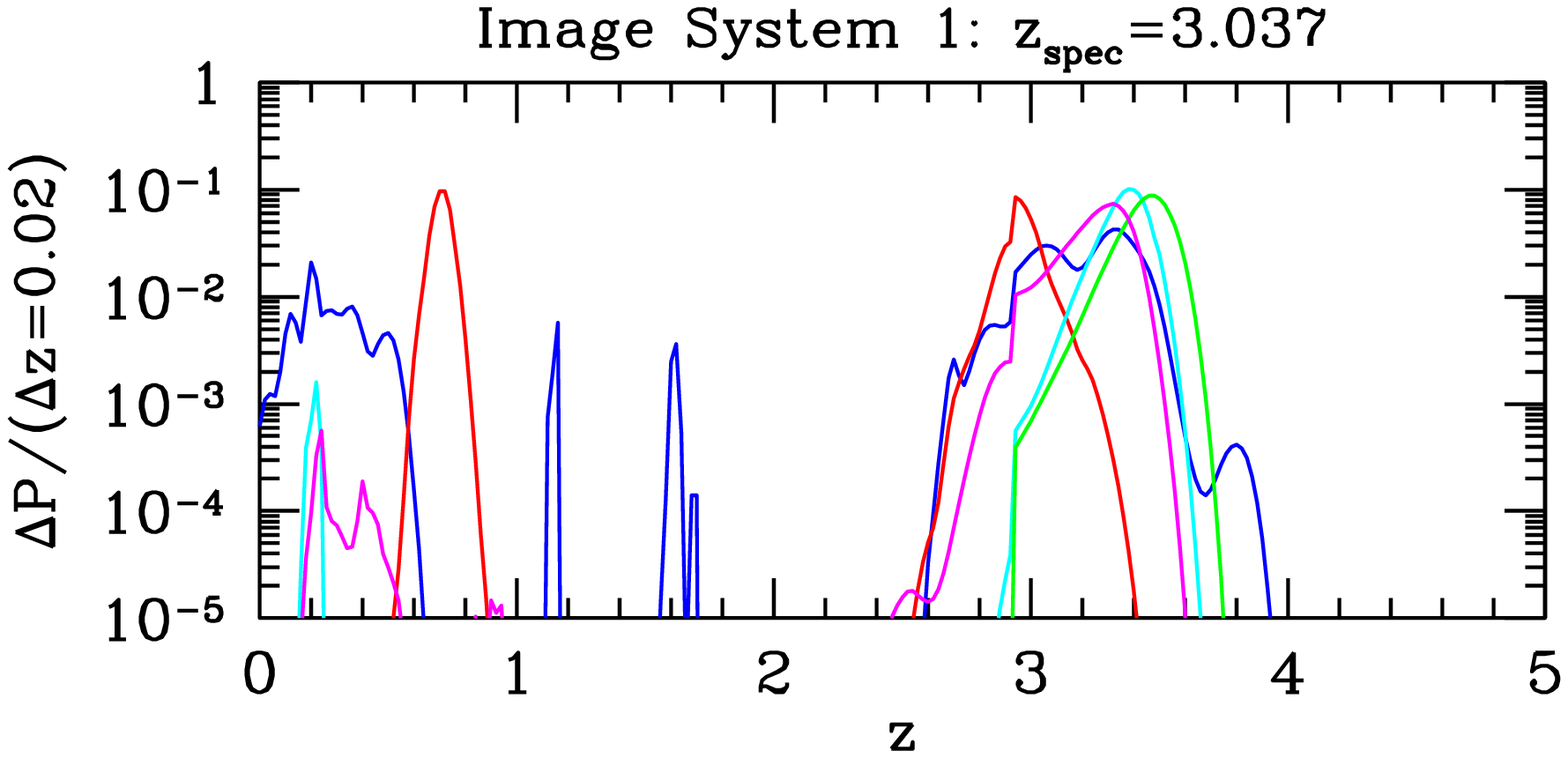}\\
  \includegraphics[width=\columnwidth,bb=18 440 592 718,clip=]{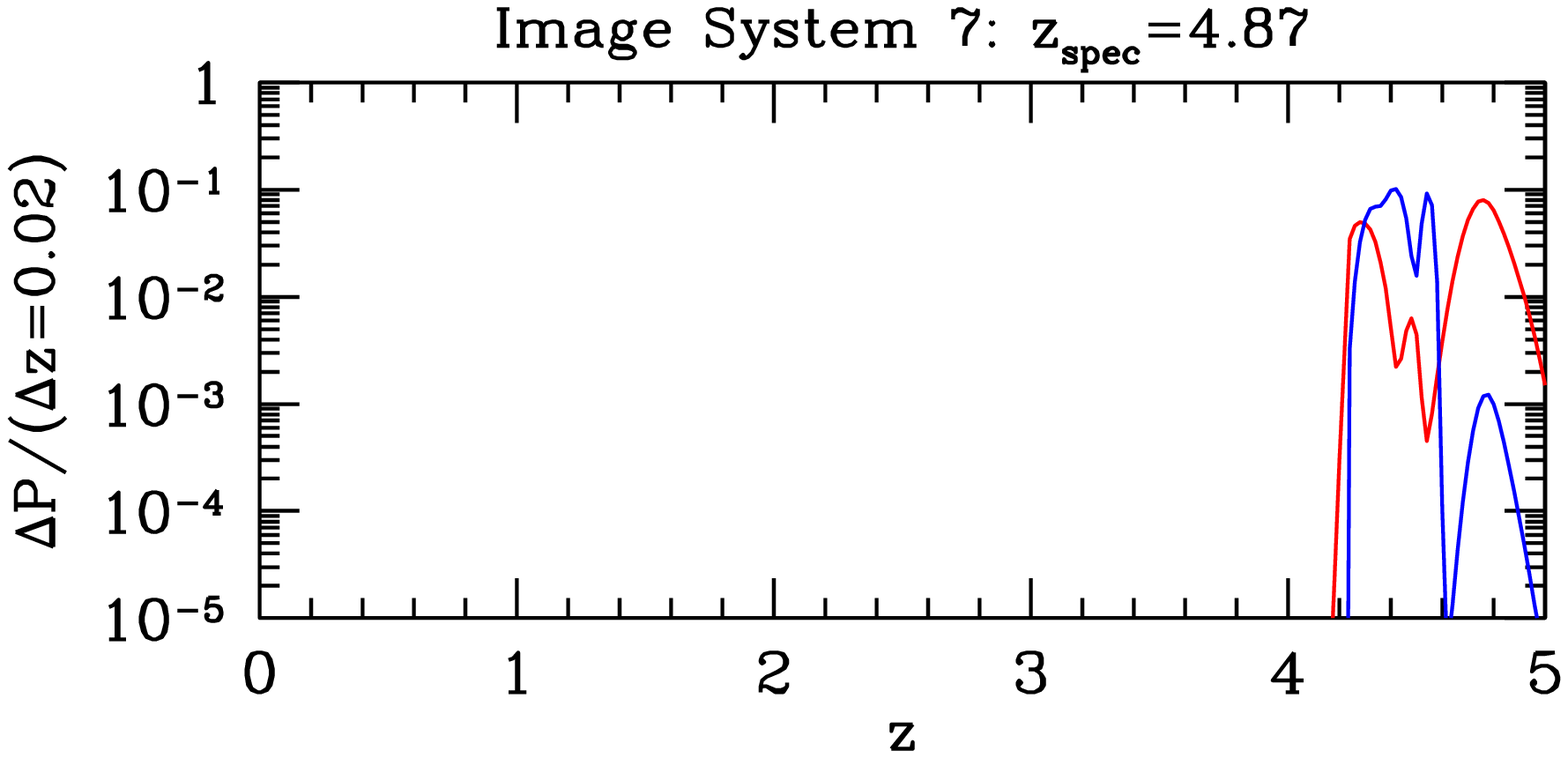}\\
  \includegraphics[width=\columnwidth,bb=18 440 592 718,clip=]{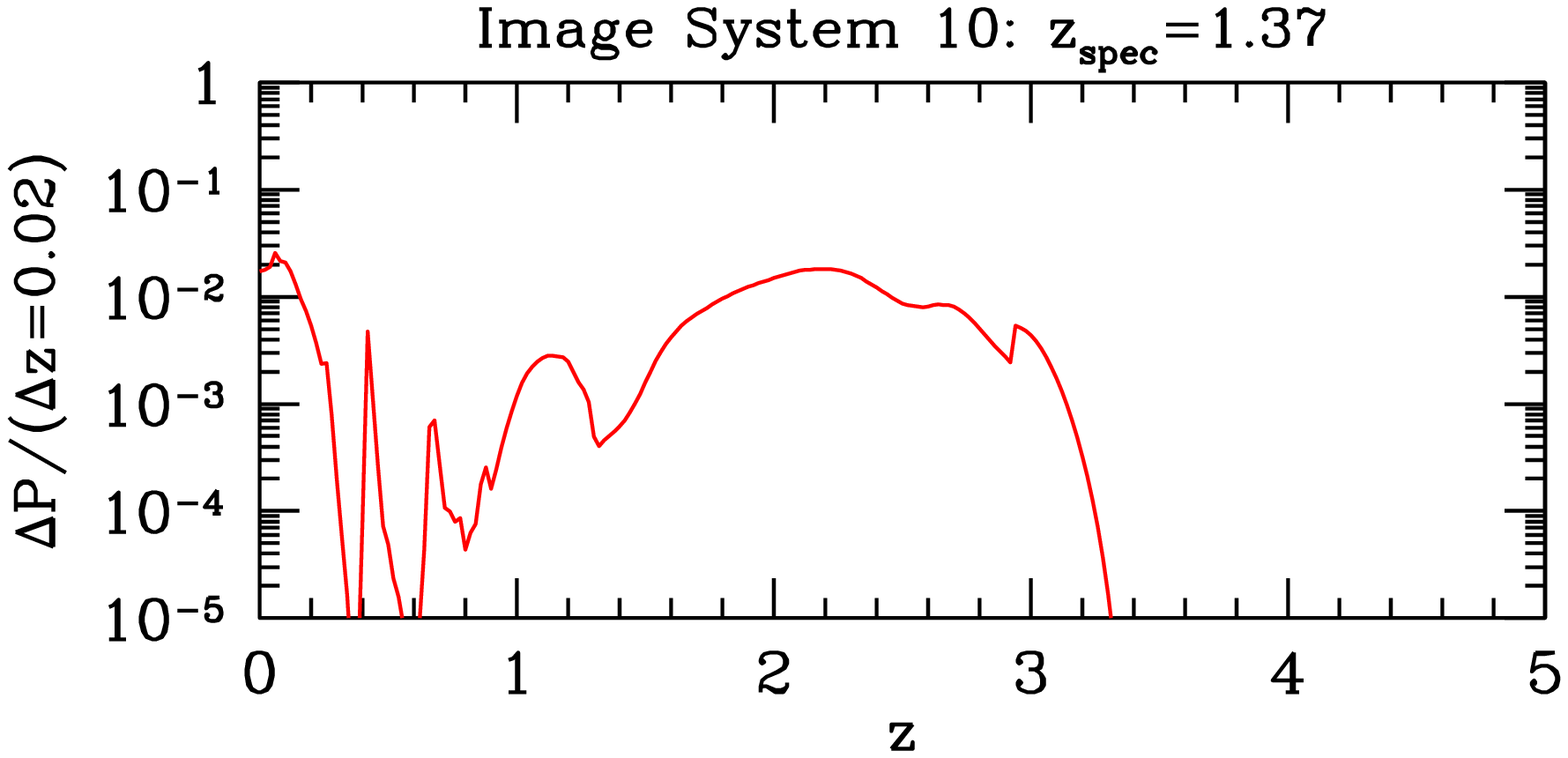}\\
  \includegraphics[width=\columnwidth,bb=18 440 592 718,clip=]{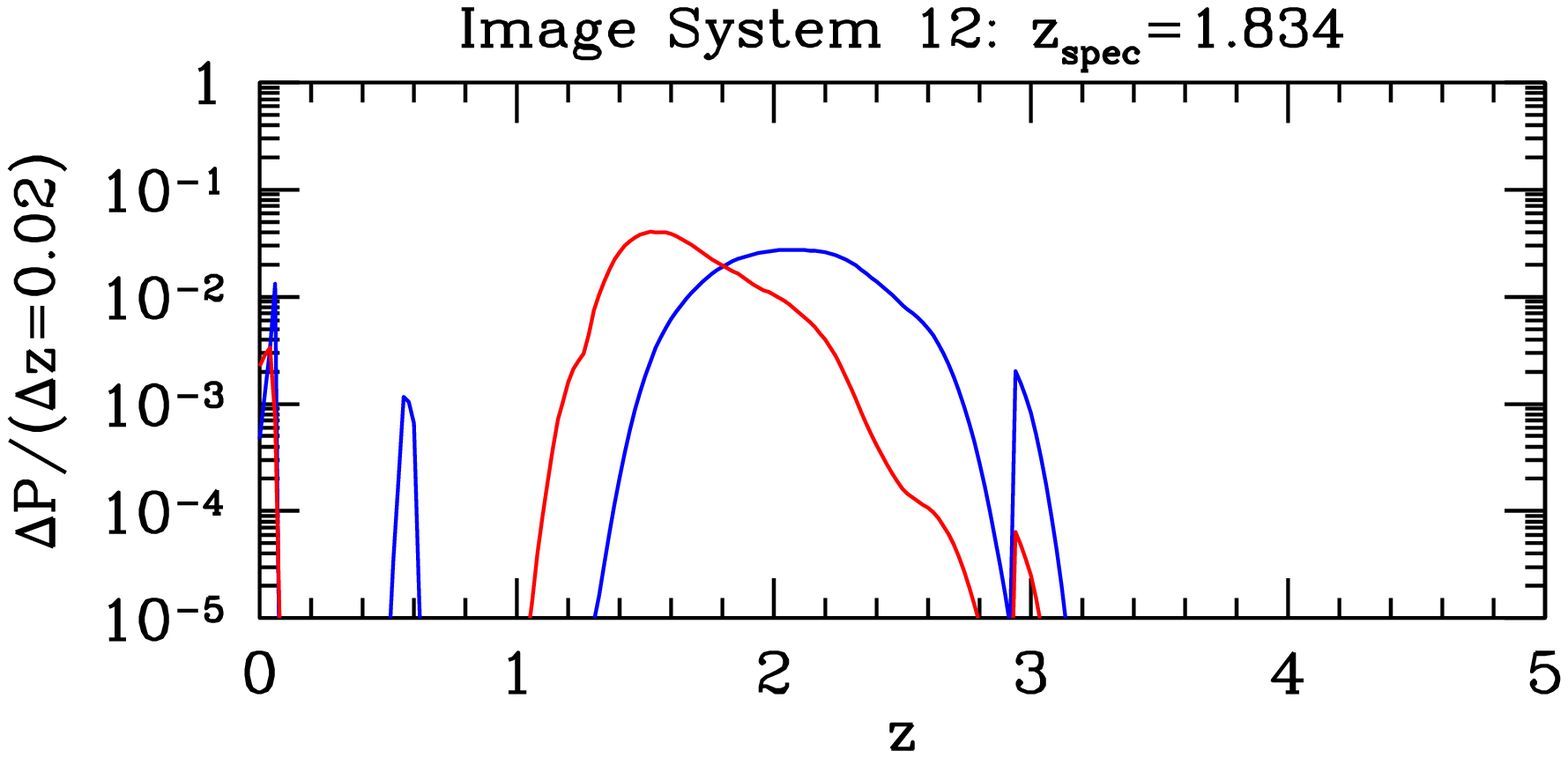}\\
  \includegraphics[width=\columnwidth,bb=18 440 592 718,clip=]{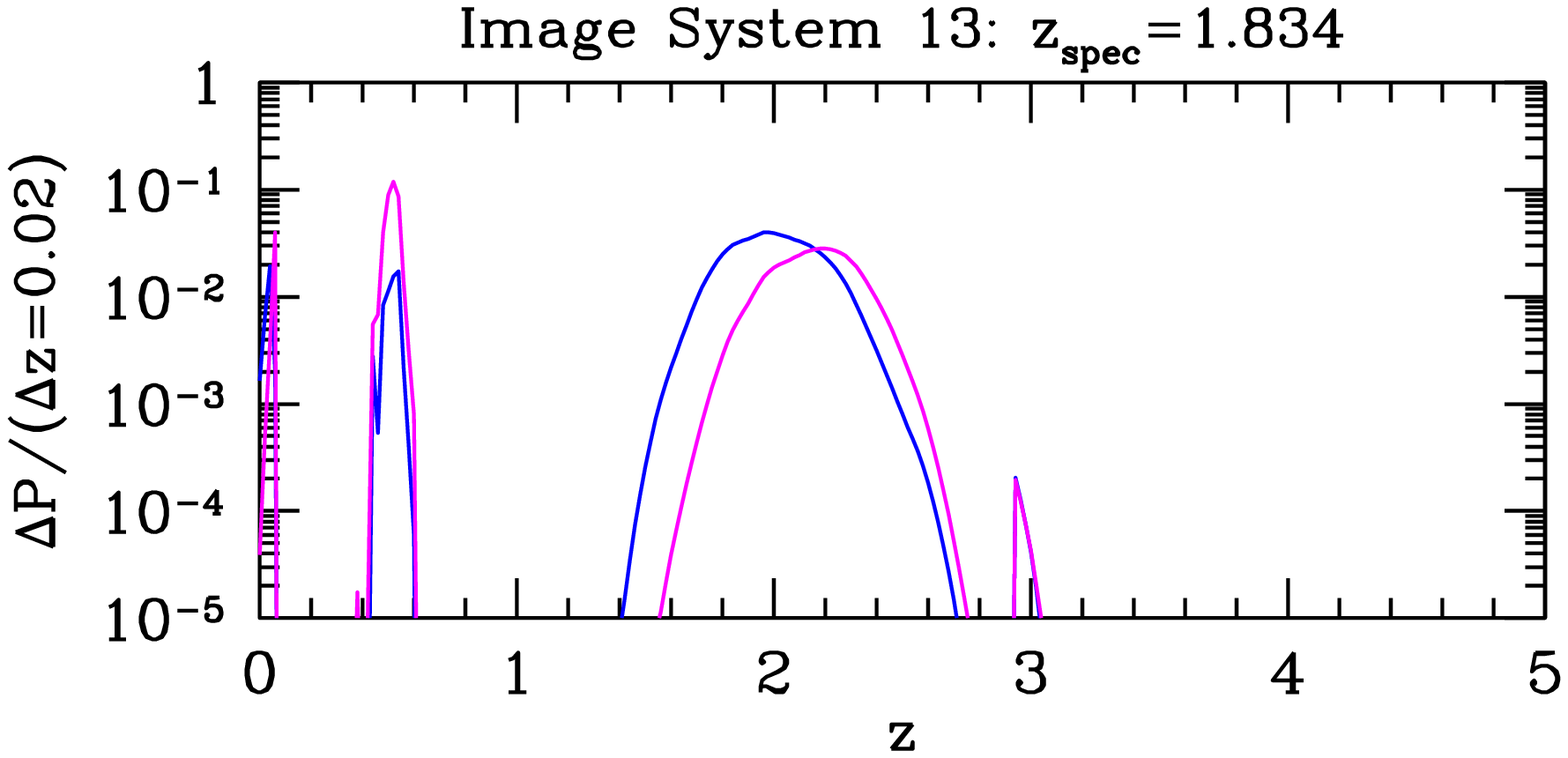}
  \caption{Photometric redshift probability density for the 5 image
    systems where spectroscopic redshift is known for at least one of
    the images in the system. In each panel the coloured thin lines
    represent individual redshift probability densities for the images
    in the system. Only image systems 10 and 12 have a broad
    photometric redshift probability density distribution, the other
    spectroscopic redshifts are recovered well.}
  \label{fig:zprob_total}
\end{figure}

In Figure \ref{fig:zprob_total} we show the photometric redshift
probability density of the 5 multiple image systems with a
spectroscopically known redshift. In the figures the different colours
represent the probability densities of individual multiple images of
the system. In most of the cases the spectroscopic and photometric
redshifts agree very well. Only image systems 10 and 12 have a broad
photometric redshift probability density distribution.

\begin{figure}
  \centering
  \includegraphics[height=\columnwidth]{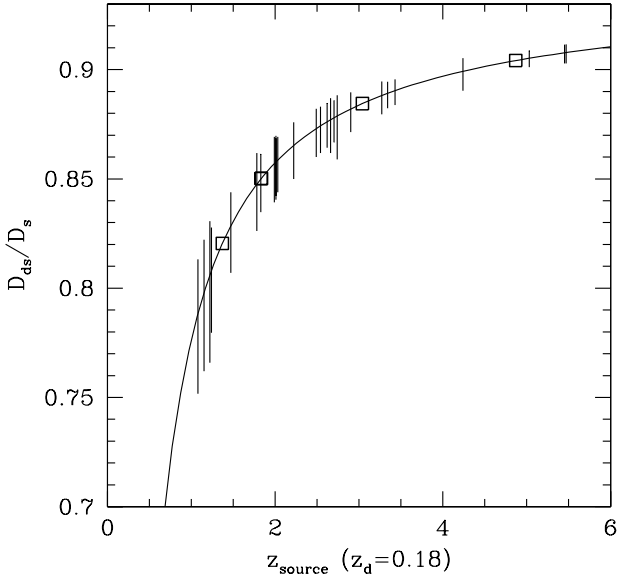}
  \caption{ The lensing power of a cluster depends on the ratio
    ${D_{ds} / D_{s}}$, where $D_{ds}$ is the angular diameter
    distance between the cluster and the source and $D_{d}$ is the
    angular diameter distance of the cluster. This ratio flattens
    rapidly for redshifts larger than 0.5 (for a cluster at redshift
    0.18).  In the figure we show the allowed ${D_{ds} / D_{s}}$
    ratios of the different multiple image systems. The squares
    indicate objects with spectroscopic redshifts. Since ${D_{ds} /
    D_{s}}$ varies only little for z$>$2 the redshifts of sources are
    not of great importance. The spectroscopic redshifts are important
    in fixing the overall mass scale.}
  \label{fig:ddsds}
\end{figure}

The lensing power of a cluster depends on the ratio ${D_{ds}/D_{s}}$,
where $D_{ds}$ is the angular diameter distance between the cluster
and the source and $D_{d}$ is the angular diameter distance of the
cluster. In Figure \ref{fig:ddsds} we show the power of a lens at
redshift 0.18 for different source redshifts. Vertical lines in Figure
\ref{fig:ddsds} show the range of allowed D$_{ds}$/D$_s$ ratios for
the image systems with photometric redshifts. The five squares mark
the image systems with known redshifts (two have the same
redshift). The D$_{ds}$/D$_s$ ratio can be well constrained by
photometric redshifts alone. With the help of the five spectroscopic
redshifts we can very accurately separate the geometric factor from
the deflection angle allowing us to constrain the cluster mass
tightly.

\subsection{Finding Optimal Model Parameters}

Goodness of fit in strong gravitational lensing can be quantified in
two ways. The proper way is to calculate a $\chi^2$ in the image
plane, i.e. how far an image predicted by a model is from the observed
one. In calculating the positions of predicted images of an image
system we assume that the images of a system originate from the
average source of the images. The expression for an image plane
$\chi^2$ is then

\begin{center}
  \begin{equation}
    {\chi}^{2}=\sum_{k}\sum_{i}{|\vec\theta_{k,i}-\vec\theta_{i}(<\vec\beta_{k,i}>)|^2
      \over \sigma_k^2},
  \end{equation}
\end{center}

where $\vec{\theta}_{k,i}$ is the position of image \textit{i} in
image system \textit{k} and $\vec{\theta}_{i}(<\vec\beta_{k,i}>)$ is
the predicted image position corresponding to image at
$\vec{\theta}_{k,i}$ from mean source of system \textit{k} at
$<\vec{\beta}_{k,i}>$ and $\sigma_k$ is the error in image positions
for system \textit{k} (estimated to be 1 pixel for all images).

Calculating image plane $\chi^2$ is unfortunately very time consuming
since the lens equation needs to be inverted numerically. An
additional complication is that for some values of the model
parameters not all observed images necessarily exist. This means that
an image plane $\chi^2$ does not necessarily converge to the optimal
parameters but is trapped in a local minimum.

Goodness of fit can also be estimated by requiring that all images of
an image system originate from the same source and hence minimise the
dispersion of the source positions. The problem in this case is that
the errors are measured in the image plane and do not necessarily
represent the errors in source positions. We take account of this by
rescaling errors in the image plane with local
magnification. Rescaling by magnification largely avoids bias towards
cluster parameters with high magnification (large core radius for the
NSIE model or small concentration for ENFW model). The source plane
$\chi^2$, $\widetilde{\chi}^2$, can be written in the following way,

\begin{center}
  \begin{equation}
    \widetilde{\chi}^{2}=\sum_{k}\sum_{i}\sum_{j>i}{|\vec\beta_{k,i}-\vec\beta_{k,j}|^2
      \over \sigma_{k,i}^2/\mu_{k,i}+\sigma_{k,j}^2/\mu_{k,j}},
    \label{eqn:source_chi2}
  \end{equation}
\end{center}

where $\vec\beta_{k,i}$ is the source position of image \textit{i} in
system \textit{k}, $\sigma_{k,i}$ is the error in the corresponding
image position and $\mu_{k,i}$ is the local image magnification.

The advantage of $\widetilde{\chi}^2$ over $\chi^2$ is that for every
image position it is always possible to calculate a corresponding
source position and so $\widetilde{\chi}^2$ can be calculated for all
values of the model parameters making $\widetilde{\chi}^2$ converge
well.

To find optimal model parameters we have first minimised
$\widetilde{\chi}^2$ to obtain model parameters close to the optimal
ones to ensure that the identified multiple images can be reproduced
by the models. The optimal model parameters were then found by
minimising $\chi^2$ properly in the image plane.


\subsection{Degeneracies}
\label{sec:degen}

Any multiple image system can only constrain the mass contained within
the images. This leads to degeneracies in the derived surface mass
profile: the so called mass sheet degeneracy states that if a given
surface mass density satisfies image constraints then a new surface
mass density can be found, by suitably rescaling this surface mass
density and by adding a constant mass sheet, which satisfies image
positions as well as relative magnifications equally well.

For haloes with variable mass profile this can also create a
degeneracy between the parameters of the profile. For the NSIE model a
high core radius can be compensated for by a larger velocity
dispersion and for NFW a higher scale radius demands a lower
concentration parameter.

These degeneracies can be broken if multiple image systems at
different redshifts and at different radii can be found. Position of a
radial critical line, and so radial arcs, depends critically on the
mass distribution in the central regions and hence the core radius. On
the other hand tangential arcs give strong constraints on the mass on
larger scales. Well defined halo parameters can be determined by
having radial arcs, a large number of multiple images at different
redshifts and by minimising in the image plane.


\section{Constructing Lensing Models}
\label{sec:result}

We have constructed in total 4 strong lensing models for the cluster.
The mass distributions are composed of two components as described in
the previous section: the cluster galaxies and a smooth dark matter
component. They give us the best fitting NSIE and ENFW parameters for
the {\it smooth} dark matter component and the mass profile of the
different mass components as well the total mass profile of the
cluster. With Models \Roman{one} and \Roman{two} we aim to establish a
well defined total mass profile for the cluster using the multiple
image positions and the photometric redshifts of the sources. The
difference between Models \Roman{one} and \Roman{two} is in the
scaling law used for the cluster galaxies. Model \Roman{one} assumes a
$s \propto \sigma^2$ law where as for Model \Roman{two} we use $s
\propto \sigma$ law. Models \mbox{\Roman{one}b} and
\mbox{\Roman{two}b} replicate the setup of \citet{broadhurst:05} and
with these we want to compare results with our parametric models to
the their more flexible kappa-in-a-grid model. For Models
\mbox{\Roman{one}b} and \mbox{\Roman{two}b} we have used images from
\citet{broadhurst:05} only and have left the photometric redshifts of
the sources free as was done in \citet{broadhurst:05}. We have kept
the spectroscopic redshifts fixed as these help to define the mass
scale of the cluster. The difference between Models
\mbox{\Roman{one}b} and \mbox{\Roman{two}b} is again in the scaling
law adopted.

In addition to the 4 strong lensing Models above we have also
constructed 2 further Models in order to derive NSIS and NFW
parameters of the total cluster mass profile and to facilitate the
comparison of our results with earlier methods used to measure cluster
masses, and numerical simulations. In Model \Roman{three} we have
fitted a NSIS and an NFW profile to the total mass obtained with
Models \Roman{one} and \Roman{two}. With Model \Roman{four} we combine
the strong lensing constraints from Model \Roman{three} and the weak
lensing constraints from \citet{broadhurst:05b} and derive accurate
NSIS and NFW parameters for the total mass profile out to $\sim$15'
($\sim$2.5~Mpc).

Most of the image systems are very well reproduced by the
Models. Image systems 8, 12, 14, 15, 30, 31 and 32 are located close
to critical lines where the image plane $\chi^2$ is difficult to
calculate due to ill determined image positions from the models. For
these image systems we have always calculated the $\chi^2$ in the
source plane.

We quantify the quality of fit by the rms distance between an observed
image position and one predicted by the models. For the images systems
mentioned above the magnification weighted source separation was used
instead. The rms distance between an image in an image system and the
image position obtained with the models are given for each image
system in table \ref{tab:app:summary} in the Appendix.

Additional information of the fit quality can be seen in appendix
\ref{app:stamps} where we show image stamps of all multiple images. In
addition to the multiple images we also show two model reproductions
for each image obtained with the two descriptions of the smooth dark
matter component (NSIE and ENFW) for Model \Roman{two}.

\begin{figure*}
  \centering
  \includegraphics[width=2\columnwidth, bb=60 60 580 580, clip=]{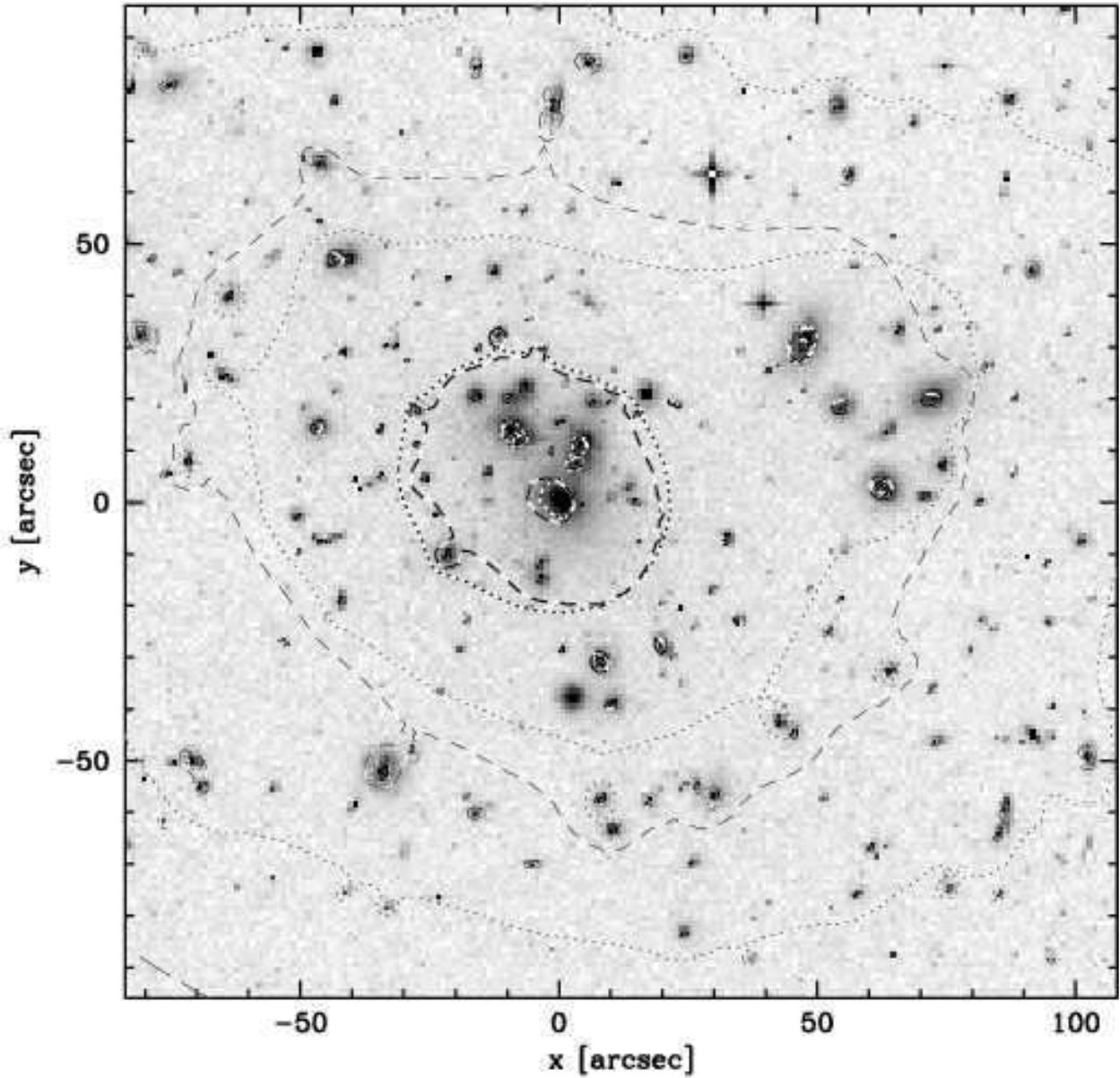}
  \caption{Surface mass density contours for NSIE (dotted) and ENFW
  (dashed) profiles for Model \Roman{one}. For each smooth DM profile
  we plot the mean $\kappa$ of the in total 4000 cluster realisations
  used in deriving the best fit parameters and errors. The contours
  are drawn at $\kappa$=0.25, 0.50, 1.00 and 2.00 levels for a source
  at redshift of 1.0. The thin contour lines are for $\kappa<$1 and
  thick lines for $\kappa\ge$1. The secondary mass concentration on
  the upper right can also be seen clearly in the surface mass density
  contours.}
  \label{fig:kappa}
\end{figure*}

Figure \ref{fig:kappa} shows the total surface-mass-density
contours obtained with Models \Roman{one} and \Roman{two} for the two
smooth DM profiles. For each smooth DM profile we plot the mean
$\kappa$ of the in total 4000 cluster realisations used in deriving
the best fit parameters and errors. The dotted contours are for the
NSIE and the dashed contours are for the ENFW profile. The contours
are drawn at $\kappa$=0.25, 0.50, 1.0 and 2.0 levels for a source
redshift of 1. The thin contour lines are for $\kappa<$1 and thick
lines for $\kappa\ge$1.\\

\begin{figure*}
  \centering
  \includegraphics[width=2\columnwidth, bb=60 60 580 580, clip=]{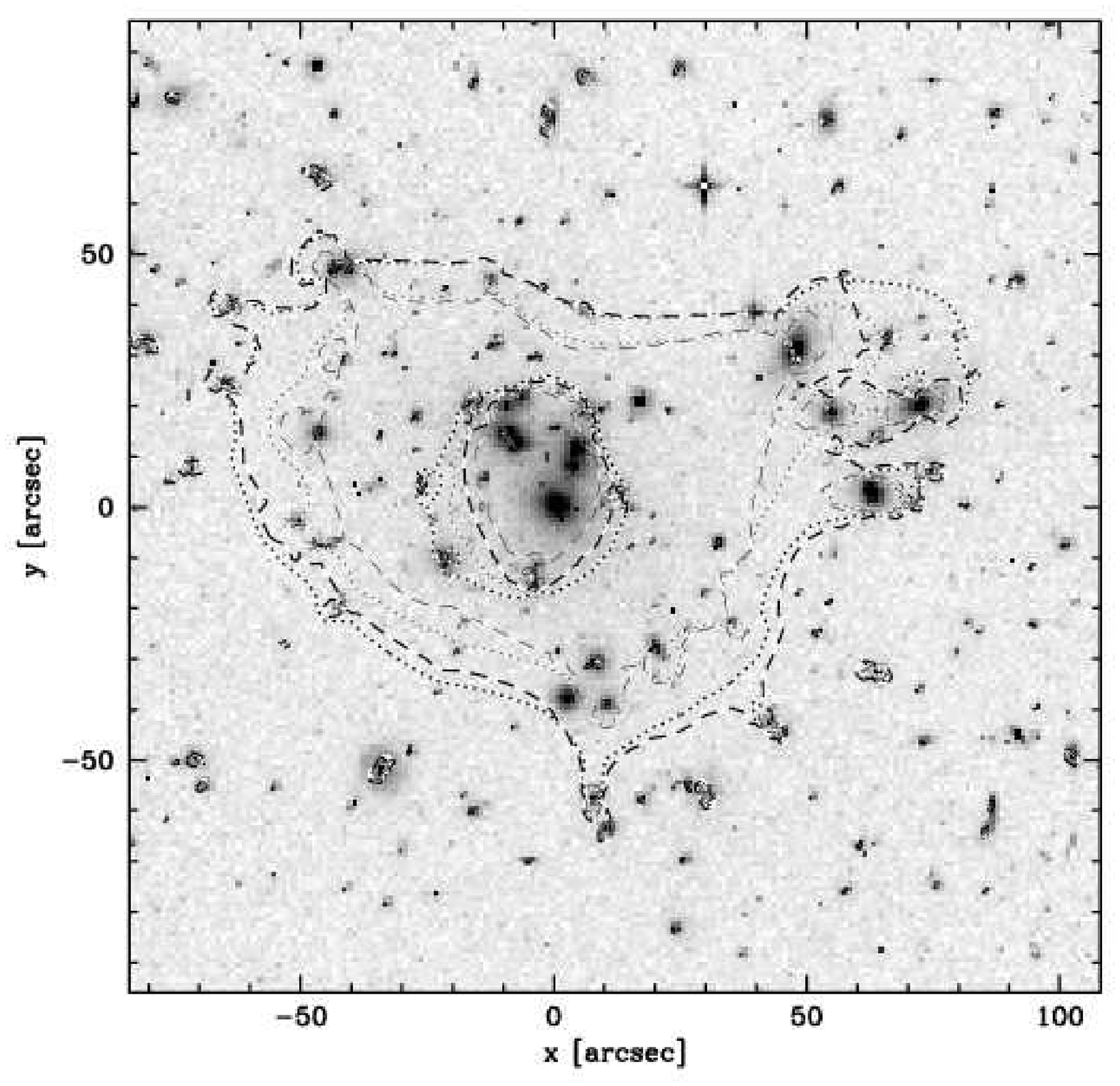}
  \caption{The critical curves obtained with Models \Roman{one} and
    \Roman{two} for the two smooth DM profiles. For each smooth DM
    profile we plot the critical curves of the average cluster of the
    cluster realisations used in deriving the best fit parameters and
    errors. The dotted contours are for the NSIE and the dashed
    contours are for the ENFW profile. The thin contours are drawn for
    a source redshift of 1 and the thick contours for a source
    redshift of 5.}
  \label{fig:crit}
\end{figure*}

Figure \ref{fig:crit} shows the critical curves obtained with Models
\Roman{one} and \Roman{two} for the two smooth DM profiles. For each
smooth DM profile we plot the critical curves of the average cluster
of the in total 4000 cluster realisations used in deriving the best
fit parameters and errors. The dotted contours are for the NSIE and
the dashed contours are for the ENFW profile. The thin contours are
drawn for a source redshift of 1 and the thick contours for a source
redshift of 5.

\subsection{Models \Roman{one} and \Roman{two}: Strong Lensing Mass 
  Reconstruction}

The first two strong lensing models aim to establish a total mass
profile for further analysis. The smooth dark matter component of the
cluster mass was modelled in exactly the same way for both models
(detailed in section \ref{sec:models}), for the galaxy component we
vary the scaling of the truncation radius of the BBS model with the
velocity dispersion. For Model \Roman{one} we have assumed that the
galaxies follow a scaling law similar to the field galaxies, namely
that the truncation radius $s$ of a galaxy scales like
$s=s_*\times(\sigma/136~km/s)^2$ with the velocity dispersion $\sigma$
of the galaxy. For Model \Roman{two} we have assumed a scaling law
expected for galaxies in cluster, $s=s_*\times(\sigma/136~km/s)$. The
normalisation of the scaling law for each Model and smooth DM profile
is shown in Table \ref{tab:results}. The haloes are strongly
truncated. This is a real effect but the actual values obtained for
$s_*$ can be affected by the optimisation process. Mass lost from
galaxies due to truncation can in part be compensated by the smooth
dark matter component, leading to possibly significant uncertainties
in the truncation radii. It is not our aim in this paper to attempt to
constrain the truncation radii of the galaxies in the cluster but
instead to reproduce the observed multiple images as accurately as
possible. The truncation of galaxies in a cluster environment will be
discussed in detail in a forth coming publication (Halkola et al.,
2006 in preparation).

The constraints for Models \Roman{one} and \Roman{two} are the
positions of the multiple images and their redshifts. The redshifts of
sources were allowed to find the optimal redshift within the 1-sigma
errors of the photometric redshifts, except sources with spectroscopic
redshifts for which we have fixed the redshift to the measured
one. The allowed ranges for the source redshifts are tabulated in
Table \ref{tab:app:summary} in appendix \ref{app:images}.

The best fit parameters for the smooth dark matter component of the
models are summarised in Table \ref{tab:results}. The errors are
caused by errors in determining the correct galaxy masses and in
measuring the multiple image position. The derivation of errors is
explained in section \ref{sec:errors}.

Our best fitting model is Model \Roman{one} with a dark matter
component described by an ENFW profile. The differences in the fit
quality between the models and smooth dark matter profiles used are
generally small as can be seen from Table \ref{tab:results} although
both Models perform better when the smooth DM is modelled with an ENFW
profile. The fit quality is ~0.5'' better than that achieved by
\citet{broadhurst:05} in their analysis of the cluster. This can be
due to better modelling of the cluster mass or to a different set of
multiple images used. If the difference is due to the changes in
multiple image systems then the $\chi^2$ in the case
\citet{broadhurst:05} of is driven by a only a few image systems since
most of the images systems are infact identical. Another difference
are the constraints imposed on the redshifts of the images in our
modelling.

\subsection{Models \mbox{\Roman{one}b} and \mbox{\Roman{two}b}: Comparison to \citet{broadhurst:05}}

In order to directly compare the performance of our parametric models
to the grid model of \citet{broadhurst:05} we have constructed two
further models that mimic their setup. The models are constrained only
by the multiple image positions from \citet{broadhurst:05}; the
photometric redshifts of the images were thus ignored and were
included as free parameters. We have fixed the spectroscopic redshifts
however since it is necessary to define an overall mass scale for the
cluster. The rest of the modelling is identical to that of Models
\Roman{one} and \Roman{two}.

The very good performance of our models relative to
\citet{broadhurst:05} is remarkable considering the large freedom in
the mass profile allowed in their modelling. This also means that the
mass profile can be very well described by parametric models making
the additional freedom allowed by non-parametric mass modelling
unnecessary, even undesirable if one is interested in comparing the
performance of different parametric mass profiles.

Assuming that the smooth mass component of \citet{broadhurst:05} is
able to reproduce both NSIE and ENFW halo profiles the other major
difference between our mass modelling and that of
\citet{broadhurst:05} is in the treatment of the galaxy component. The
assumptions needed on the properties of the cluster galaxies in our
modelling seem to be well justified based on the superior performance
of our models.\\


\subsubsection{Estimation of Errors in the Parameters of the Smooth Dark
  Matter Component}
\label{sec:errors}

Our primary source of uncertainty in the parameters of the smooth dark
matter component are the velocity dispersions of the galaxies in the
cluster.

In order to estimate the effect of measurement errors in the cluster
galaxy component on the parameters of the smooth cluster component we
have created 2000 clusters for Models \Roman{one}, \Roman{two},
\mbox{\Roman{one}b} and \mbox{\Roman{two}b} and the two profiles by
varying the velocity dispersions of cluster galaxies and positions of
multiple images by the estimated measurement errors. For each galaxy
we have assigned a new velocity dispersion from a Gaussian
distribution centred on the measured values with a spread
corresponding to the error. The truncation radii of the cluster
galaxies were adjusted accordingly. For the scaling law we have used
the normalisation as for the original cluster galaxies. New positions
for the multiple images were assigned similarly by assigning new
positions from a Gaussian distribution centred on the measured
positions.

Optimal parameters for the smooth cluster component for each cluster
were found by minimising $\widetilde{\chi}^2$ in the source plane due
to the large number of minimisations required. However, in all
subsequent analysis we have used the image plane $\chi^2$ calculated
after the optimal parameters in the source plane were found.

\begin{figure}
  \centering
  \includegraphics[height=\columnwidth]{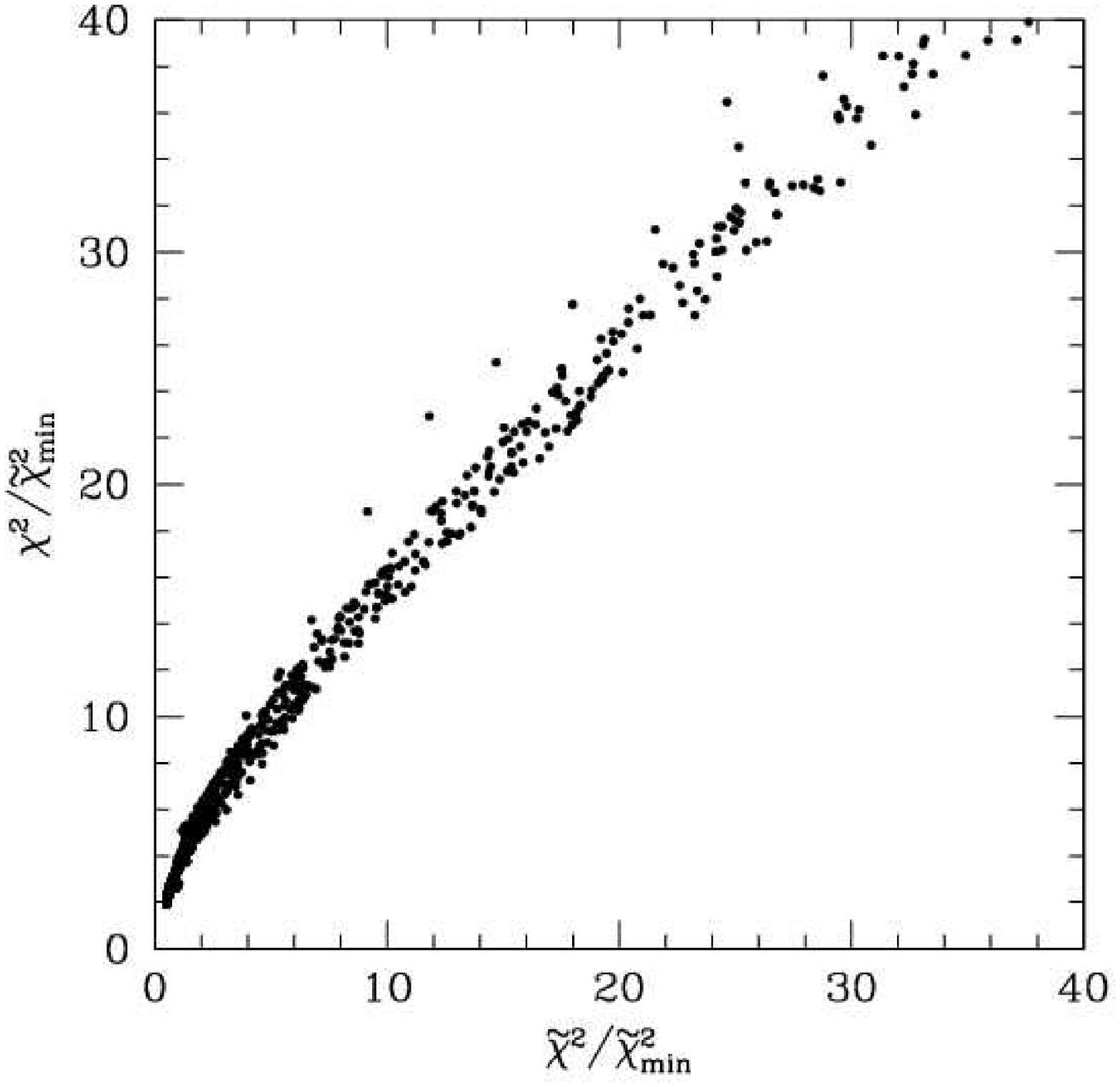}
  \caption{A comparison between $\chi^2$ (image plane $\chi^2$) and
  $\widetilde{\chi}^2$ (source plane $\chi^2$) for $\sim$3000
  different model configurations. The important property of
  $\chi^2(\widetilde{\chi}^2)$ for the minimisation is that it is a
  monotonically increasing function making $\widetilde{\chi}^2$ a
  reliable tracer for $\chi^2$ when used in finding optimal model
  parameters in the error analysis.}
  \label{fig:chi2}
\end{figure}

To justify the use of $\widetilde{\chi}^2$ instead of $\chi^2$ in the
error estimation we show in Figure \ref{fig:chi2} the final $\chi^2$
against $\widetilde{\chi}^2$ for a large number of models after
minimising $\widetilde{\chi}^2$. In the figure both $\chi^2$ and
$\widetilde{\chi}^2$ have been scaled by the minimum
$\widetilde{\chi}^2$ of the models. The good correspondence between
the two, even at high values of $\chi^2$, and that
$\chi^2(\widetilde{\chi}^2)$ is a monotonically increasing function
(unfortunately with some scatter) gives us confidence in the source
plane minimisation and our error analysis.

\begin{figure}
  \centering
  \includegraphics[height=0.7\columnwidth]{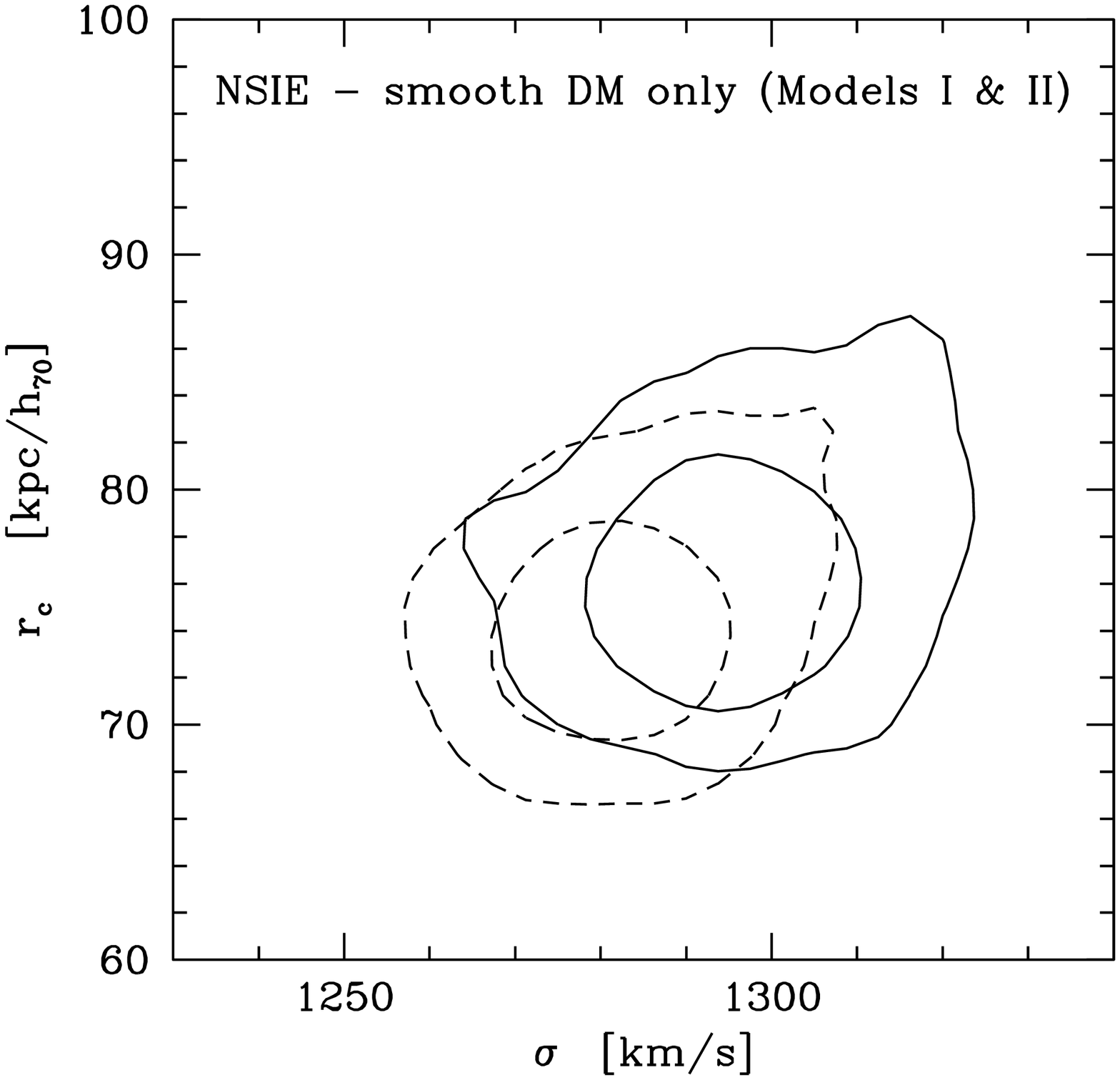}\\
  \includegraphics[height=0.7\columnwidth]{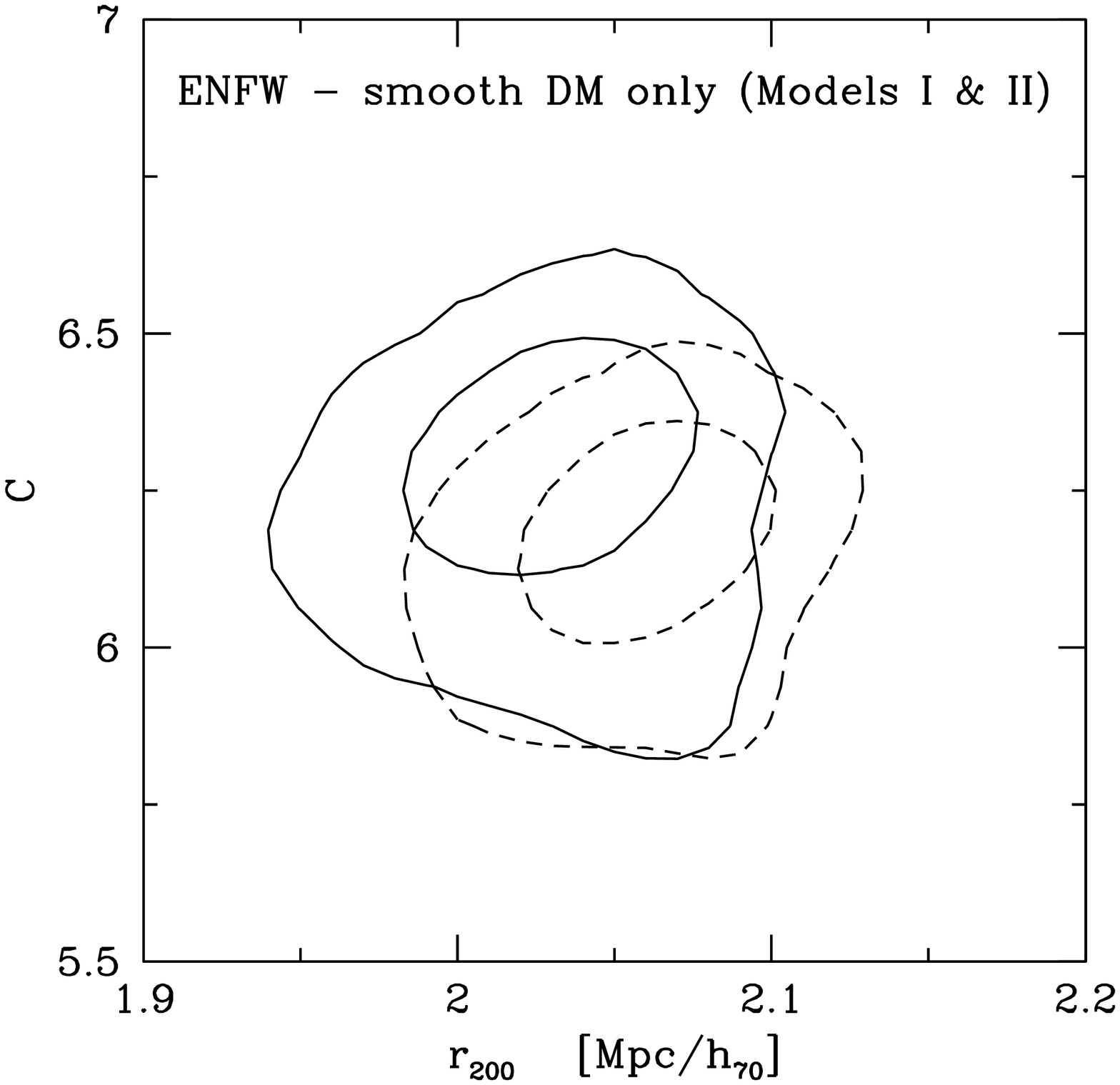}
  \caption{Estimates of the profile parameter uncertainties for the
    main halo of the smooth DM component (NSIE top, NFW bottom). In
    both panels Model \Roman{one} is represented by the solid contours
    and Model \Roman{two} by the dashed contours. Contours are drawn
    at 68\% and 95\% confidence levels. The uncertainty is mostly due
    the errors in determining the contribution of the cluster galaxies
    to the total mass.}
  \label{fig:param_errors}
\end{figure}

The optimal parameters of the generated clusters have a spread around
the best fit parameters determined for the 'real' cluster. The number
density of the optimal parameters in the parameter space cannot be
directly used to quantify the random error since areas of high density
could in fact also include a large number of relatively poor fits to
the data. To include also information of the quality of fit we weight
each realisation of the Monte-Carlo simulation with the final
1/$\chi^2$ of the realisation. In Figure \ref{fig:param_errors} we
show the number density contours for the NSIE (top panel) and ENFW
(bottom panel) halo parameters of the realisations after weighting by
the final 1/$\chi^2$. The solid lines are for Model \Roman{one} and
the dashed lines for Model \Roman{two}. The contour lines show the
regions in which 68\% and 95\% of the weighted realisations lie.\\

\begin{table*}
  \label{tab:results}
  \centering
  \caption[]{Best fit halo parameters of the smooth dark matter
  components for Models \Roman{one} and \Roman{two}. The rms error
  quoted for each model and halo type is the rms distance between the
  measured image positions and those predicted by the models expect
  for image systems where the calculations of an image plane $\chi^2$
  was not possible (8, 12, 14, 15, 30, 31 and 32) for which we have used
  the source plane $\chi^2$ defined in equation
  \ref{eqn:source_chi2}. The values given are for the best fitting
  halo optimised in the image plane. The values in brackets refer to
  the mean values obtained from the simulations (optimised in the
  source plane), the errors in the halo parameters are derived from
  the spread of the values in the simulations. For details of the
  models and error derivation see text body. }
  \begin{tabular}{lcccc}
    &\multicolumn{2}{c}{{\bf{Model \Roman{one}}}}
    &\multicolumn{2}{c}{{\bf{Model \Roman{two}}}}\\
    \hline
    \hline
    \bf{NSIE}
    &\multicolumn{2}{c}{rms error 3.17''~~(3.51$\pm$0.15'')}
    &\multicolumn{2}{c}{rms error 3.13''~~(3.45$\pm$0.16'')}\\
    &\multicolumn{2}{c}{ s$_{bbs}$~=~24~kpc*$(\sigma_{gal}$/136km/s)$^2$ }
    &\multicolumn{2}{c}{ s$_{bbs}$~=~37~kpc*$(\sigma_{gal}$/136km/s) }\\
    \hline
    \bf{Parameter}
    &\bf{Halo One}&\bf{Halo Two}
    &\bf{Halo One}&\bf{Halo Two}\\
    \hline
    $\sigma$ (km/s)
    & 1298~~(1293$^{+11}_{-10}$) &  603~~(595$\pm$20)
    & 1285~~(1281$^{+10}_{ -9}$) &  618~~(613$\pm$17)\\
    r$_c$ (kpc)
    & 77~~(76$^{+4}_{-3}$) &  75~kpc~~(72$\pm$5)
    & 75~~(74$^{+3}_{-3}$) &  75~kpc~~(73$\pm$4)\\
    \hline
    \hline
    \bf{ENFW}
    &\multicolumn{2}{c}{rms error 2.73''~~(3.12$\pm$0.20'')}
    &\multicolumn{2}{c}{rms error 2.48''~~(3.08$\pm$0.19'')}\\
    &\multicolumn{2}{c}{ s$_{bbs}$~=~21~kpc*$(\sigma_{gal}$/136km/s)$^2$ }
    &\multicolumn{2}{c}{ s$_{bbs}$~=~36~kpc*$(\sigma_{gal}$/136km/s) }\\
    \hline
    \bf{Parameter}
    &\bf{Halo One}&\bf{Halo Two}
    &\bf{Halo One}&\bf{Halo Two}\\
    \hline
    C
    & 6.5~~(6.3$\pm$0.2) & 0.5~~(0.5$\pm$0.1)
    & 6.2~~(6.2$\pm$0.1) & 0.7~~(0.7$\pm$0.1)\\
    r$_{200}$ (Mpc)
    & 2.04~~(2.03$\pm$0.03) & 2.79~~(2.81$\pm$0.06)
    & 2.07~~(2.06$\pm$0.03) & 2.52~~(2.53$\pm$0.06)\\
    \hline
    \hline
    \begin{minipage}{10mm}\bf{Model free\\parameters}\end{minipage}
    &\multicolumn{2}{c}{\begin{minipage}{35mm}\vspace{1mm}parameters of the
    smooth\\ dark matter component\vspace{1mm}\end{minipage}}
    &\multicolumn{2}{c}{\begin{minipage}{35mm}\vspace{1mm}parameters of the
    smooth dark matter component\vspace{1mm}\end{minipage}}\\
    \hline
    \begin{minipage}{10mm}\bf{Model\\constraints}\end{minipage}
    &\multicolumn{2}{c}{\begin{minipage}{35mm}\vspace{1mm}images
    from this work and\\\citet{broadhurst:05}\\z$_{phot}$ of sources\end{minipage}}
    &\multicolumn{2}{c}{\begin{minipage}{35mm}\vspace{1mm}images
    from this work and\\\citet{broadhurst:05}\\z$_{phot}$ of sources\end{minipage}}\\
    \hline
  \end{tabular}
\end{table*}

\begin{table*}
  \label{tab:results_br} 
  \centering
  \caption[]{Best fit halo parameters of the smooth dark matter
  components for Models \Roman{one}b and \Roman{two}b. The rms error
  quoted for each model and halo type is the rms distance between the
  measured image positions and those predicted by the models expect
  for image systems where the calculations of an image plane $\chi^2$
  was not possible (8, 12, 13, 14, 20 and 30) for which we have used the
  source plane $\chi^2$ defined in equation \ref{eqn:source_chi2}. The
  values given are for the best fitting halo optimised in the image
  plane. The values in brackets refer to the mean values obtained from
  the simulations (optimised in the source plane), the errors in the
  halo parameters are derived from the spread of the values in the
  simulations. For details of the models and error derivation see text
  body. }
  \begin{tabular}{lcccccccc}
    &\multicolumn{2}{c}{{\bf{Model \mbox{\Roman{one}b}}}}
    &\multicolumn{2}{c}{{\bf{Model \mbox{\Roman{two}b}}}}\\
    \hline
    \hline
    \bf{NSIE}
    &\multicolumn{2}{c}{rms error 3.03''~~(3.27$\pm$0.20'')}
    &\multicolumn{2}{c}{rms error 2.65''~~(3.29$\pm$0.21'')}\\
    &\multicolumn{2}{c}{ s$_{bbs}$~=~30~kpc*$(\sigma_{gal}$/136km/s)$^2$ }
    &\multicolumn{2}{c}{ s$_{bbs}$~=~43~kpc*$(\sigma_{gal}$/136km/s) }\\
    \hline
    \bf{Parameter}
    &\bf{Halo One}&\bf{Halo Two}
    &\bf{Halo One}&\bf{Halo Two}\\
    \hline
    $\sigma$ (km/s)
    & 1223~~(1222$\pm$13) &  647~~(645$\pm$18)
    & 1210~~(1216$\pm$11) &  658~~(660$\pm$15)\\
    r$_c$ (kpc)
    & 56~~(60$\pm$3) & 75~~(74$\pm$3)
    & 58~~(60$\pm$3) & 74~~(74$\pm$2)\\
    \hline
    \hline
    \bf{ENFW}
    &\multicolumn{2}{c}{rms error 2.74''~~(3.30$\pm$0.15'')}
    &\multicolumn{2}{c}{rms error 2.72''~~(3.31$\pm$0.15'')}\\
    &\multicolumn{2}{c}{ s$_{bbs}$~=~31~kpc*$(\sigma_{gal}$/136km/s)$^2$ }
    &\multicolumn{2}{c}{ s$_{bbs}$~=~51~kpc*$(\sigma_{gal}$/136km/s) }\\
    \hline
    \bf{Parameter}
    &\bf{Halo One}&\bf{Halo Two}
    &\bf{Halo One}&\bf{Halo Two}\\
    \hline
    C
    & 6.4~~(6.4$\pm$0.2) & 1.5~~(1.6$\pm$0.1)
    & 6.5~~(6.4$\pm$0.2) & 1.6~~(1.7$\pm$0.1)\\
    r$_{200}$ (Mpc)
    & 2.12~~(2.08$\pm$0.04) & 1.87~~(1.86$\pm$0.05)
    & 2.08~~(2.04$\pm$0.04) & 1.85~~(1.83$\pm$0.05)\\
    \hline
    \hline
    \begin{minipage}{10mm}\bf{Model free\\parameters}\end{minipage}
    &\multicolumn{2}{c}{\begin{minipage}{35mm}\vspace{1mm}parameters of the
    smooth\\ dark matter component,\\z$_{phot}$ of sources\vspace{1mm}\end{minipage}}
    &\multicolumn{2}{c}{\begin{minipage}{35mm}\vspace{1mm}parameters of the
    smooth\\dark matter component,\\z$_{phot}$ of sources\vspace{1mm}\end{minipage}}\\
    \hline
    \begin{minipage}{10mm}\bf{Model\\constraints}\end{minipage}
    &\multicolumn{2}{c}{\begin{minipage}{35mm}\vspace{1mm}images
    from \\\citet{broadhurst:05}\end{minipage}}
    &\multicolumn{2}{c}{\begin{minipage}{35mm}\vspace{1mm}images
    from \\\citet{broadhurst:05}\end{minipage}}\\
    \hline
  \end{tabular}
\end{table*}

\begin{table*}
  \label{tab:results2} 
  \centering
  \caption[]{Best fit halo parameters for the cluster profile. For
  Models \Roman{three} and \Roman{four} we have fitted the measured
  cluster parameters with a single NSIS or NFW profile. The
  constraints for the cluster parameters were mass for Model
  \Roman{three},both mass and shear for Model \Roman{four}. For
  details of the models see text body.}
  \begin{tabular}{lcc}
    &{\bf{Model \Roman{three}}}
    &{\bf{Model \Roman{four}}}\\
    \hline
    \hline
    \bf{NSIE}
    &{$\chi^2$ / dof = 10.5 / 11}
    &{$\chi^2$ / dof = 30.0 / 20}\\
    \hline
    \bf{Parameter}
    &{\bf{only one halo fitted}}
    &{\bf{only one halo fitted}}\\
    \hline
    $\sigma$
    &1514$^{+18}_{-17}$~km/s
    &1499$^{+15}_{-14}$~km/s\\
    r$_c$
    &71$\pm$5~kpc
    &66$\pm$5~kpc\\
    \hline
    \hline
    \bf{ENFW}
    &{$\chi^2$ / dof =  0.8 / 11}
    &{$\chi^2$ / dof = 31.9 / 20}\\
    \hline
    \bf{Parameter}
    &{\bf{only one halo fitted}}
    &{\bf{only one halo fitted}}\\
    \hline
    C
    &6.0$\pm$.5
    &7.6$^{+0.3}_{-0.5}$\\
    r$_{200}$
    &2.82$^{+0.11}_{-0.09}$~Mpc
    &2.55$^{+0.07}_{-0.04}$~Mpc\\
    \hline
    \hline
    \begin{minipage}{15mm}\bf{Model free\\parameters}\end{minipage}
    &\begin{minipage}{40mm}\vspace{1mm}the above parameters of \\
      the halo profiles\vspace{1mm}\end{minipage}
    &\begin{minipage}{40mm}\vspace{1mm}the above parameters of\\
      the halo profiles\vspace{1mm}\end{minipage}\\
    \hline
    \begin{minipage}{15mm}\bf{Model\\constraints}\end{minipage}
    &\begin{minipage}{40mm}\vspace{1mm}total mass obtained with\\
    Models \Roman{one} and \Roman{two}\vspace{1mm}\end{minipage}
    &\begin{minipage}{40mm}\vspace{1mm}total mass obtained with\\
    Models \Roman{one} and \Roman{two}\\shear from
    Broadhurst et al. 2005b\vspace{1mm}\end{minipage}\\ \hline
  \end{tabular}
\end{table*}

\subsection{Model \Roman{three}: Parameters for the Total Mass Profile}

It is important to realise that the multiple images constrain the
combined mass of the cluster, be it baryonic or dark. The division of
the mass to two components is done in order to take account of the
mass we can observe as accurately as possible. The uncertainty of the
description of the galaxy component is reflected in how well we can
determine the profile parameters of the smooth dark matter
component. The parameters for the total mass distribution, constrained
directly by the multiple images, can be determined significantly
better. For this reason we have also fitted single NFW and NSIS haloes
to the total mass obtained from Models \Roman{one} and \Roman{two}.

\begin{figure}
  \centering \includegraphics[width=\columnwidth]{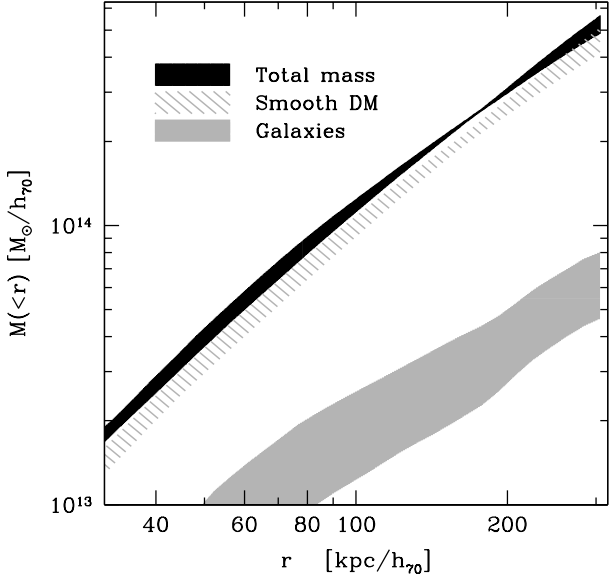}
  \caption{1-$\sigma$ confidence regions of projected mass M($<$r)
  inside radius r for two mass components and total mass. The mass
  associated with galaxies is shown as a solid grey, the smooth dark
  matter as a striped grey and the total mass as a solid black
  region. The large uncertainty in the galaxy component is well
  balanced by the smooth dark matter component to produce a tight
  M($<$r) profile for the total mass.}
  \label{fig:mass_components}
\end{figure}

We estimate the total mass profile of the cluster by combining all
mass profiles from the error analysis. In Figure
\ref{fig:mass_components} we show the 68.3\% confidence regions of
mass for the two mass components and the total mass. The galaxy
component is shown as a solid grey region, smooth dark matter as a
striped grey and the total mass as a solid black region. The regions
were determined by taking the best 68.3\% of the galaxy component
realisations from both Models \Roman{one} and \Roman{two} regardless
of the smooth DM profile used. We have decided to combine the
individual mass profiles from both models and smooth dark matter
profiles since they all provide similar fit qualities and by combining
them we allow a greater freedom in the total mass profile.

\begin{figure}
  \centering \includegraphics[width=\columnwidth]{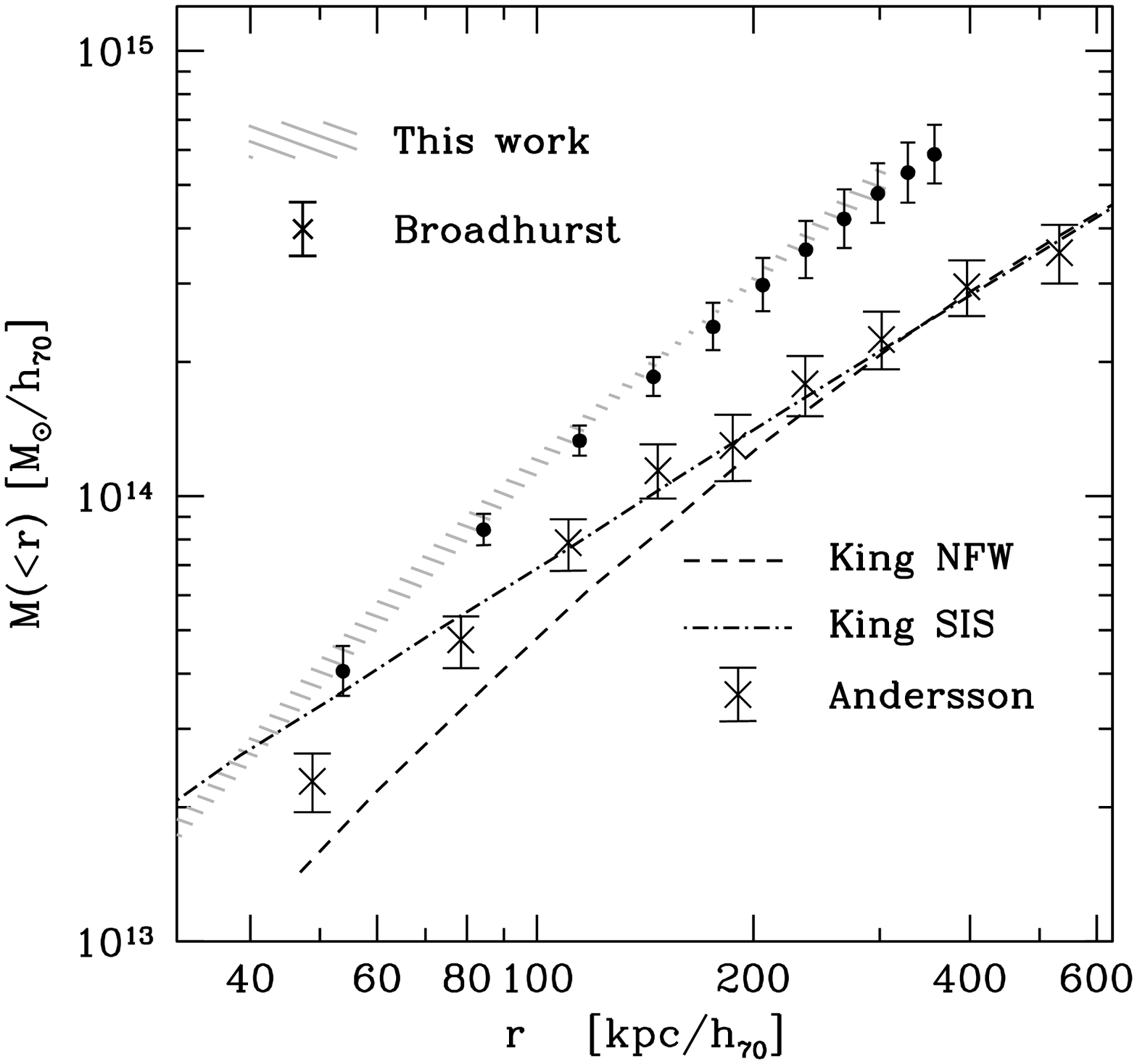}
  \caption{1-$\sigma$ confidence region of projected mass M($<$r)
  inside radius r from this work (striped grey region) compared with
  \citet{broadhurst:05} (long dashed), \citet{king:02b} (NFW dashed,
  SIS dot dashed) and \citet{andersson:04} (dashed - long dashed). The
  error bars shown for \citet{broadhurst:05} and \citet{andersson:04}
  are 1-$\sigma$~errors. The mass profile and errors for
  \citet{broadhurst:05} were obtained by integration from the surface
  mass density in their Figure 26.}
  \label{fig:mass}
\end{figure}

In Figure \ref{fig:mass} we show again the envelope of the total
masses encompassed by the best 68.3\% fits of all the model galaxies
from the error analysis in striped grey. For comparison we also show
the strong lensing mass measurement of \citet{broadhurst:05} (long
dashed), weak lensing mass from \citet{king:02b} (NFW dashed, SIS dot
dashed) and X-ray mass estimate of \citet{andersson:04} (dashed - long
dashed). For \citet{broadhurst:05} and \citet{andersson:04} points we
plot the 1-sigma errors. The \citet{broadhurst:05} mass has been
integrated from the radial surface mass density profile in their
Figure 26, and the errors have been inferred from the errors in
surface mass density. \citet{andersson:04} have also provided an
estimate of the projected X-ray mass so that the profile can be
compared with lensing mass measurements. The agreement between our
work and that of \citet{broadhurst:05} is very good, well within
1-$\sigma$ at all radii. The mass measured using strong lensing is
factor $\sim$2 larger than the mass from X-ray estimates. For a
discussion on the low mass from X-ray please refer to
\citet{andersson:04}.\\

To estimate the NSIS and NFW parameters of the total mass we can
simply fit the mass profile obtained with Models \Roman{one} and
\Roman{two} with a single NSIS or NFW halo. One should not forget that
the total mass profile was derived using the NSIE and NFW profiles
themselves. The total mass profile is composed of the mass in the
galaxies and two elliptical smooth DM haloes and hence the total mass
is no longer a pure NSIE or ENFW profile. In fitting a single halo we
also do not include ellipticity. The excellent agreement between the
total mass profile obtained in this work and that of
\citet{broadhurst:05} and the superior performance of our models to
theirs give us confidence that the parameters we have derived are
indeed representative of the total mass profile of the cluster. We do
not compare the quality of fit of NSIS and NFW haloes but instead the
obtained parameters with those from weak lensing. This should help us
to avoid problems arising from the underlying smooth DM profiles used
in obtaining the total mass profile.

The 68.3\%, 95.4\% and 99.7\% confidence contours for the NFW
parameters are shown in Figure \ref{fig:single_halo_nfw_all} (solid
black contours). Both the concentration and r$_{200}$ are well
constrained. The best fit values are C=6.0$\pm$0.5 and
r$_{200}$=$2.82^{+0.11}_{-0.09}$.

Also the NSIS parameters are well constrained. The corresponding
confidence contours are shown in Figure
\ref{fig:single_halo_nsis_all}. Both of the NSIS parameters depend on
the halo profile in the region where the multiple images have
significant constraints. Therefore the confidence contours are also
extremely tight. The best fit parameters are
$\sigma$=1514$_{-17}^{+18}$~km/s and $r_c$=71$\pm$5~kpc. The best
fitting profile parameters are summarised in Table \ref{tab:results2}.

As a comparison we have also fitted of a single isothermal sphere to
the {\it smooth} DM component only. This results in an NSIS a velocity
dispersion of 1450$^{+39}_{-31}$~km/s and a core radius of
77$^{+10}_{-8}$~kpc/h while an NFW profile has a concentration of
4.7$^{+0.6}_{-0.5}$ and a virial radius of
2.86$\pm$0.16~Mpc/h$_{70}$.


\subsection{Model \Roman{four}: Combining Information from Strong and Weak Lensing}
\label{sec:sl_wl_combined}

In this subsection we include the new weak lensing shear data from
\citet{broadhurst:05b} in our analysis to use information of the
cluster profile from larger clustercentric radii.

Strong lensing in A1689 can only constrain the mass at best to out
200-300~kpc from the cluster centre. In order to constrain the scale
radius of an NFW profile strongly it should lay within the multiple
images. Unfortunately, in the case of A1689, the scale radius seems to
be just outside the multiple images (and hence strong lensing cannot
constrain it significantly) but too small to be well constrained by
weak lensing data alone. On the other hand weak lensing can tell us
something about the total mass of the cluster and hence constrain
r$_{200}$. By combining this with information from strong lensing
(details of the profile at small radii) one should expect to have a
handle on the cluster at all radii.

In their extensive work on this cluster
\citet{broadhurst:05,broadhurst:05b} conclude that the parameters
derived from strong and weak lensing are not compatible. In the strong
lensing regime an NFW halo has only a moderate concentration
(C=6.0$\pm$0.5 in this work, \citet{broadhurst:05} find
C=6.5$^{+1.9}_{-1.6}$) whereas in the weak lensing regime a very high
concentration (\citet{broadhurst:05b} C=13.7$^{+1.4}_{-1.1}$) is
required, uncharacteristic to a halo of this size and typical to a
halo with a much lower mass.

\begin{figure}
  \centering
  \includegraphics[width=\columnwidth]{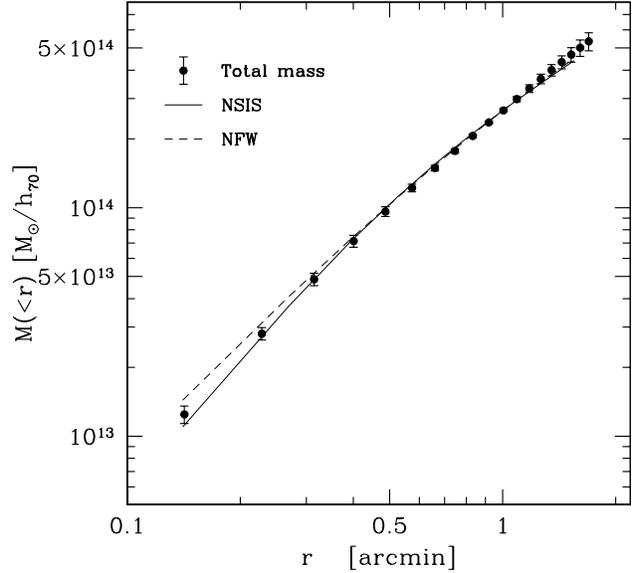}
  \caption{ Radial profile of the total mass from this work
  (circles). The lines show the best fitting NSIS (solid line) and NFW
  (dashed line) profiles to a simultaneous fit of the shear in Figure
  \ref{fig:single_halo_all_shear} and the mass in this Figure. }
  \label{fig:single_halo_all_mass}
\end{figure}

\begin{figure}
  \centering
  \includegraphics[width=\columnwidth]{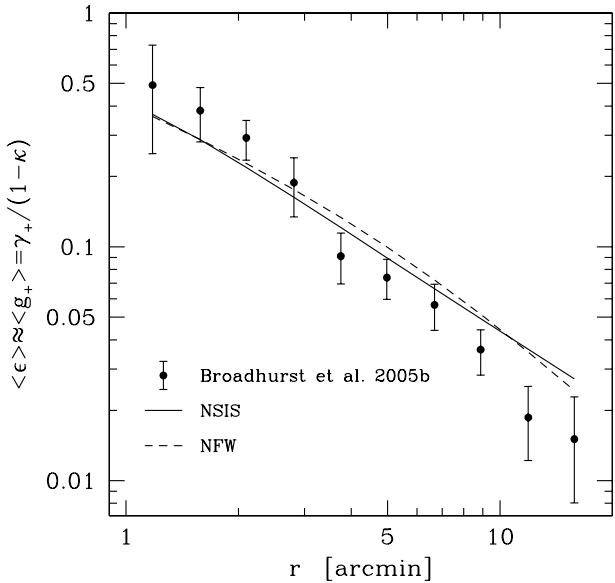}
  \caption{ Radial profile of the tangential shear from
  \citet{broadhurst:05b} (circles). The lines show the best fitting
  NSIS (solid line) and NFW (dashed line) profiles to a simultaneous
  fit of the shear in this figure and the mass in Figure
  \ref{fig:single_halo_all_mass}. }
  \label{fig:single_halo_all_shear}
\end{figure}

We have checked this inconsistency in the NFW parameters by fitting a
single halo to both the radial mass profile (Model \Roman{three}, this
work) and shear profile (\citet{broadhurst:05b}), shown in Figures
\ref{fig:single_halo_all_mass} and
\ref{fig:single_halo_all_shear}. The fit is done simultaneously to the
mass from strong lensing and reduced shear from weak lensing. The best
fitting NFW profile is plotted as a dashed black line in the two
figures.

Unlike \citet{broadhurst:05b} we do not include any prior (C$\le$30) on
the concentration in our fits since there is no obvious bias towards
NFW profiles with higher concentrations: a high quality fit with a
large concentration purely reflects the inability of shear
measurements to constrain the central cluster profile. A prior could
lead to a wrong determination of the minimum $\chi^2$ and hence favour
a smaller concentration without a physical significance.

\begin{figure}
  \centering
  \includegraphics[width=\columnwidth]{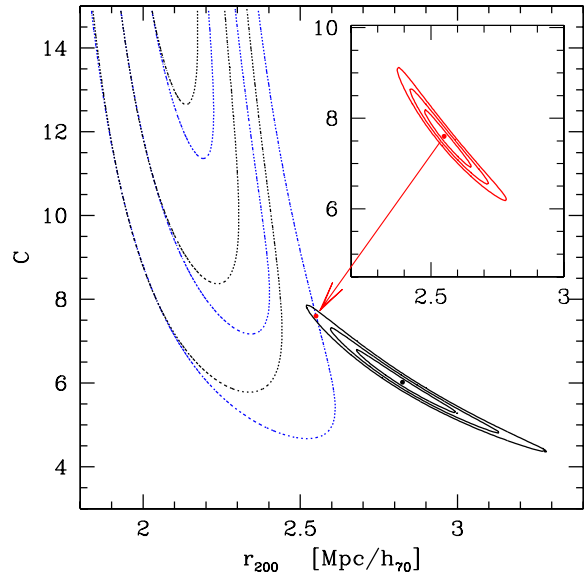}
  \caption{68.3\%, 95.4\% and 99.7\% confidence contours for the NFW
  parameters C-r$_{200}$ for a fit to reduced shear (data from
  \citet{broadhurst:05b}) only (dotted contours) and mass (data from
  this work) only (solid black contours). The dotted blue contours
  show the confidence regions from shear when the last two data points
  are excluded. The combined fit to both reduced shear and mass
  simultaneously is $\chi^2/dof\sim$~31.9~/~20 (shown in the inlet for
  clarity with the same scale). The best fitting parameters are
  C=7.6$^{+0.3}_{-0.5}$ and
  r$_{200}$=2.55$^{+0.07}_{-0.09}$~Mpc/h$_{70}$. A fit to shear only
  gives C=30.4 and r$_{200}$=1.98~Mpc/h$_{70}$.}
  \label{fig:single_halo_nfw_all}
\end{figure}

Fitting a single NFW halo to the weak lensing shear from
\citet{broadhurst:05b} gives only a lower limit for the concentration
but constrains r$_{200}$ (or equivalently virial mass M$_{200}$) to
$\sim$ 2.0-2.5Mpc/h$_{70}$ (68.3\%, 95.4\% and 99.7\% confidence
regions for C-r$_s$ are shown in Figure \ref{fig:single_halo_nfw_all}
as dotted lines). The best fit values are C=30.4 and
r$_{200}$=1.98~Mpc/h$_{70}$. The fit is excellent with $\chi^2_{WL
shear}/dof$~=~2.5~/~8. The parameters of the NFW profile from fitting
the total mass and shear independently disagree more than the
estimated 3-$\sigma$ errors.

By fitting both shear and mass simultaneously we are able to combine
the constraints from both small and large radii to obtain well defined
NFW parameters for the halo. The NFW parameters in this case become
C=7.6$^{+0.3}_{-0.5}$ and
r$_{200}$=2.55$^{+0.07}_{-0.09}$~Mpc/h$_{70}$ (confidence regions for
the combined weak and strong lensing fit are shown with solid red
contours in Fig. \ref{fig:single_halo_nfw_all}).  The mass and shear
profiles of the best fitting NFW halo are shown in Figures
\ref{fig:single_halo_all_mass} and \ref{fig:single_halo_all_shear}
respectively as dashed lines.

\begin{figure}
  \centering
  \includegraphics[width=\columnwidth]{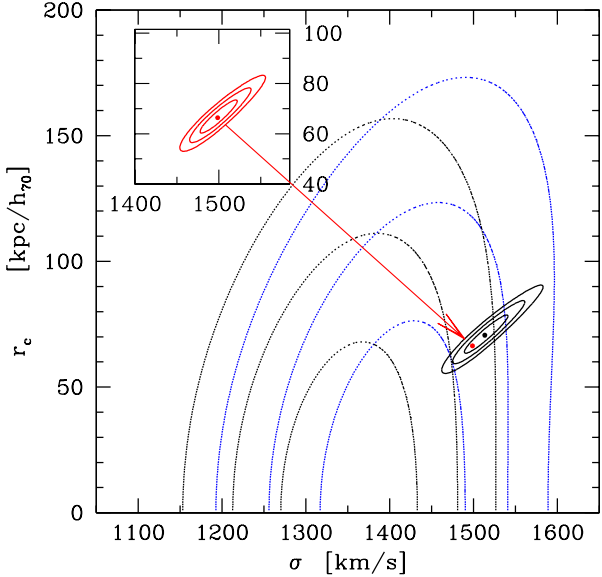}
  \caption{68.3\%, 95.4\% and 99.7\% confidence contours for the NSIS
  parameters $\sigma$-r$_c$ for a fit to reduced shear (data from
  \citet{broadhurst:05b}) only (dotted contours) and mass (data from
  this work) only (solid black contours). The dotted blue contours
  show the confidence regions from shear when the last two data points
  are excluded. The combined fit to both reduced shear and mass
  simultaneously (shown in the inlet with the same scale) is similar
  to that of NFW profile with $\chi^2/dof\sim$~30~/~27. The best
  fitting parameters are $\sigma$=1499$\pm$15~km/s and
  r$_c$=66$\pm$5~kpc/h$_{70}$.A fit to shear only gives a singular
  profile with $\sigma$=1354~km/s.}
  \label{fig:single_halo_nsis_all}
\end{figure}

We have repeated the experiment also for an NSIS halo. The
corresponding contours are shown in Figure
\ref{fig:single_halo_nsis_all}, and mass and shear profiles in Figures
\ref{fig:single_halo_all_mass} and \ref{fig:single_halo_all_shear} as
solid lines.

Like with the NFW halo the core radius of the NSIS halo is poorly
constrained by the weak lensing data alone though surprisingly a
singular profile with $\sigma$=1354~km/s has the best fit. The fit is
good with $\chi^2$~/~dof~=~10~/~8. The best fitting parameters to both
weak and strong lensing simultaneously are $\sigma$=1499$\pm$15~km/s
and r$_c$=66$\pm$5~kpc/h$_{70}$ with $\chi^2$~/~dof~=~30~/~20. The
agreement between the parameters for the NSIS halo is better than for
the NFW profile, though still only at 2-$\sigma$ level.

To see how important the last two shear data points are for the
previously derived cluster parameters we have excluded the two outer
most data points from the shear measurement. If compared to numerical
simulations the concentration of the NFW halo remains unreasonably
high although the disagreement between weak and strong lensing is
reduced to just under 3-sigma. For the NSIS halo the shear data is
still fit best with a singular profile but higher values of core
radius are allowed and the velocity dispersion is increased to make
the weak and strong lensing parameters agree at better than
2-sigma. The best fitting parameters are C=7.1$\pm$0.4 and
r$_{200}$=2.63$\pm$0.06~Mpc/h$_{70}$ for the NFW halo and
$\sigma$=1505$\pm$15~km/s and r$_c$=68$\pm$5~kpc/h$_{70}$ for the NSIS
halo. The two profiles fit well with $\chi_{nsis}^2/dof\sim$~22~/~18
compared to $\chi_{nfw}^2/dof\sim$~20~/~18. The confidence regions
with the last two shear points excluded are shown as dotted blue
contours in Figures \ref{fig:single_halo_nfw_all} and
\ref{fig:single_halo_nsis_all}.

In a recent work \citet{biviano:05} derived the mass profiles of the
different luminous and dark components of cluster masses
separately. They find that ratio of baryonic to total mass decreases
from the centre to r$\sim$0.15 virial radii and then increases
again. We see the same trend also in our work (Figure
\ref{fig:mass_components}), where the galaxy component has a minimum
contribution at around 200~kpc. This is smaller than expected
(380~kpc) if we take the r$_{200}$ of the NFW profile to be the virial
radius of the cluster.

The best fit parameters are summarised in Table \ref{tab:results2}.


\subsection{Comparison with Literature}
\label{sec:literature}

The mass of Abell 1689 has been determined in a variety of ways with
different weaknesses and strengths. Results from the three methods
used (\mbox{X-ray} temperature, line-of-sight velocity and lensing
(both weak and strong)) have disagreed considerably. This section
makes a short summary and comparison of the results using the
different methods.

Recent results are summarised in Tables \ref{tab:comparison} and
\ref{tab:comp_mass}. Parametric model fits are summarised in Table
\ref{tab:comparison} and aperture mass fits in Table
\ref{tab:comp_mass}. When comparing different mass estimates one
should bear in mind that both \mbox{X-ray} and velocity dispersion
measure a spherical mass where as lensing in the thin lens
approximation measures projected mass, i.e. mass in a cylinder of a
given radius, resulting in higher masses within a given radius. Only
lensing measures the mass directly. Both \mbox{X-ray} and velocity
dispersion rely on the cluster being a relaxed system.\\

\begin{table}
  \centering
  \caption{ Comparison between best fit parametric mass models from
    different methods for Abell 1689. \citet{andersson:04} is an
    \mbox{X-ray}, \citet{king:02} a weak lensing and
    \citet{girardi:97} a line of sight velocity study of the
    cluster. The background galaxy catalogue used by \citet{king:02}
    suffers from contamination from galaxies at low redshifts where
    lensing is inefficient (discussed in more detail in
    \citet{clowe:01}) which reduce the total measured mass of the
    cluster and hence the measured velocity dispersion.}
  \begin{tabular}{lccc}
    &\multicolumn{2}{c}{\bf{NFW Parameters}}&\\
    \hline
    \bf{Method}&C&r$_{200}$ (Mpc)&\bf{Reference}\\
    \hline
    \hline
    SL&6.0$\pm$0.5&2.82$\pm$0.11&this work\\
    \hline
    SL&6.5$^{+1.9}_{-1.6}$& 2.02 &\citealt{broadhurst:05}\\
    \hline
    X-ray&7.7$^{+1.7}_{-2.6}$&1.87$\pm$0.36&\citealt{andersson:04}\\
    \hline
    WL&4.8&1.84&\citealt{king:02}\\
    \hline
    \hline\\
    &\multicolumn{2}{c}{\bf{NSIE Parameters}}&\\
    \hline
    \bf{Method}&$\sigma$ (km/s)&r$_c$ (kpc)&\bf{Reference}\\
    \hline
    \hline
    SL&1514$\pm$18&71$\pm$5&this work\\
    \hline
    SL&1390&60&\citealt{broadhurst:05}\\
    \hline
    X-ray&918$\pm$27&SIS&\citealt{andersson:04}\\
    \hline
    X-ray&1190&27&\citealt{andersson:04}$^*$\\
    \hline
    WL&998$^{+33}_{-42}$&SIS&\citealt{king:02}\\
   \hline
    LOSVD&1429$^{+145}_{-96}$&-&\citealt{girardi:97}\\
   \hline
   \hline
    \multicolumn{4}{l}{$^*$ data from \citealt{andersson:04}, fitting
    done in this work.}\\
  \end{tabular}
  \label{tab:comparison}
\end{table}

\begin{table}
  \centering
  \caption{Comparison between mass estimates for Abell 1689 from
    different methods. The mass measured by \citet{andersson:04} are
    underestimates of the total mass if the cluster is undergoing a
    merger. For our work the mass at r=0.25~Mpc/h$_{70}$ is an
    extrapolation since the multiple images do not extend to such
    large clustercentric radii.}
  \begin{tabular}{lcl}
    \hline \multicolumn{1}{c}{\bf{M($<$r)}}&
    \multicolumn{1}{c}{\bf{r}}& \multicolumn{1}{c}{\bf{Reference}}\\
    \multicolumn{1}{c}{(10$^{15}$ M$_\odot$ h$_{100}^{-1}$)}&
    \multicolumn{1}{c}{(Mpc h$_{100}^{-1}$)}&\\ \hline \hline
    \hspace*{4mm}0.14 $\pm$ 0.01&\hspace*{4mm}0.10\hspace*{4mm}&this work, Model \Roman{three}\\
    \hline
    \hspace*{4mm}0.082 $\pm$ 0.013&0.10&\citet{andersson:04}\hspace*{1mm}\\
    \hline
    \hspace*{4mm}0.43 $\pm$ 0.02&0.24&\citet{tyson:95}\\
    \hline
    \hspace*{4mm}0.20 $\pm$ 0.03&0.25&\citet{andersson:04}\\
    \hline
    \hspace*{4mm}0.37 $\pm$ 0.06&\hspace*{4mm}0.25\hspace*{4mm}&this work, Model \Roman{three}\\
    \hline
    \hspace*{4mm}0.48 $\pm$ 0.16&0.25&\citet{dye:01}\\
    \hline
  \end{tabular}
  \label{tab:comp_mass}
\end{table}

\subsubsection{X-ray}
\label{sec:literature_xray}
\begin{figure}
  \centering
  \includegraphics[width=\columnwidth]{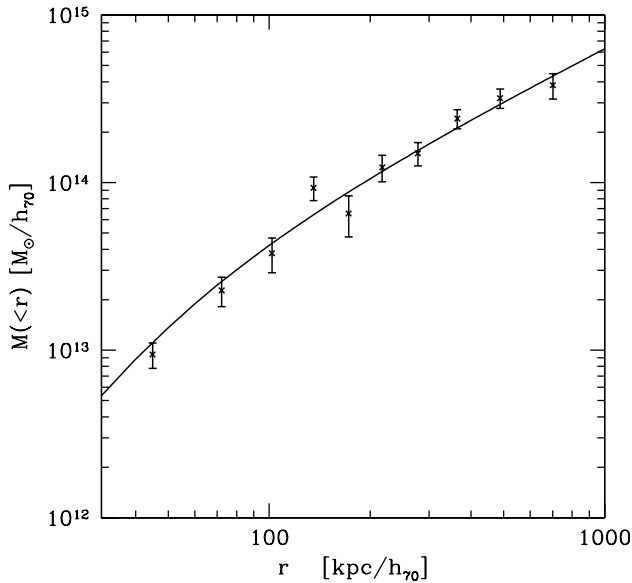}
  \caption{The spherical mass of A1689 from X-ray observations of the
  cluster by \citet{andersson:04} (points) and our best fit NSIS halo
  with $\sigma$=1190~km/s and $r_c=$27~kpc (line). The fit is very
  good with $\chi^2$ / dof = 9.2 / 8. The low mass in the centre can
  only be fitted with an isothermal sphere if a core radius is
  included.}
  \label{fig:andersson_nsis}
\end{figure}

The most recent \mbox{X-ray} measurements of the mass of A1689 are
those of \citet{xue:02} with the Chandra X-Ray Observatory and
\citet{andersson:04} with the XMM-Newton X-Ray Observatory. Both find
nearly circular \mbox{X-ray} emission centred on the cD galaxy. Best
fit NFW profile to \citet{andersson:04} data has parameters
$c=$7.7$^{+1.7}_{-2.6}$ and $r_{200}=$$\pm$0.36~Mpc/h$_{70}$. They
have also fitted a SIS profile to the data and obtain
$\sigma=$918~km/s. The NFW profile gives a much better fit to their
data.  We have also fitted an NSIS profile to \citet{andersson:04}
since it is clear that (single parameter) a SIS will not be able to
reproduce the data. We have fitted the spherical mass of an NSIS
profile with $\sigma=$1190~km/s and $r_c=$27~kpc to the data from
Figure 9 of \citet{andersson:04} and this provides a very good
fit. The NSIS halo along with the fitted points are shown in Figure
\ref{fig:andersson_nsis}. The low $\sigma$ found by
\citet{andersson:04} is mainly driven by the low central mass of the
cluster which the SIS profile can only accommodate with a low
$\sigma$. By including a core radius in the fit the mass can be
modelled very well everywhere also by an IS profile.

The total mass inside 140kpc from the cluster centre is
1.2$\times$10$^{14}$ M$_{\odot}$ and 1.9$\times$10$^{14}$ M$_{\odot}$
for \citet{andersson:04} and \citet{xue:02}
respectively. \citet{andersson:04} also discuss the effect a merger
would have on the \mbox{X-ray} mass estimates. The estimated
\mbox{X-ray} masses could increase by a factor of $\sim$2 (velocity
dispersion by factor $\sqrt{2}$) assuming that two equal mass haloes
are considered as one in the \mbox{X-ray} analysis. This would be
enough to bring \mbox{X-ray} mass of Abell 1689 in good agreement with
lensing.

\subsubsection{Spectroscopy}
\label{sec:literature_spec}
An early spectroscopic work by \citet{teague:90} found a very high
velocity dispersion of 2355km/s for Abell 1689. \citet{girardi:97}
have reanalysed the data from \citet{teague:90} and found four
different structures in A1689 with velocity dispersions of 1429km/s,
321km/s, 243km/s and 390km/s. A simple consideration of total mass in
the separate structures equals that of a single isothermal sphere with
$\sigma$$\sim$1550km/s. The separate structures are more extended than
the region of multiple images and the 2$^{nd}$ halo in this study does
not correspond to any of the spectroscopically identified groups by
\citet{girardi:97}.

\citet{czoske:04} have used VIMOS on the VLT to obtain spectra for
A1689. Their results are still preliminary but indicate a strong
gradient in the velocity dispersion from $\sim$2100km/s in the centre
to $\sim$1200km/s at larger clustercentric distances
($\gtrsim$1Mpc). The high velocity dispersion on the centre could be
due to an unrelaxed system and not an indication of a high total mass
of the cluster.

\citet{lokas:05} have recently shown that cluster mass cannot be
reliably estimated from galaxy kinematics due to the complex
kinematical structure of A1689. The obtained velocity dispersion
depends sensitively on the chosen galaxy sample.

\subsubsection{Weak Lensing}
\label{sec:literature_wl}
The mass of A1689 has been measured in a number of cases with weak
gravitational lensing. The obtained masses are always considerably
lower when compared to strong lensing masses or LOSVD measurements
with $\sigma$=1028~km/s, $\sigma$=998~km/s and $\sigma$=1030~km/s
(\citet{clowe:01}, King et al. (2002a,b) respectively).
\citet{clowe:01,king:02} use the same catalogue of lensed background
galaxies. The catalogue is very likely contaminated by unlensed
galaxies in the foreground and only weakly lensed galaxies in the
close proximity of the cluster ($z<$0.3) where lensing is
inefficient. These galaxies reduce the average shear signal leading to
lower mass estimates. The SIS velocity dispersion estimate of
\citet{clowe:01} increases to 1095~km/s if they assume that 87\% of
the faint background galaxies have $z>$0.3. Recent work by
\citet{broadhurst:05b} results in higher masses and only 1-2 $\sigma$
discrepancy between weak and (our) strong lensing models.

\citet{hoekstra:03} investigate the effect of distant (along
line-of-sight) large scale structure on the errors of derived
M$_{200}$ and concentration of an NFW halo. They conclude that the
errors could be underestimated by a factor $\sim$2.


\section{Conclusion}
\label{sec:conclusion}
We have identified 15 images systems in deep ACS images of galaxy
cluster Abell 1689. Two of these are not in the 30 image systems
identified by \citet{broadhurst:05}. By excluding one of their image
systems and splitting another in two we have constructed a new
catalogue with 107 multiple images. These from 31 image systems and we
have additionally used one long arc in the modelling. The galaxy
cluster was modelled with a two component mass model: mass associated
with cluster galaxies and an underlying smooth dark matter
component. Cluster galaxies were identified from the cluster
redsequence and their halo masses were estimated using
Fundamental-Plane and Faber-Jackson relations. The use of
Fundamental-Plane in measuring the mass for most of the galaxies used
in the cluster modelling is new and allows a very precise
determination of the (central) galaxy mass. The galaxies were modelled
with a truncated isothermal ellipsoid. The truncation of the galaxy
haloes is necessary for accurate lensing models. The smooth dark
matter component was modelled separately with two parametric
elliptical halo profiles: elliptical NFW profile and a non-singular
isothermal ellipsoid.\\

We find that both an ENFW and NSIE describe the smooth dark matter
component very well. The multiple images are reproduced extremely
well. The best fit ENFW profile of the smooth dark matter component
has a virial radius of 2.06$\pm$0.03~Mpc and a concentration parameter
of 6.2$\pm$0.1, the best fitting NSIE profile has a core radius of
74$\pm$3~kpc and a velocity dispersion of 1281$\pm$10~km/s. The
ellipticities of the two model haloes are small ($\epsilon=0.06$ in
both cases).

By fitting a single NSIS and NFW halo to the total mass we can
determine the halo parameters of the cluster as a whole very
strongly. The NFW parameters are C=6.0$\pm$0.5 and
r$_{200}$=2.82$\pm$0.11~Mpc; the NSIS parameters are
$\sigma$=1514$_{-17}^{+18}$~km/s and $r_c$=71$\pm$5~kpc.

Using the images of \citet{broadhurst:05} we obtain a fit with an rms
distance between the identified multiple images and model predictions
0.6'' better than the best model in \citet{broadhurst:05} (rms of
2.65'' compared to 3.25''). This is surprising considering the large
freedom in the mass model used by \citet{broadhurst:05} compared to
parametric models. The superior performance of our model can in part
be attributed to a careful analysis of the cluster galaxy
component. It also indicates that small scale dark matter 'mini'
haloes are not needed to explain the deflection field in A1689. The
overall mass profiles are in good agreement however. This shows that
strong gravitational lensing can be used to derive very accurate total
mass profiles; different methods and assumptions agree very well in
mass although the treatment of the cluster galaxies in particular can
be quite different.

The low masses obtained from weak lensing in the past are no longer
observed in new shear measurements by
\citet{broadhurst:05b}. According to our analysis, at least for the
NFW profile, the parameters obtained from strong and weak lensing
disagree at $\sim$3-sigma level. The high concentration of an NFW
profile fit to weak lensing data is incompatible with both the strong
lensing results presented here and in \citet{broadhurst:05}. The
discrepancy between halo parameters is present at $<$2-sigma level in
the case of an isothermal sphere dark matter halo. We do not find
support for the strong rejection of a softened isothermal sphere by
\citet{broadhurst:05b} based on the combined strong and weak lensing
mass profile.\\

The unusually high concentration (compared to numerical N-body
simulations) can be explained by a suitably aligned tri-axial halo
\citep{oguri:05} but this cannot be used to solve the discrepancy
between weak and strong lensing measurements which both measure the
same projected mass, albeit at different radii.\\


\section*{Acknowledgements}
This work was supported by the Deutsche Forschungsgemeinschaft, grant
\emph{SFB 375} ``Astroteilchenphysik''. The authors would like to
thank the referee Jean-Paul Kneib for helpful comments that have made
this paper both clearer and more comprehensive, Ralf Bender for making
his photometric redshift code available for this project, Marisa
Girardi for providing us with the list of galaxies and their redshifts
used in \citet{girardi:97}, Thomas Erben and Lindsay King for their
help with data reduction, and Peter Schneider for his comments on the
manuscript.


\bibliographystyle{mnras}
\bibliography{halkola_a1689_strong_lensing}

\clearpage
\newpage

\appendix

\section{Gravitational Lensing by NSIE, NFW and BBS profiles}
\label{app:lensing}


\subsection{Isothermal Sphere / Ellipsoid}
A model often used in gravitational lensing for galaxies and clusters
of galaxies is a singular isothermal sphere (SIS)
\citep[e.g.][]{gott:73,turner:84}. SIS naturally reproduces the
observed flat rotation curves of galaxies
\citep[e.g.][]{roberts:73}. The following equations describe a
non-singular (or softened) isothermal ellipsoid \citep{hinshaw:87}
where the singularity has been removed with a core radius, and
additionally an ellipticity has been incorporated to better model the
observed galaxy shapes \citep{seitz:98}. In the equations ellipticity
is introduced to the gravitational potential, in a similar fashion to
\citet{kochanek:89}, and not the mass distribution. This approach has
some problems with large ellipticities, when the accompanying mass
distribution can have negative values, as noted by
\citet{blandford:87}, \citet{kormann:94} and others, but is
numerically rather simple and straightforward to implement since all
parameters of interest can be calculated from the analytic derivatives
of the potential. An alternative approach is to have an elliptical
mass distribution as demonstrated by \citet{kormann:94} but the
expressions for deflection angle ($\vec{\alpha}$), surface mass
density ($\kappa$) and shear ($\vec{\gamma}$) are considerably more
complicated. \citet{kassiola:93} have done a thorough comparison
between elliptical potentials and elliptical mass distributions. We
have estimated the effect of an elliptical potential on the
ellipticity of the surface mass density, illustrated in Figure
\ref{fig:ellipticity}. This will be discussed later in more detail.

In the following equations $\psi$ is gravitational potential,
$\vec{\theta}$~is (image) position on the lens plane, $\zeta$ is a
core radius, $q$=b/a=(1-$\epsilon$)/(1+$\epsilon$) is the axis ratio
of the potential and $\theta_{E}$ is the Einstein radius of a singular
isothermal sphere.\\

The equations for the deflection potential ($\psi(\vec{\theta})$), the
deflection angle ($\vec{\alpha}$), $\kappa$ and $\vec{\gamma}$ are

\begin{equation}
  \psi(\vec{\theta})={\psi_{0}\over\zeta}\ \sqrt{\zeta^{2}+q\ \theta_1^{2}+{1\over q}\
  \theta_2^{2}}=\theta_{E}\ C(\vec{\theta})\ ,
\end{equation}

with  $\theta_{E}={\psi_{0}\over\zeta}=4\pi {D_{ds}\over D_s}({\sigma
  \over c})^2$ and $C(\vec{\theta})=\sqrt{\zeta^{2}+q\ \theta_1^{2}+{1\over q}\ \theta_2^{2}}$\ ,\\

\begin{equation}
  \vec{\alpha}(\vec{\theta})=\nabla\psi(\vec{\theta})={\theta_{E} \over C(\vec{\theta})
  }\cdot \bigg(q\ \theta_1,\ {1\over q}\ \theta_2\bigg)\ ,
\end{equation}

\begin{equation}
  \kappa(\vec{\theta}) = {1\over2}\nabla^{2} \psi(\vec{\theta})={1\over2}\ {\theta_{E}\over
    C(\vec{\theta})^3}\ \bigg(Q_{+}\ \zeta^2+\theta_1^2+\theta_2^2\bigg)\ ,
\end{equation}

with $Q_{\pm}=q\pm{1\over q}$\ ,\\

\begin{equation}
  \gamma_1(\vec{\theta})={1\over2}\ {\theta_{E}\over C(\vec{\theta})^3}\ \bigg({Q_{-}}\ \zeta^2-\theta_1^2+\theta_2^2\bigg)\ ,
\end{equation}
\begin{equation}
  \gamma_2(\vec{\theta})=-{\theta_{E}\ \theta_1\ \theta_2\over C(\vec{\theta})^3}\ ,
\end{equation}


\subsection{Universal Dark Matter Profile}

The universal dark matter profile is an analytic fit to results of
numerical N-body simulations of galactic haloes by
\citet{navarro:96}. These simulations showed that density profiles of
galactic haloes of very different sizes (two decades in radius) could
be fitted with a single 'universal' profile. At small radii (r$<$r$_s$
or x=r/r$_s<$1) the NFW-profile is flatter than isothermal with,
$\rho\propto r^{-1}$, where as for large radii (x$>$1), where $\rho\propto
r^{-3}$, it is steeper than isothermal which has $\rho\propto r^{-2}$
everywhere.

Lensing by NFW-profile has been studied in a number of papers
\citep[e.g.][]{bartelmann:96,wright:00,golse:02b}. We have implemented
an elliptical NFW-profile (ENFW) following the formalism described in
\citet{meneghetti:03}. They have introduced the ellipticity to the
deflection angle rather than the potential (or mass distribution).

For the deflection angle we are using the elliptical deflection angle
from \citet{meneghetti:03}. Here we show only the expression for the
deflection angle. For details of the derivation see
\citet{meneghetti:03}.\\

The deflection angle for a spherical NFW mass distribution at
x=r/r$_s$ is

\begin{equation}
  \alpha^{NFW}(x) = {4\kappa_{s}\over x}g(x),
\end{equation}
with
\begin{equation}
  g(x) = ln{x\over2}+\left\{ \begin{array}{lr}
    {2\over\sqrt{1 - x^2}} arctanh\sqrt{1-x\over 1+x} & ,x<1 \\
    1 & ,x=1 \\
    {2\over\sqrt{x^2 - 1}} arctan\sqrt{x-1\over x+1} & ,x>1 \\
  \end{array} \right .\\
\end{equation}\\

We approximate an elliptical mass distribution with axis ratio q by
elliptical contours of the deflection angle,

\begin{equation}
  x\rightarrow \chi=\sqrt{q x_1^2+{1\over q} x_2^2},\\
\end{equation}
\begin{equation}
  \alpha^{ENFW}_{1}=\alpha^{NFW}(\chi)\ {q\ x_1 \over \chi},\quad
  \alpha^{ENFW}_{2}=\alpha^{NFW}(\chi)\ {x_2 \over q\ \chi}\\
\end{equation}\\

The surface mass density ($\kappa$) and shear ($\vec{\gamma}$) are
calculated from the elliptical deflection angles by numerical
differentiation.


\subsection{Truncated isothermal sphere}
  
The truncated isothermal sphere has been introduced by
\citet{brainerd:96} in the framework of galaxy-galaxy lensing. The two
parameters of BBS profile are truncation radius ($s$) and central
velocity dispersion ($\sigma$). The density profile of the BBS model
is then given by

\begin{equation}
  \rho(r)={\sigma^2 \over 2\pi G r^2}{s^2 \over (r^2+s^2)}
\end{equation}

For $r<s$ the density profile is similar to a singular isothermal
sphere ($\rho(r)\propto 1/r^2$) where as for $r>s$ the density falls
off quicker ($\rho(r)\propto 1/r^4$) to avoid the infinite mass of an
isothermal sphere.\\

The deflection angle of a BBS profile is

\begin{equation}
  \alpha^{BBS}(x)={4 \pi \sigma^2 D_{ds} \over D_{s} c^2 x}\
  \bigg[1+x-\sqrt{1-x^2}\bigg]
\end{equation}
with $x=r/s$.\\

The ellipticity is included in the deflection angle in exactly the
same way as was done for the ENFW halo. The surface mass density and
shear were also calculated from elliptical deflection angles by
numerical differentiation.\\

\clearpage

\section{Fundamental Plane}
\label{app:FP}
  
\begin{table*}
  \caption{Table of galaxy properties from fitting cluster galaxies
  with a Sersic profile. The parameters of the 80 most massive
  galaxies are tabulated.}
  \begin{tabular}{ccccccccc}
    \hline
    Galaxy ID &  RA & Dec & m$_{AB}$$^1$ & $n_{ser}$& R$_e$$^2$ (kpc) & $1-b/a$& $PA$ ($^\circ$) & $\sigma_{est}$$^3$ (km/s)\\
    \hline
    \hline
         1 & +13:11:25.53 & -1:20:37.09 & $17.13\pm0.15$ & $3.1\pm0.3$ & $8.1\pm0.5$ & $0.37\pm0.02$ & $ 10\pm1$ & $178  ~~(224/141)$ \\
         2 & +13:11:25.28 & -1:19:31.12 & $19.09\pm0.10$ & $6.0\pm0.1$ & $2.2\pm0.1$ & $0.06\pm0.01$ & $ 70\pm3$ & $108  ~~(136/86)$ \\
         3 & +13:11:28.19 & -1:18:43.80 & $18.75\pm0.14$ & $2.6\pm0.1$ & $2.5\pm0.1$ & $0.57\pm0.03$ & $ 32\pm1$ & $115  ~~(144/91)$ \\
         4 & +13:11:26.09 & -1:19:51.99 & $18.39\pm0.13$ & $3.9\pm0.2$ & $2.5\pm0.1$ & $0.23\pm0.01$ & $ 79\pm1$ & $144  ~~(181/114)$ \\
         5 & +13:11:26.67 & -1:19:03.88 & $19.64\pm0.14$ & $3.2\pm0.1$ & $1.8\pm0.1$ & $0.38\pm0.01$ & $ 79\pm1$ & $ 80  ~~(100/63)$ \\
         6 & +13:11:26.38 & -1:19:56.51 & $18.33\pm0.14$ & $3.4\pm0.1$ & $1.7\pm0.1$ & $0.35\pm0.01$ & $ 34\pm1$ & $181  ~~(228/144)$ \\
         7 & +13:11:27.06 & -1:19:36.88 & $18.25\pm0.01$ & $6.0\pm0.1$ & $4.9\pm0.1$ & $0.17\pm0.01$ & $  5\pm1$ & $147  ~~(185/117)$ \\
         8 & +13:11:24.62 & -1:21:11.10 & $18.24\pm0.13$ & $4.4\pm0.3$ & $2.1\pm0.1$ & $0.55\pm0.03$ & $136\pm1$ & $170  ~~(215/135)$ \\
         9 & +13:11:25.55 & -1:20:17.25 & $19.17\pm0.10$ & $6.0\pm0.1$ & $2.6\pm0.1$ & $0.15\pm0.02$ & $ 96\pm3$ & $ 97  ~~(122/77)$ \\
        10 & +13:11:27.30 & -1:19:05.17 & $19.62\pm0.15$ & $3.8\pm0.1$ & $1.7\pm0.2$ & $0.33\pm0.01$ & $171\pm2$ & $ 81  ~~(102/64)$ \\
        11 & +13:11:28.50 & -1:18:44.81 & $18.64\pm0.09$ & $3.9\pm0.1$ & $3.0\pm0.2$ & $0.56\pm0.03$ & $179\pm1$ & $115  ~~(144/91)$ \\
        12 & +13:11:29.55 & -1:18:34.66 & $18.25\pm0.13$ & $4.1\pm0.2$ & $2.0\pm0.1$ & $0.41\pm0.01$ & $172\pm1$ & $175  ~~(221/139)$ \\
        13 & +13:11:25.27 & -1:20:02.92 & $19.65\pm0.12$ & $1.0\pm0.1$ & $1.7\pm0.1$ & $0.78\pm0.01$ & $ 93\pm1$ & $ 81  ~~(102/65)$ \\
        14 & +13:11:27.56 & -1:20:02.51 & $17.74\pm0.01$ & $8.9\pm0.1$ & $8.0\pm0.1$ & $0.15\pm0.01$ & $ 57\pm1$ & $147  ~~(185/116)$ \\
        15 & +13:11:26.62 & -1:19:47.96 & $19.69\pm0.01$ & $3.5\pm0.1$ & $2.4\pm0.1$ & $0.35\pm0.01$ & $165\pm1$ & $ 75  ~~(95/60)$ \\
        16 & +13:11:24.36 & -1:21:07.57 & $18.83\pm0.12$ & $6.1\pm0.2$ & $1.5\pm0.1$ & $0.11\pm0.02$ & $173\pm7$ & $138  ~~(174/110)$ \\
        17 & +13:11:28.27 & -1:19:31.55 & $18.38\pm0.14$ & $6.0\pm0.1$ & $2.4\pm0.2$ & $0.32\pm0.02$ & $148\pm1$ & $149  ~~(188/119)$ \\
        18 & +13:11:27.99 & -1:20:07.71 & $17.66\pm0.01$ & $5.3\pm0.1$ & $3.8\pm0.1$ & $0.06\pm0.01$ & $176\pm1$ & $205  ~~(259/163)$ \\
        19 & +13:11:28.90 & -1:19:02.55 & $19.00\pm0.14$ & $5.1\pm0.1$ & $3.6\pm0.3$ & $0.42\pm0.01$ & $ 84\pm1$ & $ 83  ~~(105/66)$ \\
        20 & +13:11:29.47 & -1:19:16.58 & $18.77\pm0.13$ & $4.0\pm0.1$ & $1.0\pm0.1$ & $0.58\pm0.02$ & $ 56\pm1$ & $174  ~~(220/139)$ \\
        21 & +13:11:28.52 & -1:19:58.47 & $18.27\pm0.14$ & $3.3\pm0.1$ & $2.6\pm0.1$ & $0.42\pm0.02$ & $159\pm1$ & $153  ~~(192/121)$ \\
        22 & +13:11:31.57 & -1:19:32.70 & $17.04\pm0.01$ & $2.3\pm0.1$ & $4.9\pm0.1$ & $0.20\pm0.01$ & $150\pm1$ & $258  ~~(325/205)$ \\
        23 & +13:11:28.38 & -1:20:43.40 & $17.75\pm0.01$ & $5.6\pm0.1$ & $7.6\pm0.1$ & $0.12\pm0.01$ & $ 17\pm1$ & $177  ~~(223/140)$ \\
        24 & +13:11:27.29 & -1:20:58.41 & $19.06\pm0.12$ & $1.3\pm0.1$ & $2.4\pm0.1$ & $0.44\pm0.03$ & $169\pm1$ & $ 97  ~~(123/77)$ \\
        25 & +13:11:29.24 & -1:19:46.93 & $19.53\pm0.14$ & $4.7\pm0.4$ & $0.6\pm0.1$ & $0.60\pm0.02$ & $142\pm1$ & $140  ~~(176/111)$ \\
        26 & +13:11:30.91 & -1:18:52.53 & $20.28\pm0.14$ & $4.3\pm0.1$ & $0.7\pm0.1$ & $0.45\pm0.01$ & $104\pm1$ & $ 84  ~~(106/67)$ \\
        27 & +13:11:31.68 & -1:19:24.65 & $18.82\pm0.10$ & $4.1\pm0.1$ & $3.4\pm0.1$ & $0.43\pm0.04$ & $ 90\pm1$ & $ 96  ~~(121/77)$ \\
        28 & +13:11:28.62 & -1:20:25.10 & $18.41\pm0.01$ & $6.0\pm0.1$ & $2.7\pm0.1$ & $0.09\pm0.01$ & $ 24\pm1$ & $136  ~~(172/108)$ \\
        29 & +13:11:30.23 & -1:20:42.74 & $17.11\pm0.01$ & $3.1\pm0.1$ & $5.7\pm0.1$ & $0.16\pm0.01$ & $ 83\pm1$ & $255  ~~(321/202)$ \\
        30 & +13:11:28.78 & -1:20:26.54 & $18.38\pm0.02$ & $4.8\pm0.1$ & $7.2\pm0.2$ & $0.44\pm0.05$ & $ 50\pm2$ & $ 94  ~~(119/75)$ \\
        31 & +13:11:30.44 & -1:20:29.13 & $17.62\pm0.02$ & $2.6\pm0.1$ & $1.9\pm0.1$ & $0.10\pm0.02$ & $ 65\pm3$ & $238  ~~(299/189)$ \\
        32 & +13:11:29.66 & -1:20:27.86 & $16.39\pm0.01$ & $1.2\pm0.1$ & $10.1\pm0.1$ & $0.14\pm0.01$ & $144\pm1$ & $303  ~~(382/241)$ \\
        33 & +13:11:28.08 & -1:21:36.68 & $18.76\pm0.16$ & $3.8\pm0.5$ & $1.9\pm0.3$ & $0.57\pm0.03$ & $175\pm1$ & $132  ~~(166/105)$ \\
        34 & +13:11:32.03 & -1:18:53.65 & $19.06\pm0.17$ & $4.5\pm0.3$ & $3.5\pm0.5$ & $0.05\pm0.01$ & $167\pm6$ & $ 82  ~~(103/65)$ \\
        35 & +13:11:32.88 & -1:19:31.48 & $16.36\pm0.02$ & $5.5\pm0.1$ & $10.1\pm0.2$ & $0.29\pm0.01$ & $ 61\pm1$ & $233  ~~(293/185)$ \\
        36 & +13:11:30.11 & -1:19:55.90 & $18.72\pm0.14$ & $2.9\pm0.1$ & $1.9\pm0.1$ & $0.07\pm0.01$ & $107\pm1$ & $135  ~~(170/107)$ \\
        37 & +13:11:28.02 & -1:21:12.90 & $18.90\pm0.18$ & $4.7\pm0.4$ & $2.6\pm0.4$ & $0.20\pm0.01$ & $ 95\pm1$ & $104  ~~(131/83)$ \\
        38 & +13:11:30.15 & -1:20:40.11 & $17.66\pm0.01$ & $2.5\pm0.1$ & $5.3\pm0.1$ & $0.36\pm0.01$ & $ 52\pm1$ & $180  ~~(227/143)$ \\
        39 & +13:11:30.40 & -1:20:51.69 & $17.64\pm0.01$ & $6.0\pm0.1$ & $6.0\pm0.1$ & $0.15\pm0.01$ & $ 89\pm1$ & $192  ~~(241/152)$ \\
        40 & +13:11:32.28 & -1:19:46.72 & $17.59\pm0.01$ & $5.2\pm0.1$ & $4.4\pm0.1$ & $0.04\pm0.01$ & $144\pm1$ & $215  ~~(271/171)$ \\
        41 & +13:11:32.15 & -1:19:24.22 & $18.26\pm0.13$ & $4.4\pm0.4$ & $2.5\pm0.1$ & $0.19\pm0.01$ & $ 24\pm3$ & $156  ~~(196/124)$ \\
        42 & +13:11:30.20 & -1:20:28.41 & $17.54\pm0.01$ & $2.5\pm0.1$ & $3.2\pm0.1$ & $0.13\pm0.01$ & $125\pm6$ & $197  ~~(248/156)$ \\
        43 & +13:11:29.18 & -1:21:16.67 & $17.99\pm0.01$ & $4.6\pm0.1$ & $4.6\pm0.1$ & $0.10\pm0.01$ & $178\pm1$ & $176  ~~(222/140)$ \\
        44 & +13:11:30.04 & -1:20:15.10 & $19.02\pm0.01$ & $3.8\pm0.1$ & $2.7\pm0.1$ & $0.37\pm0.02$ & $116\pm1$ & $ 86  ~~(109/69)$ \\
        45 & +13:11:30.18 & -1:20:17.29 & $19.50\pm0.01$ & $6.0\pm0.1$ & $1.3\pm0.1$ & $0.41\pm0.01$ & $ 59\pm1$ & $108  ~~(136/86)$ \\
        46 & +13:11:29.17 & -1:20:53.83 & $18.95\pm0.12$ & $2.7\pm0.1$ & $1.0\pm0.1$ & $0.19\pm0.01$ & $178\pm1$ & $160  ~~(201/127)$ \\
        47 & +13:11:32.83 & -1:19:58.55 & $16.15\pm0.01$ & $4.6\pm0.1$ & $8.3\pm0.1$ & $0.31\pm0.01$ & $ 22\pm1$ & $292  ~~(370/230)$ \\
        48 & +13:11:29.93 & -1:21:00.55 & $18.84\pm0.01$ & $5.2\pm0.1$ & $1.9\pm0.1$ & $0.12\pm0.01$ & $ 25\pm1$ & $114  ~~(143/90)$ \\
        49 & +13:11:32.26 & -1:19:36.44 & $18.92\pm0.01$ & $4.8\pm0.1$ & $3.5\pm0.1$ & $0.36\pm0.01$ & $ 39\pm6$ & $ 81  ~~(102/64)$ \\
        50 & +13:11:30.75 & -1:20:43.62 & $17.82\pm0.01$ & $3.5\pm0.1$ & $4.2\pm0.1$ & $0.12\pm0.01$ & $174\pm1$ & $185  ~~(233/147)$ \\
    \hline\\
    \multicolumn{9}{l}{$^1$ Total F775W AB magnitude obtained from the surface brightness profile fitting}\\
    \multicolumn{9}{l}{$^2$ Circularized physical half light radius in units of kpc}\\
    \multicolumn{9}{l}{$^3$ Estimated galaxy velocity dispersion, see text for details, and the corresponding 1$\sigma$ range}\\
  \end{tabular}
  \label{tab:app:FP}
\end{table*}

\clearpage

\begin{table*}
  \captcont*{Table of galaxy properties continued...}
  \begin{tabular}{ccccccccc}
    \hline
    Galaxy ID &  RA & Dec & m$_{AB}$$^1$ & $n_{ser}$& R$_e$$^2$ (kpc) & $1-b/a$& $PA$ ($^\circ$) & $\sigma_{est}$$^3$ (km/s)\\
    \hline
    \hline
        51 & +13:11:30.56 & -1:20:34.81 & $18.97\pm0.02$ & $3.5\pm0.1$ & $1.4\pm0.1$ & $0.35\pm0.01$ & $ 70\pm6$ & $124  ~~(153/100)$ \\
        52 & +13:11:30.56 & -1:20:45.35 & $18.13\pm0.01$ & $4.3\pm0.1$ & $2.0\pm0.1$ & $0.12\pm0.02$ & $ 92\pm3$ & $173  ~~(213/135)$ \\
        53 & +13:11:29.49 & -1:21:05.48 & $19.82\pm0.01$ & $3.0\pm0.1$ & $0.8\pm0.1$ & $0.24\pm0.01$ & $179\pm1$ & $ 77  ~~(98/62)$ \\
        54 & +13:11:29.26 & -1:21:37.37 & $18.77\pm0.09$ & $3.1\pm0.1$ & $1.9\pm0.1$ & $0.53\pm0.03$ & $118\pm1$ & $131  ~~(165/104)$ \\
        55 & +13:11:29.30 & -1:21:55.18 & $18.45\pm0.01$ & $4.1\pm0.1$ & $3.1\pm0.1$ & $0.28\pm0.01$ & $144\pm1$ & $146  ~~(184/116)$ \\
        56 & +13:11:33.36 & -1:19:16.81 & $20.81\pm0.02$ & $2.7\pm0.1$ & $0.5\pm0.1$ & $0.06\pm0.01$ & $167\pm5$ & $ 77  ~~(96/61)$ \\
        57 & +13:11:33.49 & -1:19:42.82 & $18.76\pm0.15$ & $5.3\pm0.1$ & $2.9\pm0.3$ & $0.10\pm0.03$ & $ 17\pm2$ & $107  ~~(135/85)$ \\
        58 & +13:11:31.31 & -1:21:25.05 & $17.79\pm0.01$ & $5.0\pm0.1$ & $6.3\pm0.2$ & $0.09\pm0.01$ & $ 65\pm2$ & $145  ~~(179/116)$ \\
        59 & +13:11:30.21 & -1:21:18.09 & $19.40\pm0.01$ & $3.3\pm0.1$ & $0.9\pm0.1$ & $0.27\pm0.01$ & $171\pm1$ & $135  ~~(170/107)$ \\
        60 & +13:11:31.27 & -1:21:27.71 & $18.25\pm0.01$ & $3.5\pm0.1$ & $1.7\pm0.1$ & $0.25\pm0.03$ & $ 61\pm2$ & $175  ~~(220/139)$ \\
        61 & +13:11:31.26 & -1:20:52.44 & $18.52\pm0.16$ & $5.5\pm0.1$ & $1.8\pm0.3$ & $0.07\pm0.01$ & $ 41\pm8$ & $156  ~~(196/124)$ \\
        62 & +13:11:31.32 & -1:20:44.07 & $19.58\pm0.02$ & $5.7\pm0.2$ & $0.9\pm0.1$ & $0.04\pm0.07$ & $ 18\pm2$ & $121  ~~(149/97)$ \\
        63 & +13:11:31.17 & -1:21:27.72 & $19.16\pm0.01$ & $3.2\pm0.1$ & $1.6\pm0.1$ & $0.17\pm0.02$ & $138\pm1$ & $102  ~~(130/82)$ \\
        64 & +13:11:30.22 & -1:21:42.98 & $18.65\pm0.15$ & $5.3\pm0.1$ & $1.4\pm0.2$ & $0.51\pm0.02$ & $ 24\pm1$ & $165  ~~(207/131)$ \\
        65 & +13:11:34.93 & -1:19:24.36 & $18.69\pm0.15$ & $5.1\pm0.1$ & $2.6\pm0.2$ & $0.13\pm0.01$ & $  2\pm1$ & $117  ~~(148/93)$ \\
        66 & +13:11:31.97 & -1:20:58.57 & $18.54\pm0.01$ & $5.8\pm0.1$ & $5.1\pm0.2$ & $0.06\pm0.01$ & $ 24\pm2$ & $ 86  ~~(109/68)$ \\
        67 & +13:11:34.23 & -1:21:01.72 & $18.01\pm0.07$ & $2.3\pm0.2$ & $3.1\pm0.3$ & $0.61\pm0.02$ & $177\pm2$ & $164  ~~(207/130)$ \\
        68 & +13:11:35.76 & -1:20:12.09 & $18.16\pm0.02$ & $1.9\pm0.1$ & $5.1\pm0.2$ & $0.13\pm0.01$ & $ 15\pm2$ & $111  ~~(140/88)$ \\
        69 & +13:11:35.03 & -1:20:04.29 & $18.75\pm0.03$ & $4.6\pm0.1$ & $2.9\pm0.2$ & $0.31\pm0.05$ & $ 11\pm3$ & $100  ~~(127/79)$ \\
        70 & +13:11:32.28 & -1:21:37.97 & $18.26\pm0.02$ & $4.4\pm0.1$ & $3.4\pm0.1$ & $0.27\pm0.01$ & $120\pm1$ & $146  ~~(183/116)$ \\
        71 & +13:11:32.38 & -1:22:10.64 & $18.11\pm0.01$ & $5.5\pm0.1$ & $4.1\pm0.1$ & $0.20\pm0.01$ & $ 63\pm3$ & $146  ~~(180/115)$ \\
        72 & +13:11:34.26 & -1:21:18.50 & $19.03\pm0.12$ & $3.1\pm0.1$ & $1.8\pm0.1$ & $0.53\pm0.02$ & $ 90\pm1$ & $113  ~~(143/90)$ \\
        73 & +13:11:35.37 & -1:21:18.87 & $18.85\pm0.14$ & $3.9\pm0.1$ & $1.7\pm0.1$ & $0.24\pm0.01$ & $ 66\pm1$ & $129  ~~(163/103)$ \\
        74 & +13:11:35.72 & -1:21:09.01 & $19.00\pm0.15$ & $6.1\pm0.3$ & $2.8\pm0.3$ & $0.19\pm0.01$ & $167\pm2$ & $ 95  ~~(119/75)$ \\
        75 & +13:11:34.94 & -1:20:58.99 & $18.29\pm0.13$ & $3.8\pm0.2$ & $2.5\pm0.1$ & $0.31\pm0.01$ & $117\pm1$ & $152  ~~(192/121)$ \\
        76 & +13:11:36.79 & -1:19:42.49 & $19.17\pm0.13$ & $2.8\pm0.1$ & $3.0\pm0.1$ & $0.15\pm0.05$ & $ 43\pm6$ & $ 82  ~~(104/65)$ \\
        77 & +13:11:36.01 & -1:19:57.25 & $19.72\pm0.11$ & $3.1\pm0.3$ & $1.3\pm0.1$ & $0.38\pm0.01$ & $170\pm1$ & $ 86  ~~(109/69)$ \\
        78 & +13:11:35.55 & -1:20:42.52 & $18.58\pm0.16$ & $5.0\pm0.2$ & $2.2\pm0.2$ & $0.05\pm0.01$ & $ 90\pm1$ & $137  ~~(173/109)$ \\
        79 & +13:11:33.45 & -1:21:53.28 & $18.12\pm0.01$ & $2.8\pm0.1$ & $1.4\pm0.1$ & $0.32\pm0.01$ & $ 15\pm1$ & $ 73  ~~(92/58)$ \\
        80 & +13:11:35.34 & -1:21:12.50 & $19.33\pm0.14$ & $2.9\pm0.2$ & $1.8\pm0.1$ & $0.25\pm0.01$ & $ 24\pm1$ & $ 94  ~~(118/75)$ \\
    \hline\\
    \multicolumn{9}{l}{$^1$ Total F775W AB magnitude obtained from the surface brightness profile fitting}\\
    \multicolumn{9}{l}{$^2$ Circularized physical half light radius in units of kpc}\\
    \multicolumn{9}{l}{$^3$ Estimated galaxy velocity dispersion, see text for details, and the corresponding 1$\sigma$ range}\\
  \end{tabular}
  \label{tab:app:FP2}
\end{table*}

\clearpage


\section{List of Images}
\label{app:images}

\begin{table*}
  \centering
  \caption[]{A summary of image systems used in this study. For the
 columns with redshifts and rms from lensing the values are mean
 values from Models \Roman{one} and \Roman{two} optimised in the image
 plane, in brackets the mean of simulations (optimised in the source
 plane) is given.}
  \begin{tabular}{cccccccccc}
    \cline{1-5}\cline{7-10}\\[-2mm]
    Image  & No. of  \\
    system &  images & $z_{sys}^1$ & $z_{min}$ - $z_{max} $$^2$ & $z_s^3$ && $z_{nsie}^6$  & $z_{enfw}^7$ & $rms_{nsie}^8$ & $rms_{enfw}^9$ \\[1mm]
    \cline{1-5}\cline{7-10}\\[-2.8mm]
    \cline{1-5}\cline{7-10}\\[-2mm]
       1 &  7 & 3.04 &     3.04    &  3.04$^{4,5}$ &&  3.04\quad(3.04) &  3.04\quad(3.04) &    1.72      &  2.54      \\ 
       2 &  5 & 2.46 & 1.88 - 3.03 &       -       &&  2.27\quad(2.28) &  2.12\quad(2.23) &    1.59      &  2.04      \\ 
       3 &  3 & 5.46 & 5.09 - 5.98 &       -       &&  5.74\quad(5.88) &  5.88\quad(5.92) &    0.70      &  1.01      \\ 
       4 &  5 & 1.21 & 0.98 - 1.43 &       -       &&  1.07\quad(1.06) &  1.29\quad(1.29) &    1.64      &  1.80      \\ 
       5 &  3 & 2.87 & 2.34 - 3.39 &       -       &&  2.35\quad(2.37) &  2.52\quad(2.56) &    1.31      &  2.48      \\ 
       6 &  4 & 1.08 & 0.77 - 1.39 &       -       &&  0.98\quad(0.97) &  1.36\quad(1.33) &    0.77      &  1.24      \\ 
       7 &  3 & 4.87 &     4.87    &  4.87$^{5}$   &&  4.87\quad(4.87) &  4.87\quad(4.87) &   10.38      &  6.40      \\ 
       8 &  5 & 2.84 & 1.31 - 3.35 &       -       &&  2.26\quad(2.25) &  2.30\quad(2.36) &    2.90$^{10}$ &  3.51$^{10}$ \\ 
       9 &  4 & 3.98 & 3.50 - 4.46 &       -       &&  3.52\quad(3.55) &  4.46\quad(4.41) &    1.83      &  2.77      \\ 
      10 &  3 & 1.37 &     1.37    &  1.37$^{5}$   &&  1.37\quad(1.37) &  1.37\quad(1.37) &    2.50      &  2.79      \\ 
      11 &  3 & 2.44 & 1.78 - 3.11 &       -       &&  2.48\quad(2.56) &  2.47\quad(2.53) &    1.23      &  2.15      \\ 
      12 &  2 & 1.83 &     1.83    &  1.83$^{4,5}$ &&  1.83\quad(1.83) &  1.83\quad(1.83) &    0.60$^{10}$ &  0.35$^{10}$ \\ 
      13 &  4 & 1.83 &     1.83    &  1.83$^{4,5}$ &&  1.83\quad(1.83) &  1.83\quad(1.83) &    2.01      &  2.88      \\ 
      14 &  3 & 0.99 & 0.77 - 1.21 &       -       &&  1.21\quad(1.20) &  1.21\quad(1.18) &    0.89$^{10}$ &  1.74$^{10}$ \\ 
      15 &  2 & 3.54 & 3.04 - 4.04 &       -       &&  3.04\quad(3.50) &  3.04\quad(3.47) &    1.16$^{10}$ &  1.46$^{10}$ \\ 
      16 &  3 & 1.99 & 1.59 - 2.38 &       -       &&  1.66\quad(1.66) &  1.78\quad(1.83) &    2.28      &  1.48      \\ 
      17 &  3 & 2.04 & 1.65 - 2.42 &       -       &&  1.71\quad(1.65) &  1.76\quad(1.89) &    2.79      &  1.35      \\ 
      18 &  3 & 2.40 & 2.03 - 2.77 &       -       &&  2.77\quad(2.76) &  2.77\quad(2.77) &    1.18      &  2.42      \\ 
      19 &  3 & 2.13 & 1.84 - 2.41 &       -       &&  1.84\quad(1.84) &  1.86\quad(1.86) &    1.90      &  1.56      \\ 
      20 &  5 & 2.70 & 2.15 - 3.25 &       -       &&  3.25\quad(3.25) &  3.25\quad(3.25) &    4.54      &  4.31      \\ 
      21 &  3 & 1.74 & 1.36 - 2.13 &       -       &&  1.50\quad(1.51) &  1.56\quad(1.63) &    1.48      &  1.64      \\ 
      22 &  3 & 2.04 & 1.68 - 2.41 &       -       &&  1.98\quad(1.84) &  1.92\quad(1.86) &    1.52      &  1.28      \\ 
      23 &  3 & 2.01 & 1.62 - 2.40 &       -       &&  2.02\quad(1.90) &  1.90\quad(1.86) &    1.32      &  1.40      \\ 
      24 &  5 & 3.09 & 2.47 - 3.71 &       -       &&  2.47\quad(2.47) &  2.47\quad(2.53) &    5.36      &  2.22      \\ 
      25 &  2 & 4.25 & 3.41 - 5.08 &       -       &&  3.41\quad(3.41) &  3.41\quad(3.41) &    6.90      &  4.81      \\ 
      26 &  3 & 1.02 & 0.33 - 1.71 &       -       &&  1.71\quad(1.71) &  1.71\quad(1.71) &    3.89      &  4.98      \\ 
      27 &  3 & 2.58 & 1.91 - 3.25 &       -       &&  1.91\quad(2.11) &  1.93\quad(2.53) &    4.38      &  5.37      \\ 
      28 &  2 & 1.58 & 0.02 - 3.15 &       -       &&  3.15\quad(3.15) &  3.15\quad(3.14) &    2.15      &  1.64      \\ 
      29 &  5 & 3.26 & 2.60 - 3.93 &       -       &&  2.60\quad(2.60) &  2.60\quad(2.71) &    6.85      &  4.34      \\ 
      30 &  3 & 3.51 & 3.08 - 3.94 &       -       &&  3.94\quad(3.89) &  3.08\quad(3.56) &    1.57$^{10}$ &  3.63$^{10}$ \\ 
      31 &  2 & 2.07 & 1.58 - 2.57 &       -       &&  1.92\quad(1.98) &  2.57\quad(2.45) &    0.35$^{10}$ &  0.72$^{10}$ \\ 
      32 & 19 & 1.50 & 0.50 - 8.00 &       -       &&  1.29\quad(1.32) &  1.20\quad(1.26) &    0.16$^{10}$ &  0.13$^{10}$ \\ 
    \cline{1-5}\cline{7-10}\\[-2mm]
    \multicolumn{7}{l}{$^1$ Redshift of the image system}\\
    \multicolumn{7}{l}{$^2$ Redshift range allowed in modelling}\\
    \multicolumn{7}{l}{$^3$ Spectroscopic redshift}\\
    \multicolumn{7}{l}{$^4$ Spectroscopic redshift \citealt{golse:phd}}\\
    \multicolumn{7}{l}{$^5$ Spectroscopic redshift \citealt{broadhurst:05}}\\
    \multicolumn{7}{l}{$^6$ Mean redshift for the NSIE-profile from Models \Roman{one} and \Roman{two}}\\
    \multicolumn{7}{l}{$^7$ Mean redshift for the ENFW-profile from Models \Roman{one} and \Roman{two}}\\
    \multicolumn{7}{l}{$^8$ Mean image rms for the NSIE-profile from Models \Roman{one} and \Roman{two}}\\
    \multicolumn{7}{l}{$^9$ Mean image rms for the ENFW-profile from Models \Roman{one} and \Roman{two}}\\
    \multicolumn{7}{l}{$^{10}$ Source plane $\chi^2$}\\
  \end{tabular}
  \label{tab:app:summary}
\end{table*}

\clearpage

\begin{table*}
  \centering
  \caption[]{Images used in this study.}
  \begin{tabular}{cccccccc}
    \hline
    \multicolumn{2}{c}{ID} & ID B05 & RA & Dec & $z_p^1$ & $z_{br}^2$ & $z_s^3$ \\
    \hline
    \hline
        1 & a &  1.1 & +13:11:26.60 & -1:19:56.73 &      -        & 3.03$^{+0.53}_{-0.53}$ & 3.04$^{4,5}$ \\
          & b &  1.2 & +13:11:26.43 & -1:20:00.29 & 3.41$\pm$0.22 & 3.04$^{+0.53}_{-0.53}$ &       -      \\
          & c &  1.3 & +13:11:29.92 & -1:21:07.49 & 3.40$\pm$0.30 & 3.27$^{+0.56}_{-0.56}$ &       -      \\
          & d &  1.4 & +13:11:33.21 & -1:20:27.42 & 3.34$\pm$0.20 & 2.94$^{+0.52}_{-0.52}$ &  3.0$^{5}$   \\
          & e &  1.5 & +13:11:32.08 & -1:20:06.01 & 3.48$\pm$0.11 & 3.35$^{+0.57}_{-0.57}$ &       -      \\
          & f &  1.6 & +13:11:30.00 & -1:20:38.43 & 3.32$\pm$1.26 & 1.06$^{+0.27}_{-1.91}$ &       -      \\
          & g &   -  & +13:11:26.58 & -1:19:57.35 &      -        &          -             &       -      \\
     \hline	 						   			  					  
        2 & a &  2.1 & +13:11:26.67 & -1:19:55.47 & 3.02$\pm$1.94 & 2.62$^{+0.48}_{-0.47}$ &       -      \\
          & b &  2.2 & +13:11:33.11 & -1:20:25.51 & 2.92$\pm$0.73 & 2.57$^{+0.47}_{-0.47}$ &       -      \\
          & c &  2.3 & +13:11:32.12 & -1:20:07.17 & 2.90$\pm$0.63 & 2.64$^{+0.48}_{-0.48}$ &       -      \\
          & d &  2.4 & +13:11:29.96 & -1:21:06.03 &      -        & 2.36$^{+0.44}_{-0.44}$ &       -      \\
          & e &  2.5 & +13:11:30.03 & -1:20:39.38 &      -        & 1.59$^{+0.34}_{-0.86}$ &       -      \\
     \hline	 						   			  					  
        3 & a &  3.1 & +13:11:32.19 & -1:20:27.54 & 1.44$\pm$0.07 & 5.48$^{+0.85}_{-0.85}$ &       -      \\
          & b &  3.2 & +13:11:32.32 & -1:20:33.30 & 0.78$\pm$0.13 & 5.45$^{+0.85}_{-0.85}$ &       -      \\
          & c &  3.3 & +13:11:31.83 & -1:20:56.06 &      -        &          -             &       -      \\
     \hline	 						   			  					  
        4 & a &  4.1 & +13:11:32.32 & -1:20:57.37 & 1.20$\pm$0.40 & 1.06$^{+0.27}_{-0.27}$ &       -      \\
          & b &  4.2 & +13:11:30.67 & -1:21:12.05 & 1.12$\pm$0.34 & 1.32$^{+0.31}_{-0.30}$ &       -      \\
          & c &  4.3 & +13:11:30.90 & -1:20:08.34 & 1.14$\pm$0.14 & 1.47$^{+0.33}_{-0.32}$ &       -      \\
          & d &  4.4 & +13:11:26.43 & -1:20:35.45 & 1.00$\pm$0.35 & 1.33$^{+0.31}_{-0.31}$ &       -      \\
          & e &  4.5 & +13:11:29.99 & -1:20:29.38 &      -        &          -             &       -      \\
     \hline	 						   			  					  
        5 & a &  5.1 & +13:11:29.21 & -1:20:48.79 & 3.28$\pm$0.80 & 3.29$^{+0.56}_{-0.56}$ &       -      \\
          & b &  5.2 & +13:11:29.37 & -1:20:44.17 &      -        & 3.16$^{+0.55}_{-0.55}$ &       -      \\
          & c &  5.3 & +13:11:34.27 & -1:20:20.93 &      -        & 2.15$^{+0.41}_{-0.67}$ &       -      \\
     \hline	 						   			  					  
        6 & a &  6.1 & +13:11:30.90 & -1:19:38.01 & 1.08$\pm$0.33 & 1.22$^{+0.29}_{-0.29}$ &       -      \\
          & b &  6.2 & +13:11:33.50 & -1:20:12.19 & 0.96$\pm$0.94 & 1.31$^{+0.30}_{-0.30}$ &       -      \\
          & c &  6.3 & +13:11:32.90 & -1:19:54.52 &      -        & 0.94$^{+0.25}_{-0.26}$ &       -      \\
          & d &  6.4 & +13:11:32.63 & -1:19:58.88 &      -        & 1.09$^{+0.27}_{-0.27}$ &       -      \\
     \hline	 						   			  					  
        7 & a &  7.1 & +13:11:25.60 & -1:20:51.86 & 4.28$\pm$0.41 & 4.92$^{+0.78}_{-0.78}$ &  4.87$^{5}$  \\
          & b &  7.2 & +13:11:30.82 & -1:20:13.92 & 4.42$\pm$0.14 & 5.20$^{+0.81}_{-0.81}$ &       -      \\
          & c &  7.3 & +13:11:29.97 & -1:20:24.89 &      -        & 0.77$^{+0.23}_{-4.01}$ &       -      \\
     \hline	 						   			  					  
        8 & a &  8.1 & +13:11:32.44 & -1:20:50.93 & 3.10$\pm$0.89 & 2.63$^{+0.48}_{-0.48}$ &       -      \\
          & b &  8.2 & +13:11:31.55 & -1:21:05.56 & 2.94$\pm$1.13 & 2.77$^{+0.50}_{-0.50}$ &       -      \\
          & c &  8.3 & +13:11:31.65 & -1:20:14.10 &      -        & 2.75$^{+0.89}_{-0.49}$ &       -      \\
          & d &  8.4 & +13:11:25.68 & -1:20:20.18 &      -        & 0.70$^{+0.22}_{-0.22}$ &       -      \\
          & e &  8.5 & +13:11:30.48 & -1:20:30.51 &      -        & 0.77$^{+0.23}_{-2.53}$ &       -      \\
     \hline	 						   			  					  
        9 & a &  9.1 & +13:11:30.45 & -1:19:48.67 & 4.30$\pm$0.18 & 4.97$^{+0.78}_{-0.78}$ &       -      \\
          & b &  9.2 & +13:11:33.67 & -1:20:50.35 &      -        & 1.06$^{+0.27}_{-0.27}$ &       -      \\
          & c &  9.3 & +13:11:28.90 & -1:21:15.83 &      -        & 5.16$^{+0.81}_{-0.81}$ &       -      \\
          & d &  9.4 & +13:11:26.42 & -1:20:26.95 & 4.98$\pm$0.30 & 5.17$^{+0.81}_{-0.81}$ &       -      \\
     \hline							   			  					  
       10 & a & 10.1 & +13:11:34.13 & -1:20:50.87 & 2.16$\pm$0.98 & 1.75$^{+0.36}_{-0.74}$ & 1.37$^{5}$   \\
          & b & 10.2 & +13:11:28.20 & -1:20:12.50 &      -        & 1.54$^{+0.33}_{-0.33}$ &       -      \\
          & c & 10.3 & +13:11:29.46 & -1:20:27.76 &      -        & 2.57$^{+0.63}_{-0.47}$ &       -      \\
     \hline							   			  					  
       11 & a & 11.1 & +13:11:33.49 & -1:21:06.77 & 2.96$\pm$0.92 & 2.91$^{+0.51}_{-0.51}$ &       -      \\
          & b & 11.2 & +13:11:29.20 & -1:20:01.31 & 2.76$\pm$2.18 & 2.87$^{+0.51}_{-0.51}$ &       -      \\
          & c & 11.3 & +13:11:29.64 & -1:20:26.40 &      -        & 1.58$^{+0.52}_{-0.73}$ &       -      \\
     \hline							   			  					  
       12 & a & 12.2 & +13:11:27.51 & -1:20:54.90 & 1.45$\pm$0.33 & 1.99$^{+0.39}_{-0.39}$ & 1.82$^{4,5}$ \\
          & b &   -  & +13:11:27.36 & -1:20:51.85 & 2.13$\pm$0.39 &          -             &       -      \\
    \hline\\
    \multicolumn{7}{l}{$^1$ Photometric redshift, this work}\\
    \multicolumn{7}{l}{$^2$ Photometric redshift, \citealt{broadhurst:05}}\\
    \multicolumn{7}{l}{$^3$ Spectroscopic redshift}\\
    \multicolumn{7}{l}{$^4$ Spectroscopic redshift, \citealt{golse:phd}}\\
    \multicolumn{7}{l}{$^5$ Spectroscopic redshift, \citealt{broadhurst:05}}\\
  \end{tabular}
  \label{tab:app:details}
\end{table*}

\begin{table*}
  \centering
  \captcont[]{...continued...}
  \begin{tabular}{cccccccc}
    \hline
    \multicolumn{2}{c}{ID} & ID B05 & RA & Dec & $z_p^1$ & $z_{br}^2$ & $z_s^3$ \\
    \hline
    \hline
       13 & a &   -  & +13:11:33.42 & -1:20:44.40 & 2.40$\pm$0.66 &          -             &       -      \\
          & b &   -  & +13:11:26.65 & -1:20:22.12 & 0.52$\pm$1.27 &          -             &       -      \\
          & c & 12.1 & +13:11:30.50 & -1:19:51.45 &      -        & 1.87$^{+0.38}_{-0.38}$ & 1.82$^{4,5}$ \\
          & d & 12.4 & +13:11:29.11 & -1:21:10.31 & 1.98$\pm$0.51 & 1.92$^{+0.38}_{-0.38}$ &       -      \\
     \hline							    		    	    				
       14 & a & 13.1 & +13:11:32.97 & -1:19:24.39 &      -        & 1.02$^{+0.27}_{-0.28}$ &       -      \\
          & b & 13.2 & +13:11:33.13 & -1:19:25.85 &      -        & 0.72$^{+0.23}_{-0.23}$ &       -      \\
          & c & 13.3 & +13:11:33.54 & -1:19:31.15 & 1.36$\pm$0.18 & 1.10$^{+0.28}_{-0.28}$ &       -      \\
     \hline							    		    	    				
       15 & a & 14.1 & +13:11:29.18 & -1:21:41.82 & 3.50$\pm$0.35 & 3.37$^{+0.82}_{-0.57}$ &       -      \\
          & b & 14.2 & +13:11:29.60 & -1:21:42.65 &      -        & 3.64$^{+0.61}_{-0.61}$ &       -      \\
     \hline							    		    	    				
       16 & a & 15.1 & +13:11:28.22 & -1:20:15.21 &      -        & 1.99$^{+0.39}_{-0.39}$ &       -      \\
          & b & 15.2 & +13:11:34.22 & -1:20:51.33 &      -        & 2.00$^{+0.39}_{-0.39}$ &       -      \\
          & c & 15.3 & +13:11:29.38 & -1:20:27.59 &      -        & 1.97$^{+0.43}_{-0.39}$ &       -      \\
     \hline							    		    	    				
       17 & a & 16.1 & +13:11:28.13 & -1:20:25.34 & 2.28$\pm$0.48 & 1.81$^{+0.37}_{-0.37}$ &       -      \\
          & b & 16.2 & +13:11:29.06 & -1:20:28.57 &      -        & 2.26$^{+0.43}_{-0.43}$ &       -      \\
          & c & 16.3 & +13:11:34.54 & -1:20:46.42 &      -        & 1.80$^{+0.37}_{-0.71}$ &       -      \\
     \hline							    		    	    				
       18 & a & 17.1 & +13:11:30.80 & -1:20:24.91 &      -        & 2.74$^{+0.49}_{-0.49}$ &       -      \\
          & b & 17.2 & +13:11:30.54 & -1:20:27.79 &      -        & 2.02$^{+0.40}_{-0.40}$ &       -      \\
          & c & 17.3 & +13:11:25.13 & -1:20:41.89 & 2.64$\pm$0.67 & 2.25$^{+0.43}_{-0.43}$ &       -      \\
     \hline							    		    	    				
       19 & a & 18.1 & +13:11:28.39 & -1:20:09.56 & 2.44$\pm$0.55 & 2.56$^{+0.47}_{-0.47}$ &       -      \\
          & b & 18.2 & +13:11:33.97 & -1:20:54.56 & 2.30$\pm$0.48 &  -                     &       -      \\
          & c & 18.3 & +13:11:29.51 & -1:20:27.41 &      -        & 1.58$^{+0.52}_{-0.73}$ &       -      \\
     \hline							    		    	    				
       20 & a & 19.1 & +13:11:31.78 & -1:20:22.61 &      -        & 1.72$^{+0.36}_{-0.36}$ &       -      \\
          & b & 19.2 & +13:11:25.39 & -1:20:20.03 &      -        & 2.74$^{+0.49}_{-0.49}$ &       -      \\
          & c & 19.3 & +13:11:32.10 & -1:20:59.33 &      -        & 1.57$^{+0.34}_{-0.34}$ &       -      \\
          & d & 19.4 & +13:11:32.20 & -1:20:57.15 & 3.28$\pm$0.52 & 2.58$^{+0.47}_{-0.47}$ &       -      \\
          & e & 19.5 & +13:11:30.36 & -1:20:33.98 &      -        & 4.54$^{+0.73}_{-1.66}$ &       -      \\
     \hline							    		    	    				
       21 & a & 21.1 & +13:11:31.17 & -1:20:45.80 & 1.94$\pm$0.37 & 1.79$^{+0.37}_{-0.37}$ &       -      \\
          & b & 21.2 & +13:11:30.95 & -1:20:44.76 &      -        & 1.59$^{+0.34}_{-0.34}$ &       -      \\
          & c & 21.3 & +13:11:25.40 & -1:20:11.23 & 1.78$\pm$1.04 & 1.78$^{+0.36}_{-0.36}$ &       -      \\
     \hline							    		    	    				
       22 & a & 22.1 & +13:11:29.83 & -1:20:08.81 & 1.97$\pm$0.44 & 1.99$^{+0.39}_{-0.39}$ &       -      \\
          & b & 22.2 & +13:11:29.76 & -1:20:23.78 &      -        & 1.99$^{+0.39}_{-0.59}$ &       -      \\
          & c & 22.3 & +13:11:32.56 & -1:21:15.93 & 2.37$\pm$0.49 & 1.96$^{+0.39}_{-0.39}$ &       -      \\
     \hline							    		    	    				
       23 & a & 23.1 & +13:11:29.68 & -1:20:10.04 & 2.03$\pm$0.37 & 2.03$^{+0.40}_{-0.40}$ &       -      \\
          & b & 23.2 & +13:11:29.70 & -1:20:22.91 &      -        & 1.99$^{+0.39}_{-0.62}$ &       -      \\
          & c & 23.3 & +13:11:32.80 & -1:21:15.22 &      -        & 2.00$^{+0.39}_{-0.39}$ &       -      \\
     \hline							    		    	    				
       24 & a & 24.1 & +13:11:29.34 & -1:20:56.20 & 3.04$\pm$1.18 & 2.63$^{+0.48}_{-0.48}$ &       -      \\
          & b & 24.2 & +13:11:32.21 & -1:19:50.58 &      -        & 2.50$^{+0.46}_{-0.46}$ &       -      \\
          & c & 24.3 & +13:11:30.44 & -1:19:34.16 & 3.04$\pm$0.89 & 2.43$^{+0.45}_{-0.45}$ &       -      \\
          & d & 24.4 & +13:11:33.87 & -1:20:19.88 & 2.84$\pm$1.35 & 2.81$^{+0.69}_{-0.50}$ &       -      \\
          & e & 24.5 & +13:11:29.78 & -1:20:37.02 &      -        & 4.55$^{+0.73}_{-0.80}$ &       -      \\
     \hline							    		    	    				
       25 & a & 25.1 & +13:11:28.64 & -1:20:35.01 &      -        & 4.59$^{+0.73}_{-0.73}$ &       -      \\
          & b & 25.2 & +13:11:34.80 & -1:20:33.59 & 3.38$\pm$1.73 & 4.42$^{+0.71}_{-0.71}$ &       -      \\
    \hline\\
    \multicolumn{7}{l}{$^1$ Photometric redshift, this work}\\
    \multicolumn{7}{l}{$^2$ Photometric redshift, \citealt{broadhurst:05}}\\
    \multicolumn{7}{l}{$^3$ Spectroscopic redshift}\\
    \multicolumn{7}{l}{$^4$ Spectroscopic redshift, \citealt{golse:phd}}\\
    \multicolumn{7}{l}{$^5$ Spectroscopic redshift, \citealt{broadhurst:05}}\\
  \end{tabular}
  \label{tab:app:details2}
\end{table*}

\begin{table*}
  \centering
  \captcont[]{...continued}
  \begin{tabular}{cccccccc}
    \hline
    \multicolumn{2}{c}{ID} & ID B05 & RA & Dec & $z_p^1$ & $z_{br}^2$ & $z_s^3$ \\
    \hline
    \hline
       26 & a & 26.1 & +13:11:25.30 & -1:20:32.78 & 1.42$\pm$0.75 & 1.08$^{+0.27}_{-0.39}$ &       -      \\
          & b & 26.2 & +13:11:31.47 & -1:20:25.26 &      -        & 1.04$^{+0.27}_{-0.27}$ &       -      \\
          & c & 26.3 & +13:11:30.39 & -1:20:32.61 &      -        & 0.77$^{+0.23}_{-2.53}$ &       -      \\
     \hline							    		    	    				
       27 & a & 27.1 & +13:11:25.32 & -1:20:33.13 & 1.42$\pm$0.75 & 1.81$^{+0.37}_{-0.37}$ &       -      \\
          & b & 27.2 & +13:11:31.51 & -1:20:24.66 &      -        & 1.58$^{+0.34}_{-0.48}$ &       -      \\
          & c & 27.3 & +13:11:30.34 & -1:20:32.92 &      -        & 4.55$^{+0.73}_{-1.63}$ &       -      \\
     \hline							    		    	    				
       28 & a & 28.1 & +13:11:28.45 & -1:20:10.93 &      -        & 1.17$^{+0.29}_{-4.29}$ &       -      \\
          & b & 28.2 & +13:11:34.41 & -1:21:00.02 &      -        & 2.00$^{+1.23}_{-0.43}$ &       -      \\
     \hline							    		    	    				
       29 & a & 29.1 & +13:11:29.37 & -1:20:57.93 &      -        & 2.47$^{+0.46}_{-0.57}$ &       -      \\
          & b & 29.2 & +13:11:30.18 & -1:19:34.23 &      -        & 3.40$^{+0.58}_{-0.58}$ &       -      \\
          & c & 29.3 & +13:11:32.29 & -1:19:52.58 &      -        & 2.50$^{+0.46}_{-0.46}$ &       -      \\
          & d & 29.4 & +13:11:33.77 & -1:20:20.83 &      -        & 3.35$^{+0.57}_{-0.57}$ &       -      \\
          & e & 29.5 & +13:11:29.88 & -1:20:36.62 &      -        & 4.59$^{+0.73}_{-1.66}$ &       -      \\
     \hline							    		    	    				
       30 & a & 30.1 & +13:11:32.57 & -1:19:19.84 & 3.39$\pm$0.13 & 4.49$^{+0.72}_{-0.72}$ &       -      \\
          & b & 30.2 & +13:11:33.33 & -1:19:26.08 & 3.16$\pm$1.05 & 3.23$^{+0.76}_{-0.56}$ &       -      \\
          & c & 30.3 & +13:11:33.80 & -1:19:32.71 & 3.50$\pm$0.21 & 3.30$^{+0.56}_{-0.56}$ &       -      \\
     \hline
       31 & a &   -  & +13:11:31.82 & -1:19:47.34 & 2.62$\pm$0.63 &          -             &       -      \\
          & b &   -  & +13:11:31.71 & -1:19:45.97 & 1.52$\pm$0.36 &          -             &       -      \\
     \hline
       32 & a &   -  & +13:11:33.59 & -1:20:05.99 &      -        &          -             &       -      \\
          & b &   -  & +13:11:33.58 & -1:20:05.49 &      -        &          -             &       -      \\
          & c &   -  & +13:11:33.56 & -1:20:04.93 &      -        &          -             &       -      \\
          & d &   -  & +13:11:33.55 & -1:20:04.40 &      -        &          -             &       -      \\
          & e &   -  & +13:11:33.53 & -1:20:03.64 &      -        &          -             &       -      \\
          & f &   -  & +13:11:33.54 & -1:20:04.00 &      -        &          -             &       -      \\
          & g &   -  & +13:11:33.52 & -1:20:03.16 &      -        &          -             &       -      \\
          & h &   -  & +13:11:33.51 & -1:20:02.72 &      -        &          -             &       -      \\
          & i &   -  & +13:11:33.51 & -1:20:02.29 &      -        &          -             &       -      \\
          & j &   -  & +13:11:33.50 & -1:20:01.80 &      -        &          -             &       -      \\
          & k &   -  & +13:11:33.48 & -1:20:00.80 &      -        &          -             &       -      \\
          & l &   -  & +13:11:33.47 & -1:20:00.31 &      -        &          -             &       -      \\
          & m &   -  & +13:11:33.41 & -1:19:57.01 &      -        &          -             &       -      \\
          & n &   -  & +13:11:33.42 & -1:19:57.44 &      -        &          -             &       -      \\
          & o &   -  & +13:11:33.43 & -1:19:57.88 &      -        &          -             &       -      \\
          & p &   -  & +13:11:33.44 & -1:19:58.37 &      -        &          -             &       -      \\
          & q &   -  & +13:11:33.45 & -1:19:58.75 &      -        &          -             &       -      \\
          & r &   -  & +13:11:33.45 & -1:19:59.25 &      -        &          -             &       -      \\
          & s &   -  & +13:11:33.47 & -1:19:59.81 &      -        &          -             &       -      \\
    \hline\\
    \multicolumn{7}{l}{$^1$ Photometric redshift, this work}\\
    \multicolumn{7}{l}{$^2$ Photometric redshift, \citealt{broadhurst:05}}\\
    \multicolumn{7}{l}{$^3$ Spectroscopic redshift}\\
  \end{tabular}
  \label{tab:app:detail3}
\end{table*}

\clearpage

\section{multiple images}
\label{app:stamps}

In this section of the appendix we show image stamps for all multiple
images in the different image systems. Observed images are shown on
the top row of each figure. Model predictions are shown in the second
and third rows for NSIE and ENFW Models \Roman{two} respectively
except for the source image for which we show the unlensed image
instead. The images used as sources are indicated with a \# on the
observed image. The thumbnails for the galaxies in the image plane
have a box width of 2'', those in the source plane have a box width of
1''. The scale of the images are also marked on the bottom right
corner of the images used as sources. To make the model predictions of
the images we have found a mapping from one region of the image plane
(i) to another (s) via the source plane,
$\vec{\theta_s}(\vec{\beta}(\vec{\theta_i},z),z)$. Region i is where
we expect to see an image from region s. In general a pixel from
region i is mapped to a quadrilateral in region s which will overlap
several pixels. We have redistributed the flux from pixels in region s
to the pixel mapped from region i in a way which preserves surface
brightness. This allows us to create an image of region s in region i
which has mesh of square pixels in region i. We have used a three
colour image composed of F850LP (red), F625W (green) and F475W (blue)
for the mapping. For all images the colour cuts are the same expect
for images near bright sources for which we have used only a single
filter (F775W) with the bright source subtracted in order to show the
multiple image more clearly. Image systems 3, 9 and 28 that have very
red colours we show as a grey scale images in filter F850LP.

\begin{table*}
  \caption{Image system 1:}\vspace{0mm}
  \begin{tabular}{cccccccc}
    \multicolumn{1}{m{1cm}}{{\Large A1689}}
    & \multicolumn{1}{m{1.7cm}}{\includegraphics[height=2.00cm,clip]{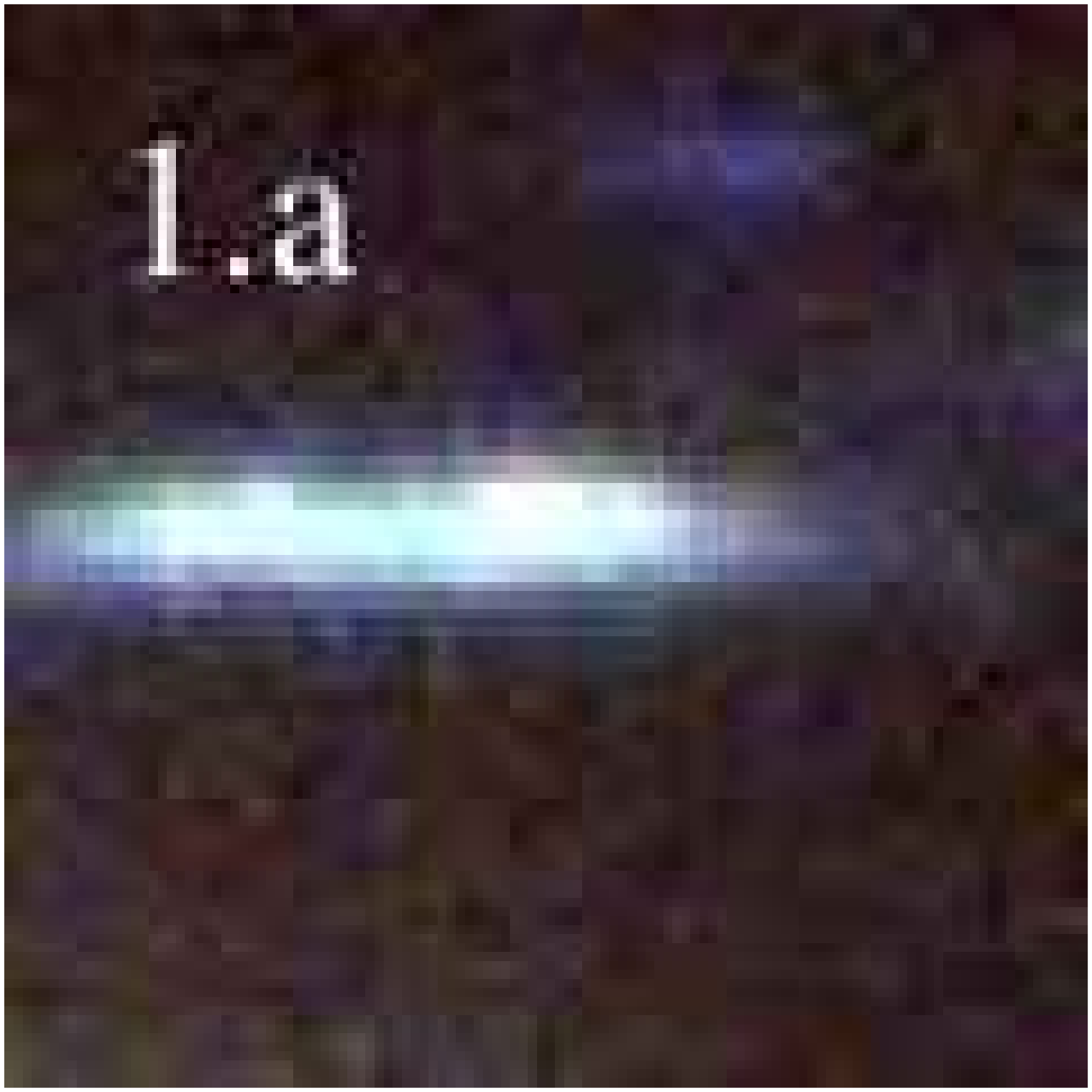}}
    & \multicolumn{1}{m{1.7cm}}{\includegraphics[height=2.00cm,clip]{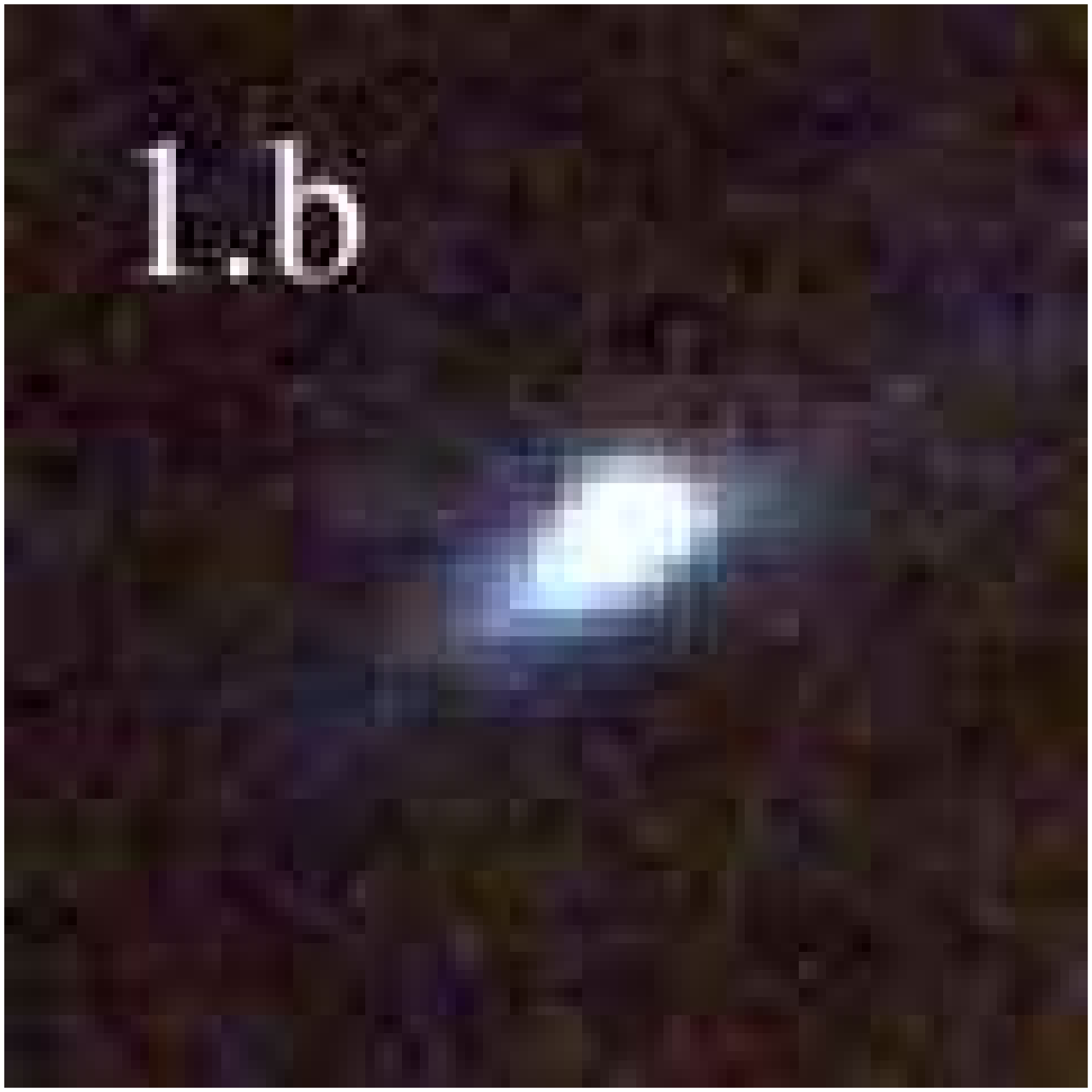}}
    & \multicolumn{1}{m{1.7cm}}{\includegraphics[height=2.00cm,clip]{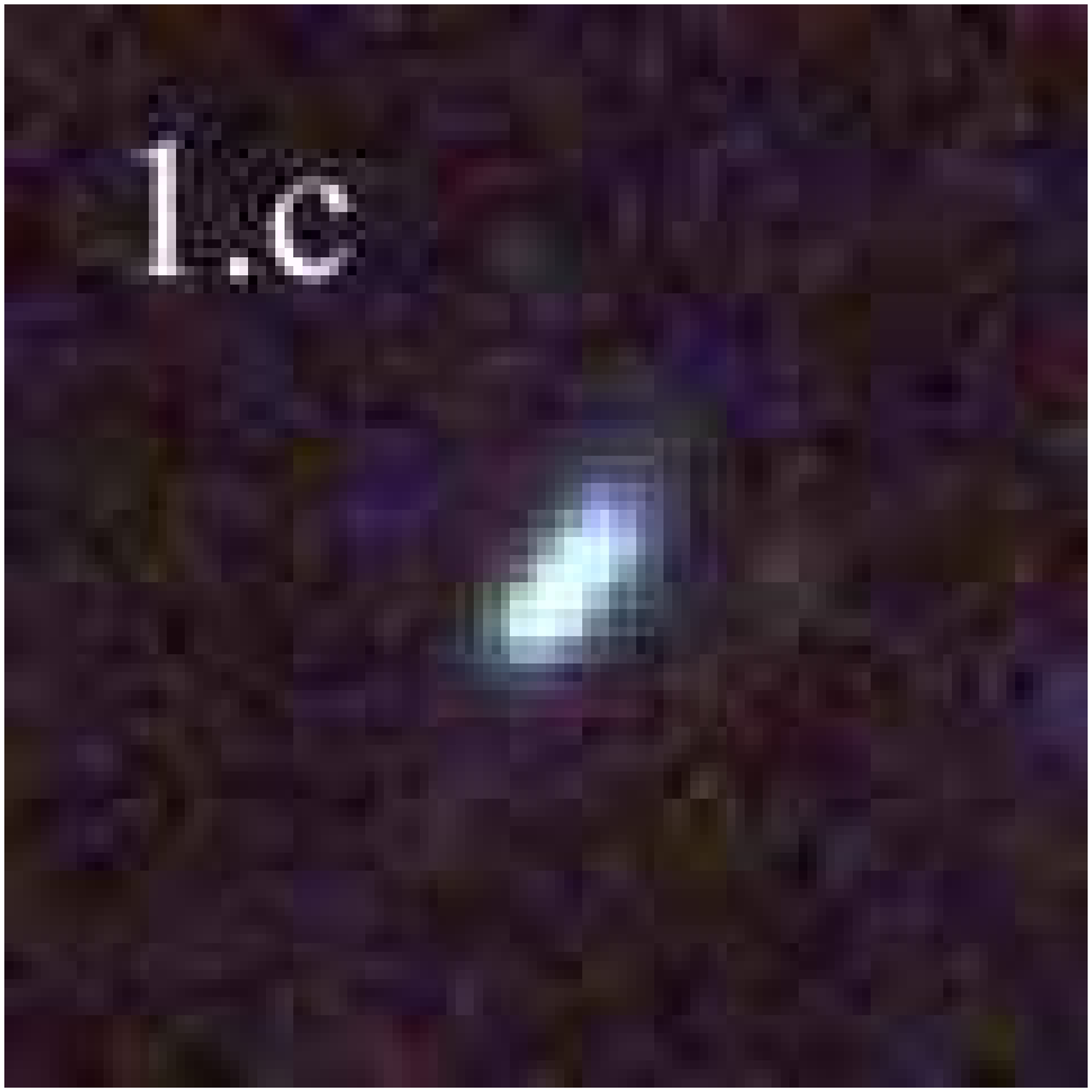}}
    & \multicolumn{1}{m{1.7cm}}{\includegraphics[height=2.00cm,clip]{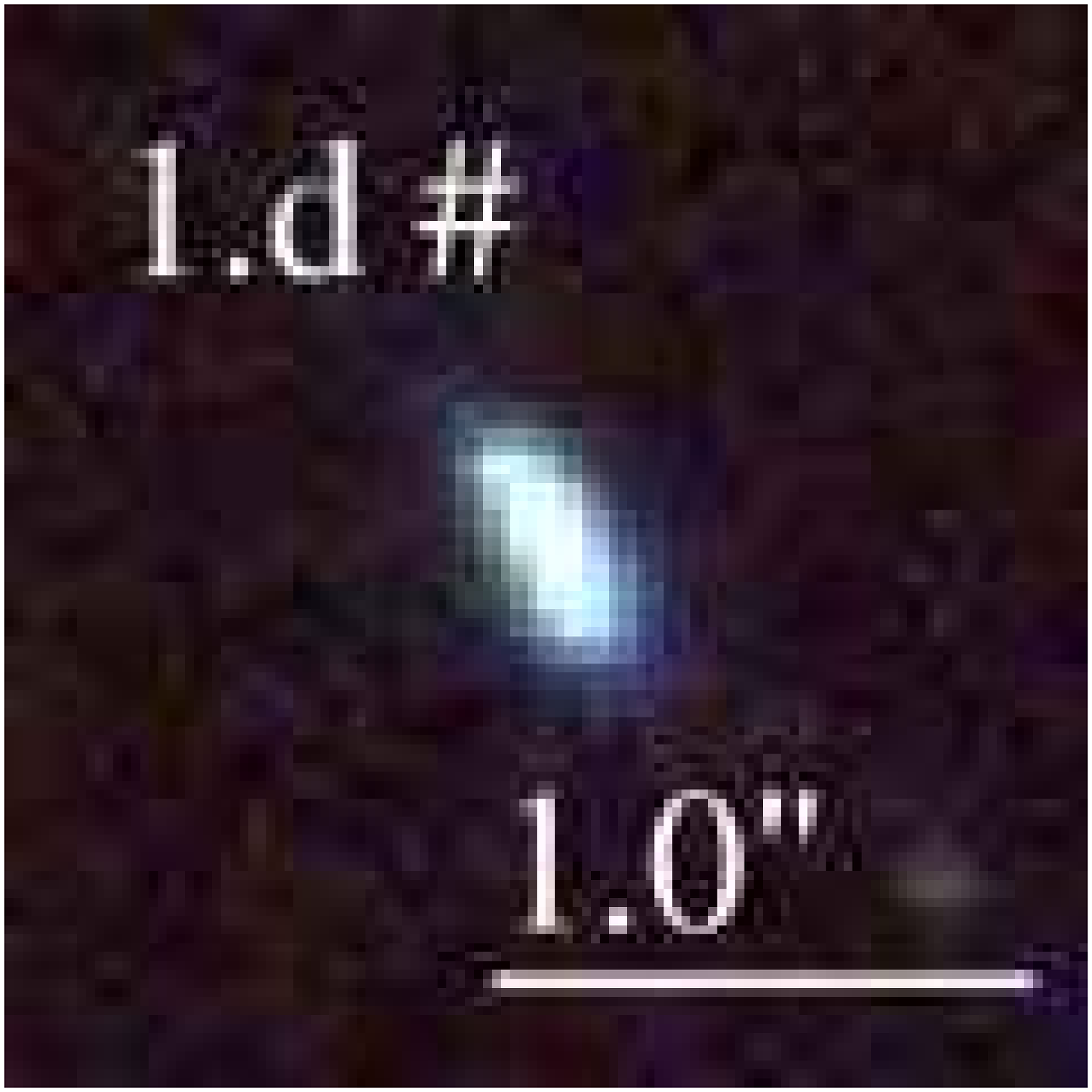}}
    & \multicolumn{1}{m{1.7cm}}{\includegraphics[height=2.00cm,clip]{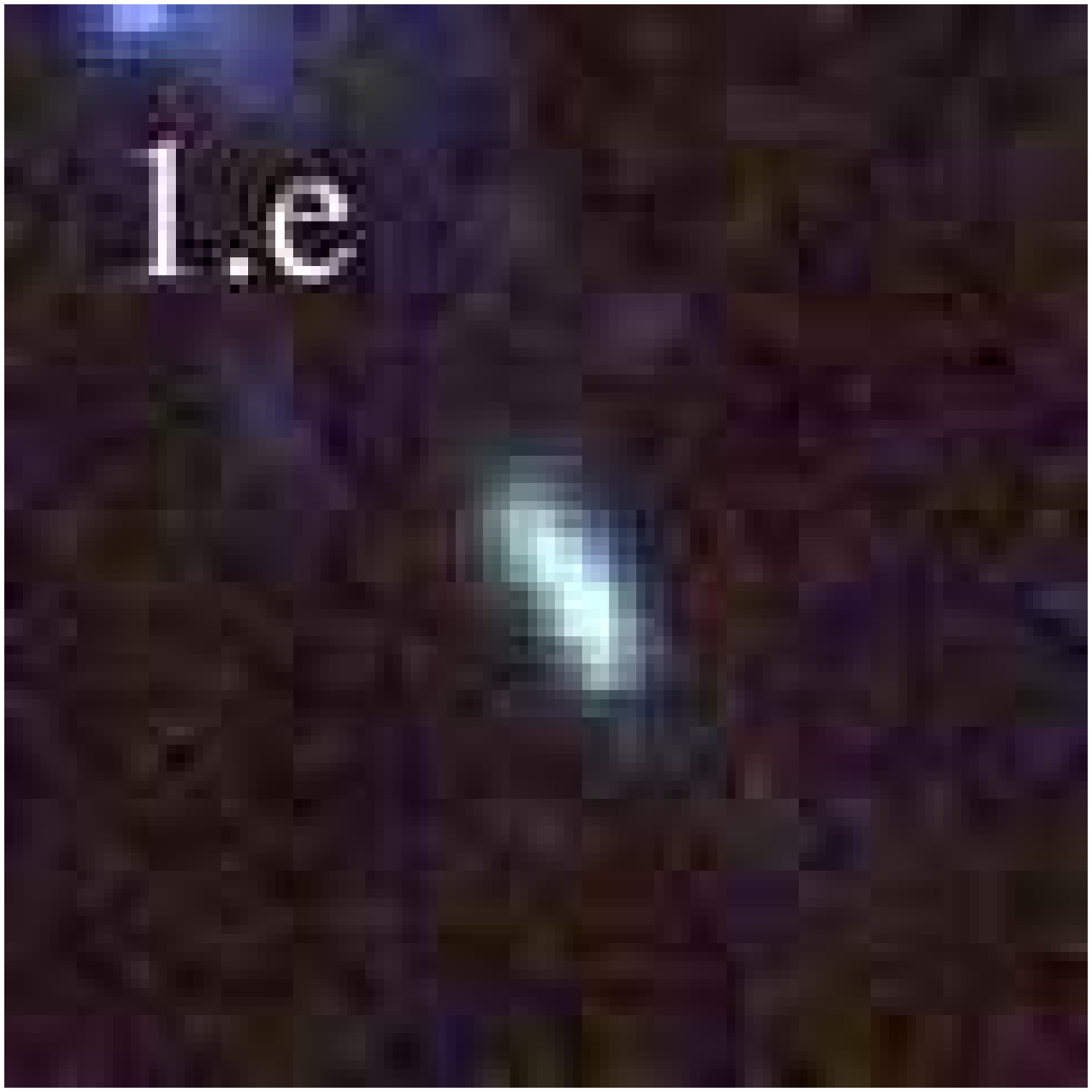}}
    & \multicolumn{1}{m{1.7cm}}{\includegraphics[height=2.00cm,clip]{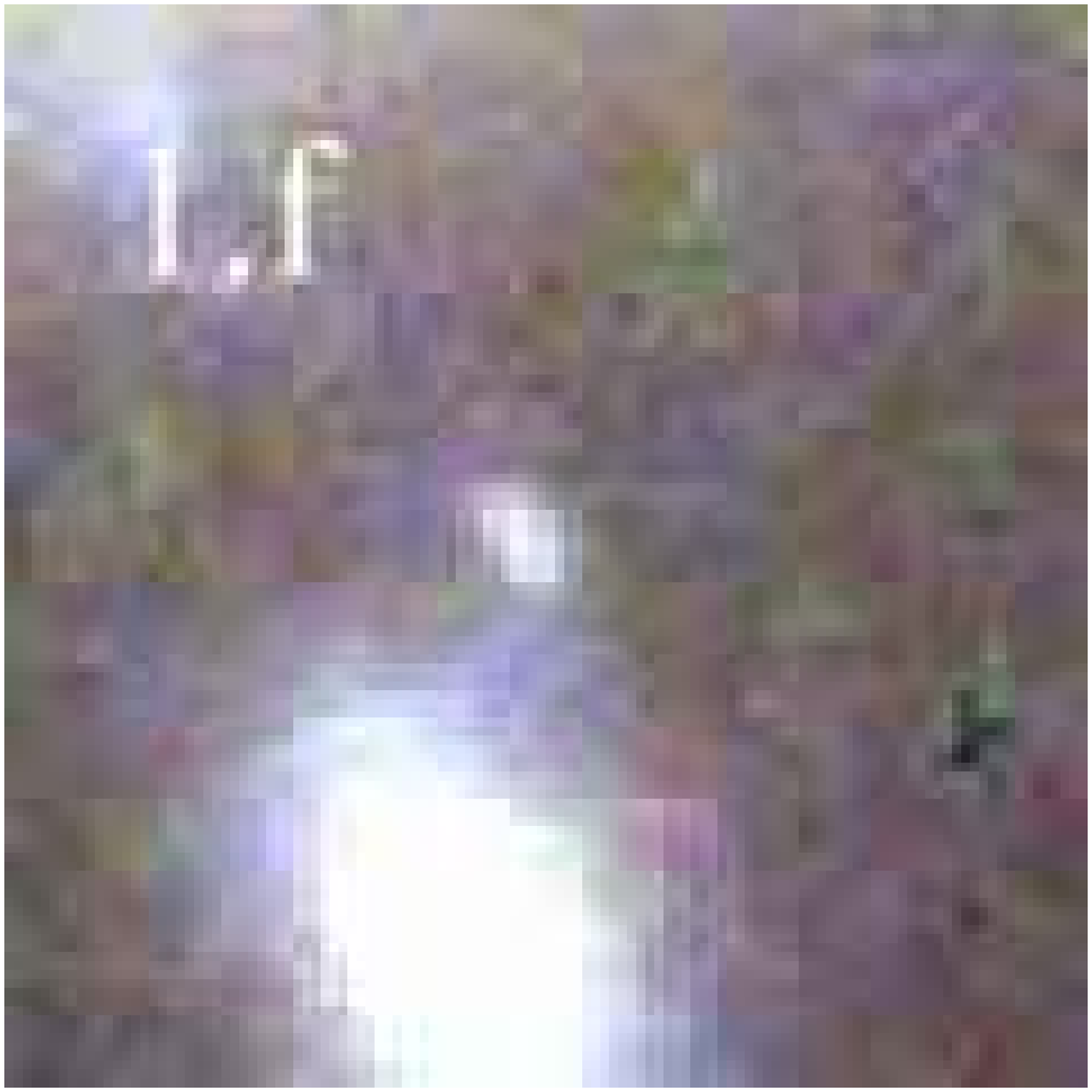}}
    & \multicolumn{1}{m{1.7cm}}{\includegraphics[height=2.00cm,clip]{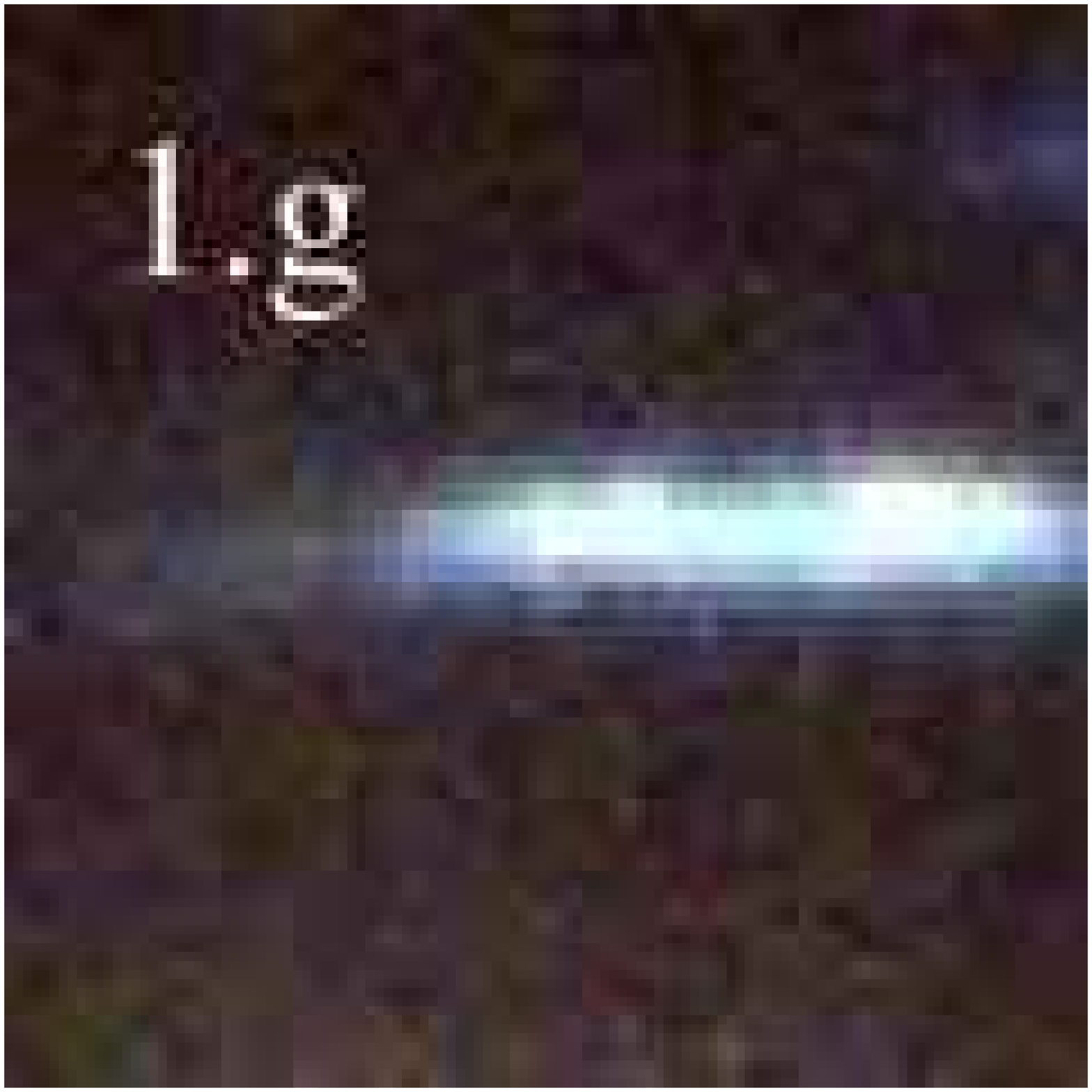}} \\
    \multicolumn{1}{m{1cm}}{{\Large NSIE}}
    & \multicolumn{1}{m{1.7cm}}{\includegraphics[height=2.00cm,clip]{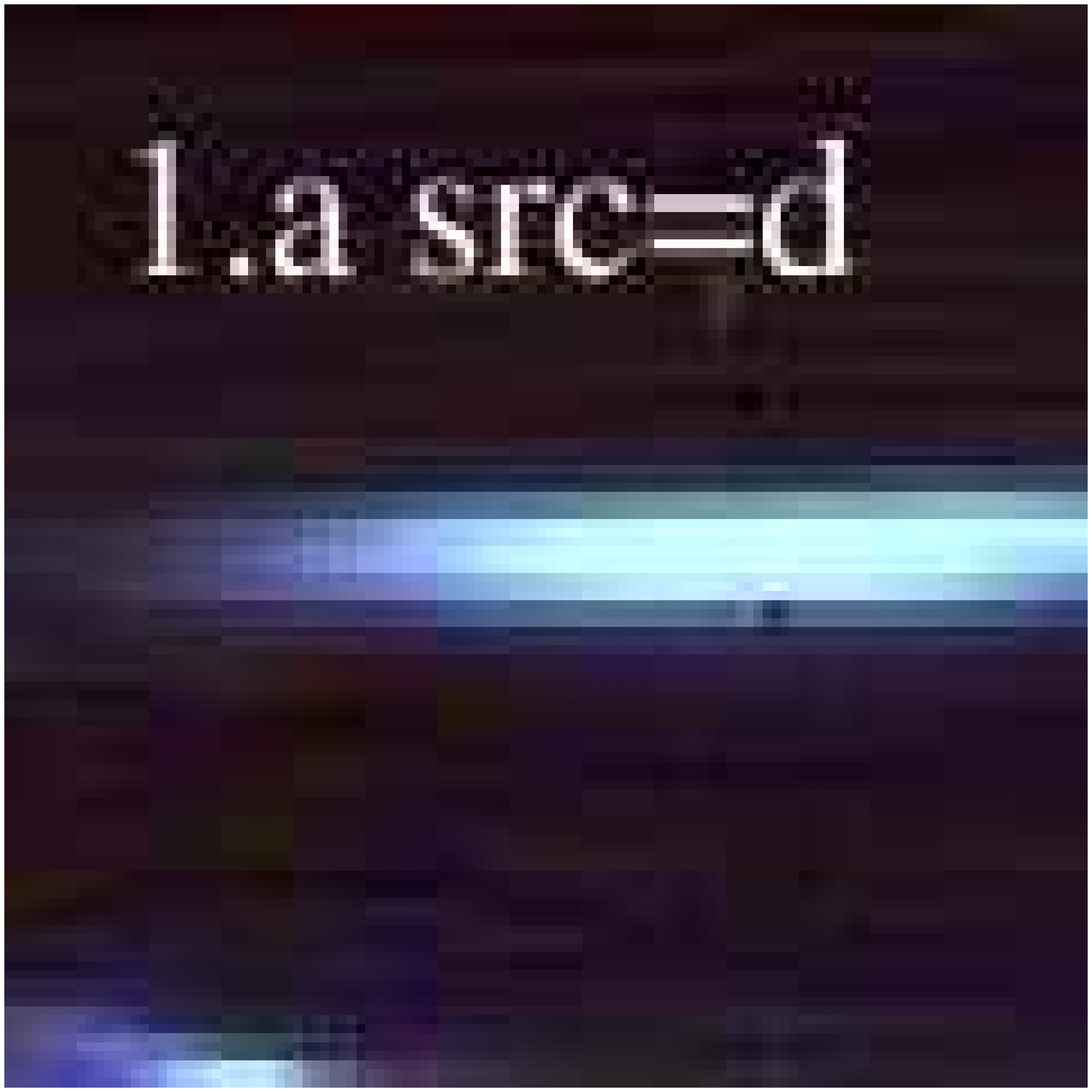}}
    & \multicolumn{1}{m{1.7cm}}{\includegraphics[height=2.00cm,clip]{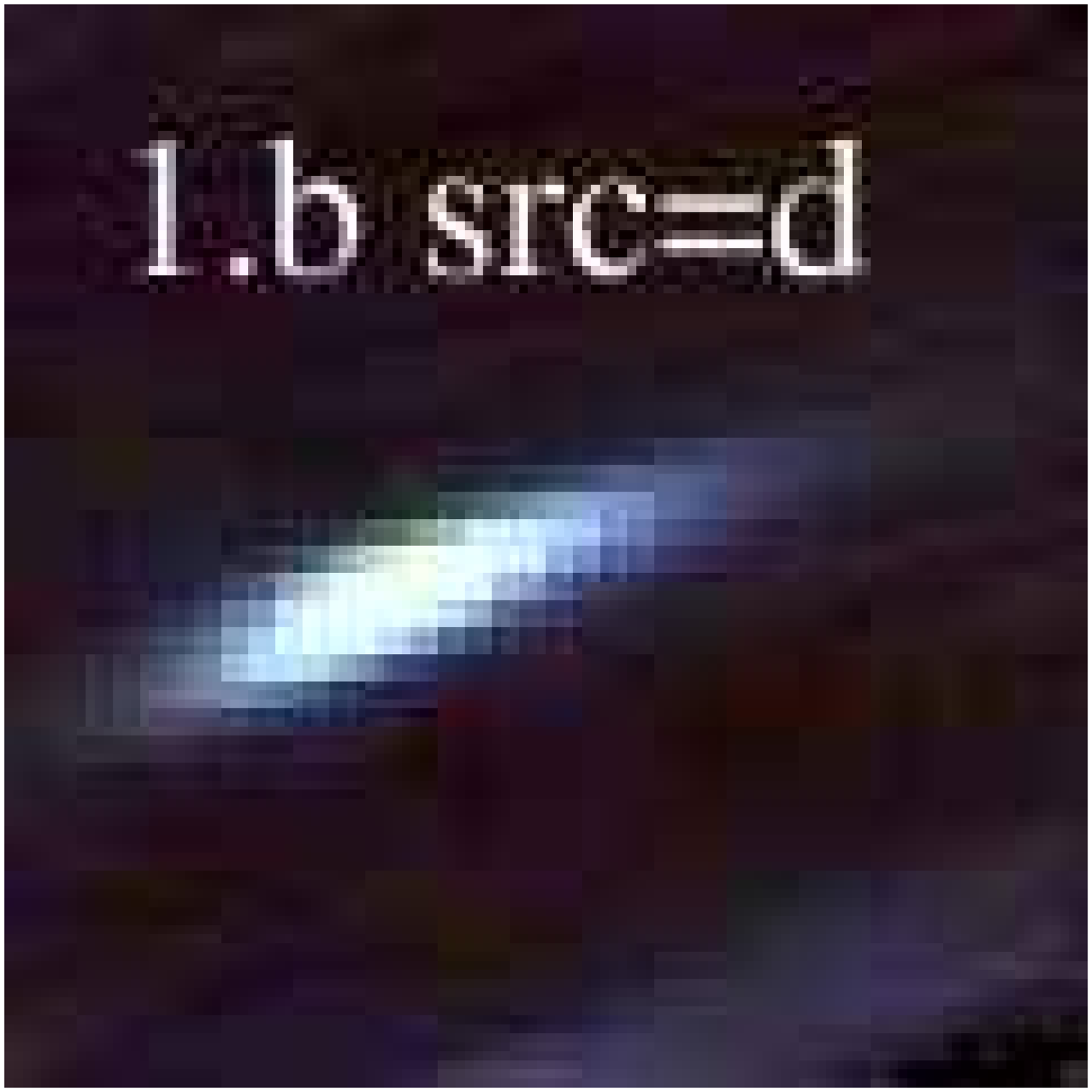}}
    & \multicolumn{1}{m{1.7cm}}{\includegraphics[height=2.00cm,clip]{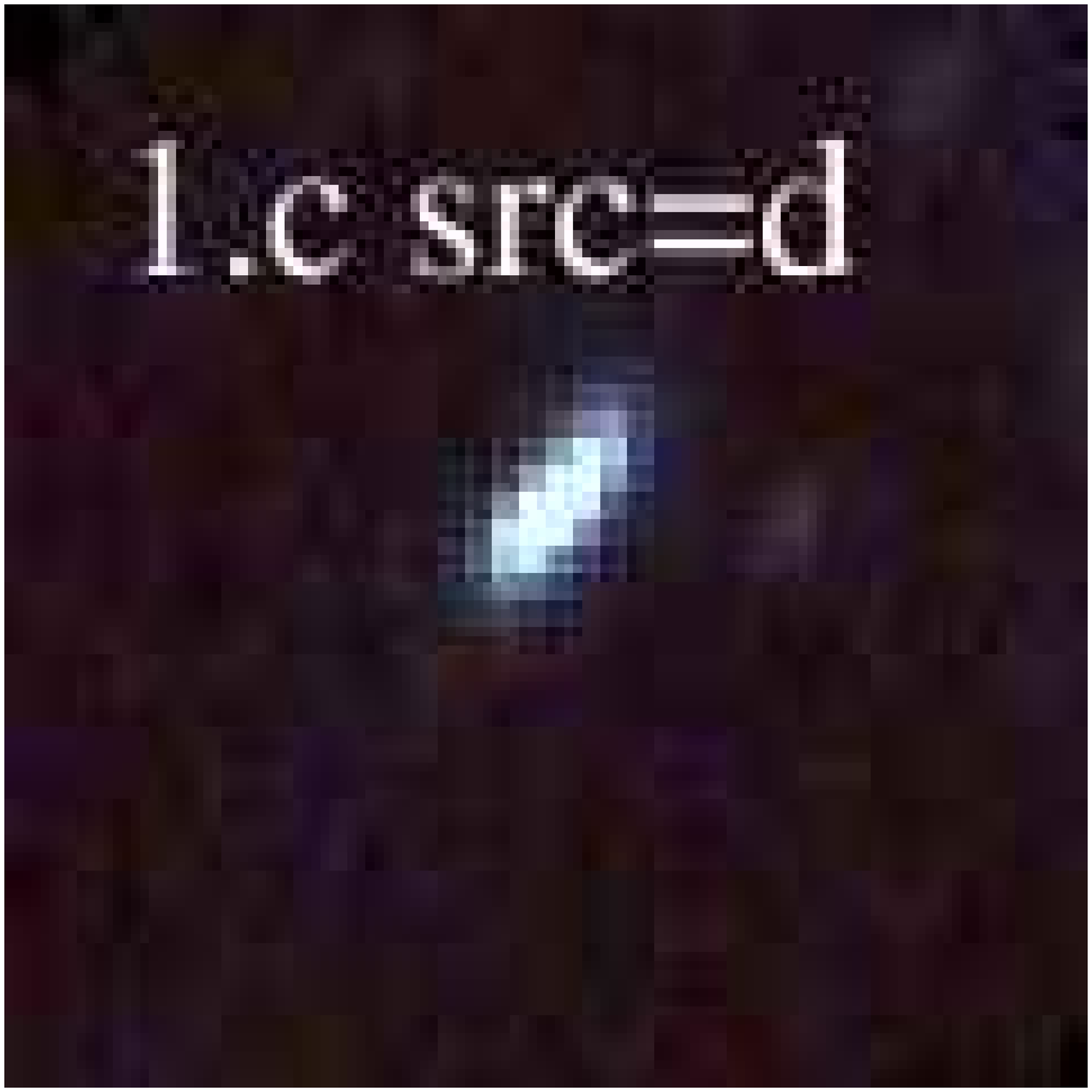}}
    & \multicolumn{1}{m{1.7cm}}{\includegraphics[height=2.00cm,clip]{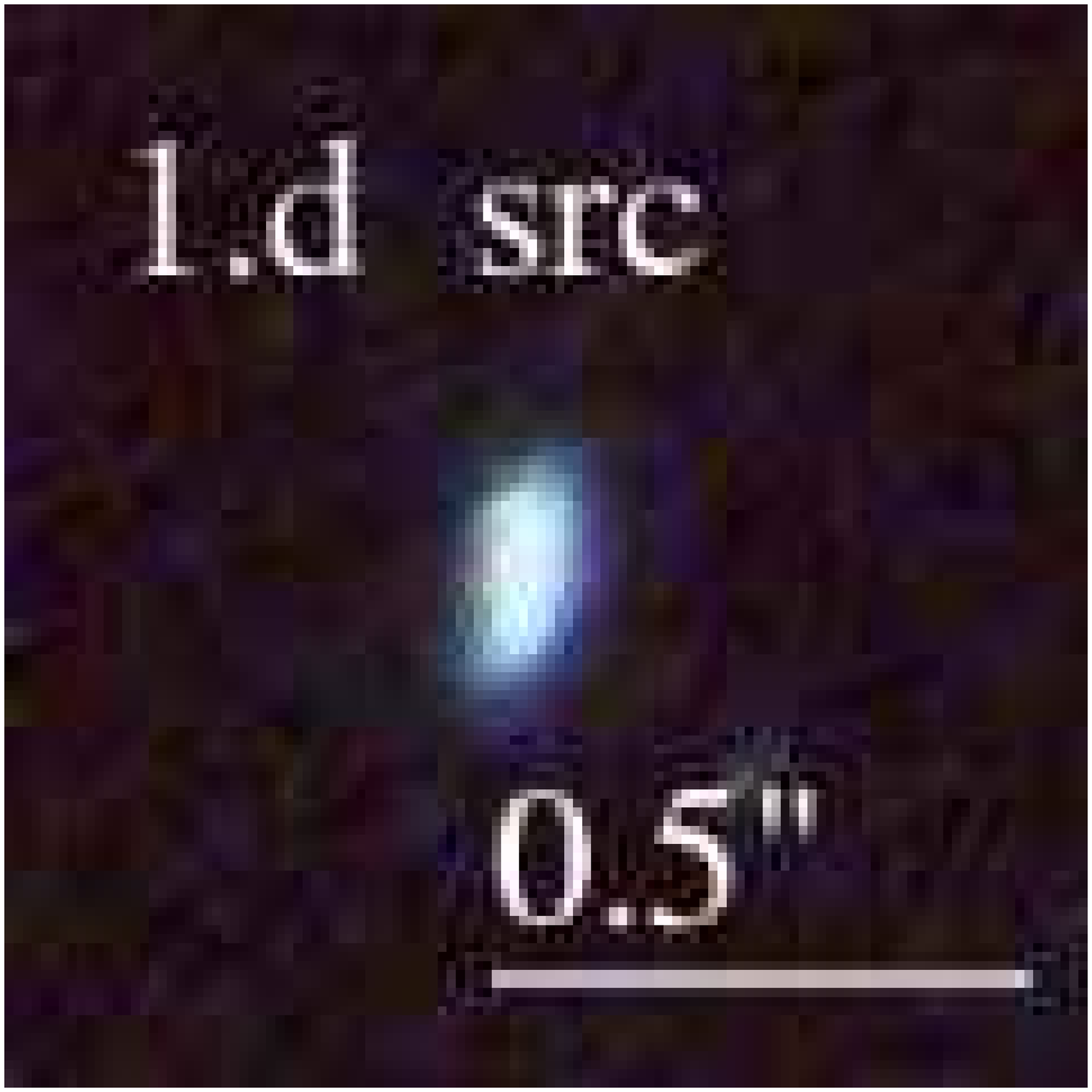}}
    & \multicolumn{1}{m{1.7cm}}{\includegraphics[height=2.00cm,clip]{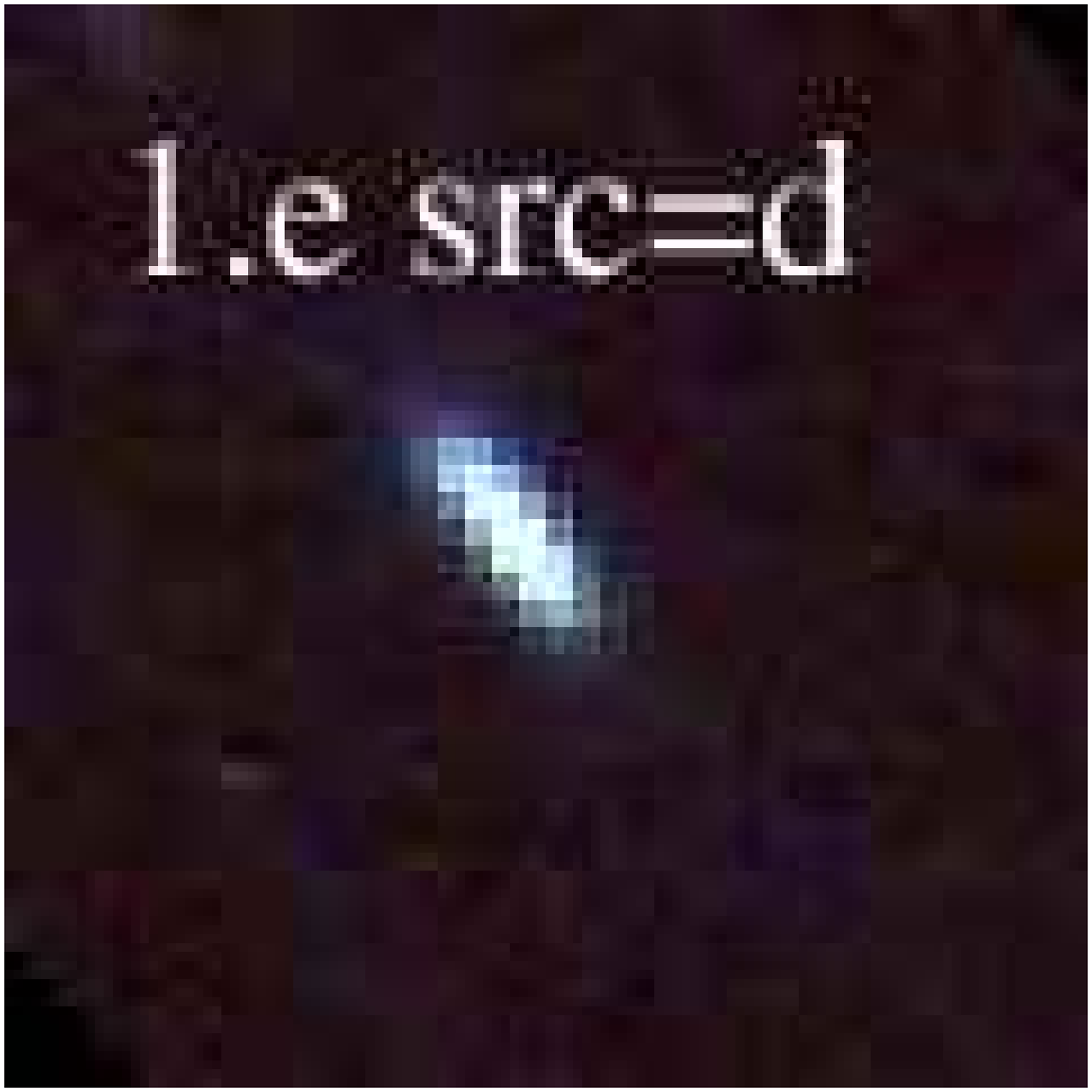}}
    & \multicolumn{1}{m{1.7cm}}{\includegraphics[height=2.00cm,clip]{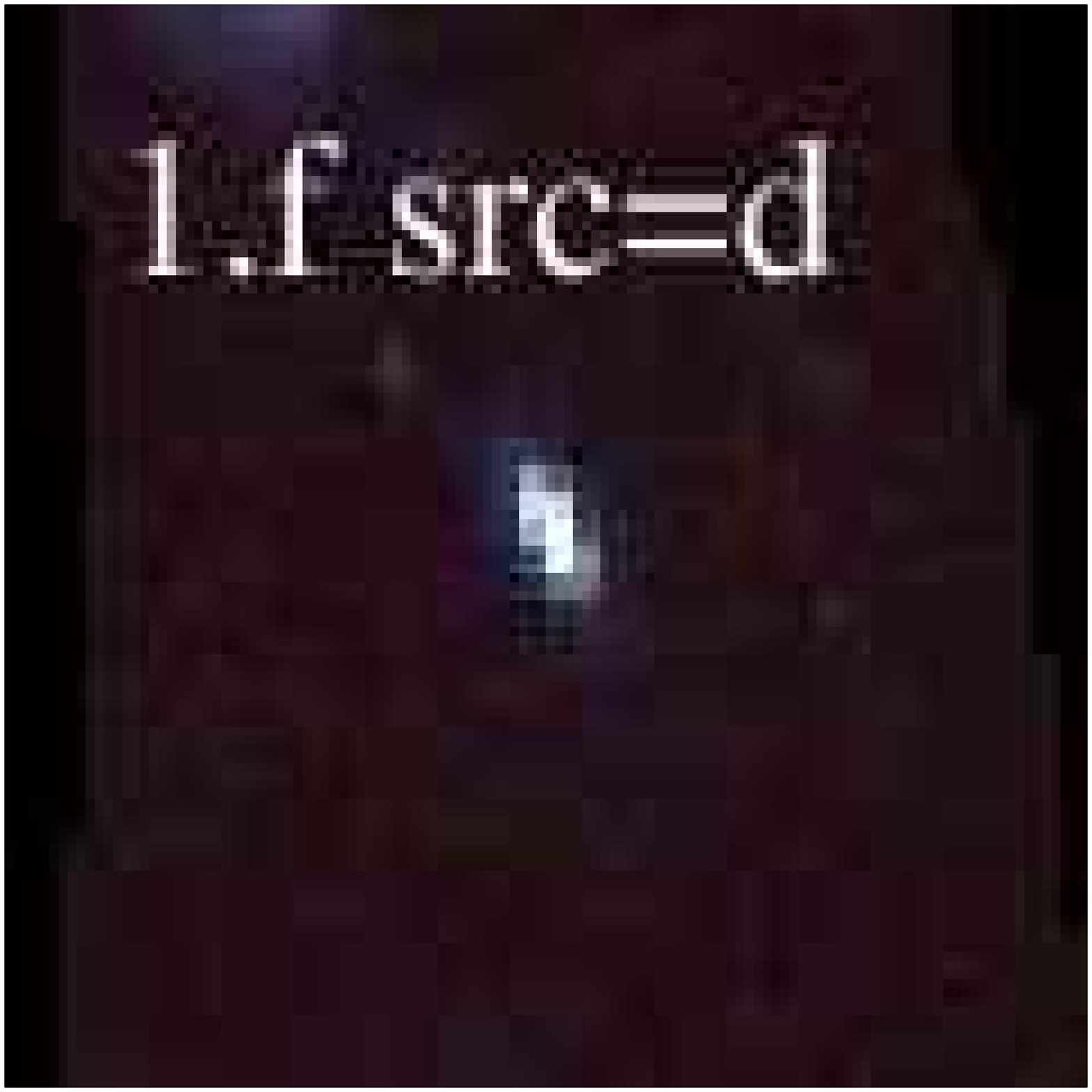}}
    & \multicolumn{1}{m{1.7cm}}{\includegraphics[height=2.00cm,clip]{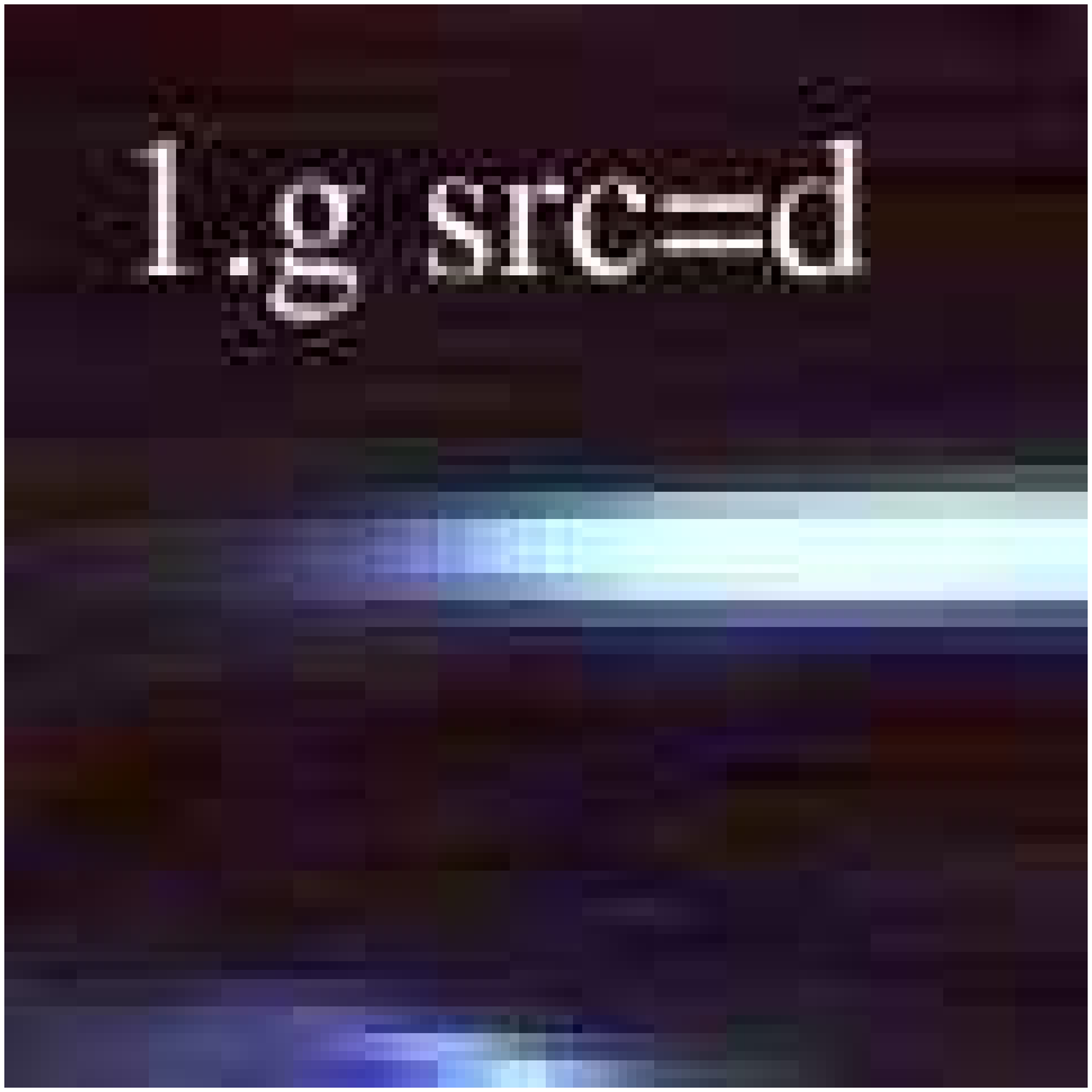}} \\
    \multicolumn{1}{m{1cm}}{{\Large ENFW}}
    & \multicolumn{1}{m{1.7cm}}{\includegraphics[height=2.00cm,clip]{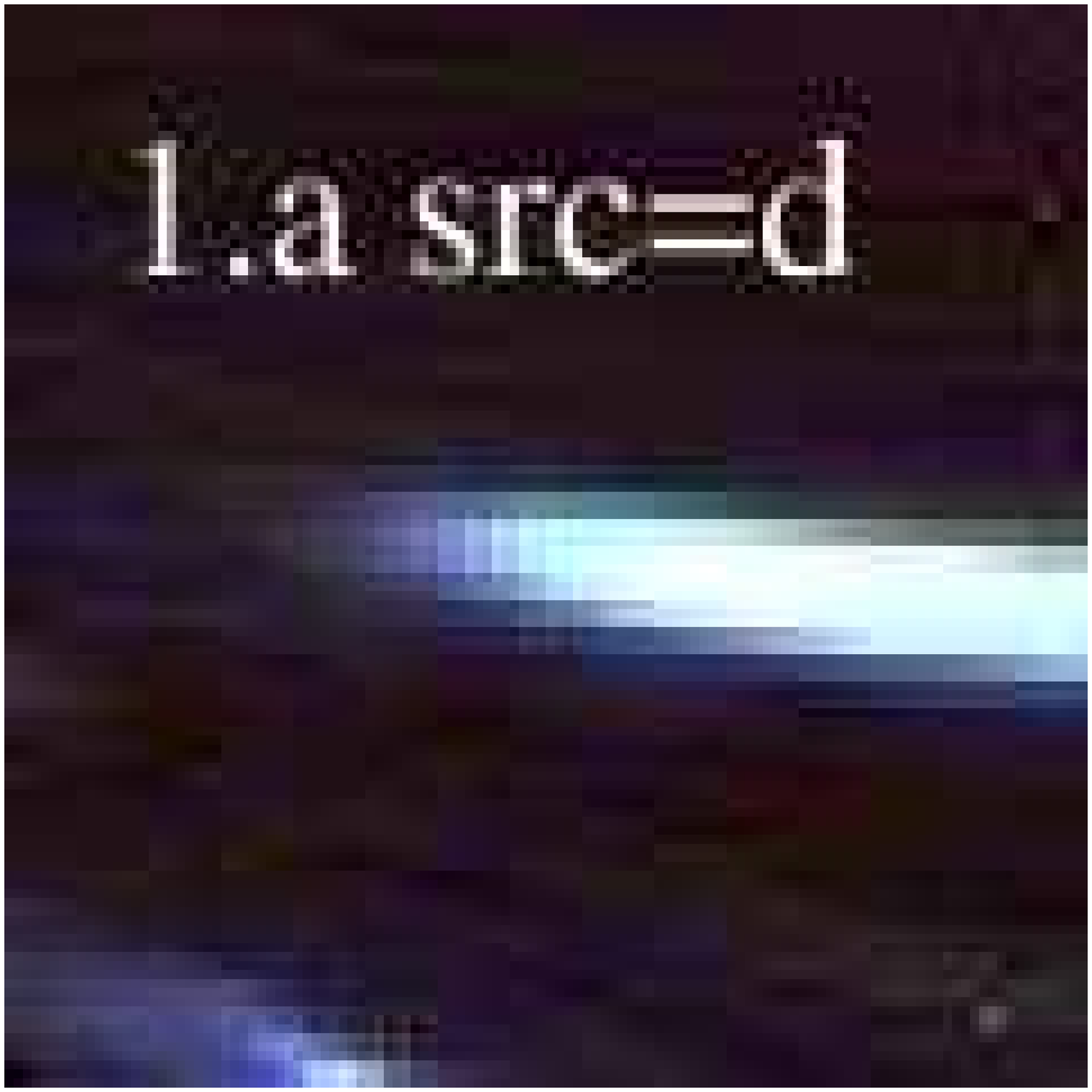}}
    & \multicolumn{1}{m{1.7cm}}{\includegraphics[height=2.00cm,clip]{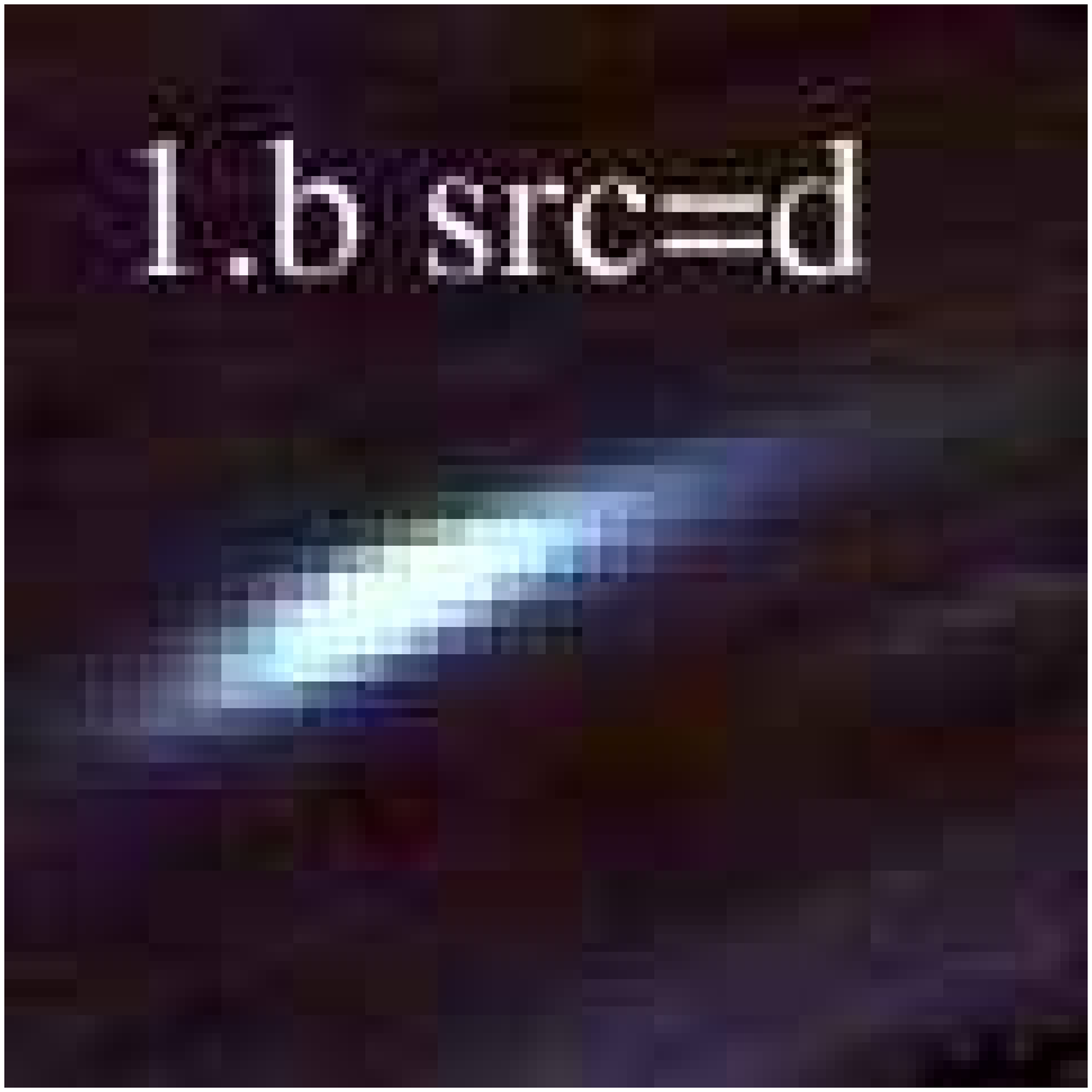}}
    & \multicolumn{1}{m{1.7cm}}{\includegraphics[height=2.00cm,clip]{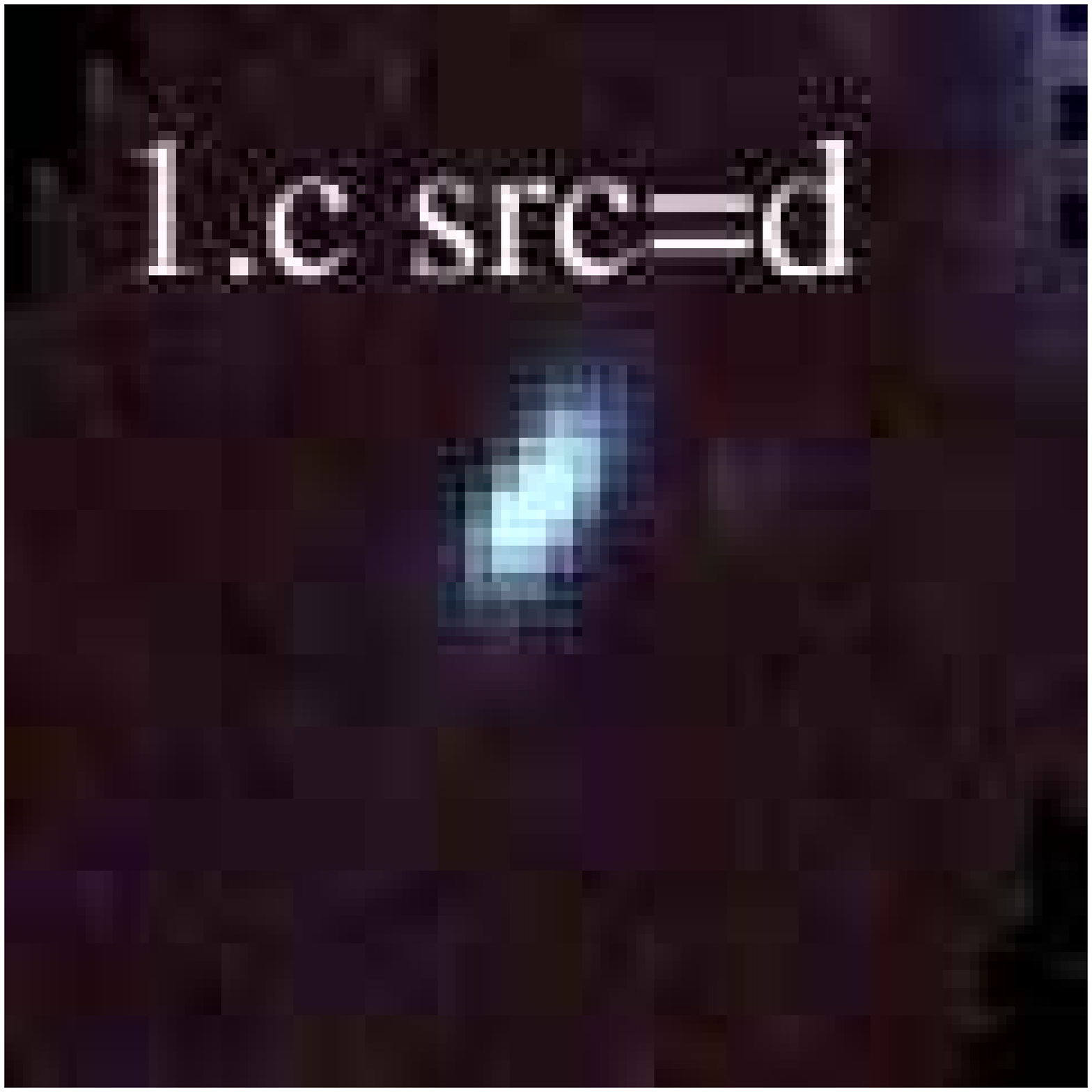}}
    & \multicolumn{1}{m{1.7cm}}{\includegraphics[height=2.00cm,clip]{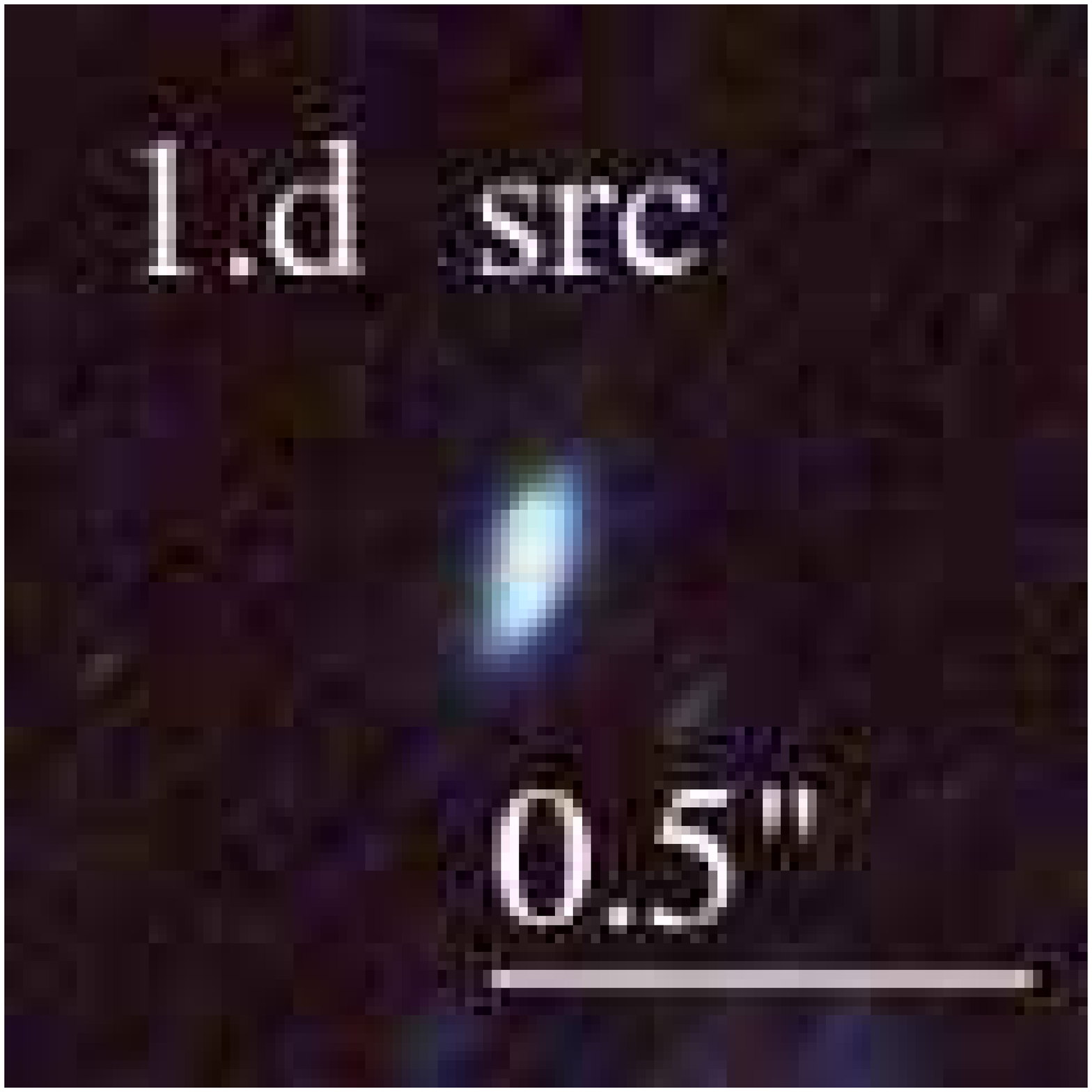}}
    & \multicolumn{1}{m{1.7cm}}{\includegraphics[height=2.00cm,clip]{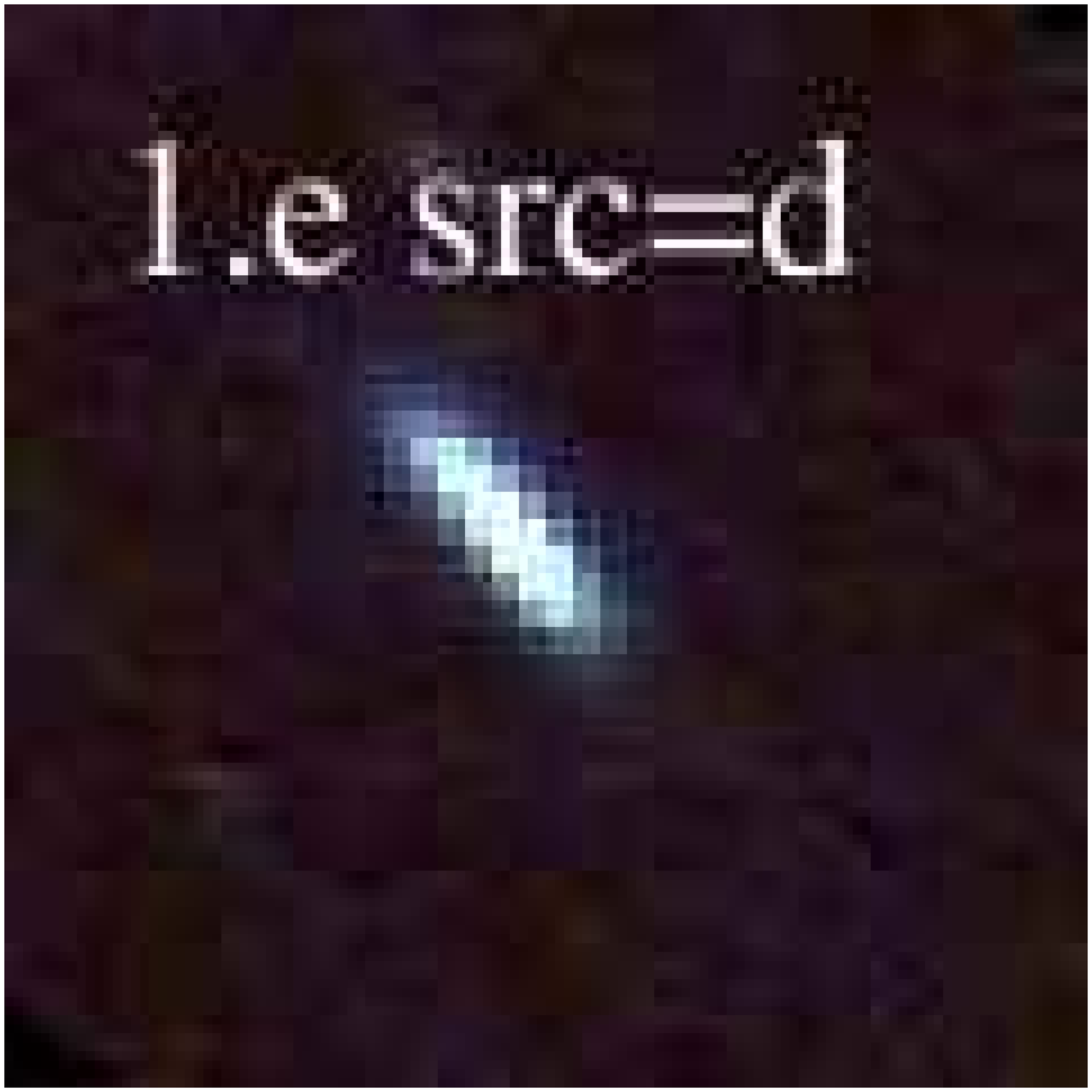}}
    & \multicolumn{1}{m{1.7cm}}{\includegraphics[height=2.00cm,clip]{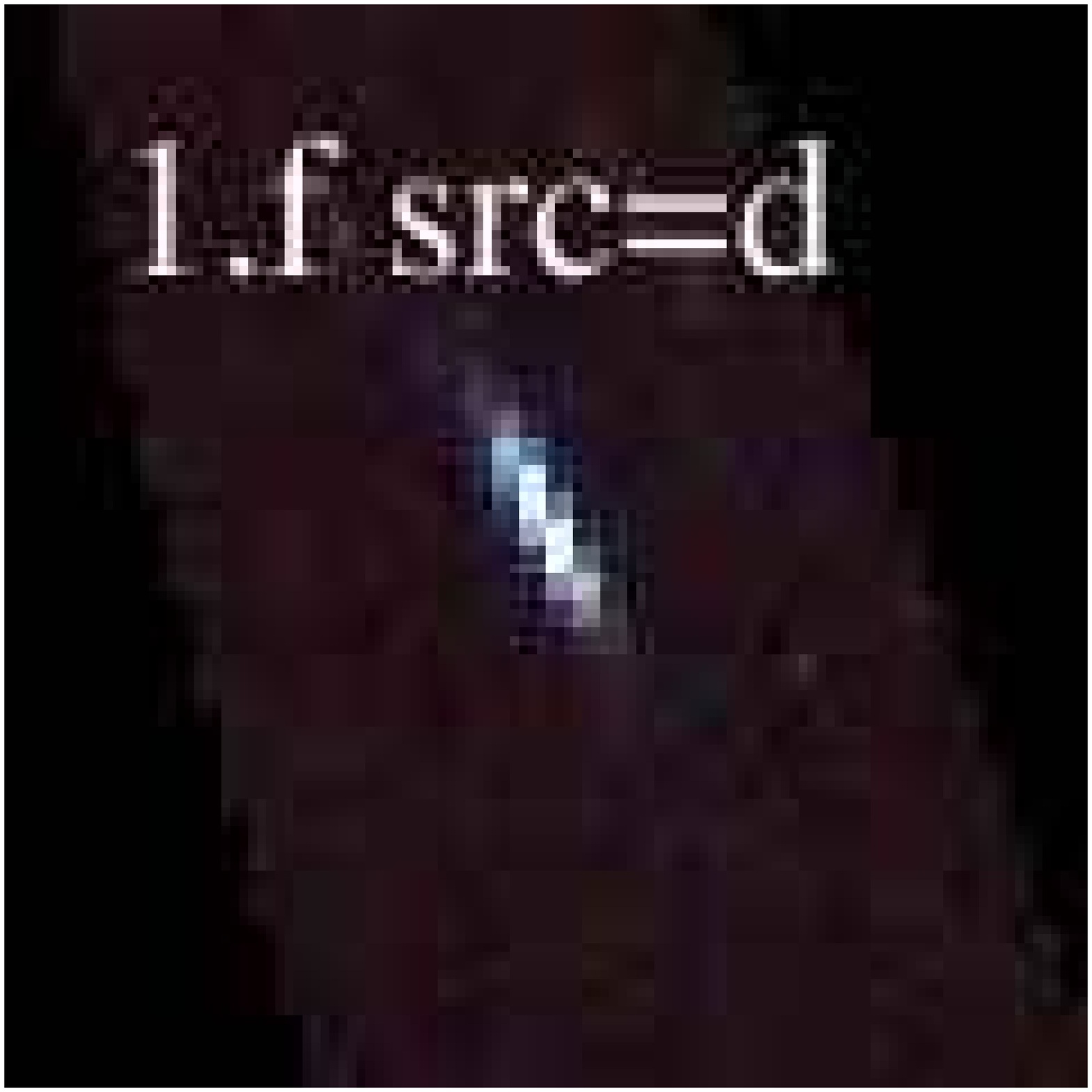}}
    & \multicolumn{1}{m{1.7cm}}{\includegraphics[height=2.00cm,clip]{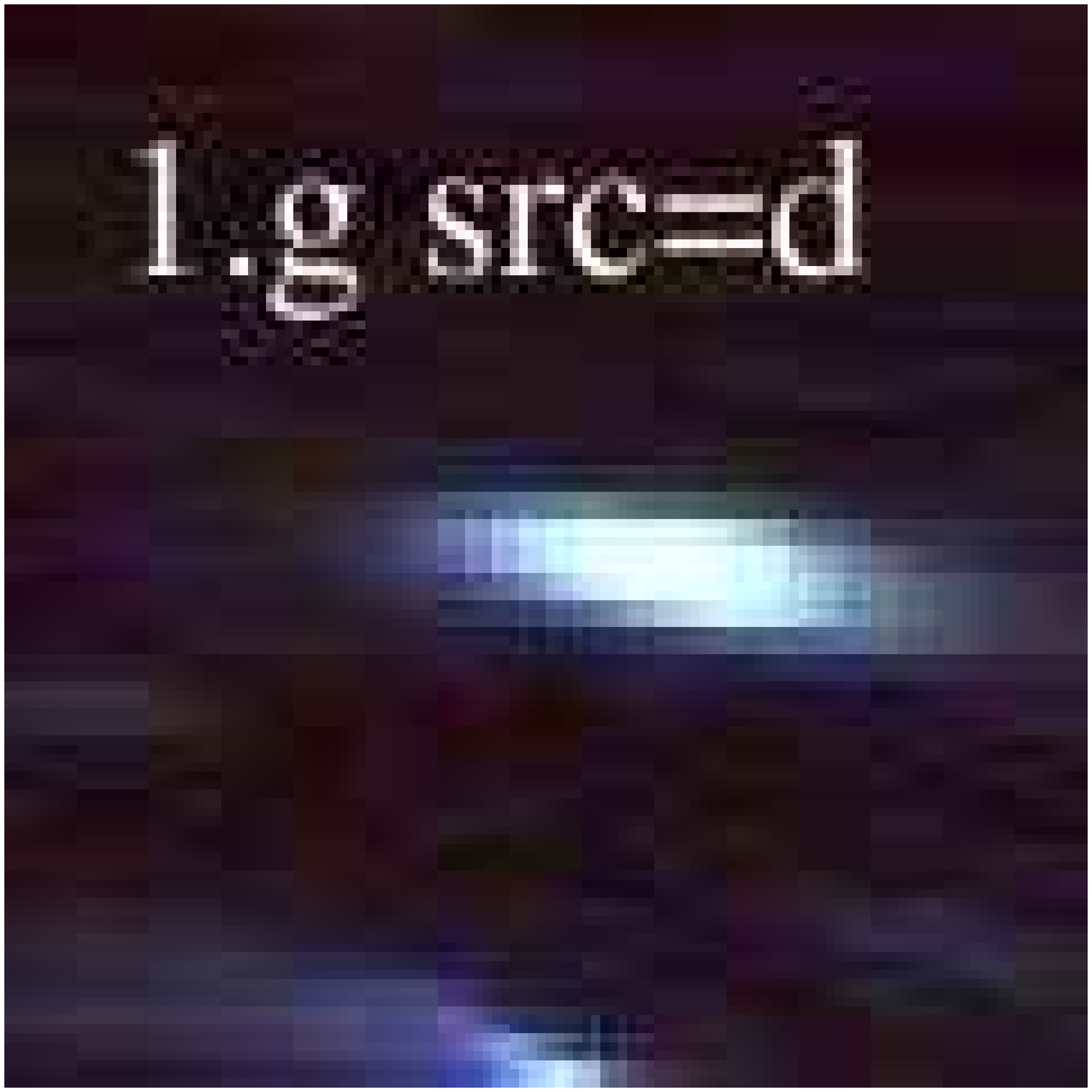}} \\
  \end{tabular}

\end{table*}

\begin{table*}
  \caption{Image system 2:}\vspace{0mm}
  \begin{tabular}{cccccc}
    \multicolumn{1}{m{1cm}}{{\Large A1689}}
    & \multicolumn{1}{m{1.7cm}}{\includegraphics[height=2.00cm,clip]{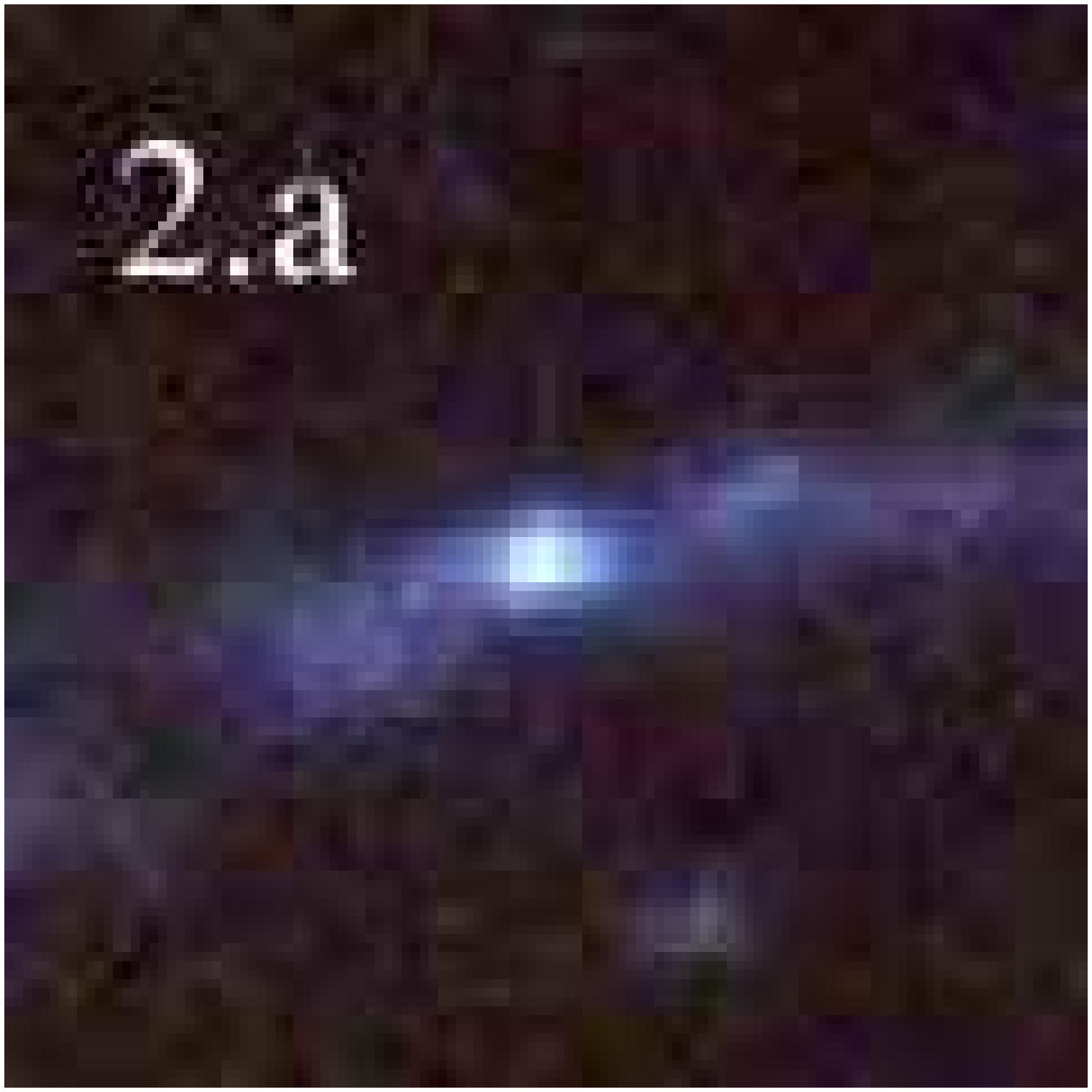}}
    & \multicolumn{1}{m{1.7cm}}{\includegraphics[height=2.00cm,clip]{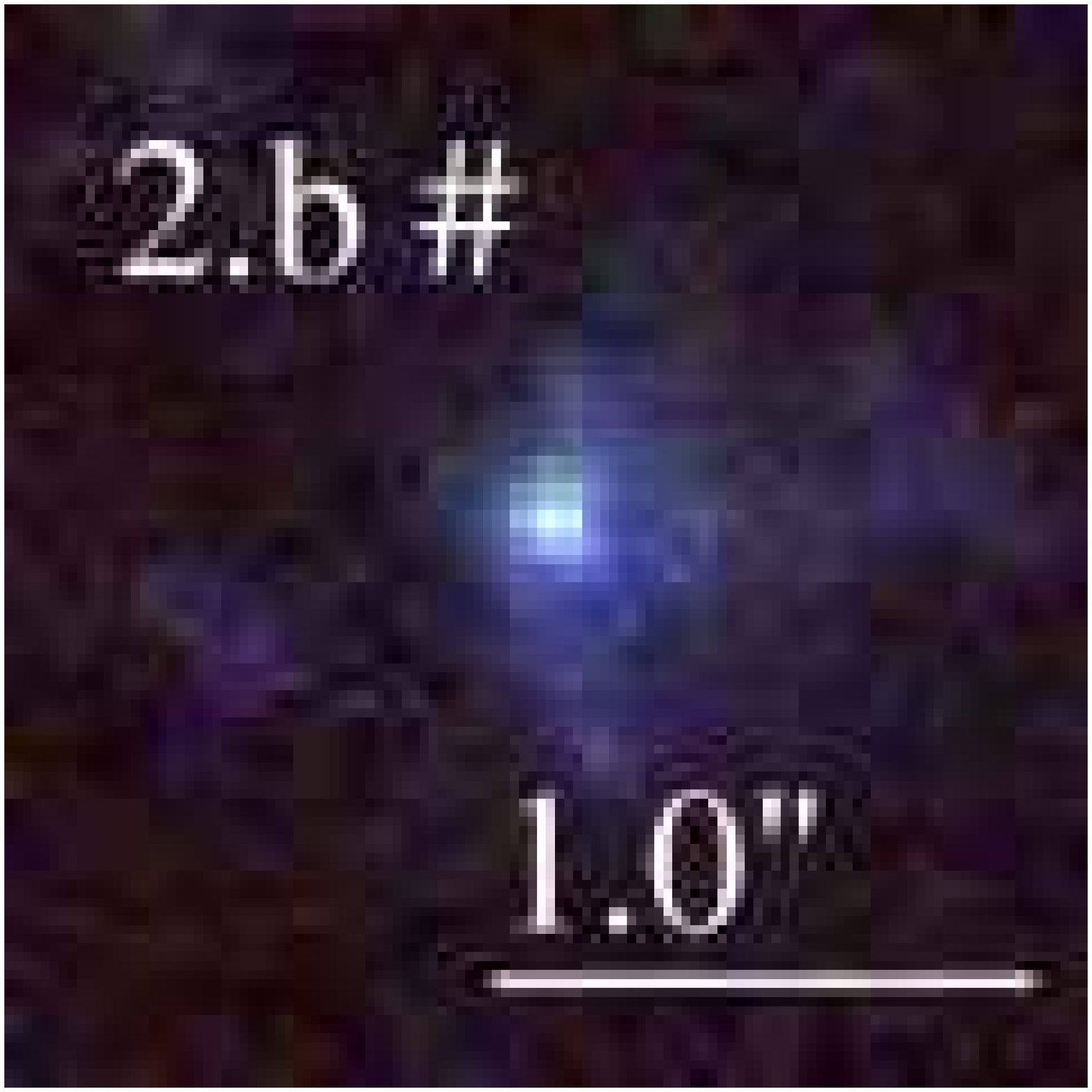}}
    & \multicolumn{1}{m{1.7cm}}{\includegraphics[height=2.00cm,clip]{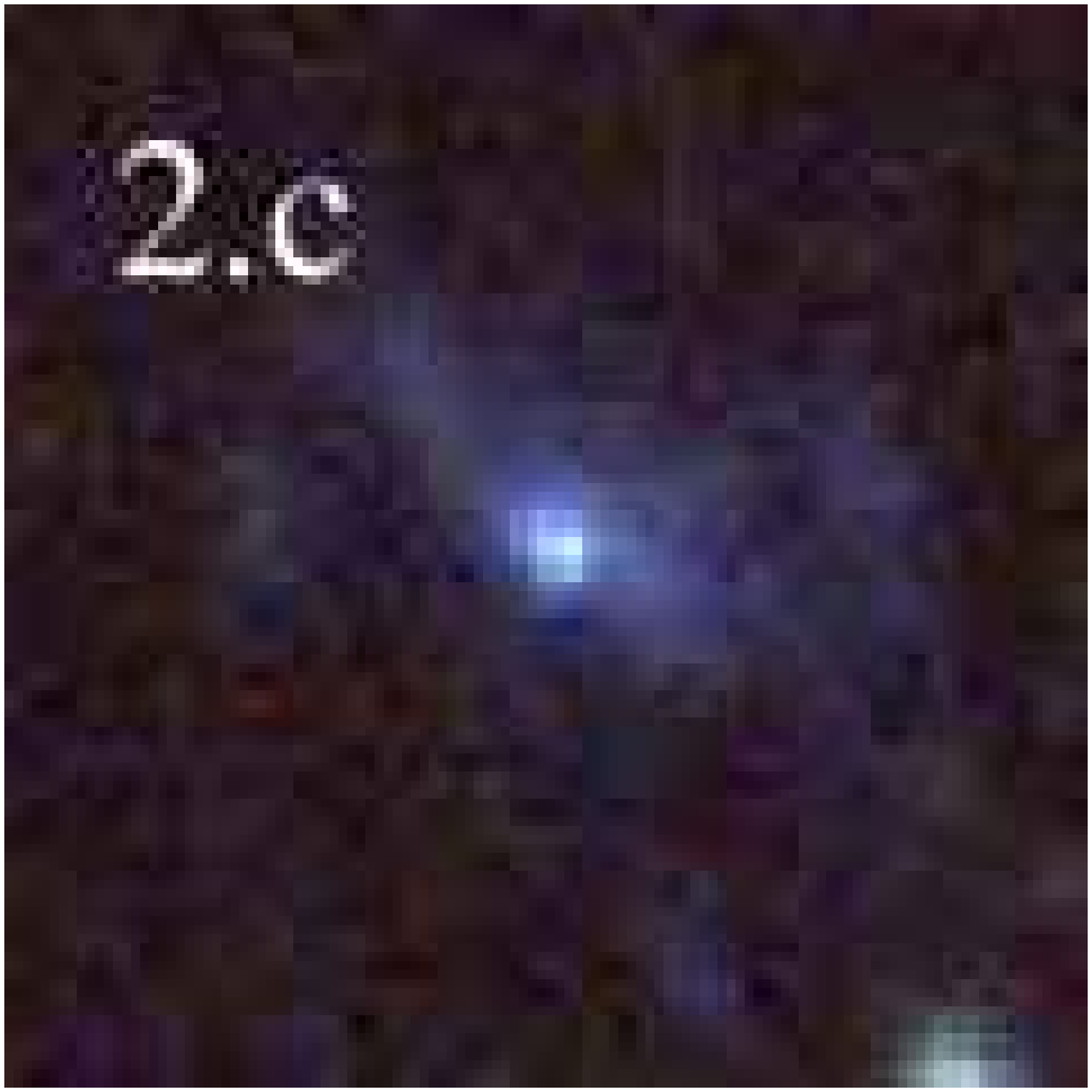}}
    & \multicolumn{1}{m{1.7cm}}{\includegraphics[height=2.00cm,clip]{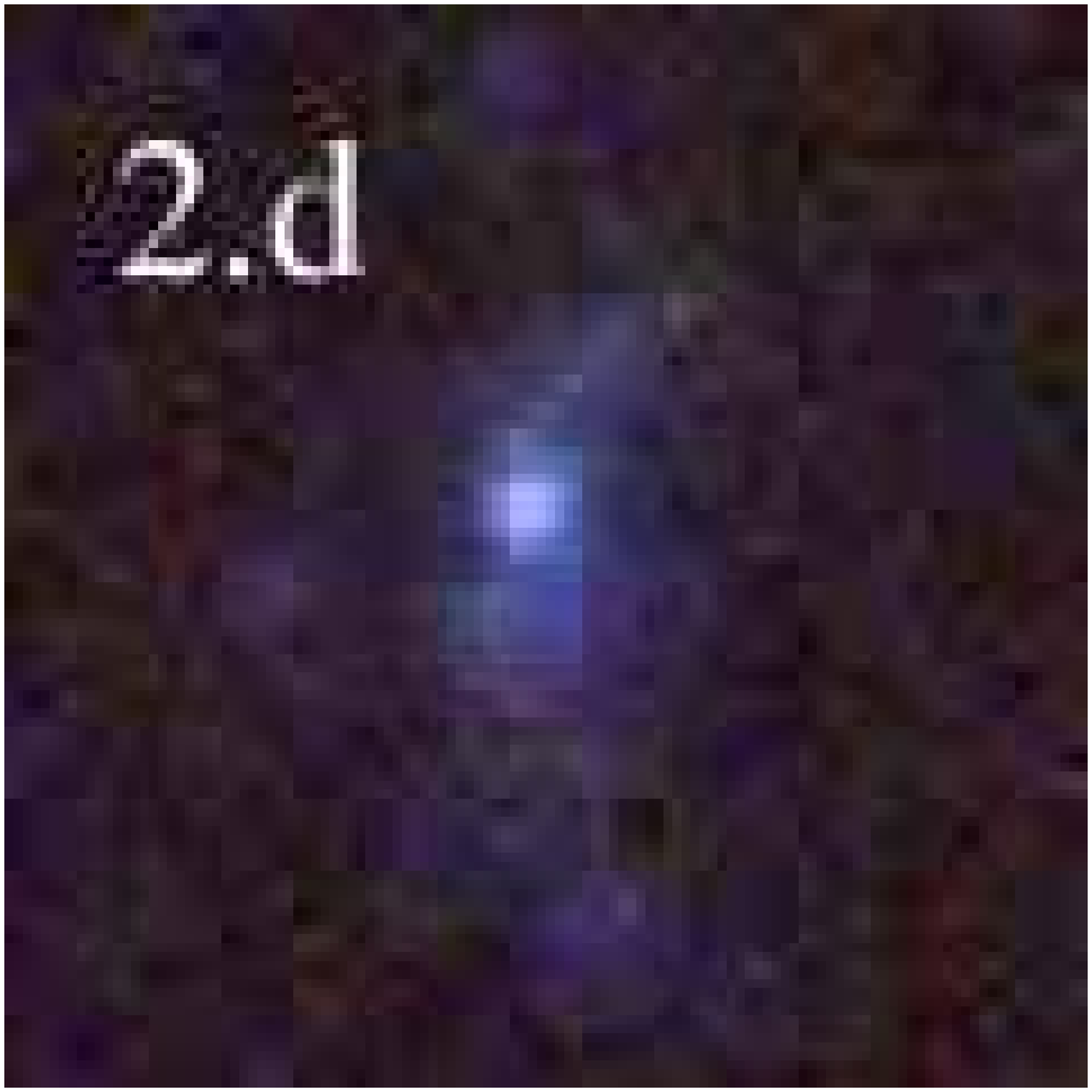}}
    & \multicolumn{1}{m{1.7cm}}{\includegraphics[height=2.00cm,clip]{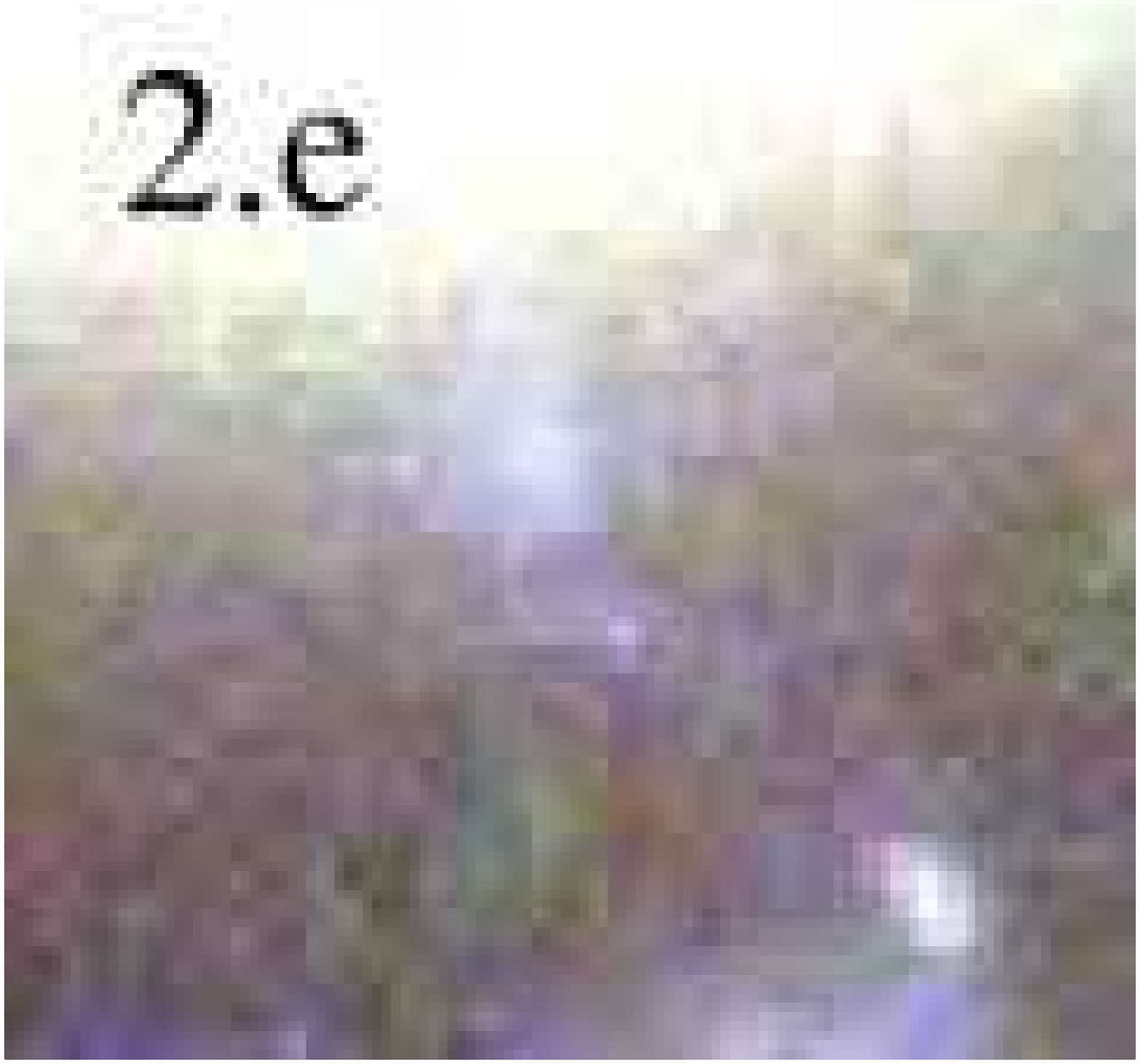}} \\
    \multicolumn{1}{m{1cm}}{{\Large NSIE}}
    & \multicolumn{1}{m{1.7cm}}{\includegraphics[height=2.00cm,clip]{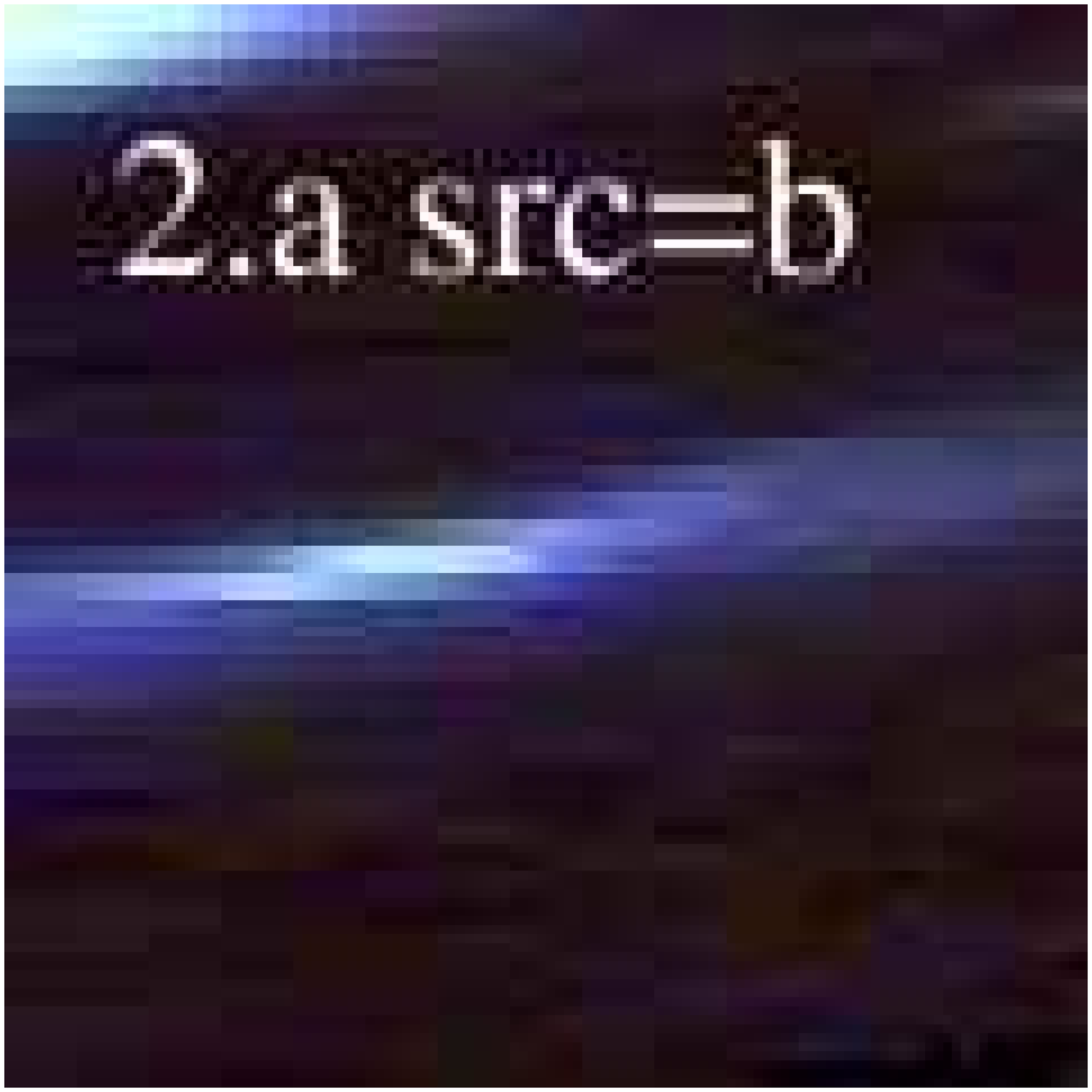}}
    & \multicolumn{1}{m{1.7cm}}{\includegraphics[height=2.00cm,clip]{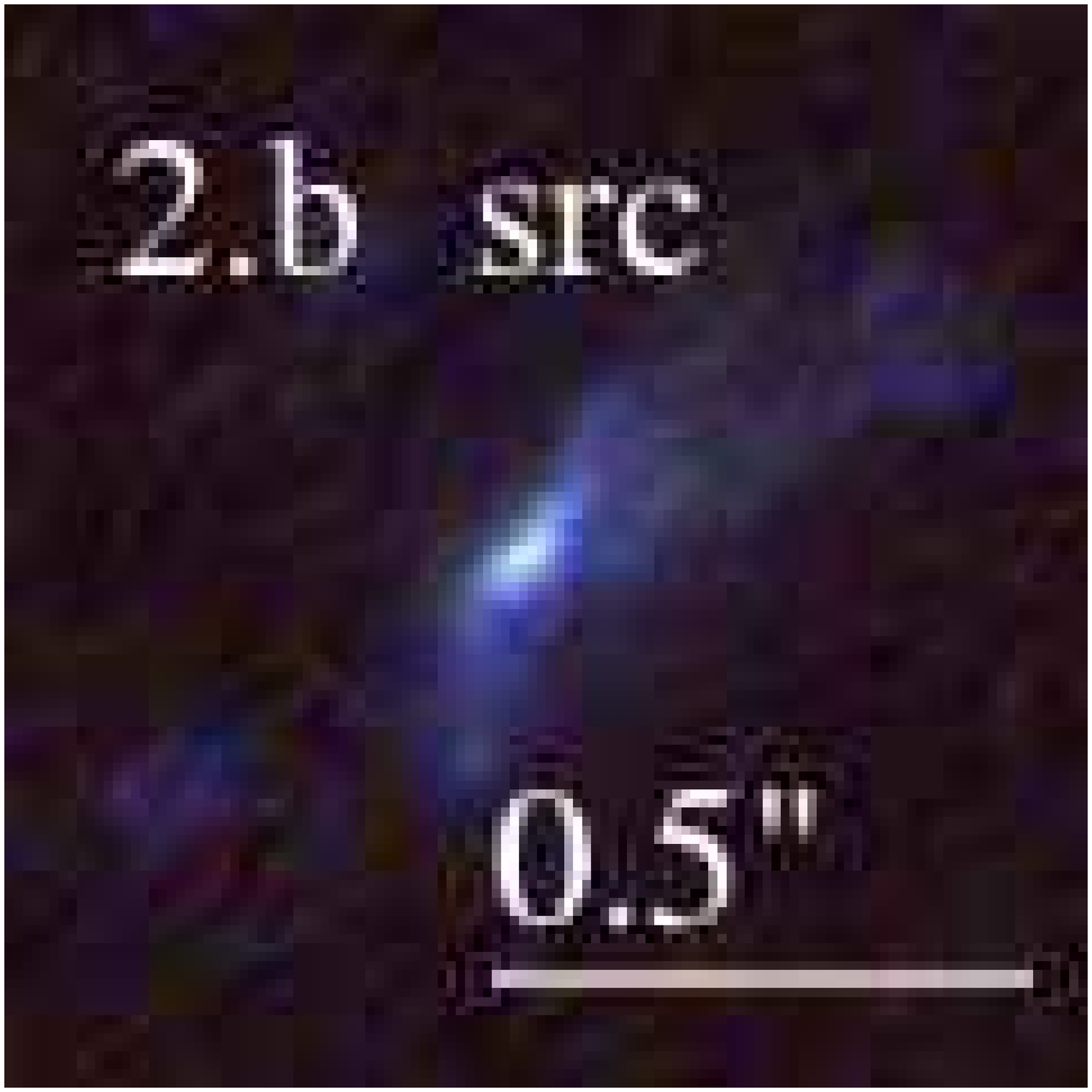}}
    & \multicolumn{1}{m{1.7cm}}{\includegraphics[height=2.00cm,clip]{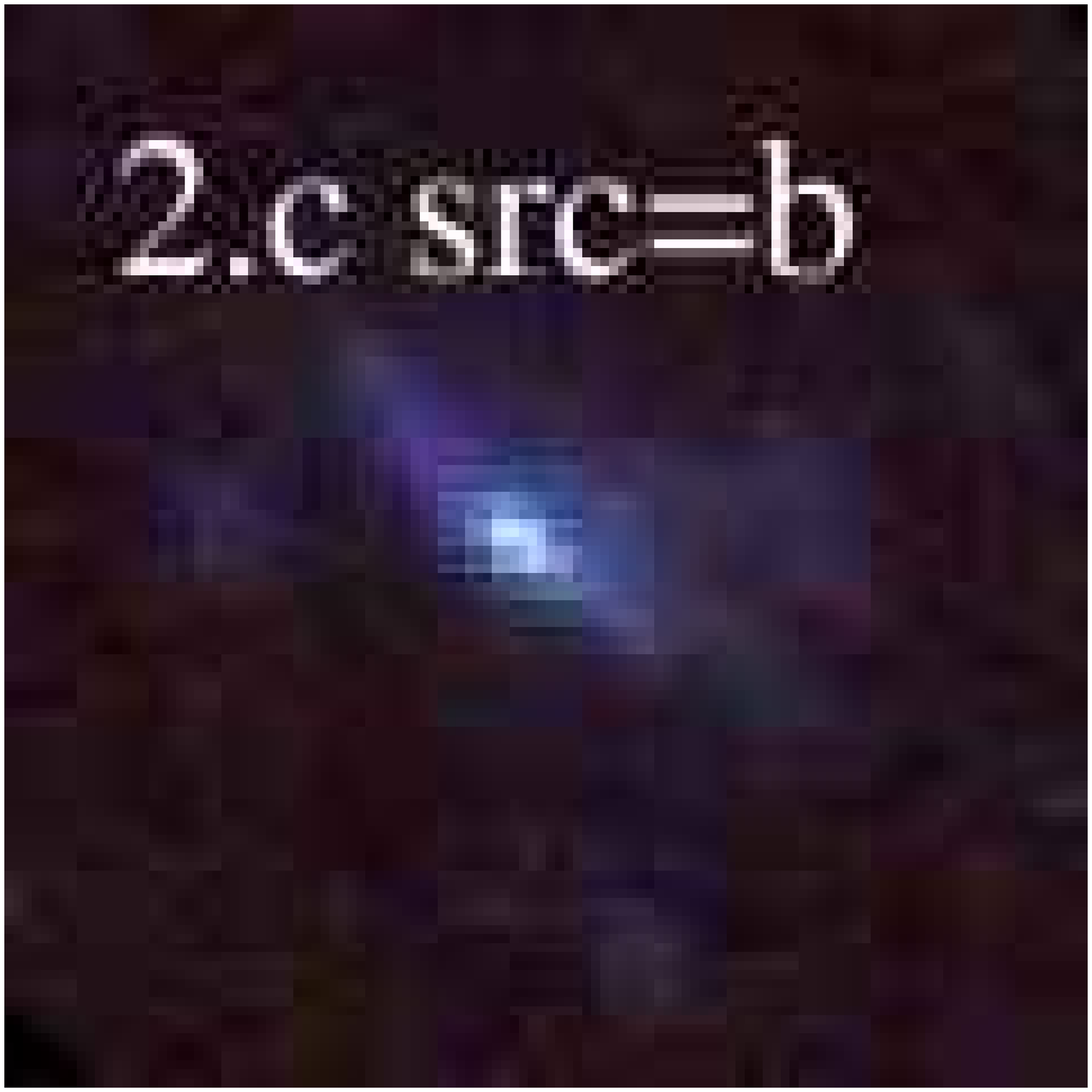}}
    & \multicolumn{1}{m{1.7cm}}{\includegraphics[height=2.00cm,clip]{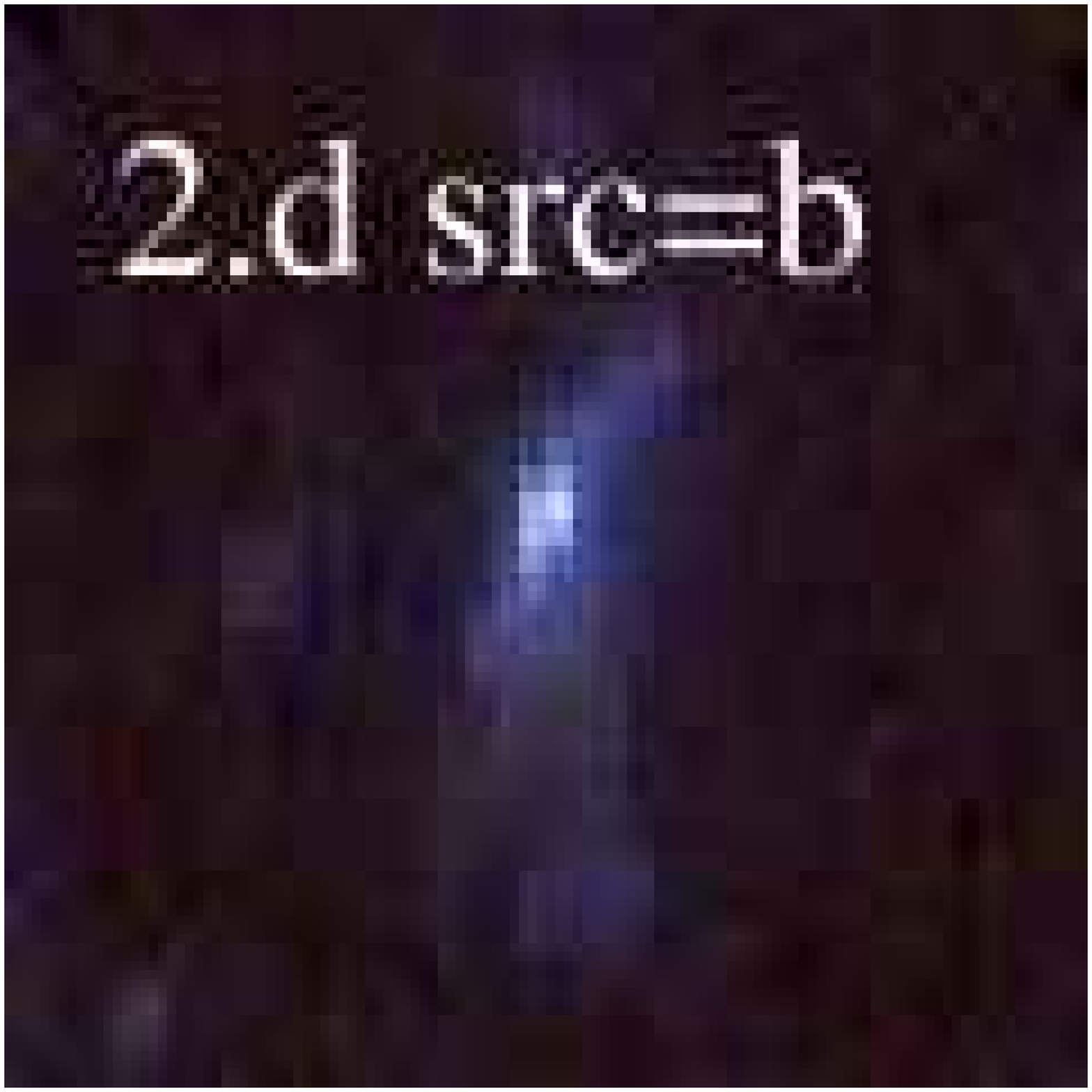}}
    & \multicolumn{1}{m{1.7cm}}{\includegraphics[height=2.00cm,clip]{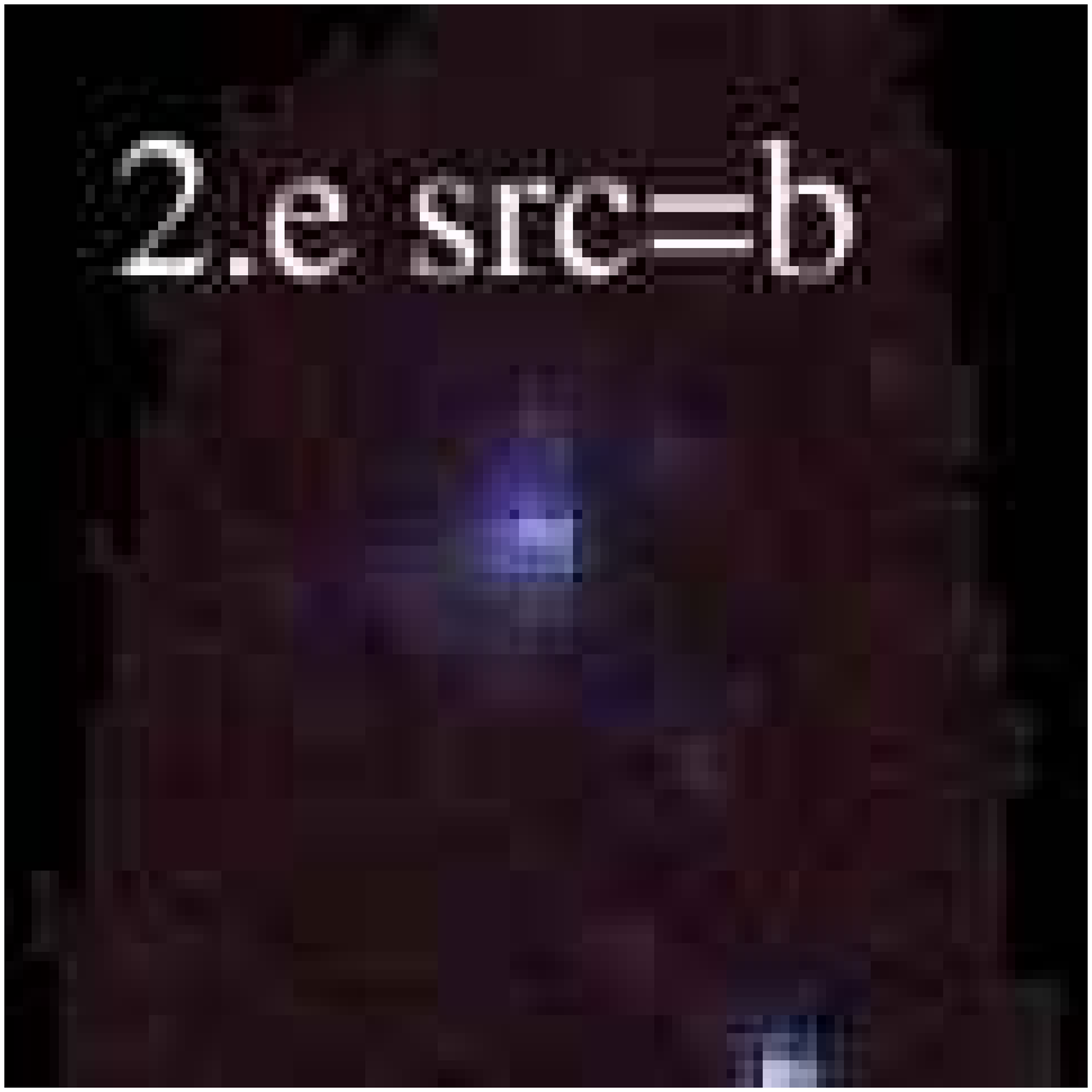}} \\
    \multicolumn{1}{m{1cm}}{{\Large ENFW}}
    & \multicolumn{1}{m{1.7cm}}{\includegraphics[height=2.00cm,clip]{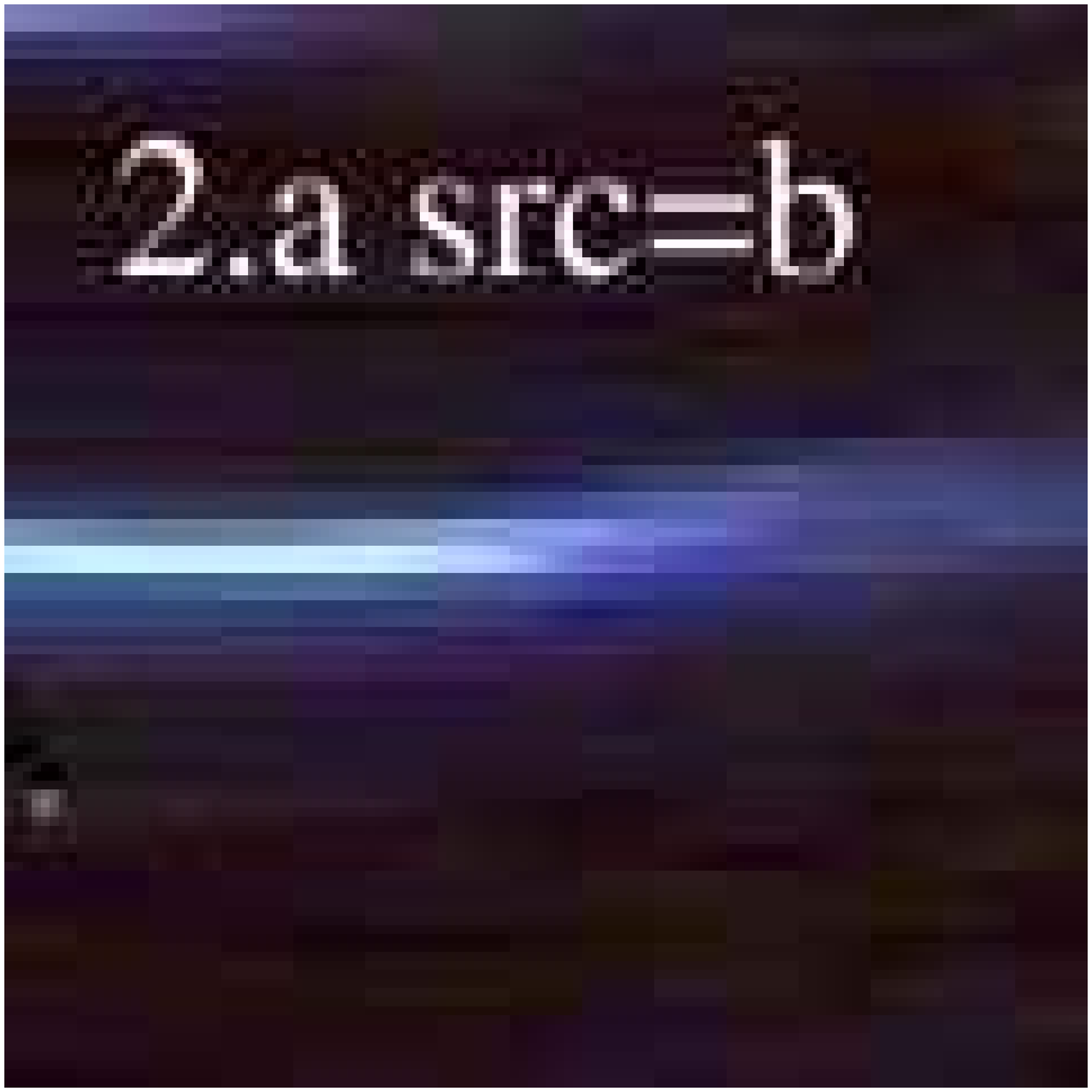}}
    & \multicolumn{1}{m{1.7cm}}{\includegraphics[height=2.00cm,clip]{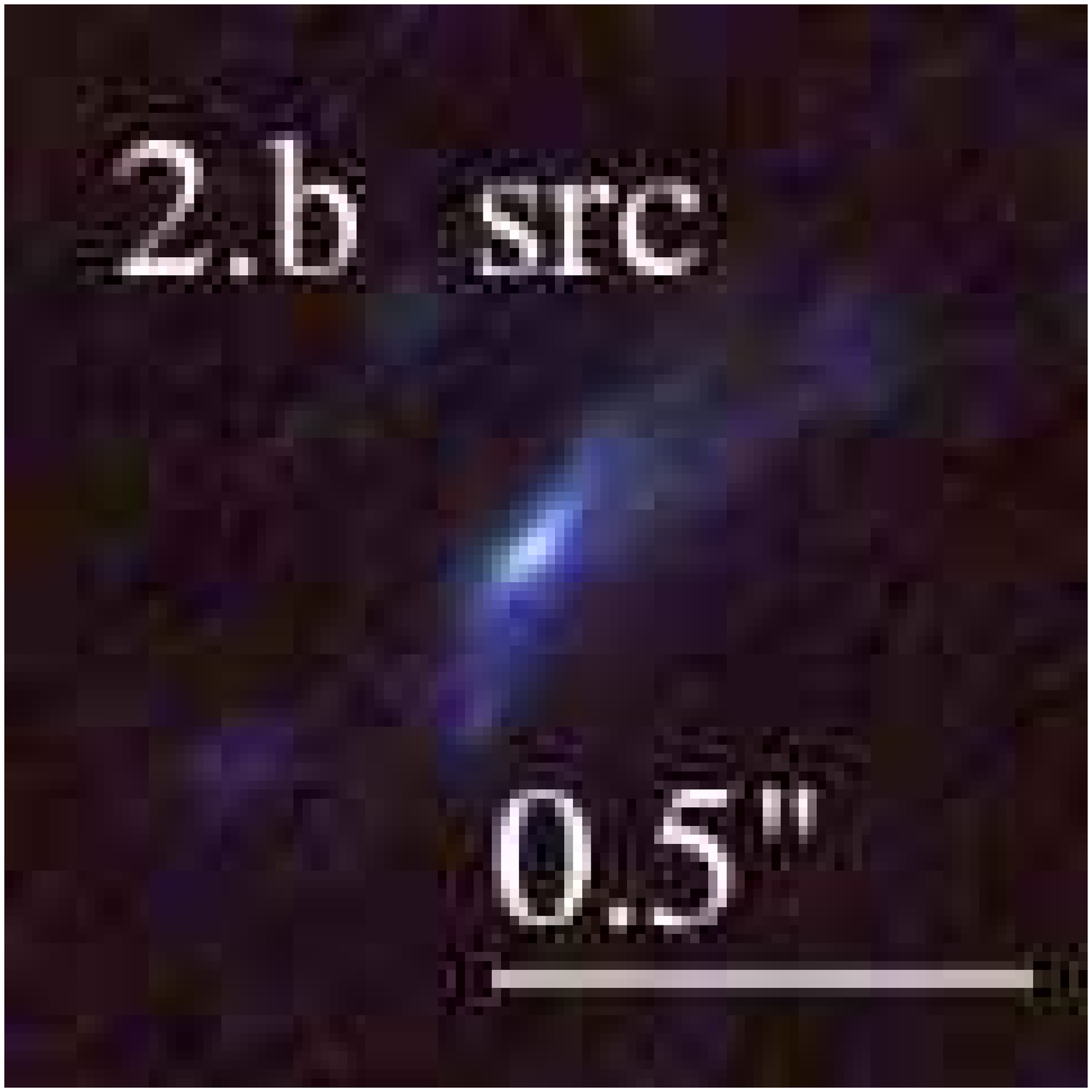}}
    & \multicolumn{1}{m{1.7cm}}{\includegraphics[height=2.00cm,clip]{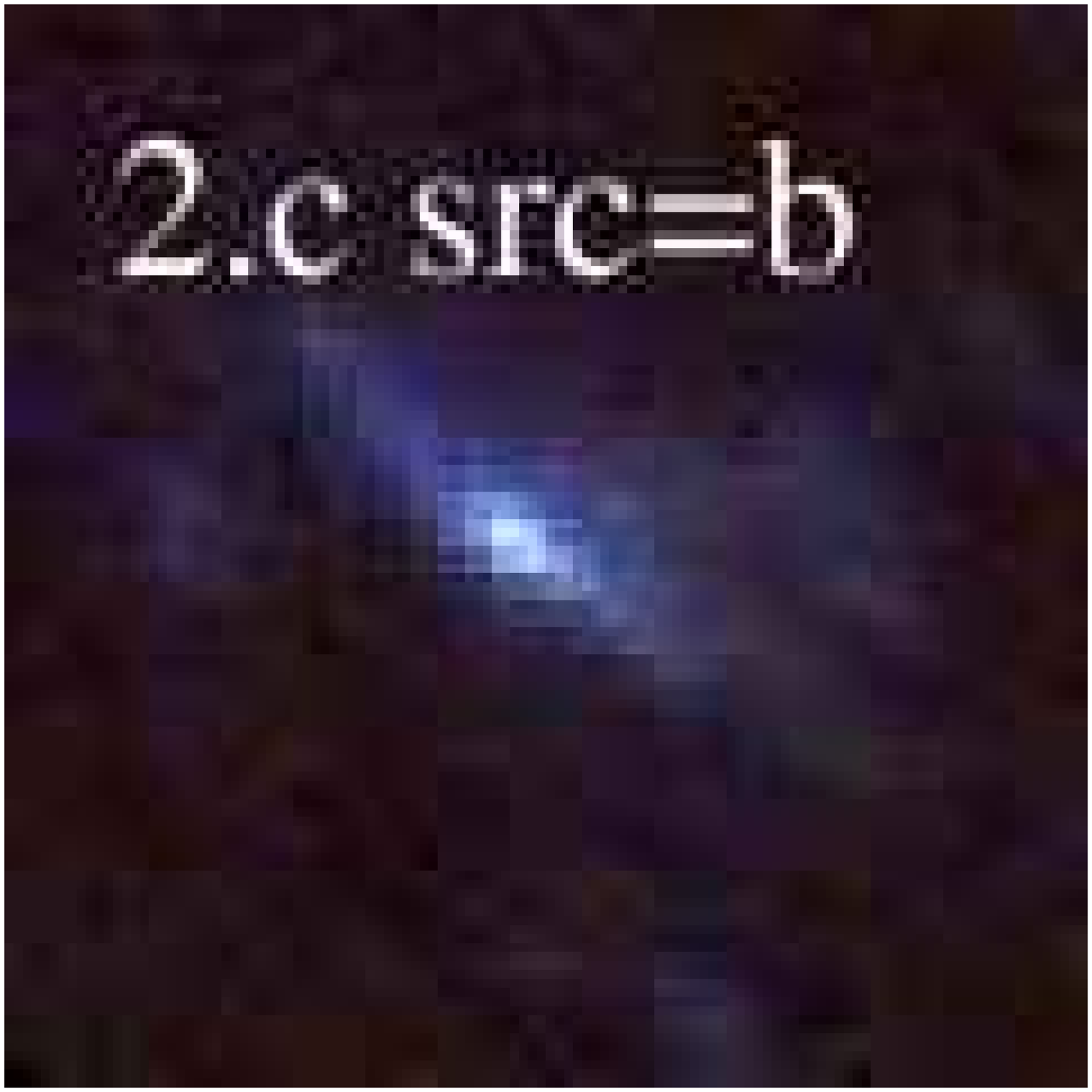}}
    & \multicolumn{1}{m{1.7cm}}{\includegraphics[height=2.00cm,clip]{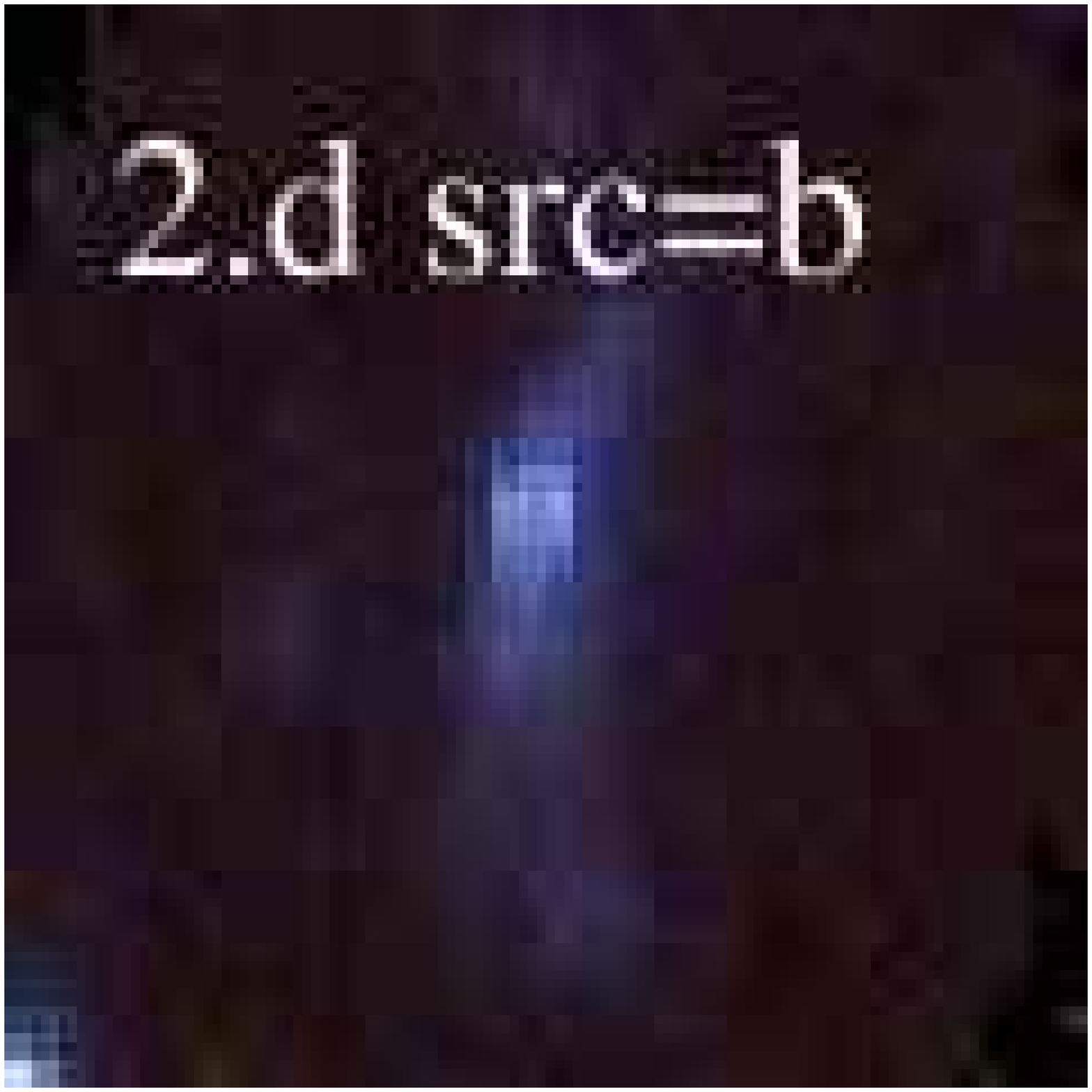}}
    & \multicolumn{1}{m{1.7cm}}{\includegraphics[height=2.00cm,clip]{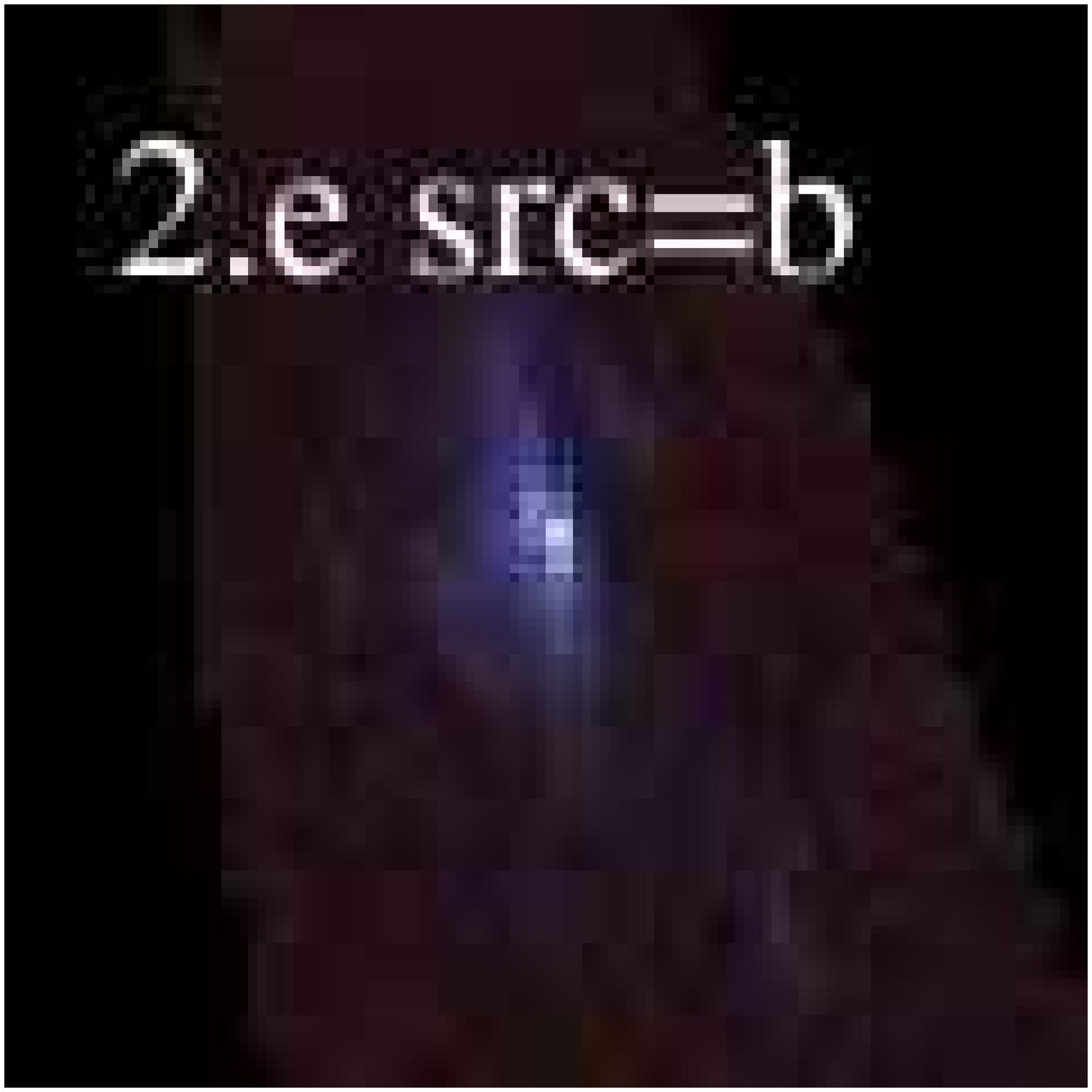}} \\
  \end{tabular}

\end{table*}

\begin{table*}
  \caption{Image system 3:}\vspace{0mm}
  \begin{tabular}{cccc}
    \multicolumn{1}{m{1cm}}{{\Large A1689}}
    & \multicolumn{1}{m{1.7cm}}{\includegraphics[height=2.00cm,clip]{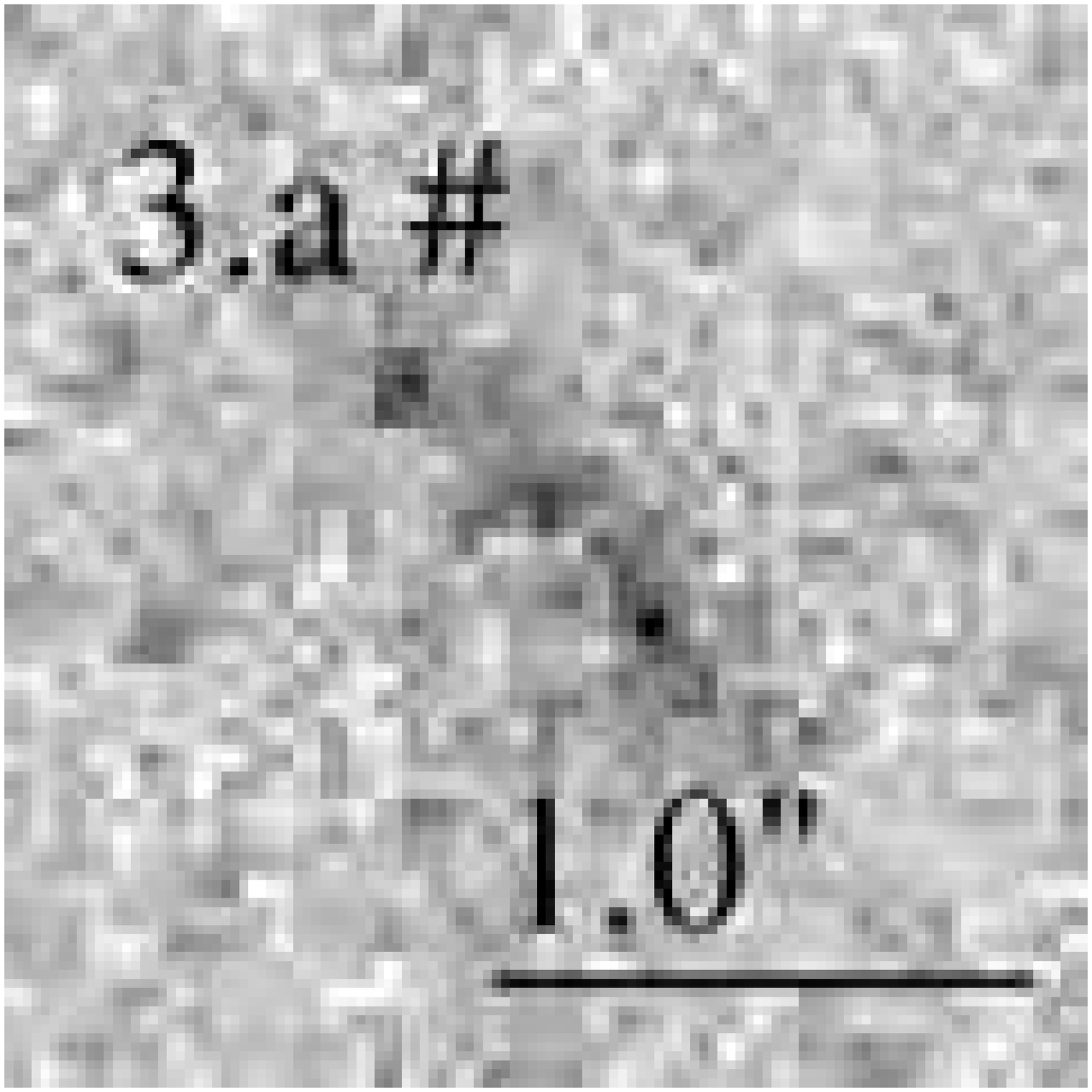}}
    & \multicolumn{1}{m{1.7cm}}{\includegraphics[height=2.00cm,clip]{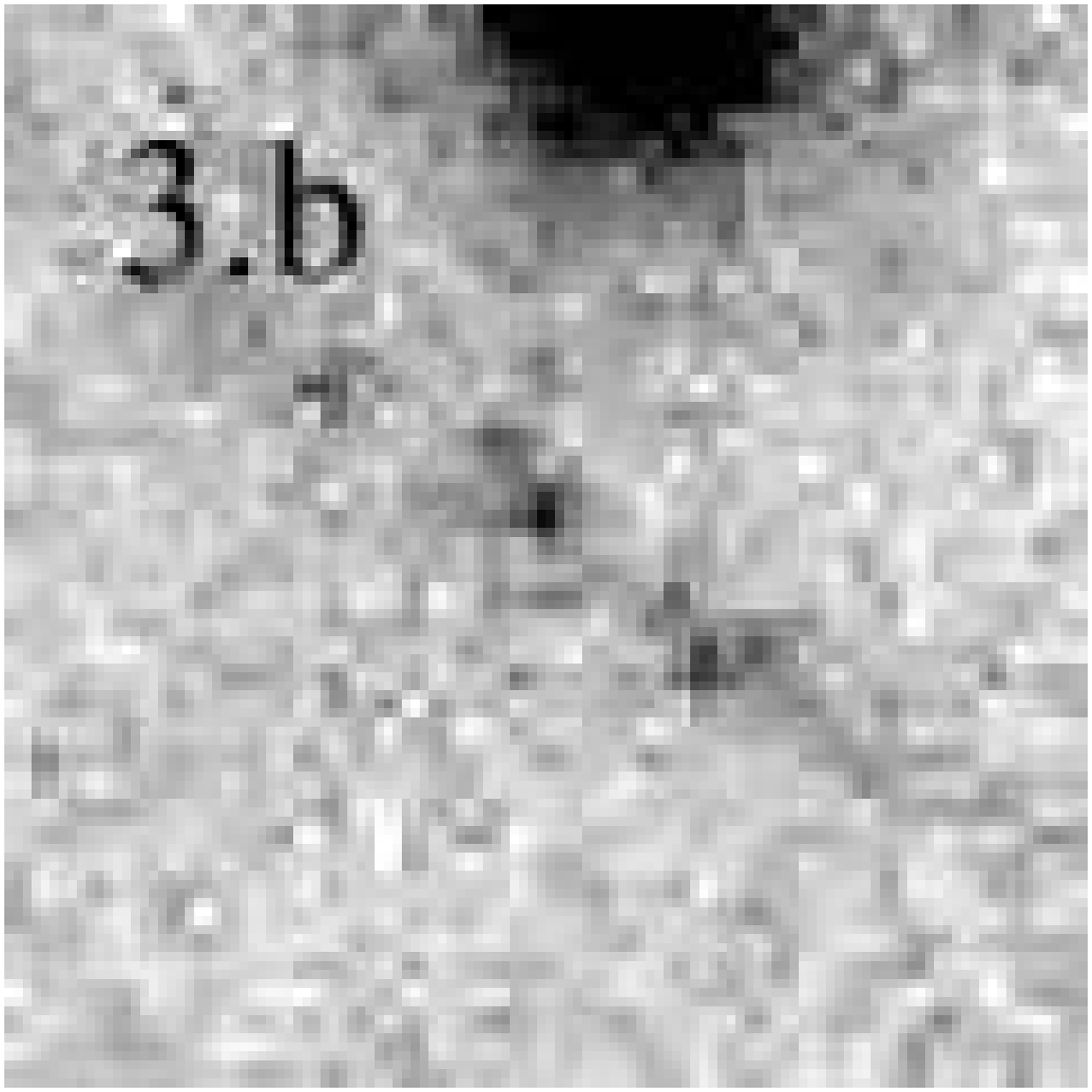}}
    & \multicolumn{1}{m{1.7cm}}{\includegraphics[height=2.00cm,clip]{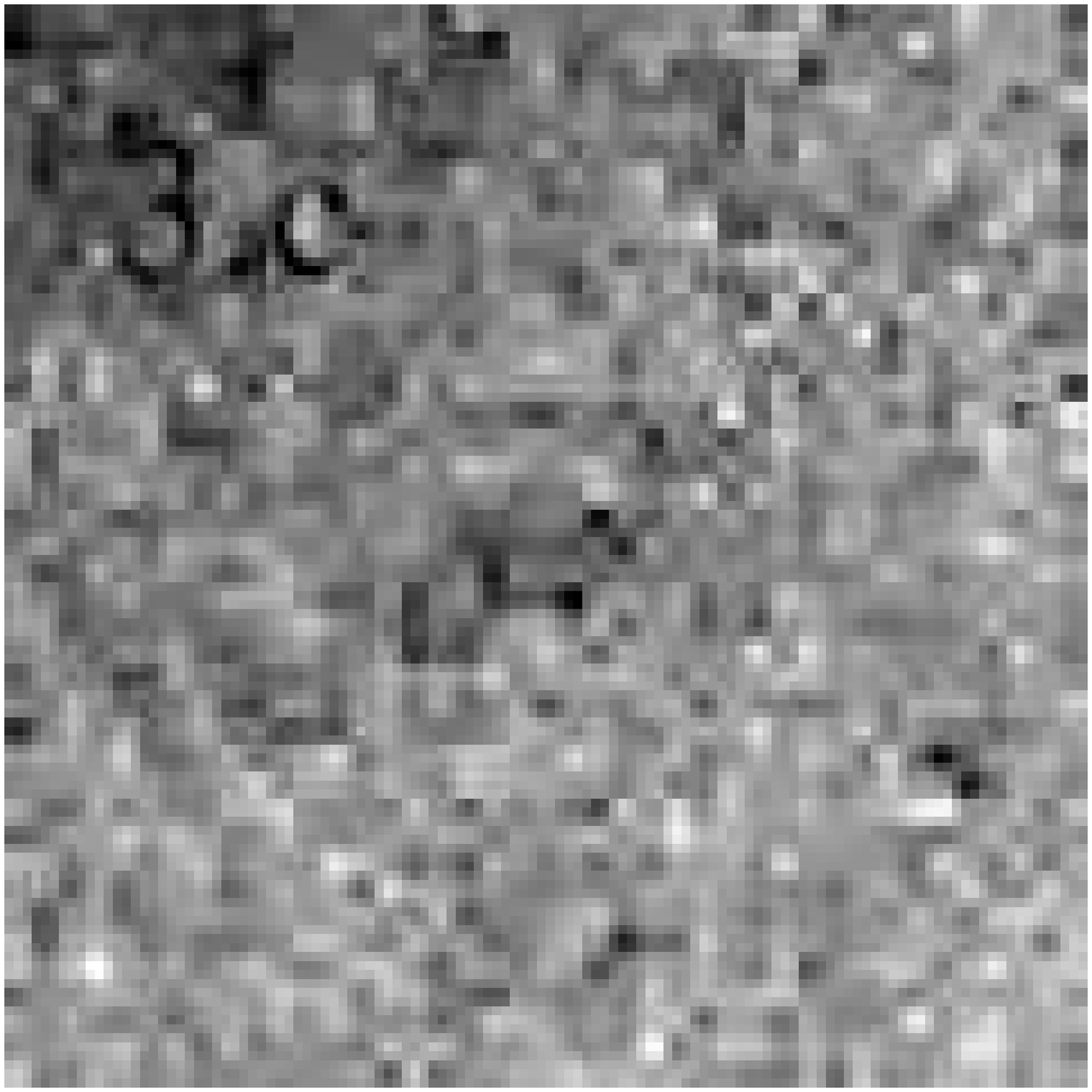}} \\
    \multicolumn{1}{m{1cm}}{{\Large NSIE}}
    & \multicolumn{1}{m{1.7cm}}{\includegraphics[height=2.00cm,clip]{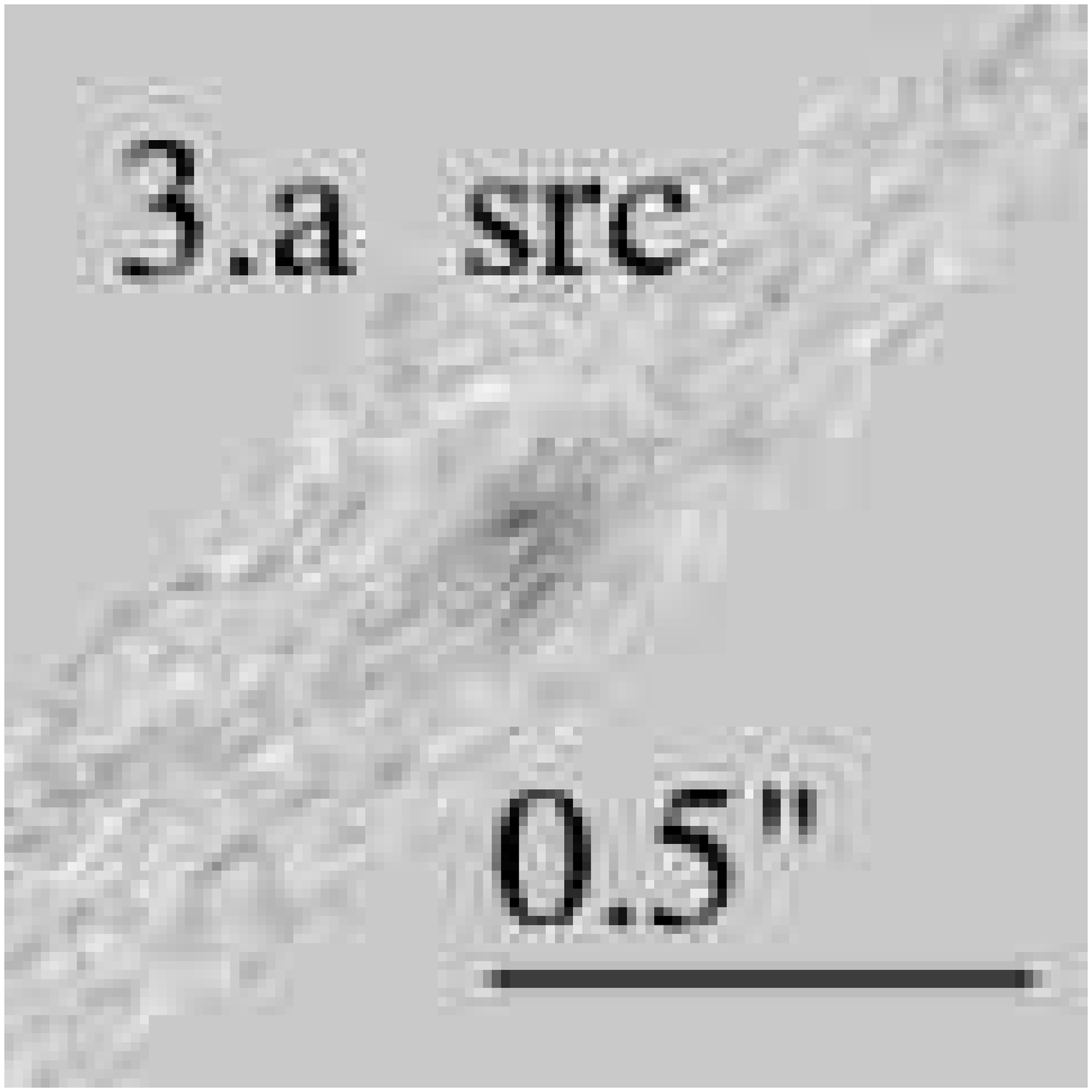}}
    & \multicolumn{1}{m{1.7cm}}{\includegraphics[height=2.00cm,clip]{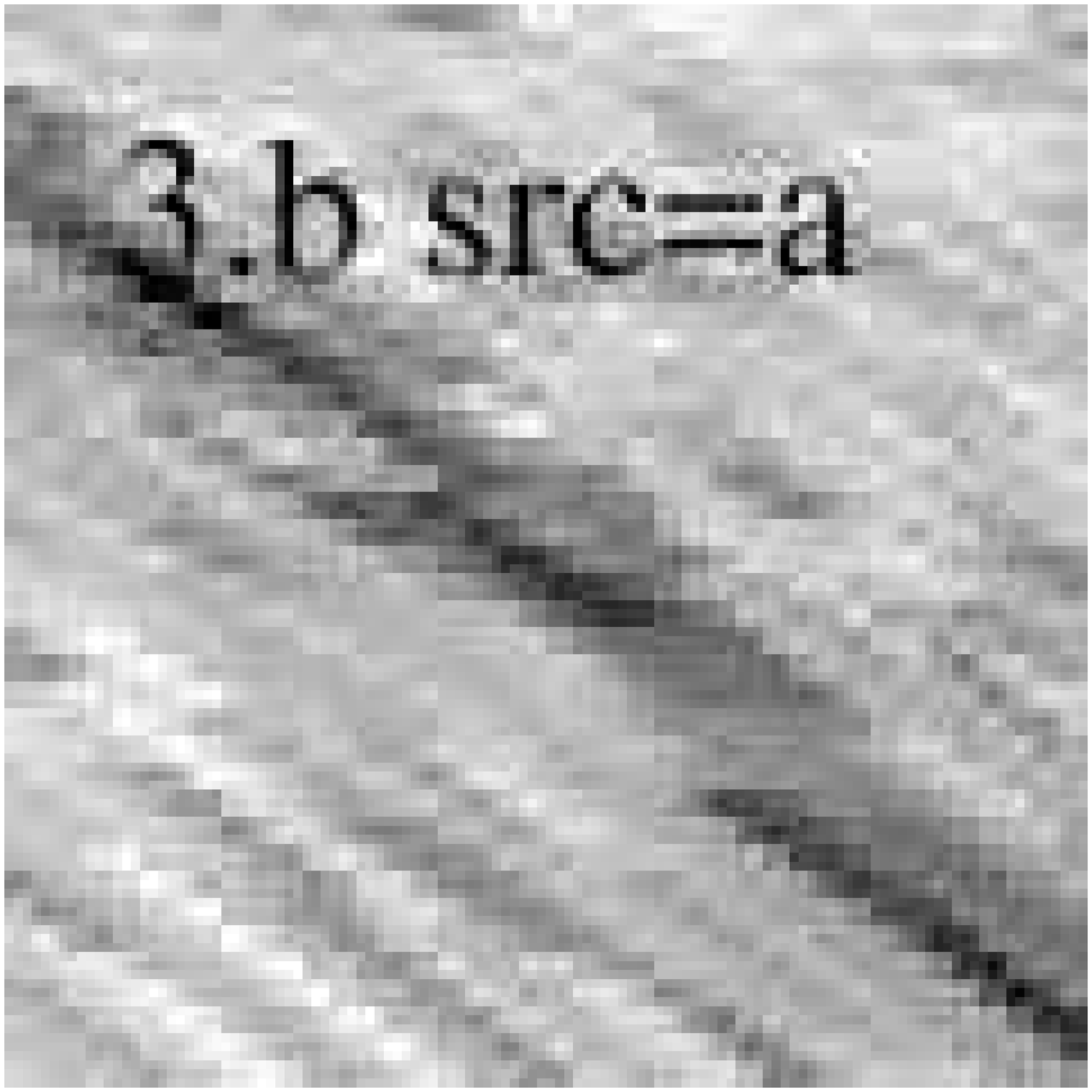}}
    & \multicolumn{1}{m{1.7cm}}{\includegraphics[height=2.00cm,clip]{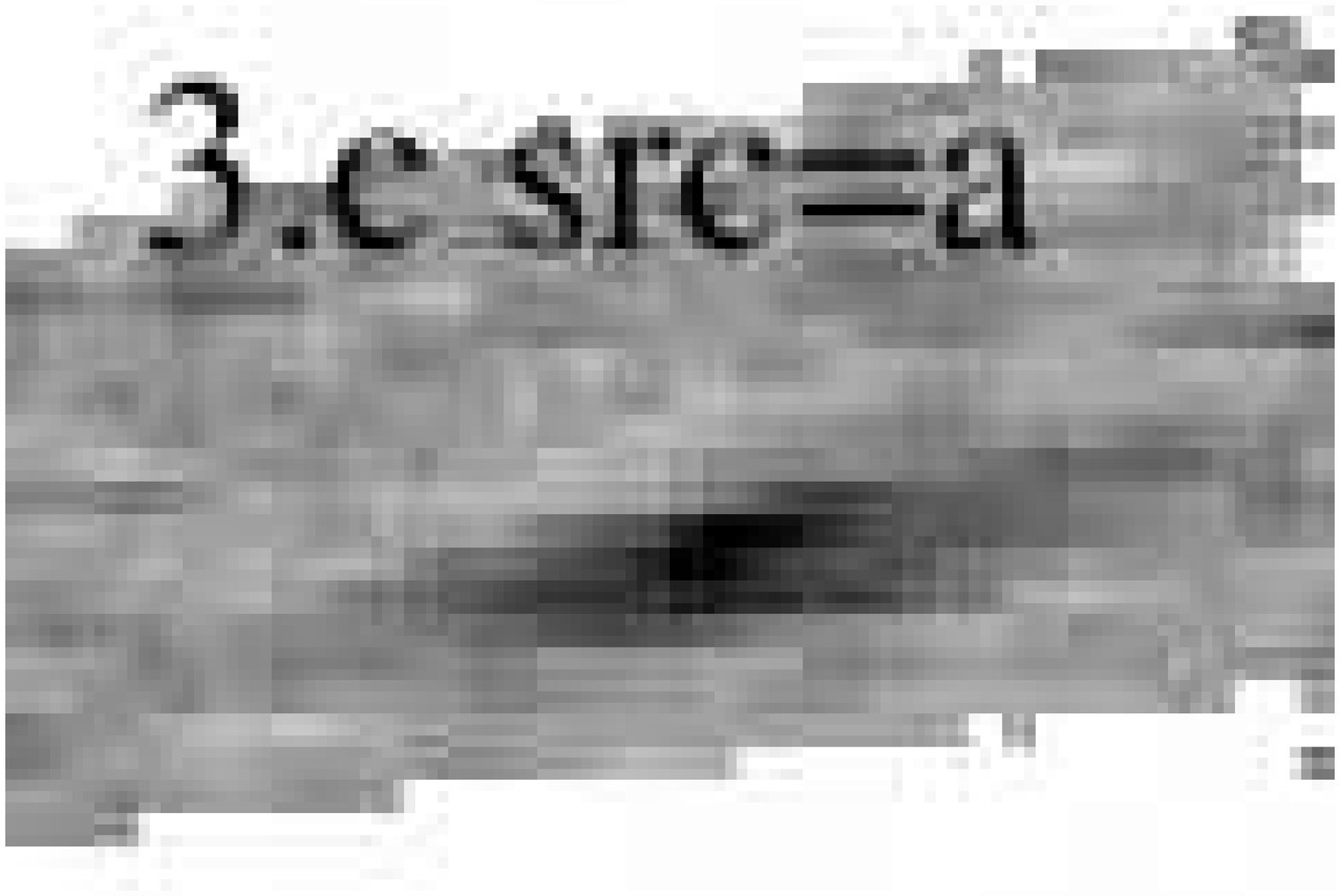}} \\
    \multicolumn{1}{m{1cm}}{{\Large ENFW}}
    & \multicolumn{1}{m{1.7cm}}{\includegraphics[height=2.00cm,clip]{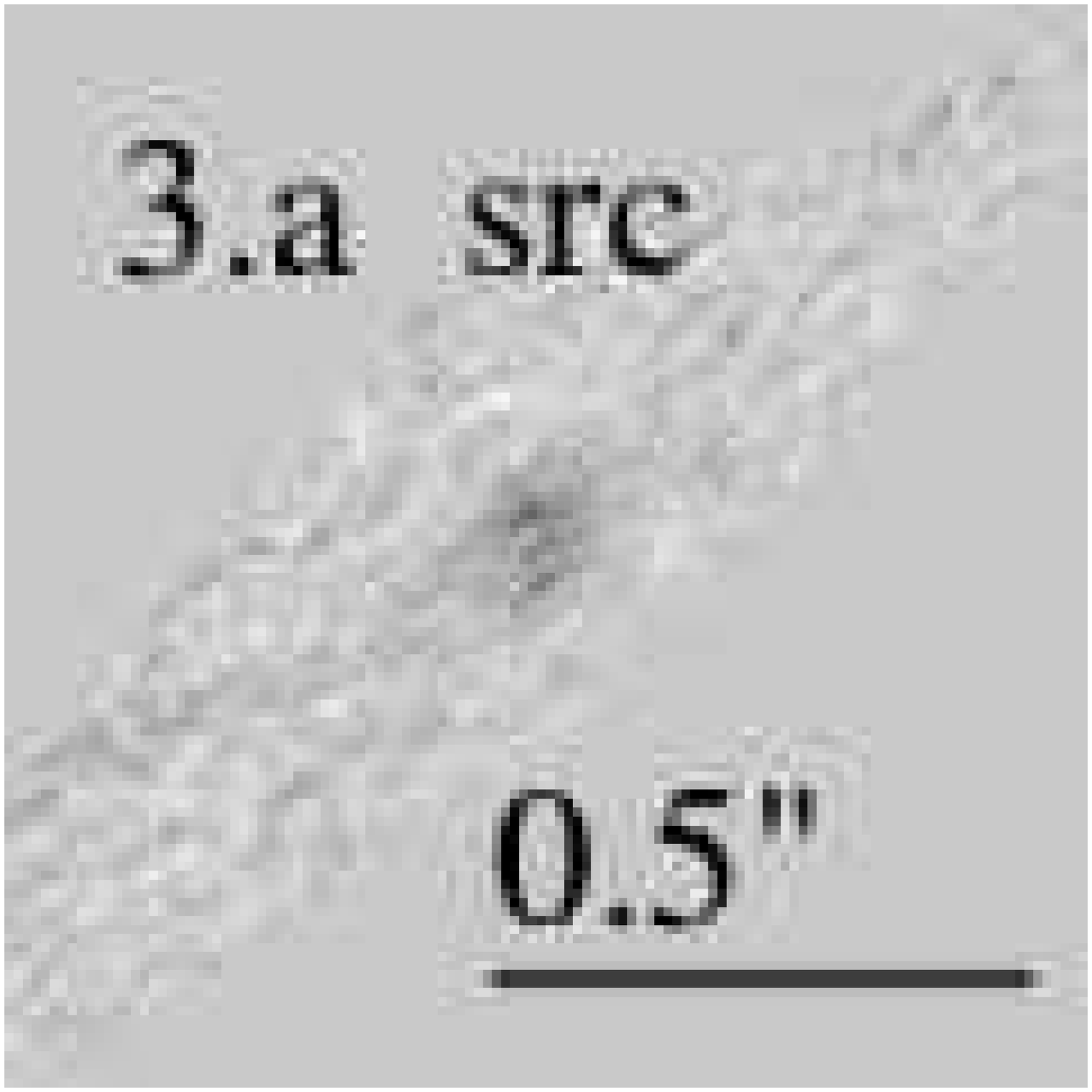}}
    & \multicolumn{1}{m{1.7cm}}{\includegraphics[height=2.00cm,clip]{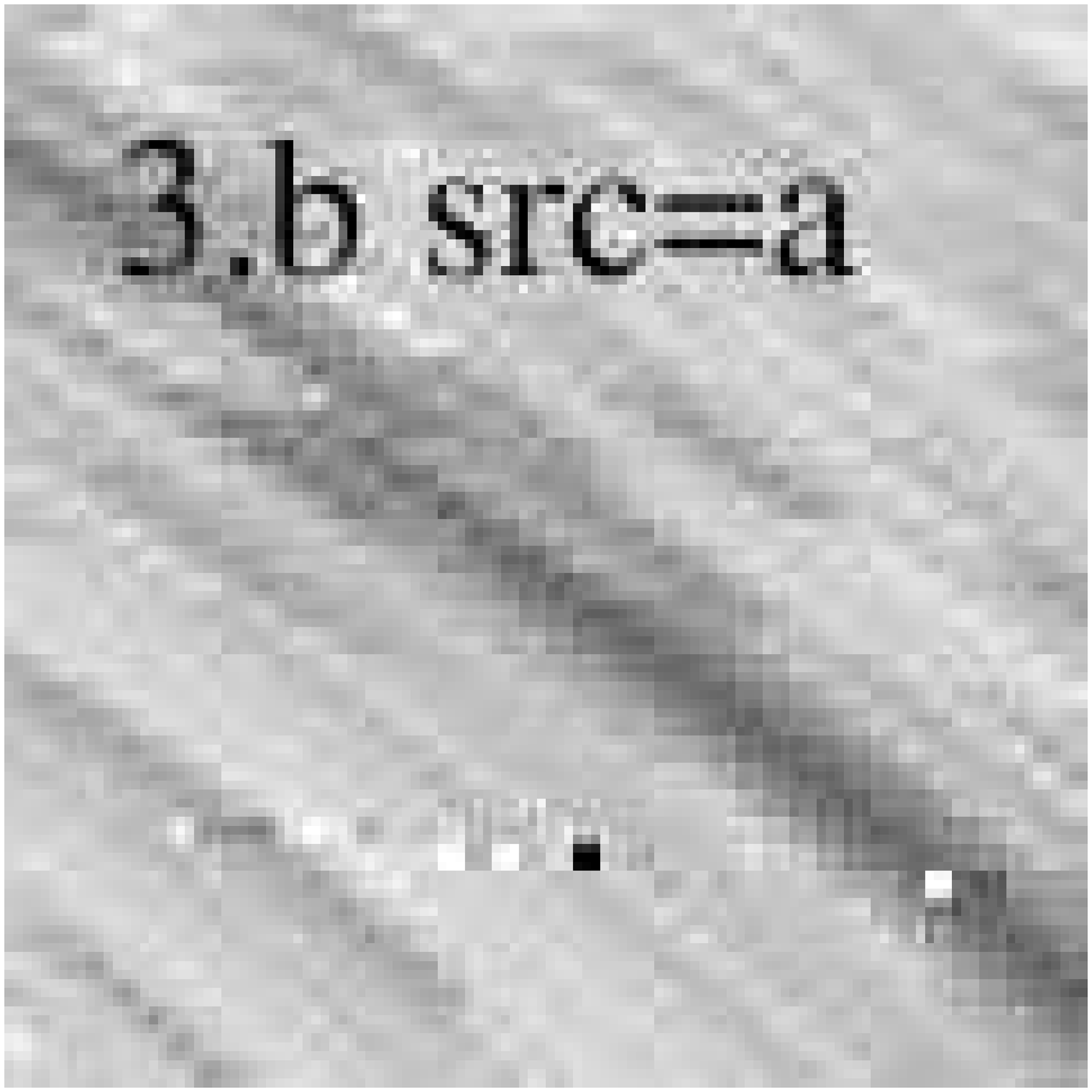}}
    & \multicolumn{1}{m{1.7cm}}{\includegraphics[height=2.00cm,clip]{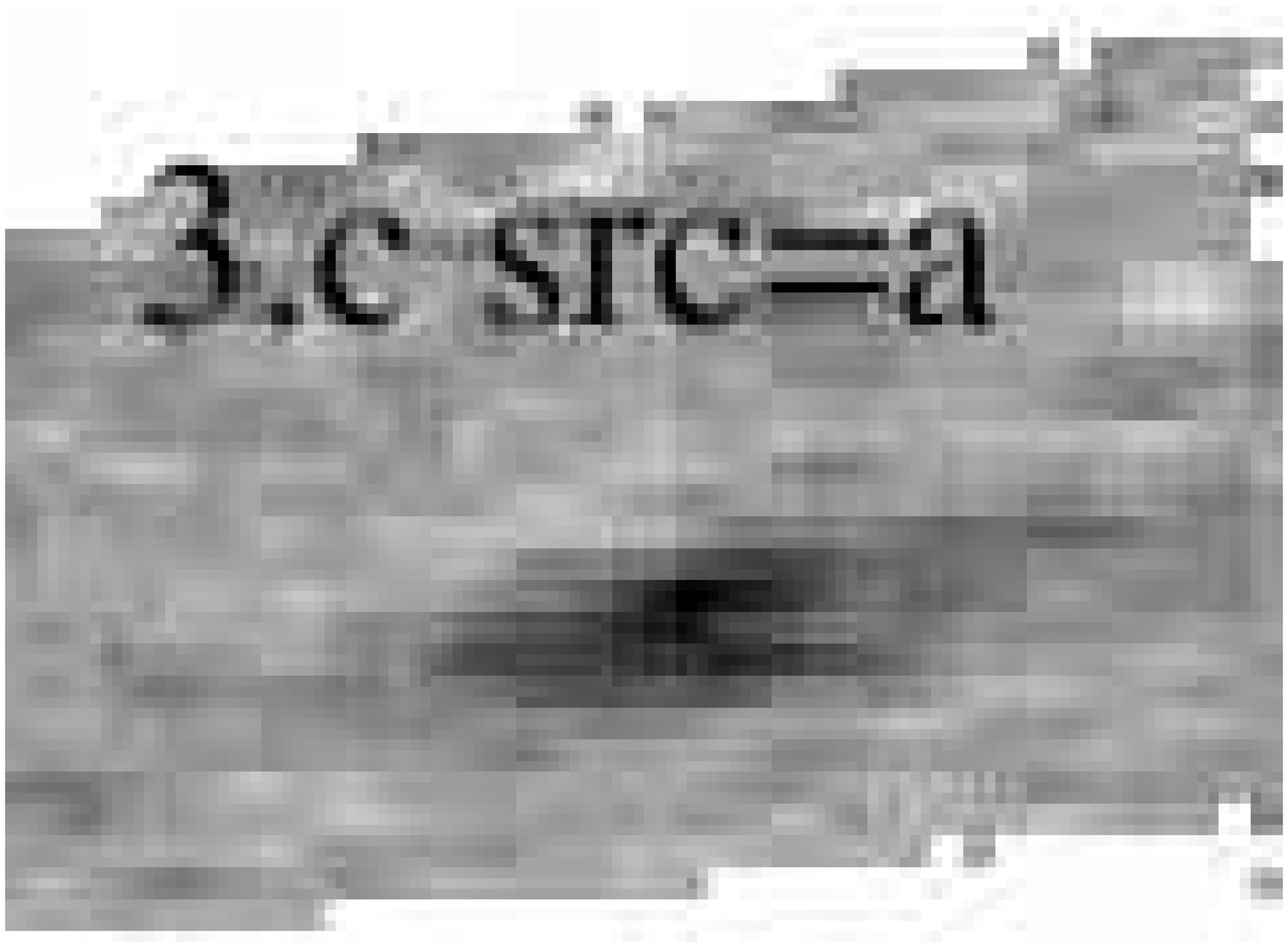}} \\
  \end{tabular}

\end{table*}

\clearpage

\begin{table*}
  \caption{Image system 4:}\vspace{0mm}
  \begin{tabular}{cccccc}
    \multicolumn{1}{m{1cm}}{{\Large A1689}}
    & \multicolumn{1}{m{1.7cm}}{\includegraphics[height=2.00cm,clip]{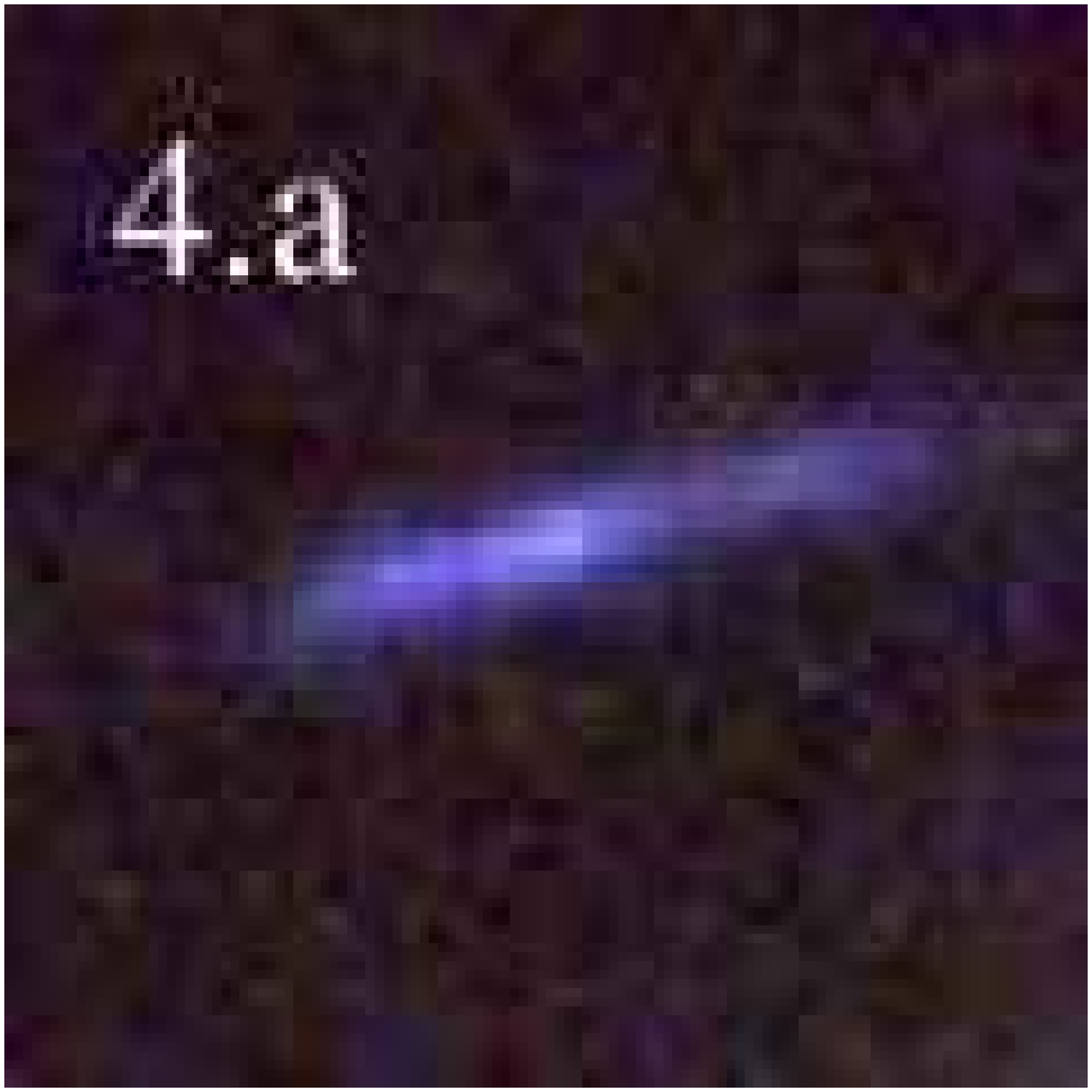}}
    & \multicolumn{1}{m{1.7cm}}{\includegraphics[height=2.00cm,clip]{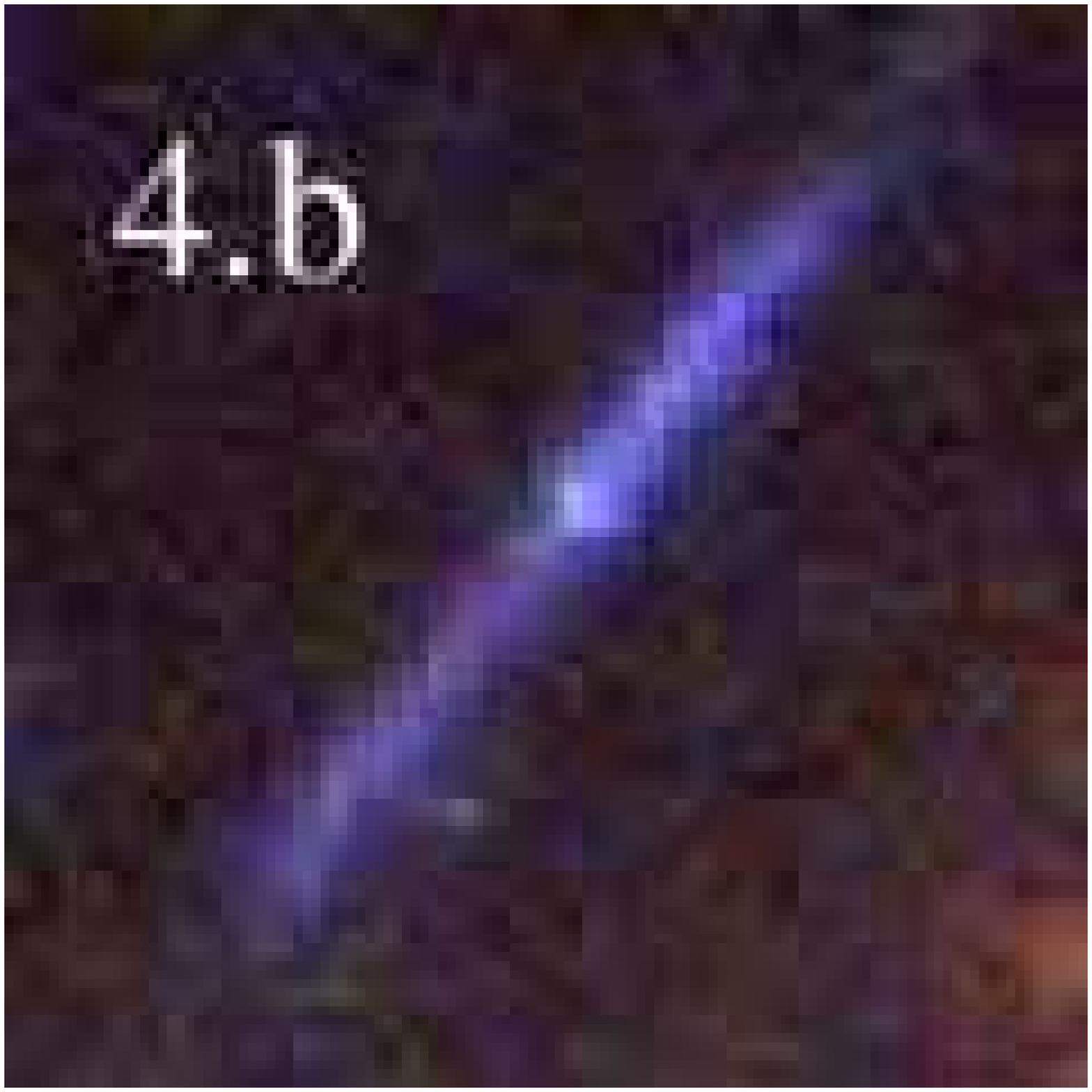}}
    & \multicolumn{1}{m{1.7cm}}{\includegraphics[height=2.00cm,clip]{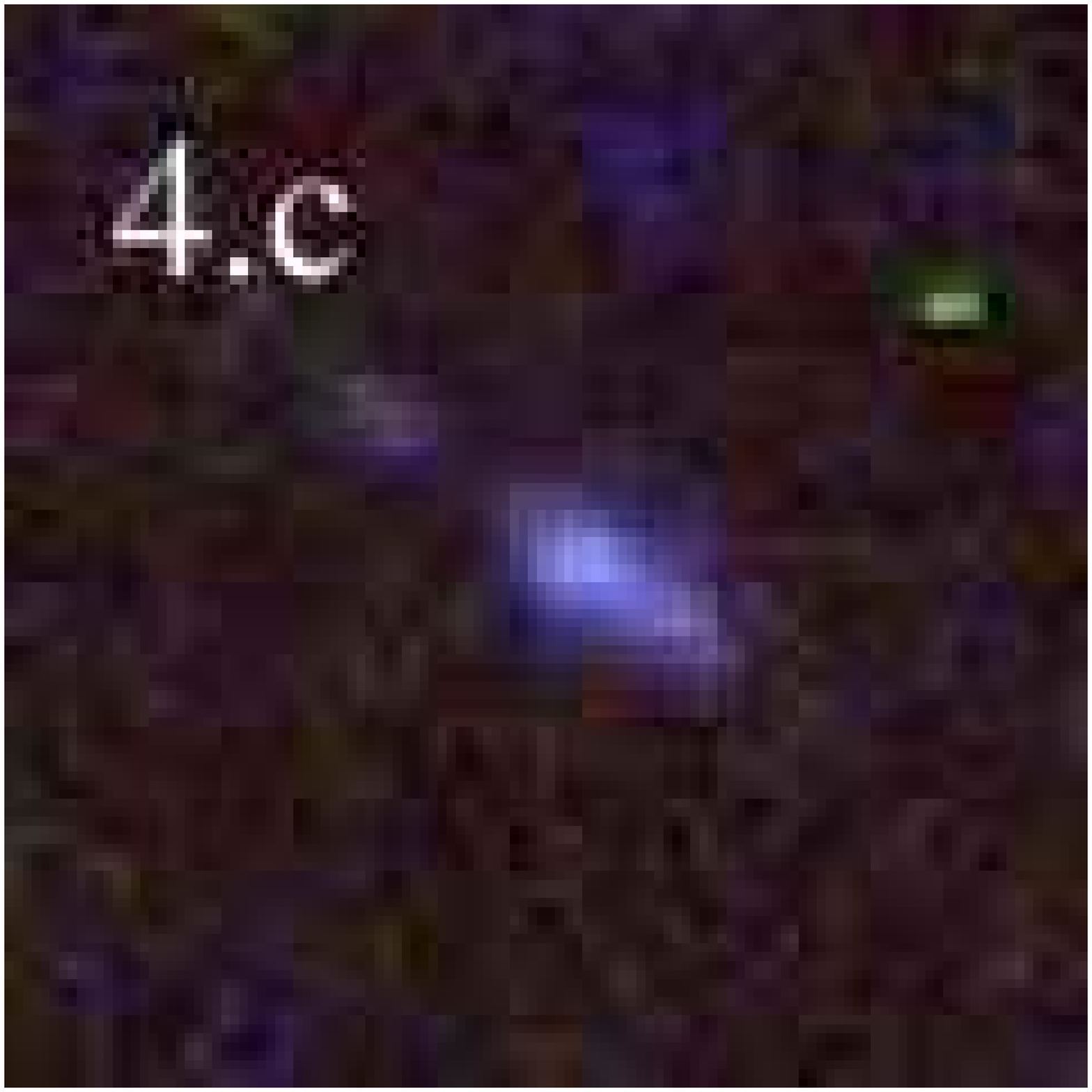}}
    & \multicolumn{1}{m{1.7cm}}{\includegraphics[height=2.00cm,clip]{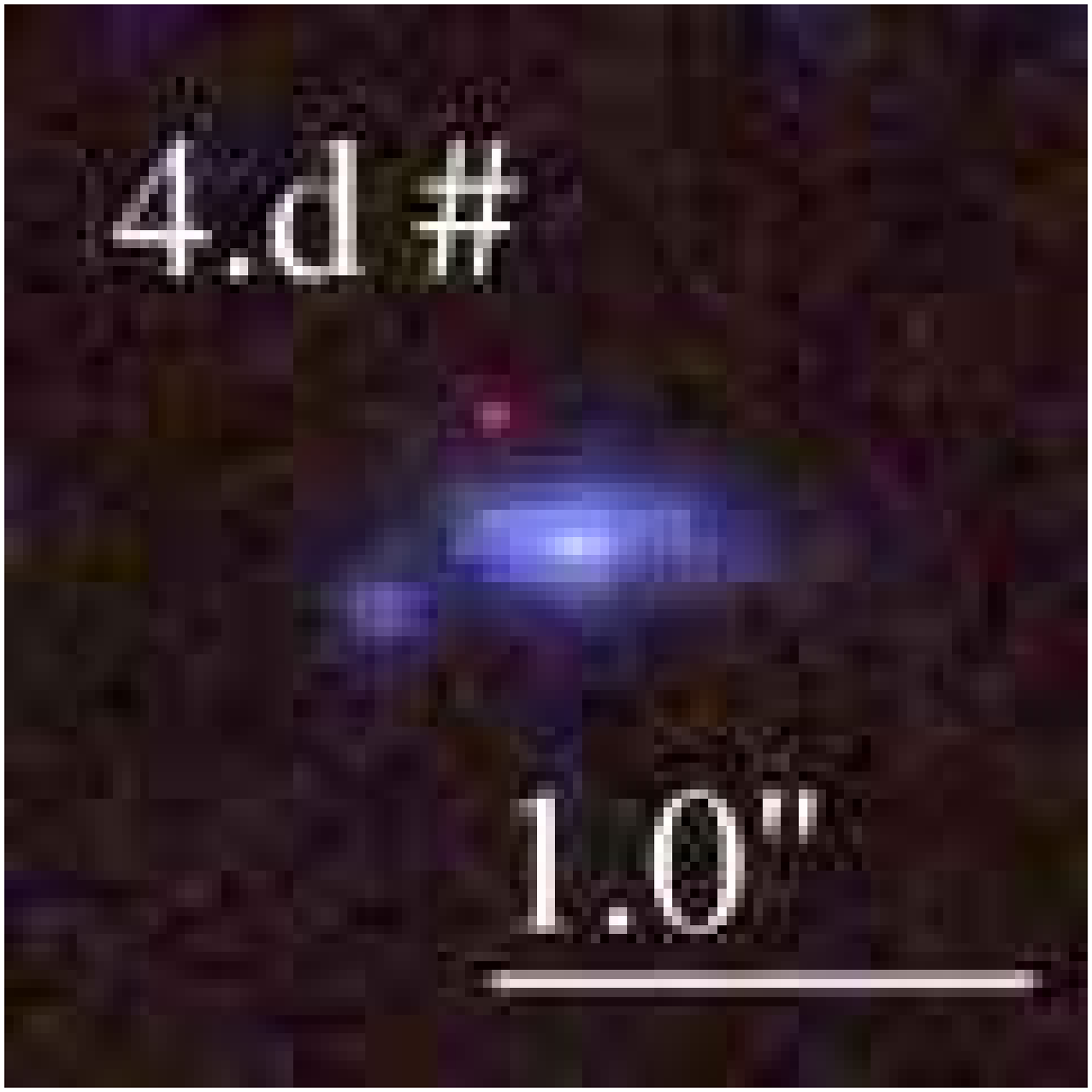}}
    & \multicolumn{1}{m{1.7cm}}{\includegraphics[height=2.00cm,clip]{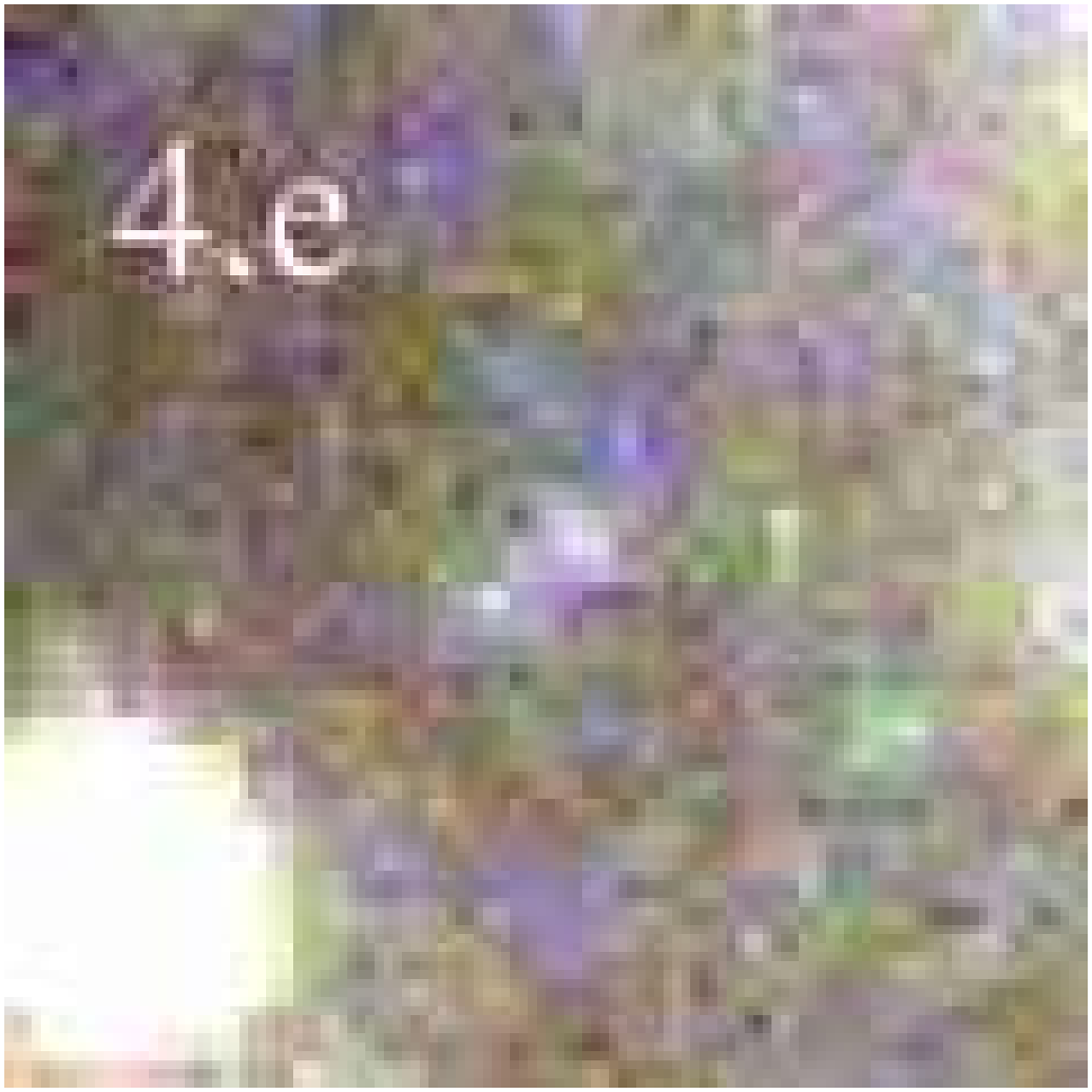}} \\
    \multicolumn{1}{m{1cm}}{{\Large NSIE}}
    & \multicolumn{1}{m{1.7cm}}{\includegraphics[height=2.00cm,clip]{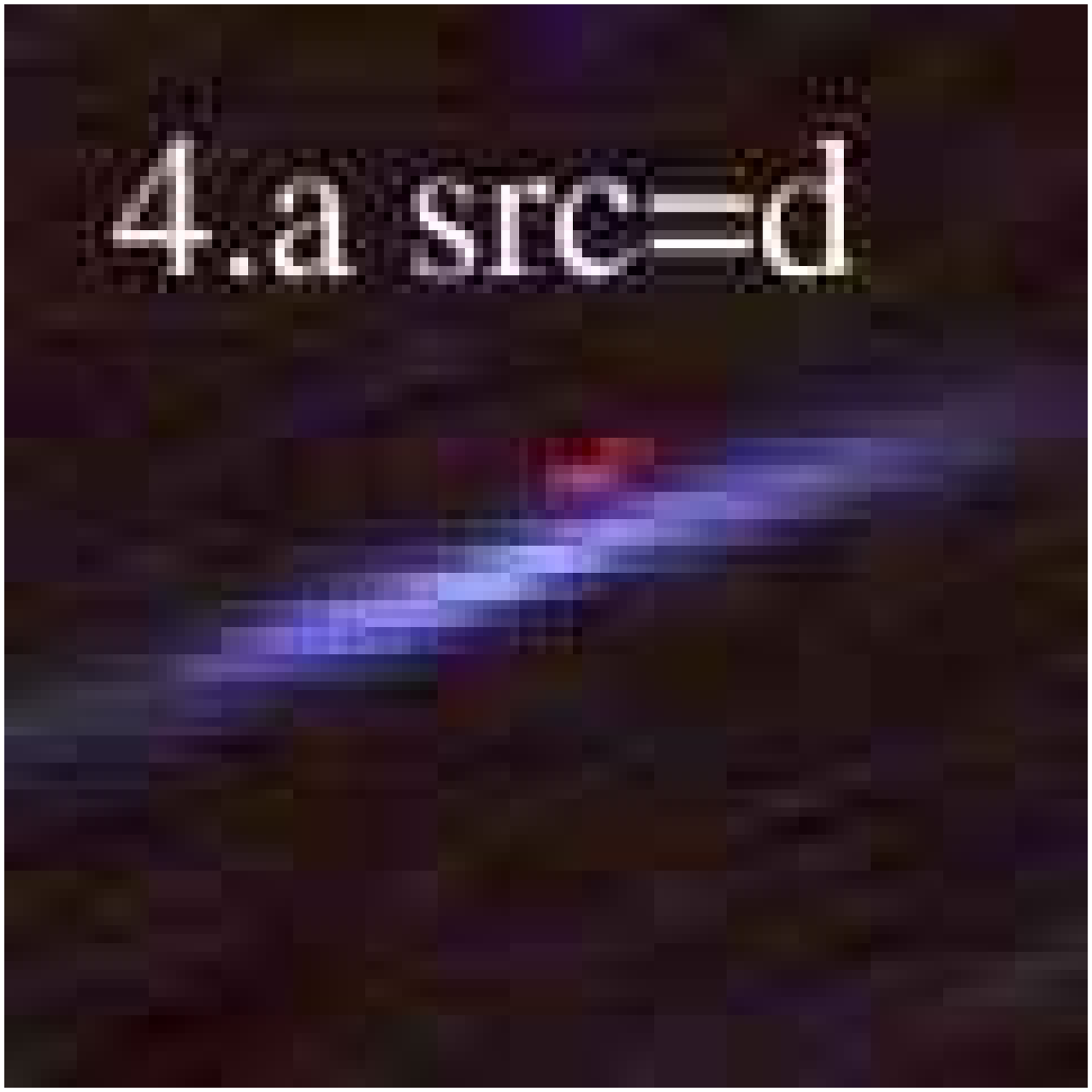}}
    & \multicolumn{1}{m{1.7cm}}{\includegraphics[height=2.00cm,clip]{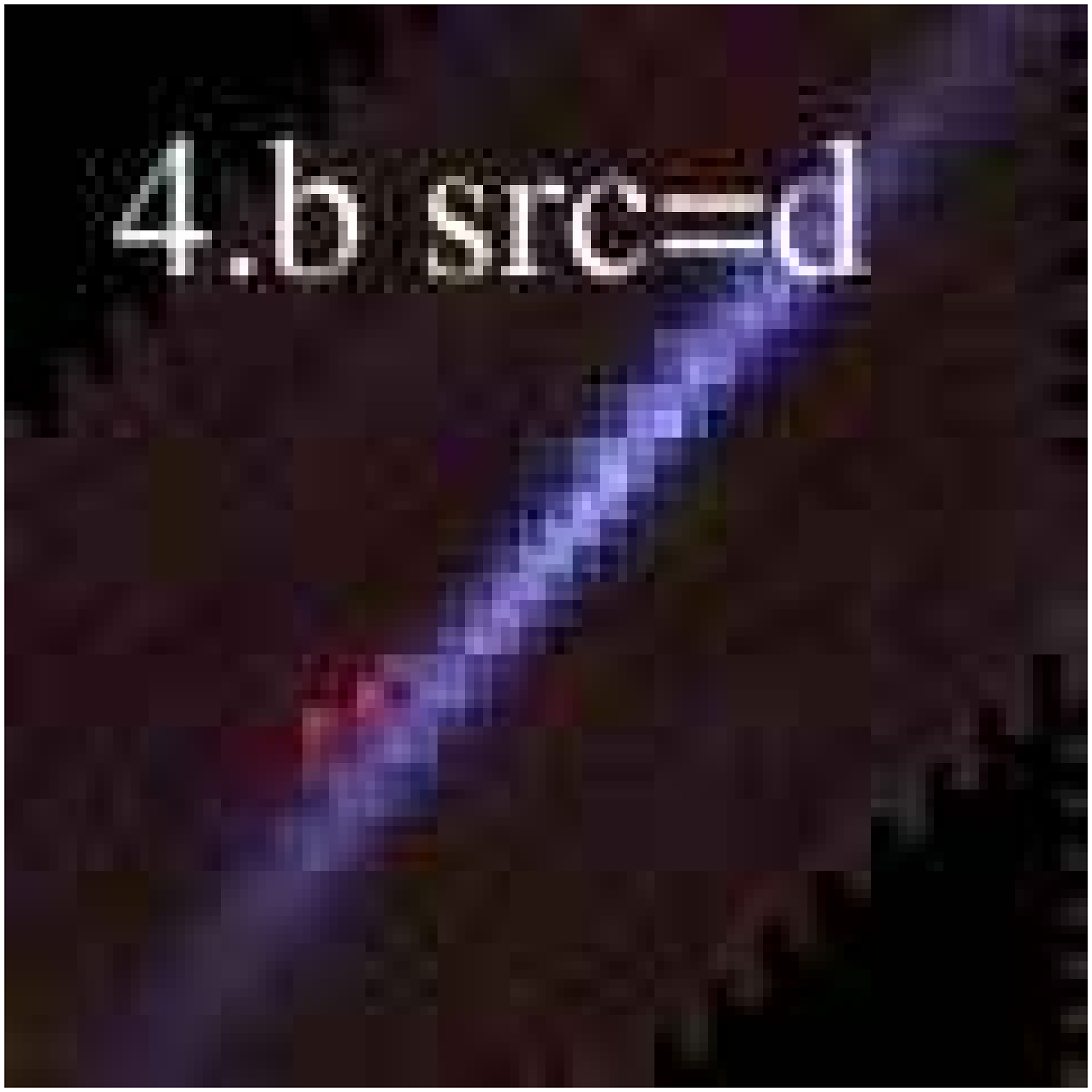}}
    & \multicolumn{1}{m{1.7cm}}{\includegraphics[height=2.00cm,clip]{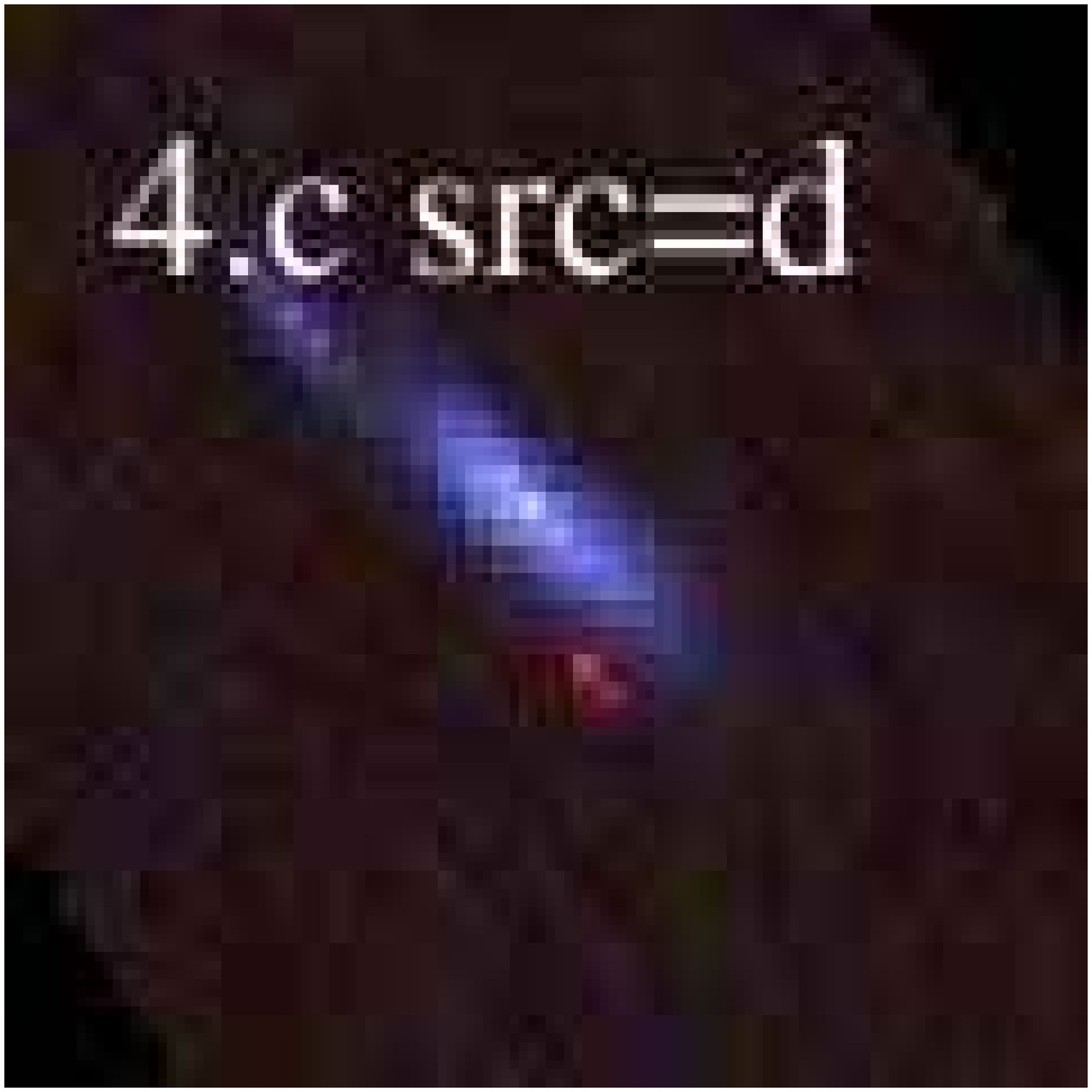}}
    & \multicolumn{1}{m{1.7cm}}{\includegraphics[height=2.00cm,clip]{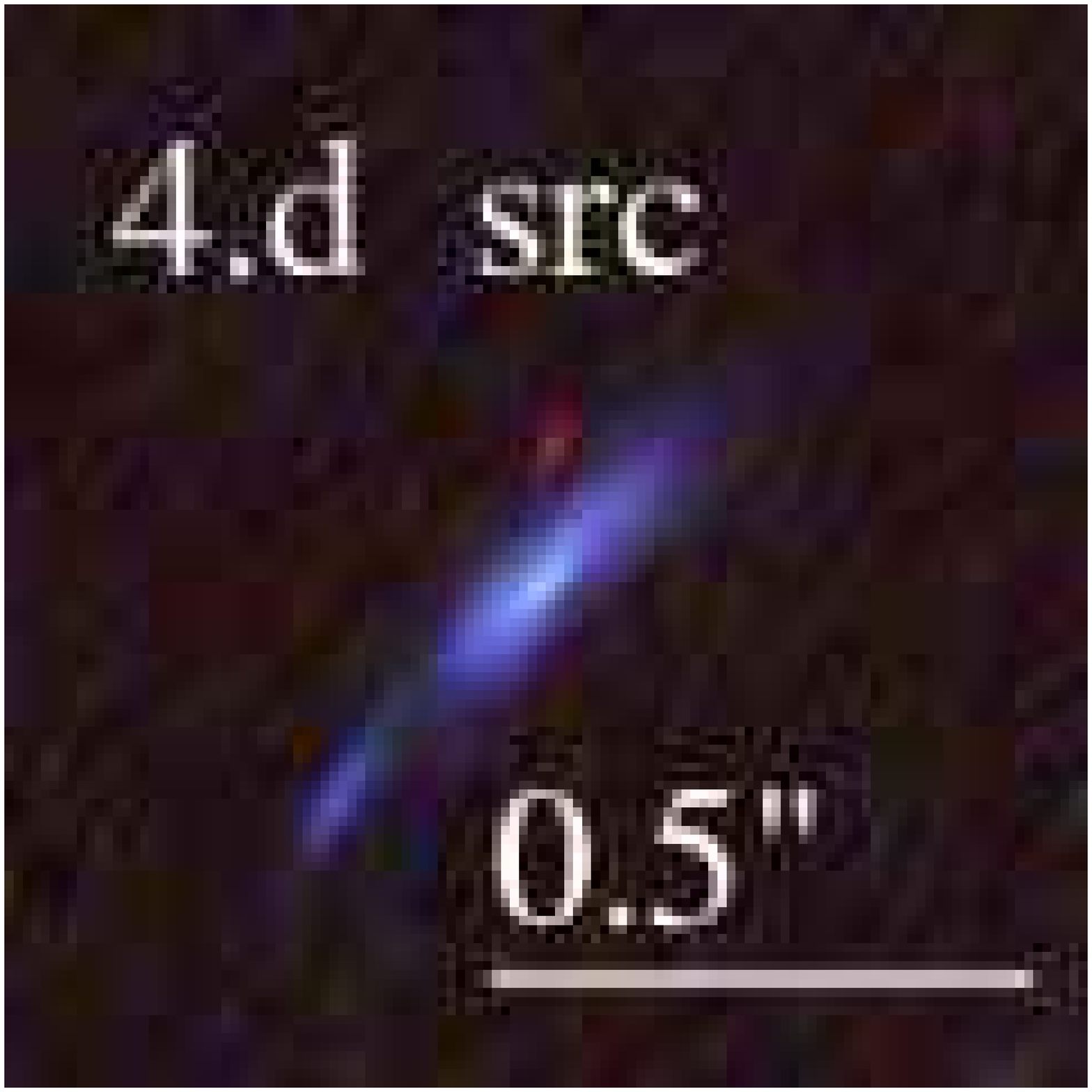}}
    & \multicolumn{1}{m{1.7cm}}{\includegraphics[height=2.00cm,clip]{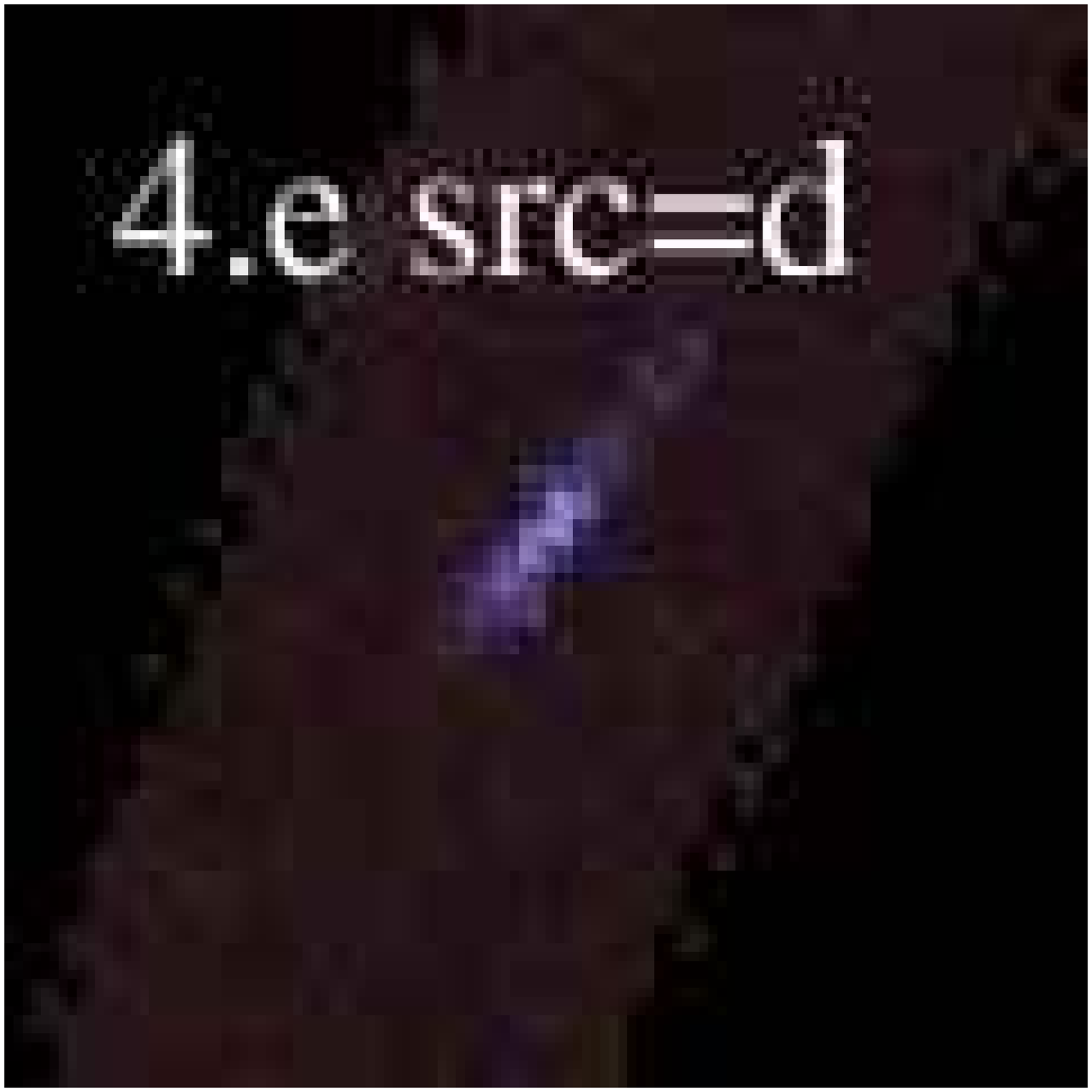}} \\
    \multicolumn{1}{m{1cm}}{{\Large ENFW}}
    & \multicolumn{1}{m{1.7cm}}{\includegraphics[height=2.00cm,clip]{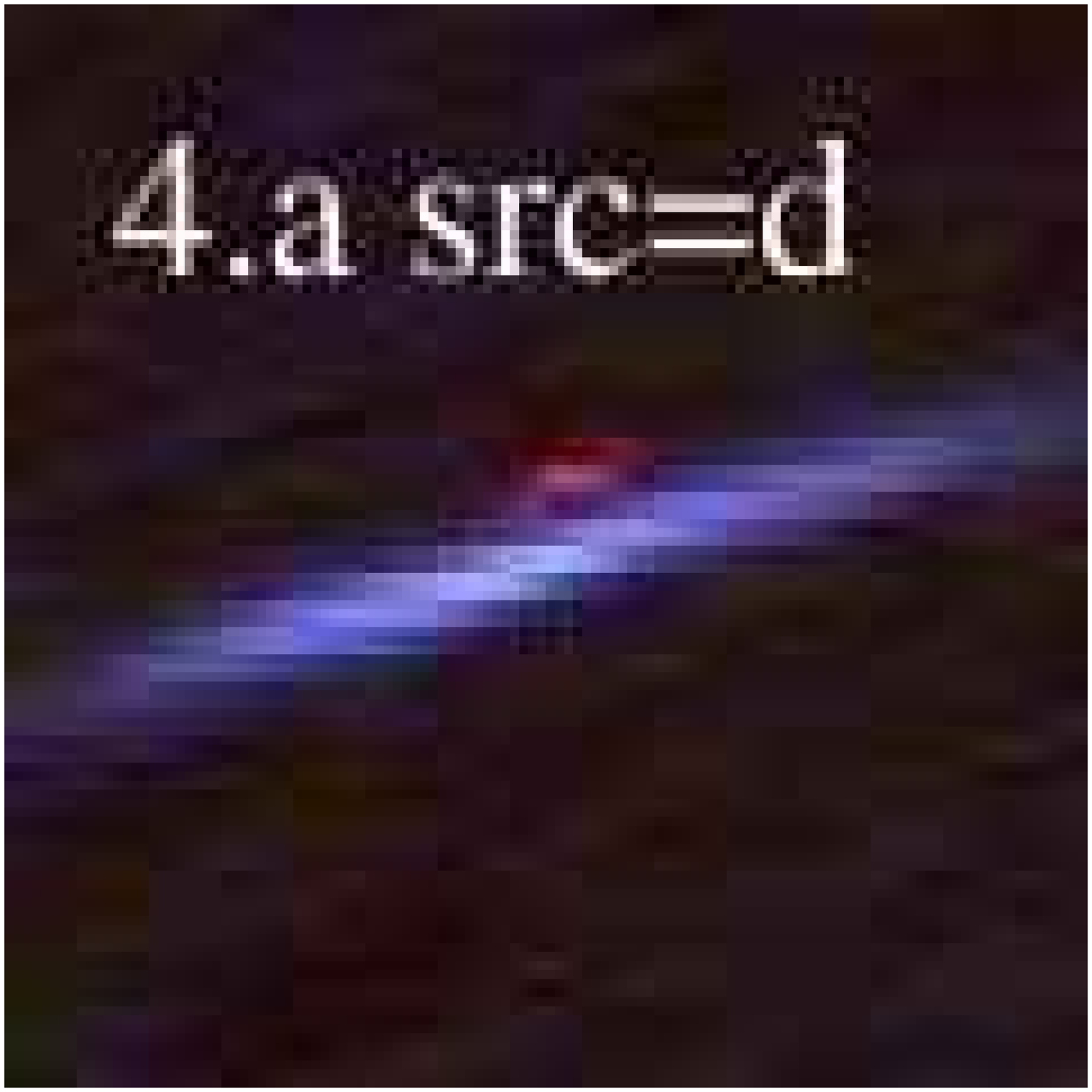}}
    & \multicolumn{1}{m{1.7cm}}{\includegraphics[height=2.00cm,clip]{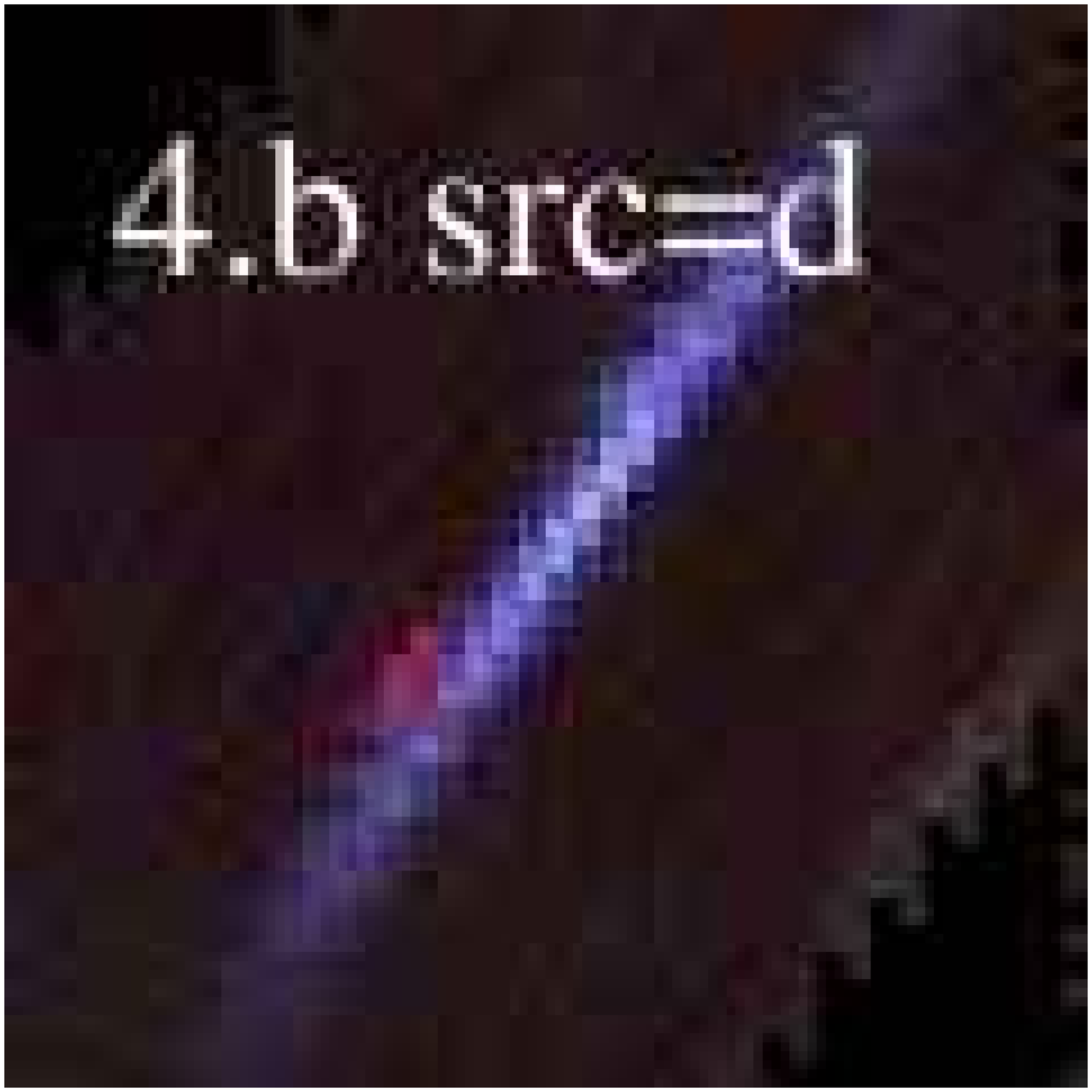}}
    & \multicolumn{1}{m{1.7cm}}{\includegraphics[height=2.00cm,clip]{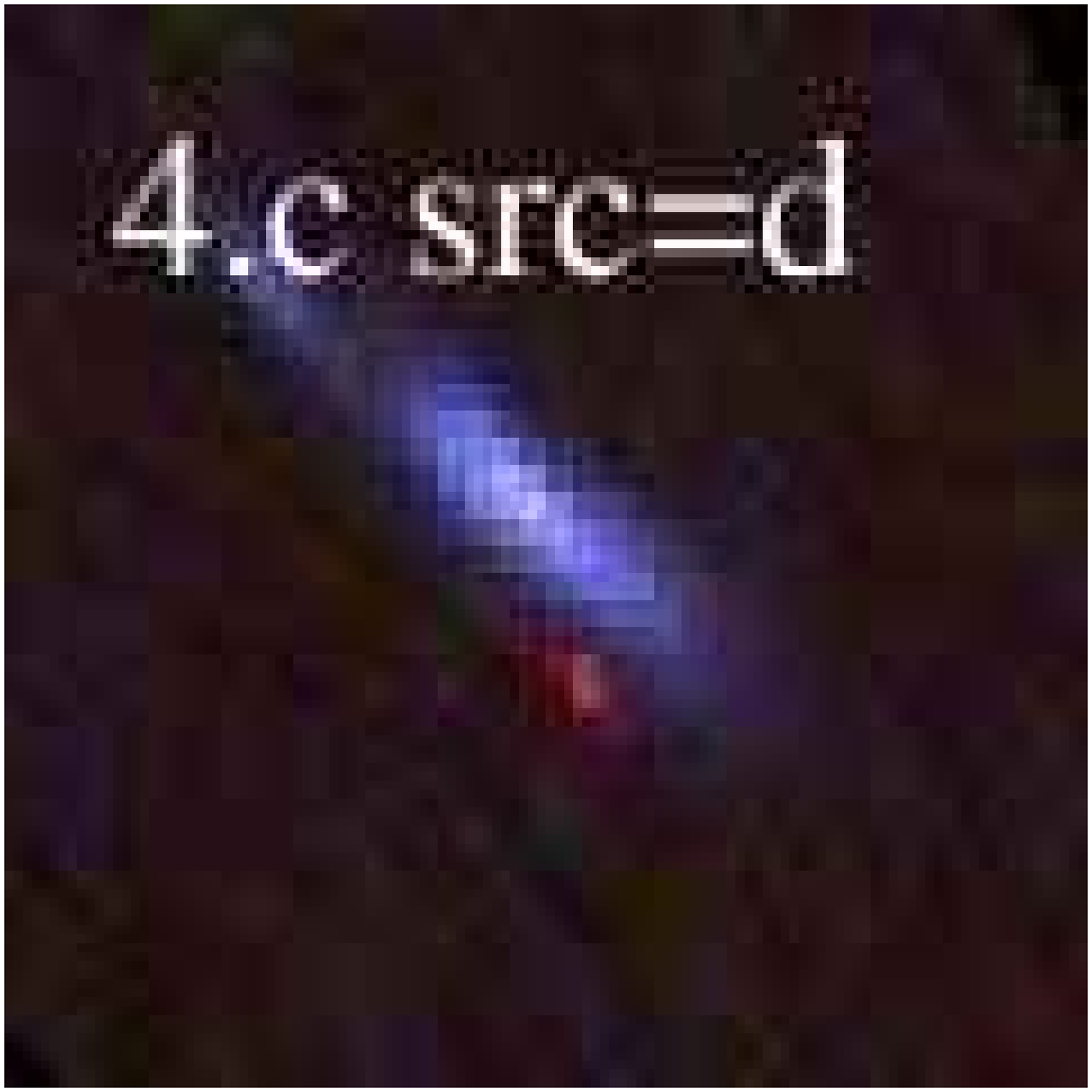}}
    & \multicolumn{1}{m{1.7cm}}{\includegraphics[height=2.00cm,clip]{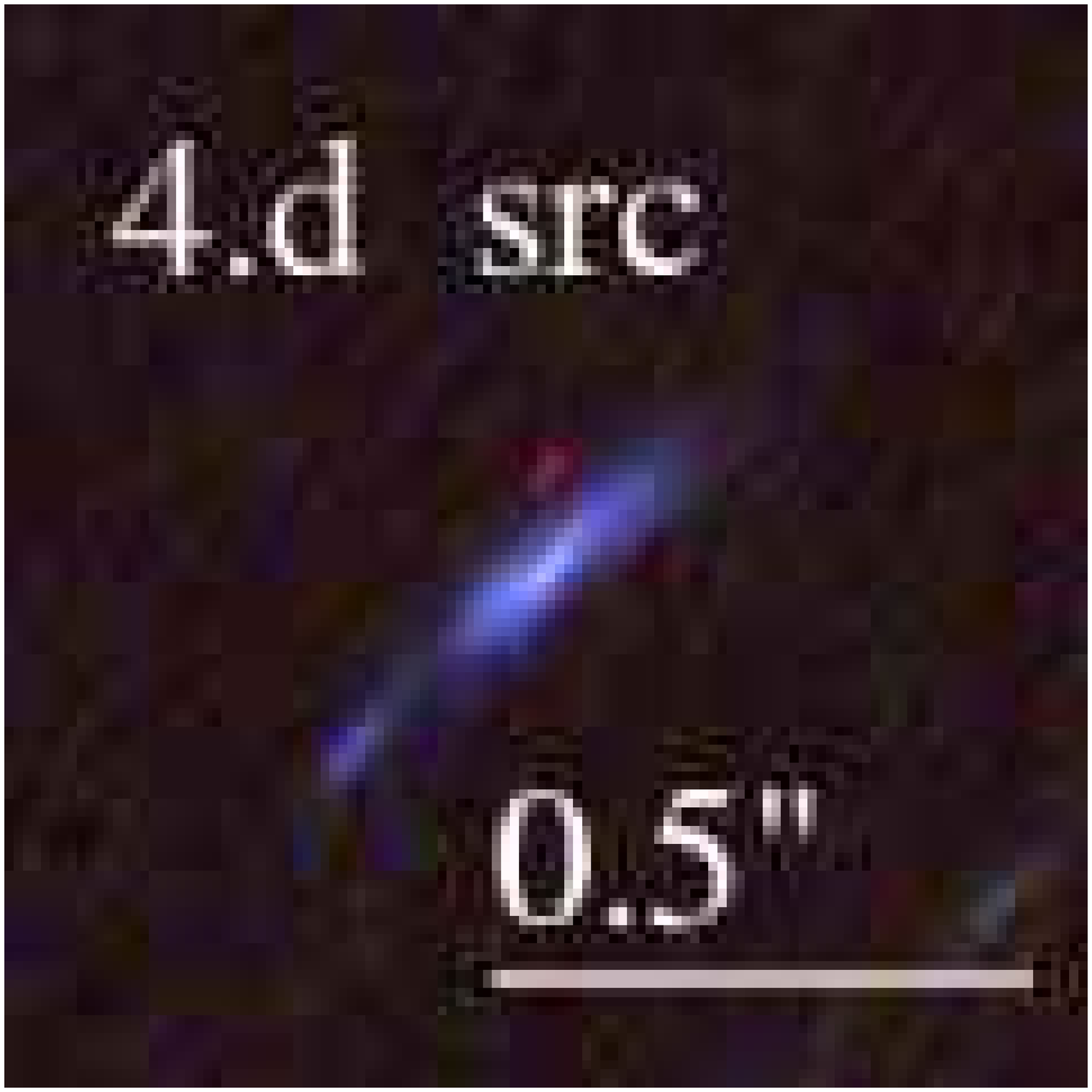}}
    & \multicolumn{1}{m{1.7cm}}{\includegraphics[height=2.00cm,clip]{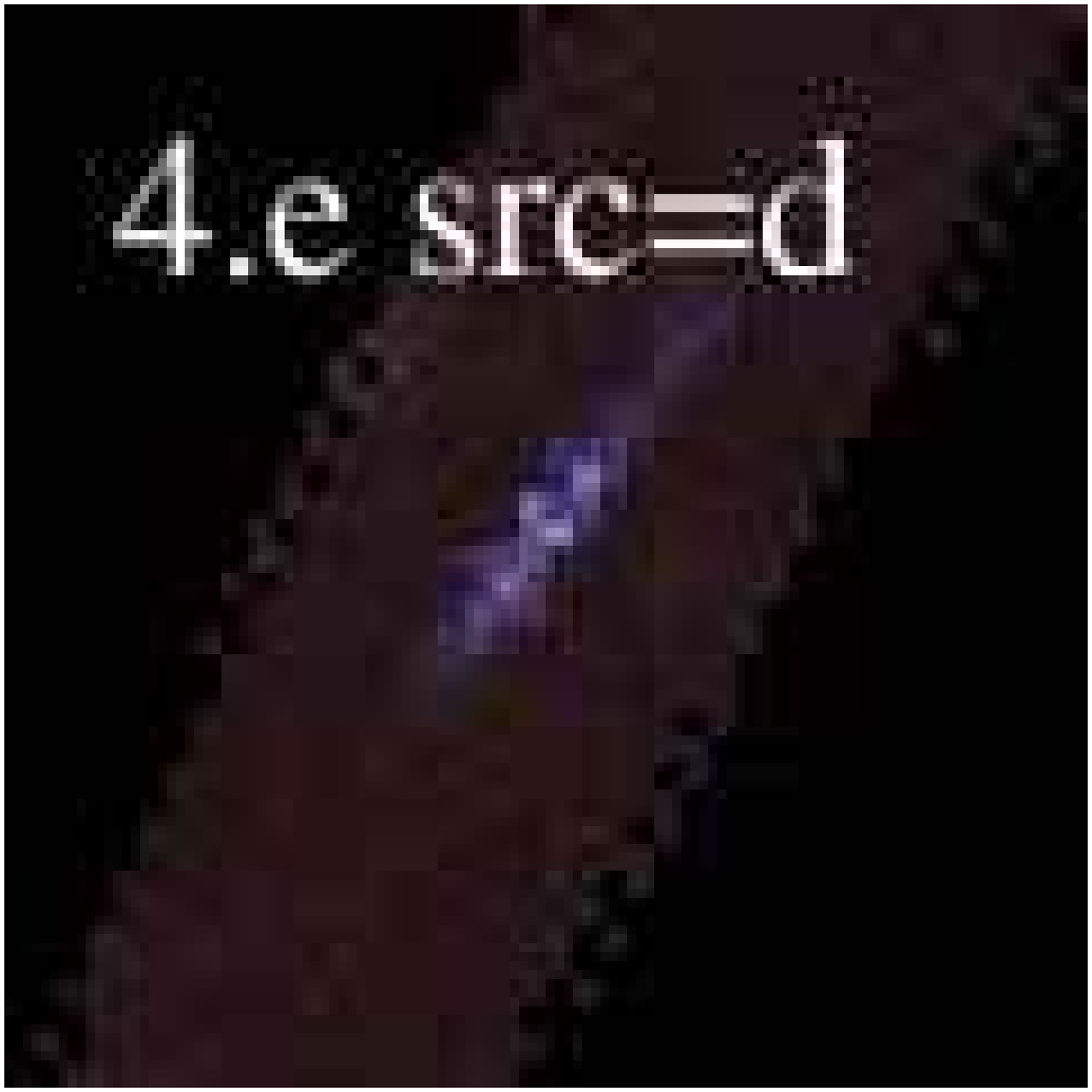}} \\
  \end{tabular}

\end{table*}

\begin{table*}
  \caption{Image system 5:}\vspace{0mm}
  \begin{tabular}{cccc}
    \multicolumn{1}{m{1cm}}{{\Large A1689}}
    & \multicolumn{1}{m{1.7cm}}{\includegraphics[height=2.00cm,clip]{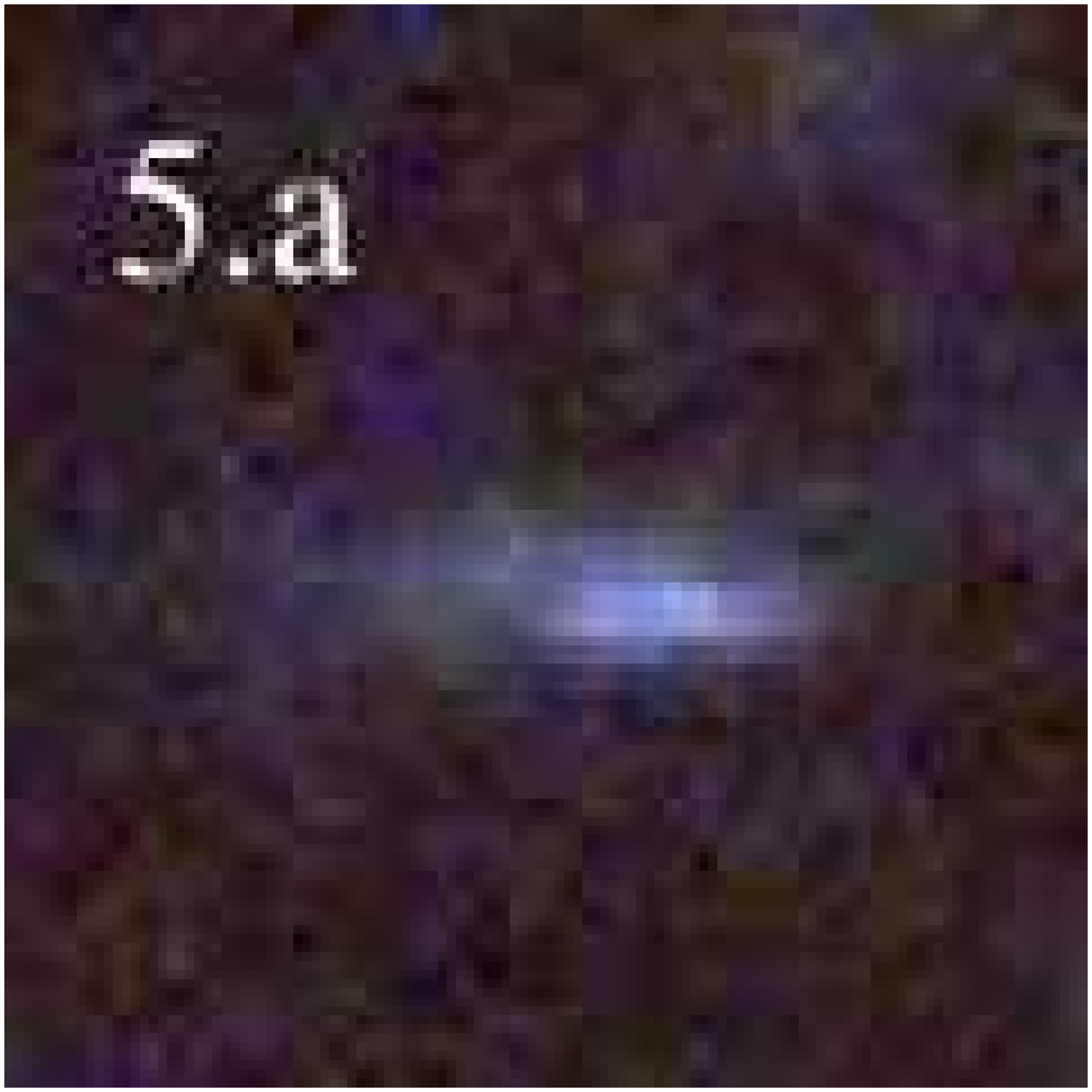}}
    & \multicolumn{1}{m{1.7cm}}{\includegraphics[height=2.00cm,clip]{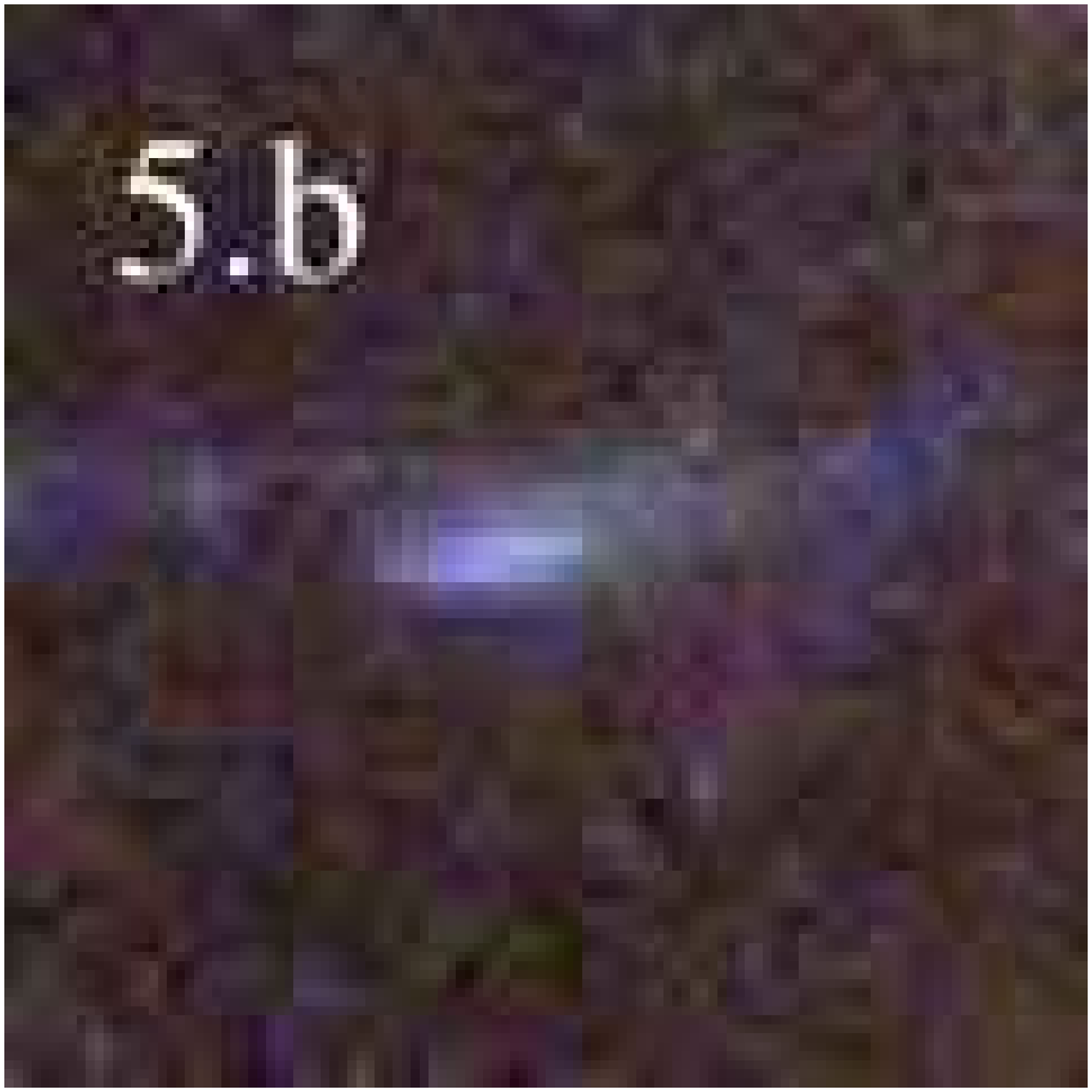}}
    & \multicolumn{1}{m{1.7cm}}{\includegraphics[height=2.00cm,clip]{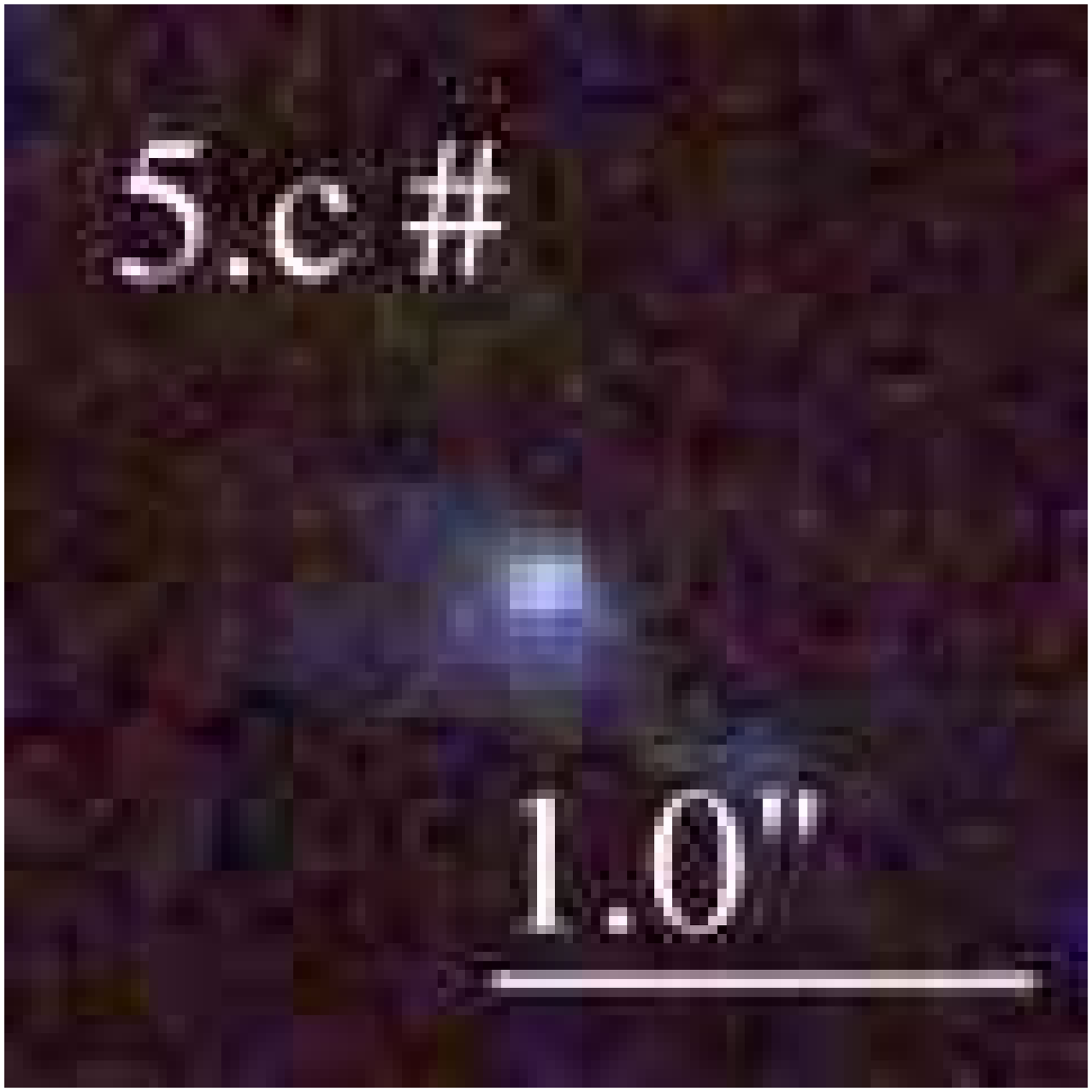}} \\
    \multicolumn{1}{m{1cm}}{{\Large NSIE}}
    & \multicolumn{1}{m{1.7cm}}{\includegraphics[height=2.00cm,clip]{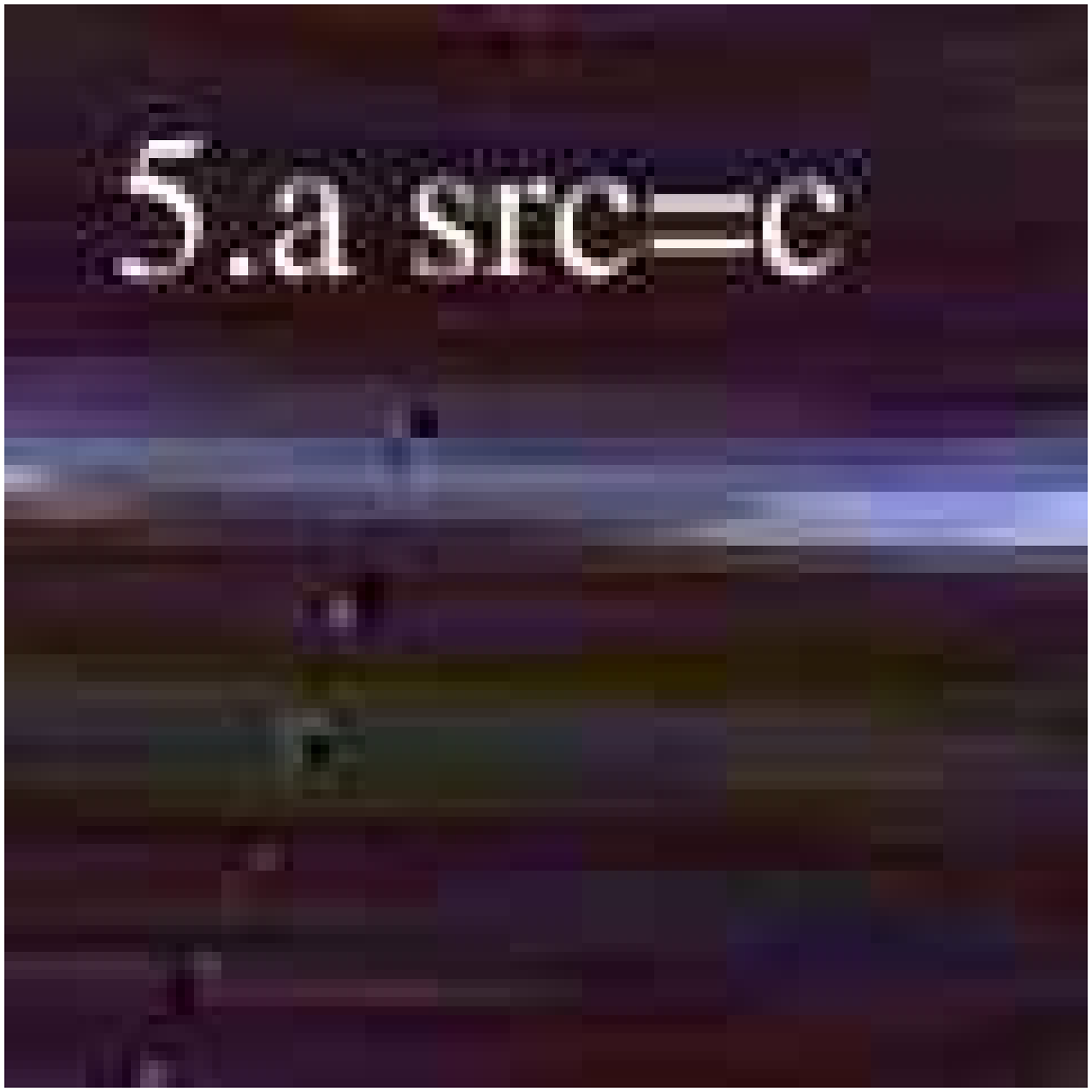}}
    & \multicolumn{1}{m{1.7cm}}{\includegraphics[height=2.00cm,clip]{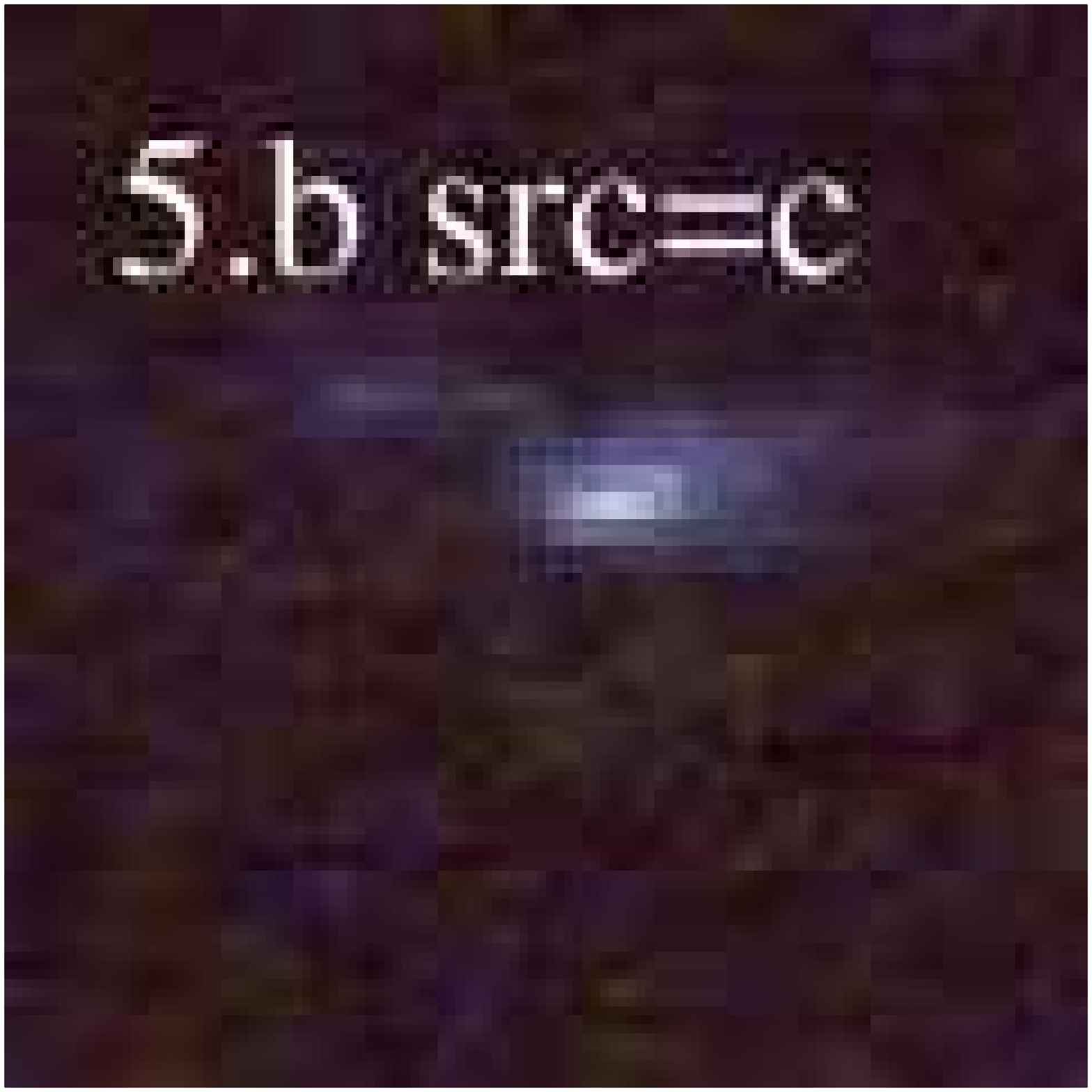}}
    & \multicolumn{1}{m{1.7cm}}{\includegraphics[height=2.00cm,clip]{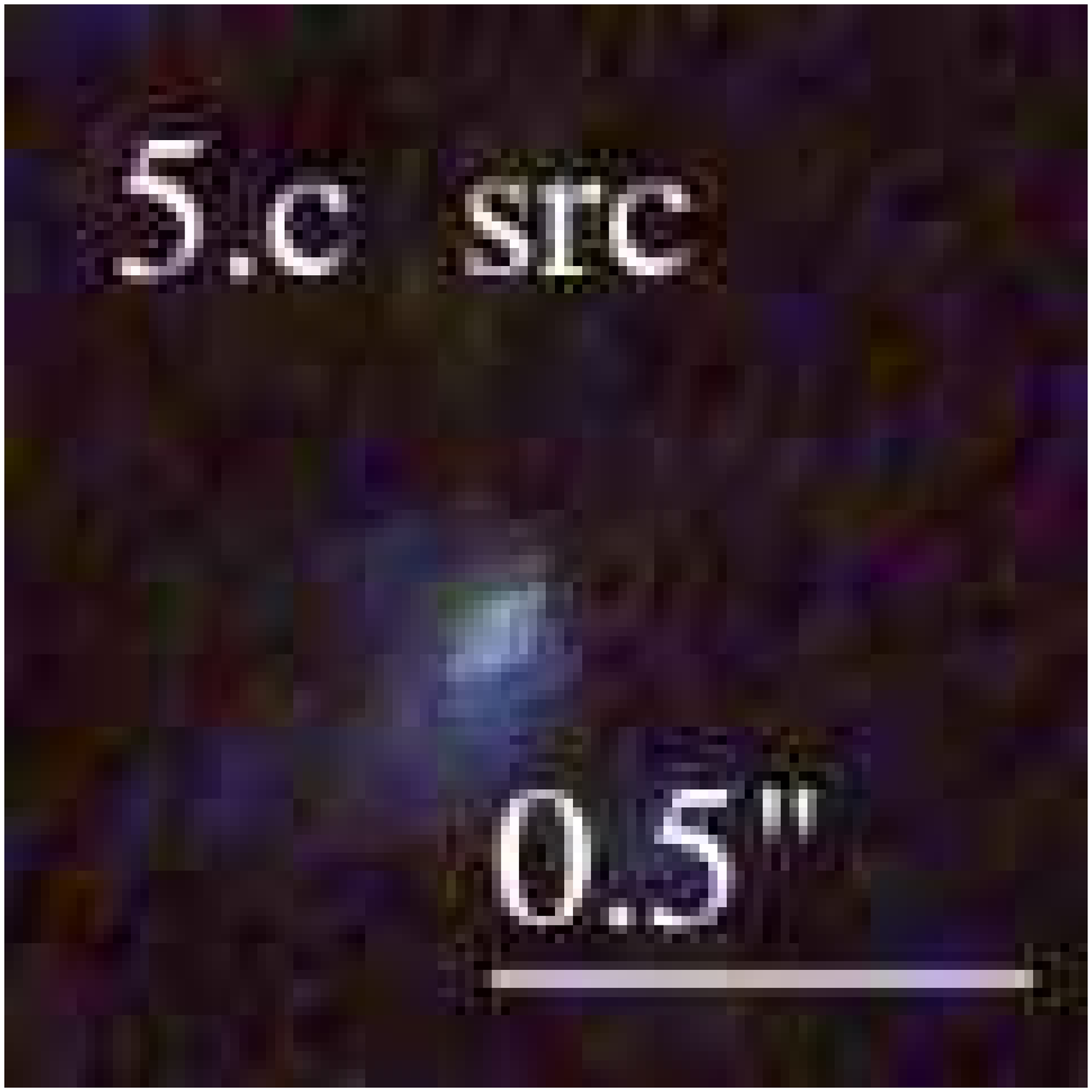}} \\
    \multicolumn{1}{m{1cm}}{{\Large ENFW}}
    & \multicolumn{1}{m{1.7cm}}{\includegraphics[height=2.00cm,clip]{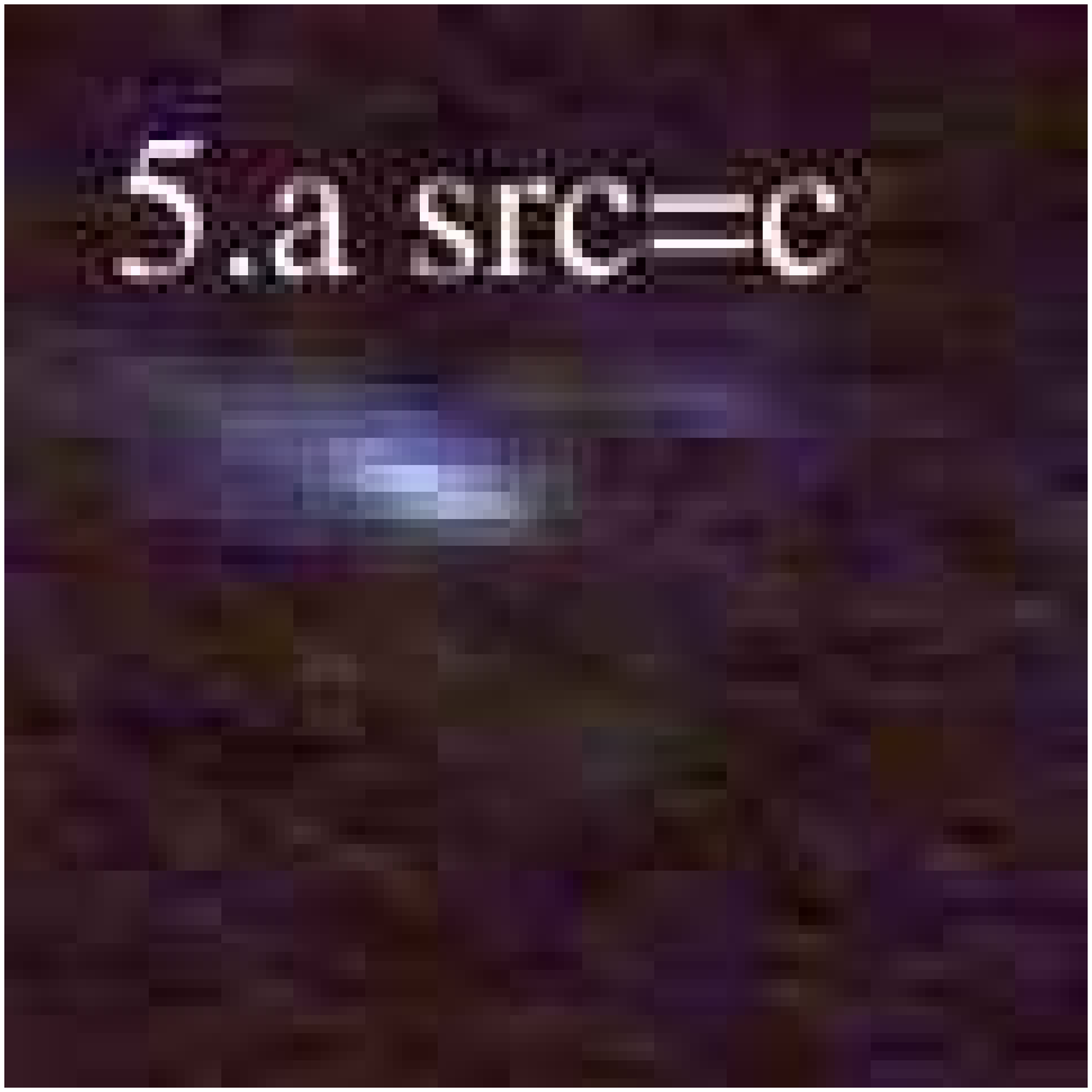}}
    & \multicolumn{1}{m{1.7cm}}{\includegraphics[height=2.00cm,clip]{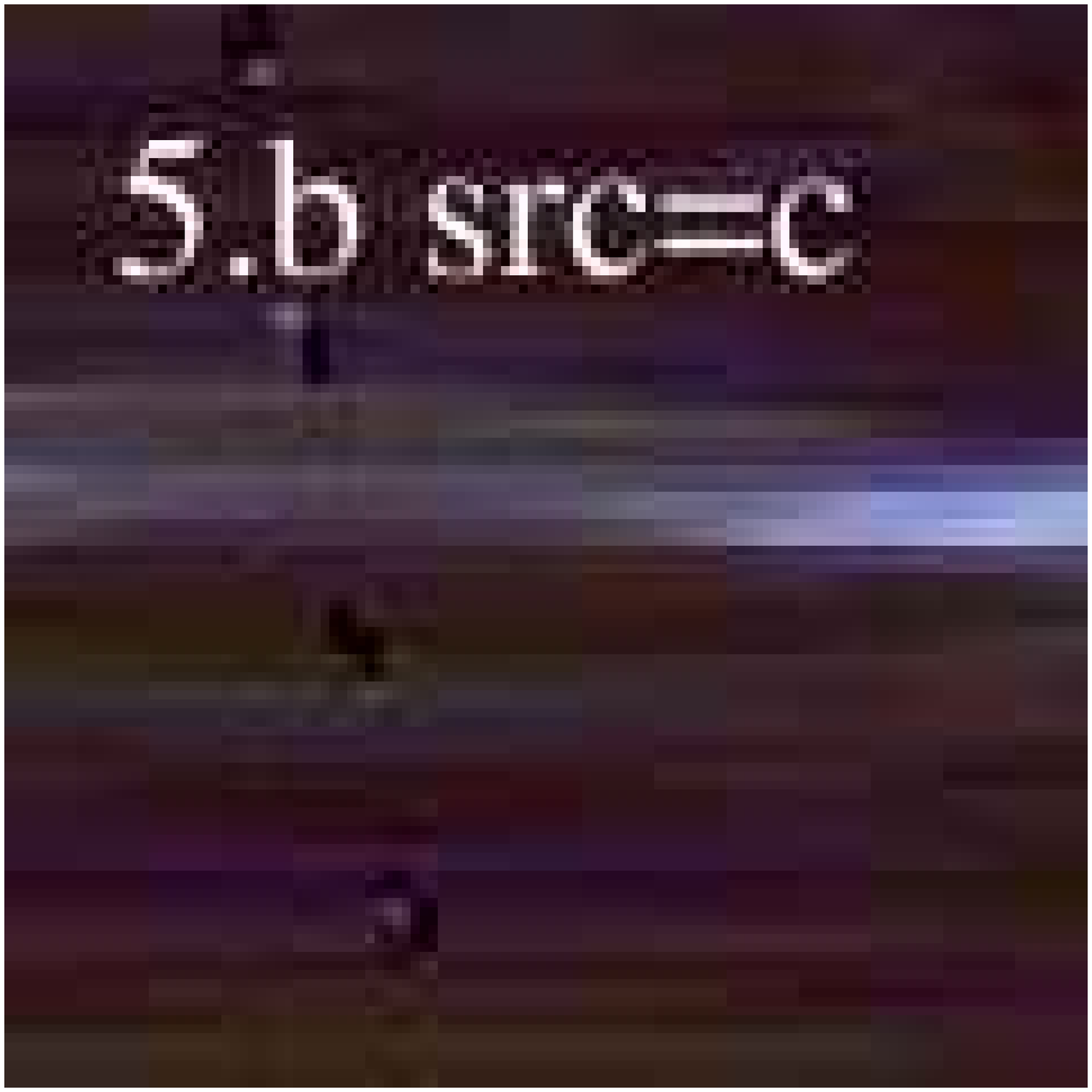}}
    & \multicolumn{1}{m{1.7cm}}{\includegraphics[height=2.00cm,clip]{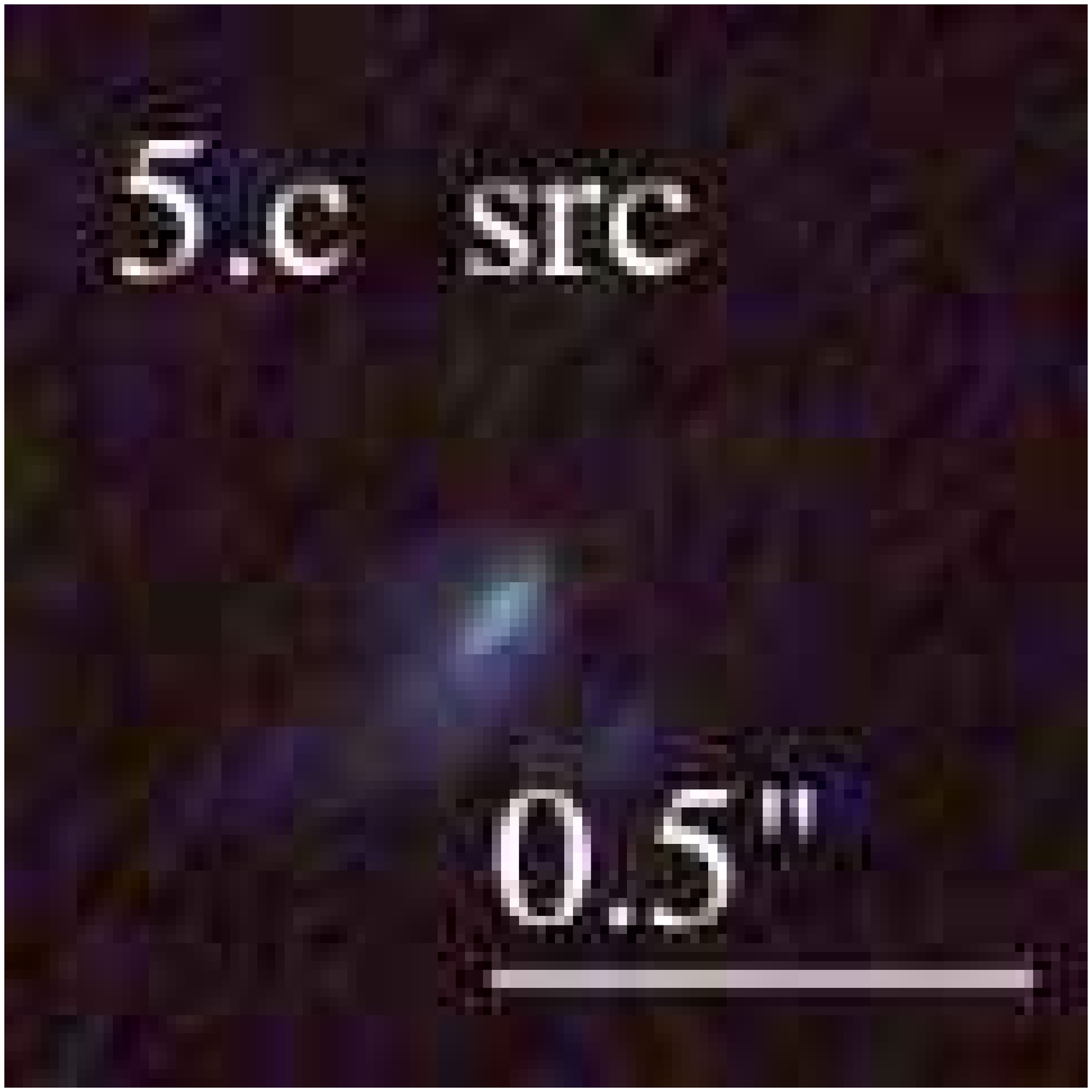}} \\
  \end{tabular}

\end{table*}

\begin{table*}
  \caption{Image system 6:}\vspace{0mm}
  \begin{tabular}{ccccc}
    \multicolumn{1}{m{1cm}}{{\Large A1689}}
    & \multicolumn{1}{m{1.7cm}}{\includegraphics[height=2.00cm,clip]{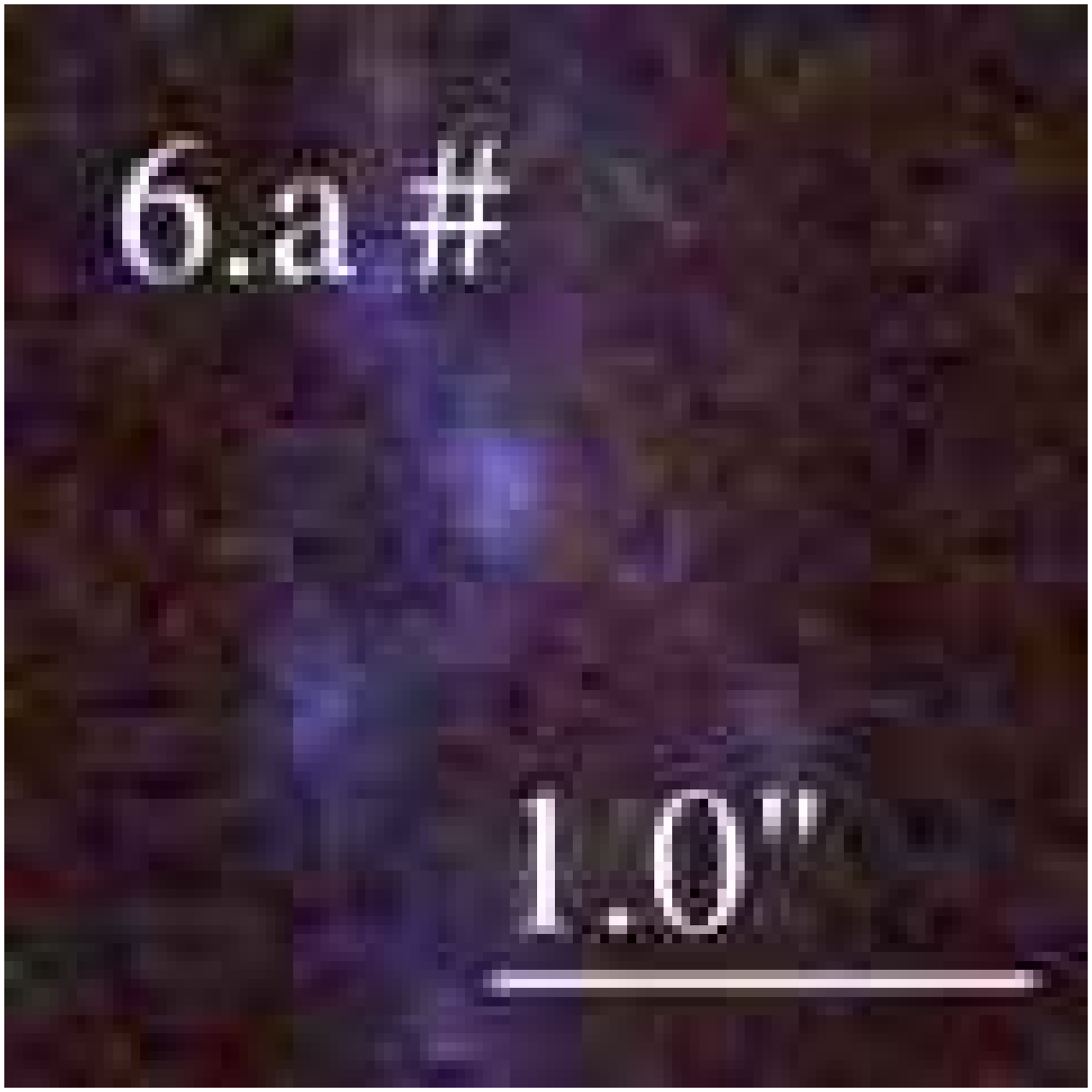}}
    & \multicolumn{1}{m{1.7cm}}{\includegraphics[height=2.00cm,clip]{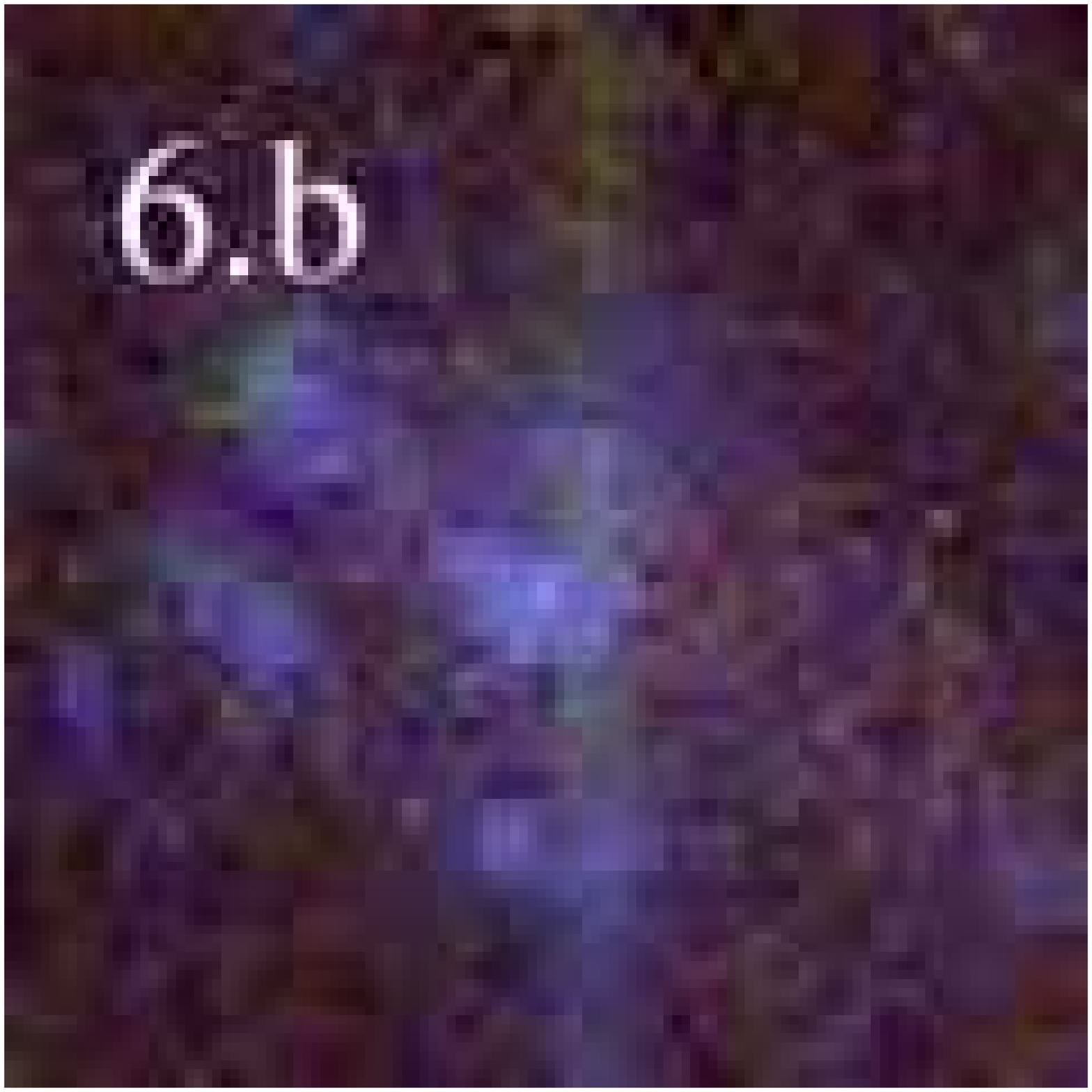}}
    & \multicolumn{1}{m{1.7cm}}{\includegraphics[height=2.00cm,clip]{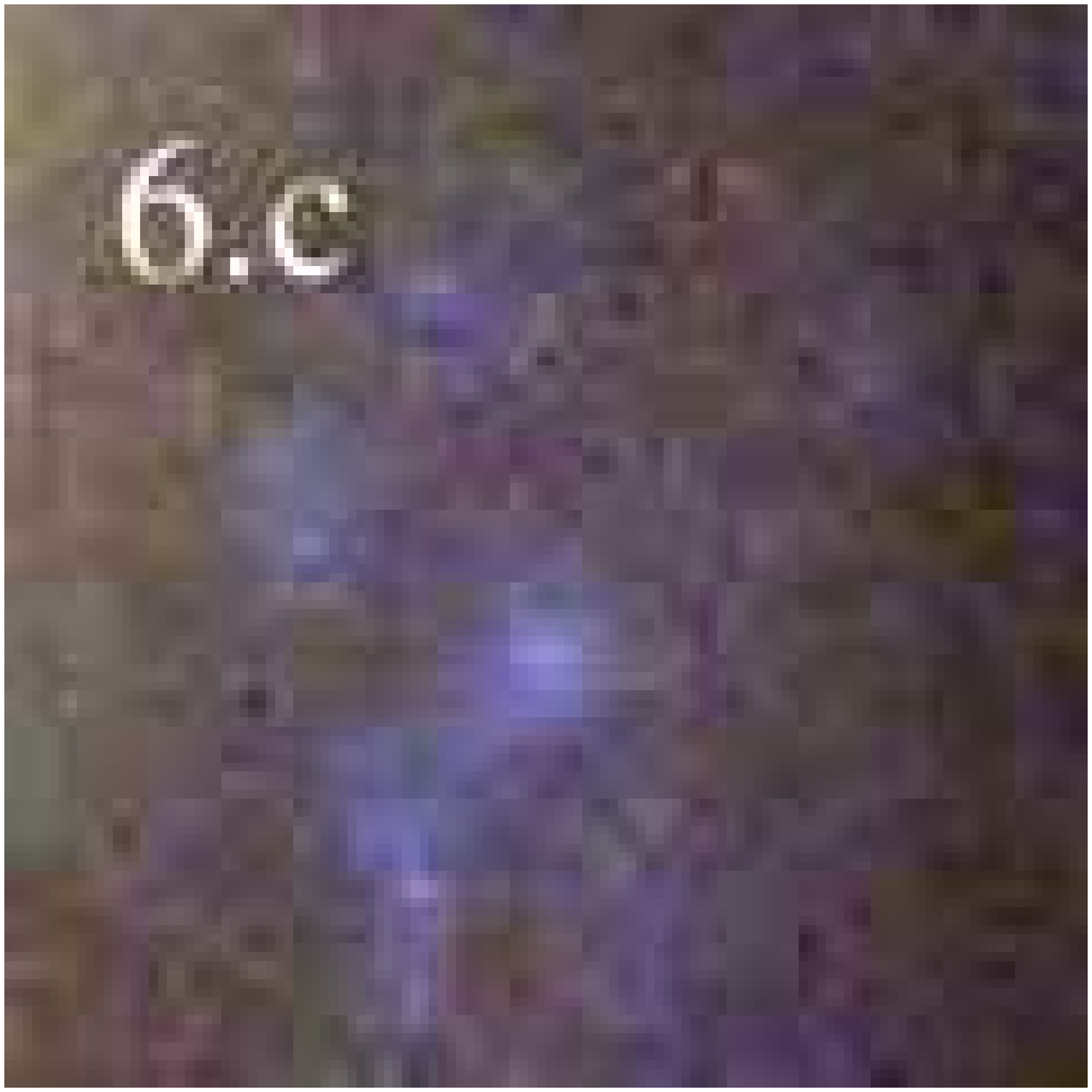}}
    & \multicolumn{1}{m{1.7cm}}{\includegraphics[height=2.00cm,clip]{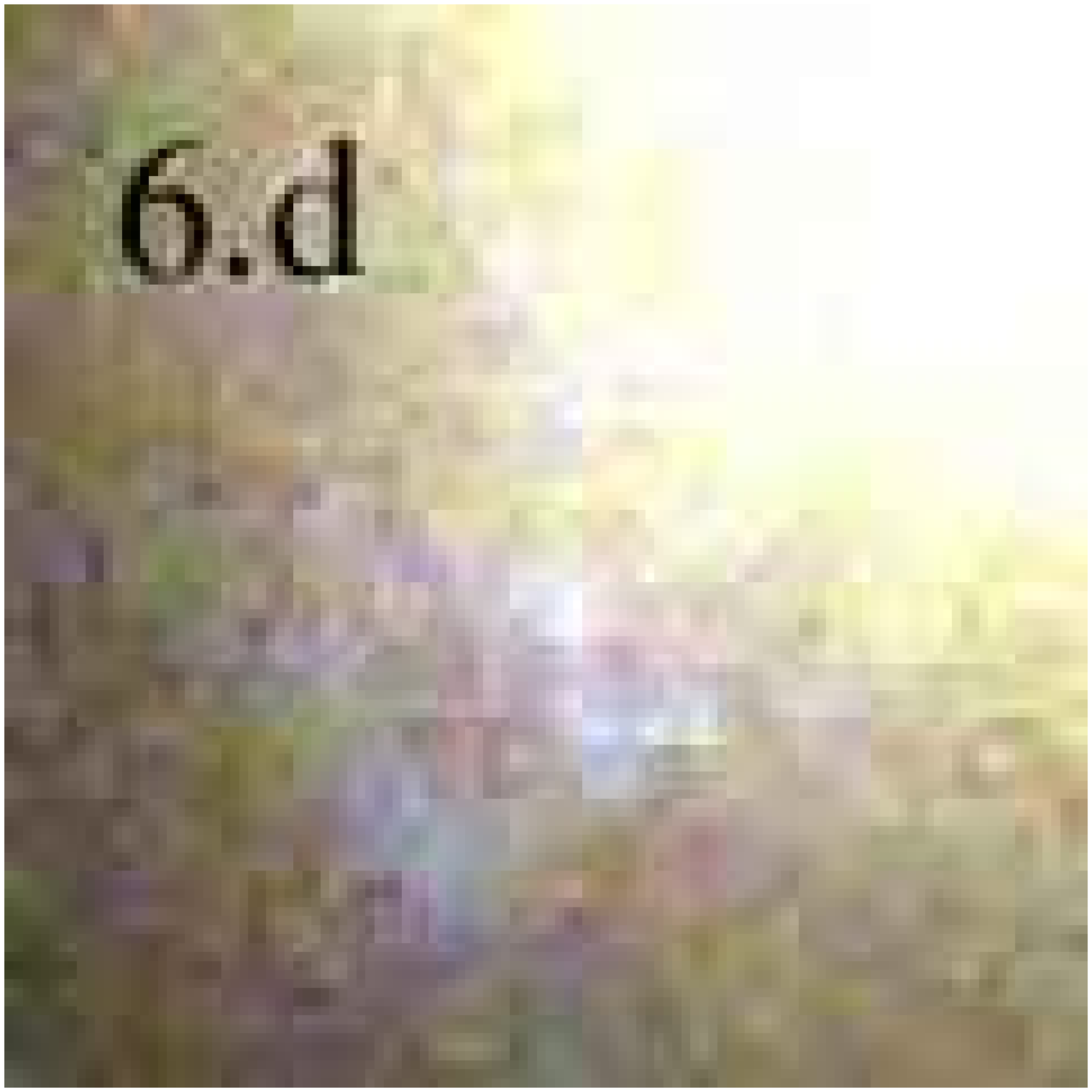}} \\
    \multicolumn{1}{m{1cm}}{{\Large NSIE}}
    & \multicolumn{1}{m{1.7cm}}{\includegraphics[height=2.00cm,clip]{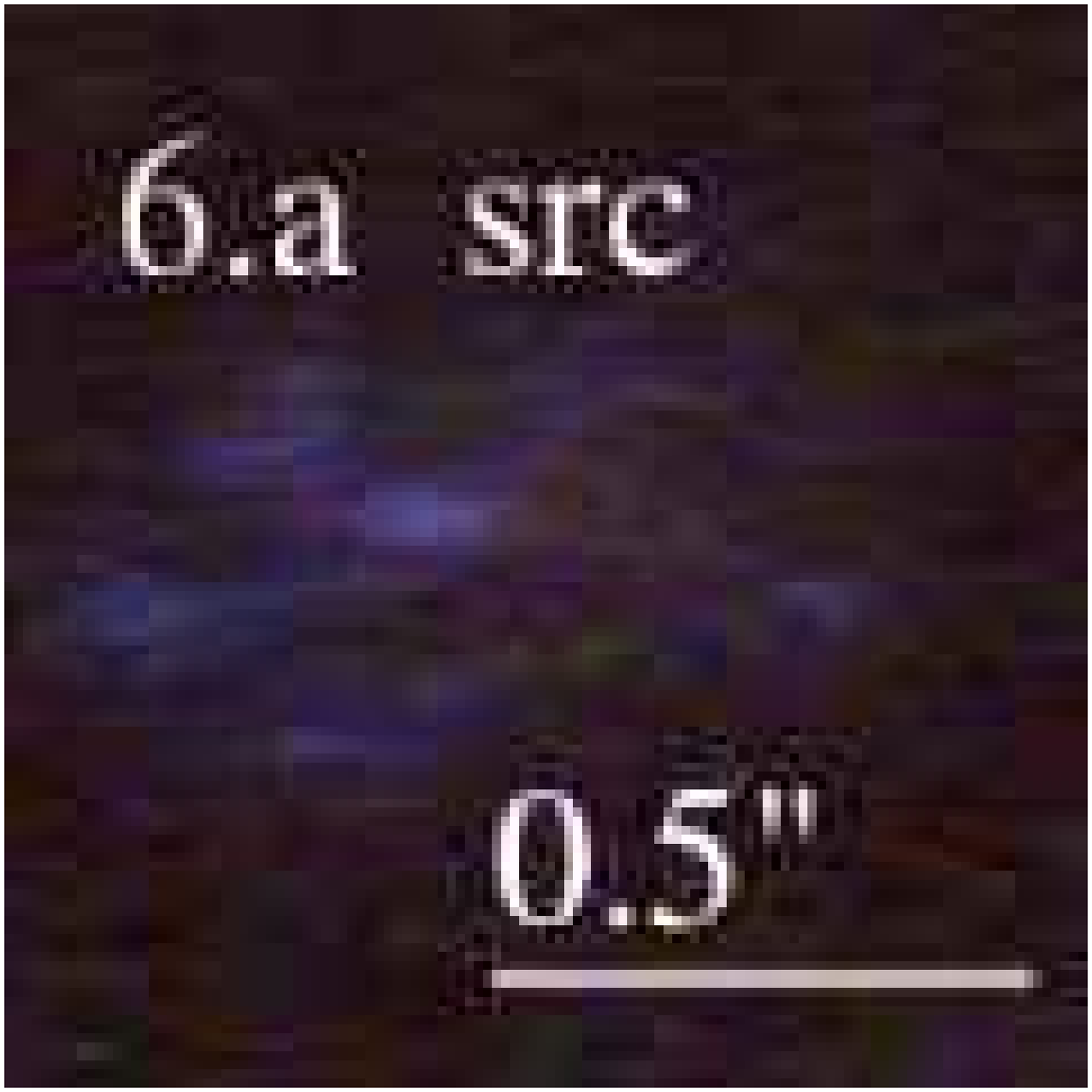}}
    & \multicolumn{1}{m{1.7cm}}{\includegraphics[height=2.00cm,clip]{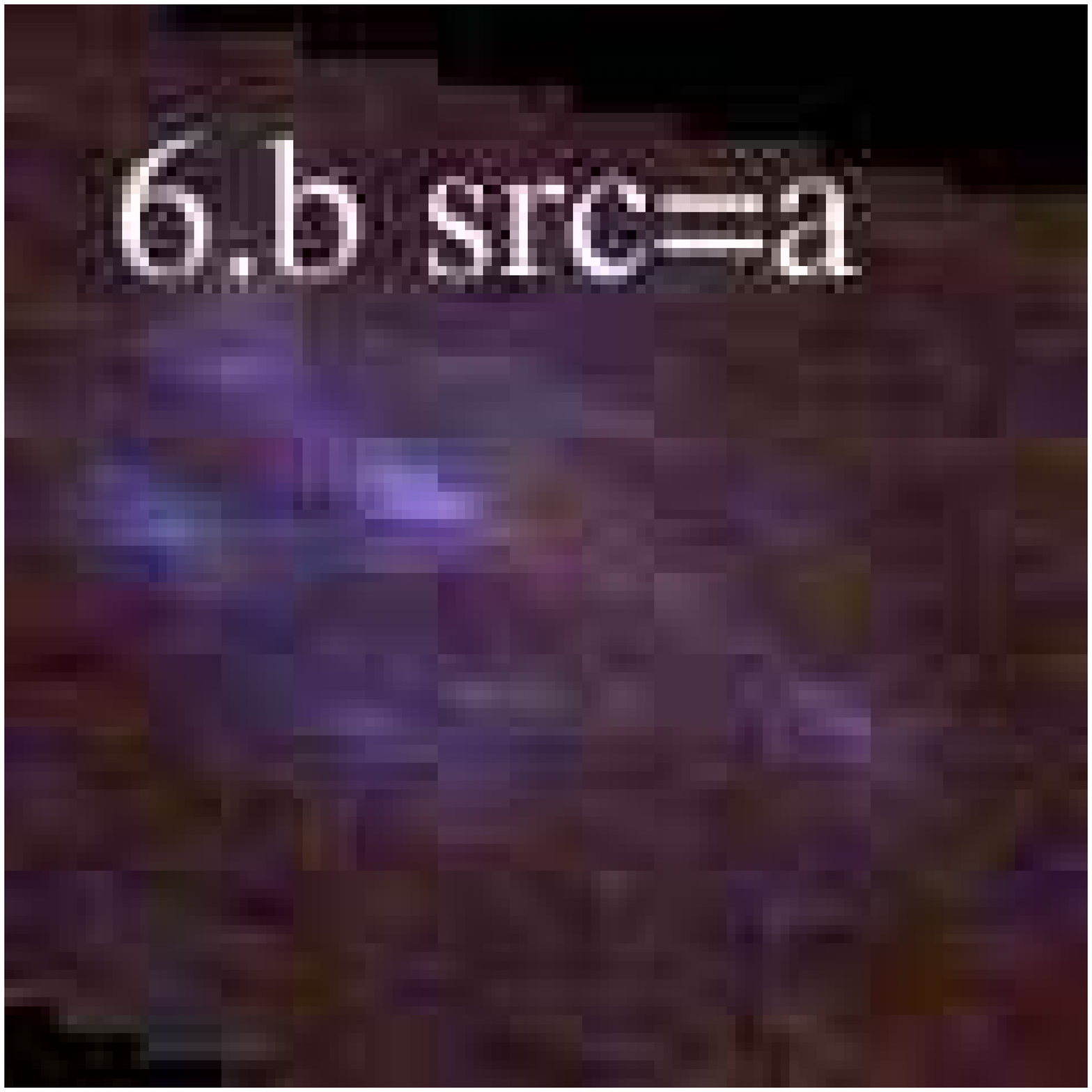}}
    & \multicolumn{1}{m{1.7cm}}{\includegraphics[height=2.00cm,clip]{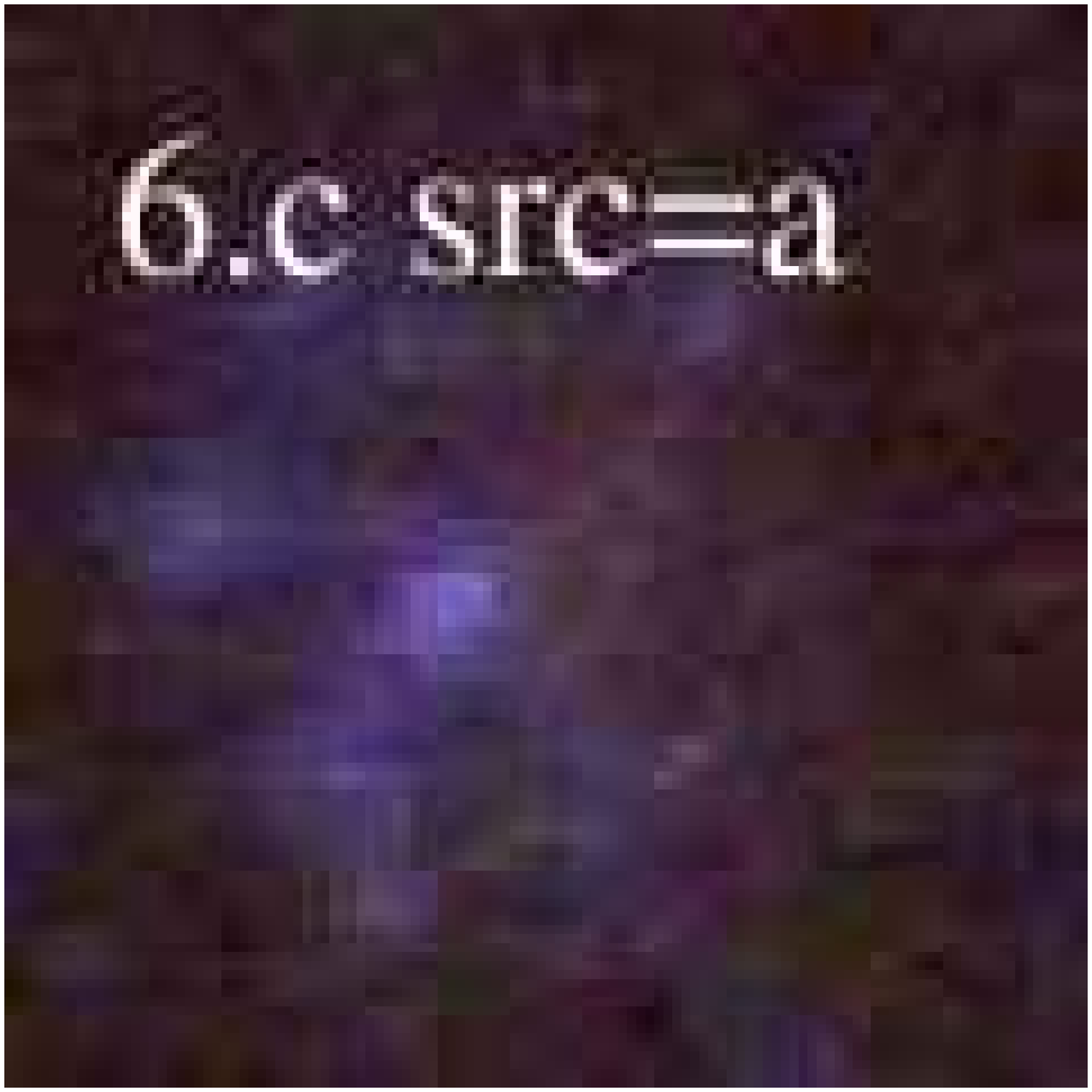}}
    & \multicolumn{1}{m{1.7cm}}{\includegraphics[height=2.00cm,clip]{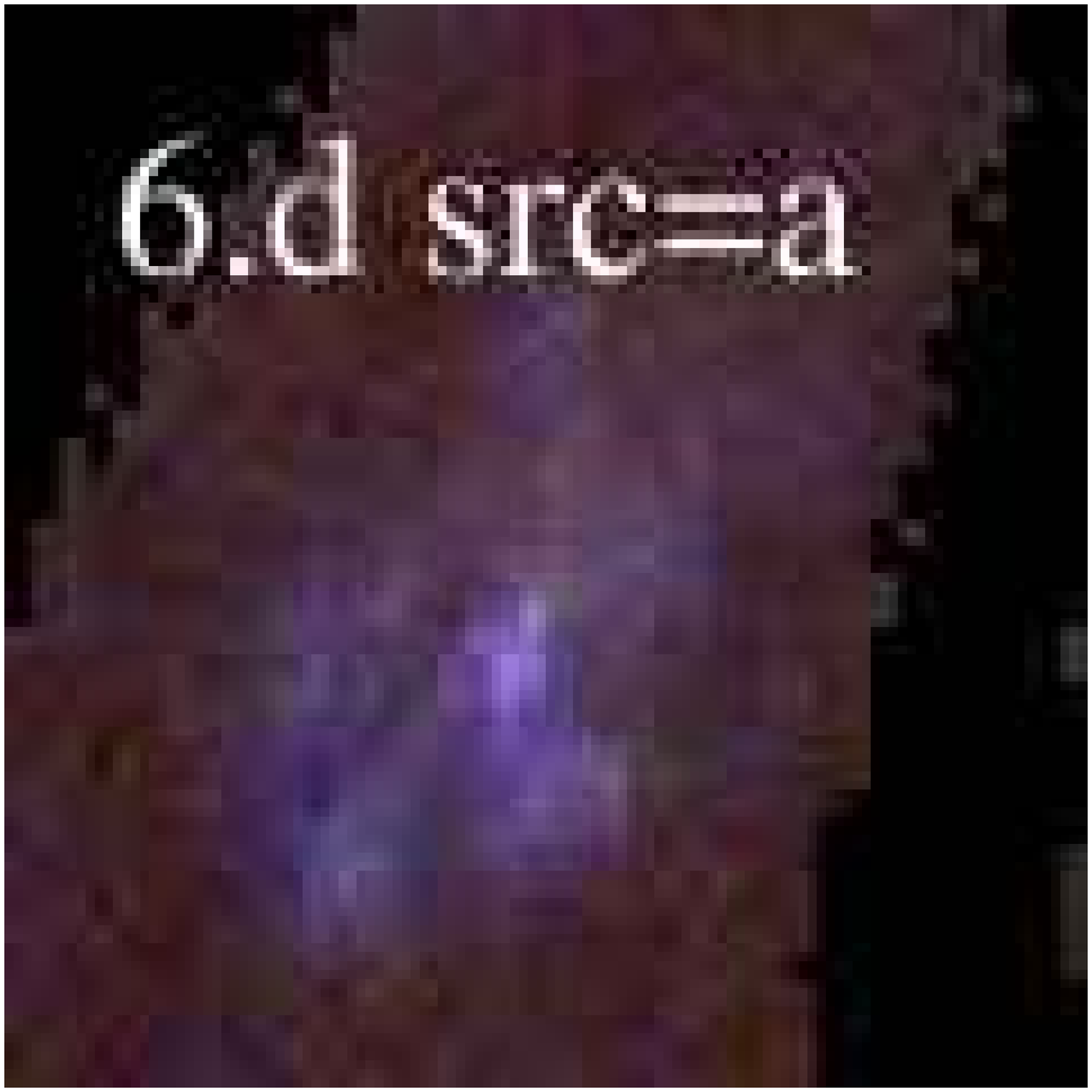}} \\
    \multicolumn{1}{m{1cm}}{{\Large ENFW}}
    & \multicolumn{1}{m{1.7cm}}{\includegraphics[height=2.00cm,clip]{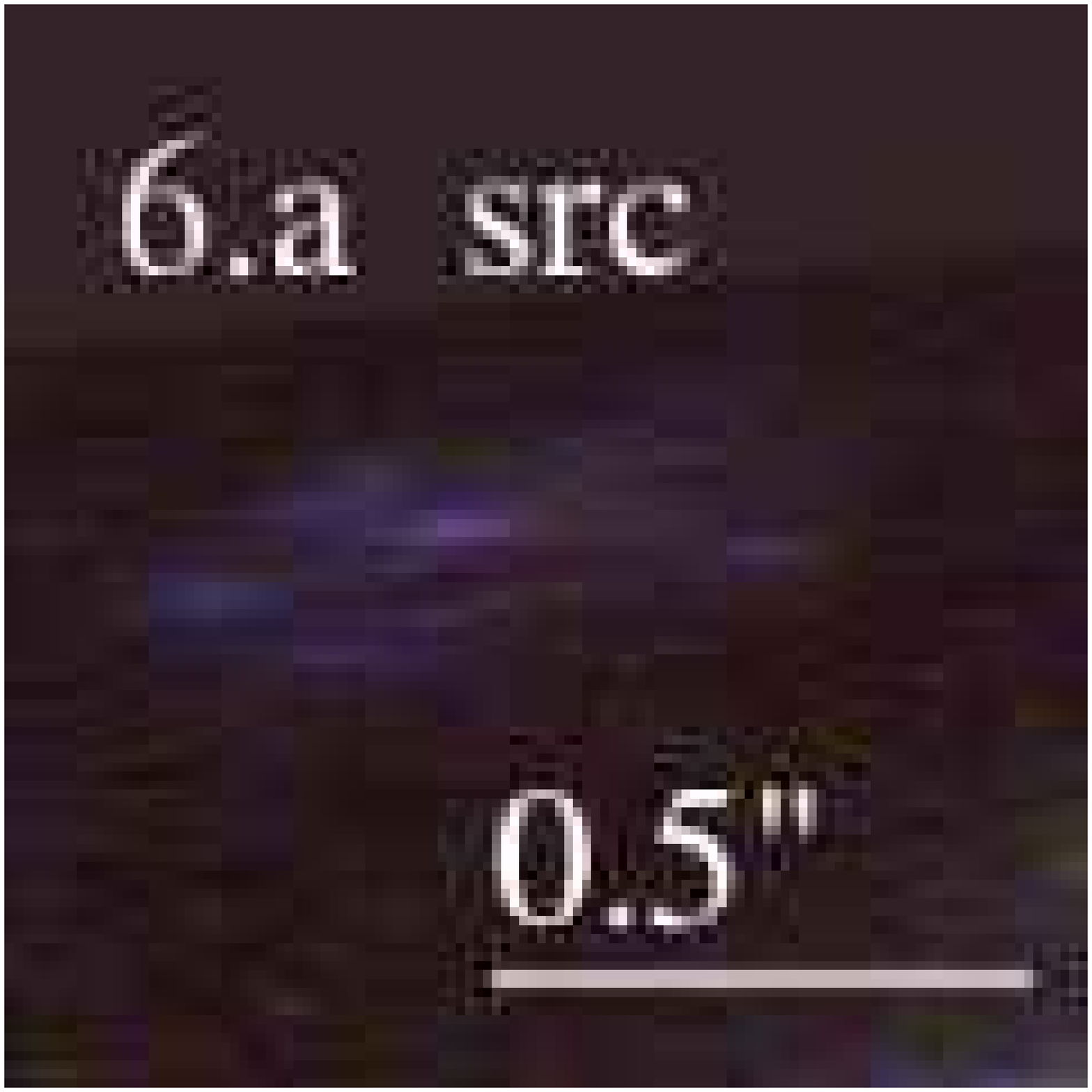}}
    & \multicolumn{1}{m{1.7cm}}{\includegraphics[height=2.00cm,clip]{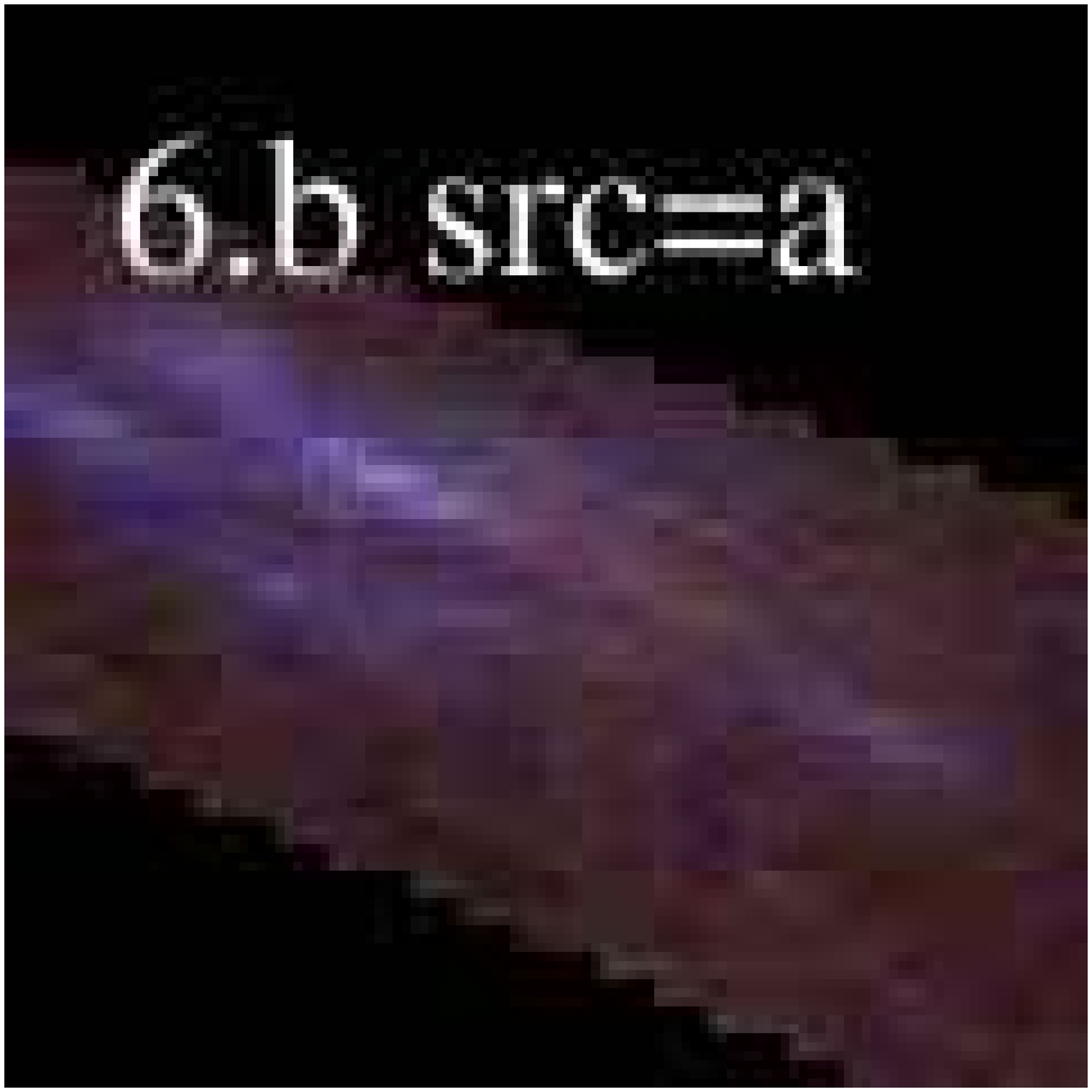}}
    & \multicolumn{1}{m{1.7cm}}{\includegraphics[height=2.00cm,clip]{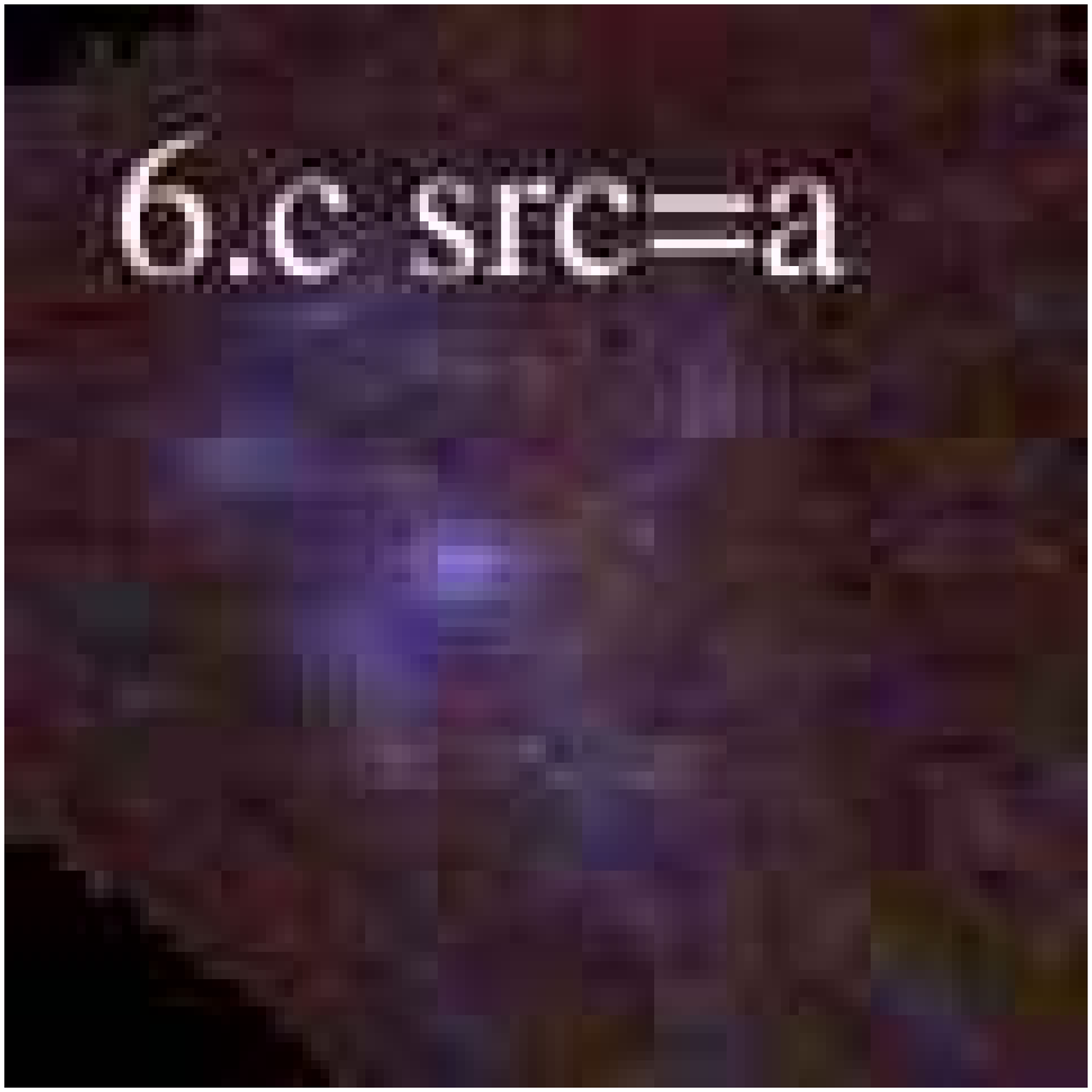}}
    & \multicolumn{1}{m{1.7cm}}{\includegraphics[height=2.00cm,clip]{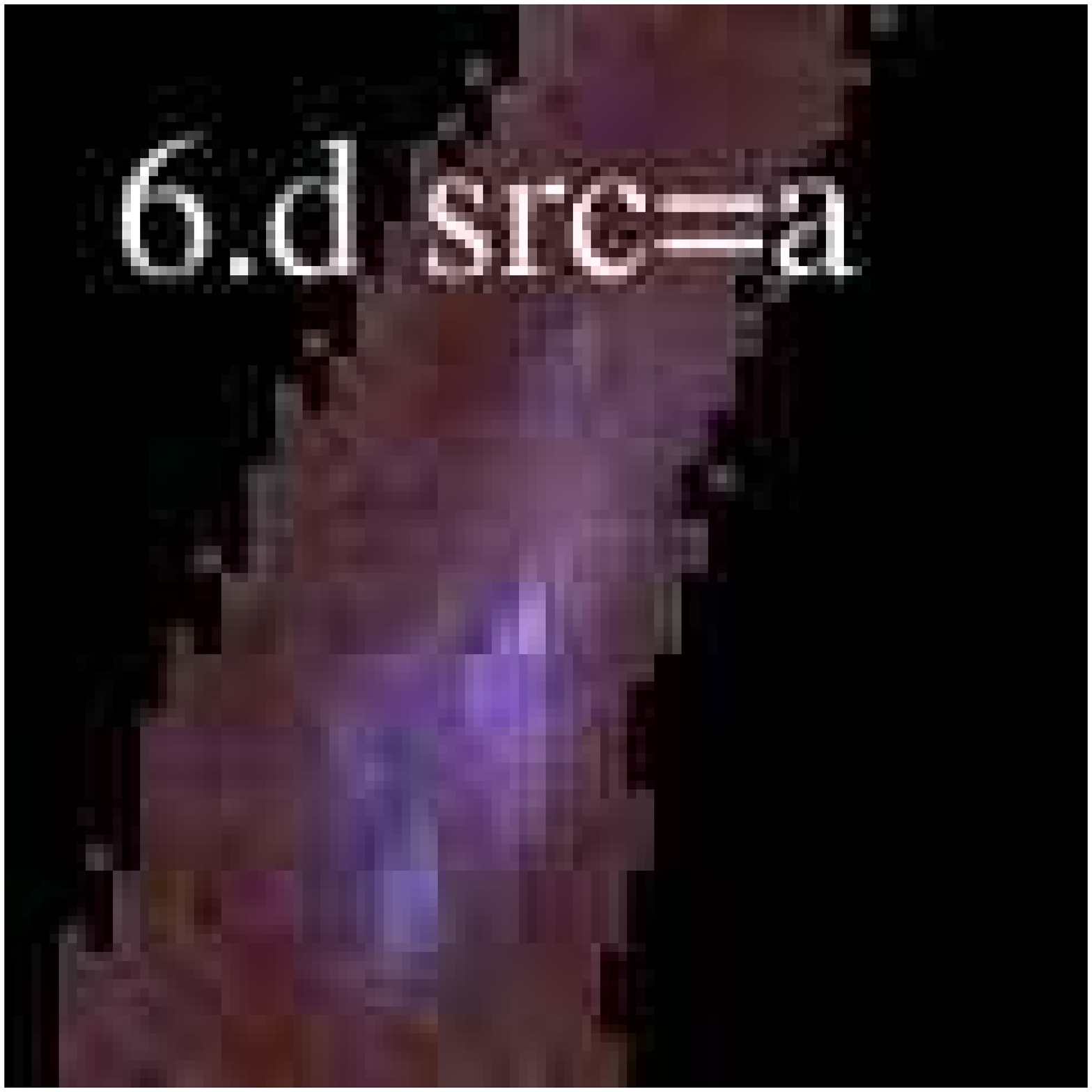}} \\
  \end{tabular}

\end{table*}

\clearpage

\begin{table*}
  \caption{Image system 7:}\vspace{0mm}
  \begin{tabular}{cccc}
    \multicolumn{1}{m{1cm}}{{\Large A1689}}
    & \multicolumn{1}{m{1.7cm}}{\includegraphics[height=2.00cm,clip]{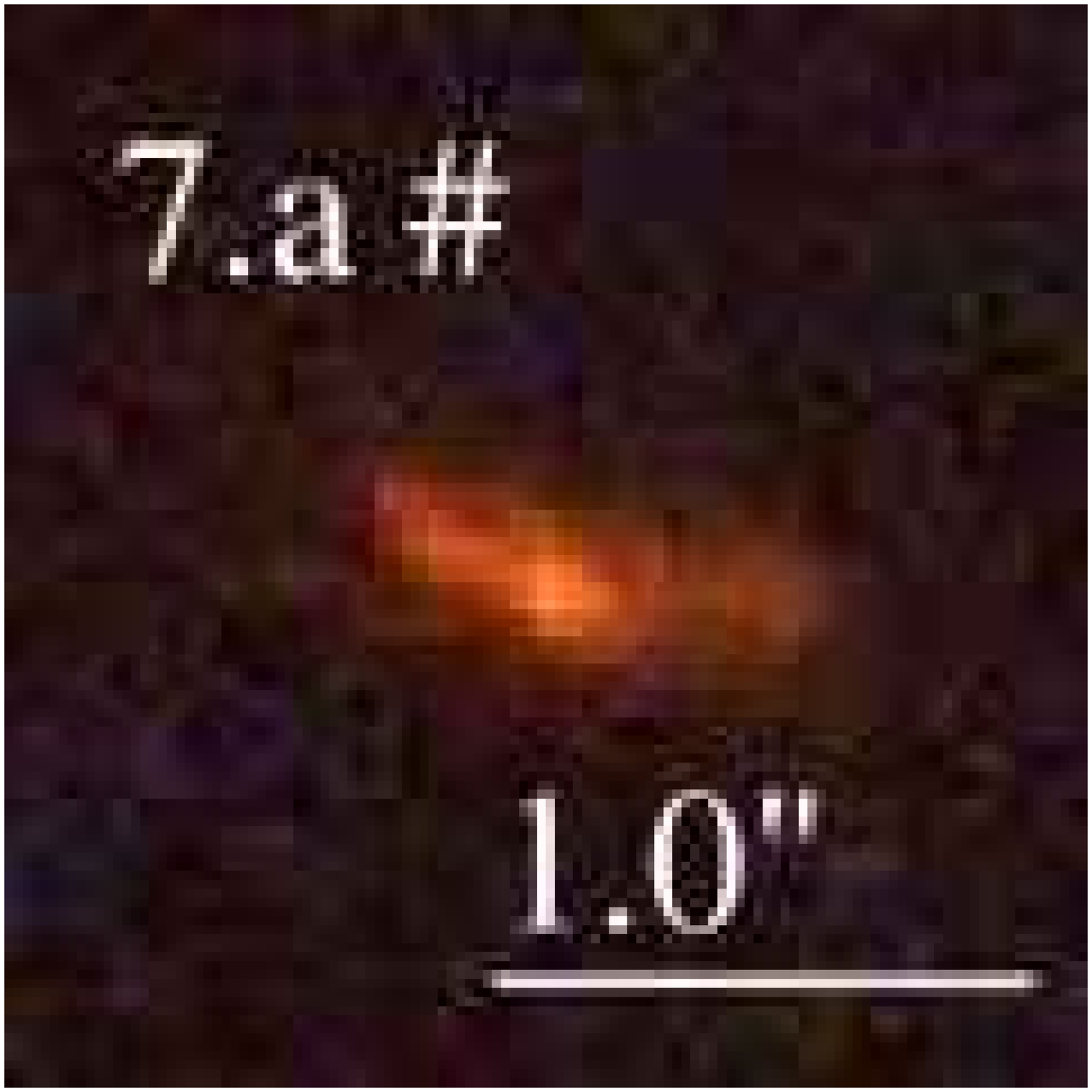}}
    & \multicolumn{1}{m{1.7cm}}{\includegraphics[height=2.00cm,clip]{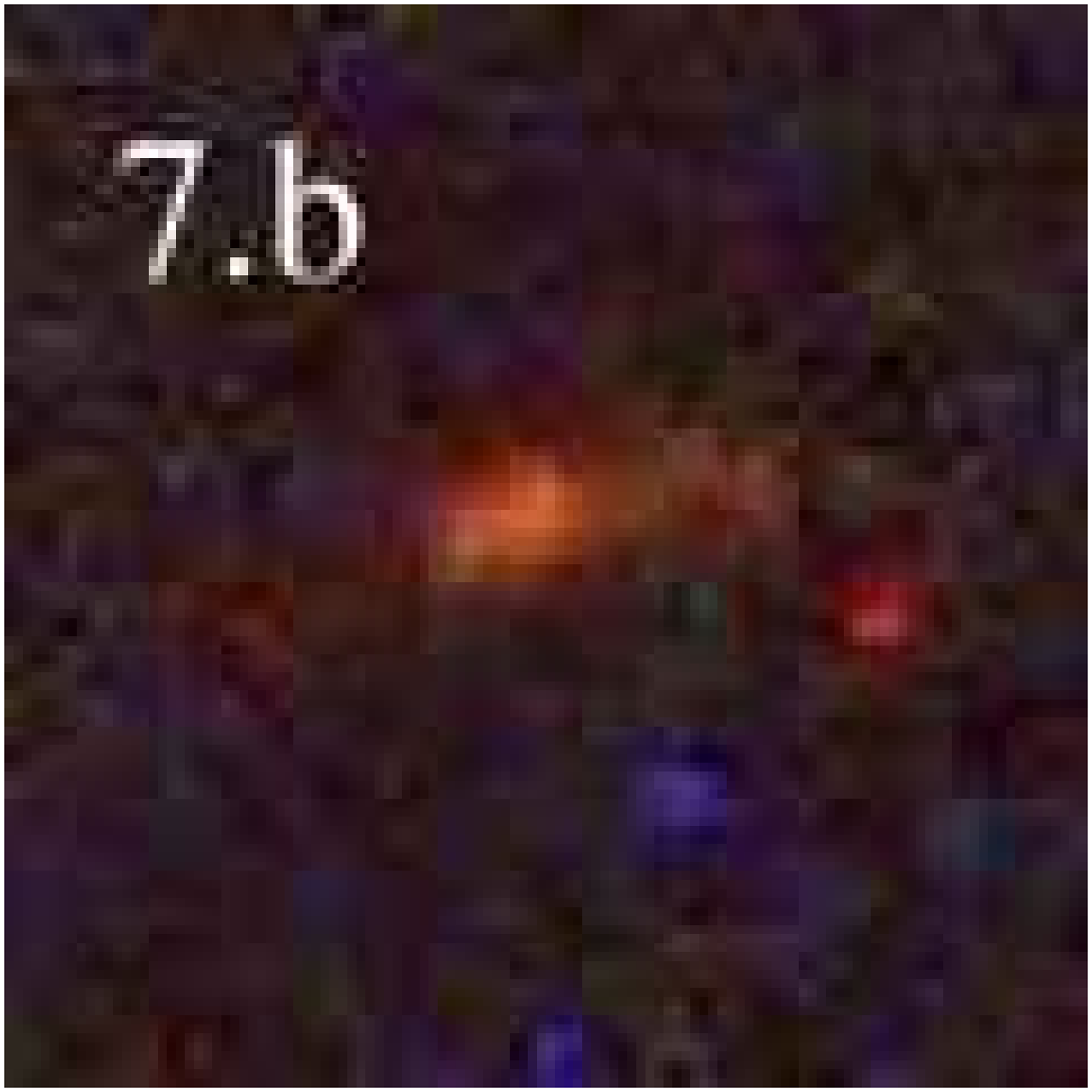}}
    & \multicolumn{1}{m{1.7cm}}{\includegraphics[height=2.00cm,clip]{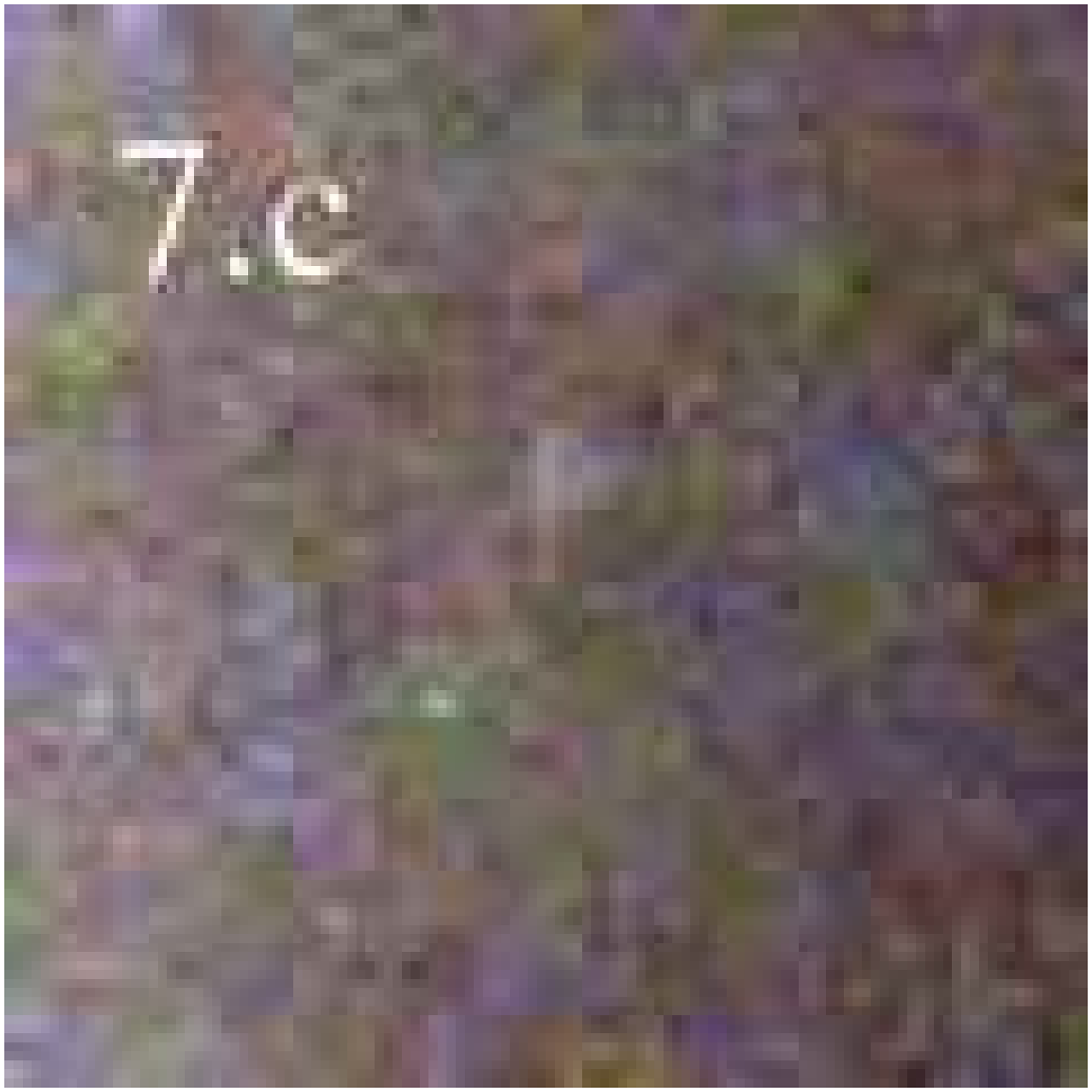}} \\
    \multicolumn{1}{m{1cm}}{{\Large NSIE}}
    & \multicolumn{1}{m{1.7cm}}{\includegraphics[height=2.00cm,clip]{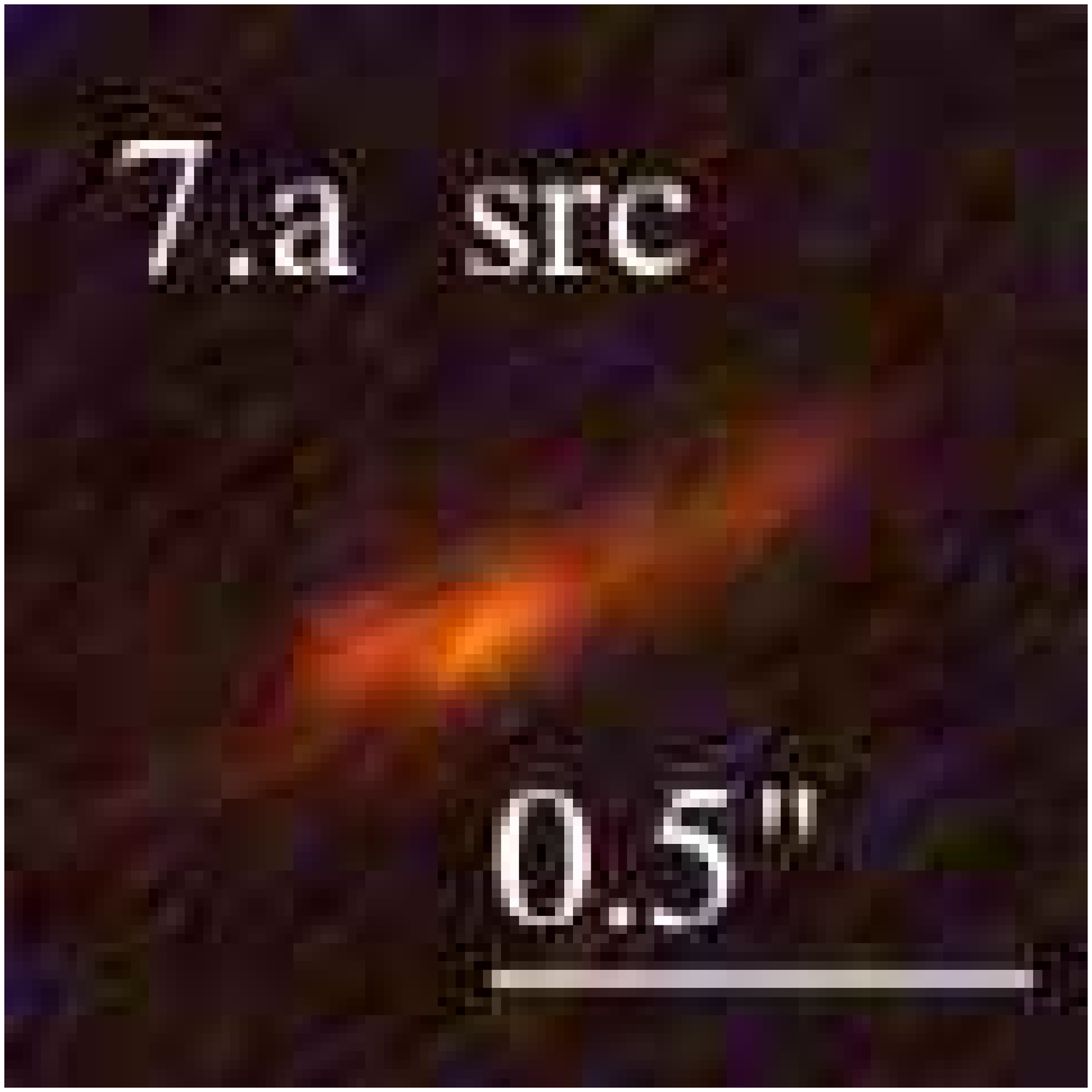}}
    & \multicolumn{1}{m{1.7cm}}{\includegraphics[height=2.00cm,clip]{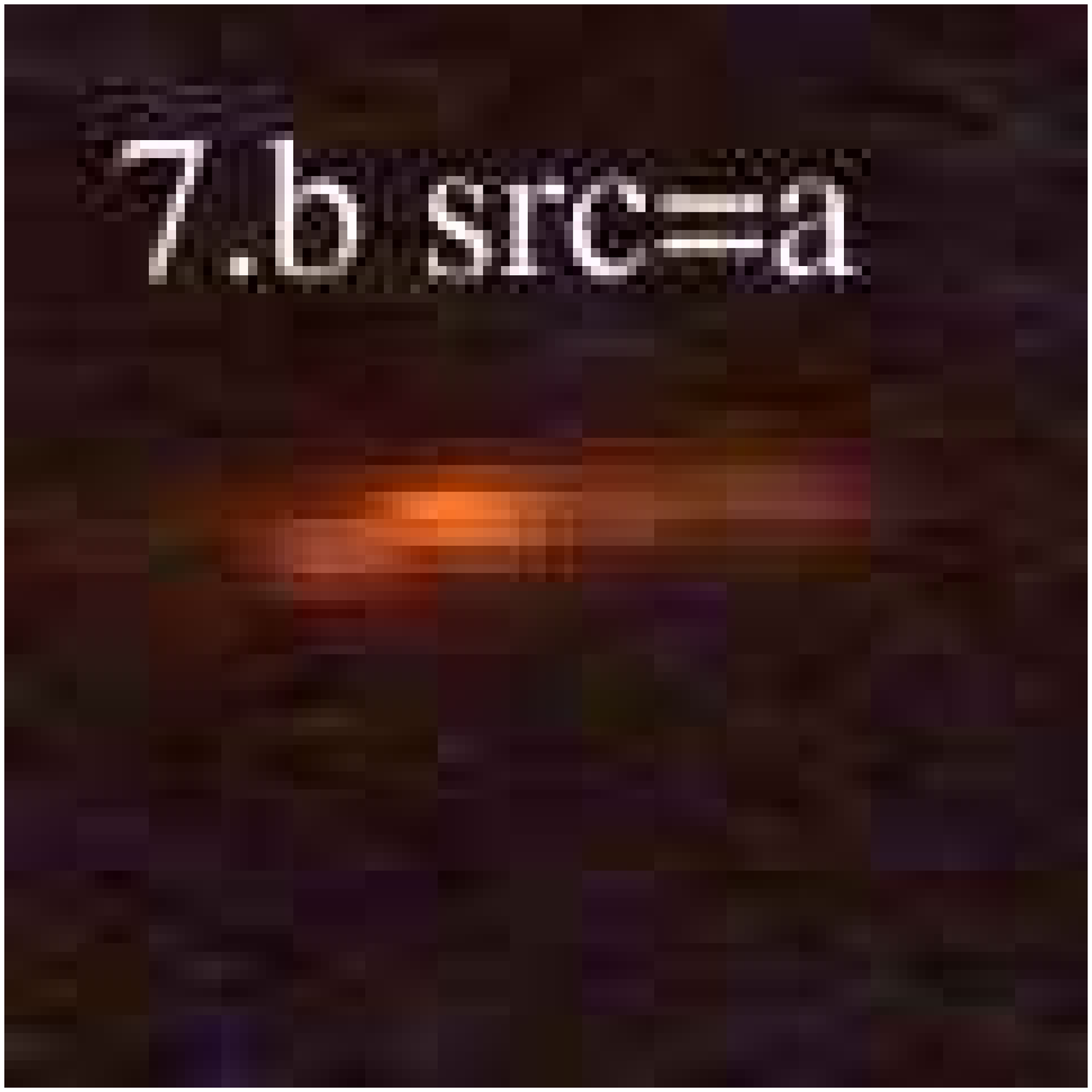}}
    & \multicolumn{1}{m{1.7cm}}{\includegraphics[height=2.00cm,clip]{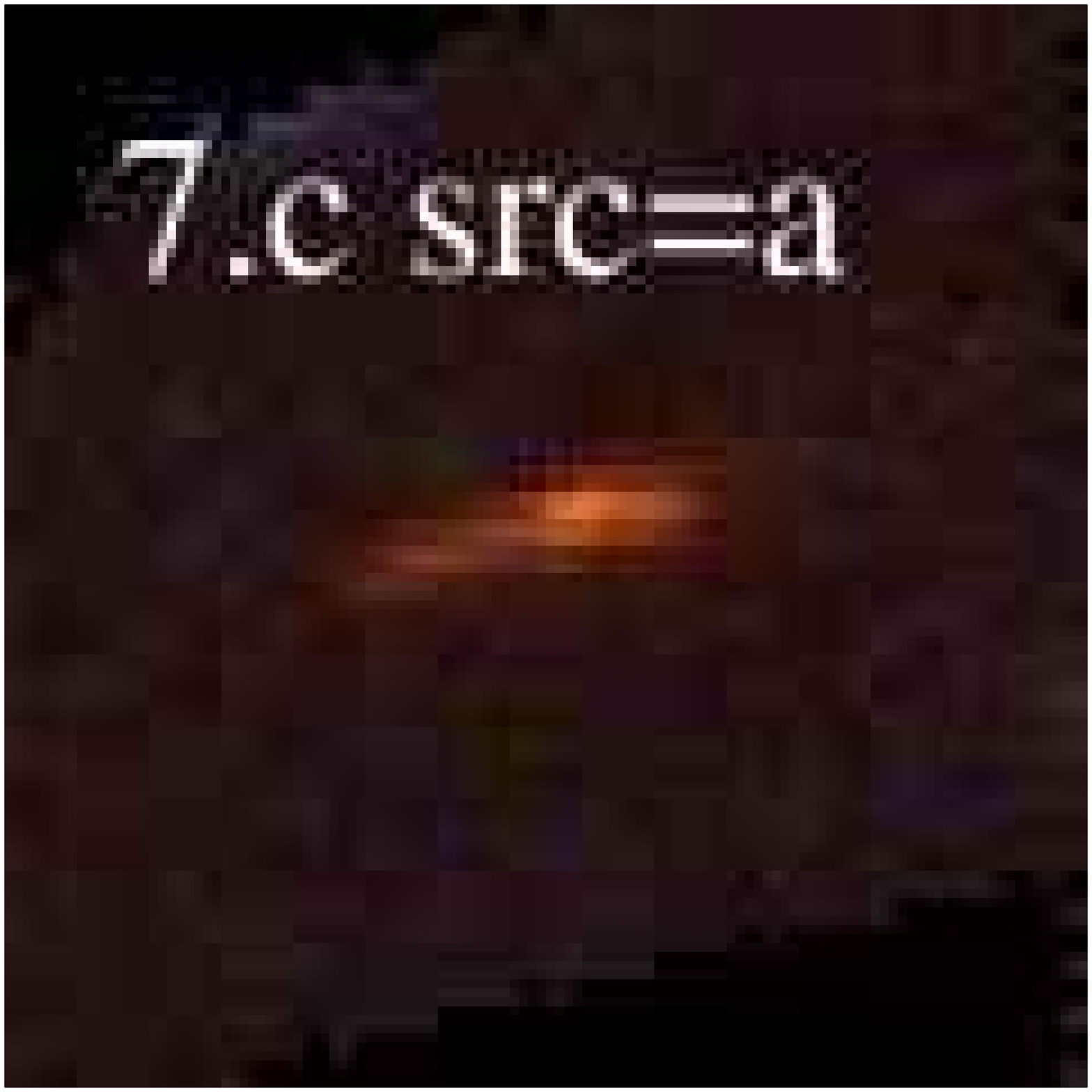}} \\
    \multicolumn{1}{m{1cm}}{{\Large ENFW}}
    & \multicolumn{1}{m{1.7cm}}{\includegraphics[height=2.00cm,clip]{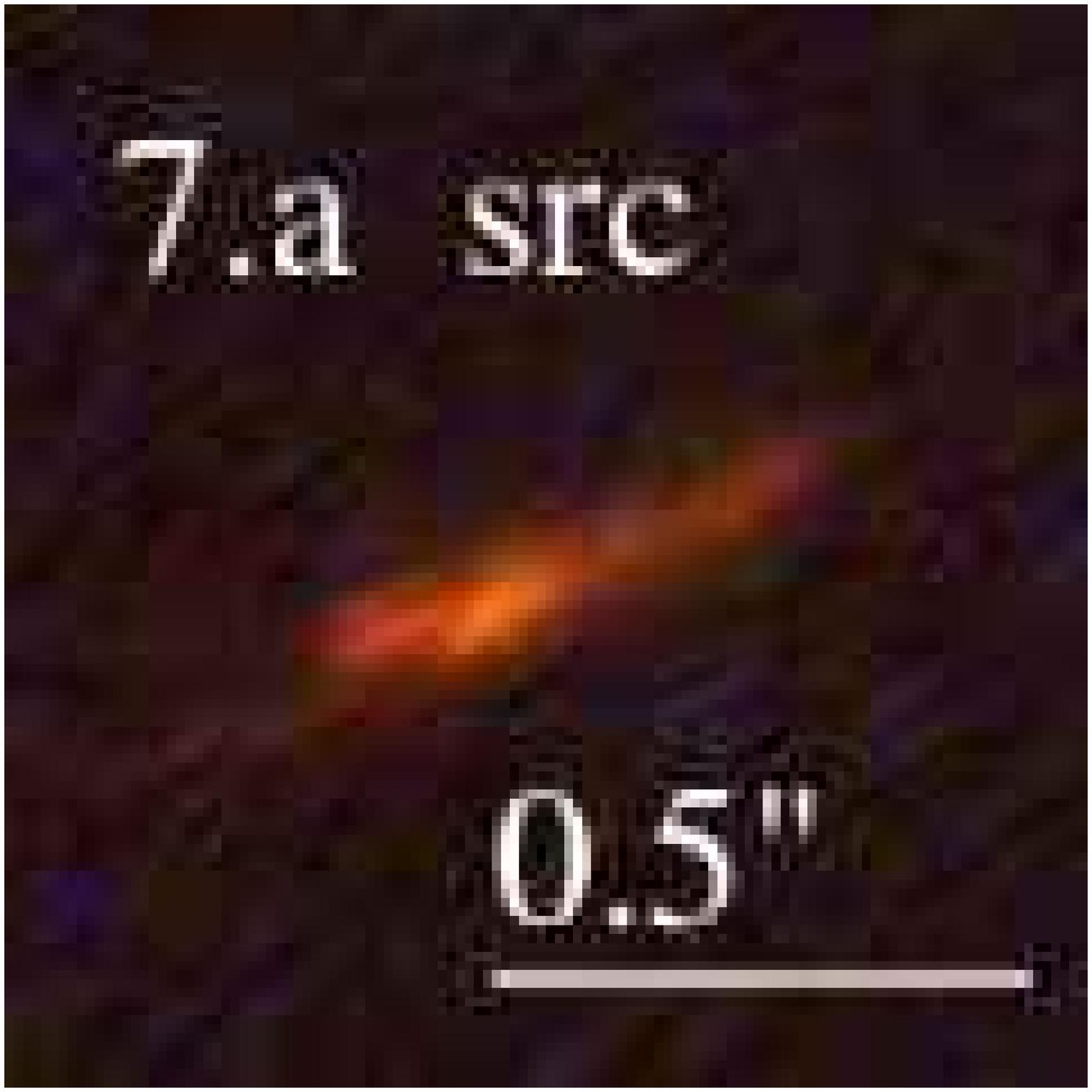}}
    & \multicolumn{1}{m{1.7cm}}{\includegraphics[height=2.00cm,clip]{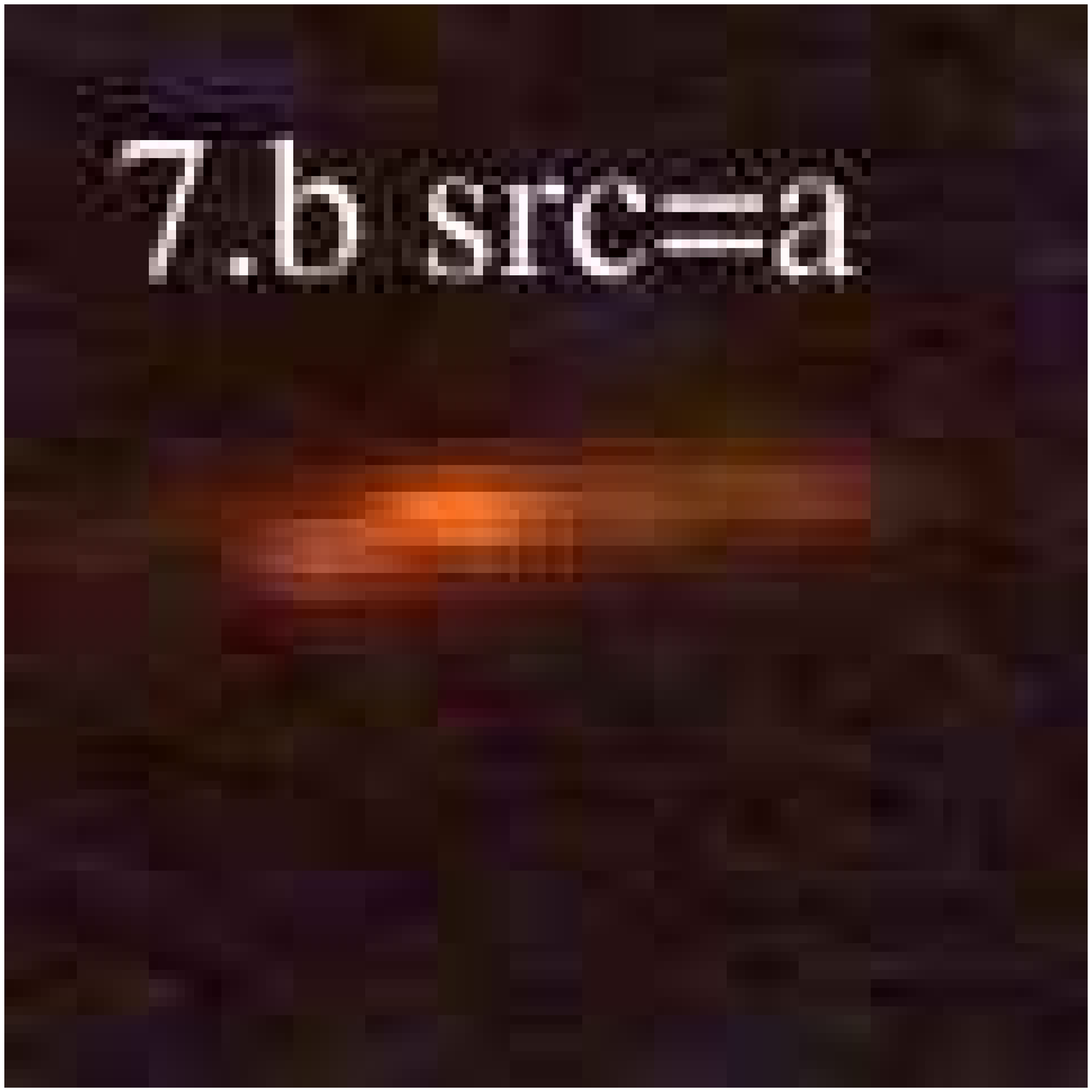}}
    & \multicolumn{1}{m{1.7cm}}{\includegraphics[height=2.00cm,clip]{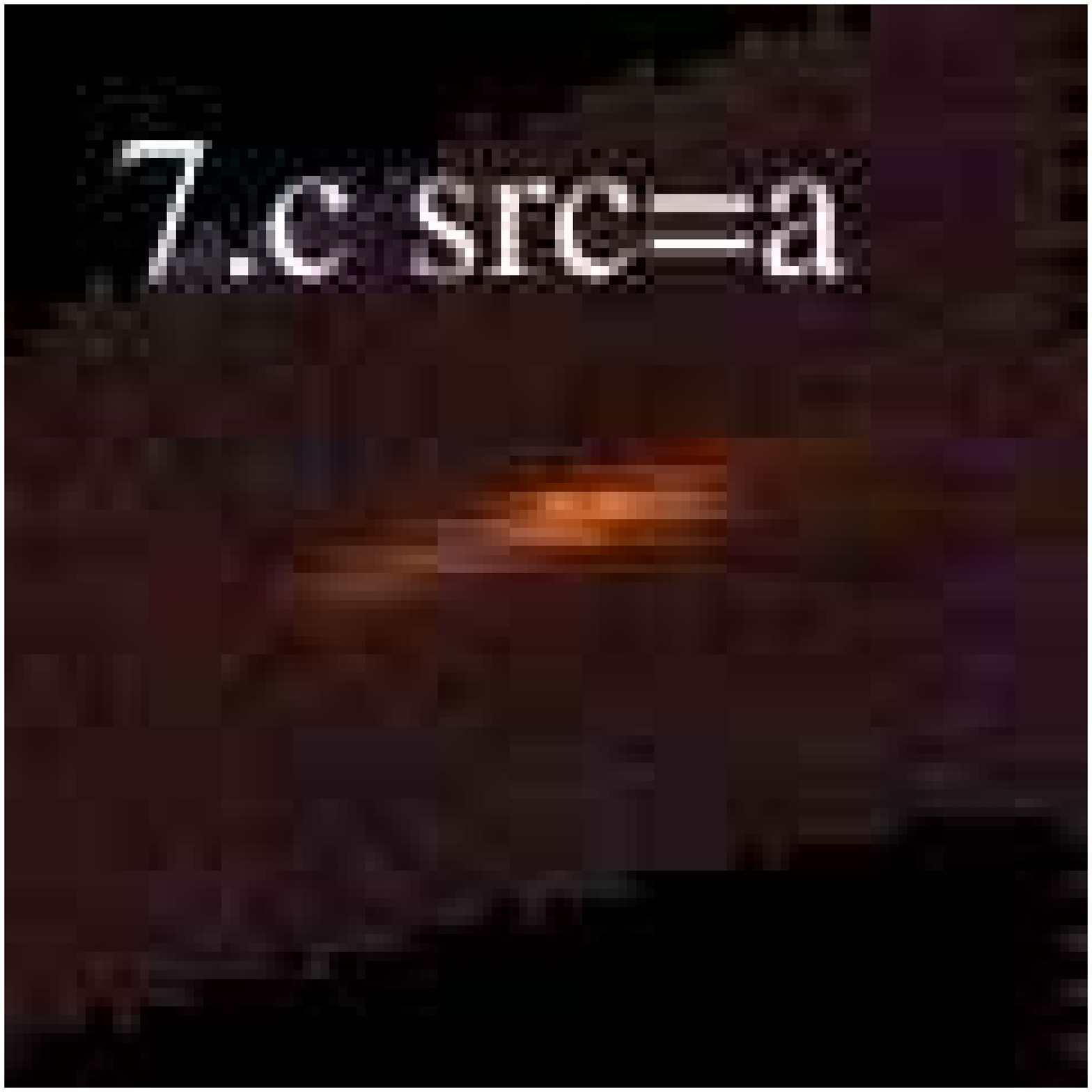}} \\
  \end{tabular}

\end{table*}

\begin{table*}
  \caption{Image system 8:}\vspace{0mm}
  \begin{tabular}{cccccc}
    \multicolumn{1}{m{1cm}}{{\Large A1689}}
    & \multicolumn{1}{m{1.7cm}}{\includegraphics[height=2.00cm,clip]{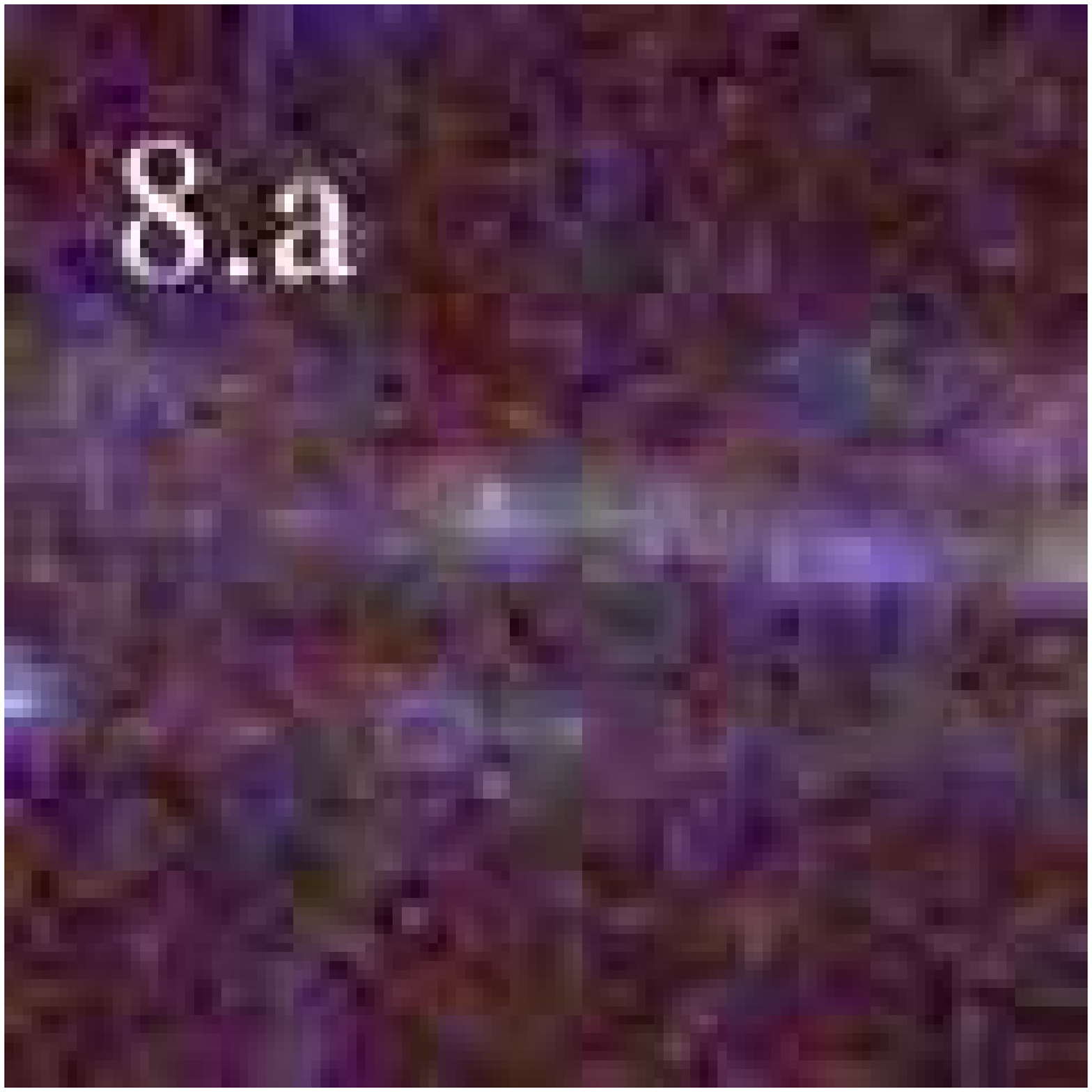}}
    & \multicolumn{1}{m{1.7cm}}{\includegraphics[height=2.00cm,clip]{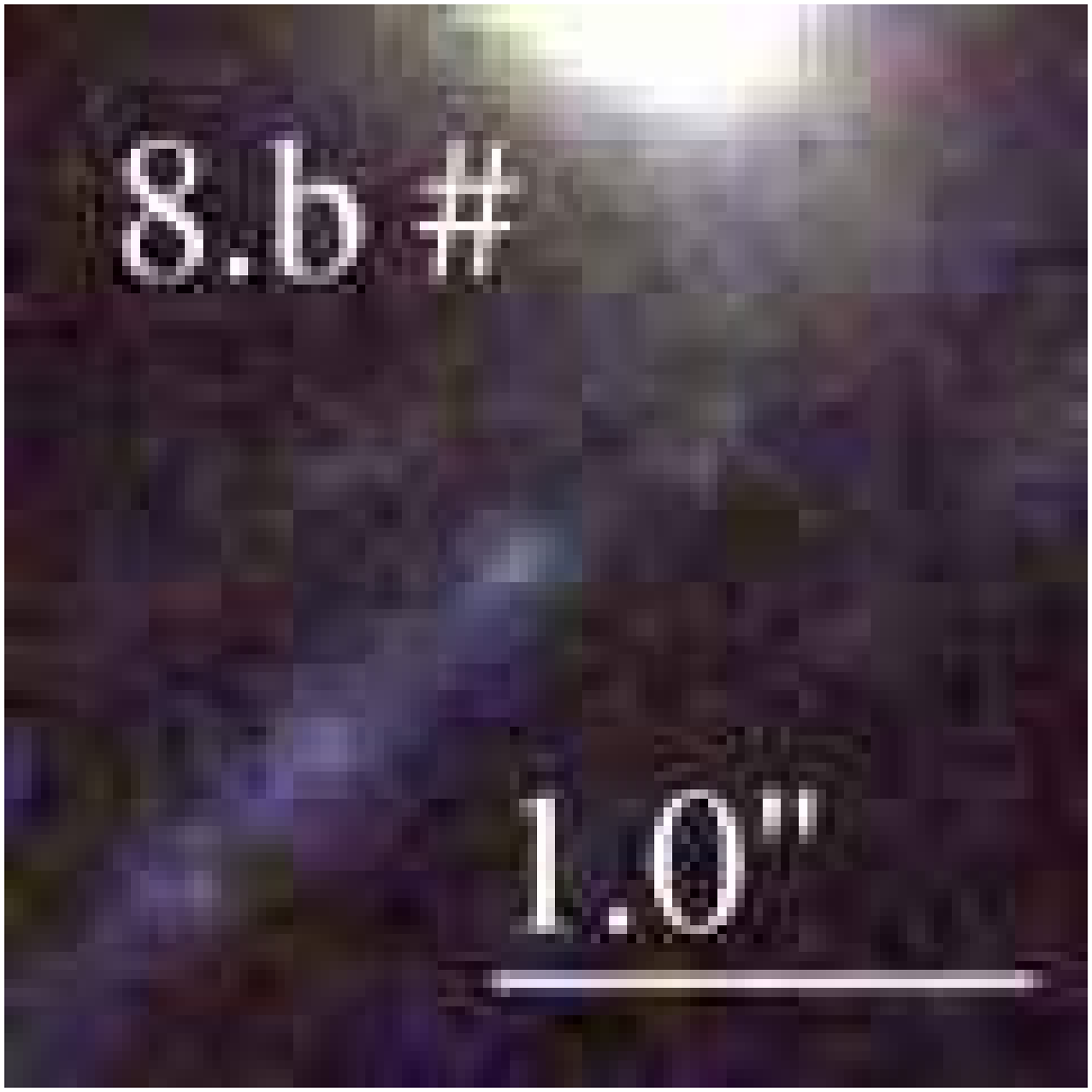}}
    & \multicolumn{1}{m{1.7cm}}{\includegraphics[height=2.00cm,clip]{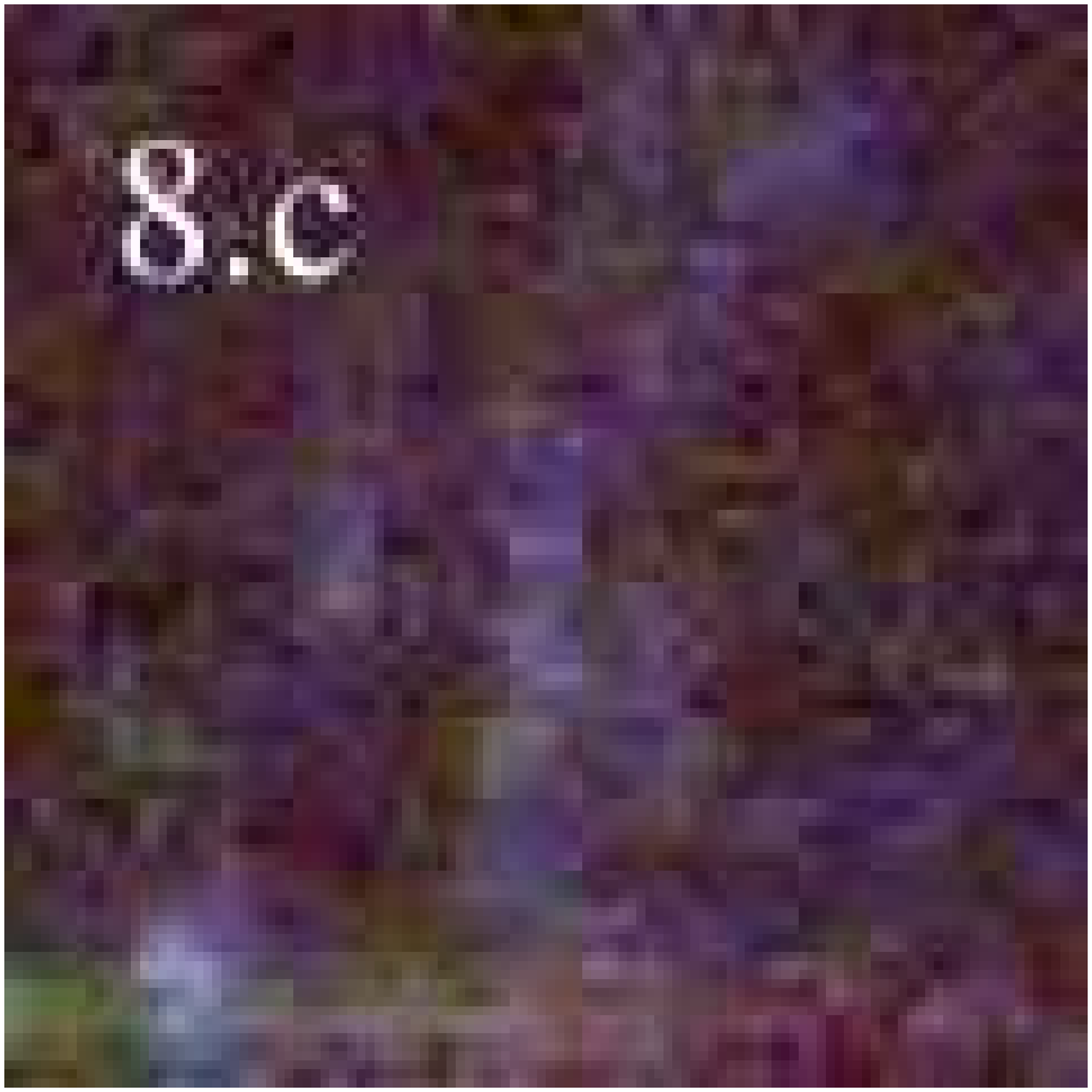}}
    & \multicolumn{1}{m{1.7cm}}{\includegraphics[height=2.00cm,clip]{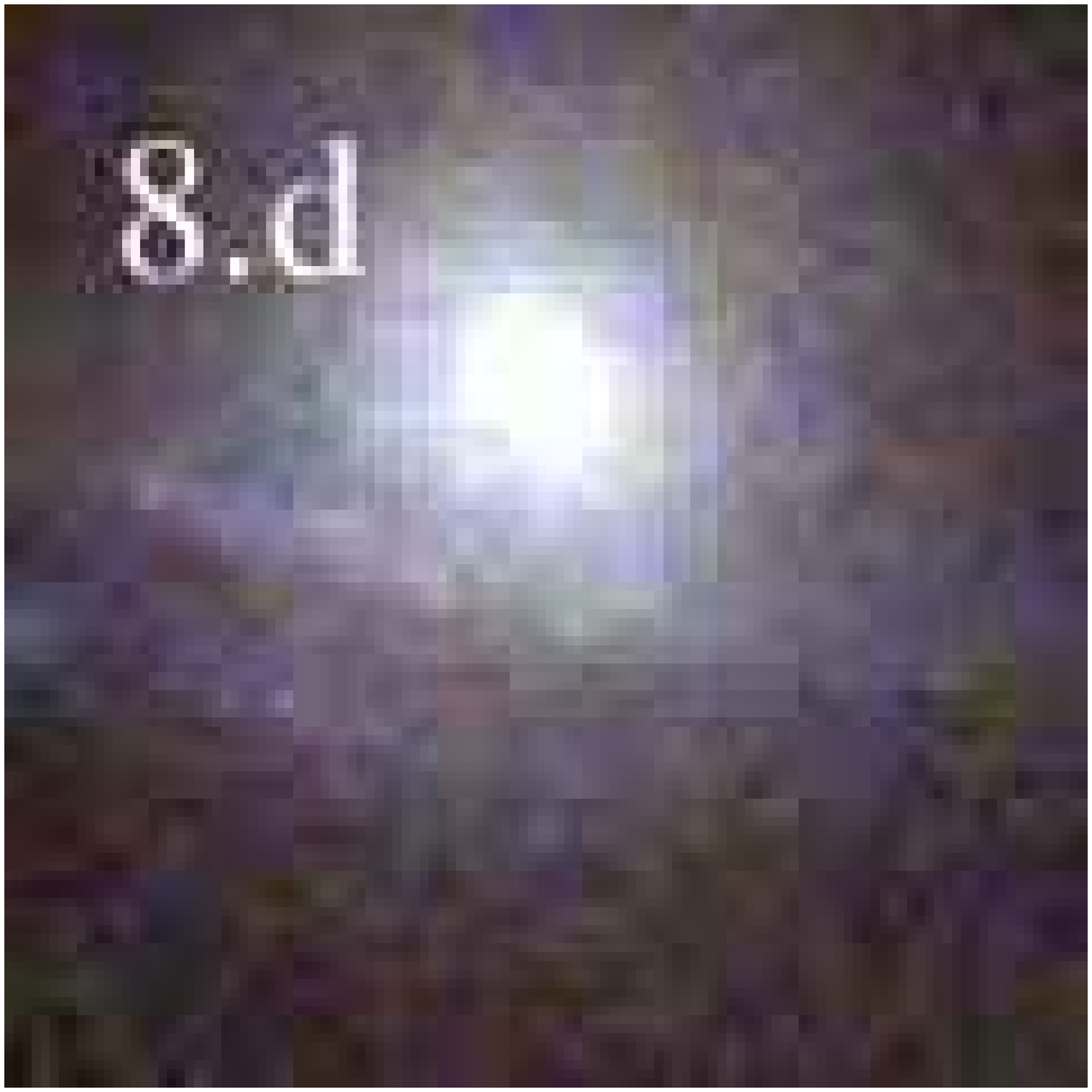}}
    & \multicolumn{1}{m{1.7cm}}{\includegraphics[height=2.00cm,clip]{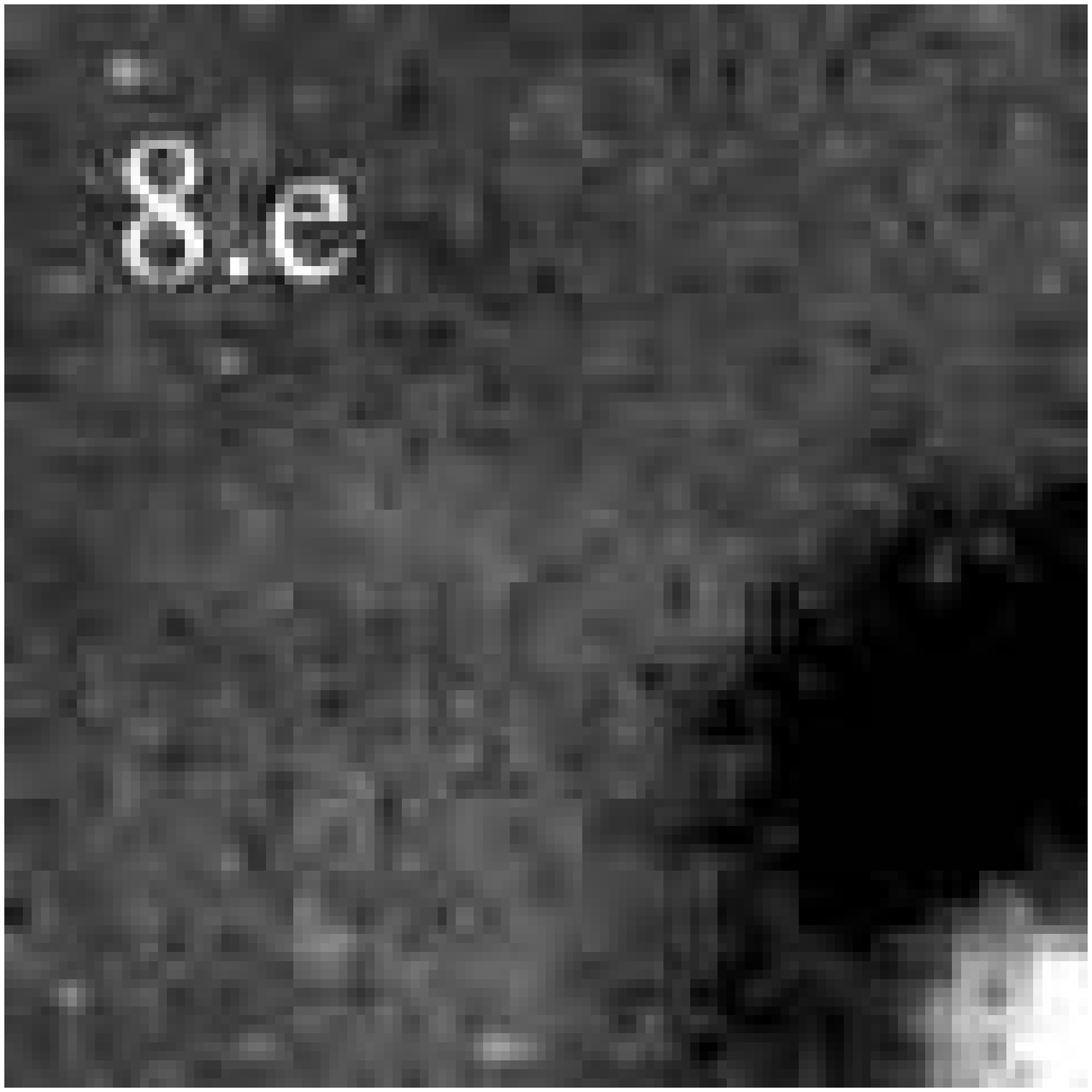}} \\
    \multicolumn{1}{m{1cm}}{{\Large NSIE}}
    & \multicolumn{1}{m{1.7cm}}{\includegraphics[height=2.00cm,clip]{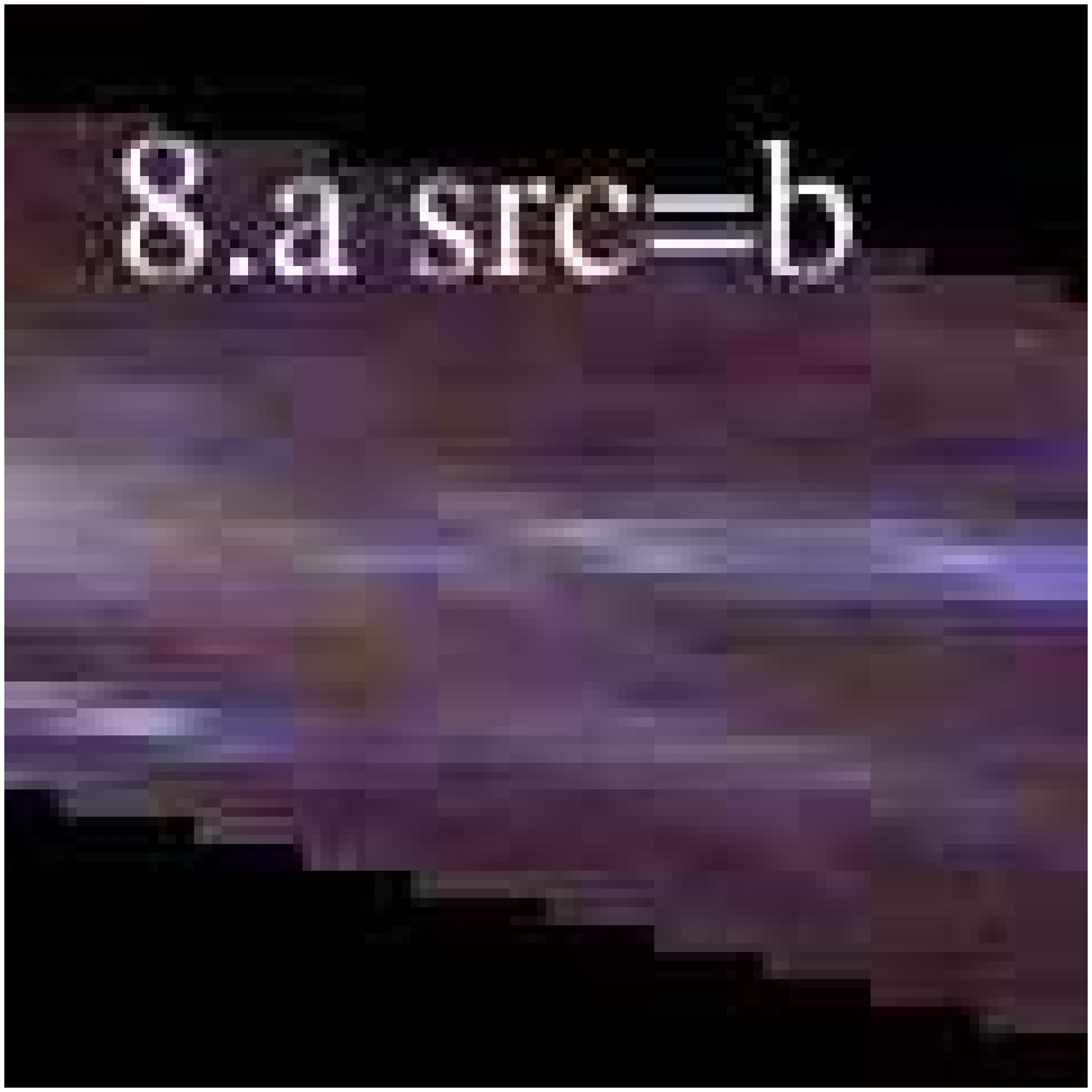}}
    & \multicolumn{1}{m{1.7cm}}{\includegraphics[height=2.00cm,clip]{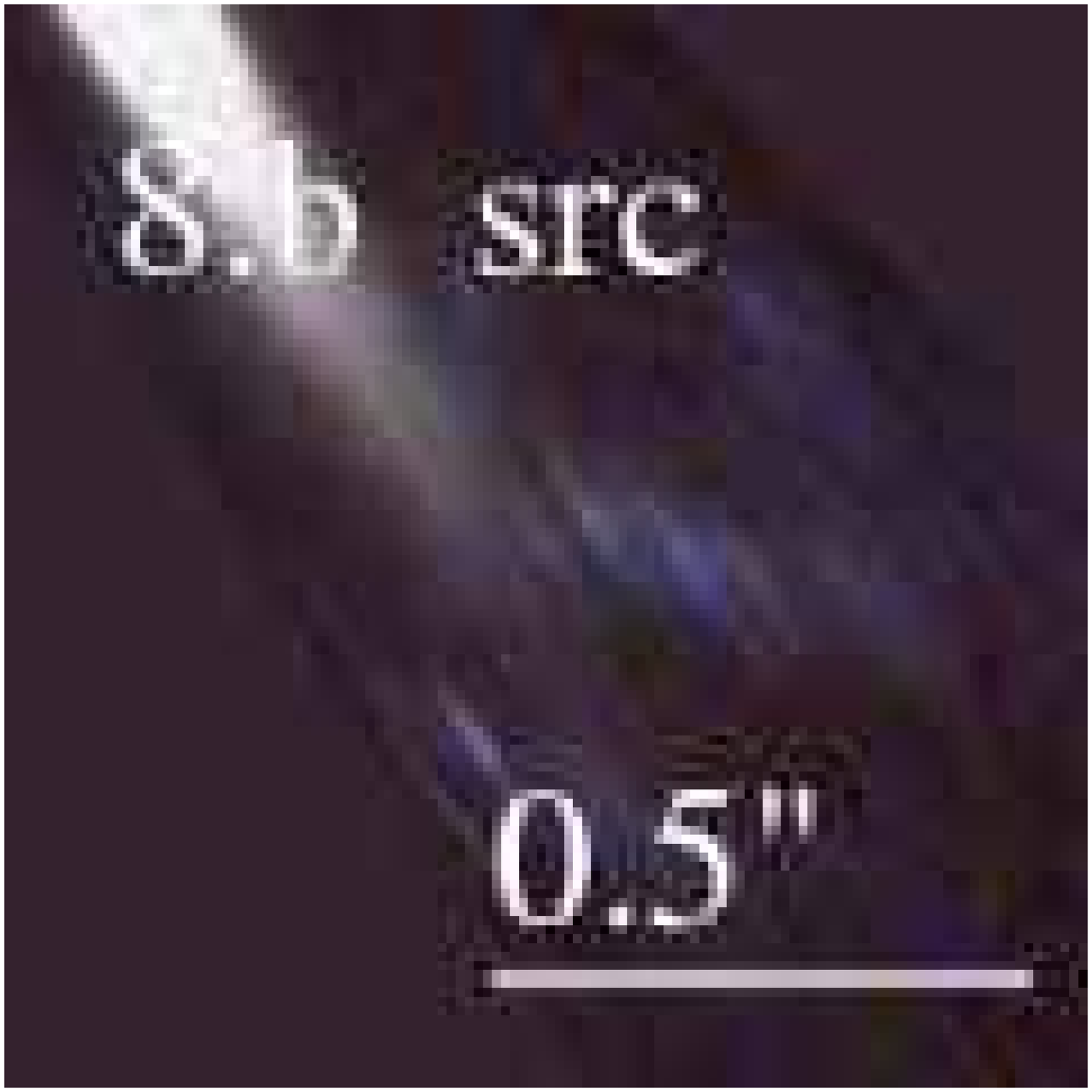}}
    & \multicolumn{1}{m{1.7cm}}{\includegraphics[height=2.00cm,clip]{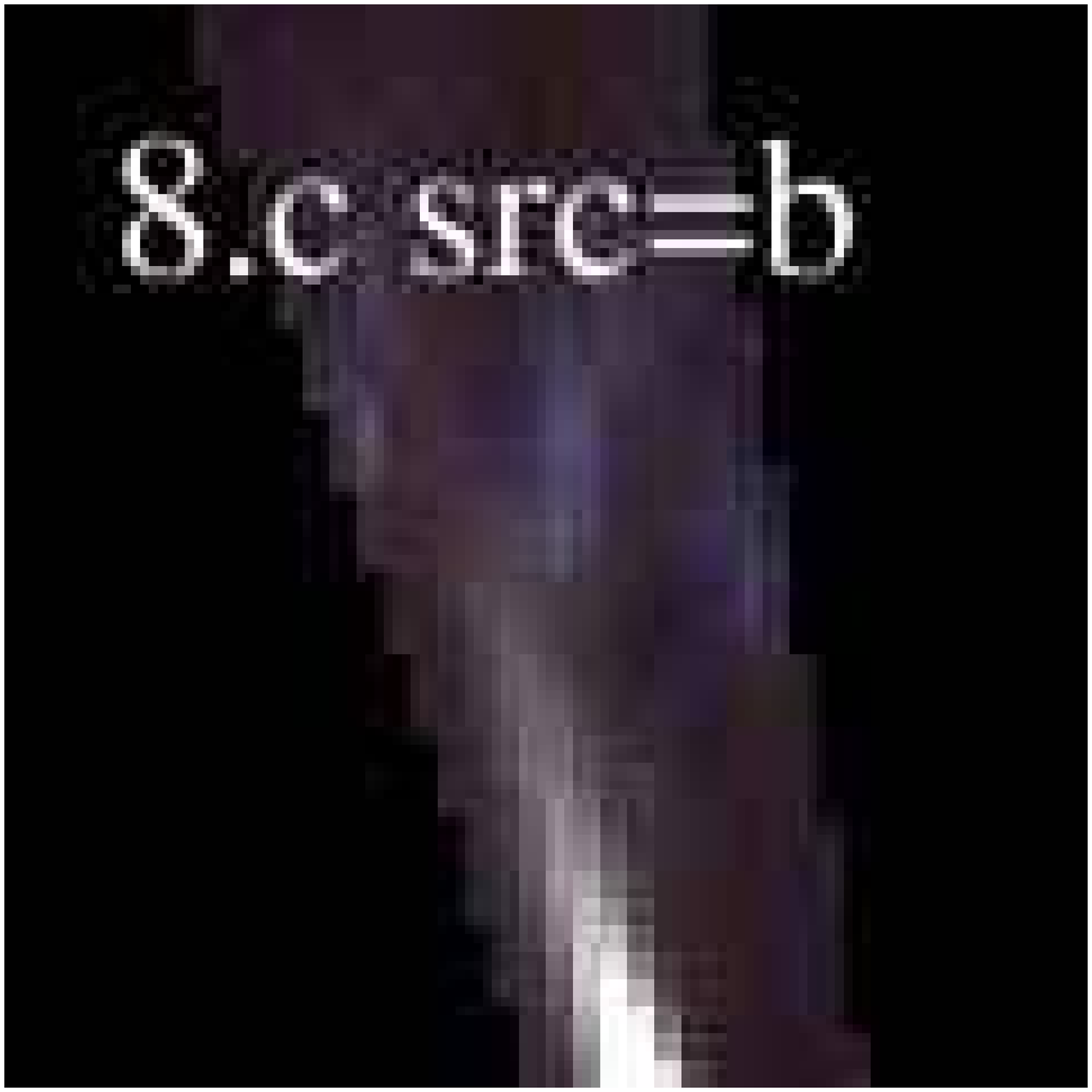}}
    & \multicolumn{1}{m{1.7cm}}{\includegraphics[height=2.00cm,clip]{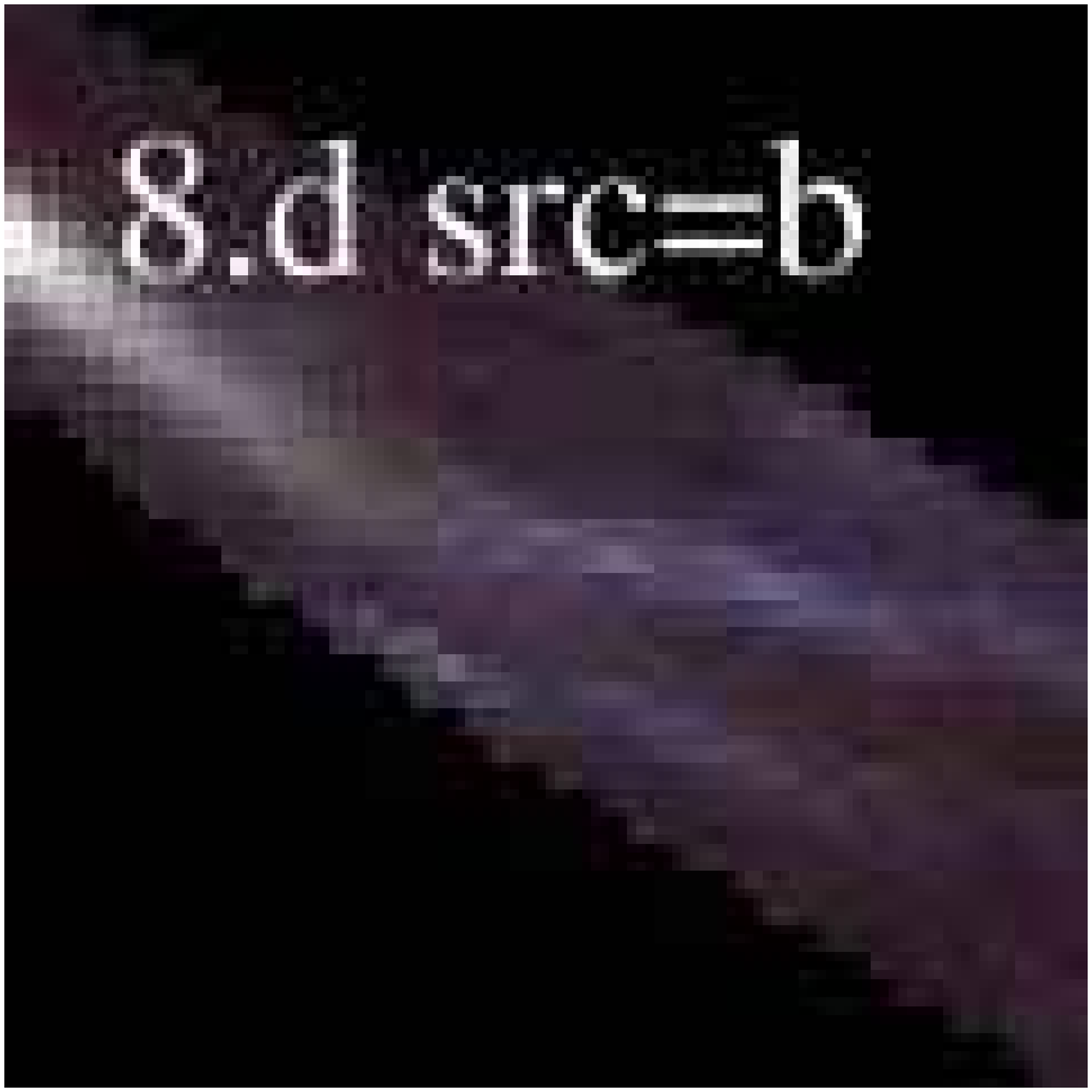}}
    & \multicolumn{1}{m{1.7cm}}{\includegraphics[height=2.00cm,clip]{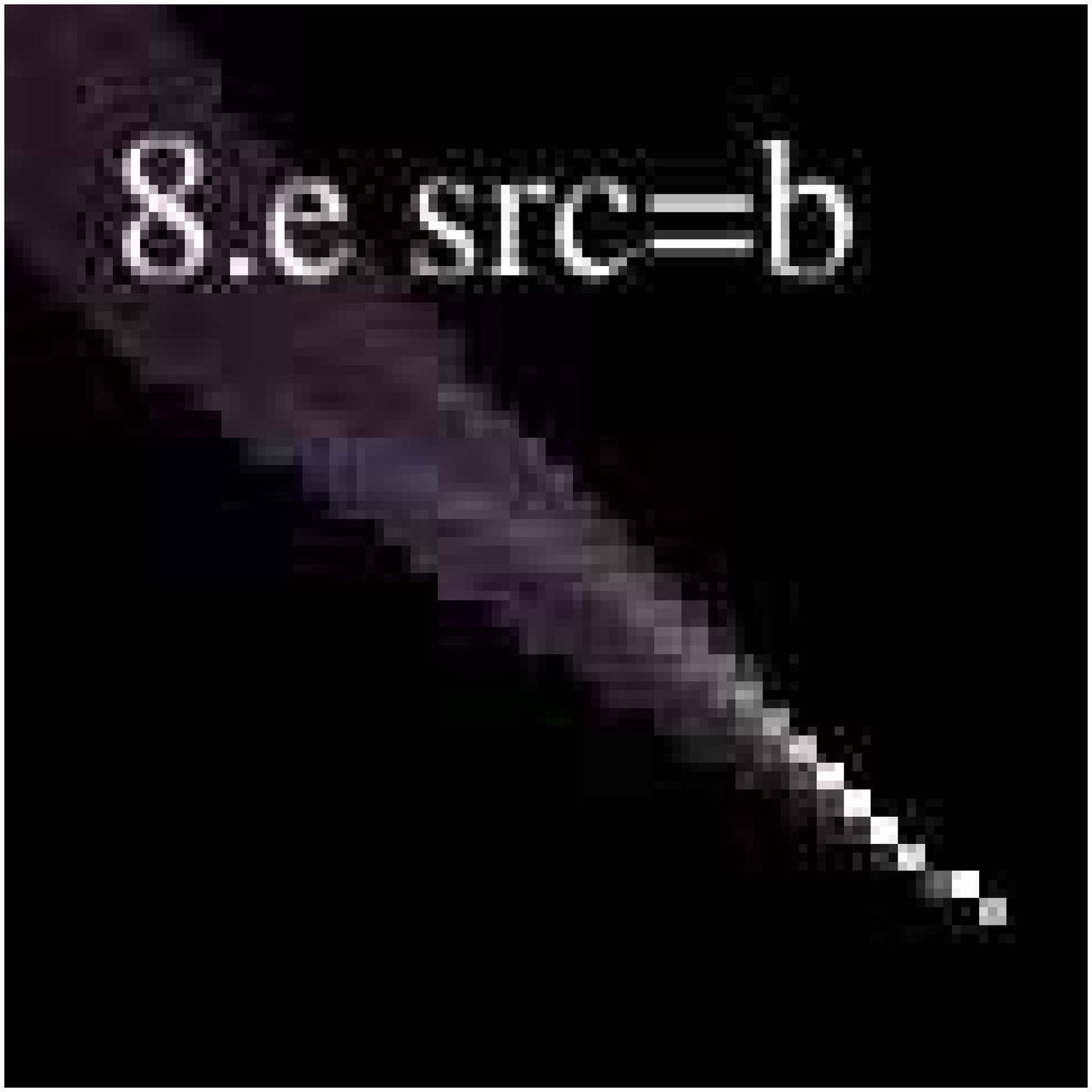}} \\
    \multicolumn{1}{m{1cm}}{{\Large ENFW}}
    & \multicolumn{1}{m{1.7cm}}{\includegraphics[height=2.00cm,clip]{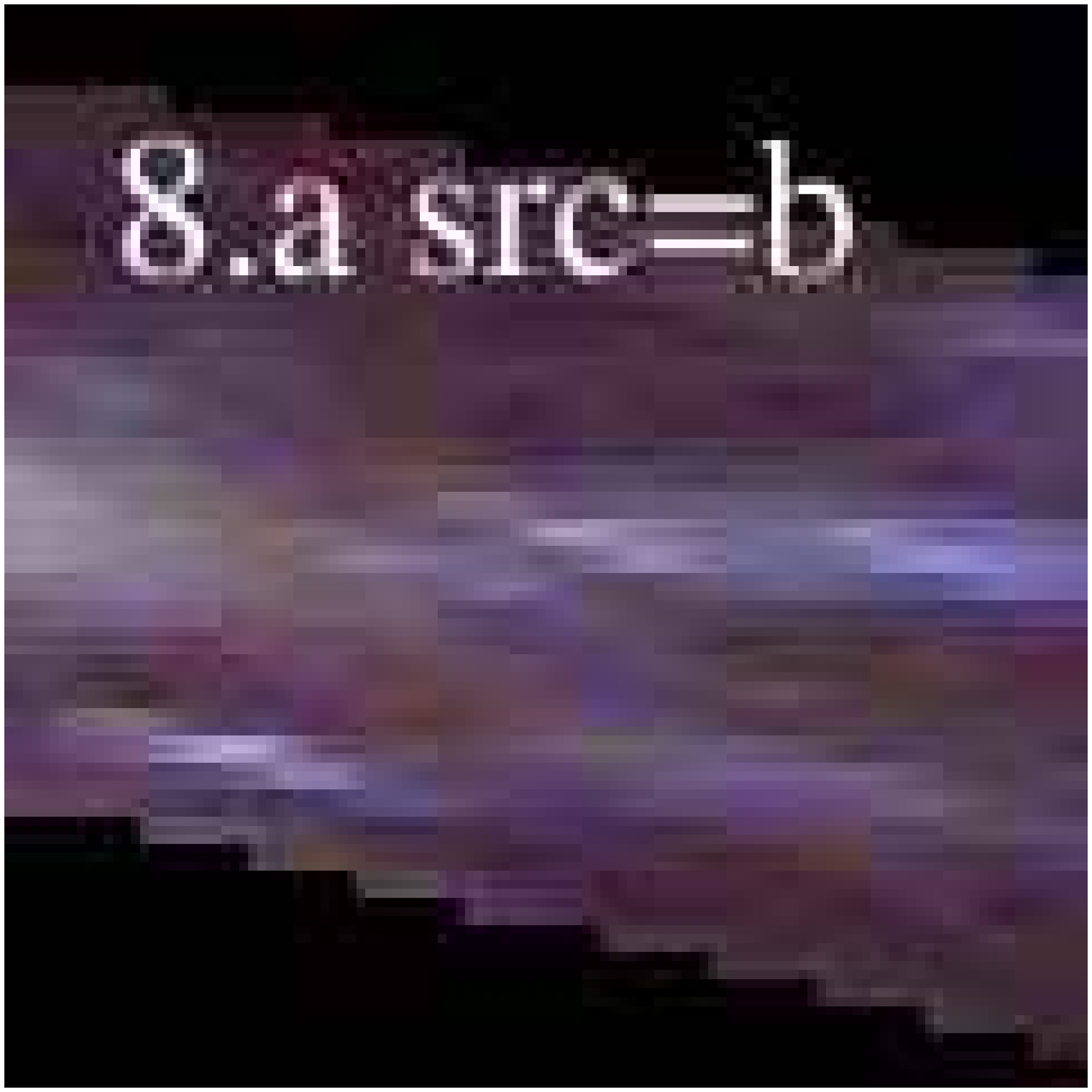}}
    & \multicolumn{1}{m{1.7cm}}{\includegraphics[height=2.00cm,clip]{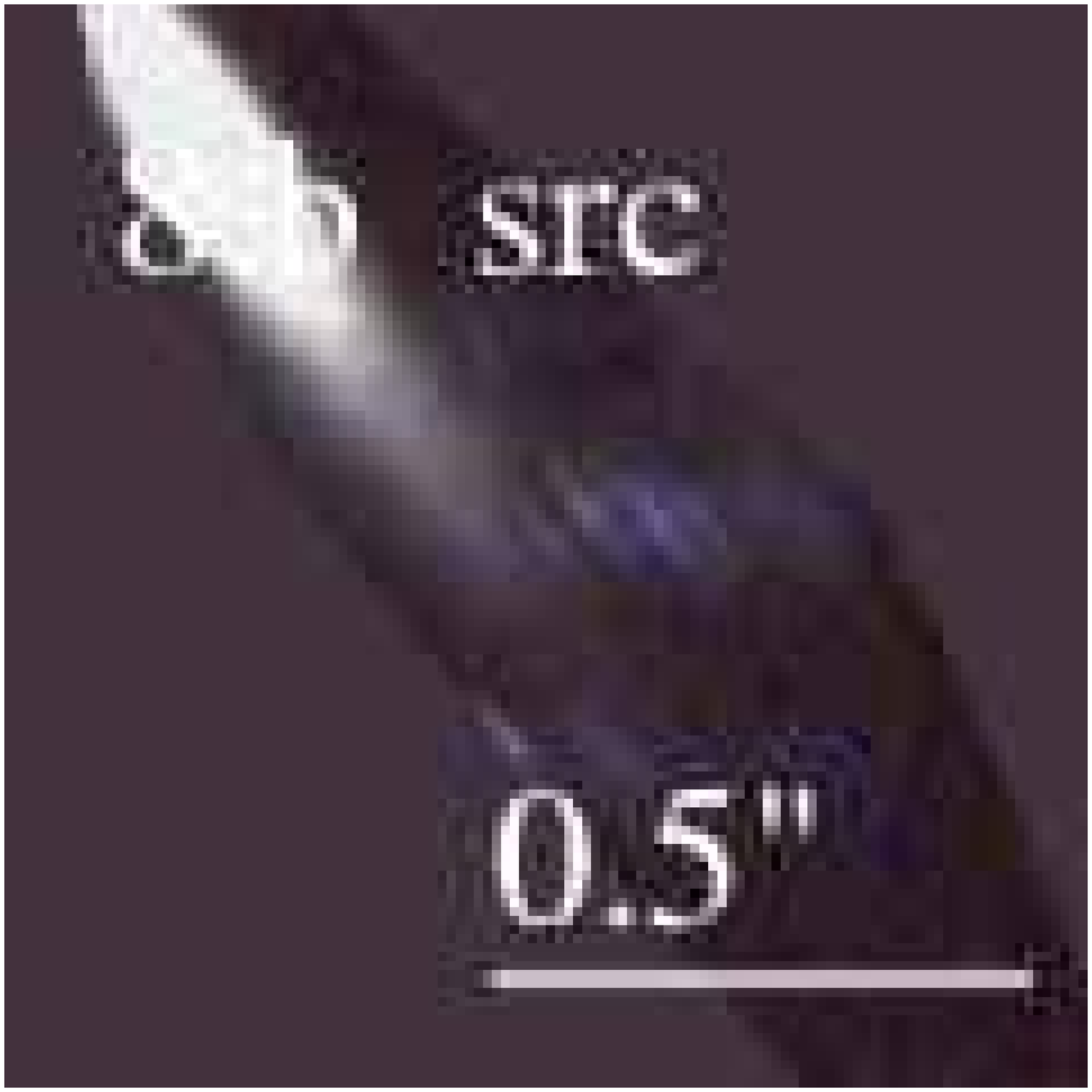}}
    & \multicolumn{1}{m{1.7cm}}{\includegraphics[height=2.00cm,clip]{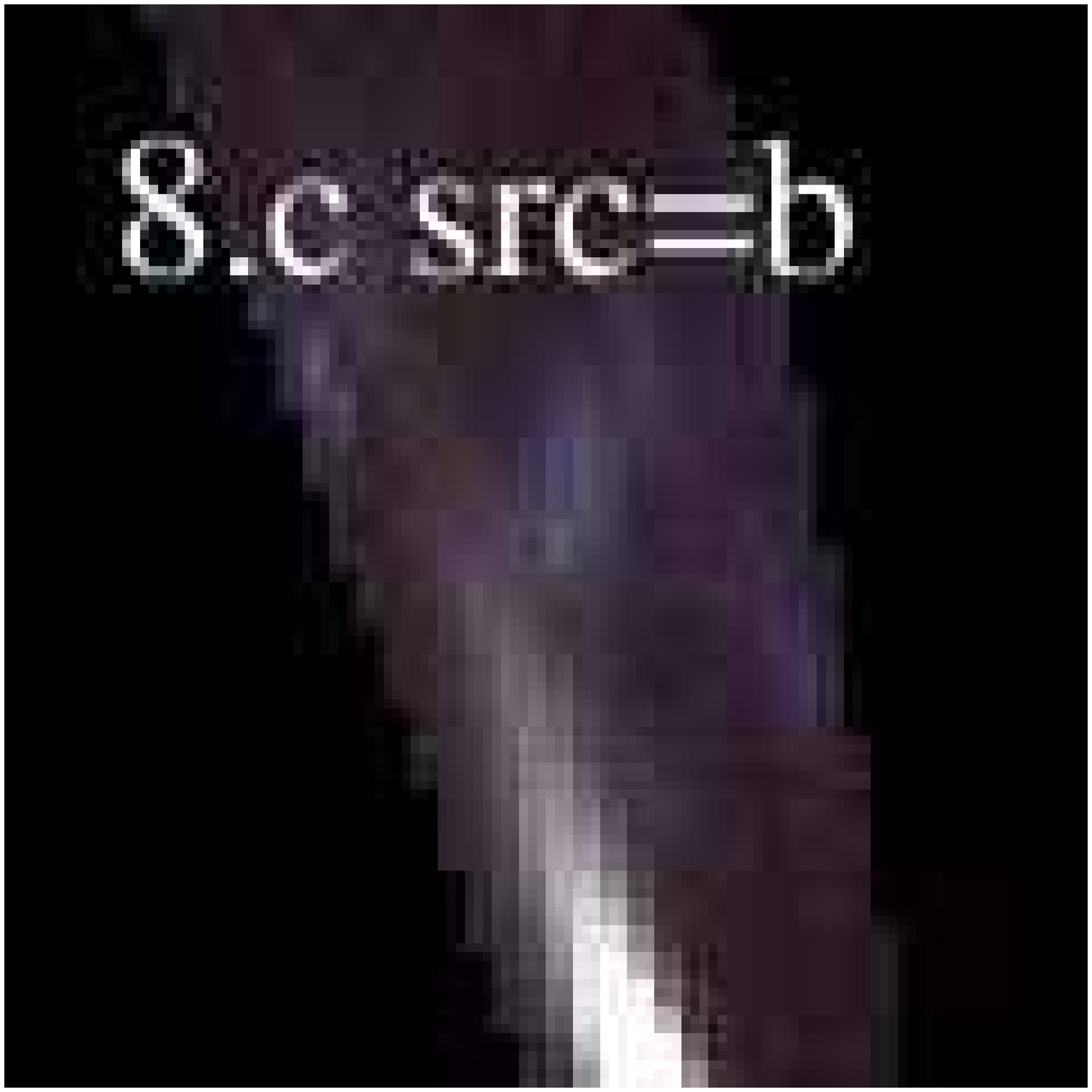}}
    & \multicolumn{1}{m{1.7cm}}{\includegraphics[height=2.00cm,clip]{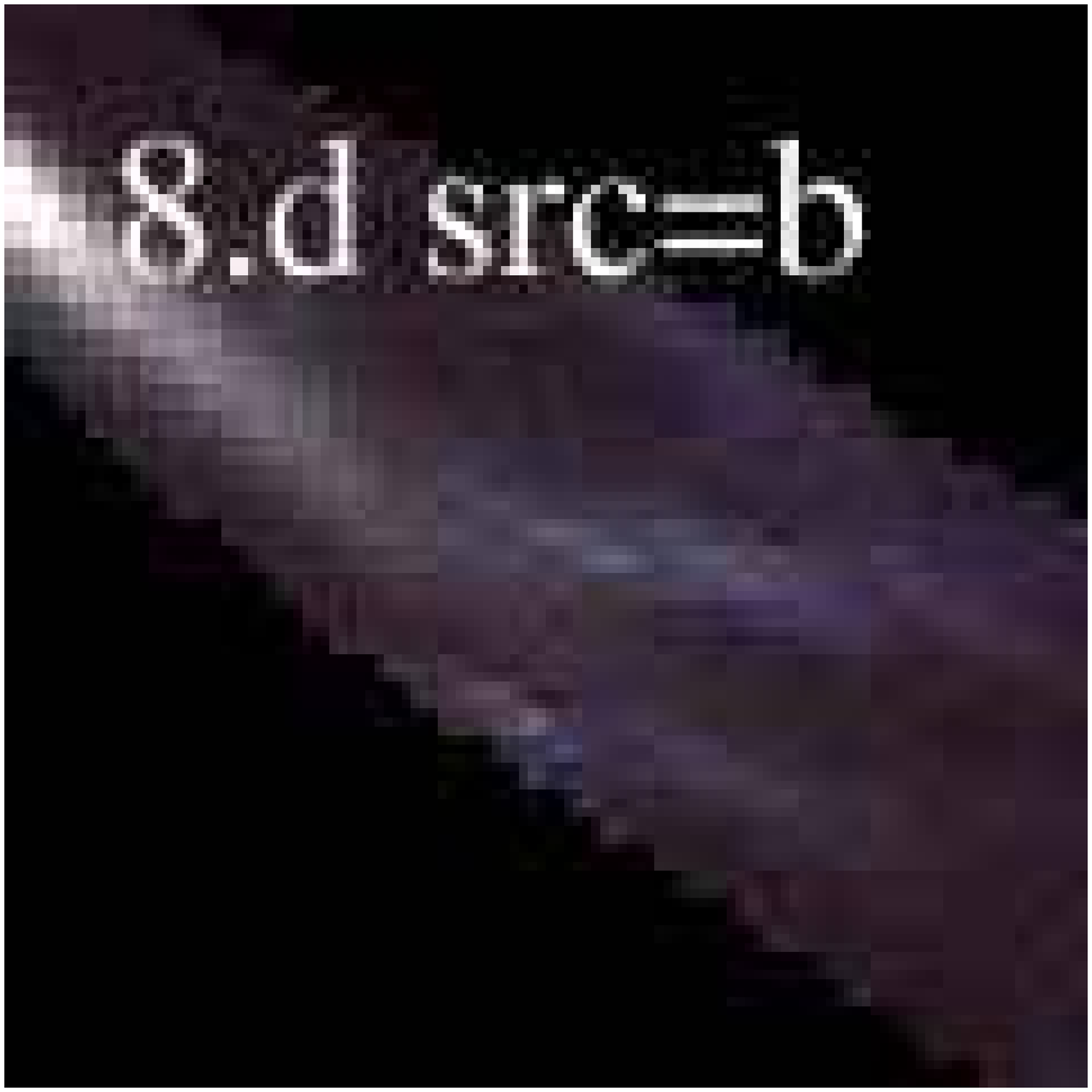}}
    & \multicolumn{1}{m{1.7cm}}{\includegraphics[height=2.00cm,clip]{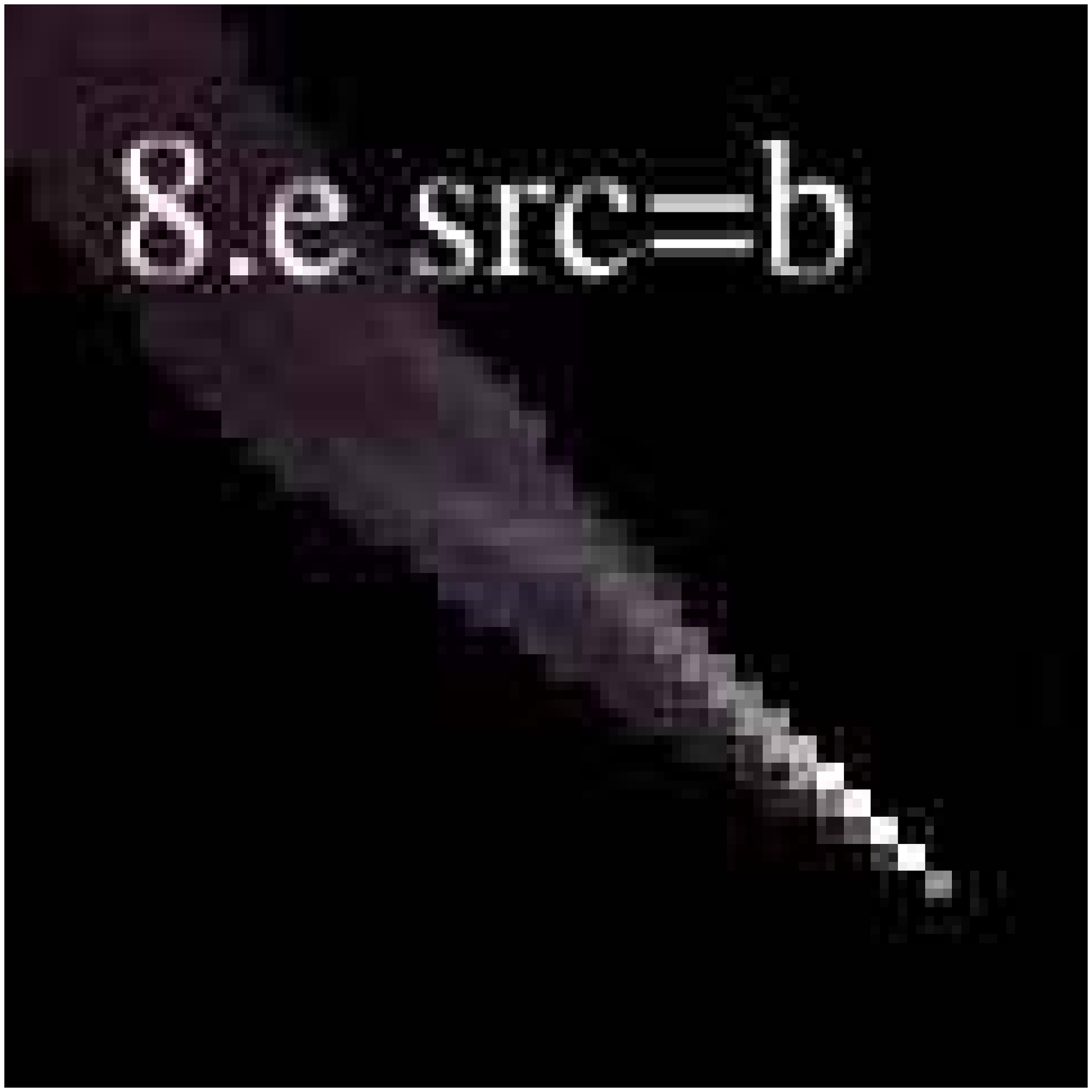}} \\
  \end{tabular}

\end{table*}

\begin{table*}
  \caption{Image system 9:}\vspace{0mm}
  \begin{tabular}{ccccc}
    \multicolumn{1}{m{1cm}}{{\Large A1689}}
    & \multicolumn{1}{m{1.7cm}}{\includegraphics[height=2.00cm,clip]{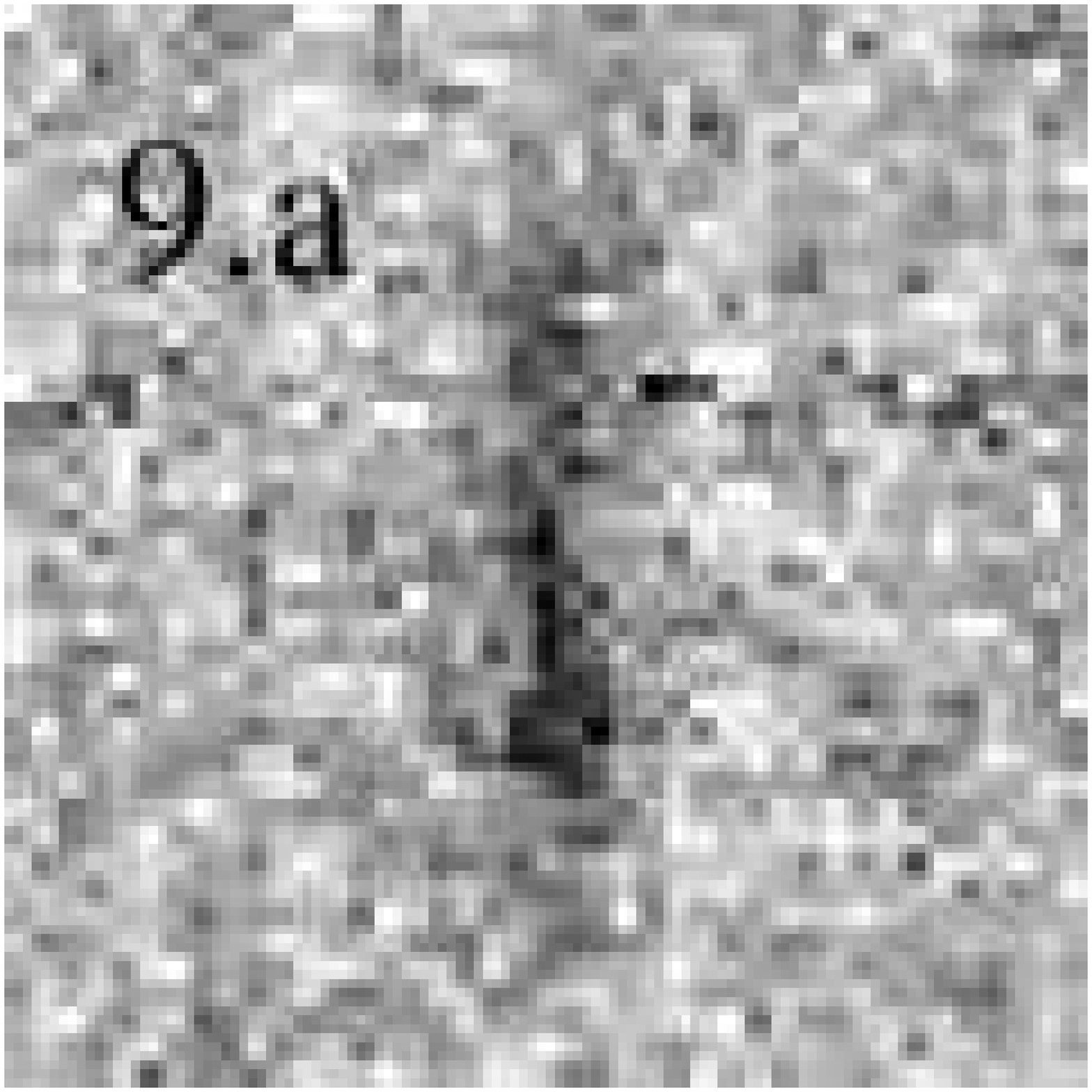}}
    & \multicolumn{1}{m{1.7cm}}{\includegraphics[height=2.00cm,clip]{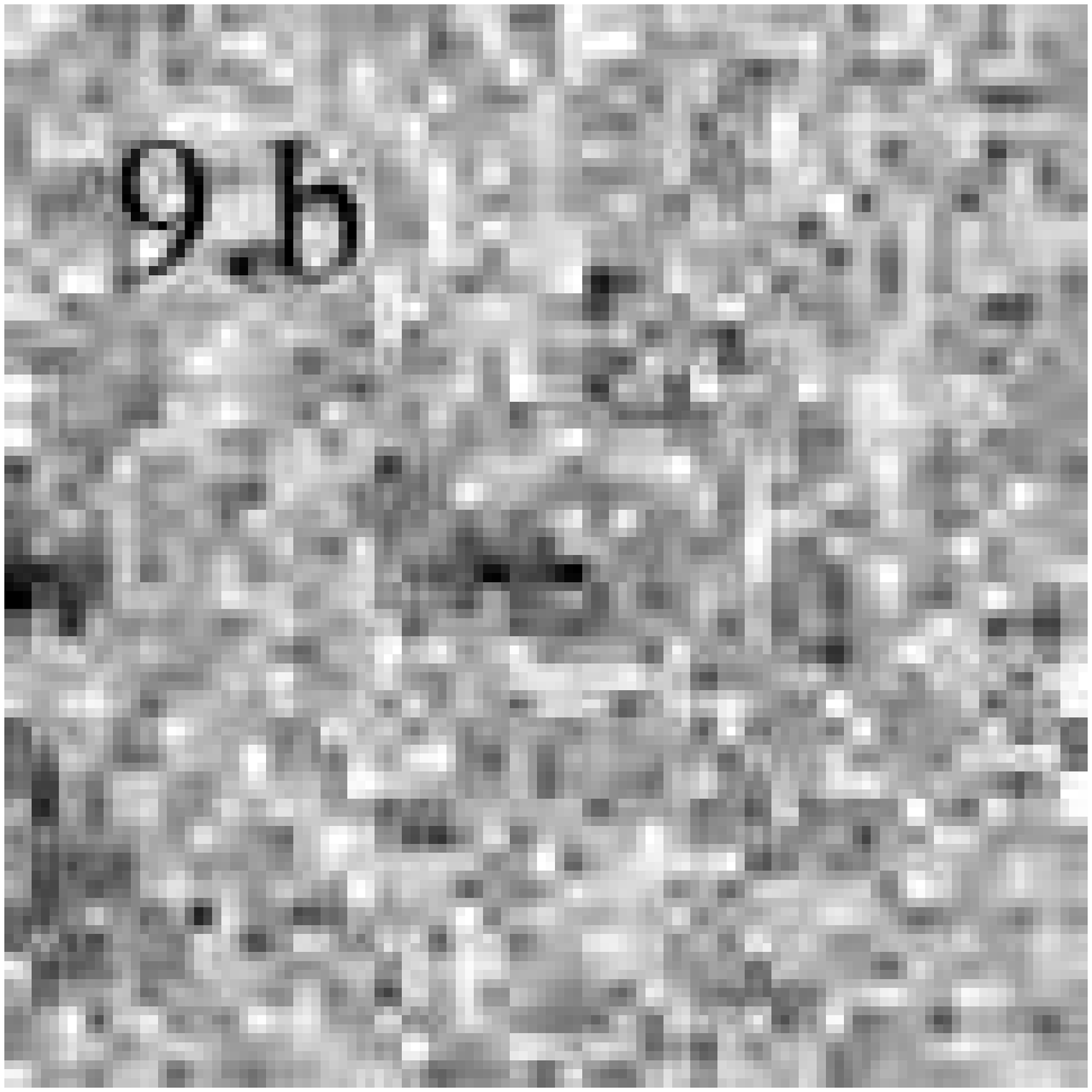}}
    & \multicolumn{1}{m{1.7cm}}{\includegraphics[height=2.00cm,clip]{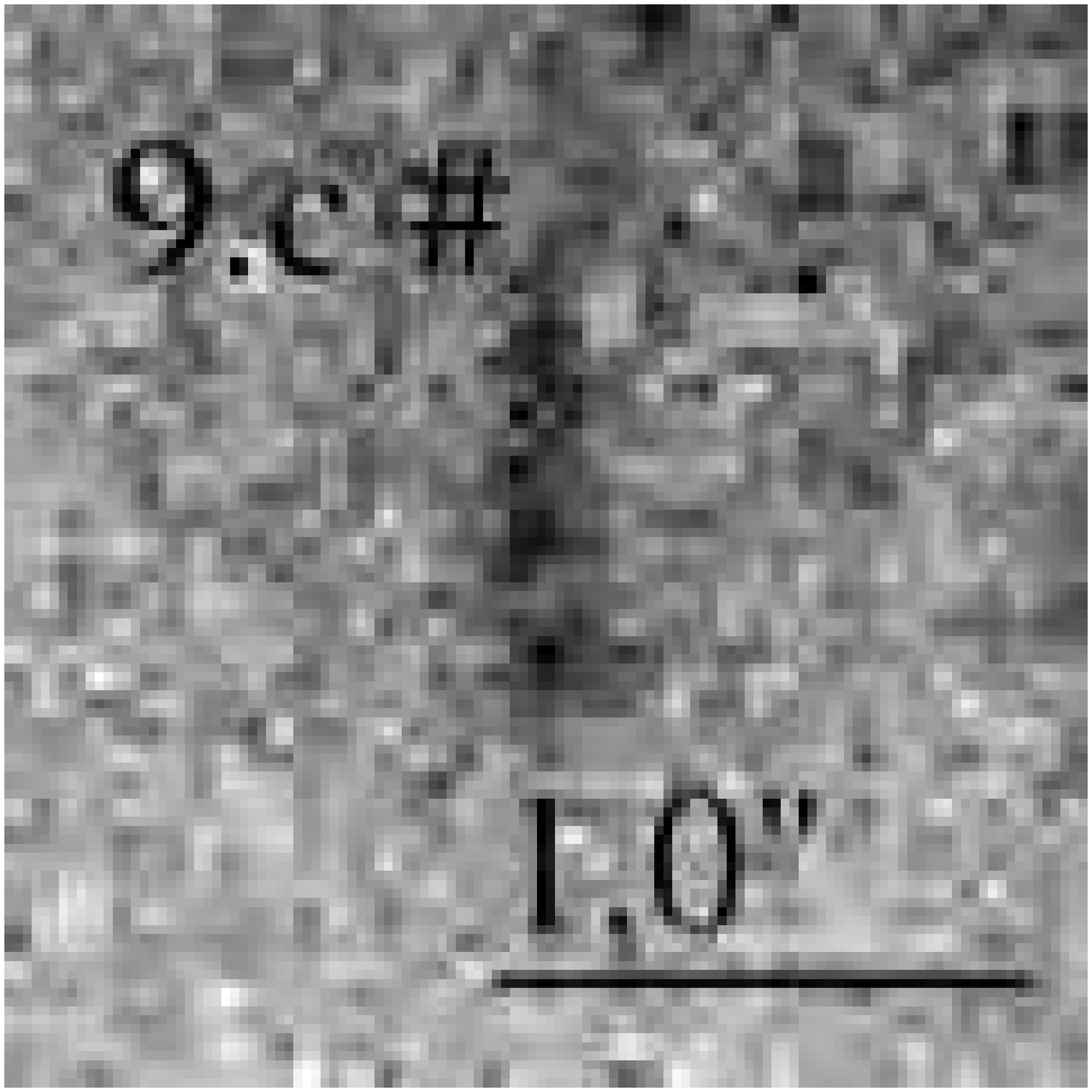}}
    & \multicolumn{1}{m{1.7cm}}{\includegraphics[height=2.00cm,clip]{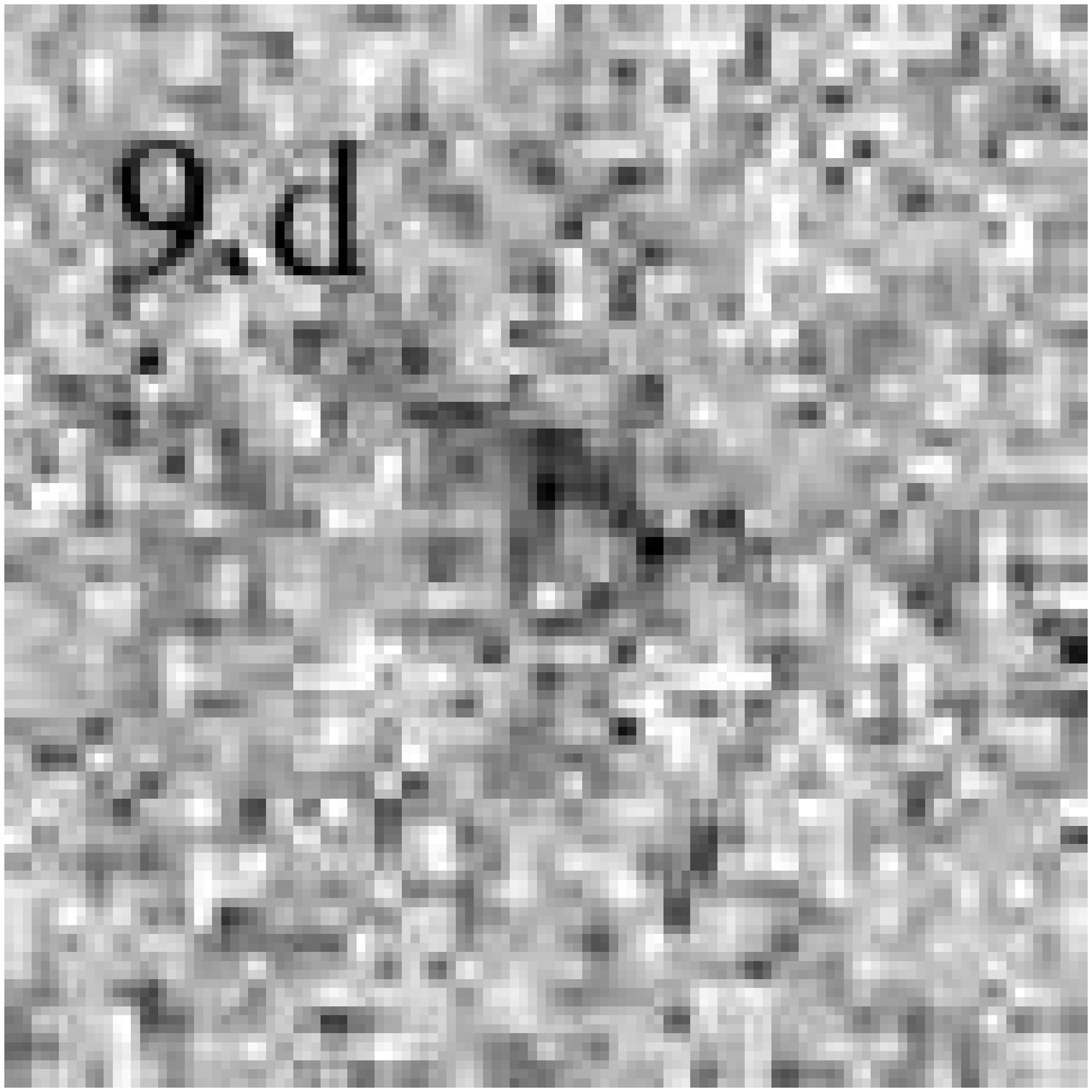}} \\
    \multicolumn{1}{m{1cm}}{{\Large NSIE}}
    & \multicolumn{1}{m{1.7cm}}{\includegraphics[height=2.00cm,clip]{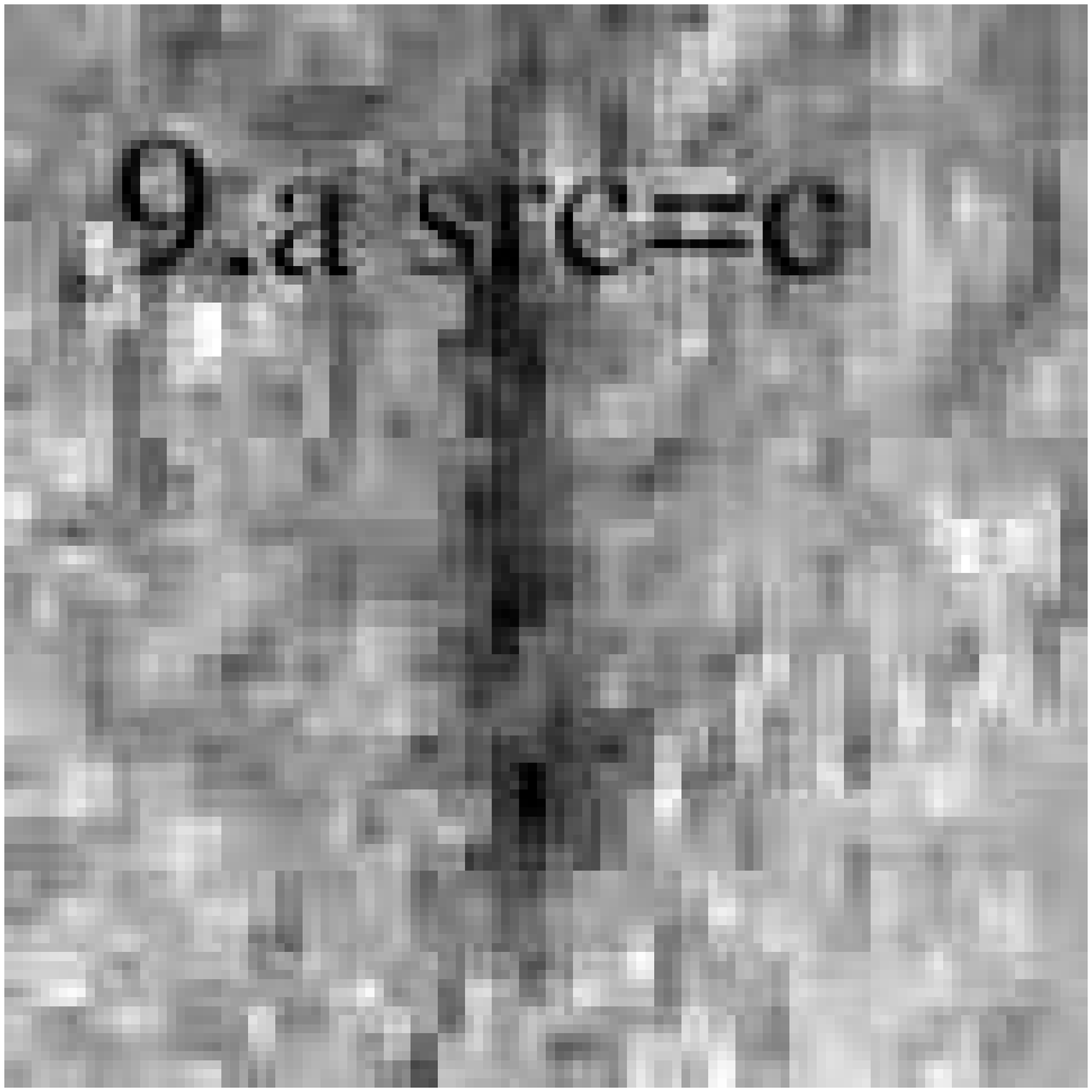}}
    & \multicolumn{1}{m{1.7cm}}{\includegraphics[height=2.00cm,clip]{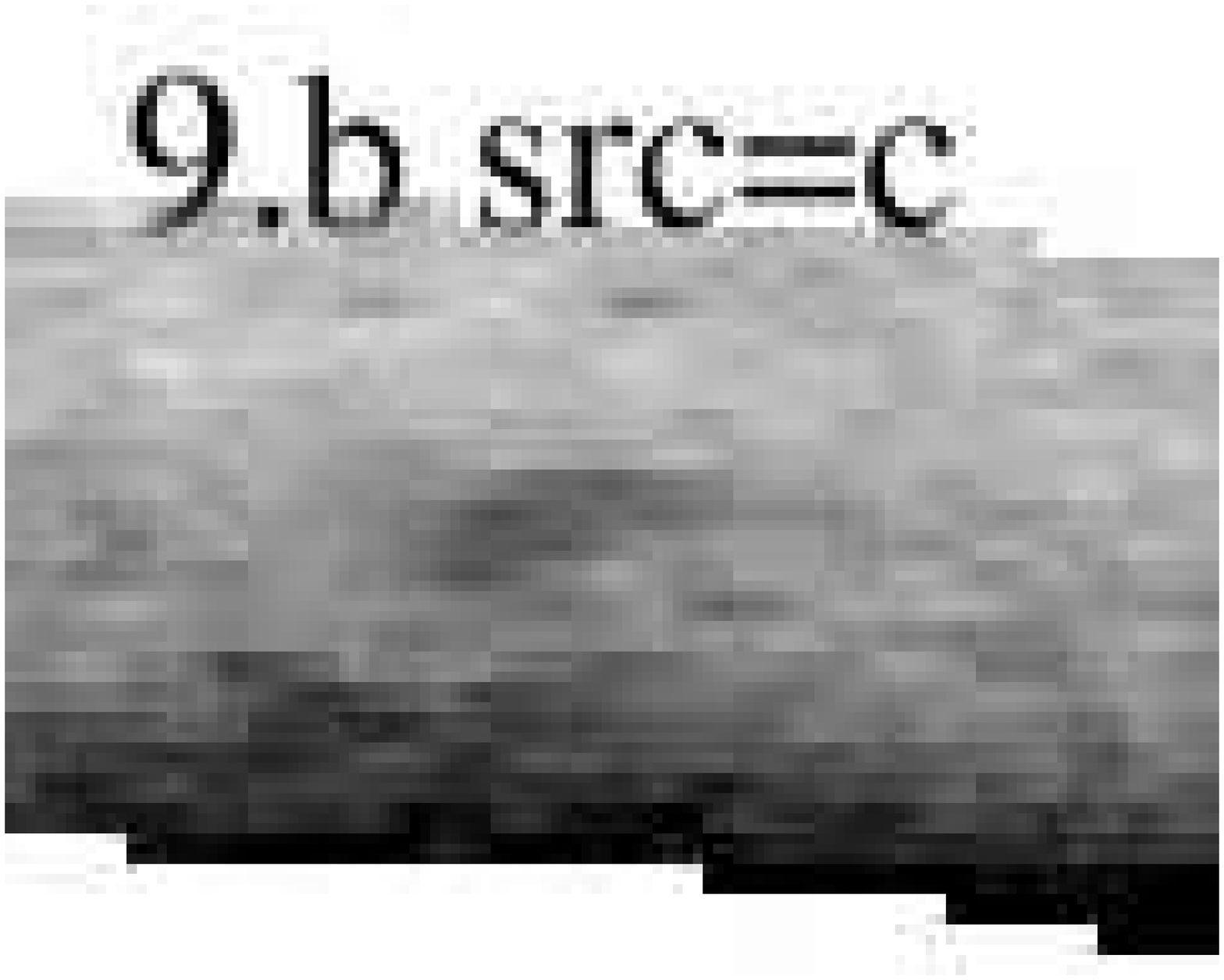}}
    & \multicolumn{1}{m{1.7cm}}{\includegraphics[height=2.00cm,clip]{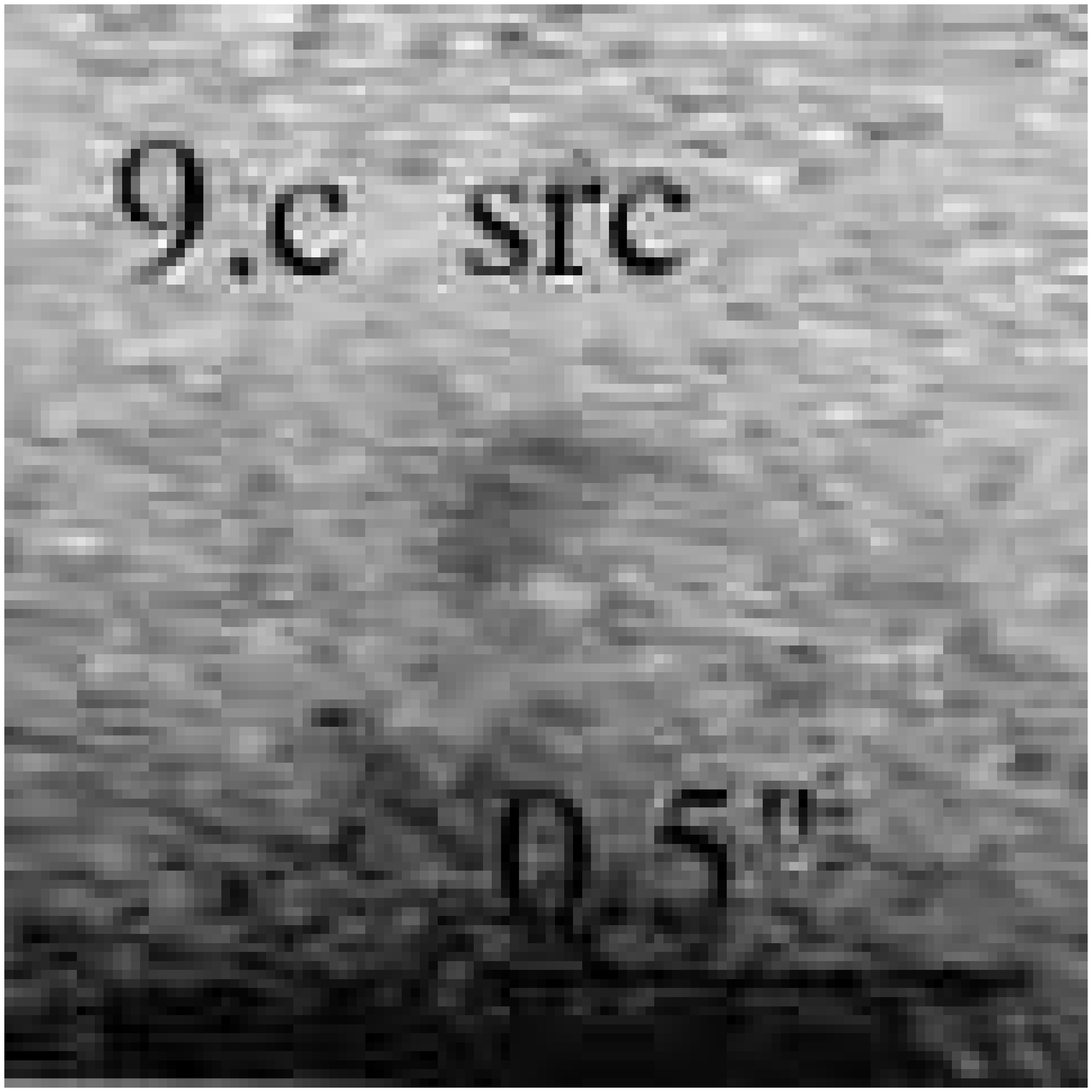}}
    & \multicolumn{1}{m{1.7cm}}{\includegraphics[height=2.00cm,clip]{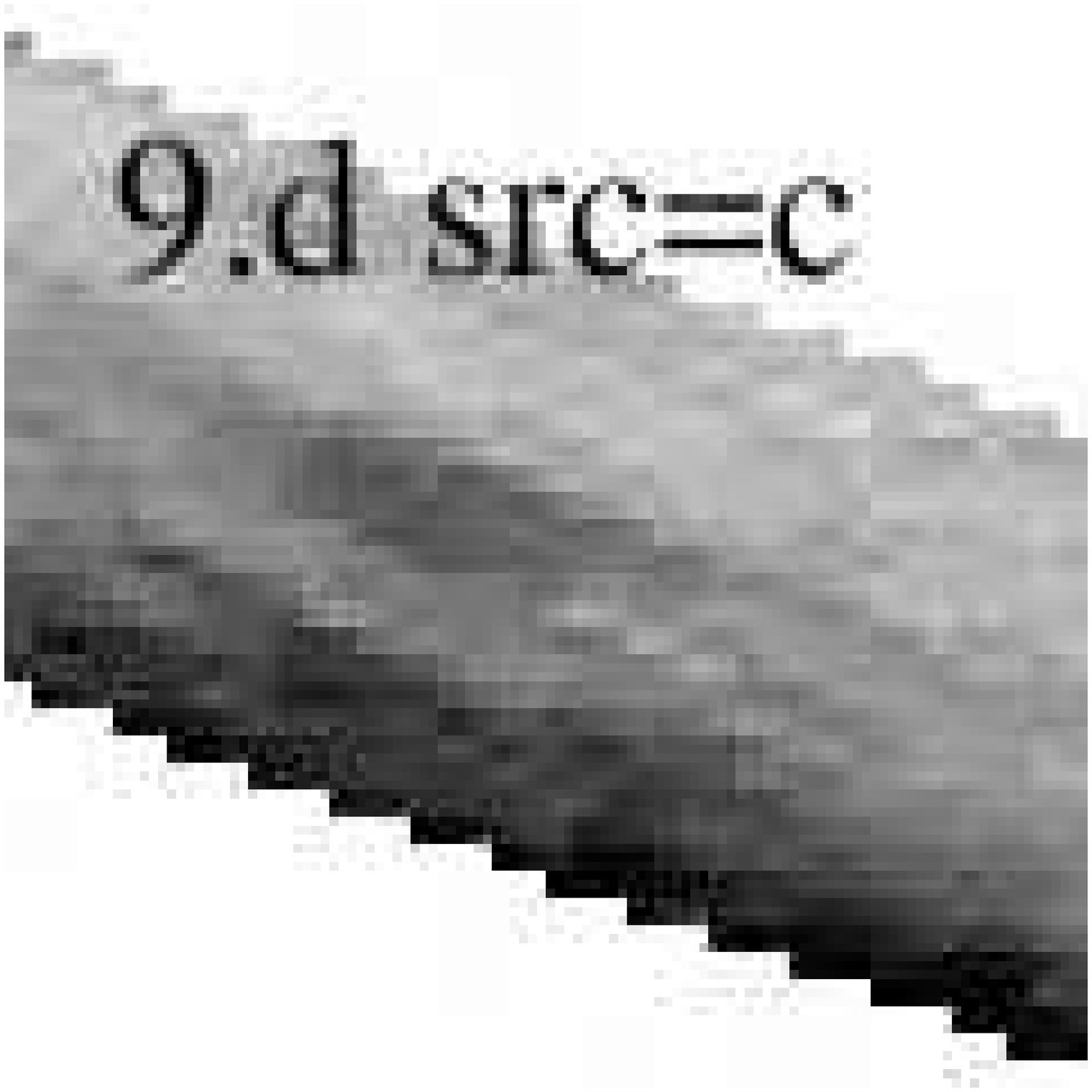}} \\
    \multicolumn{1}{m{1cm}}{{\Large ENFW}}
    & \multicolumn{1}{m{1.7cm}}{\includegraphics[height=2.00cm,clip]{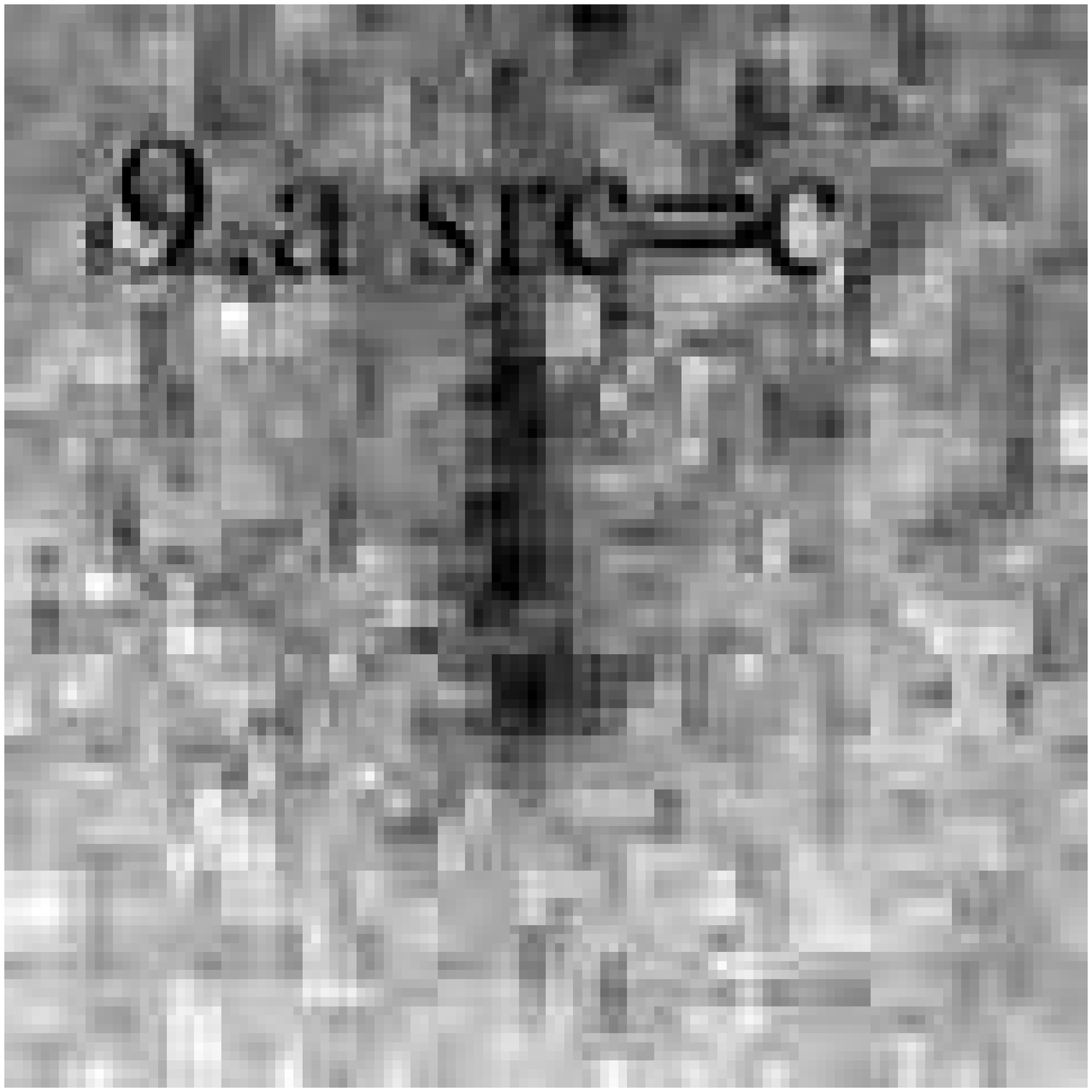}}
    & \multicolumn{1}{m{1.7cm}}{\includegraphics[height=2.00cm,clip]{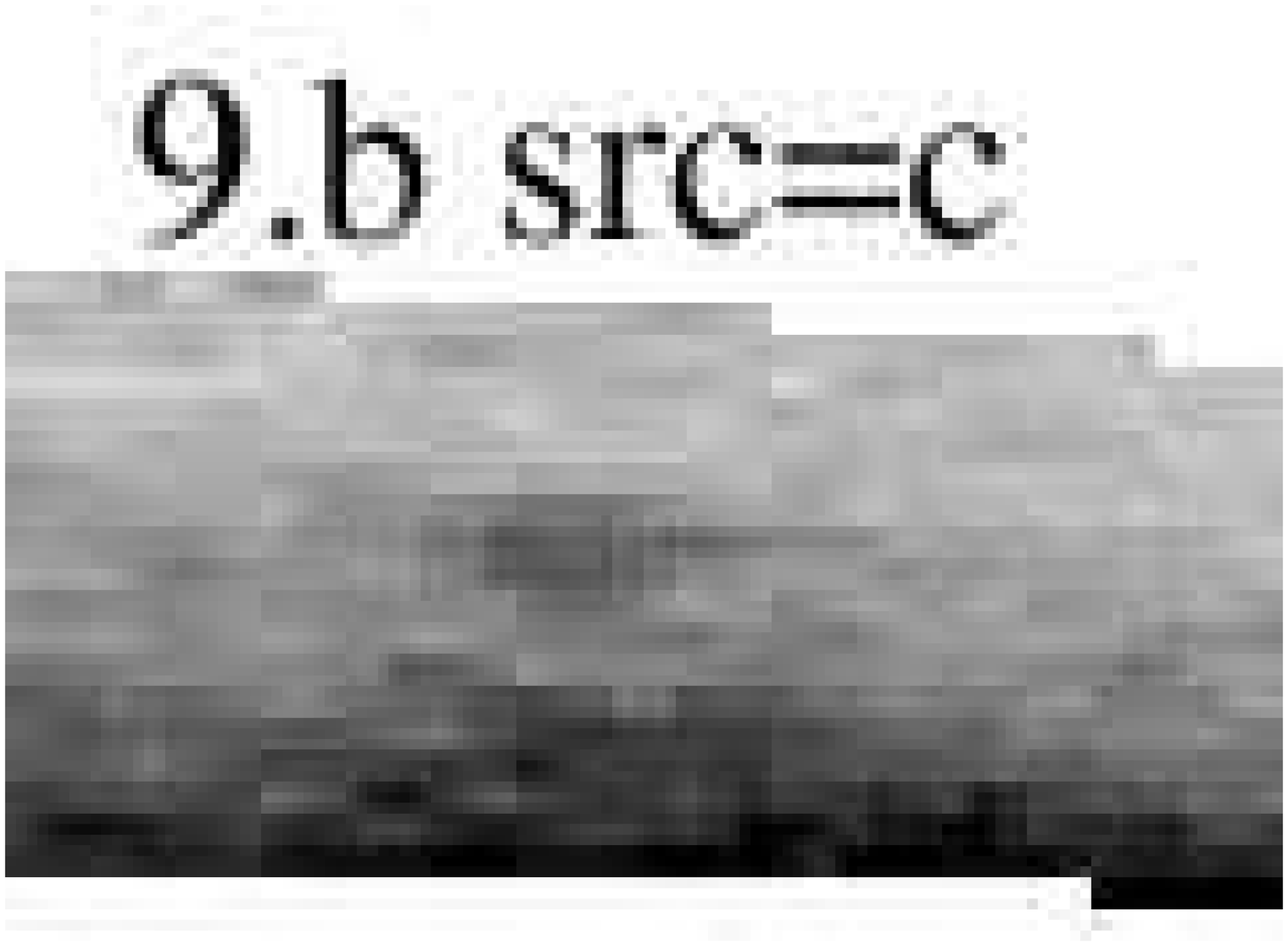}}
    & \multicolumn{1}{m{1.7cm}}{\includegraphics[height=2.00cm,clip]{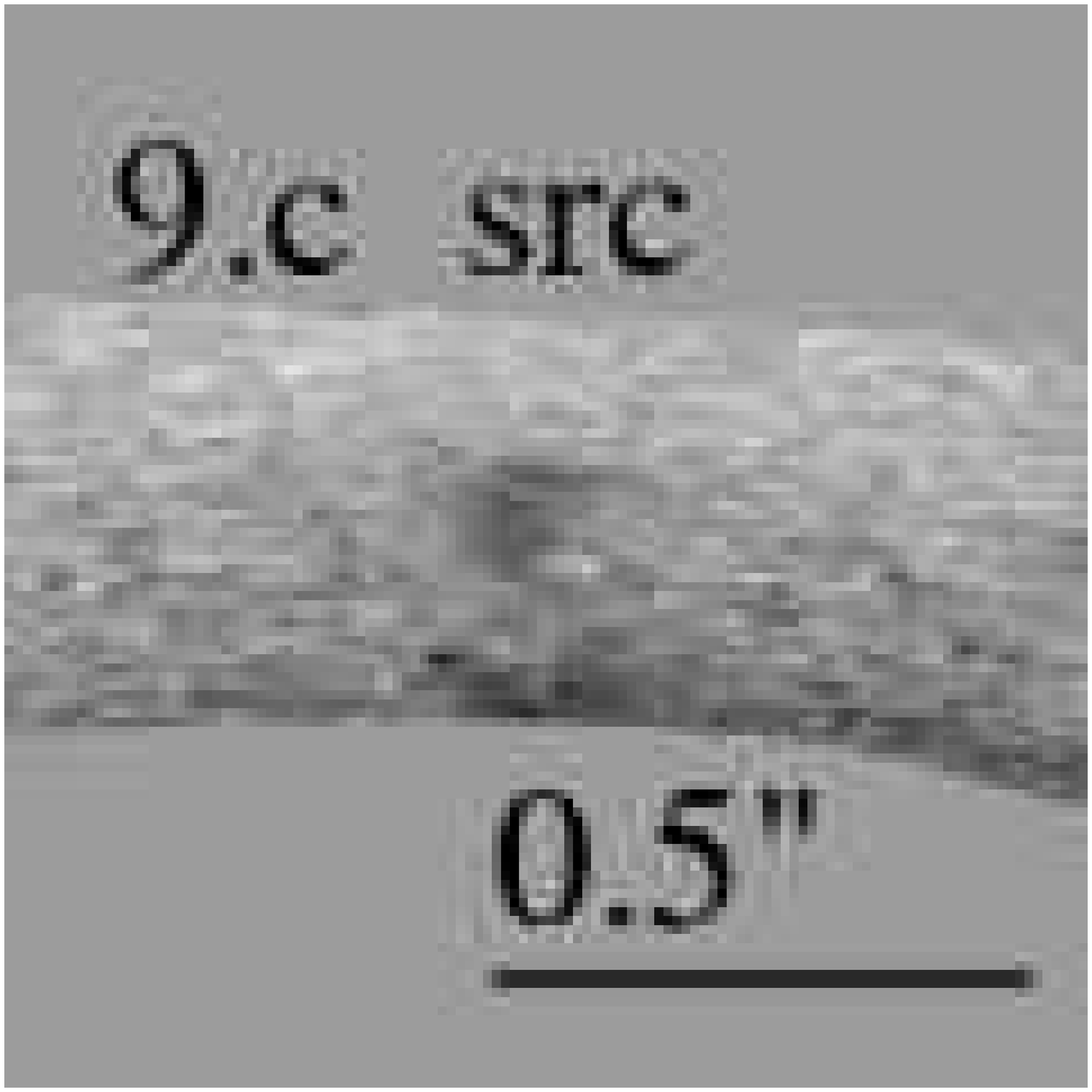}}
    & \multicolumn{1}{m{1.7cm}}{\includegraphics[height=2.00cm,clip]{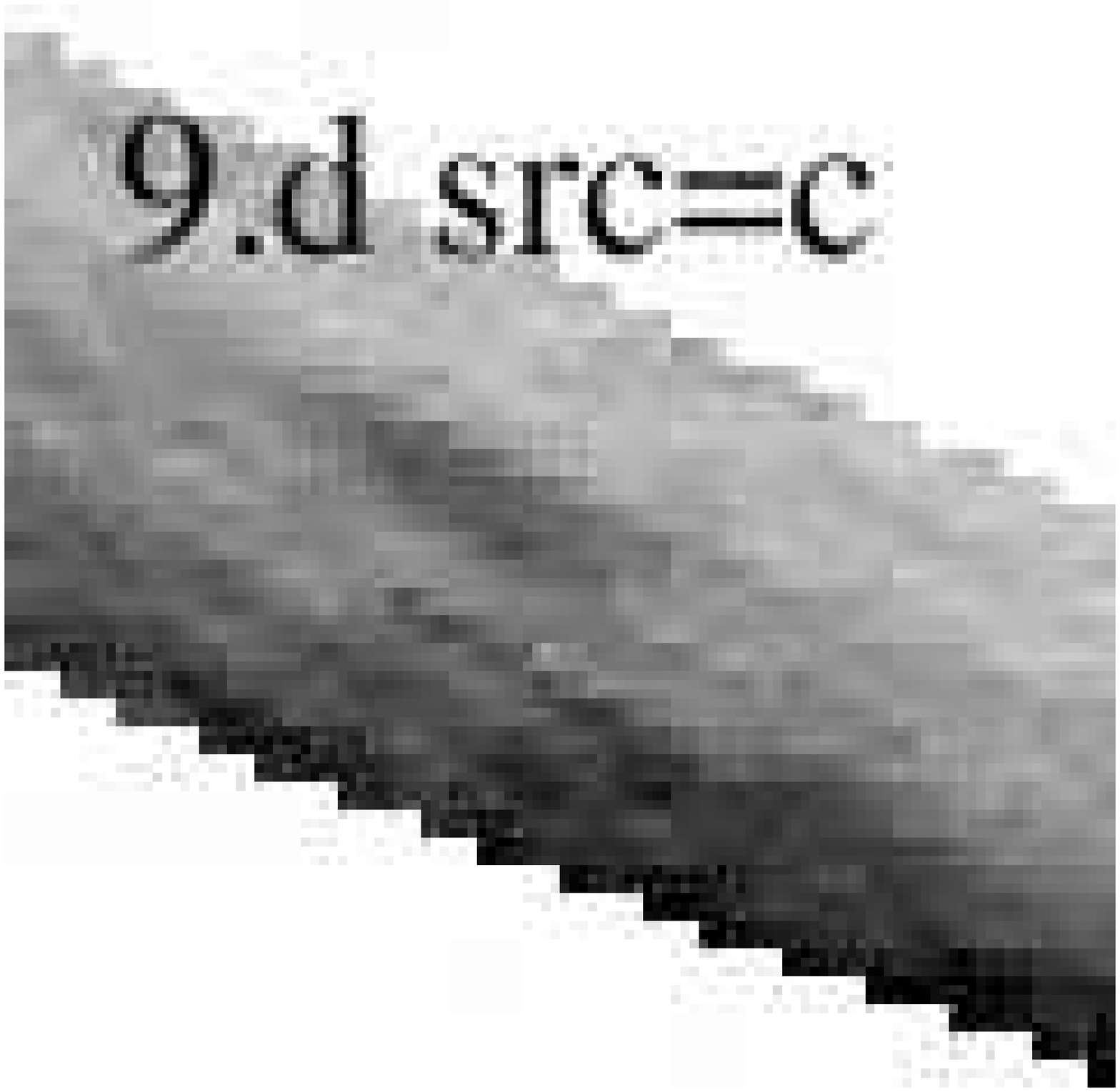}} \\
  \end{tabular}

\end{table*}

\clearpage

\begin{table*}
  \caption{Image system 10:}\vspace{0mm}
  \begin{tabular}{cccc}
    \multicolumn{1}{m{1cm}}{{\Large A1689}}
    & \multicolumn{1}{m{1.7cm}}{\includegraphics[height=2.00cm,clip]{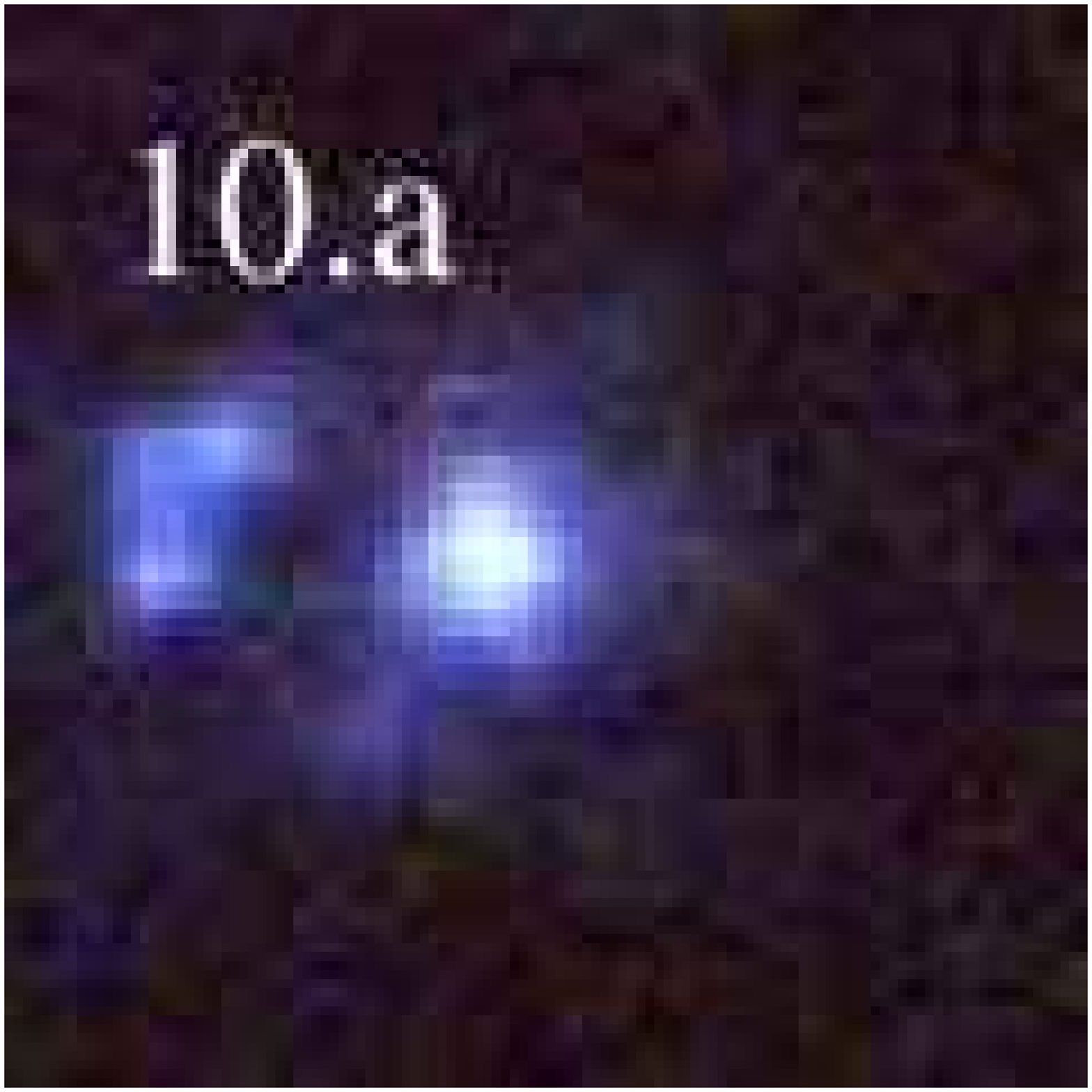}}
    & \multicolumn{1}{m{1.7cm}}{\includegraphics[height=2.00cm,clip]{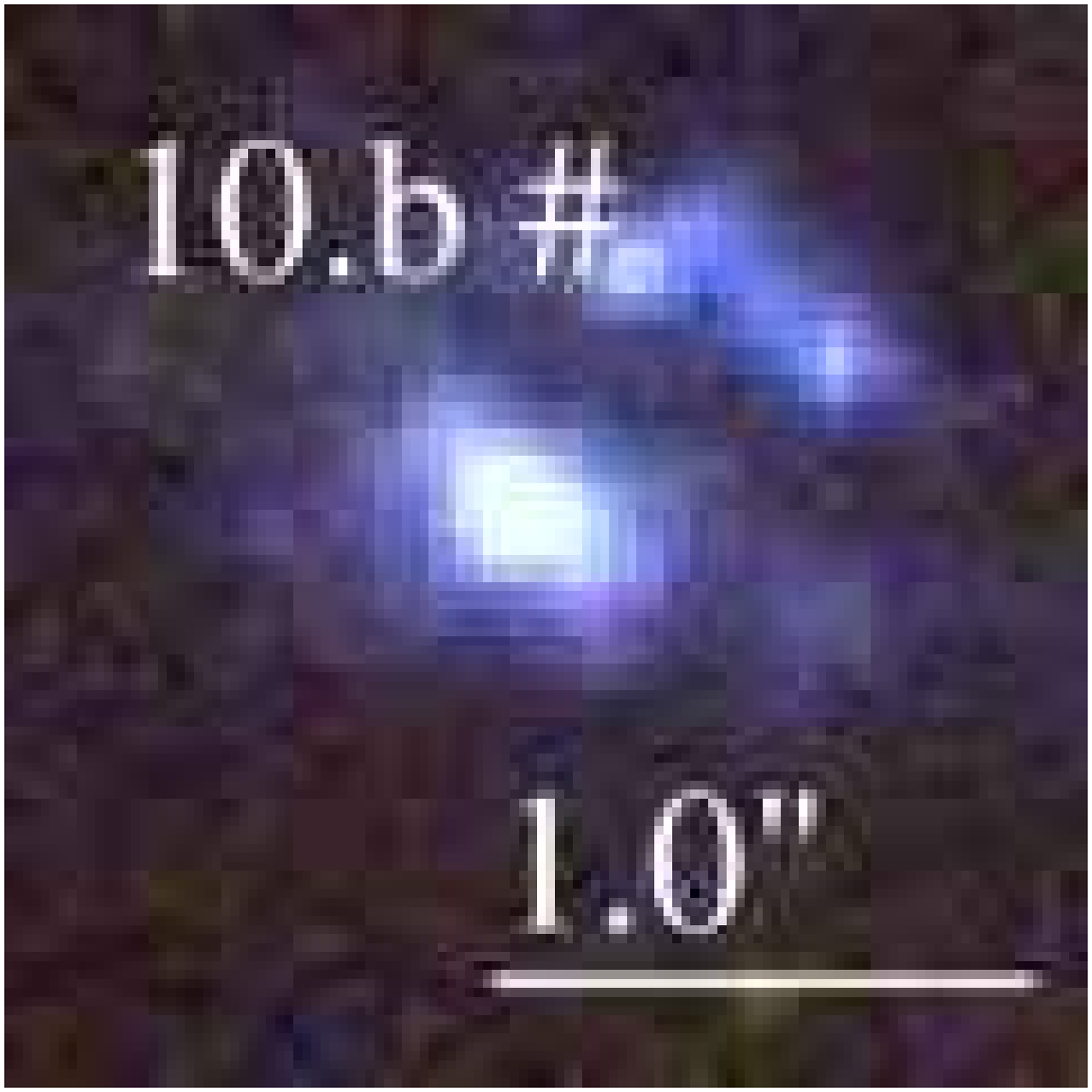}}
    & \multicolumn{1}{m{1.7cm}}{\includegraphics[height=2.00cm,clip]{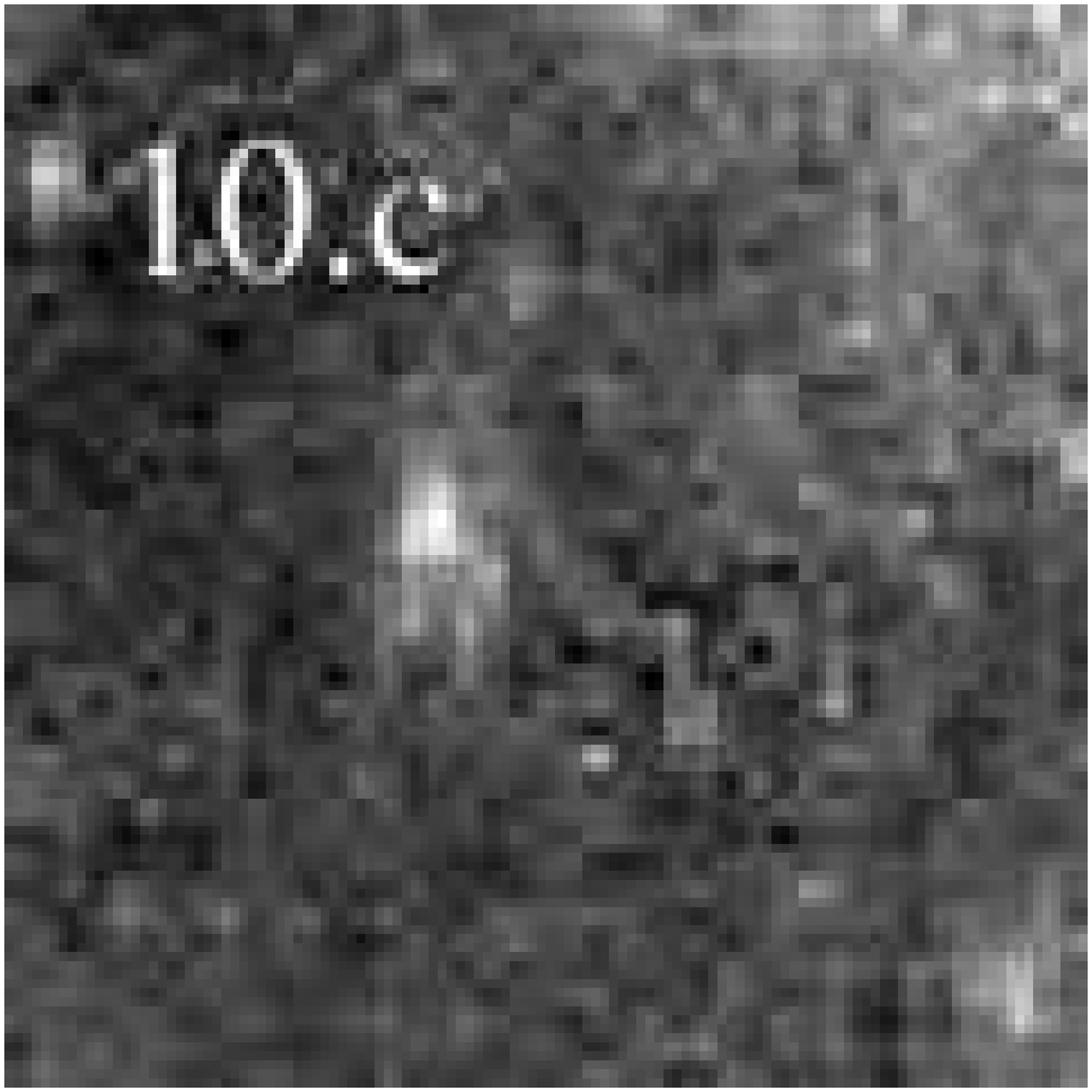}} \\
    \multicolumn{1}{m{1cm}}{{\Large NSIE}}
    & \multicolumn{1}{m{1.7cm}}{\includegraphics[height=2.00cm,clip]{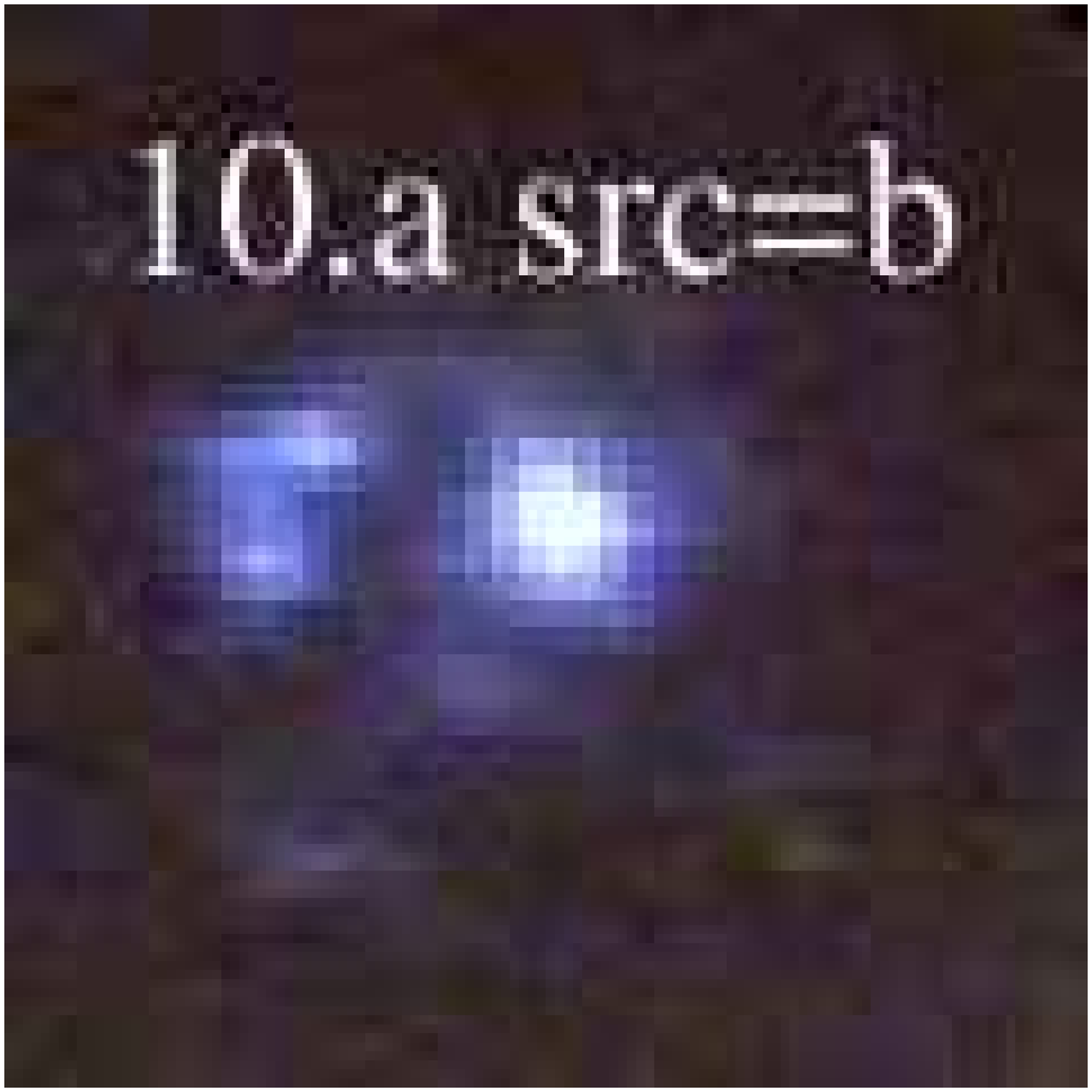}}
    & \multicolumn{1}{m{1.7cm}}{\includegraphics[height=2.00cm,clip]{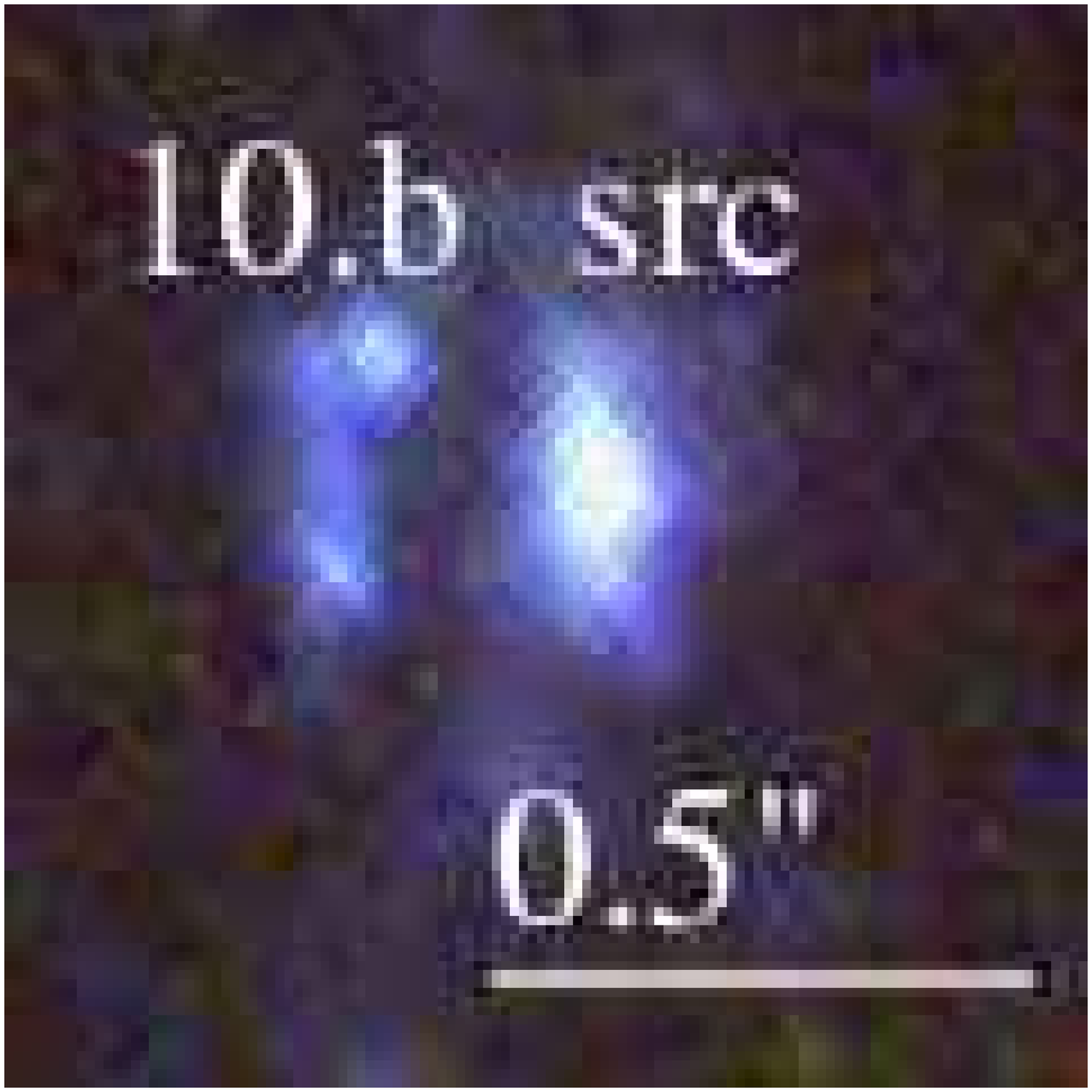}}
    & \multicolumn{1}{m{1.7cm}}{\includegraphics[height=2.00cm,clip]{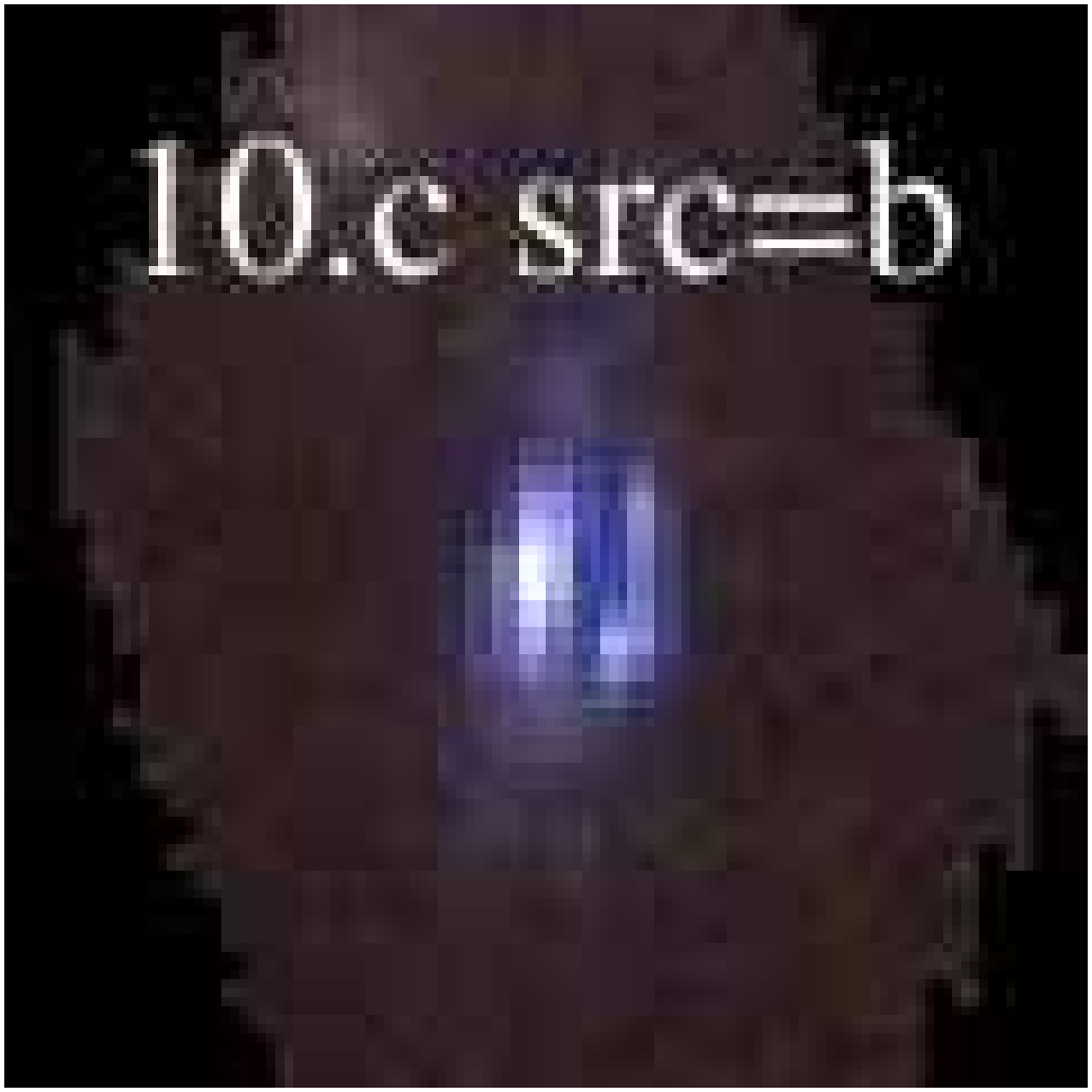}} \\
    \multicolumn{1}{m{1cm}}{{\Large ENFW}}
    & \multicolumn{1}{m{1.7cm}}{\includegraphics[height=2.00cm,clip]{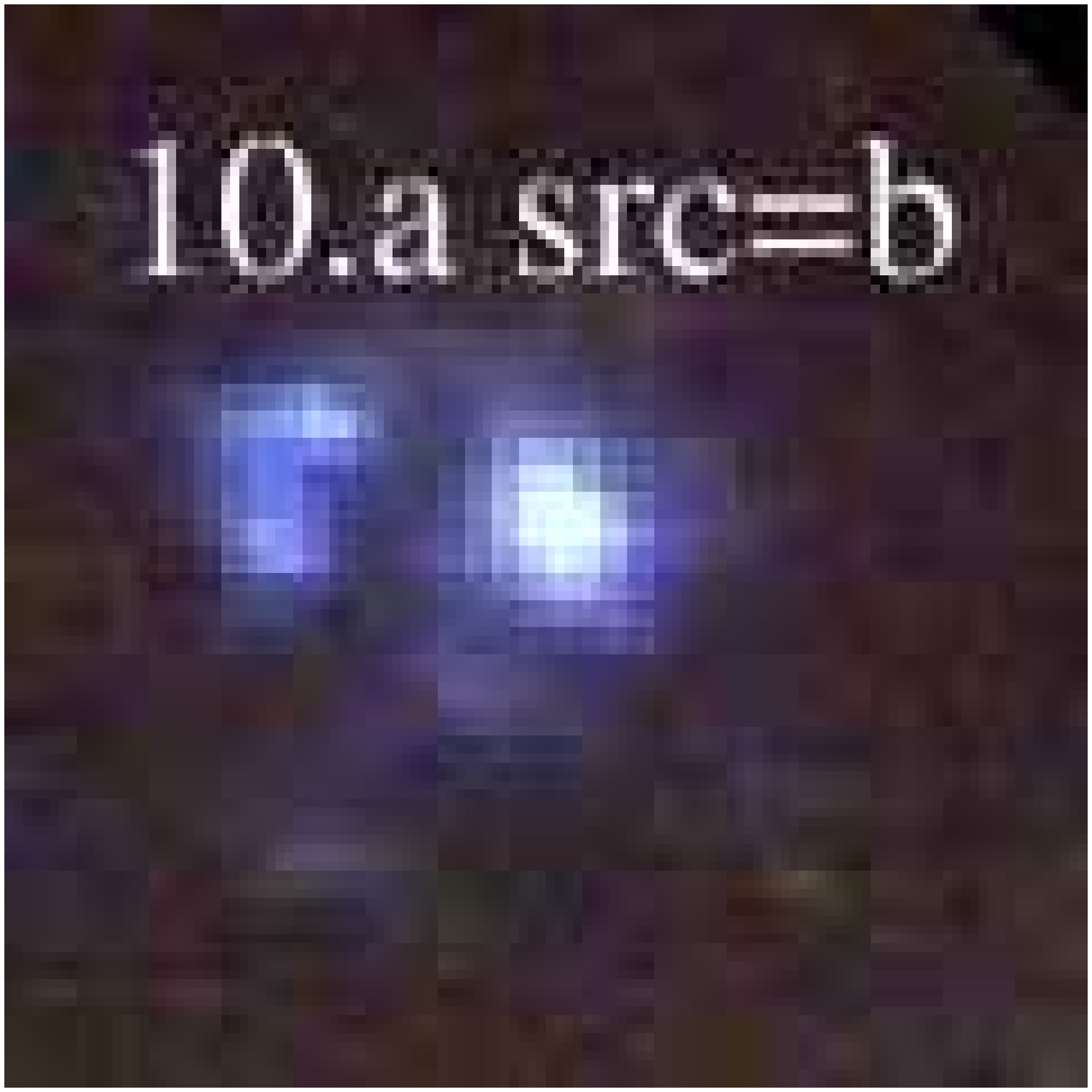}}
    & \multicolumn{1}{m{1.7cm}}{\includegraphics[height=2.00cm,clip]{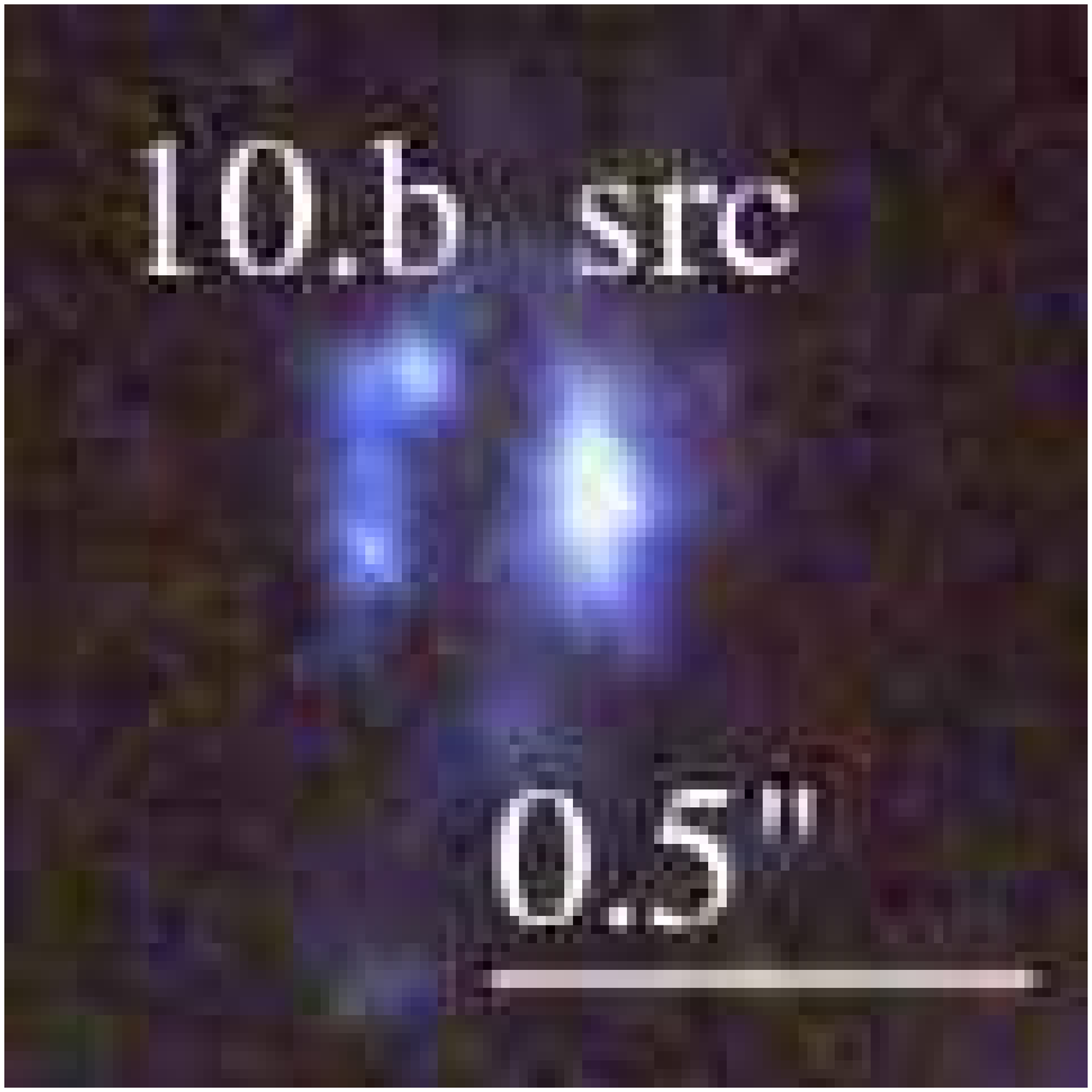}}
    & \multicolumn{1}{m{1.7cm}}{\includegraphics[height=2.00cm,clip]{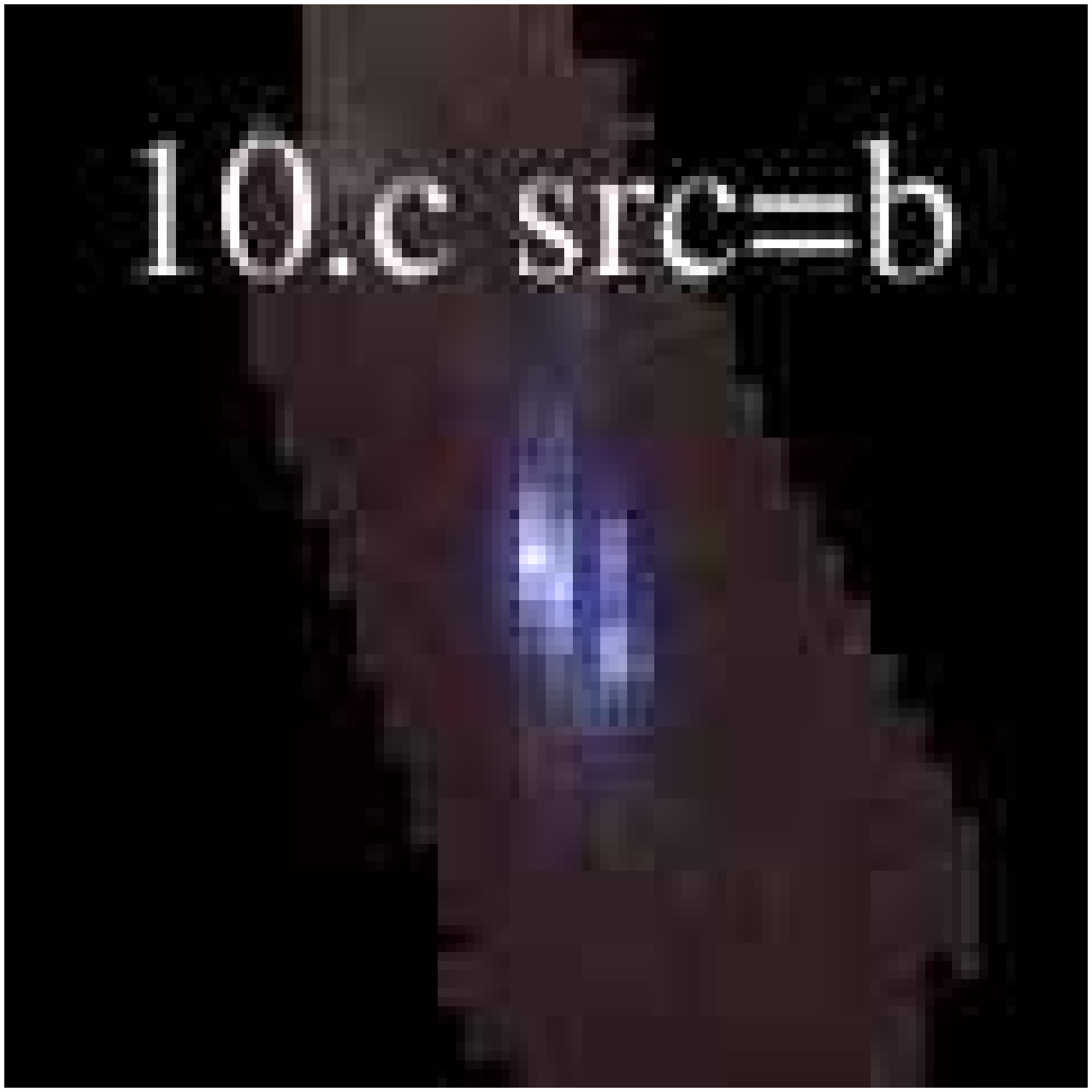}} \\
  \end{tabular}

\end{table*}

\begin{table*}
  \caption{Image system 11:}\vspace{0mm}
  \begin{tabular}{cccc}
    \multicolumn{1}{m{1cm}}{{\Large A1689}}
    & \multicolumn{1}{m{1.7cm}}{\includegraphics[height=2.00cm,clip]{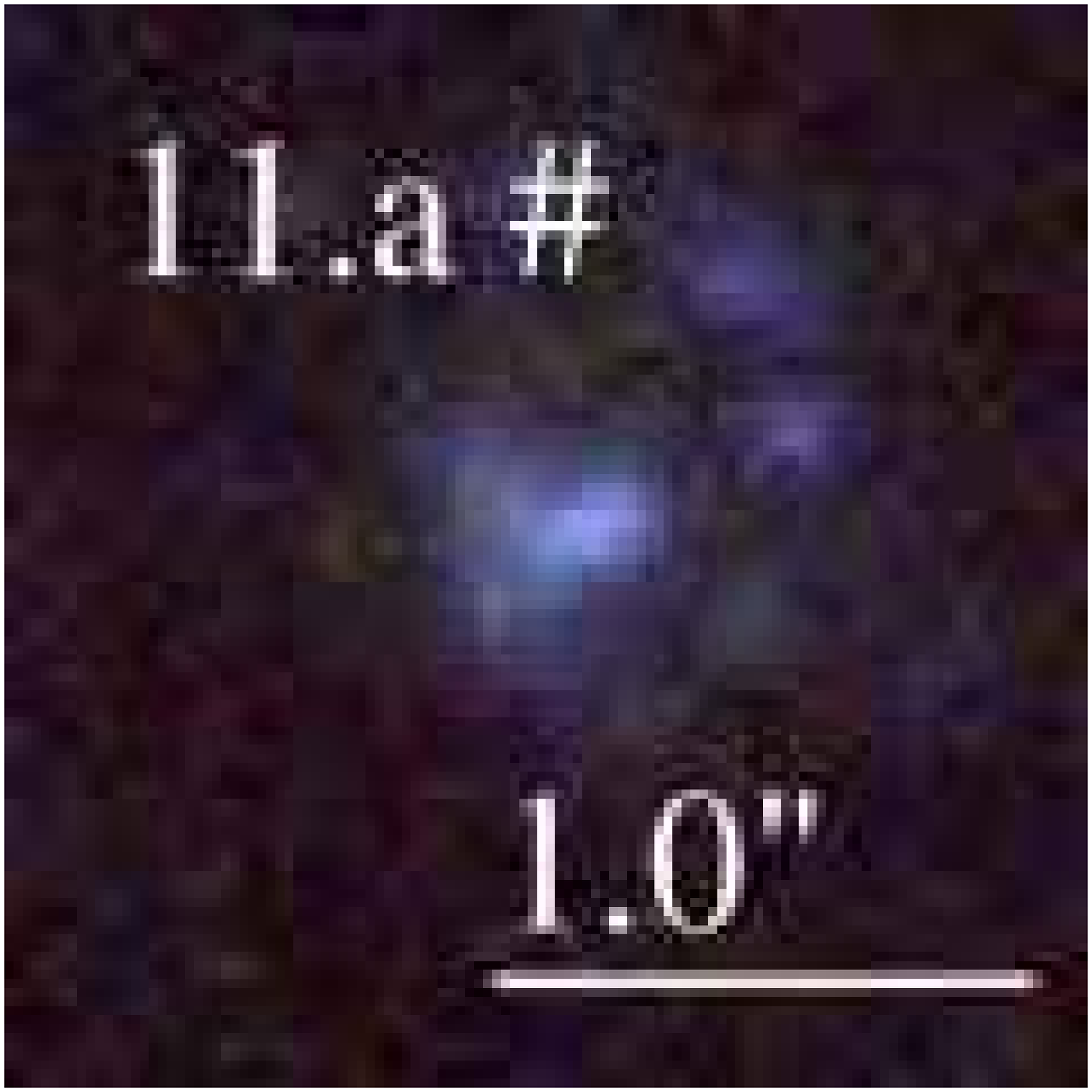}}
    & \multicolumn{1}{m{1.7cm}}{\includegraphics[height=2.00cm,clip]{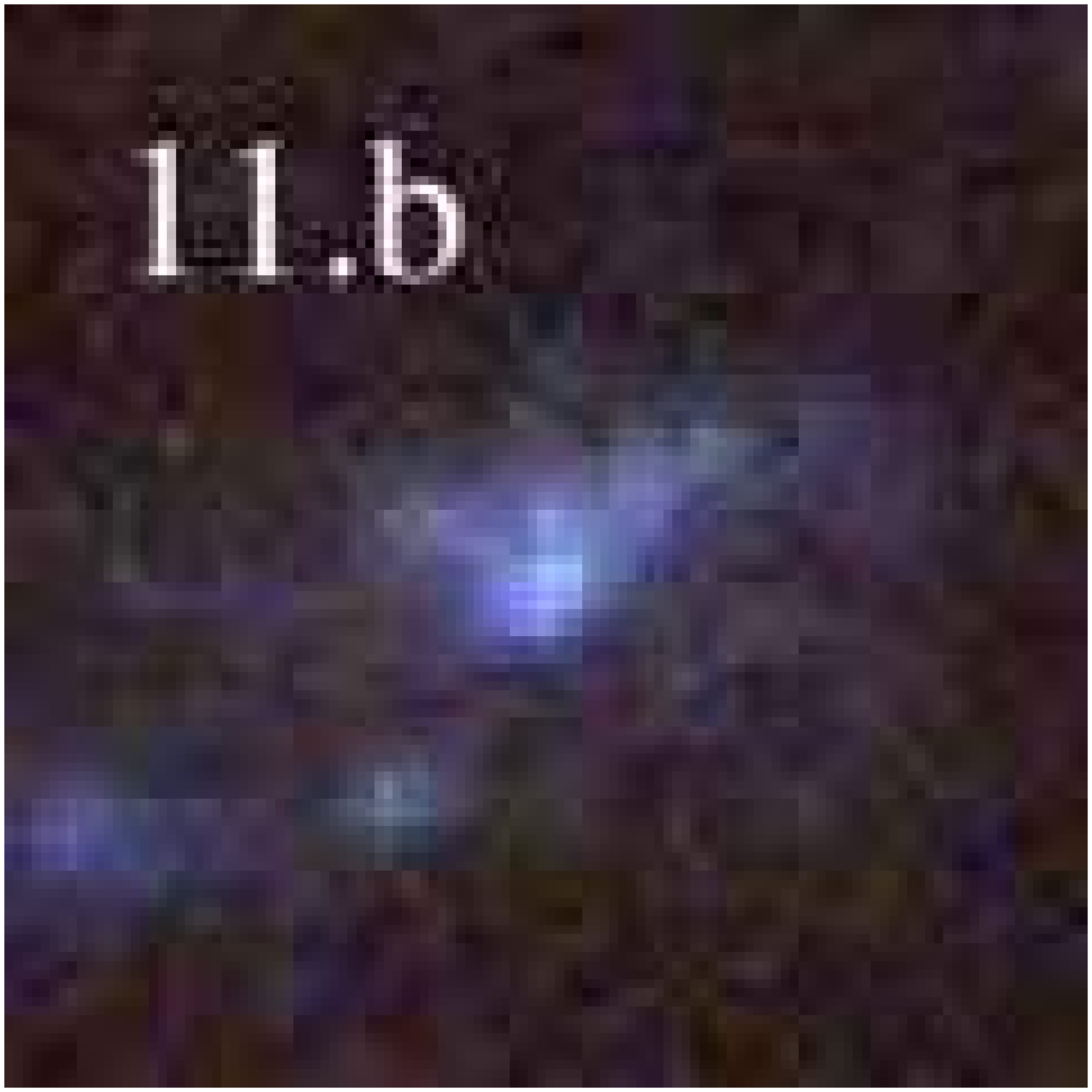}}
    & \multicolumn{1}{m{1.7cm}}{\includegraphics[height=2.00cm,clip]{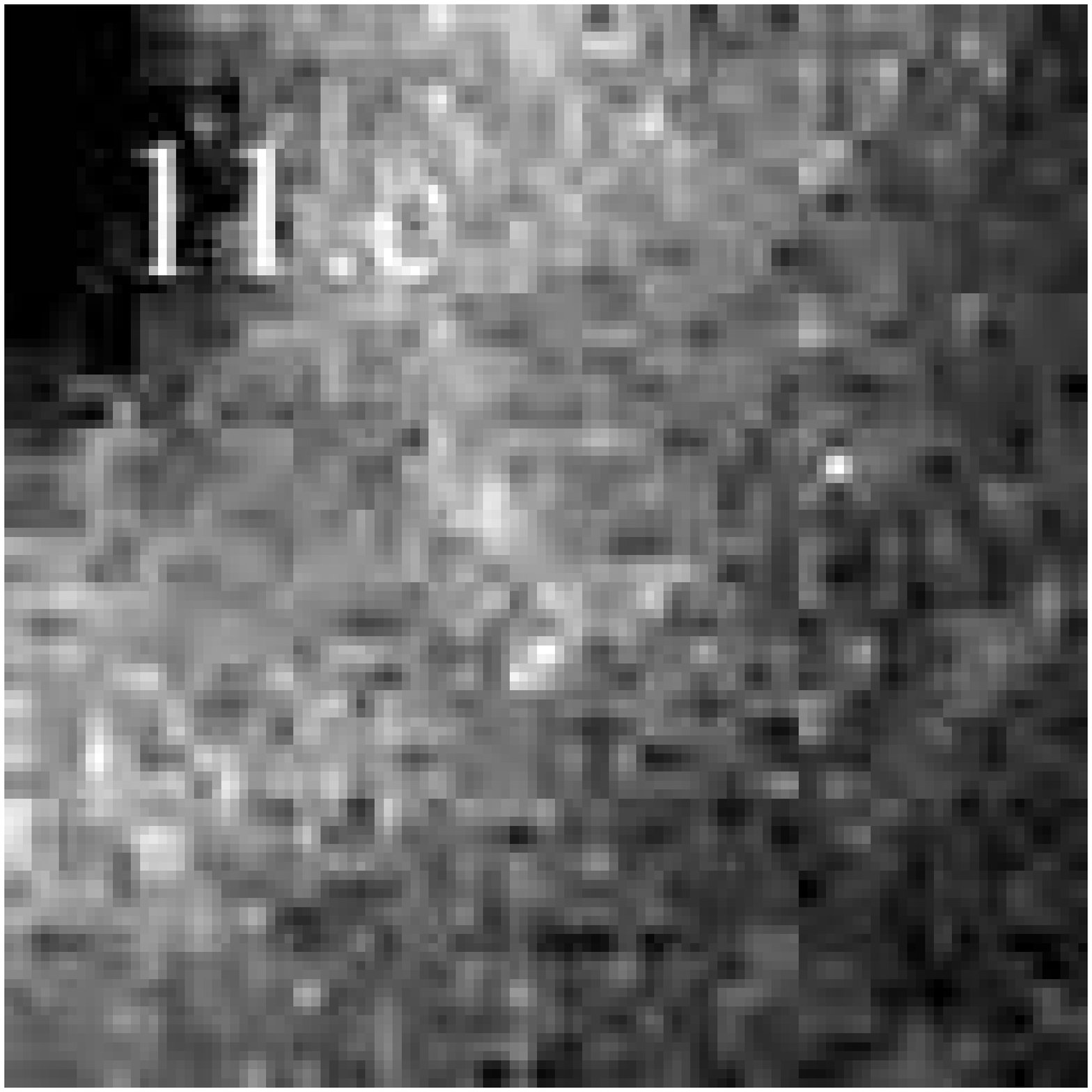}} \\
    \multicolumn{1}{m{1cm}}{{\Large NSIE}}
    & \multicolumn{1}{m{1.7cm}}{\includegraphics[height=2.00cm,clip]{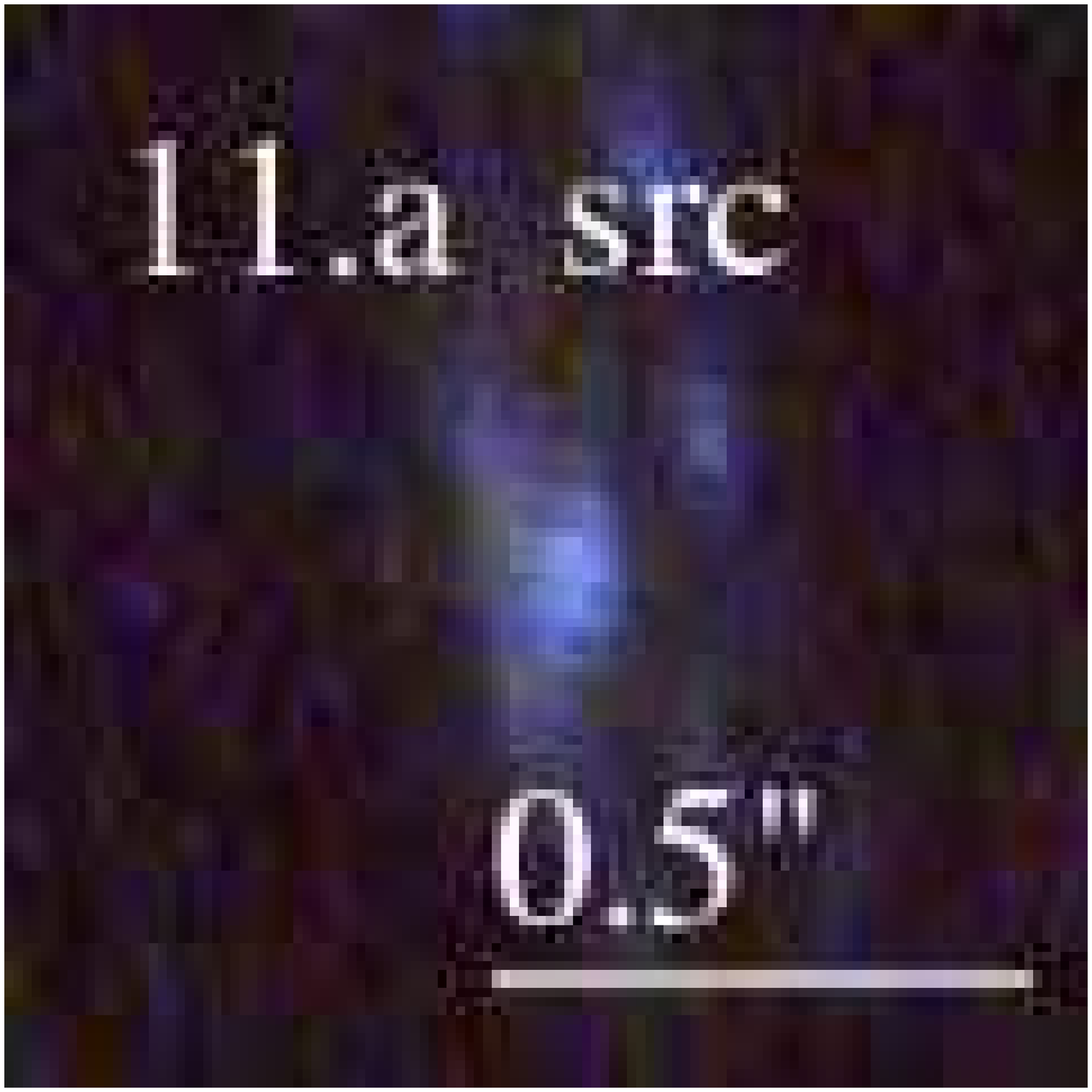}}
    & \multicolumn{1}{m{1.7cm}}{\includegraphics[height=2.00cm,clip]{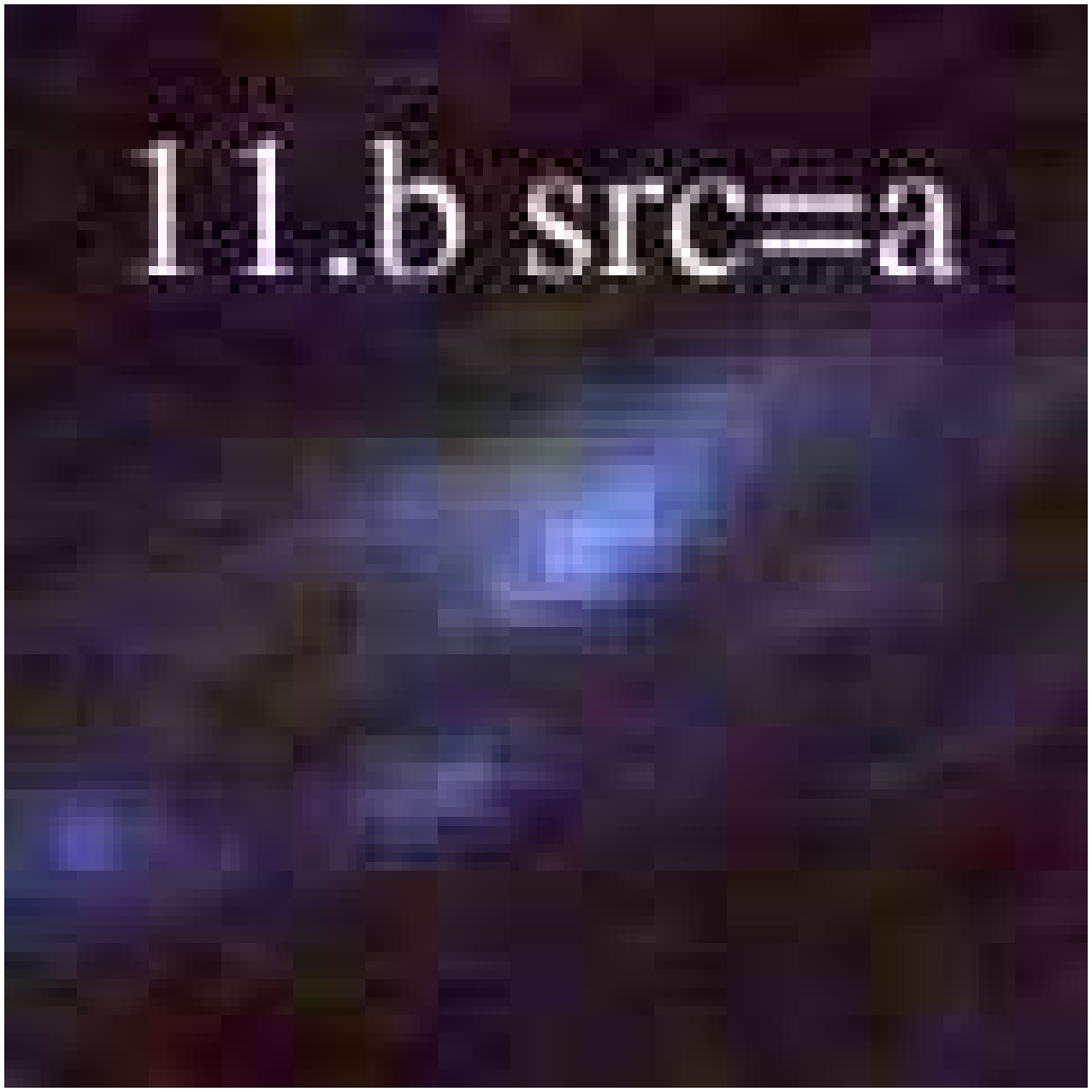}}
    & \multicolumn{1}{m{1.7cm}}{\includegraphics[height=2.00cm,clip]{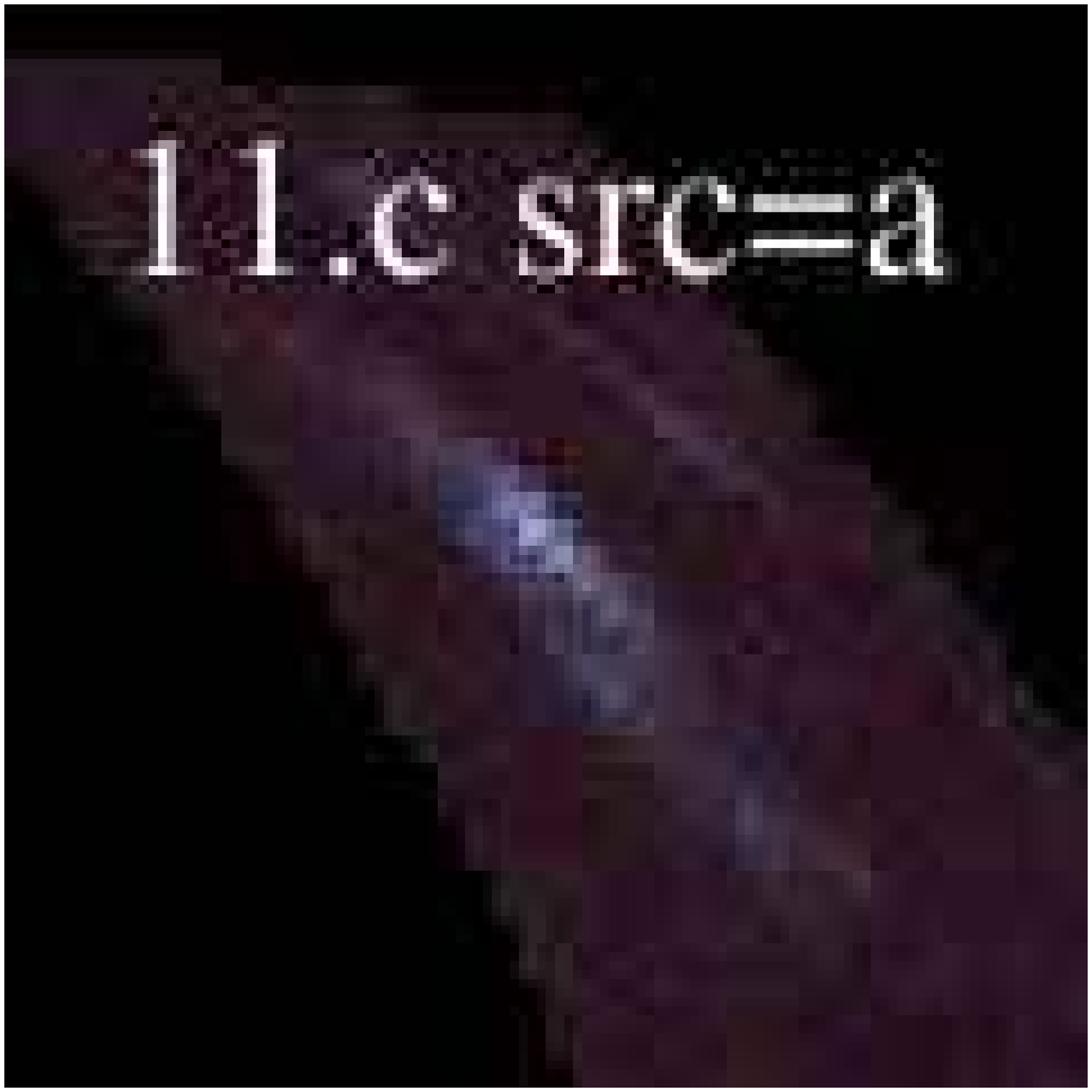}} \\
    \multicolumn{1}{m{1cm}}{{\Large ENFW}}
    & \multicolumn{1}{m{1.7cm}}{\includegraphics[height=2.00cm,clip]{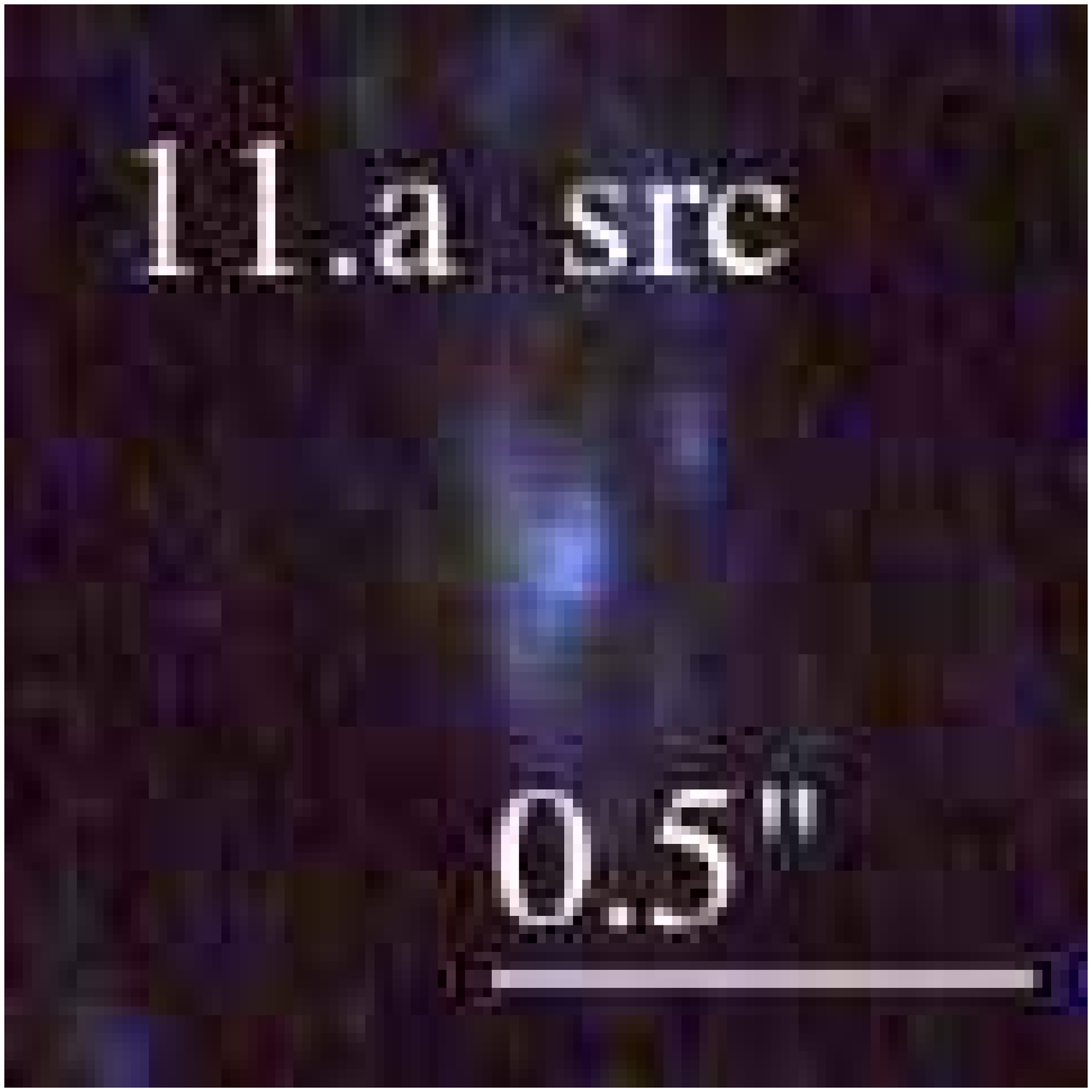}}
    & \multicolumn{1}{m{1.7cm}}{\includegraphics[height=2.00cm,clip]{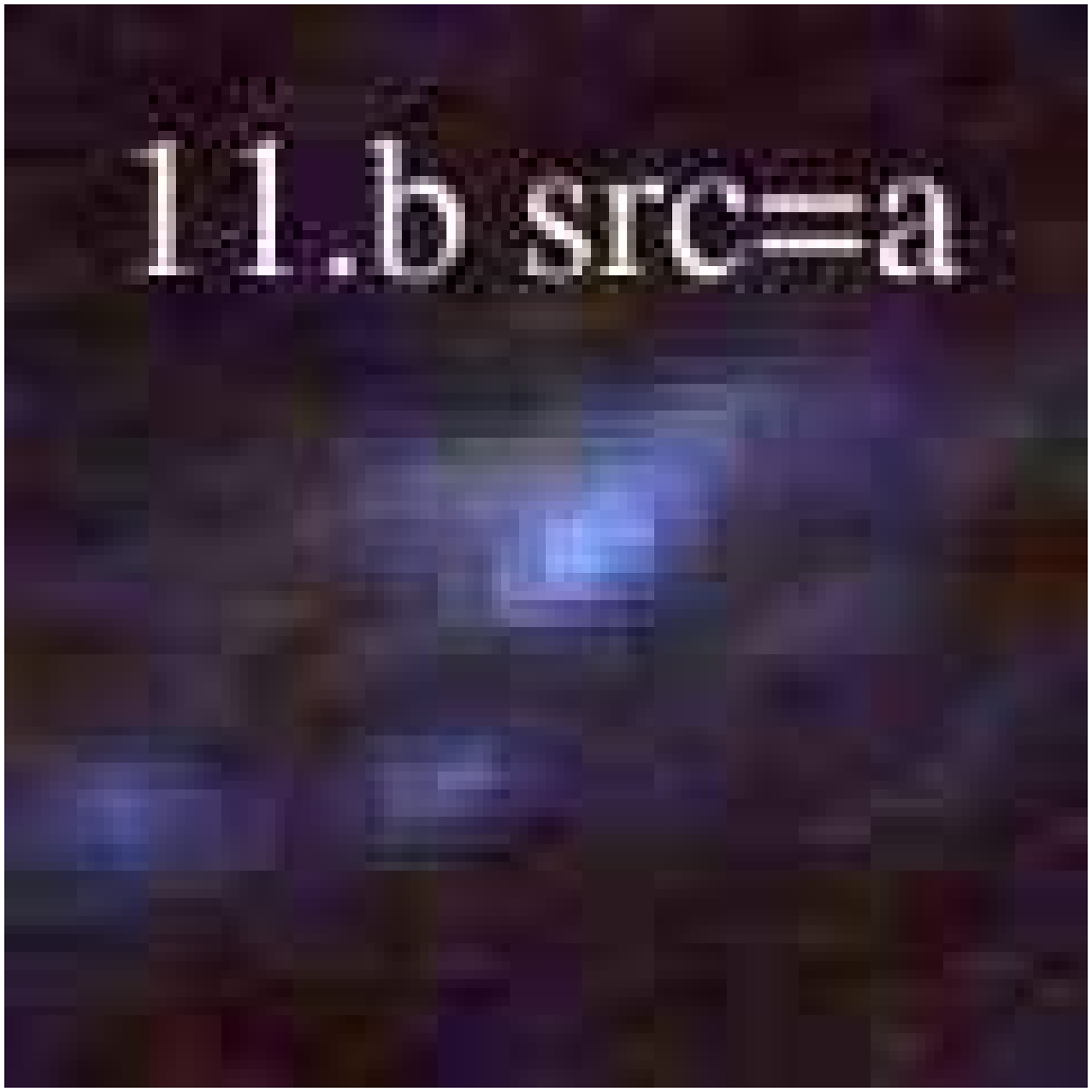}}
    & \multicolumn{1}{m{1.7cm}}{\includegraphics[height=2.00cm,clip]{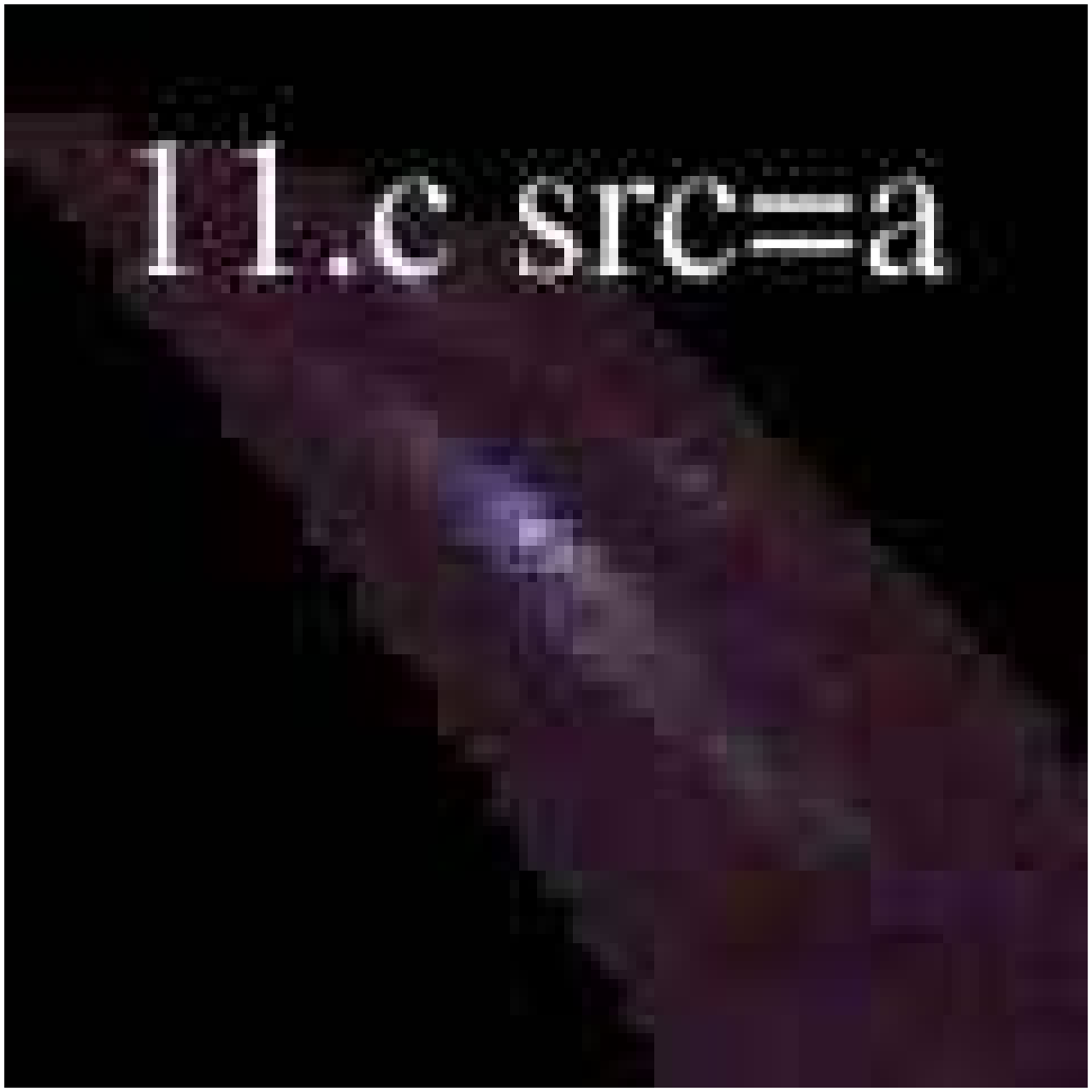}} \\
  \end{tabular}

\end{table*}

\begin{table*}
  \caption{Image system 12:}\vspace{0mm}
  \begin{tabular}{ccc}
    \multicolumn{1}{m{1cm}}{{\Large A1689}}
    & \multicolumn{1}{m{1.7cm}}{\includegraphics[height=2.00cm,clip]{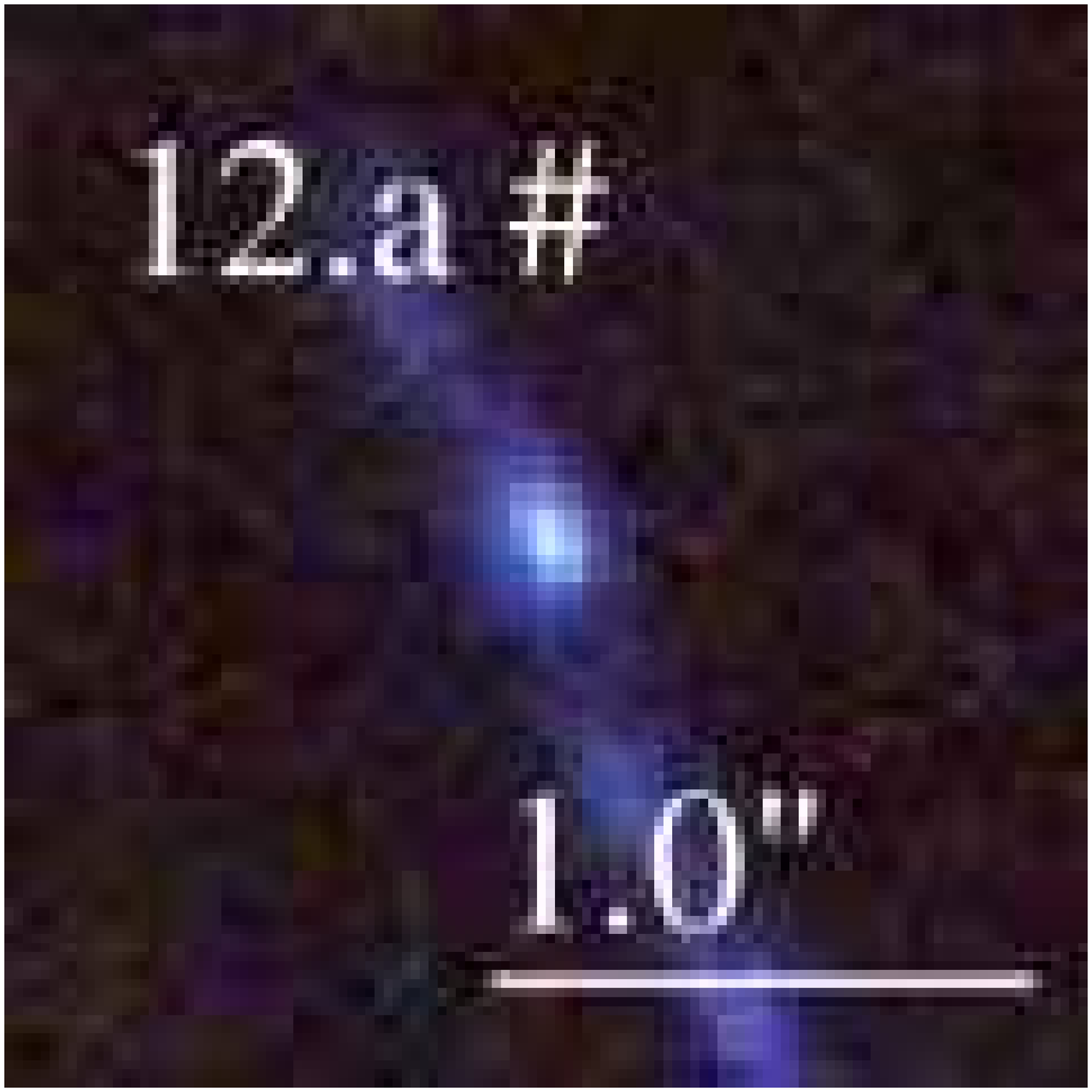}}
    & \multicolumn{1}{m{1.7cm}}{\includegraphics[height=2.00cm,clip]{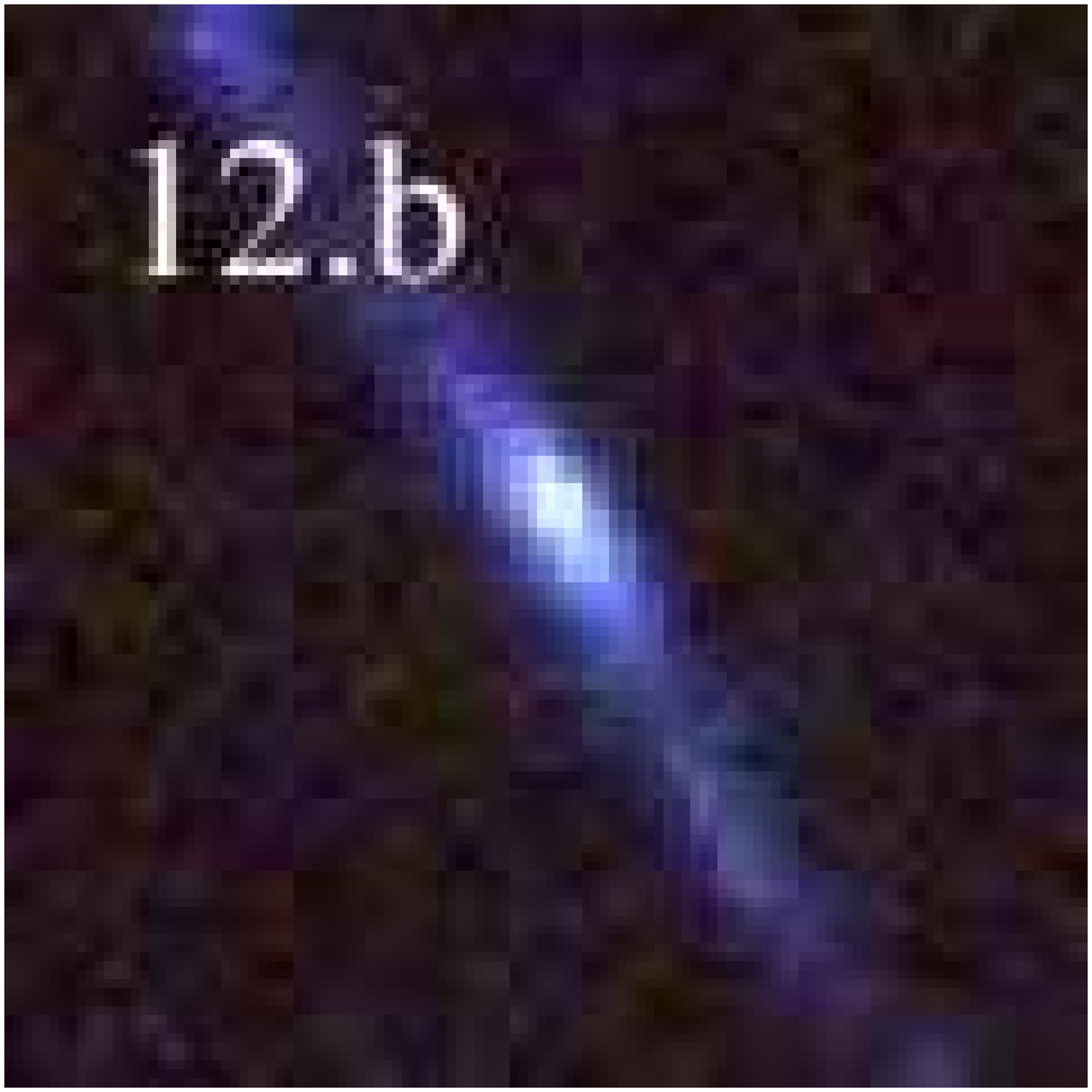}} \\
    \multicolumn{1}{m{1cm}}{{\Large NSIE}}
    & \multicolumn{1}{m{1.7cm}}{\includegraphics[height=2.00cm,clip]{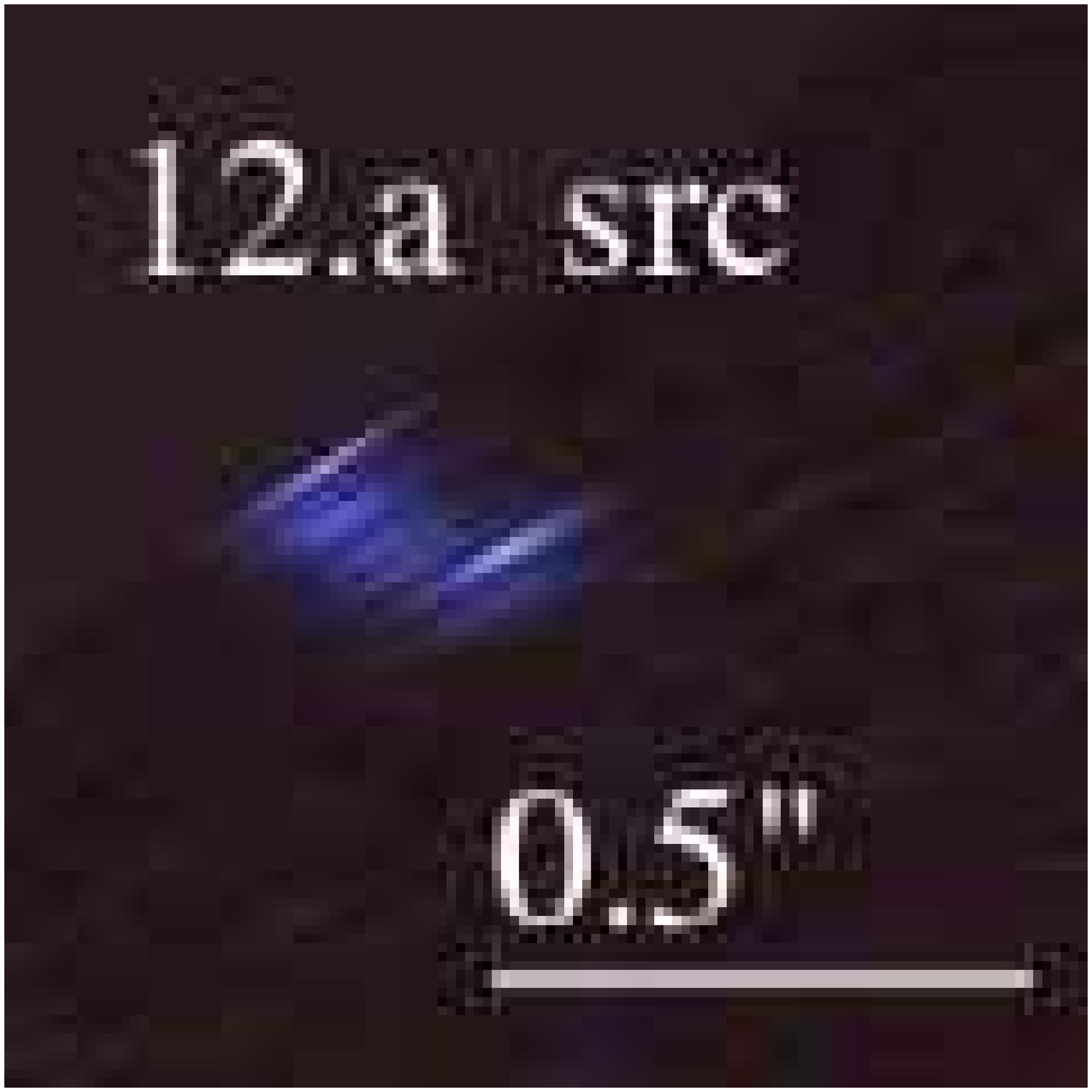}}
    & \multicolumn{1}{m{1.7cm}}{\includegraphics[height=2.00cm,clip]{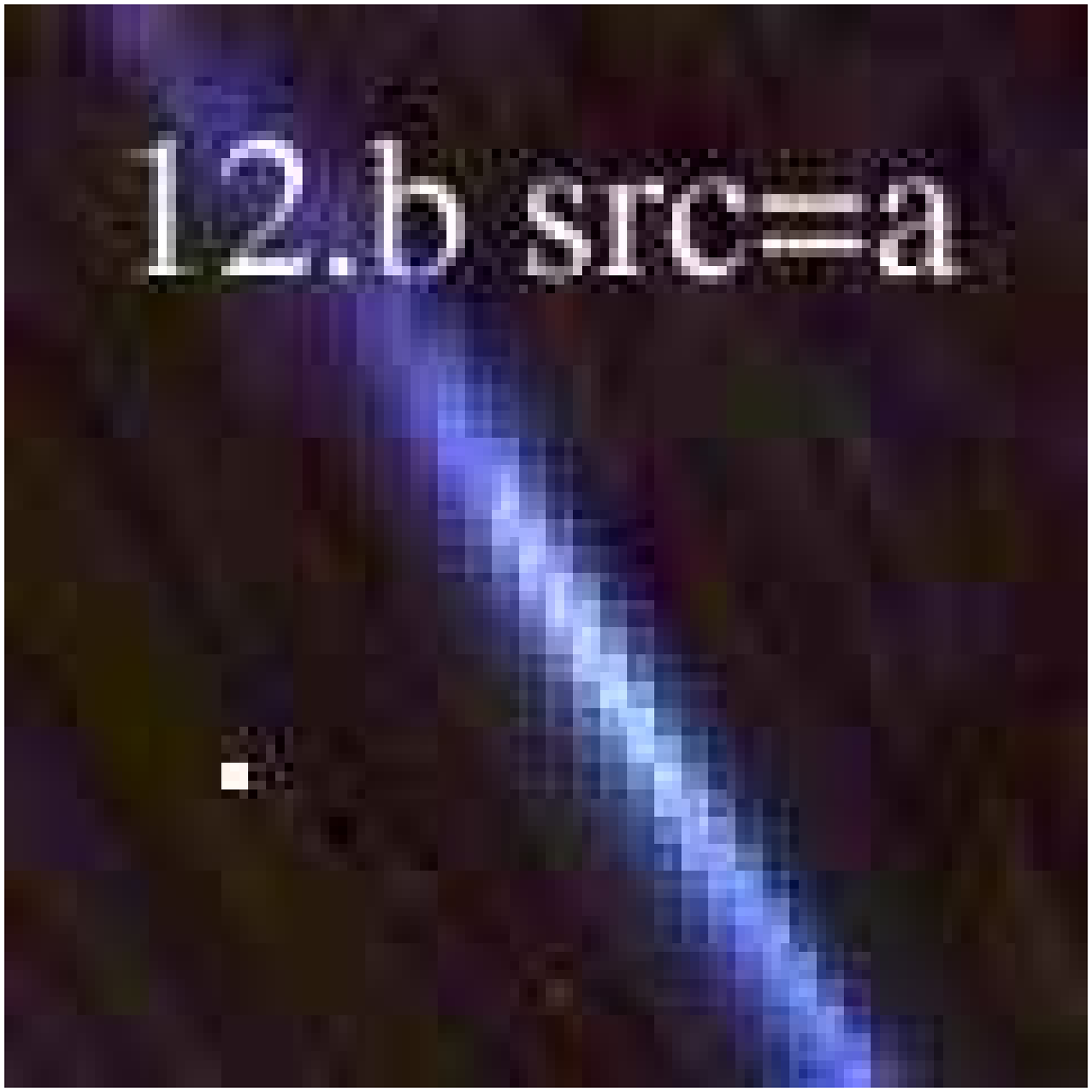}} \\
    \multicolumn{1}{m{1cm}}{{\Large ENFW}}
    & \multicolumn{1}{m{1.7cm}}{\includegraphics[height=2.00cm,clip]{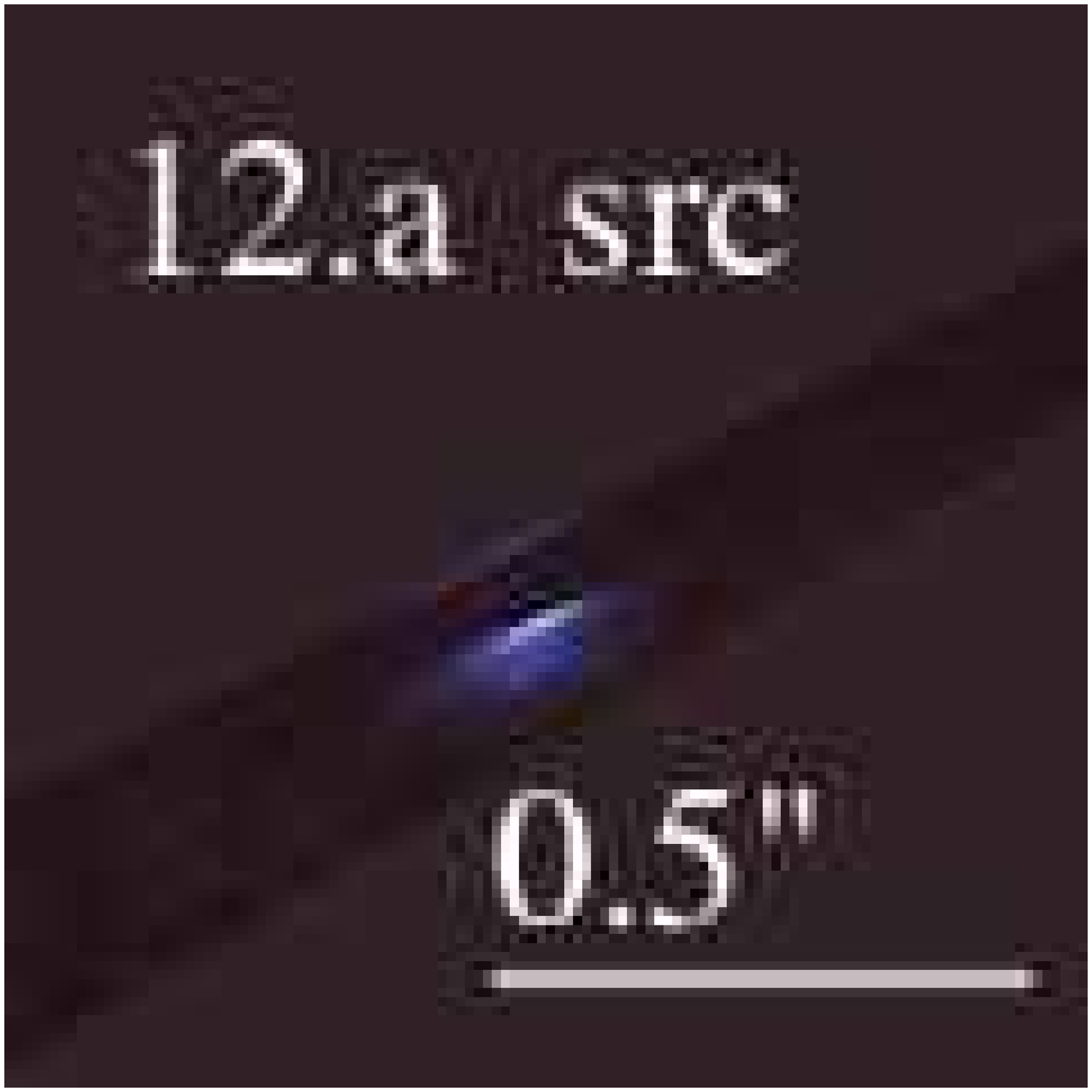}}
    & \multicolumn{1}{m{1.7cm}}{\includegraphics[height=2.00cm,clip]{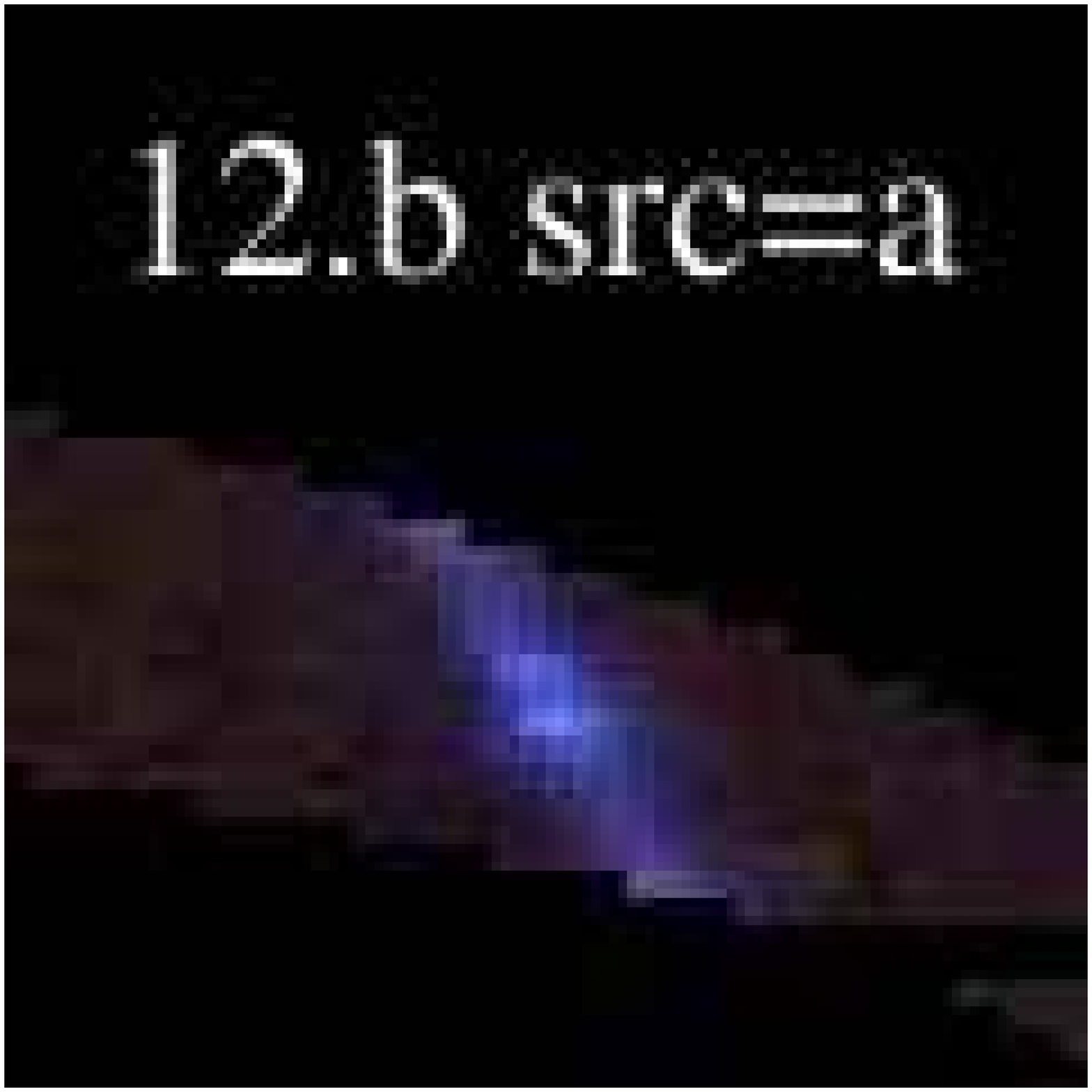}} \\
  \end{tabular}

\end{table*}

\clearpage

\begin{table*}
  \caption{Image system 13:}\vspace{0mm}
  \begin{tabular}{ccccc}
    \multicolumn{1}{m{1cm}}{{\Large A1689}}
    & \multicolumn{1}{m{1.7cm}}{\includegraphics[height=2.00cm,clip]{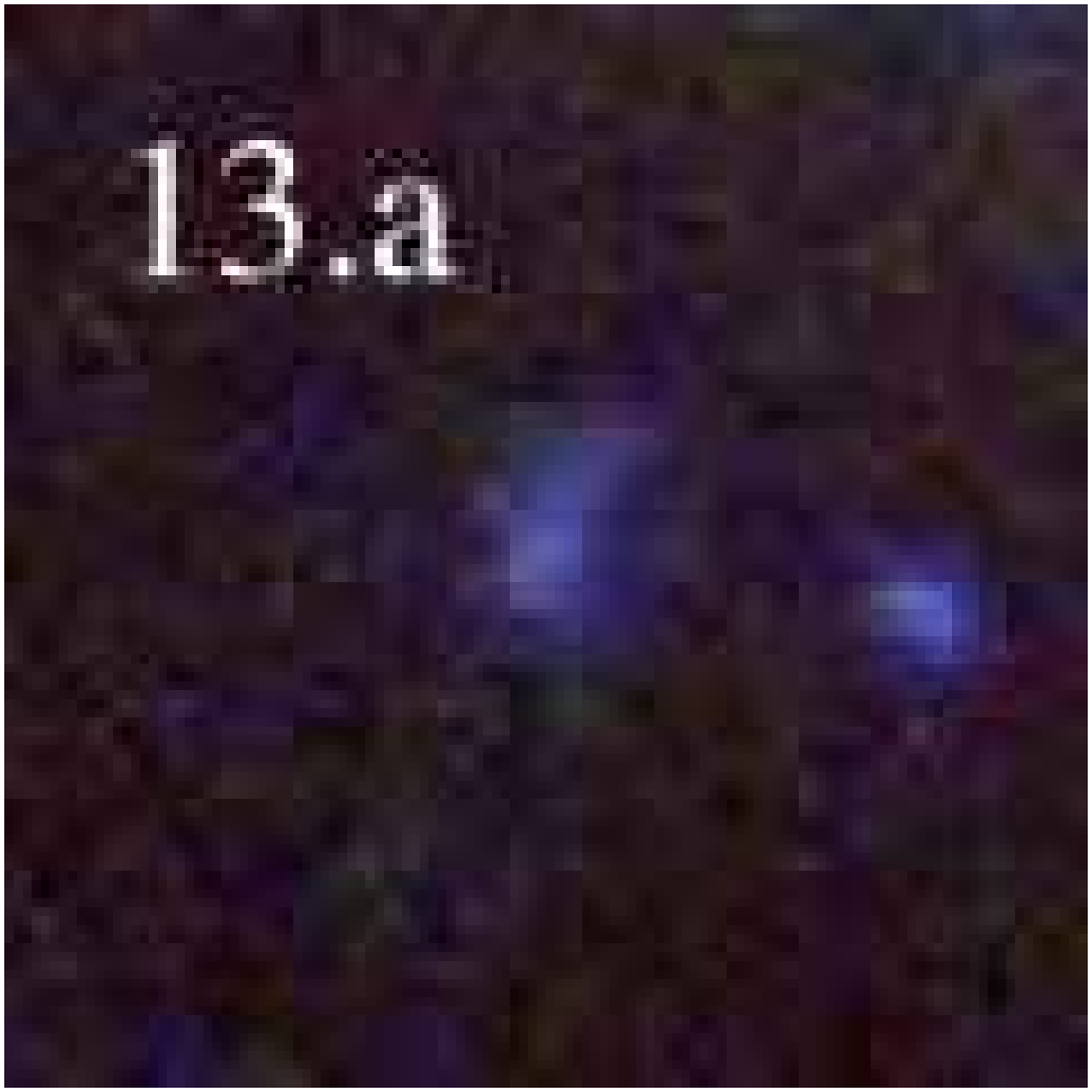}}
    & \multicolumn{1}{m{1.7cm}}{\includegraphics[height=2.00cm,clip]{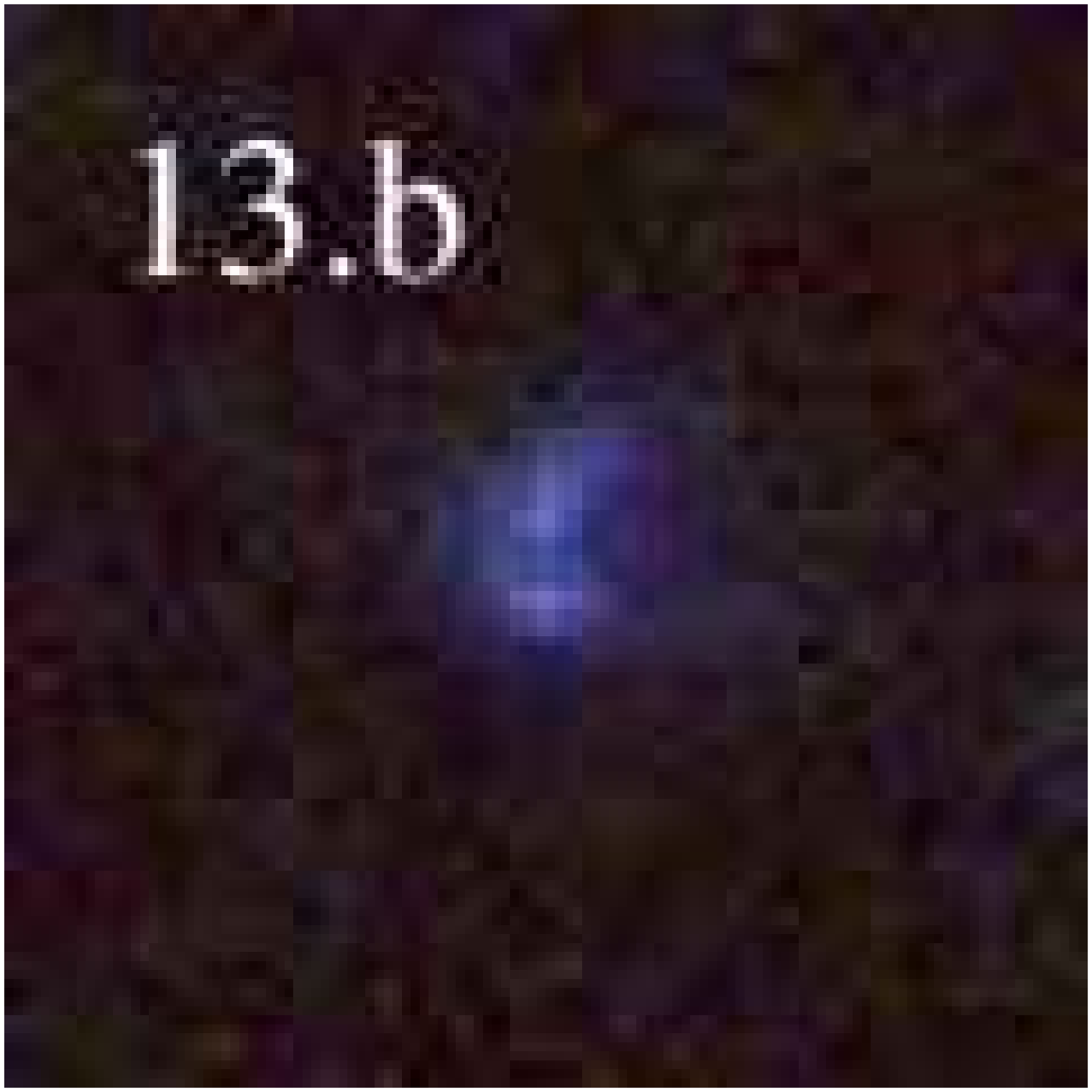}}
    & \multicolumn{1}{m{1.7cm}}{\includegraphics[height=2.00cm,clip]{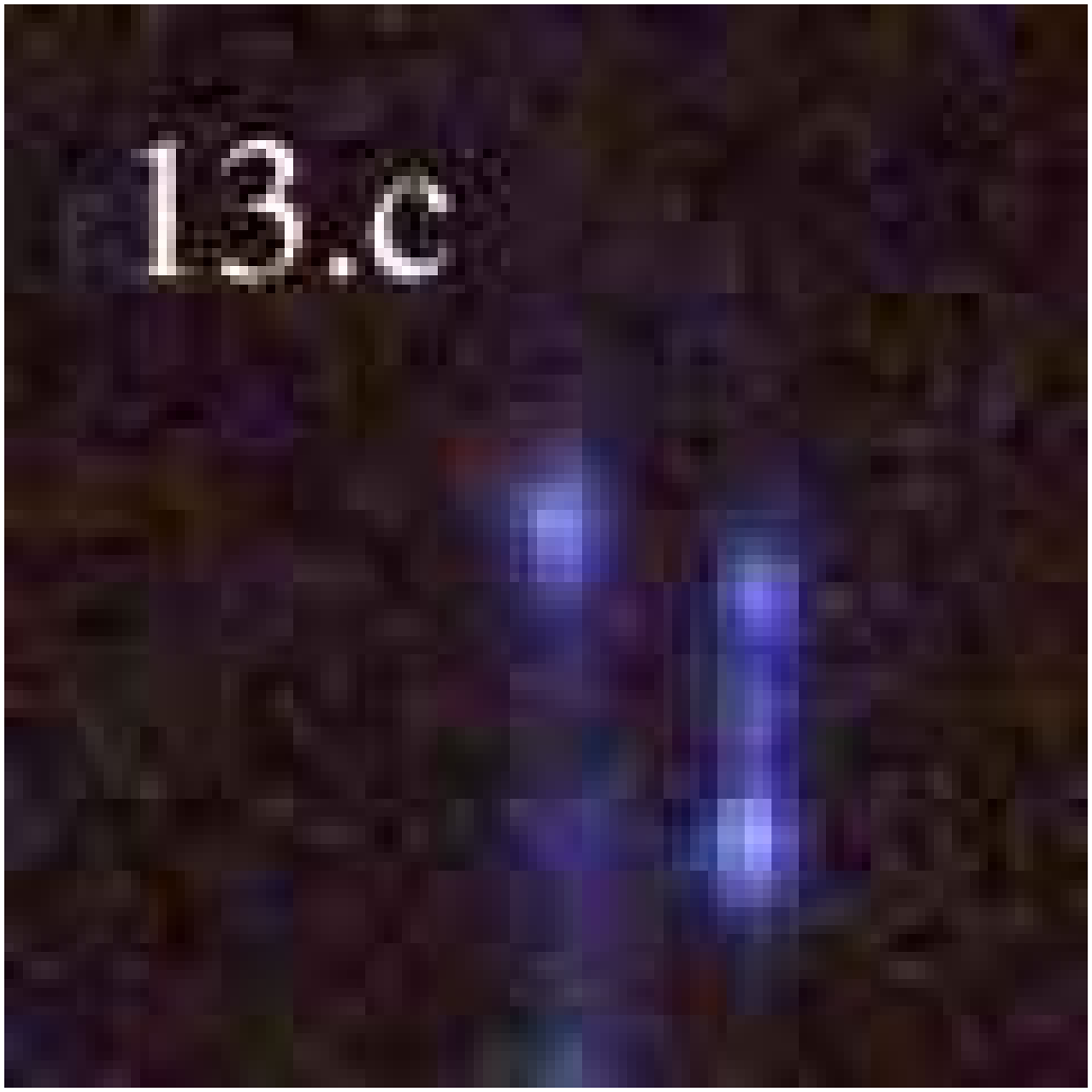}}
    & \multicolumn{1}{m{1.7cm}}{\includegraphics[height=2.00cm,clip]{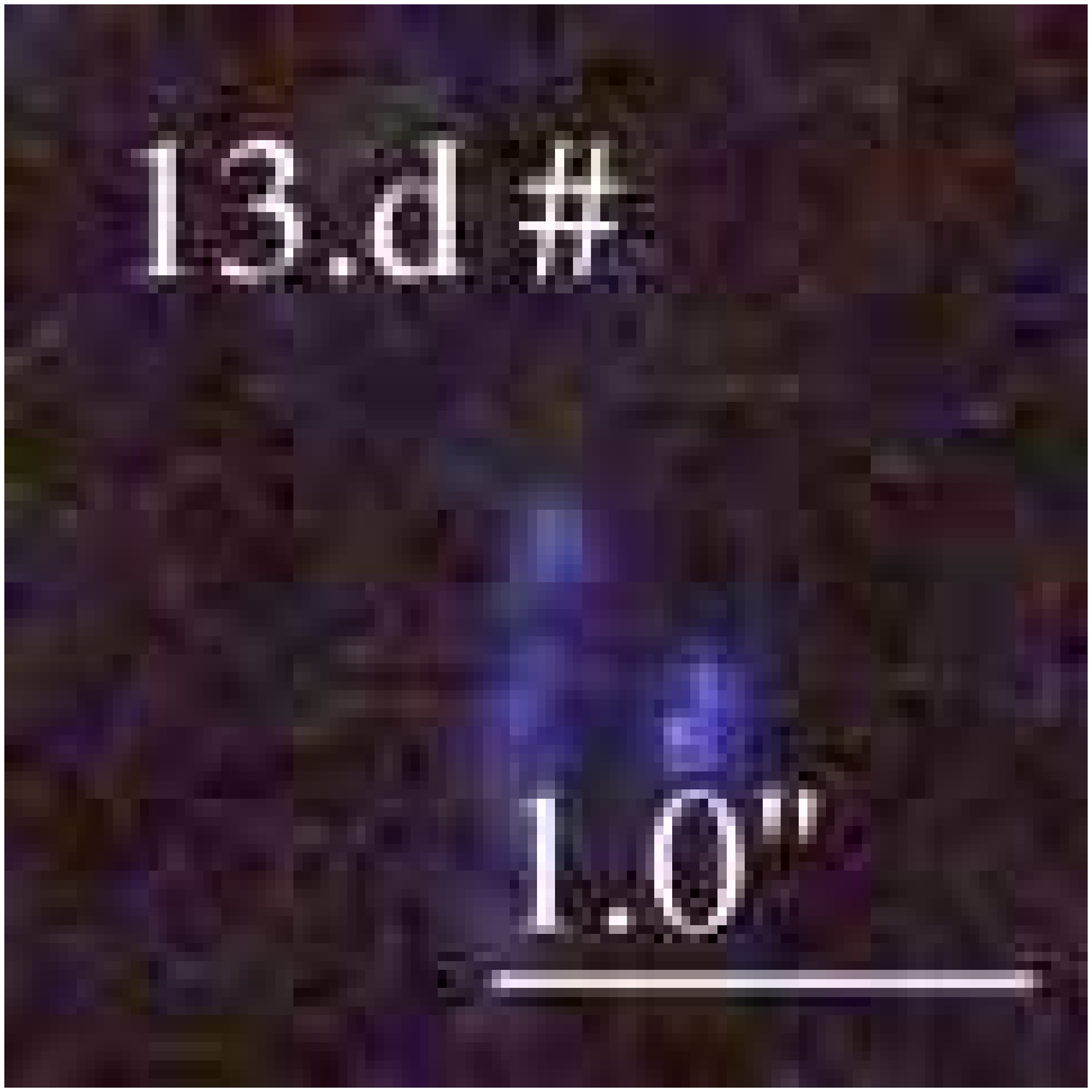}} \\
    \multicolumn{1}{m{1cm}}{{\Large NSIE}}
    & \multicolumn{1}{m{1.7cm}}{\includegraphics[height=2.00cm,clip]{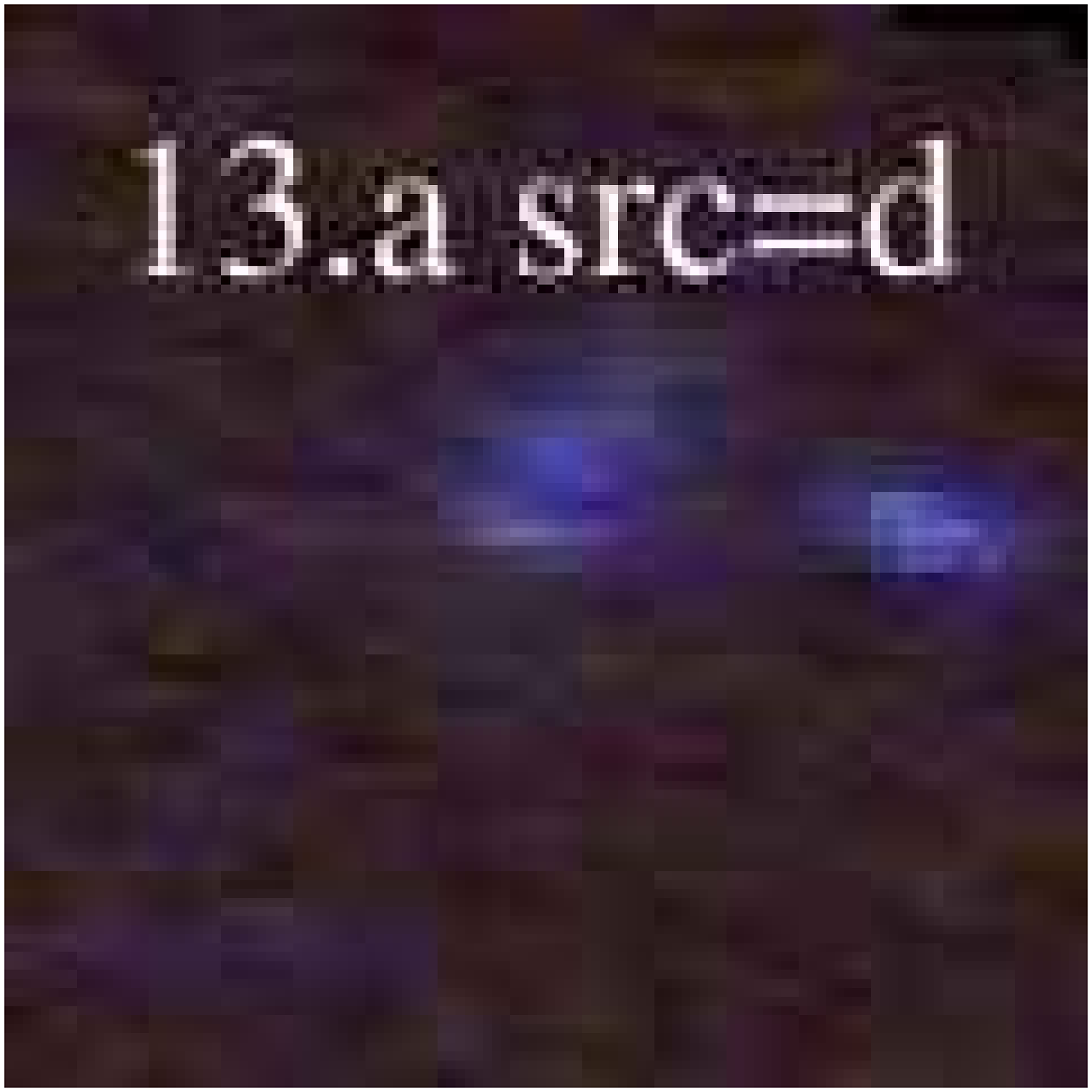}}
    & \multicolumn{1}{m{1.7cm}}{\includegraphics[height=2.00cm,clip]{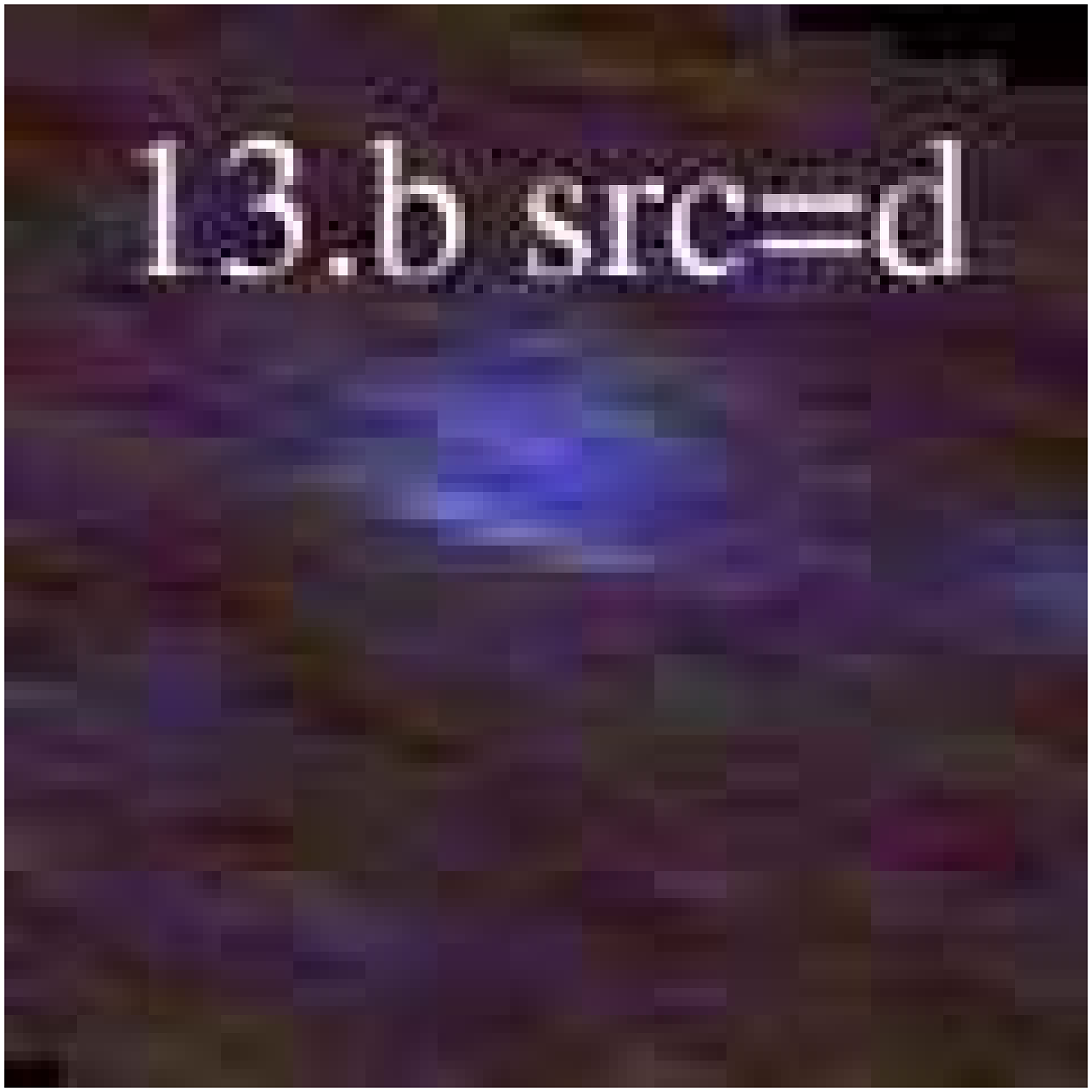}}
    & \multicolumn{1}{m{1.7cm}}{\includegraphics[height=2.00cm,clip]{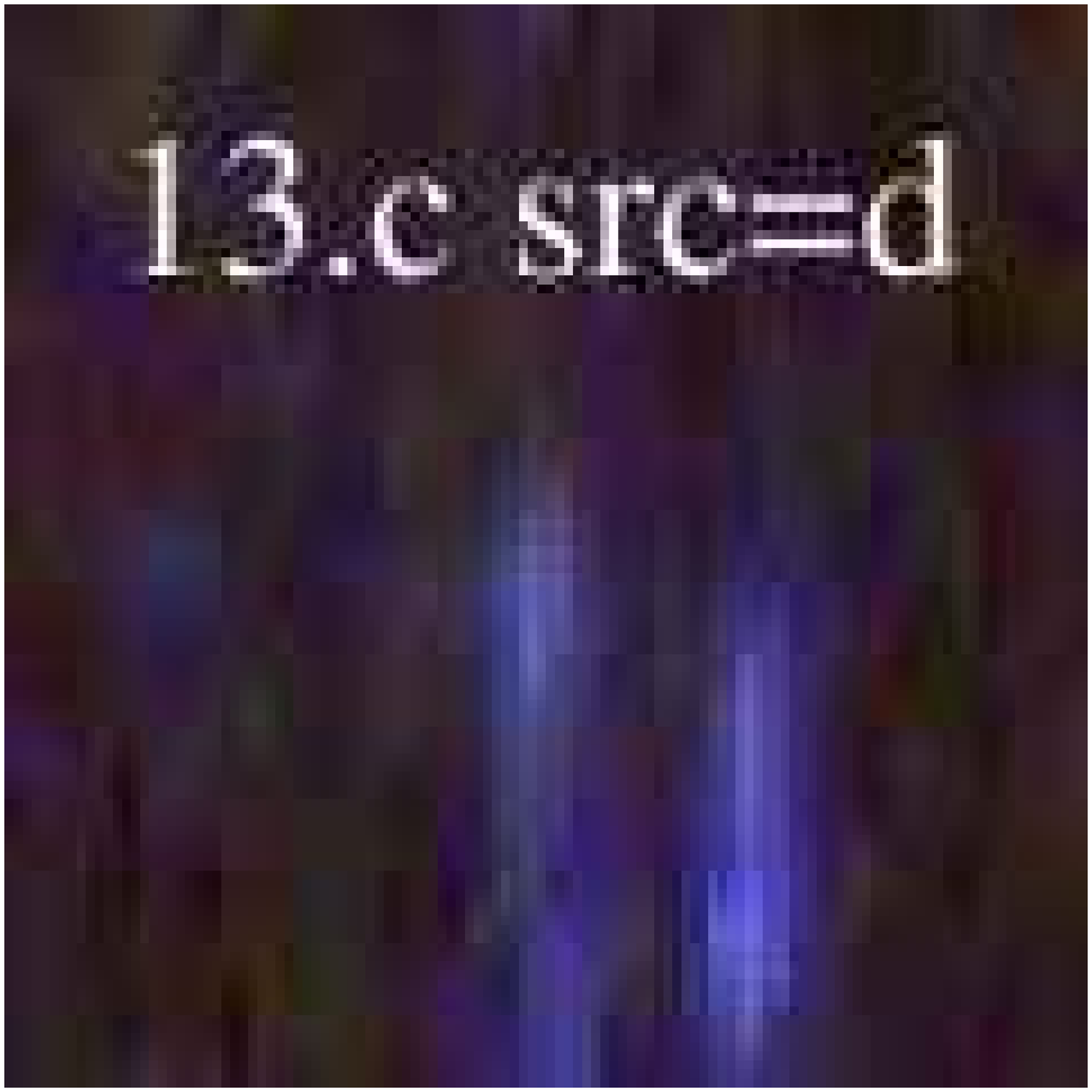}}
    & \multicolumn{1}{m{1.7cm}}{\includegraphics[height=2.00cm,clip]{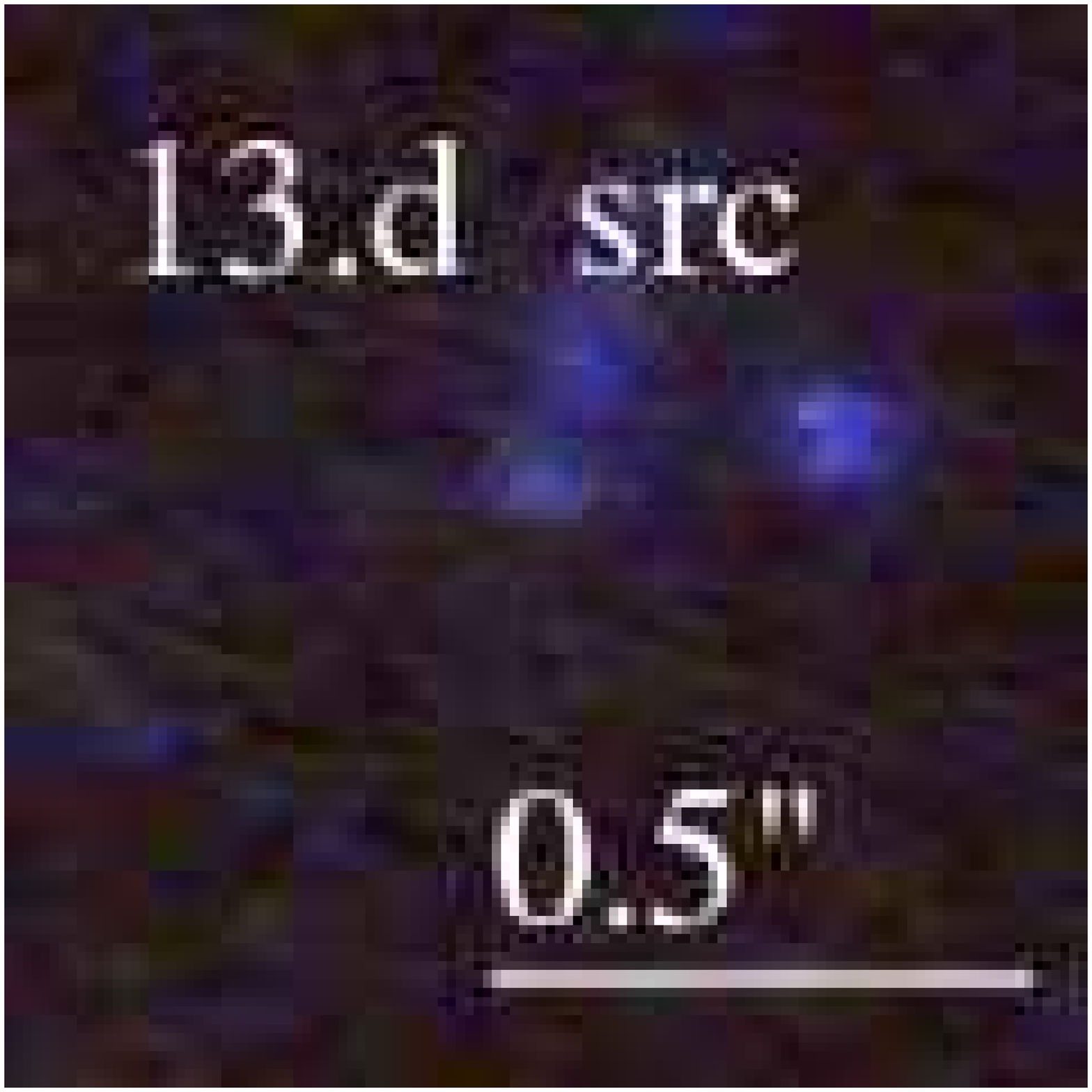}} \\
    \multicolumn{1}{m{1cm}}{{\Large ENFW}}
    & \multicolumn{1}{m{1.7cm}}{\includegraphics[height=2.00cm,clip]{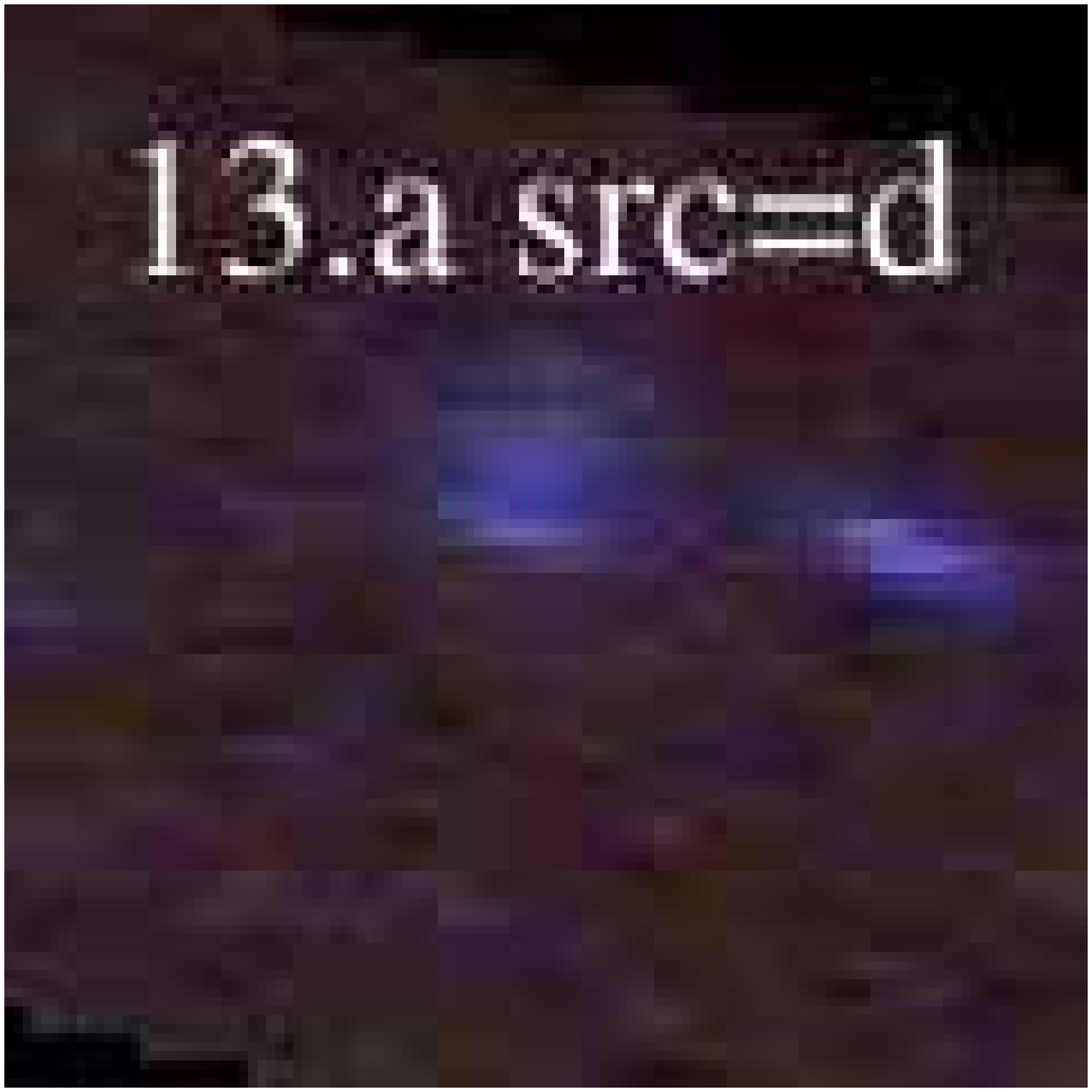}}
    & \multicolumn{1}{m{1.7cm}}{\includegraphics[height=2.00cm,clip]{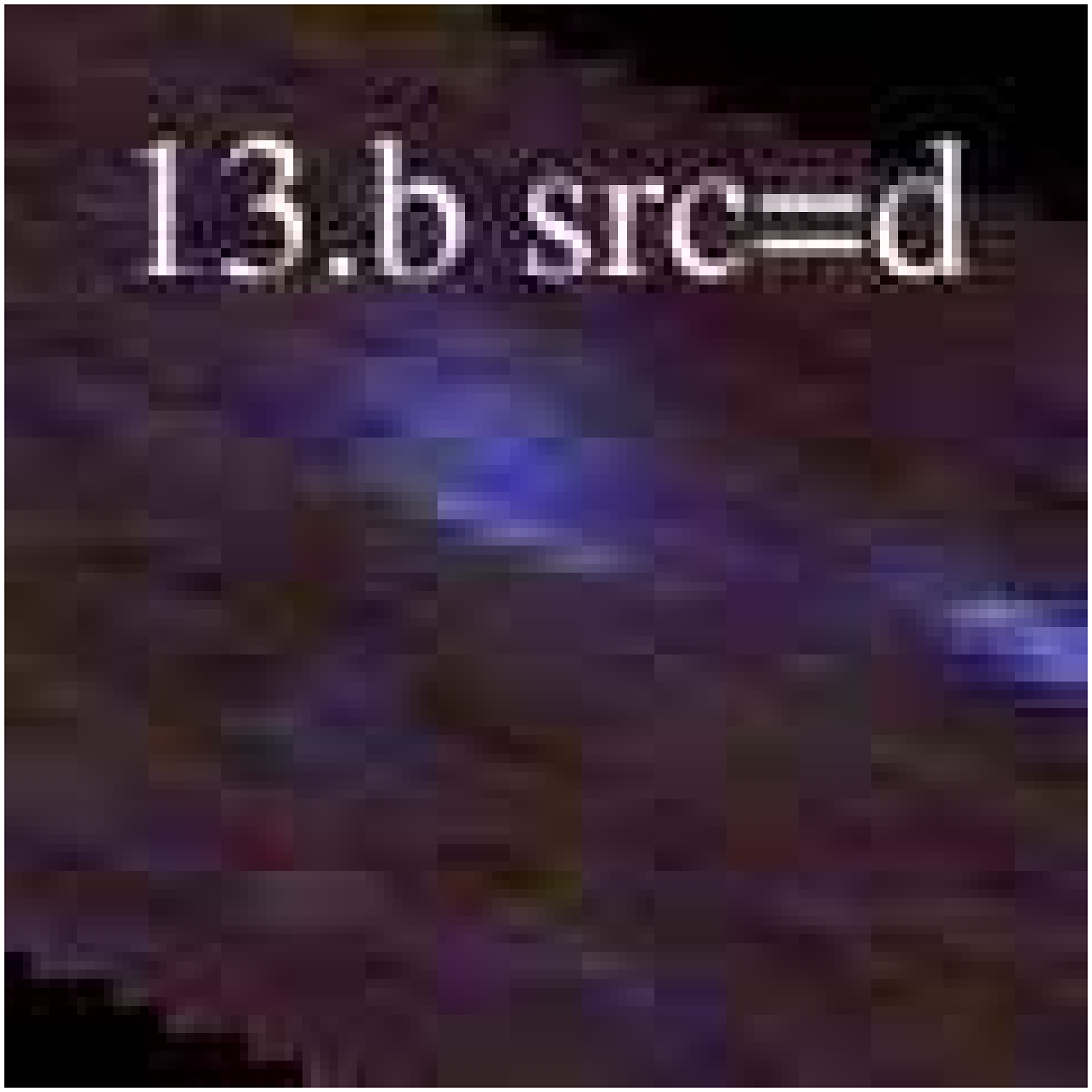}}
    & \multicolumn{1}{m{1.7cm}}{\includegraphics[height=2.00cm,clip]{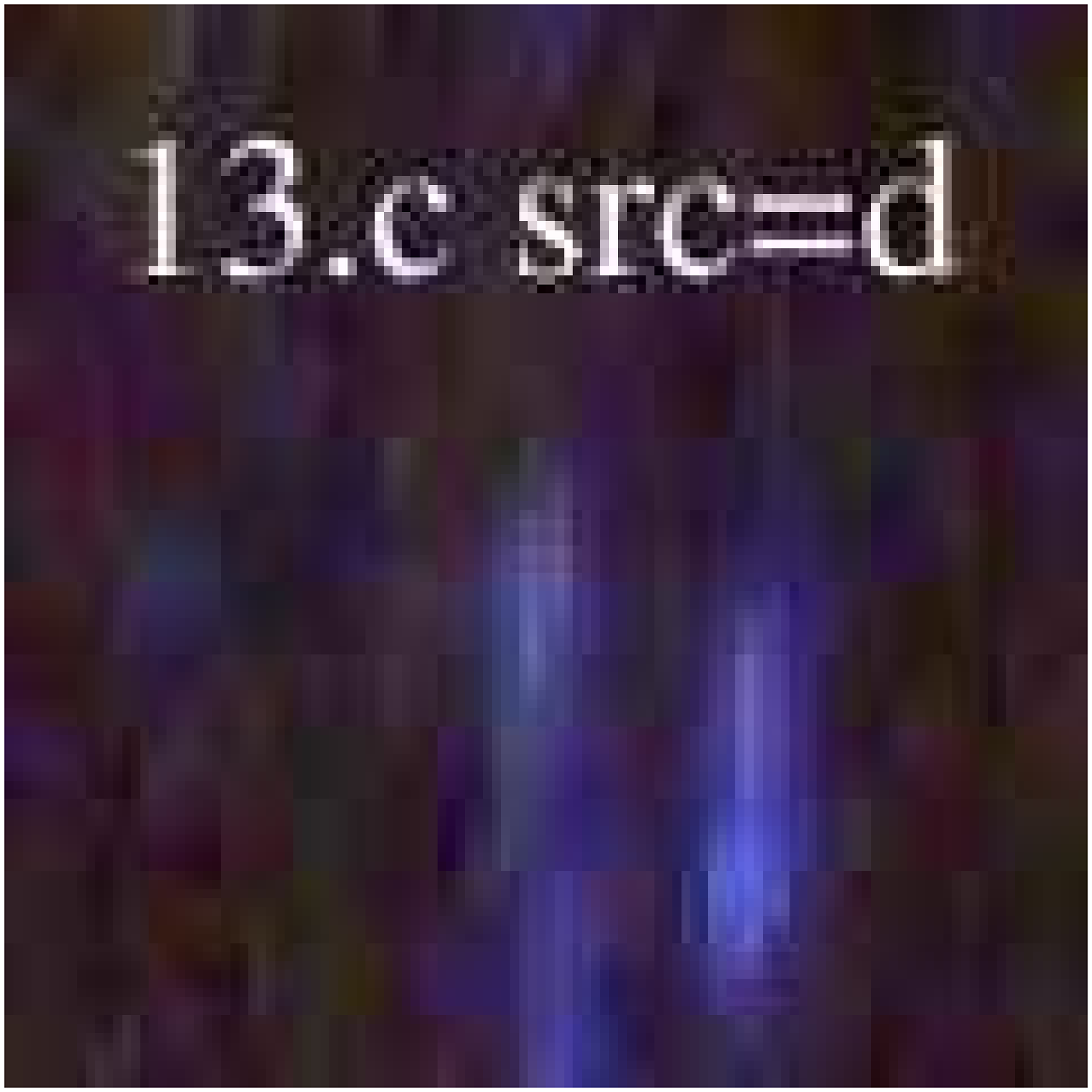}}
    & \multicolumn{1}{m{1.7cm}}{\includegraphics[height=2.00cm,clip]{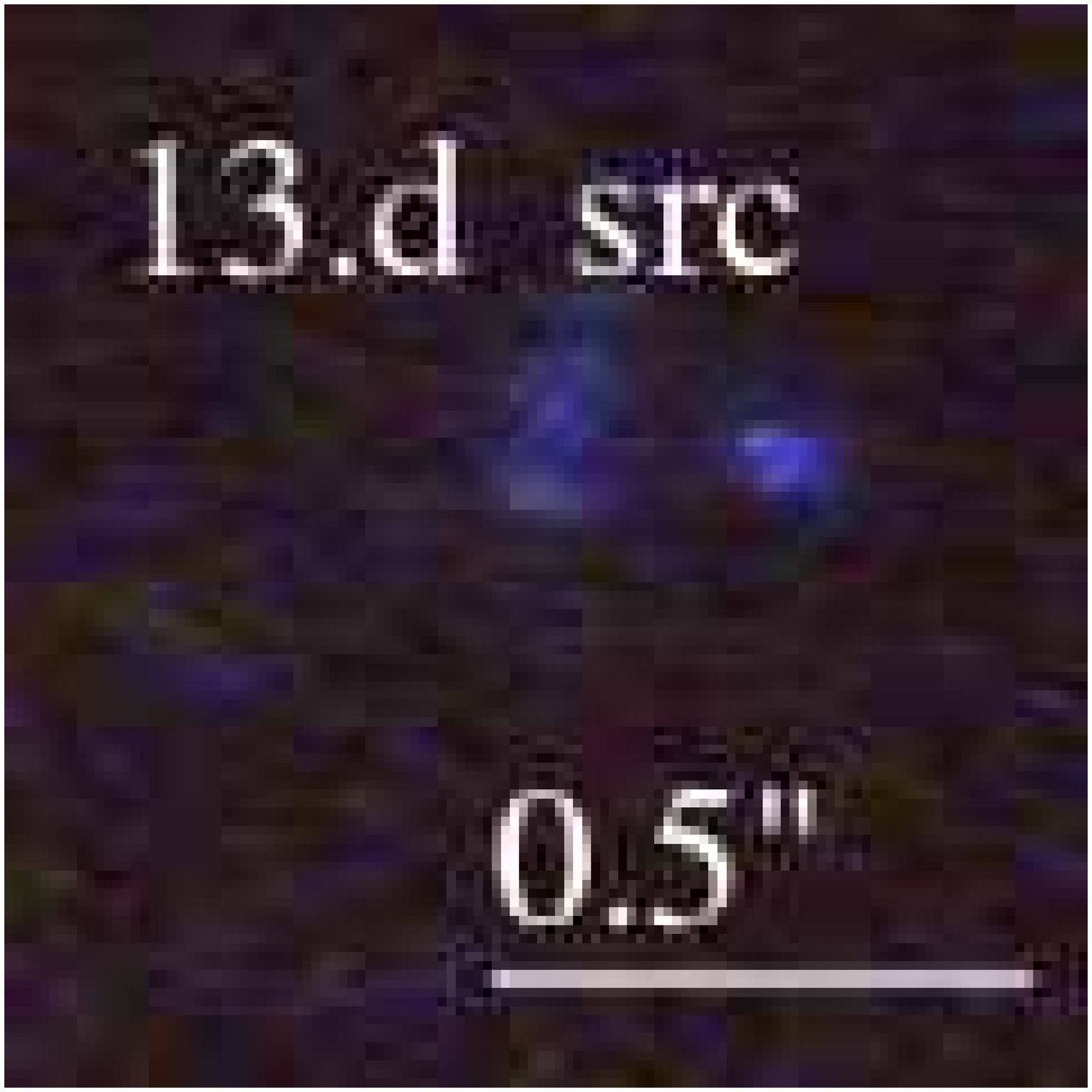}} \\
  \end{tabular}

\end{table*}

\begin{table*}
  \caption{Image system 14:}\vspace{0mm}
  \begin{tabular}{cccc}
    \multicolumn{1}{m{1cm}}{{\Large A1689}}
    & \multicolumn{1}{m{1.7cm}}{\includegraphics[height=2.00cm,clip]{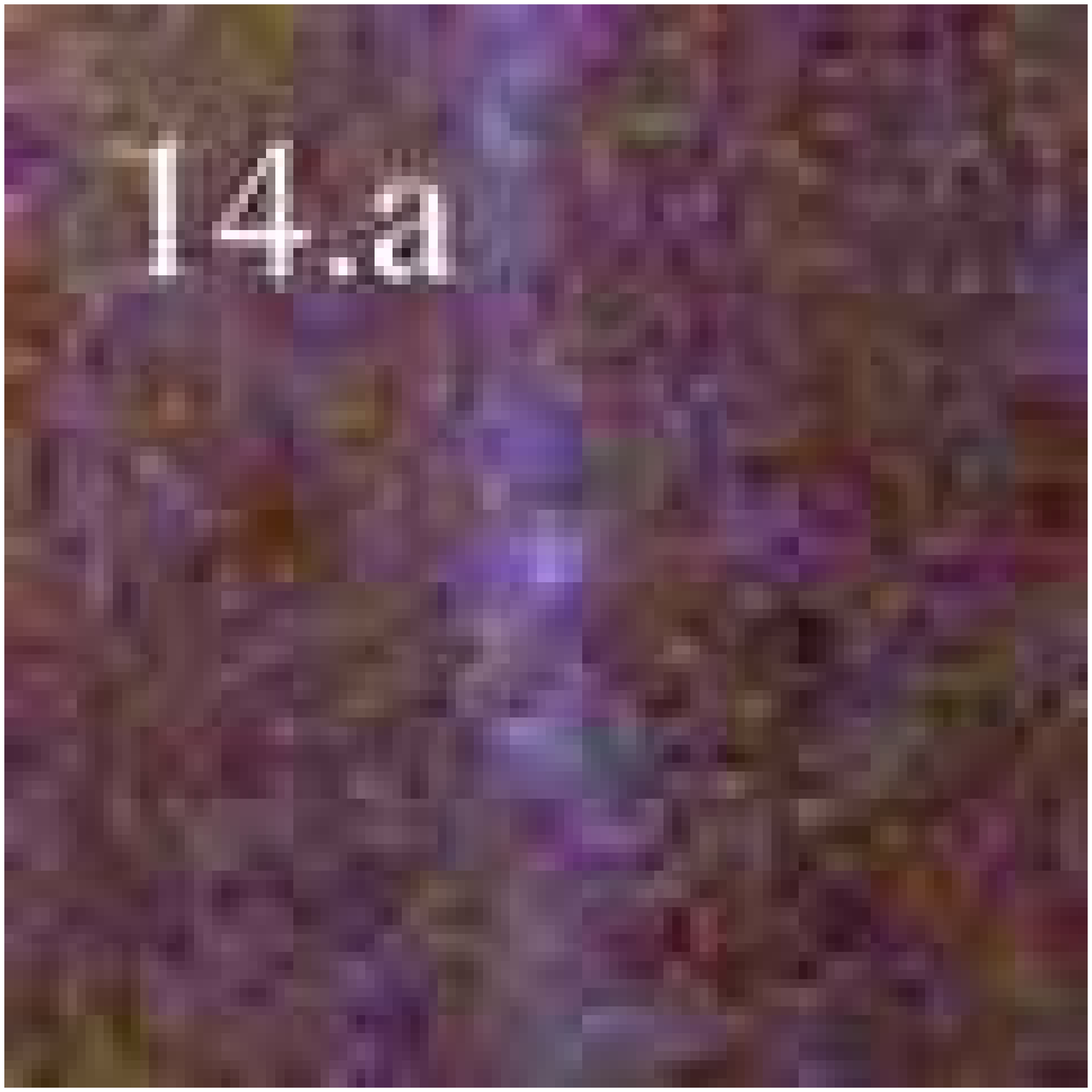}}
    & \multicolumn{1}{m{1.7cm}}{\includegraphics[height=2.00cm,clip]{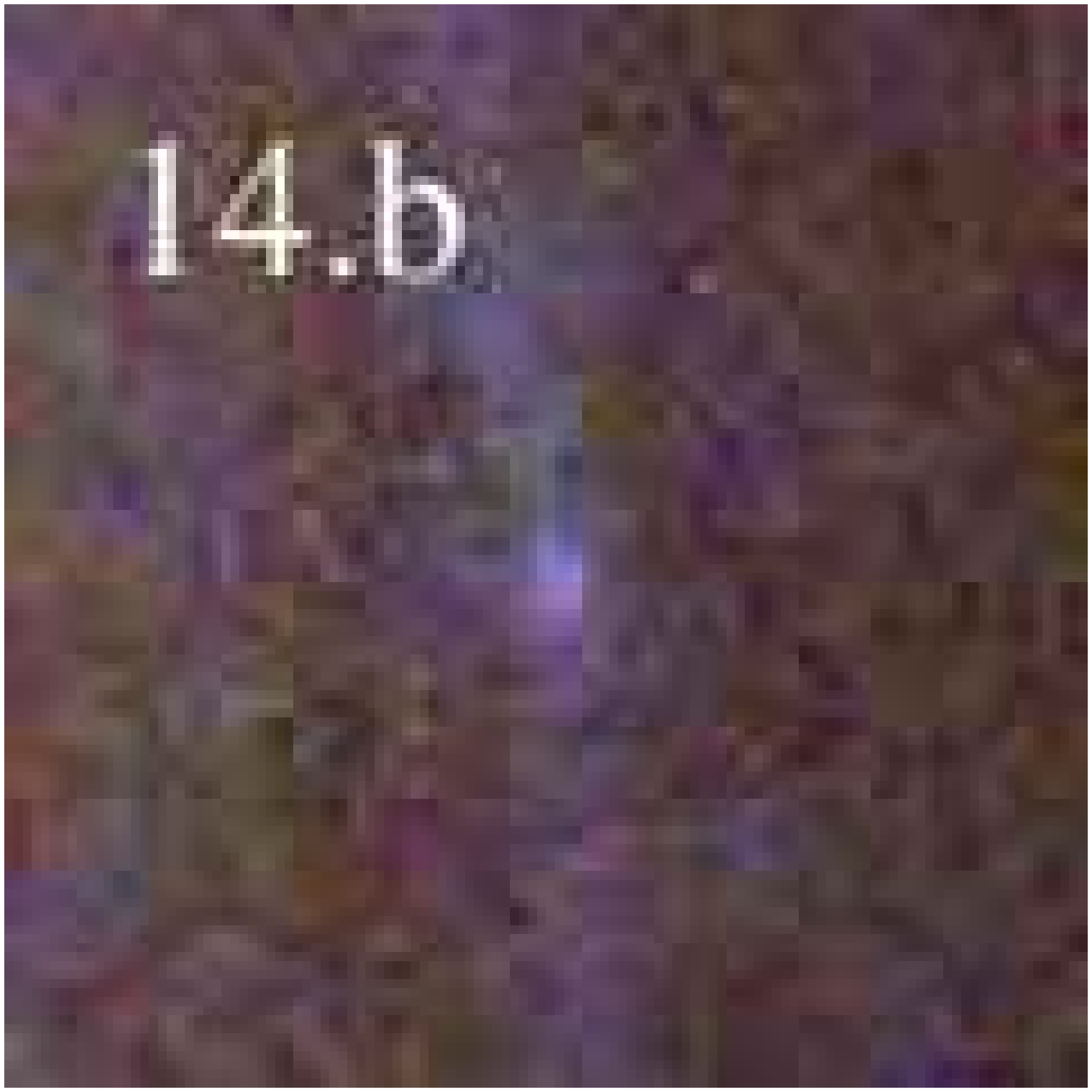}}
    & \multicolumn{1}{m{1.7cm}}{\includegraphics[height=2.00cm,clip]{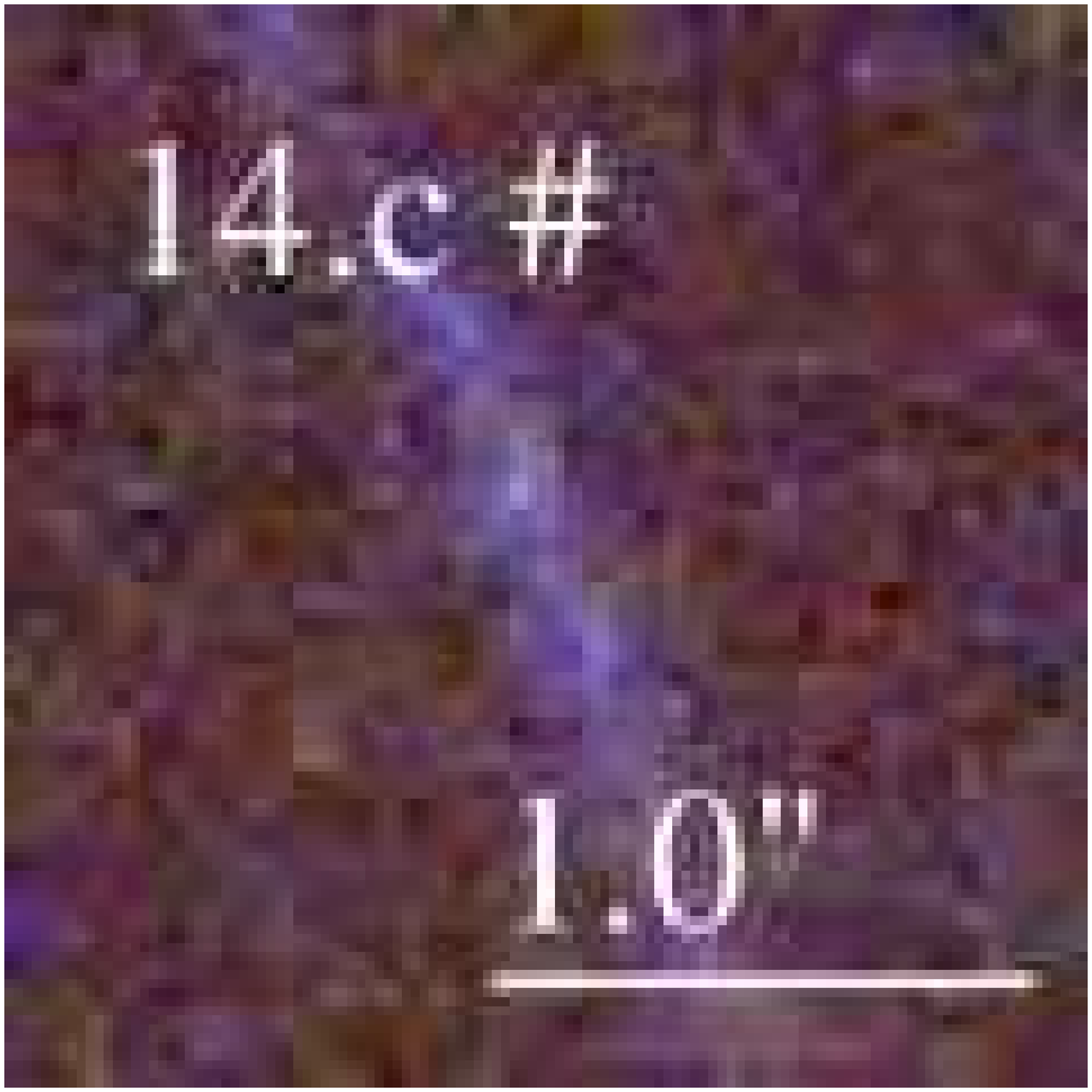}} \\
    \multicolumn{1}{m{1cm}}{{\Large NSIE}}
    & \multicolumn{1}{m{1.7cm}}{\includegraphics[height=2.00cm,clip]{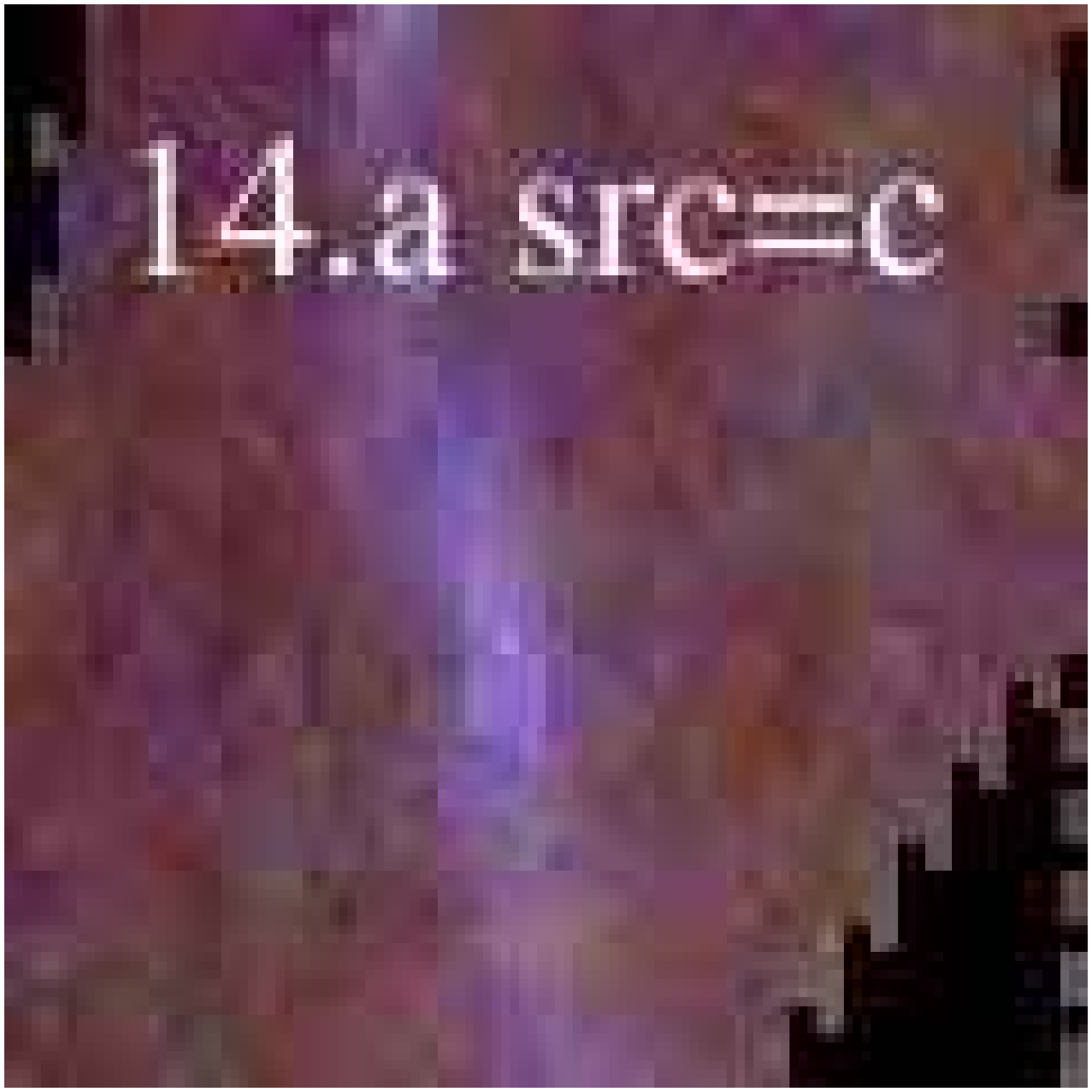}}
    & \multicolumn{1}{m{1.7cm}}{\includegraphics[height=2.00cm,clip]{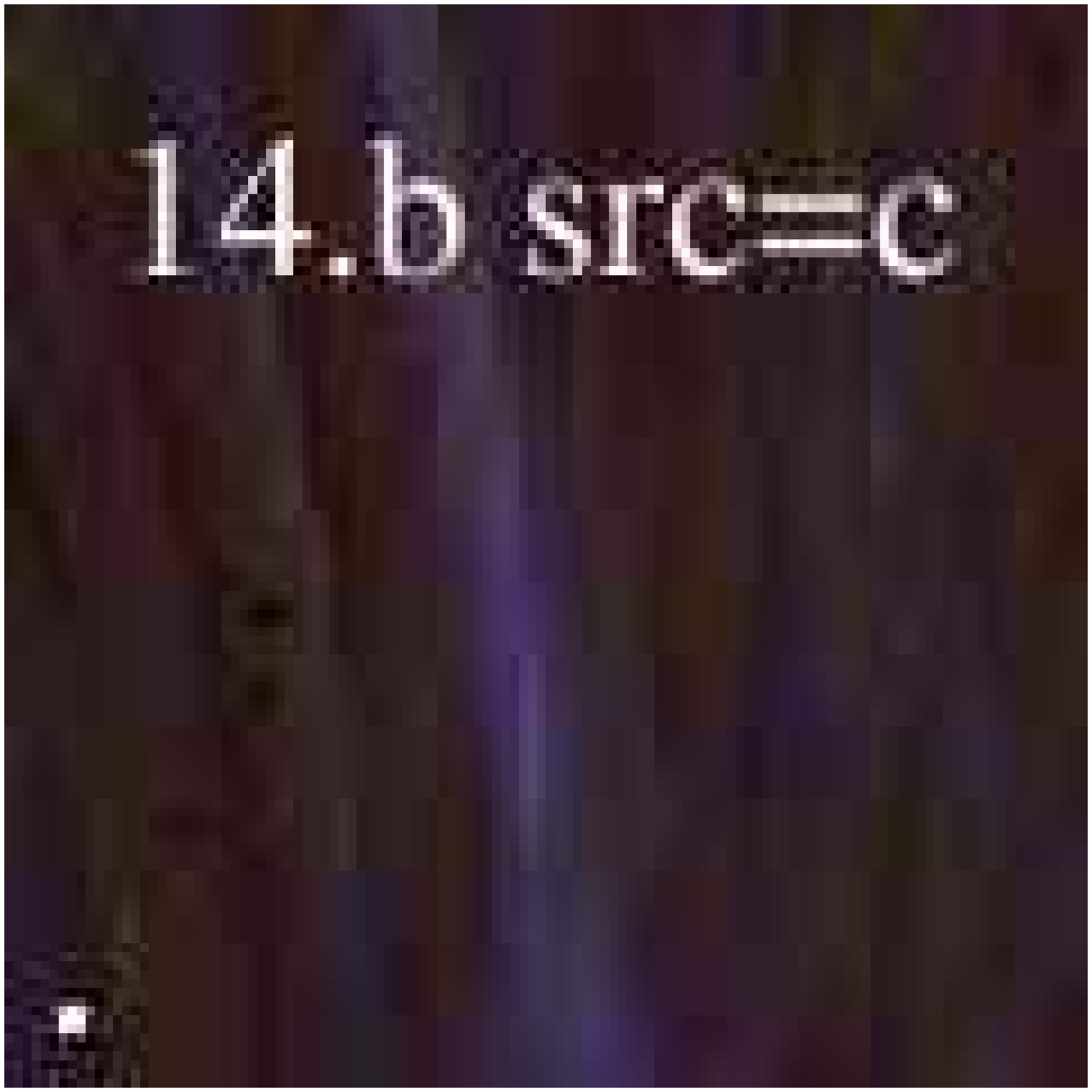}}
    & \multicolumn{1}{m{1.7cm}}{\includegraphics[height=2.00cm,clip]{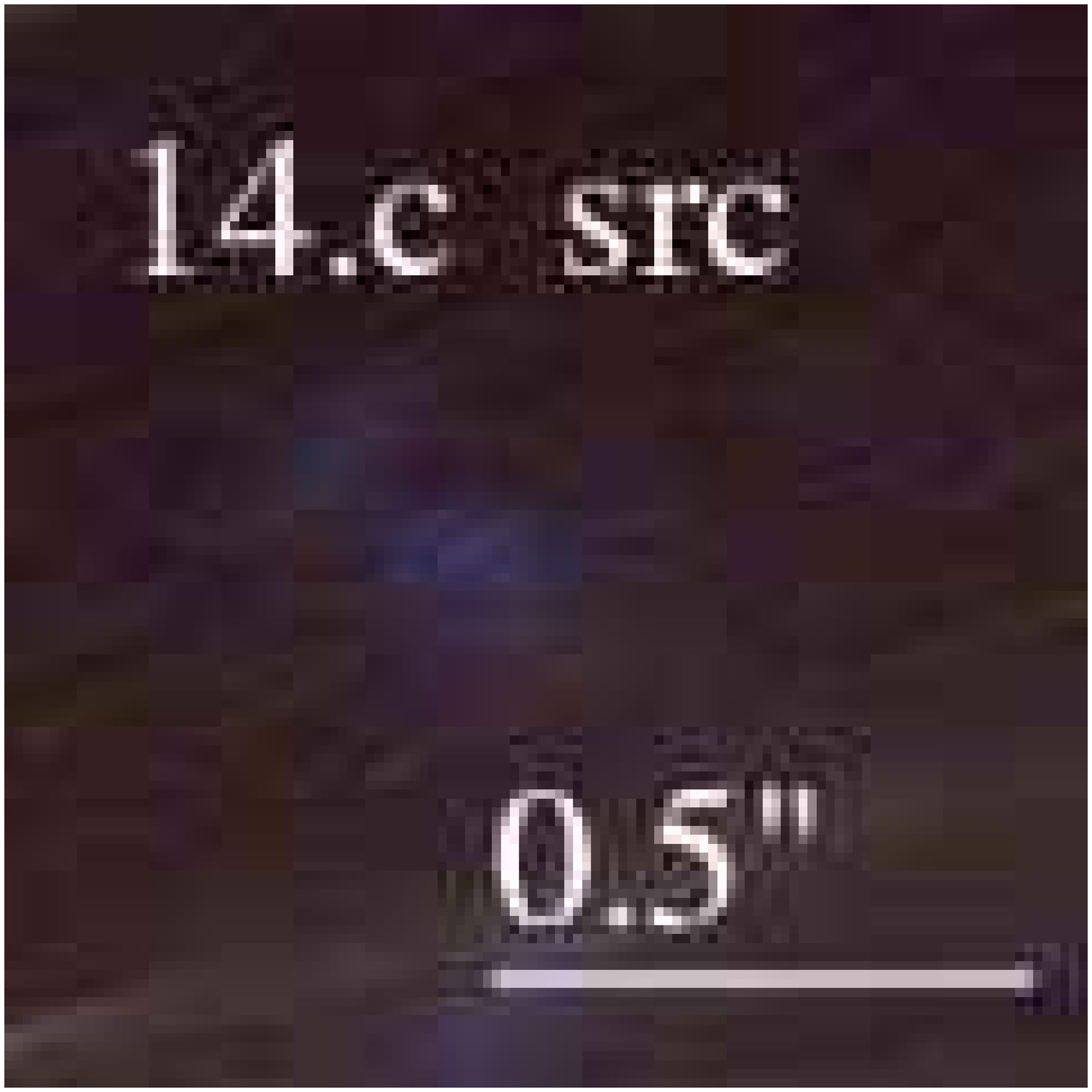}} \\
    \multicolumn{1}{m{1cm}}{{\Large ENFW}}
    & \multicolumn{1}{m{1.7cm}}{\includegraphics[height=2.00cm,clip]{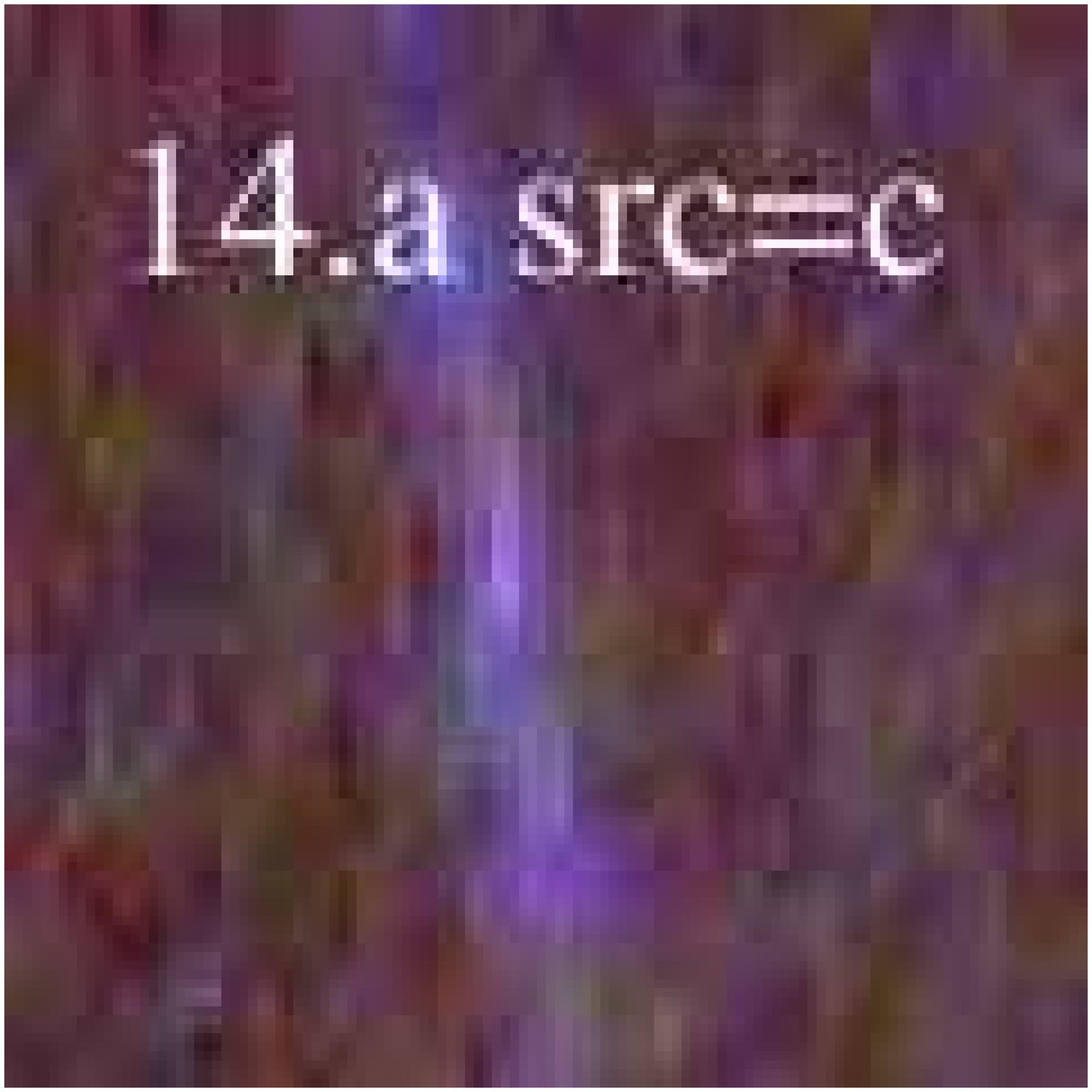}}
    & \multicolumn{1}{m{1.7cm}}{\includegraphics[height=2.00cm,clip]{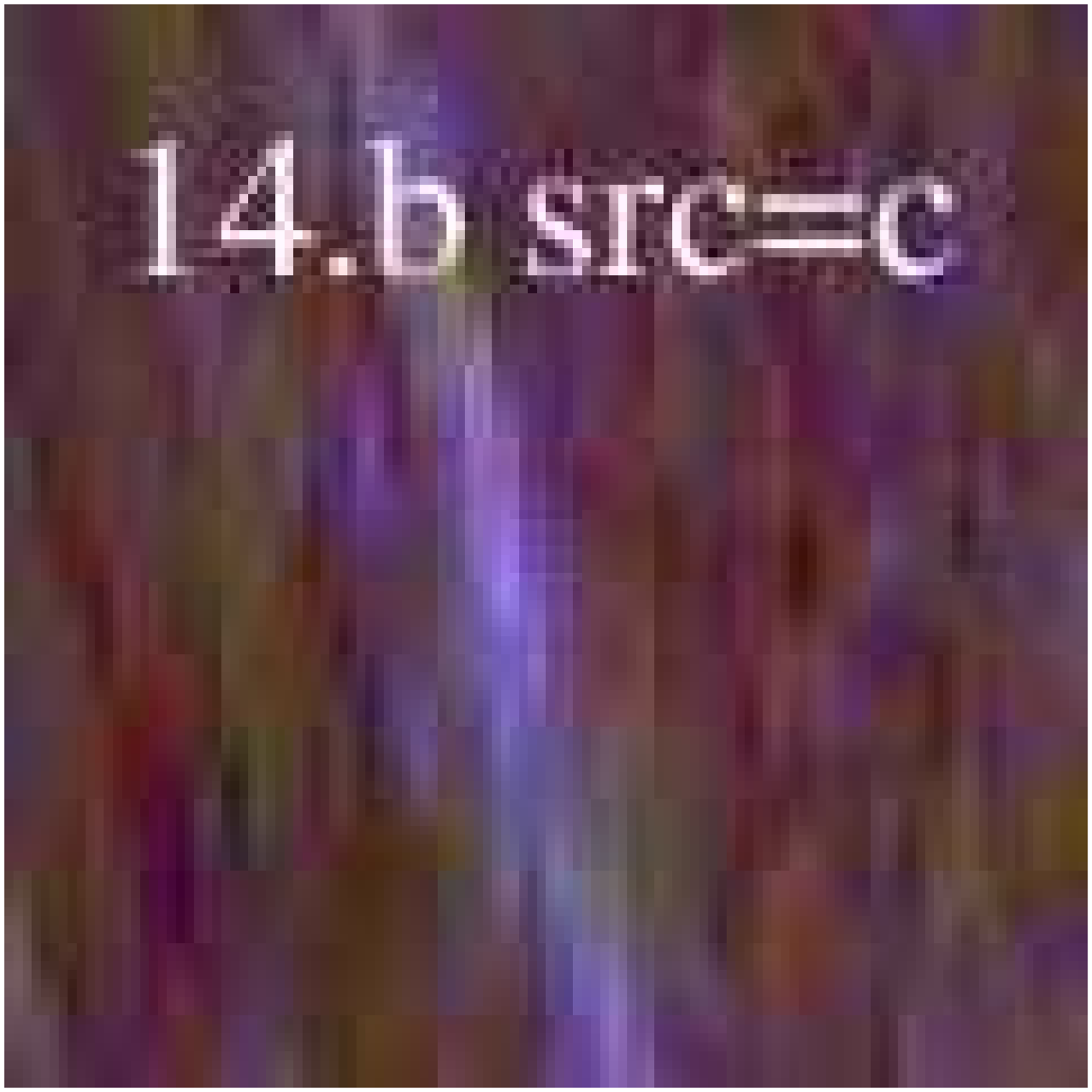}}
    & \multicolumn{1}{m{1.7cm}}{\includegraphics[height=2.00cm,clip]{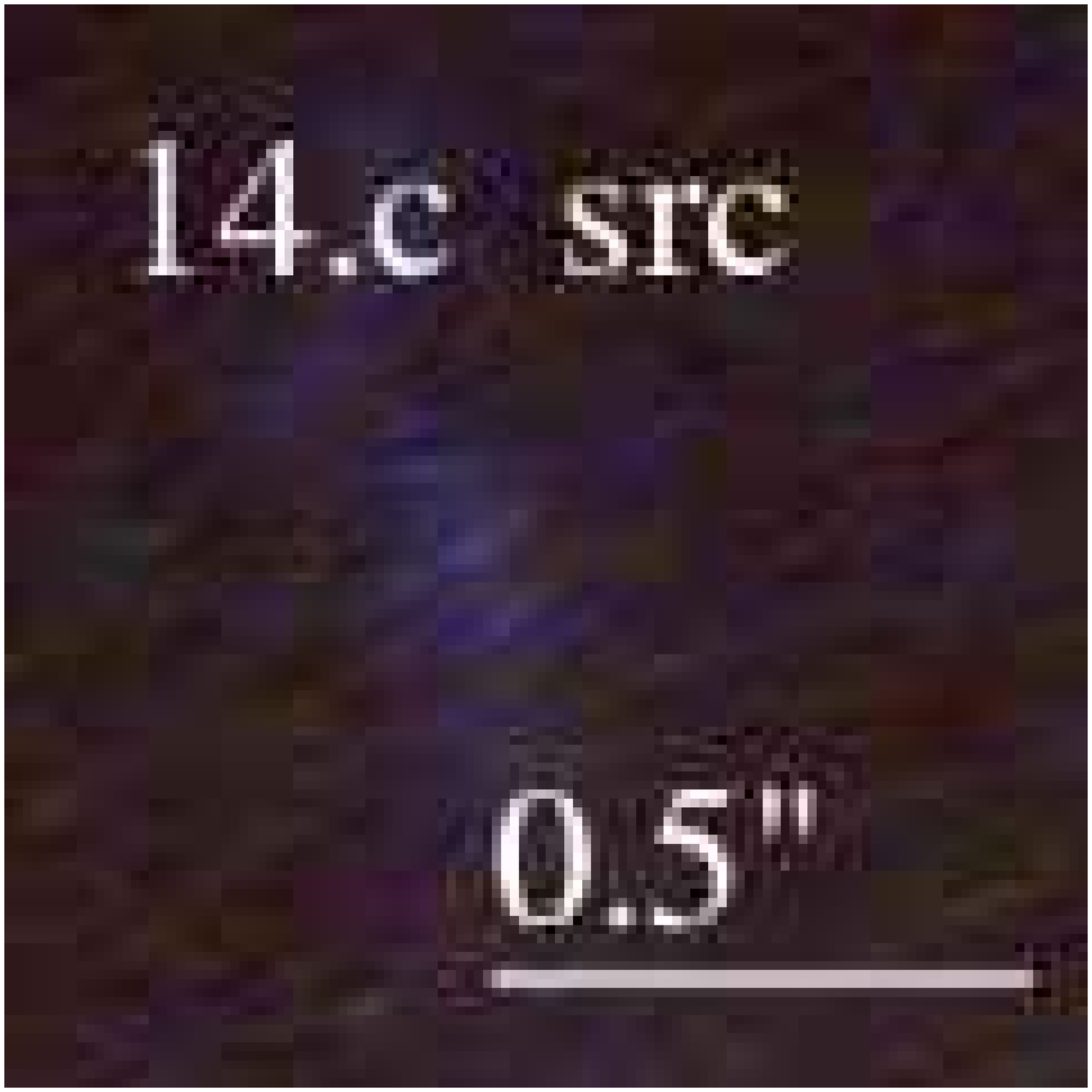}} \\
  \end{tabular}

\end{table*}

\begin{table*}
  \caption{Image system 15:}\vspace{0mm}
  \begin{tabular}{ccc}
    \multicolumn{1}{m{1cm}}{{\Large A1689}}
    & \multicolumn{1}{m{1.7cm}}{\includegraphics[height=2.00cm,clip]{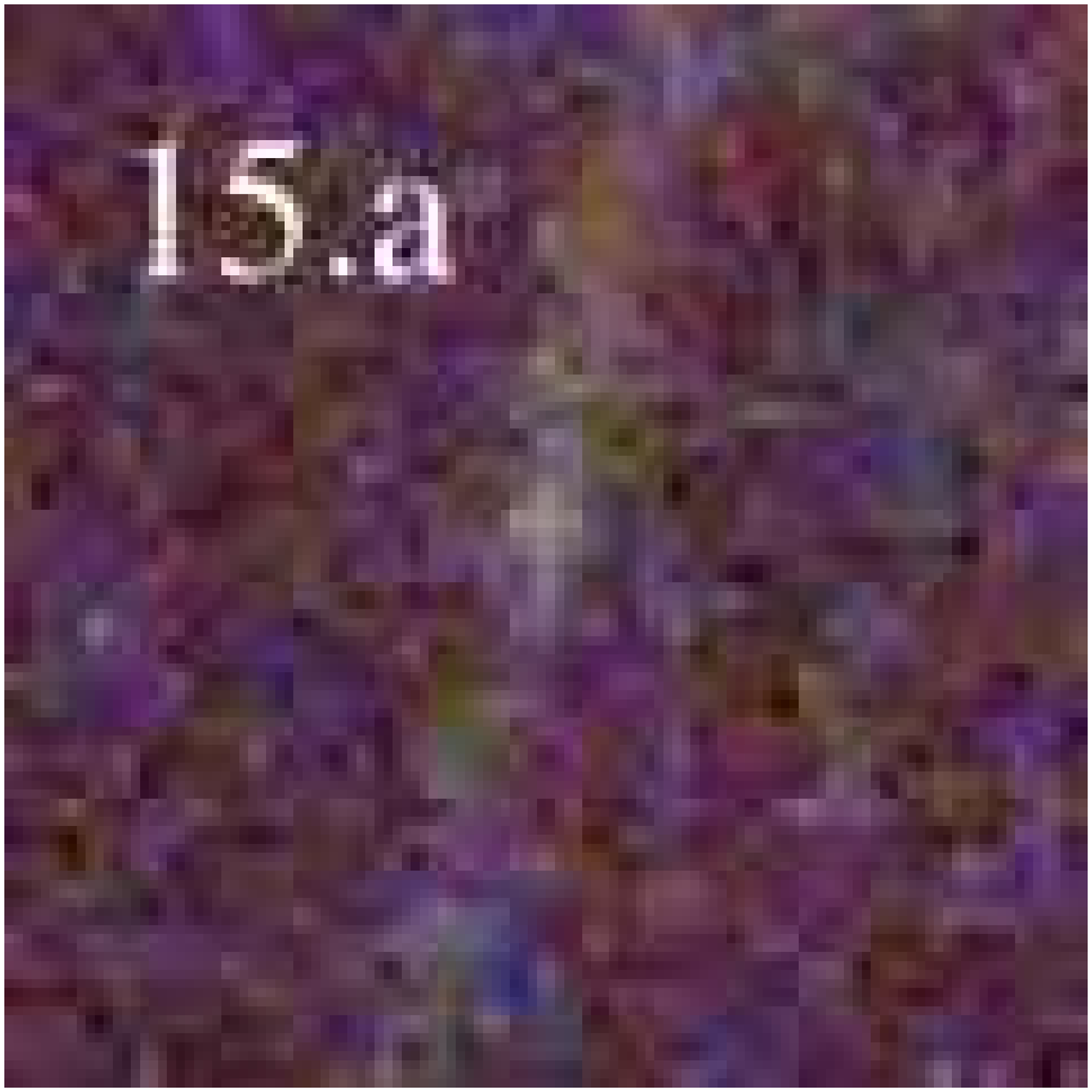}}
    & \multicolumn{1}{m{1.7cm}}{\includegraphics[height=2.00cm,clip]{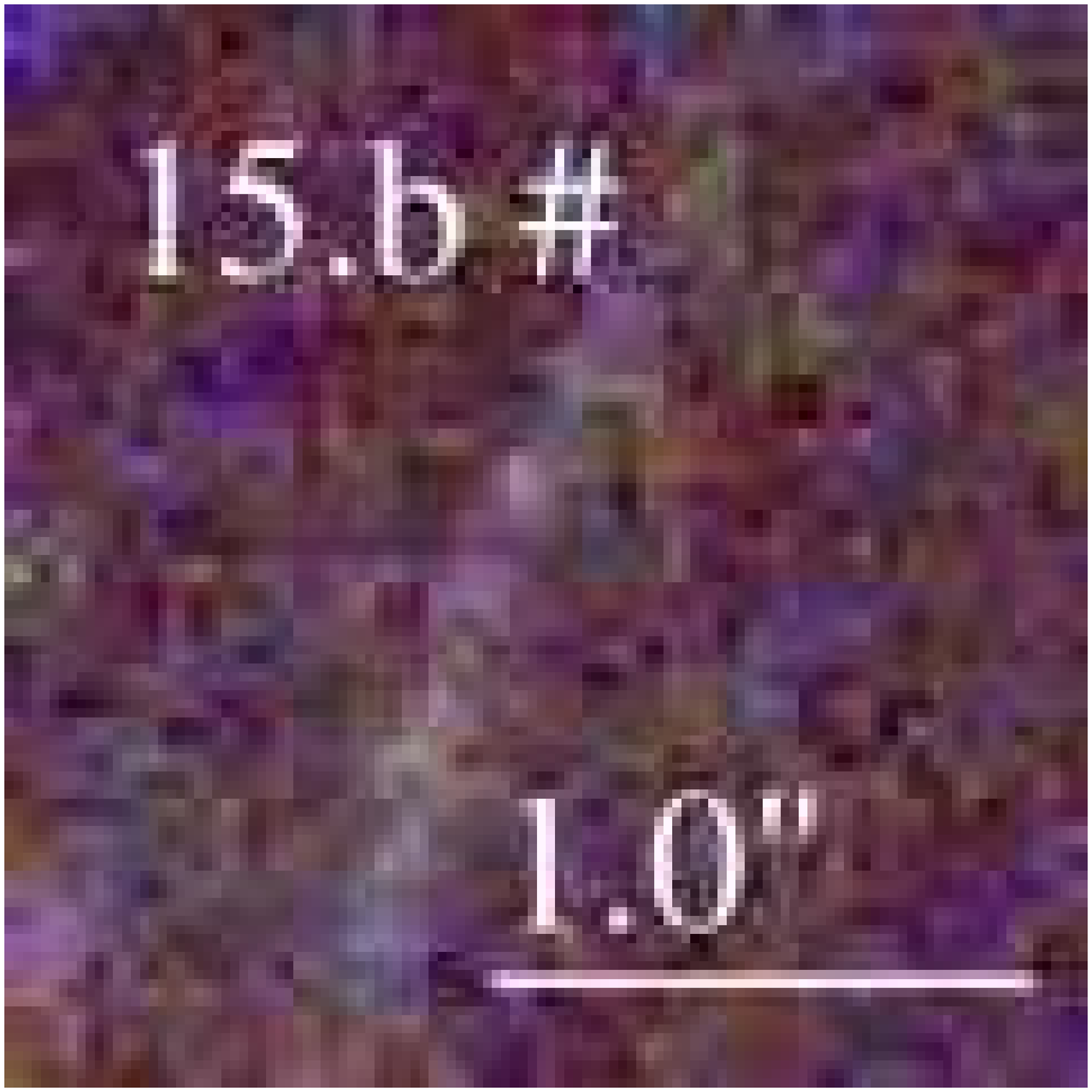}} \\
    \multicolumn{1}{m{1cm}}{{\Large NSIE}}
    & \multicolumn{1}{m{1.7cm}}{\includegraphics[height=2.00cm,clip]{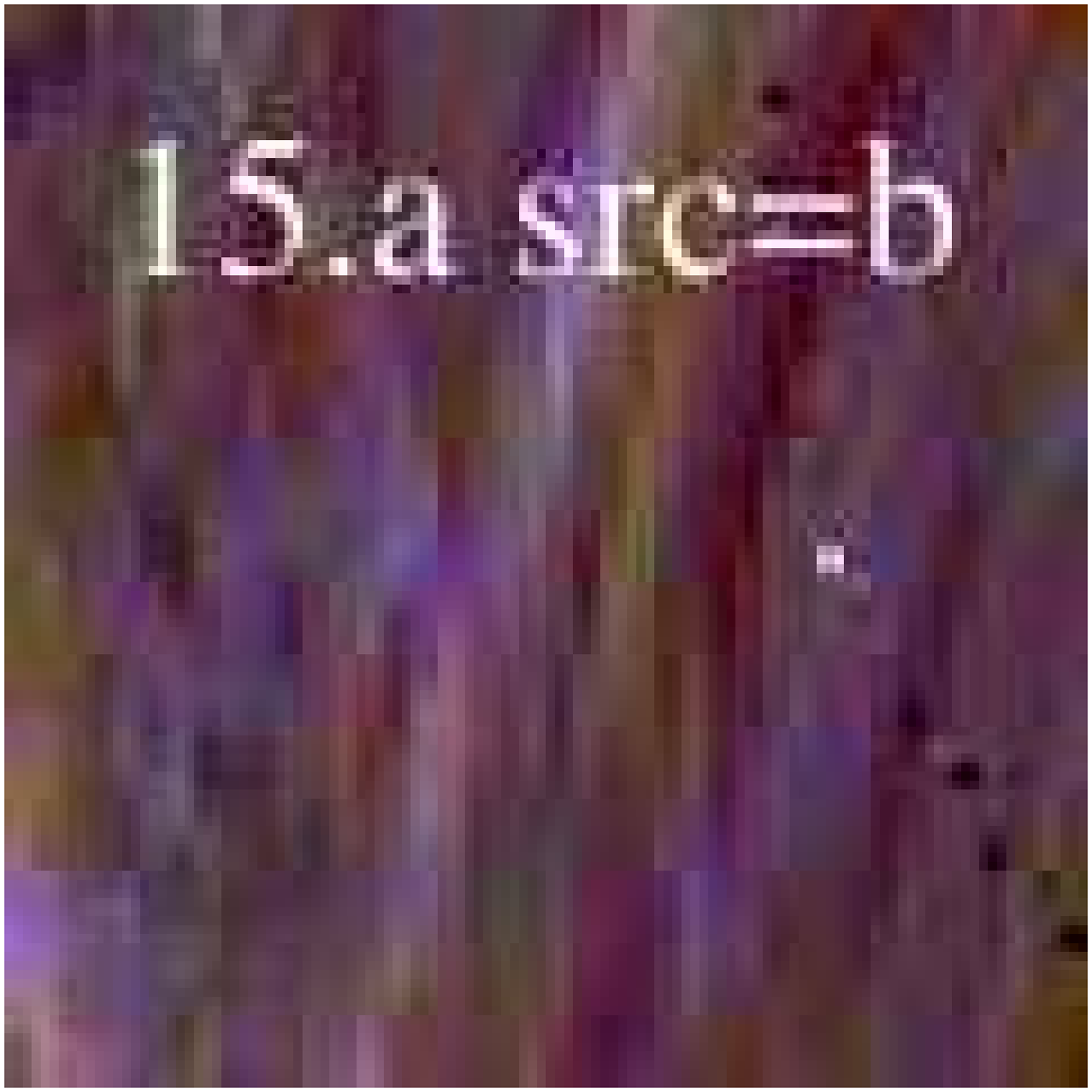}}
    & \multicolumn{1}{m{1.7cm}}{\includegraphics[height=2.00cm,clip]{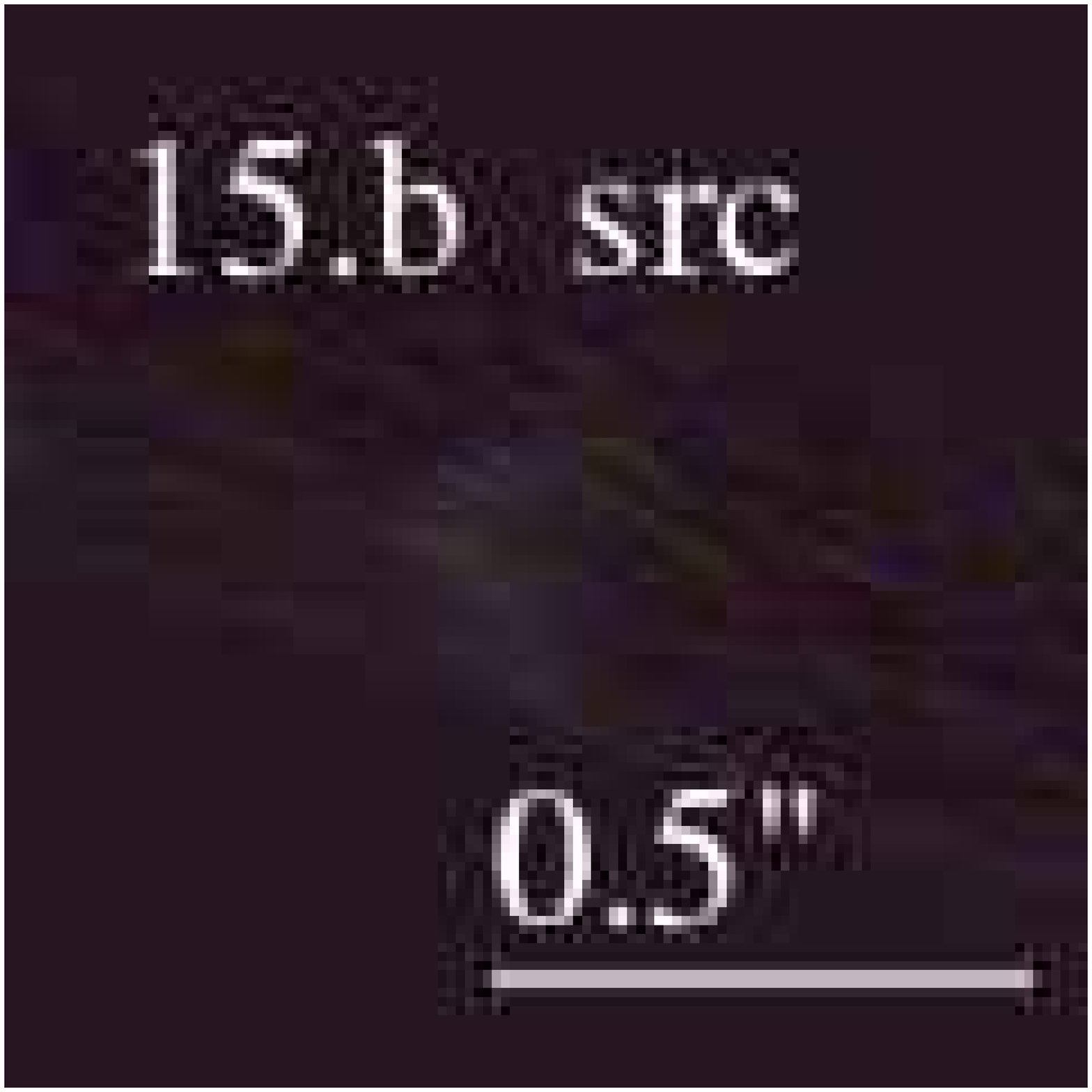}} \\
    \multicolumn{1}{m{1cm}}{{\Large ENFW}}
    & \multicolumn{1}{m{1.7cm}}{\includegraphics[height=2.00cm,clip]{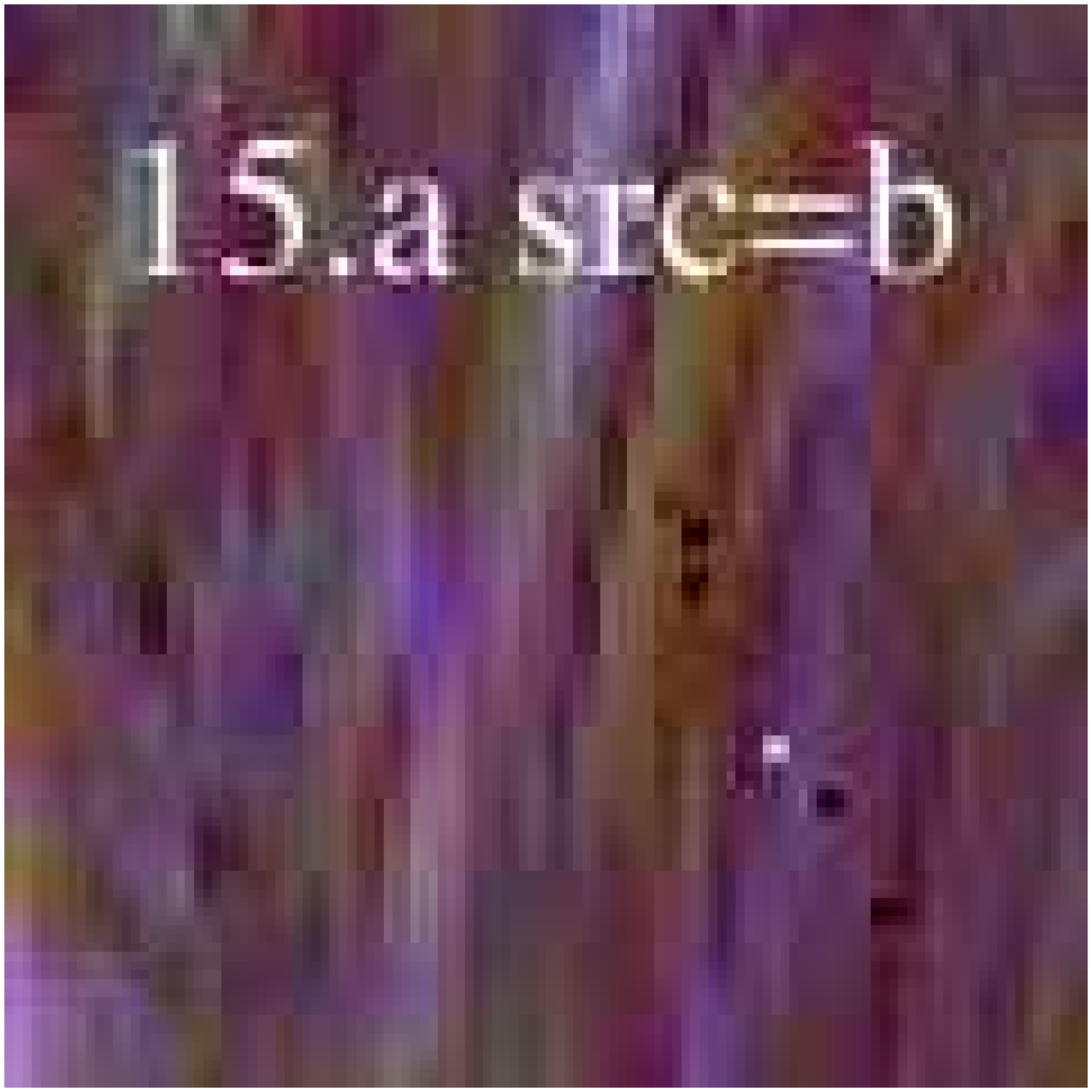}}
    & \multicolumn{1}{m{1.7cm}}{\includegraphics[height=2.00cm,clip]{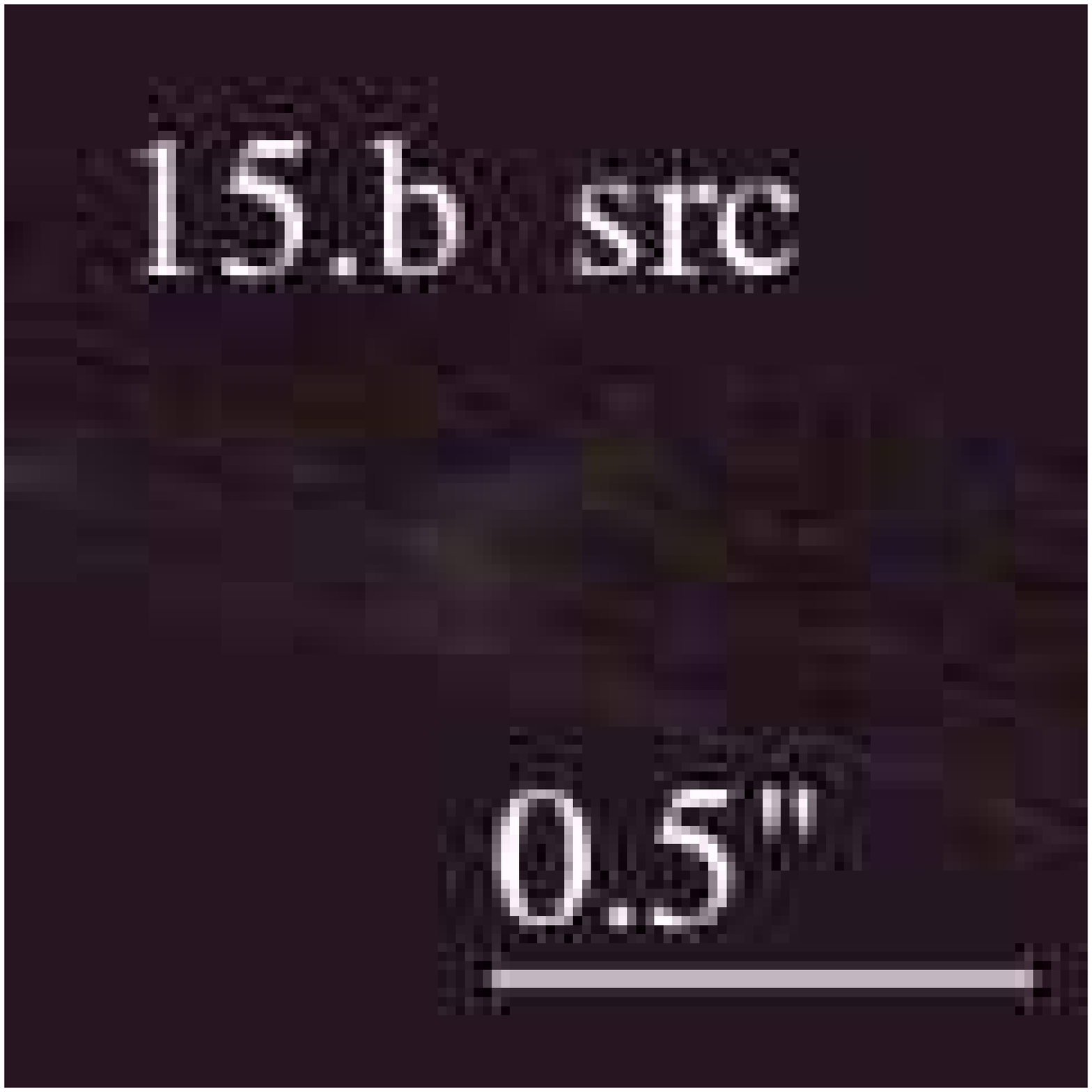}} \\
  \end{tabular}

\end{table*}

\clearpage

\begin{table*}
  \caption{Image system 16:}\vspace{0mm}
  \begin{tabular}{cccc}
    \multicolumn{1}{m{1cm}}{{\Large A1689}}
    & \multicolumn{1}{m{1.7cm}}{\includegraphics[height=2.00cm,clip]{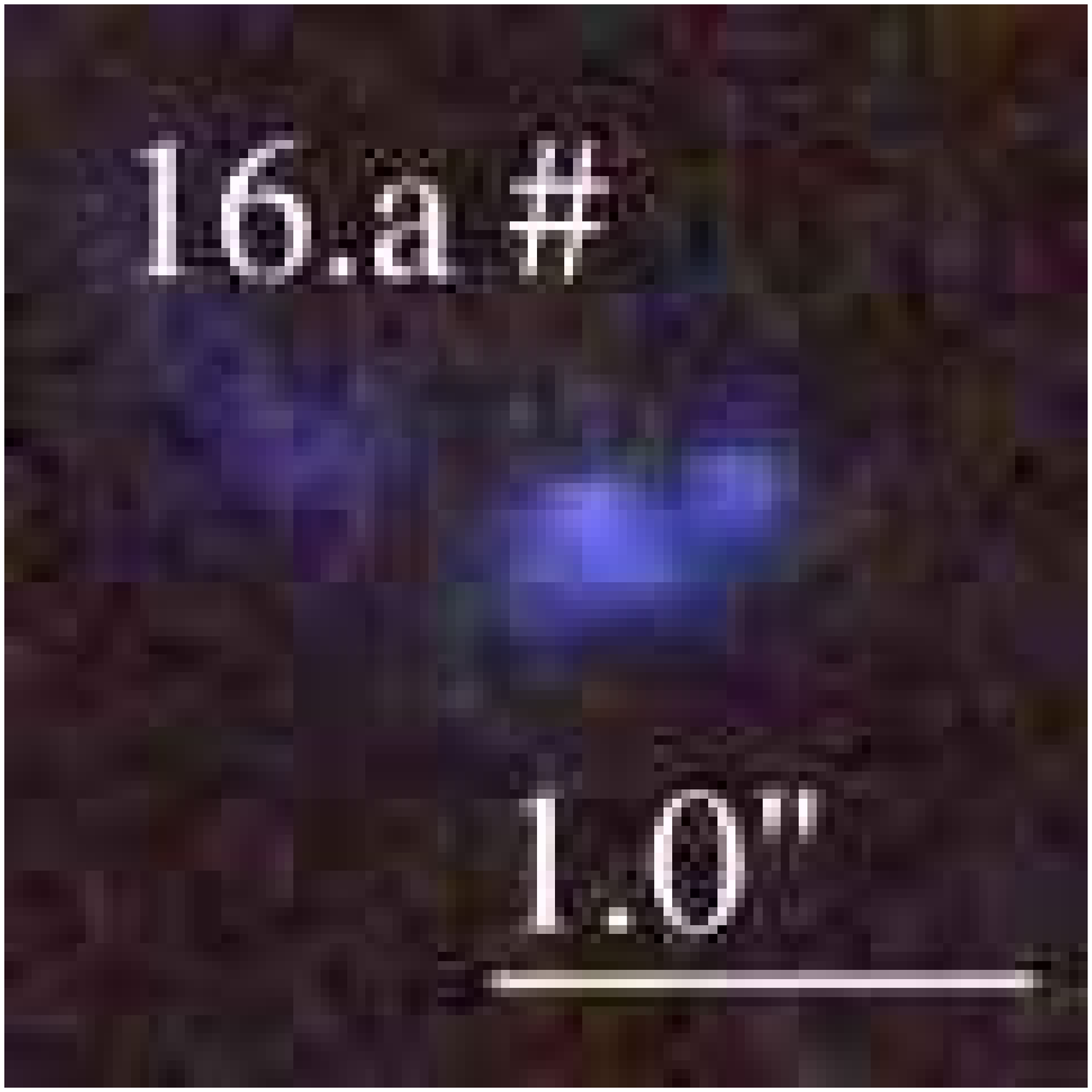}}
    & \multicolumn{1}{m{1.7cm}}{\includegraphics[height=2.00cm,clip]{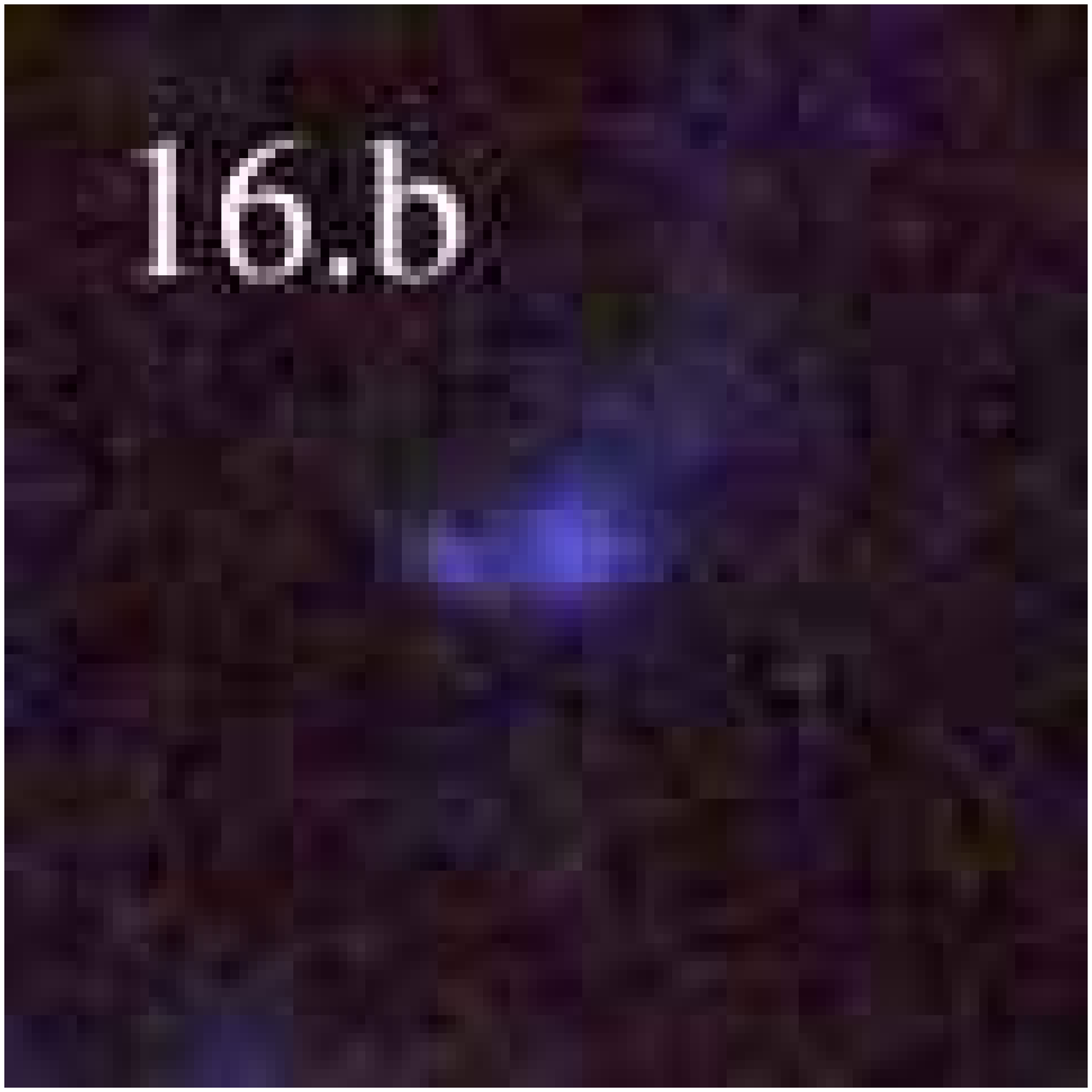}}
    & \multicolumn{1}{m{1.7cm}}{\includegraphics[height=2.00cm,clip]{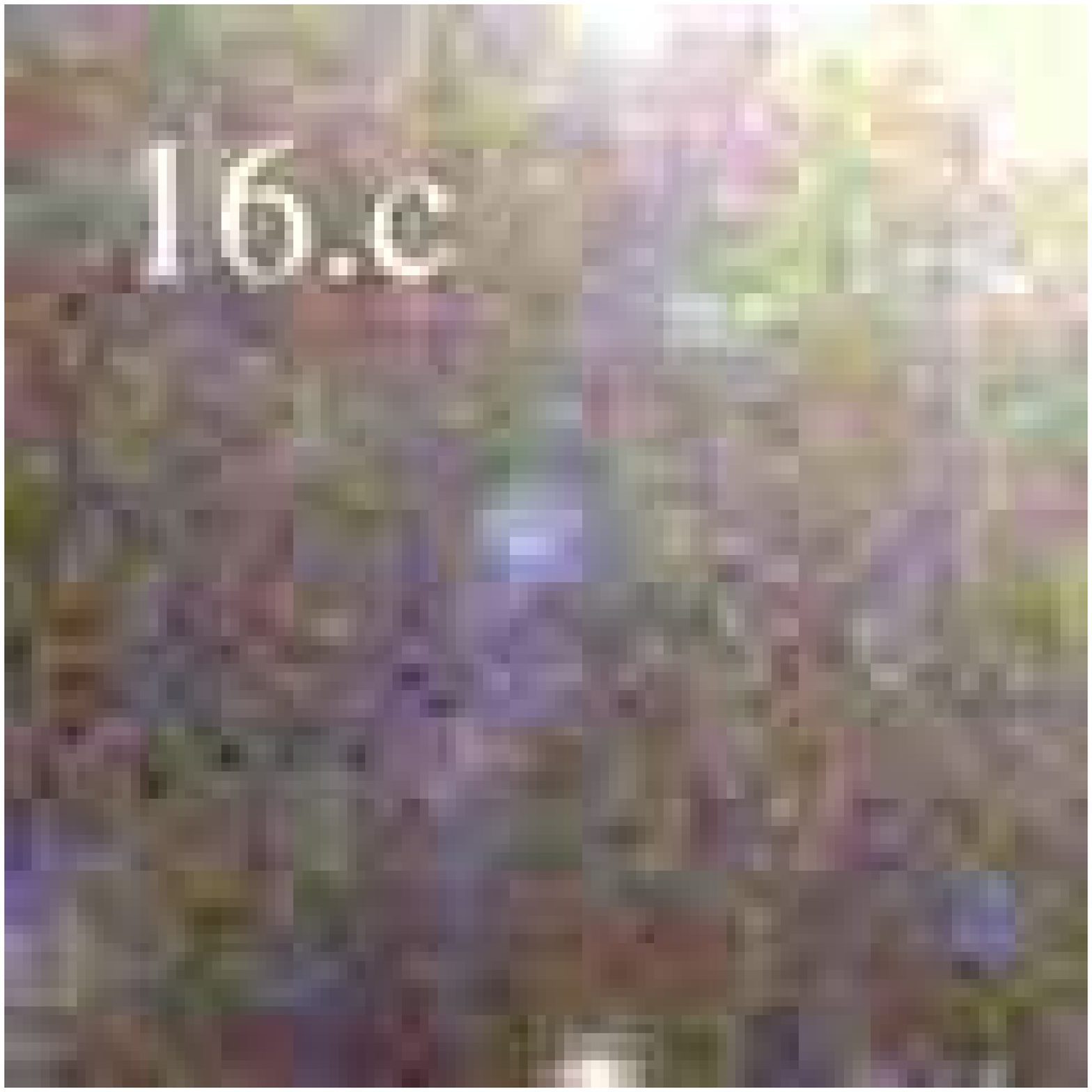}} \\
    \multicolumn{1}{m{1cm}}{{\Large NSIE}}
    & \multicolumn{1}{m{1.7cm}}{\includegraphics[height=2.00cm,clip]{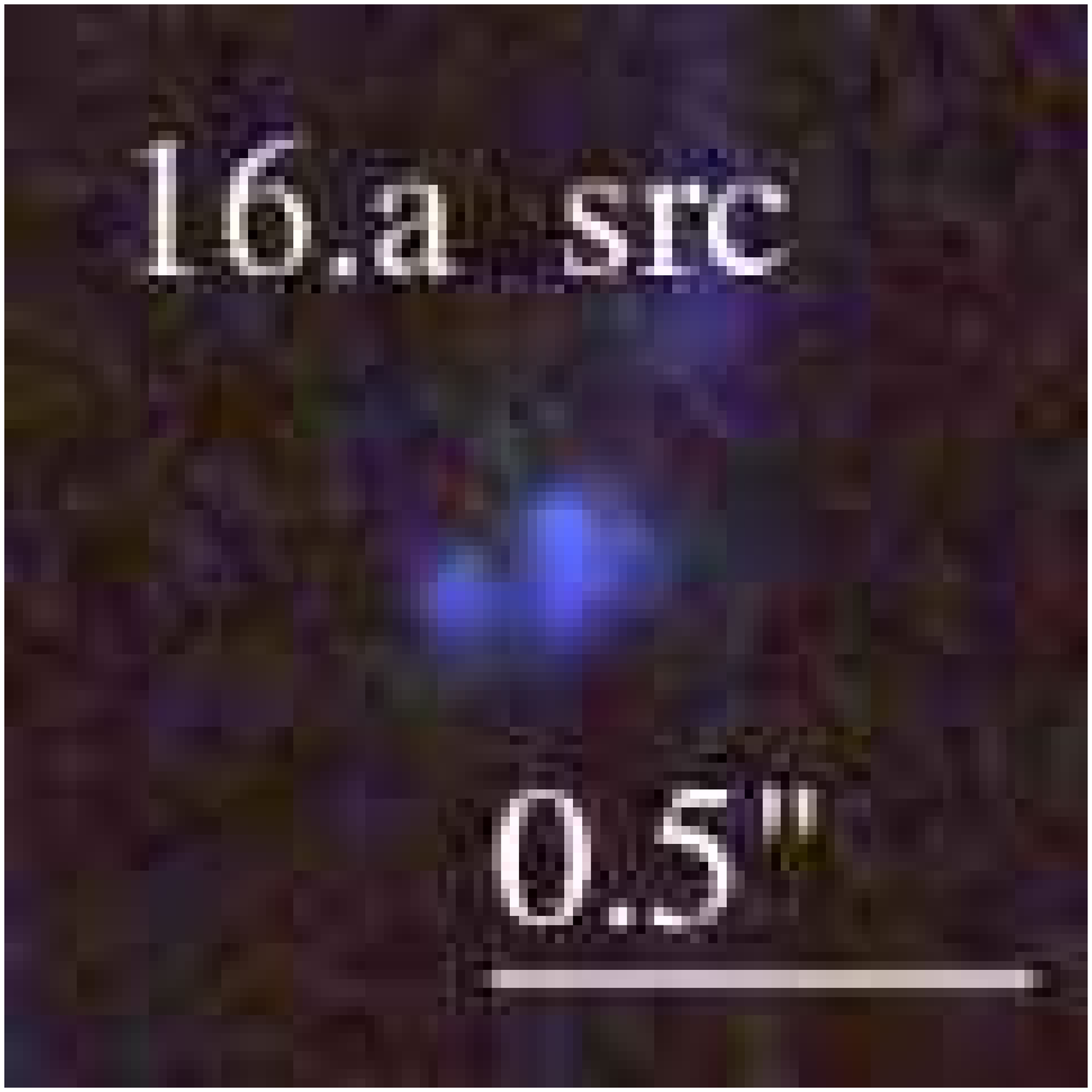}}
    & \multicolumn{1}{m{1.7cm}}{\includegraphics[height=2.00cm,clip]{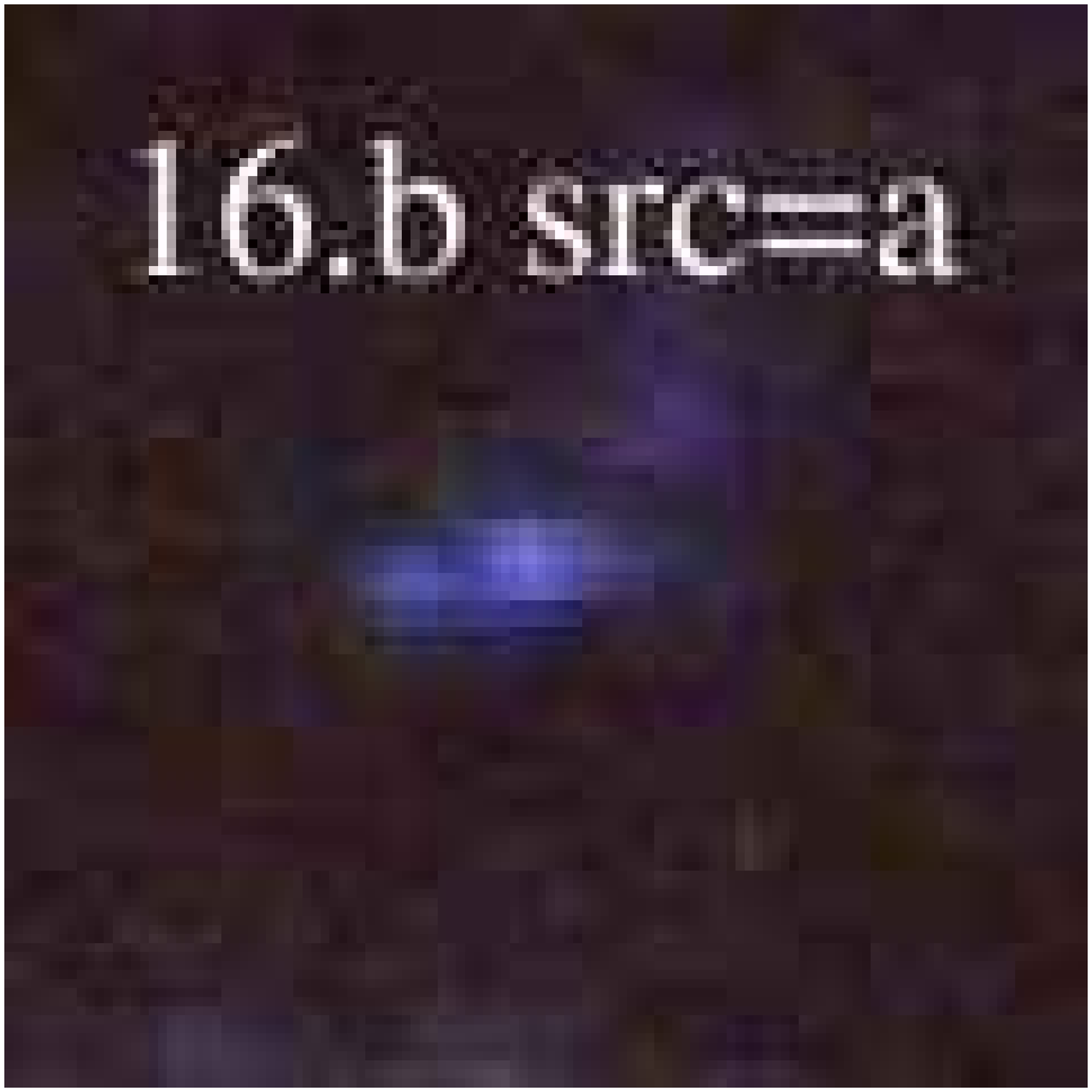}}
    & \multicolumn{1}{m{1.7cm}}{\includegraphics[height=2.00cm,clip]{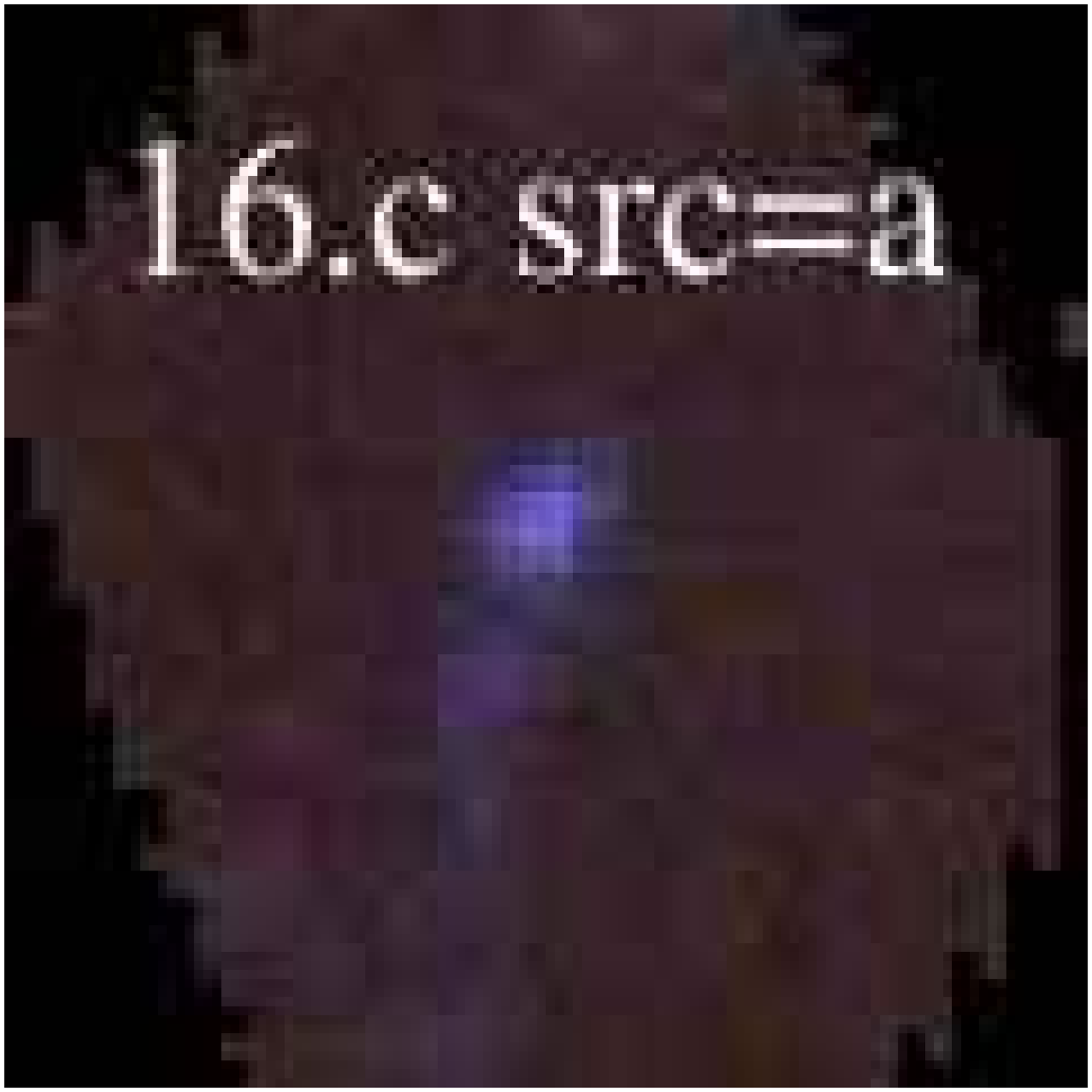}} \\
    \multicolumn{1}{m{1cm}}{{\Large ENFW}}
    & \multicolumn{1}{m{1.7cm}}{\includegraphics[height=2.00cm,clip]{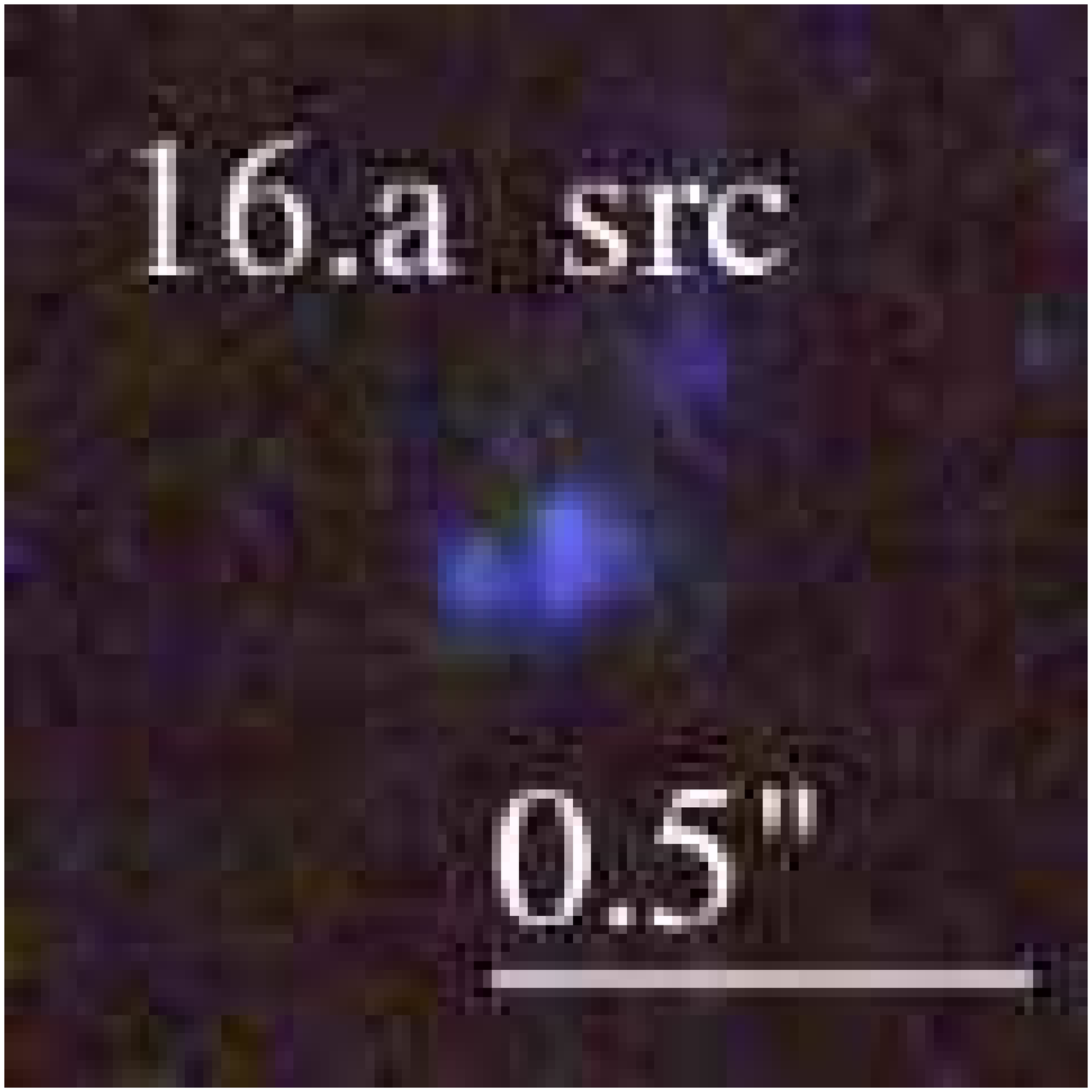}}
    & \multicolumn{1}{m{1.7cm}}{\includegraphics[height=2.00cm,clip]{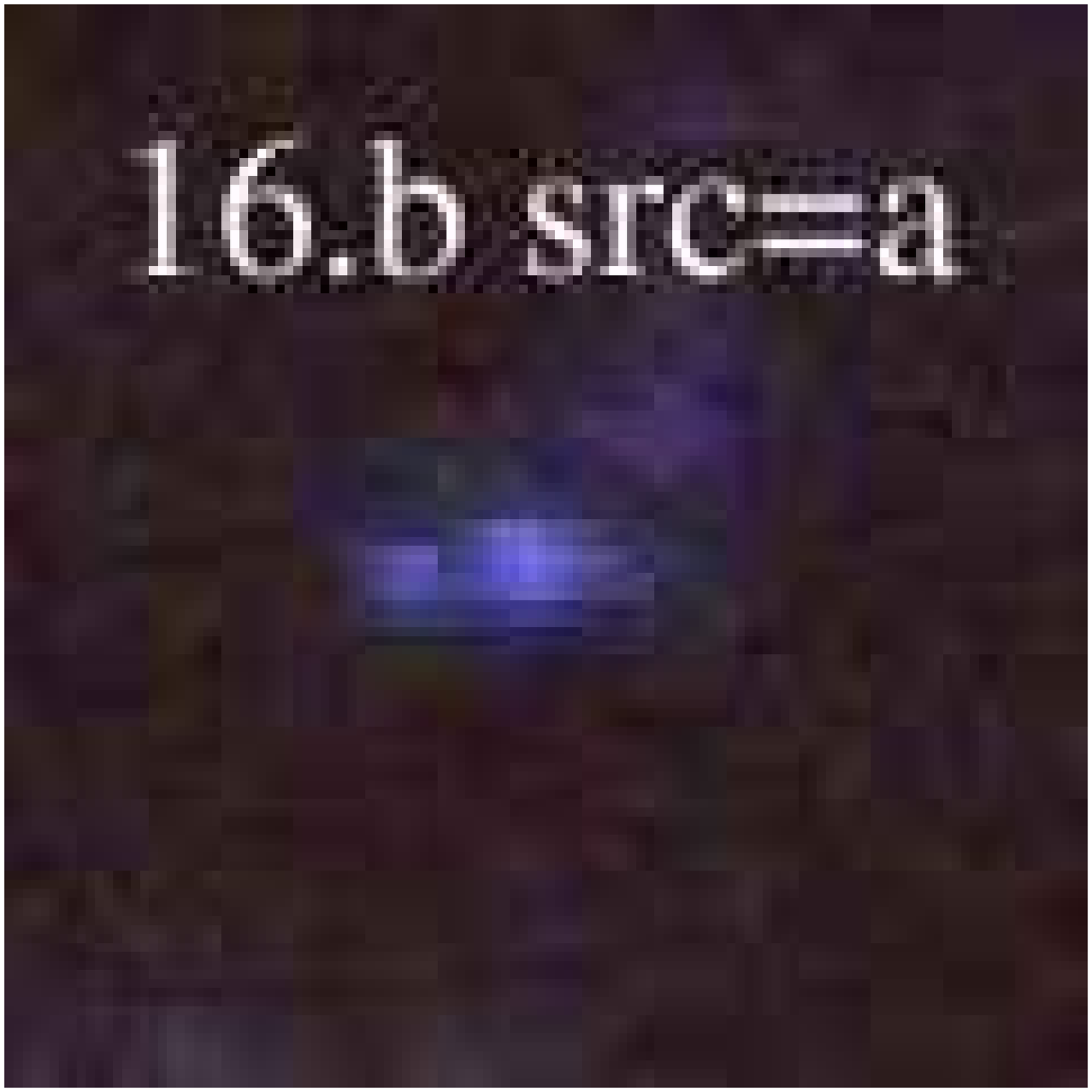}}
    & \multicolumn{1}{m{1.7cm}}{\includegraphics[height=2.00cm,clip]{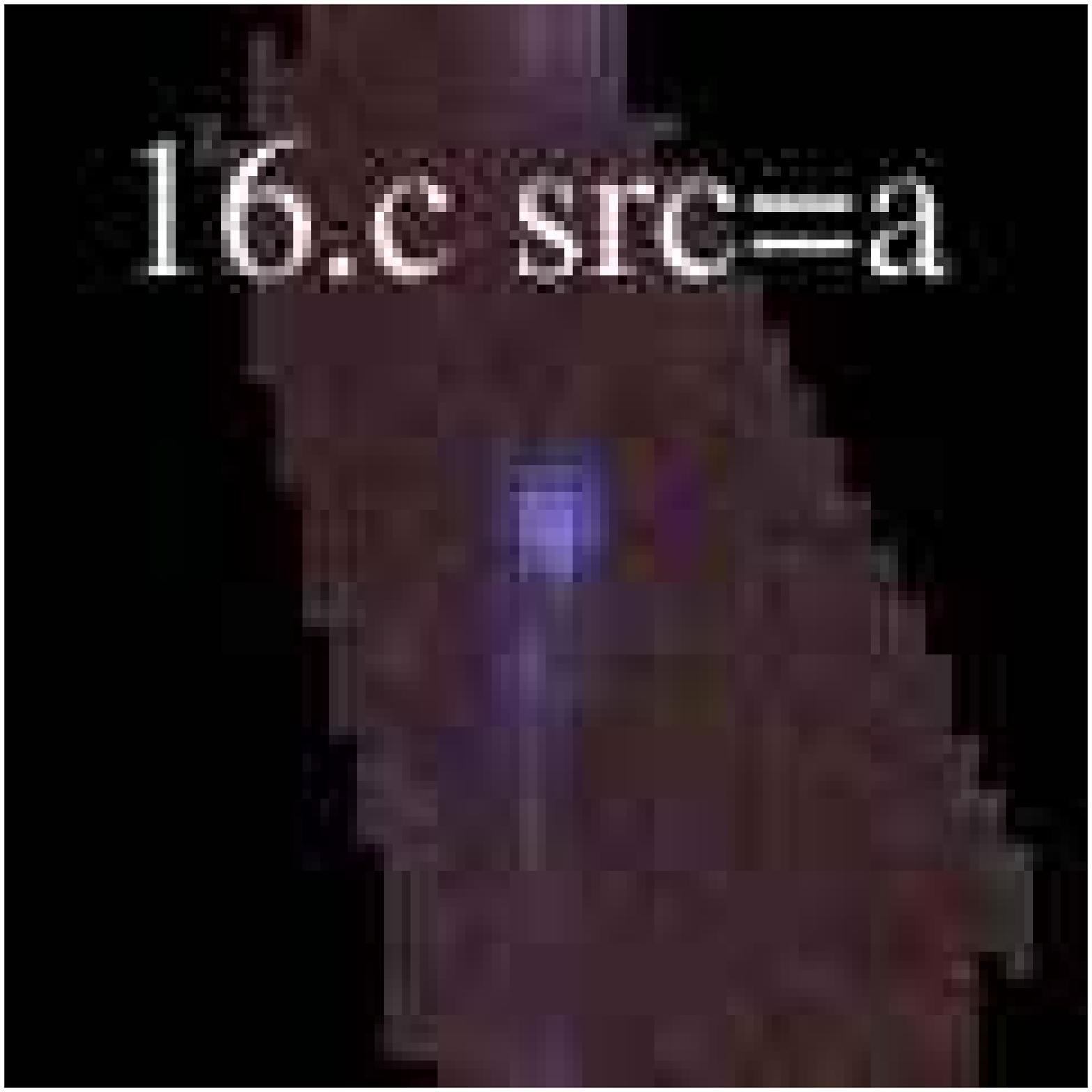}} \\
  \end{tabular}

\end{table*}

\begin{table*}
  \caption{Image system 17:}\vspace{0mm}
  \begin{tabular}{cccc}
    \multicolumn{1}{m{1cm}}{{\Large A1689}}
    & \multicolumn{1}{m{1.7cm}}{\includegraphics[height=2.00cm,clip]{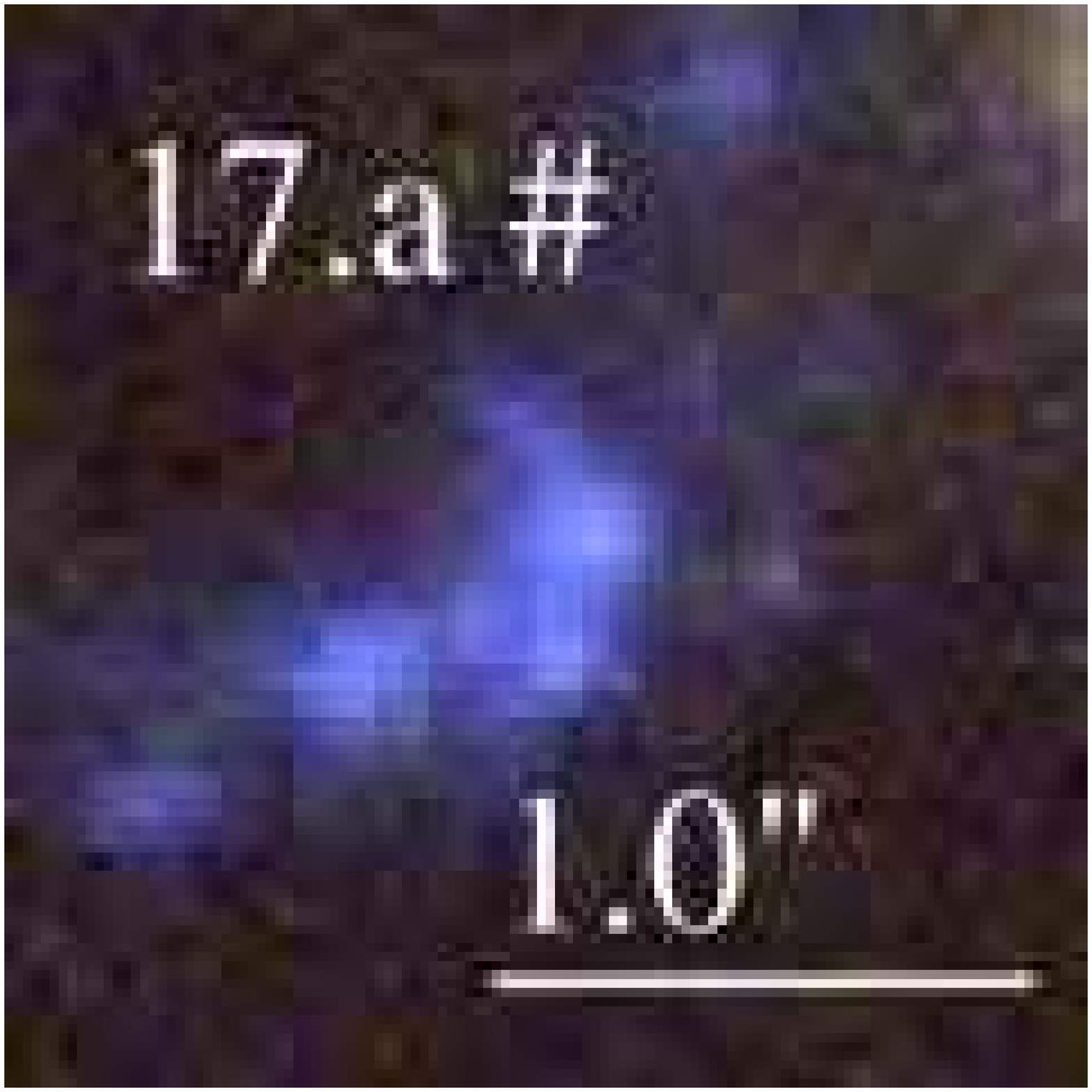}}
    & \multicolumn{1}{m{1.7cm}}{\includegraphics[height=2.00cm,clip]{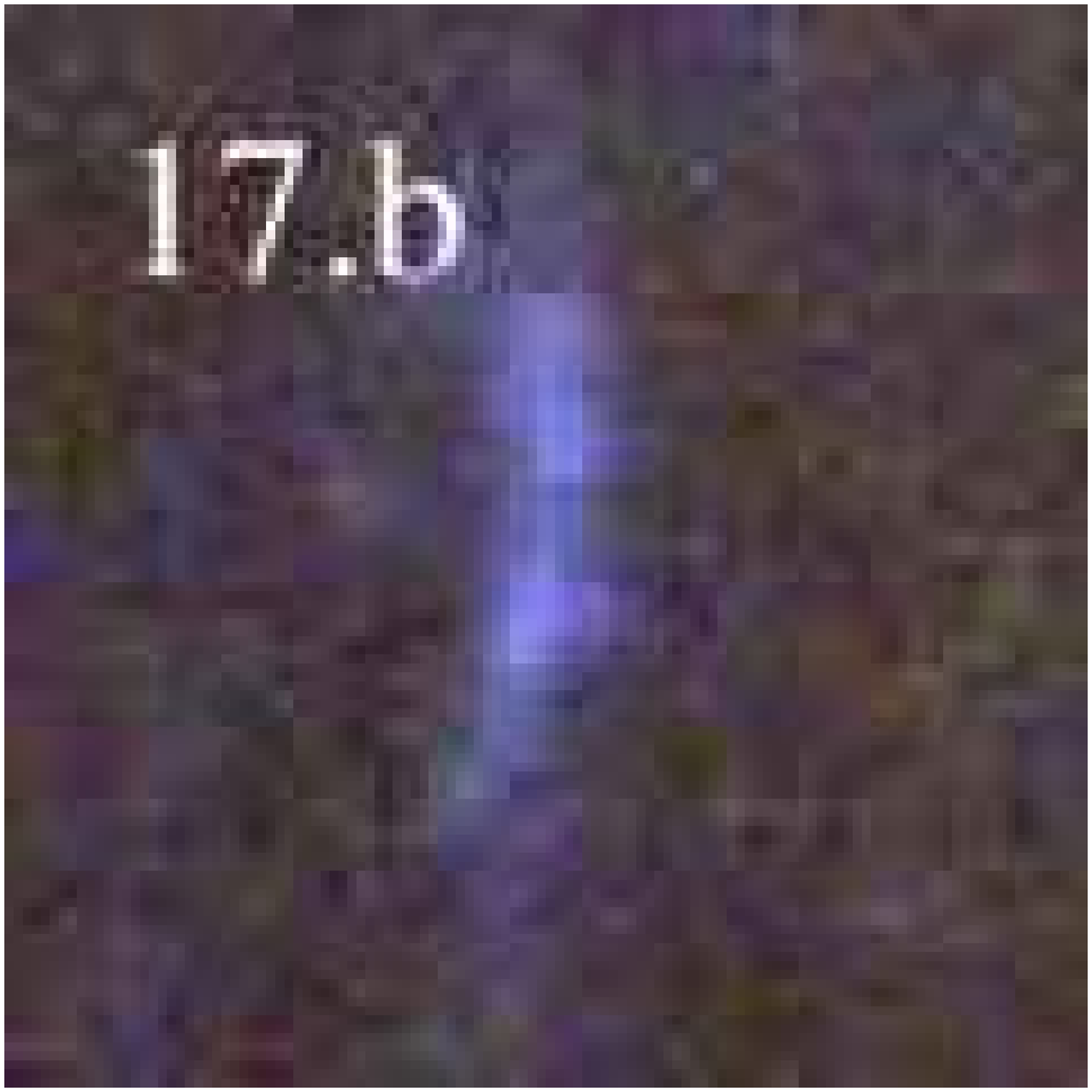}}
    & \multicolumn{1}{m{1.7cm}}{\includegraphics[height=2.00cm,clip]{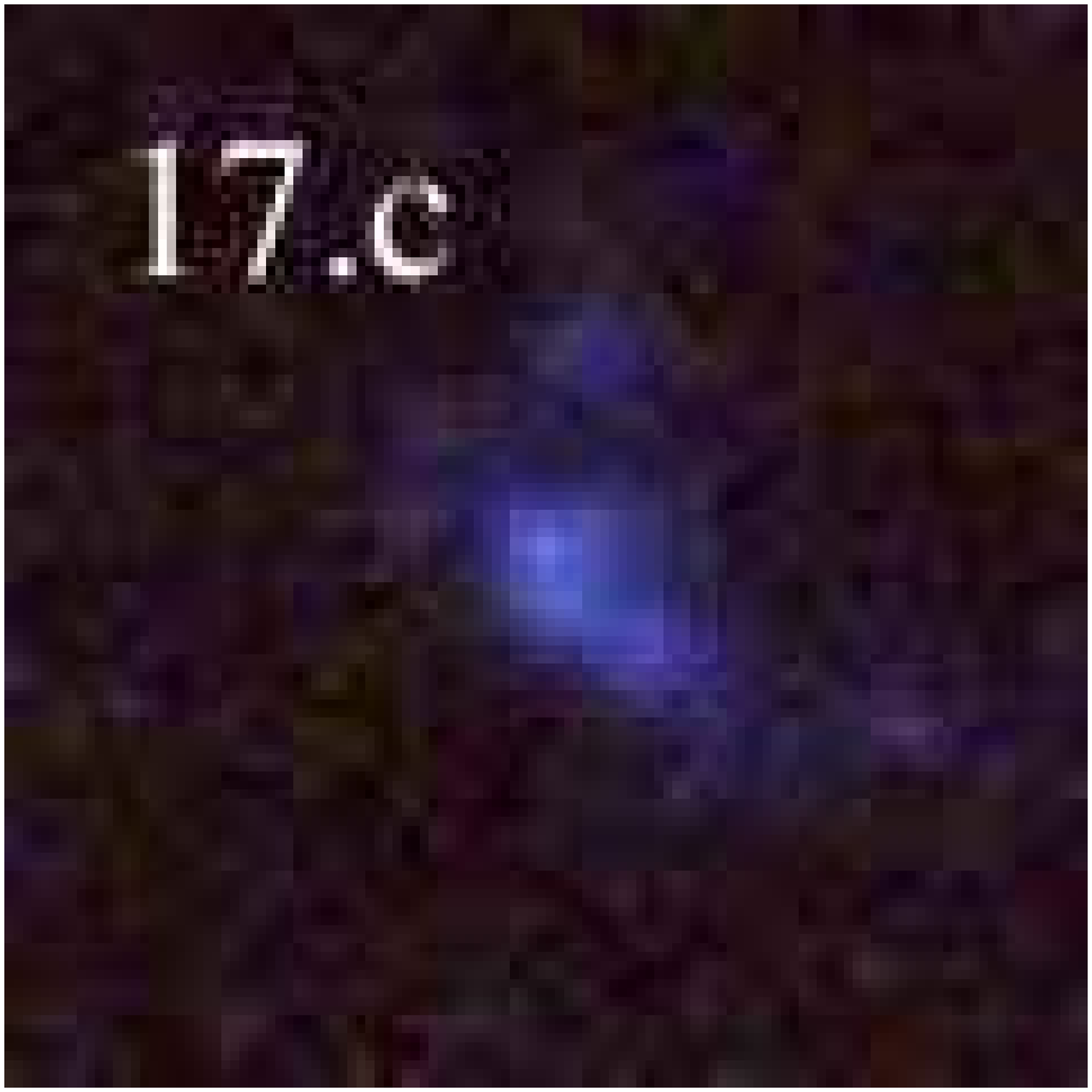}} \\
    \multicolumn{1}{m{1cm}}{{\Large NSIE}}
    & \multicolumn{1}{m{1.7cm}}{\includegraphics[height=2.00cm,clip]{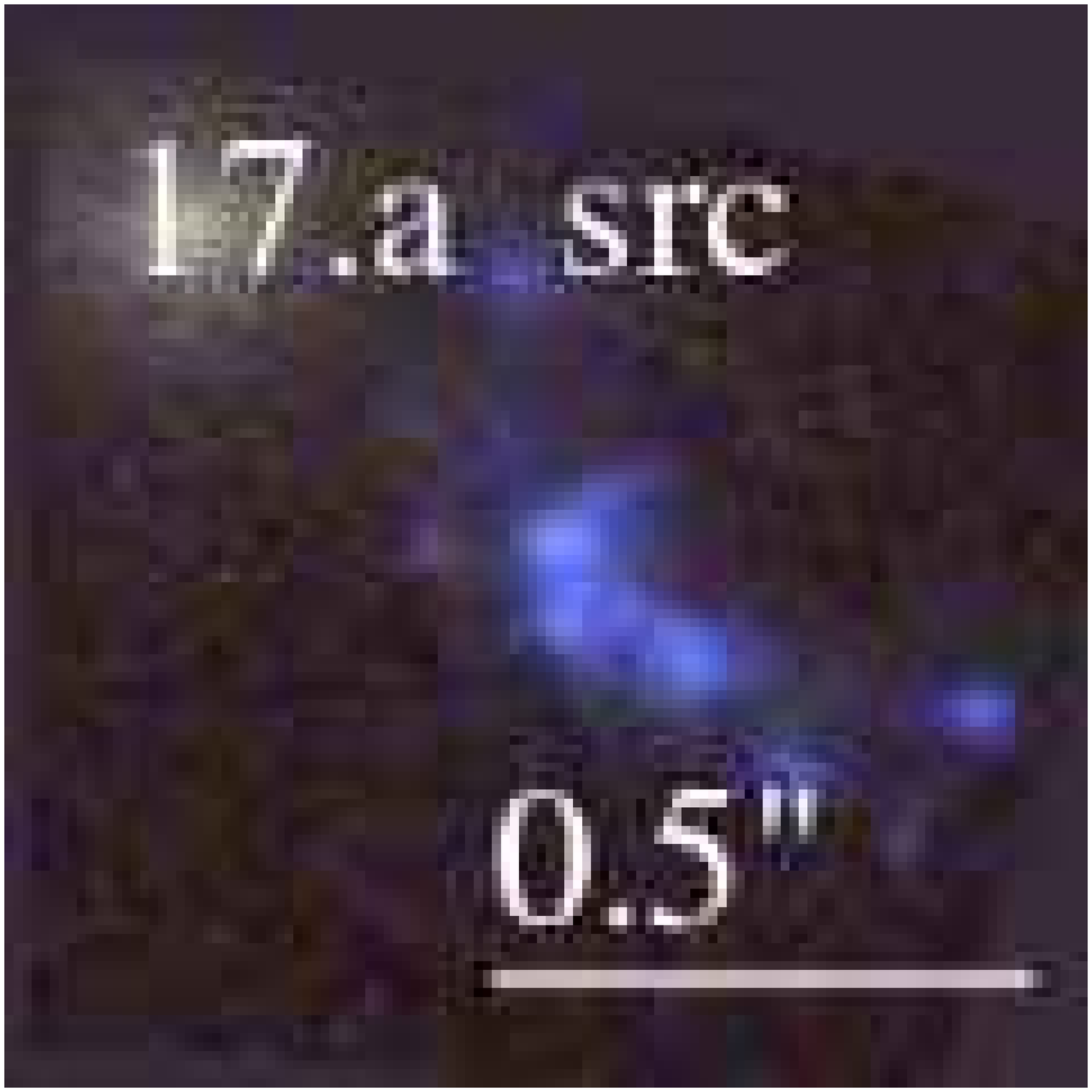}}
    & \multicolumn{1}{m{1.7cm}}{\includegraphics[height=2.00cm,clip]{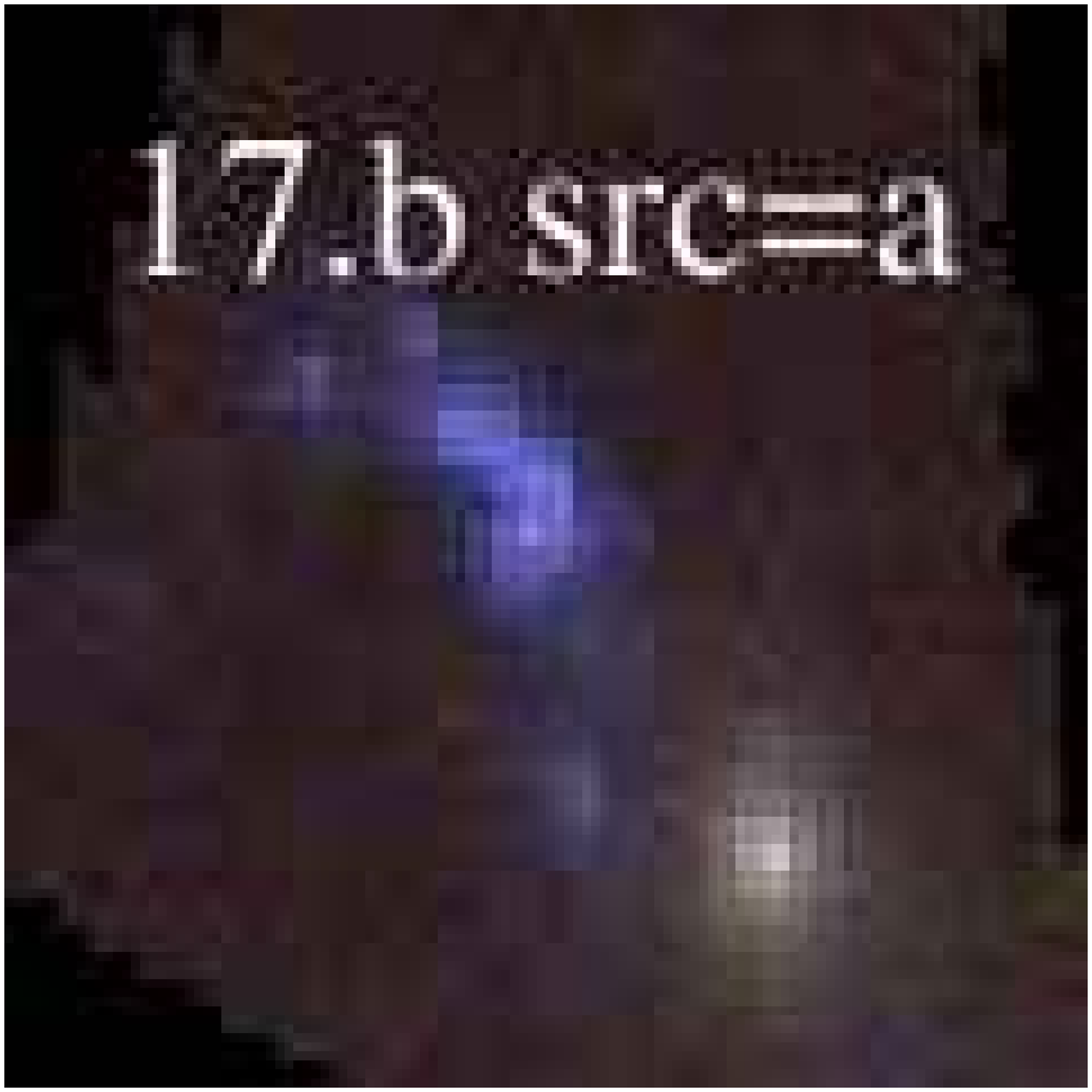}}
    & \multicolumn{1}{m{1.7cm}}{\includegraphics[height=2.00cm,clip]{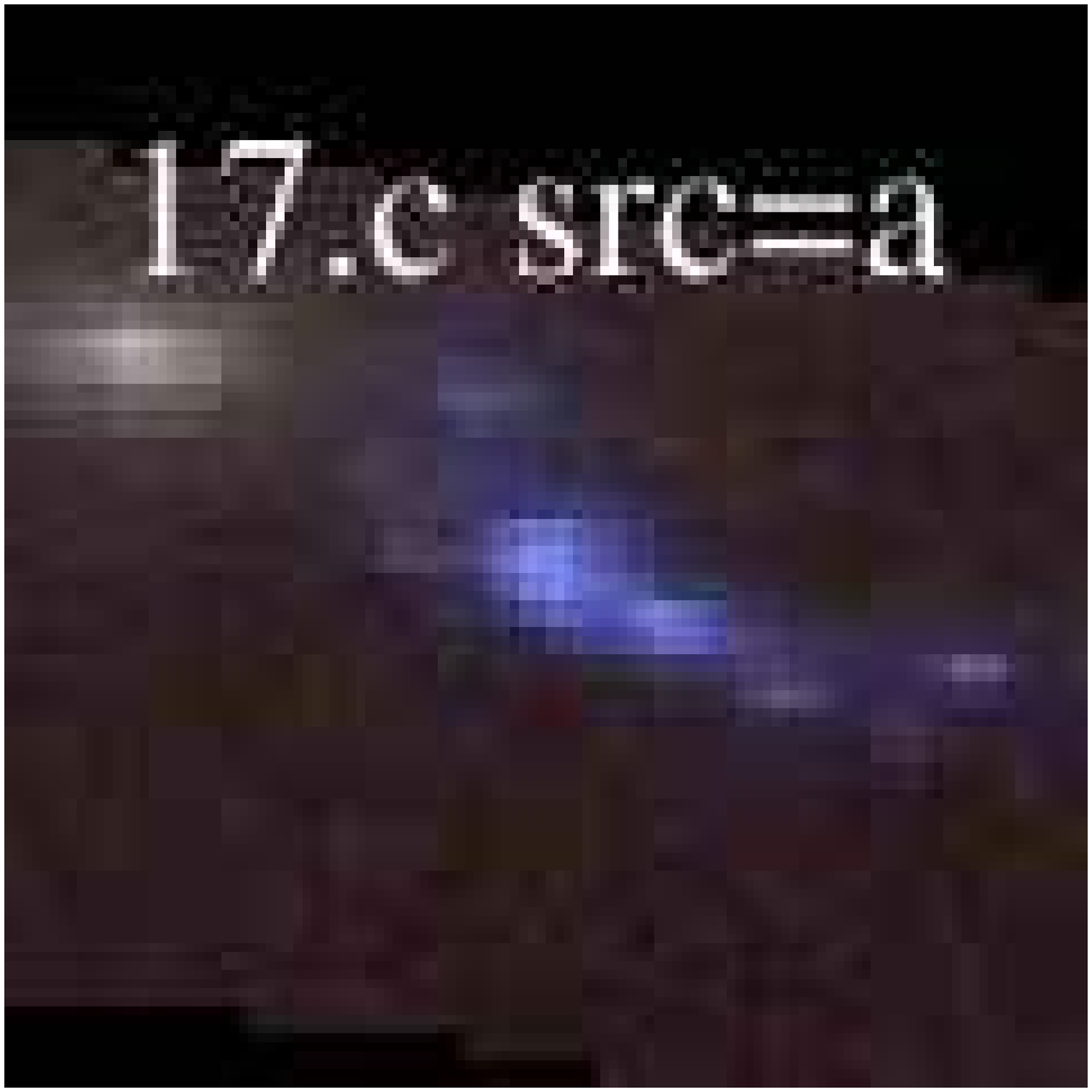}} \\
    \multicolumn{1}{m{1cm}}{{\Large ENFW}}
    & \multicolumn{1}{m{1.7cm}}{\includegraphics[height=2.00cm,clip]{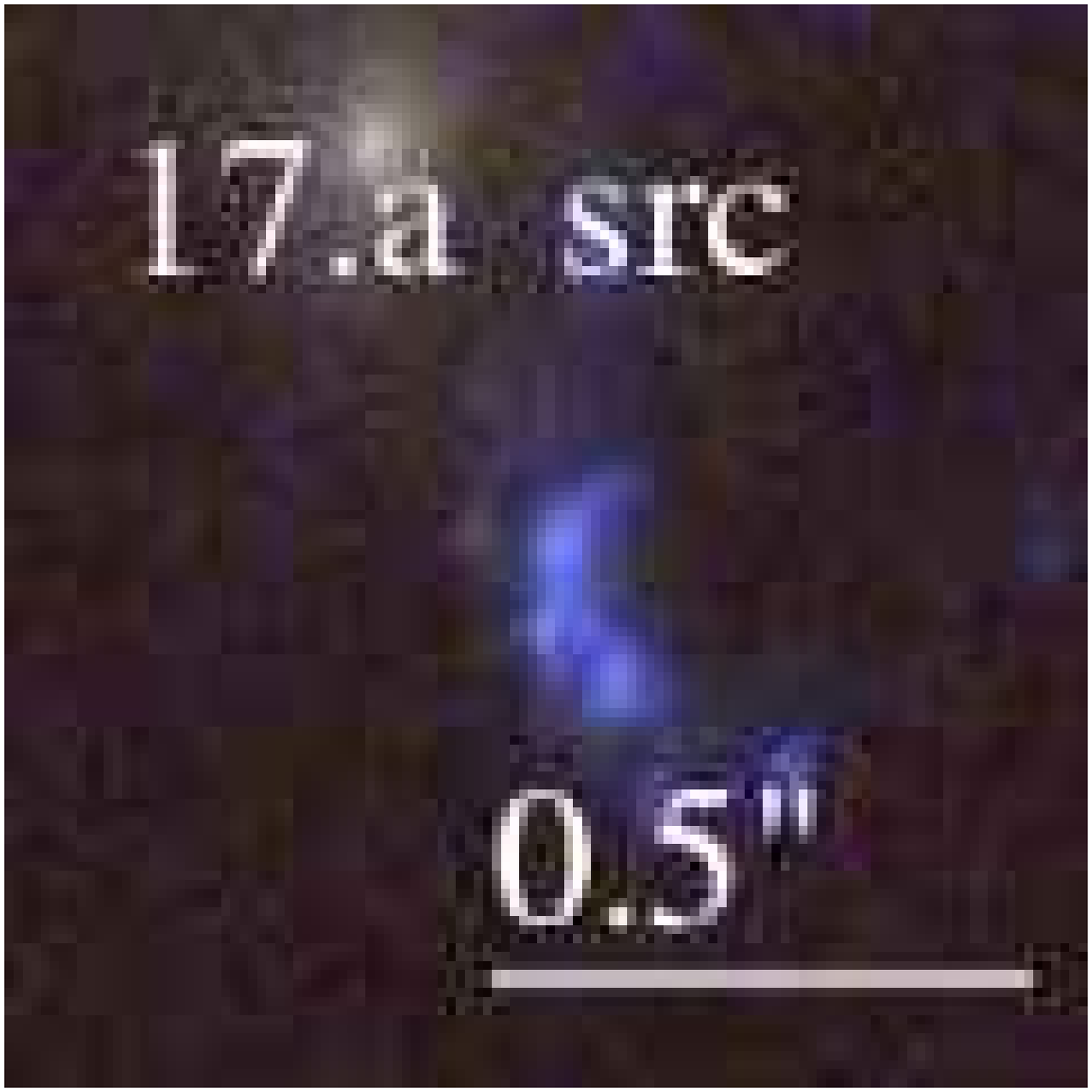}}
    & \multicolumn{1}{m{1.7cm}}{\includegraphics[height=2.00cm,clip]{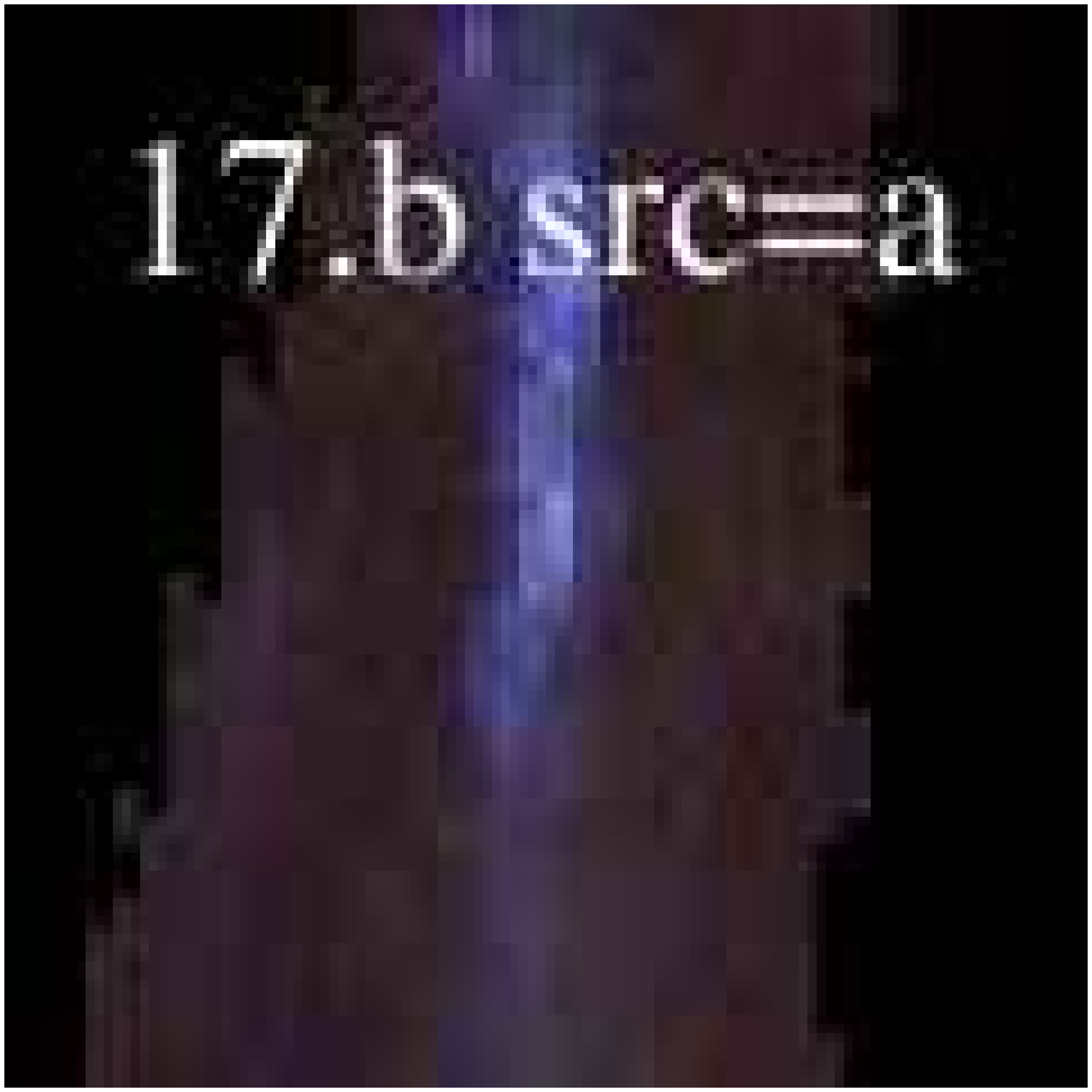}}
    & \multicolumn{1}{m{1.7cm}}{\includegraphics[height=2.00cm,clip]{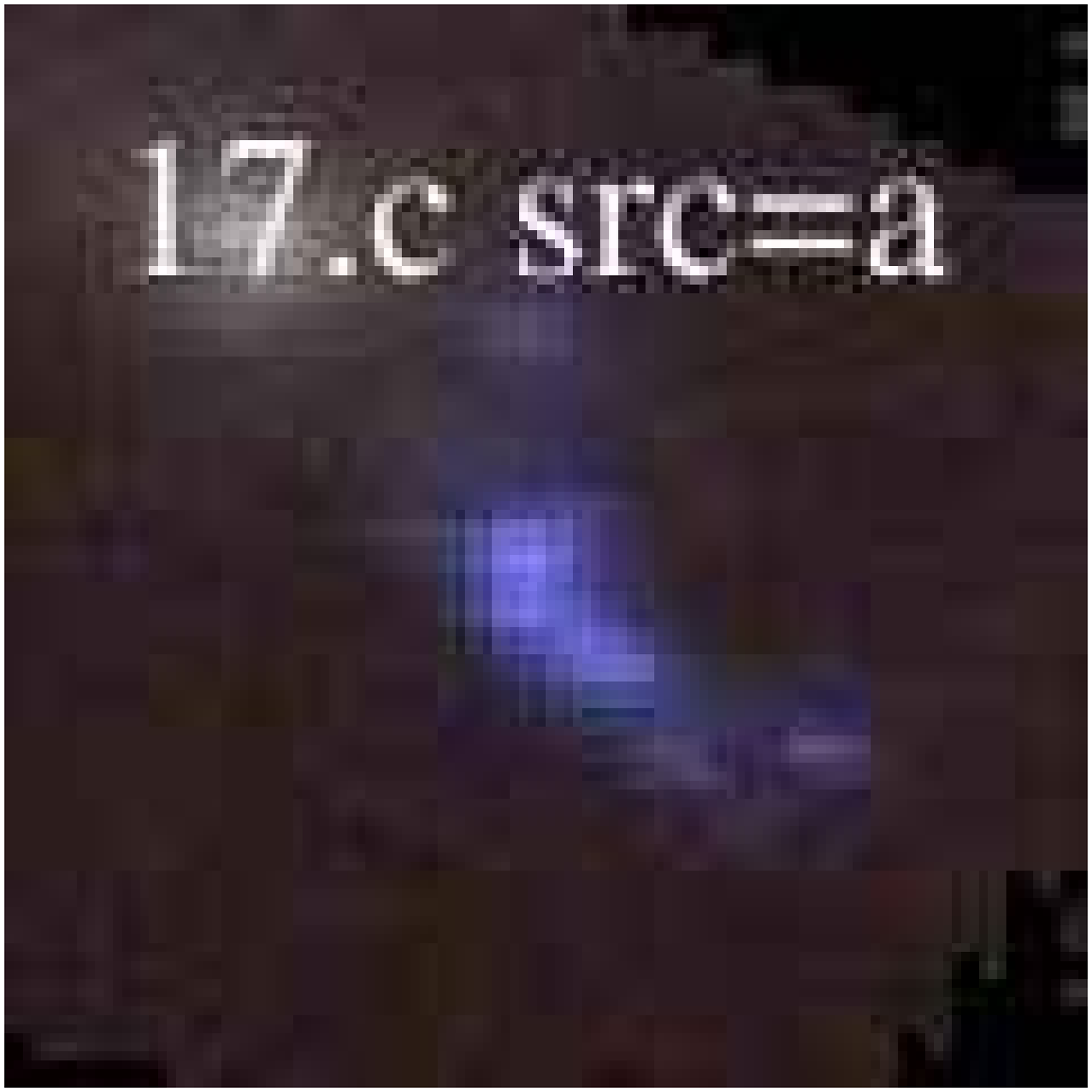}} \\
  \end{tabular}

\end{table*}

\begin{table*}
  \caption{Image system 18:}\vspace{0mm}
  \begin{tabular}{cccc}
    \multicolumn{1}{m{1cm}}{{\Large A1689}}
    & \multicolumn{1}{m{1.7cm}}{\includegraphics[height=2.00cm,clip]{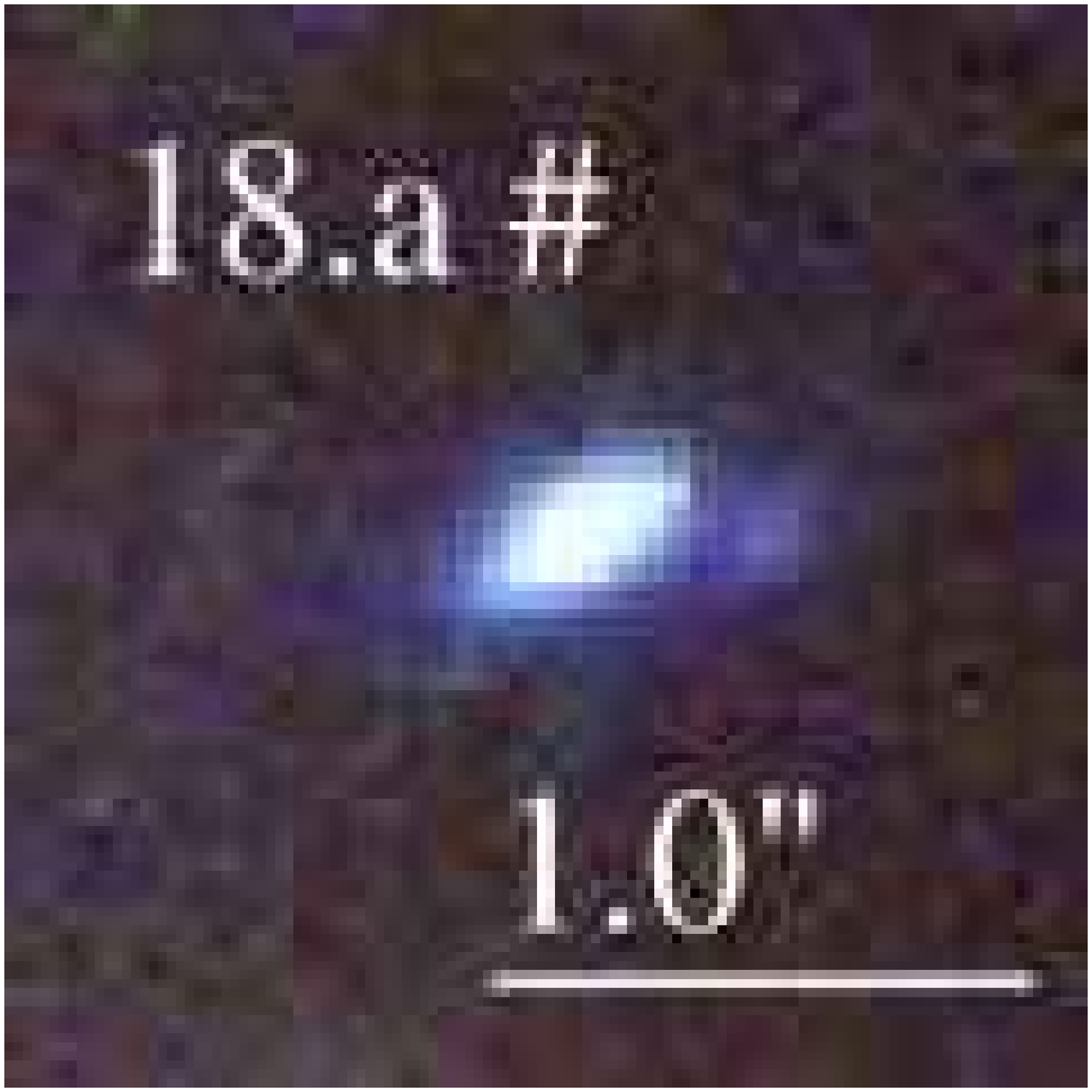}}
    & \multicolumn{1}{m{1.7cm}}{\includegraphics[height=2.00cm,clip]{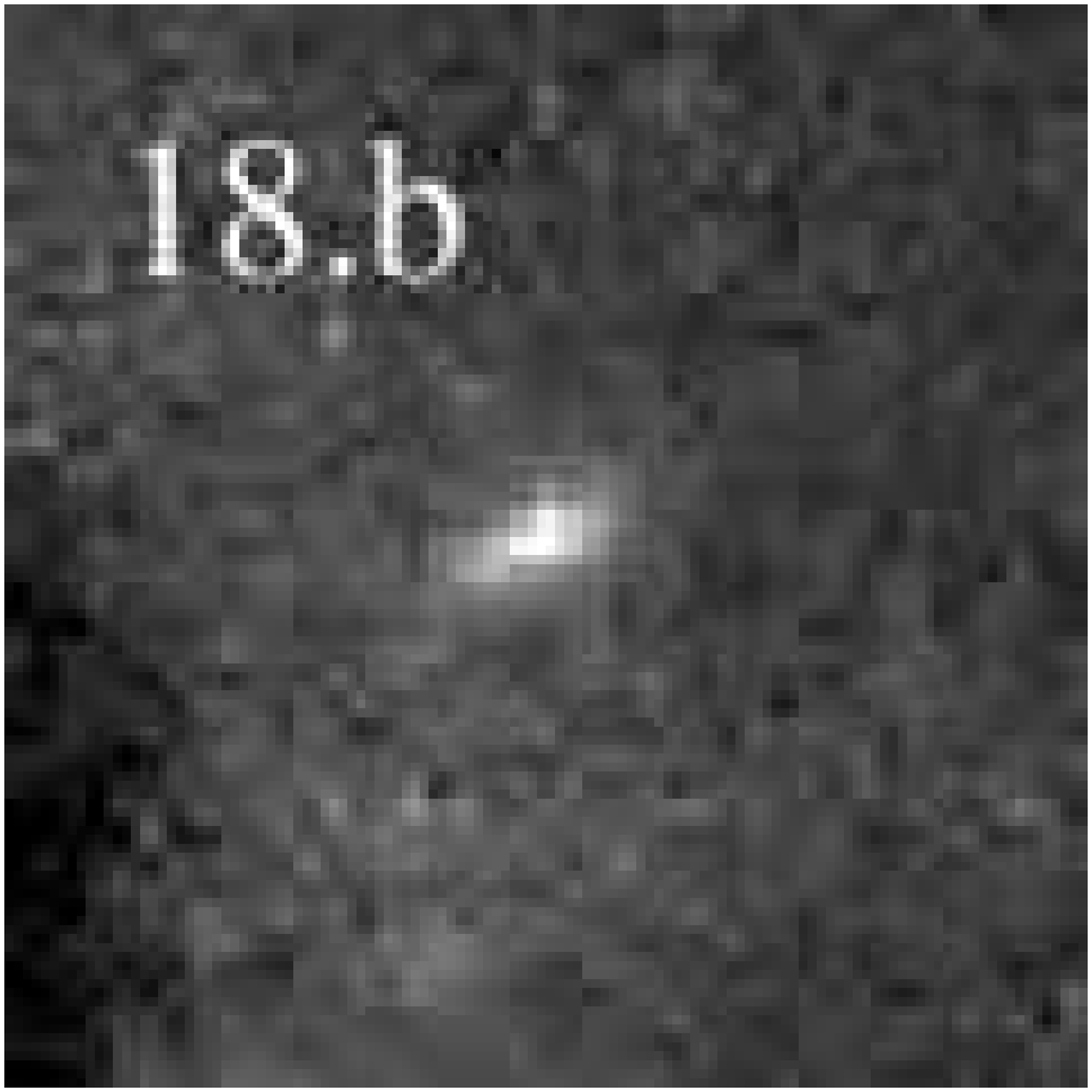}}
    & \multicolumn{1}{m{1.7cm}}{\includegraphics[height=2.00cm,clip]{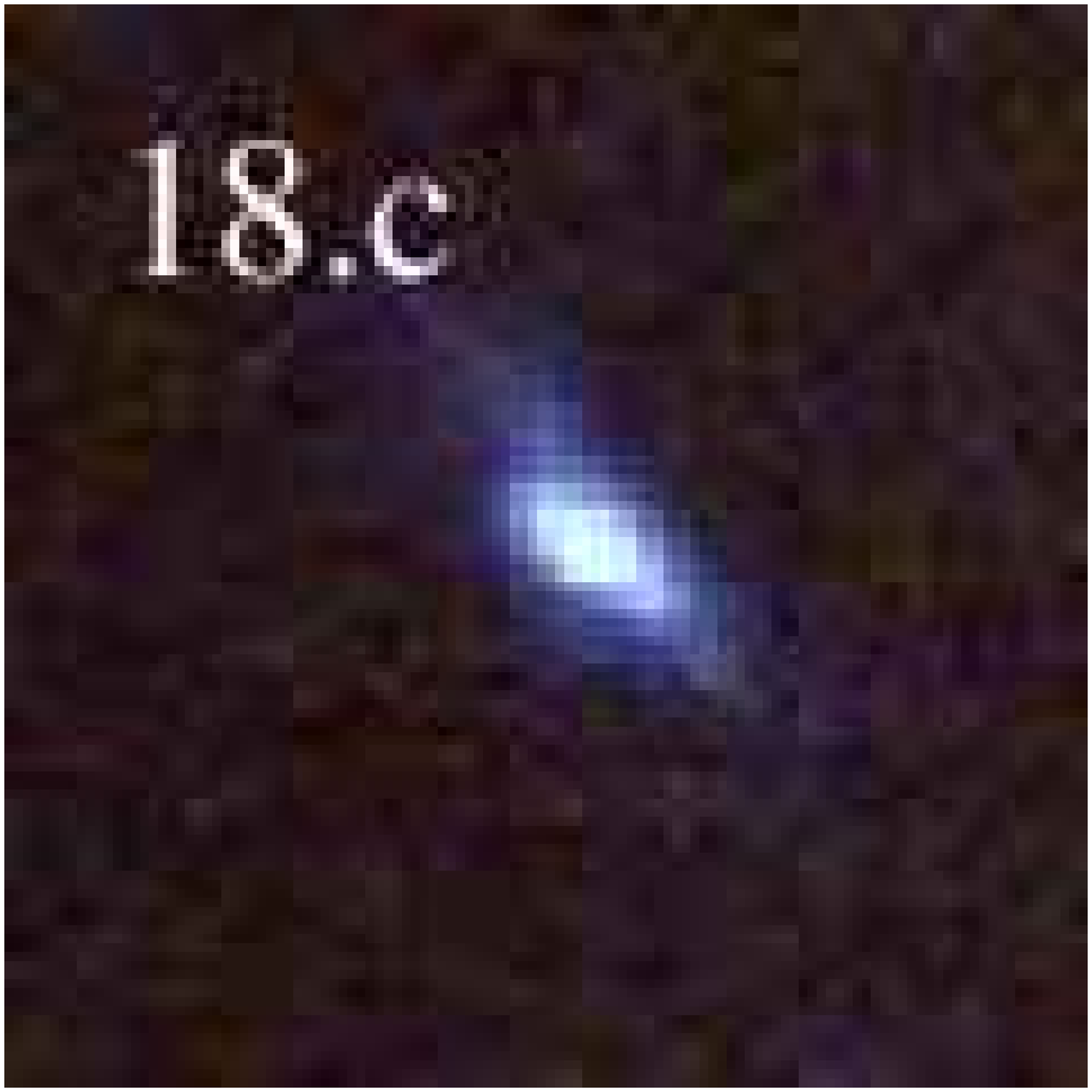}} \\
    \multicolumn{1}{m{1cm}}{{\Large NSIE}}
    & \multicolumn{1}{m{1.7cm}}{\includegraphics[height=2.00cm,clip]{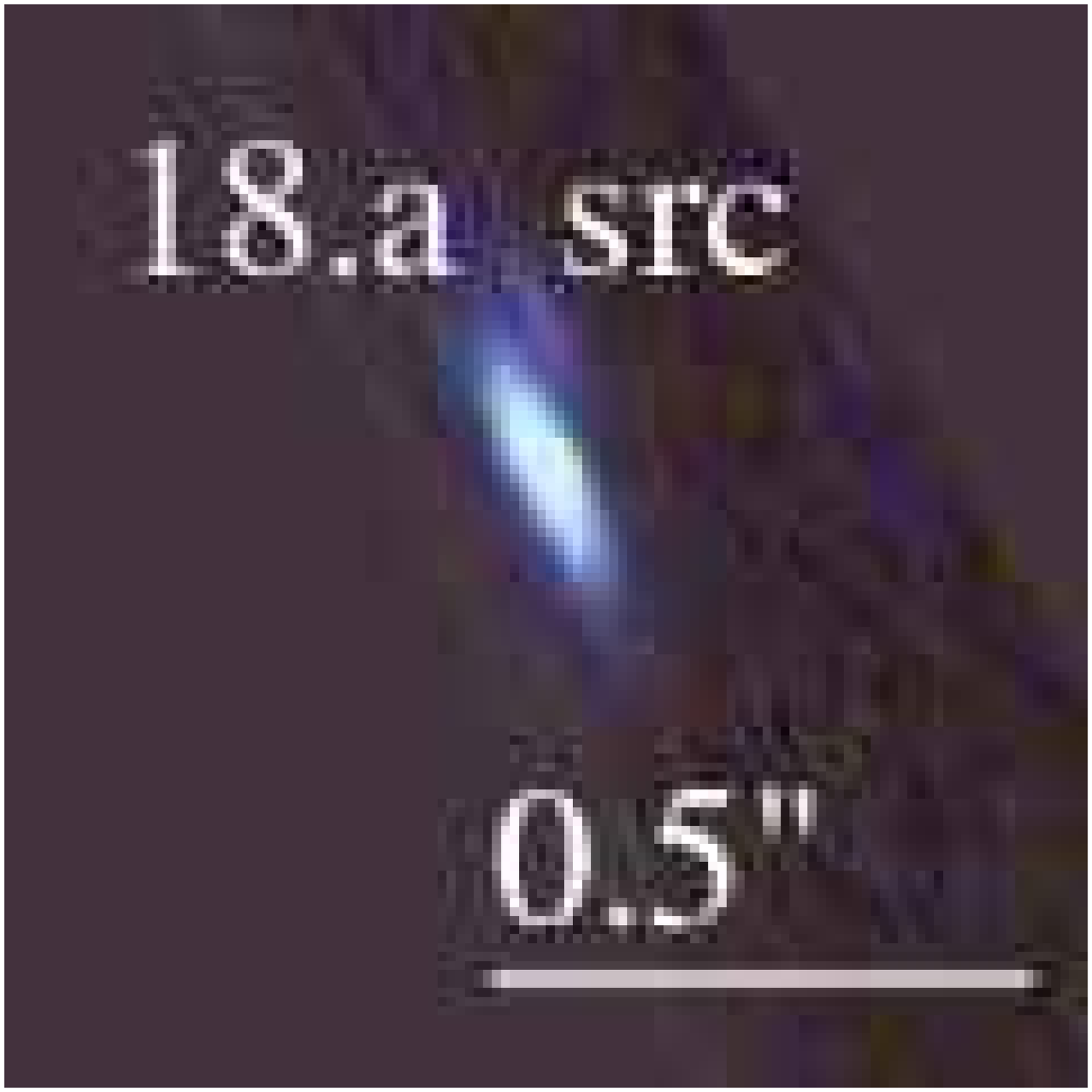}}
    & \multicolumn{1}{m{1.7cm}}{\includegraphics[height=2.00cm,clip]{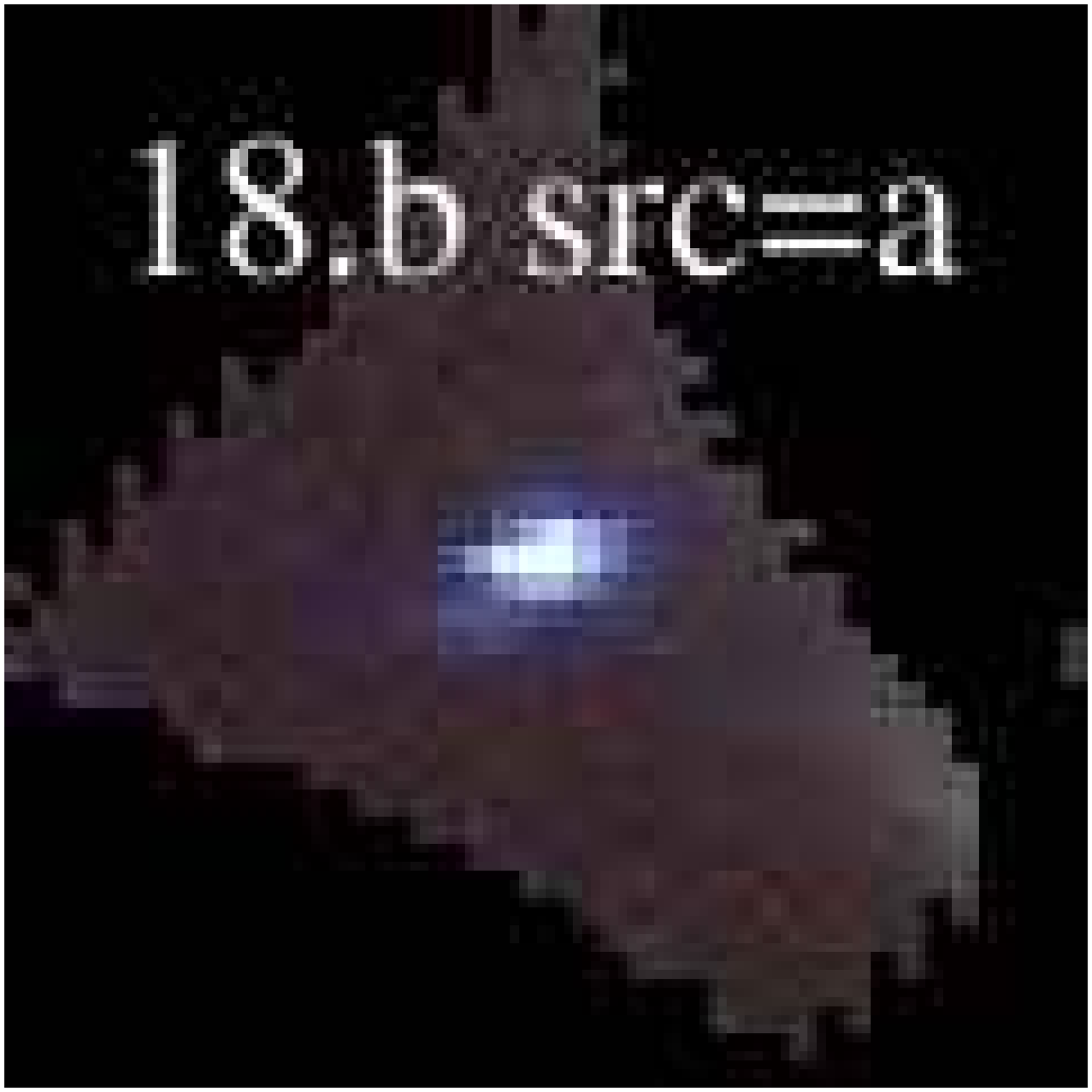}}
    & \multicolumn{1}{m{1.7cm}}{\includegraphics[height=2.00cm,clip]{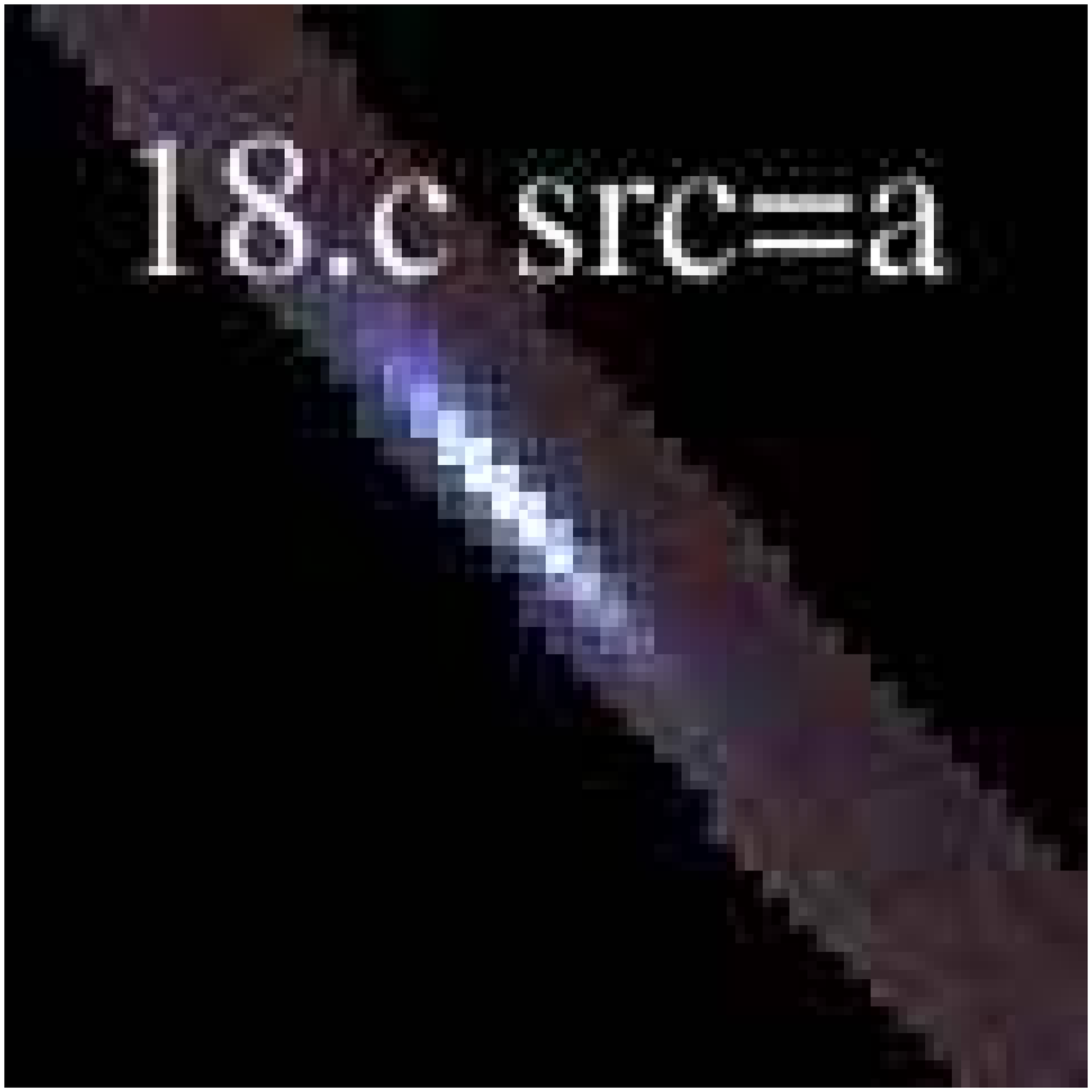}} \\
    \multicolumn{1}{m{1cm}}{{\Large ENFW}}
    & \multicolumn{1}{m{1.7cm}}{\includegraphics[height=2.00cm,clip]{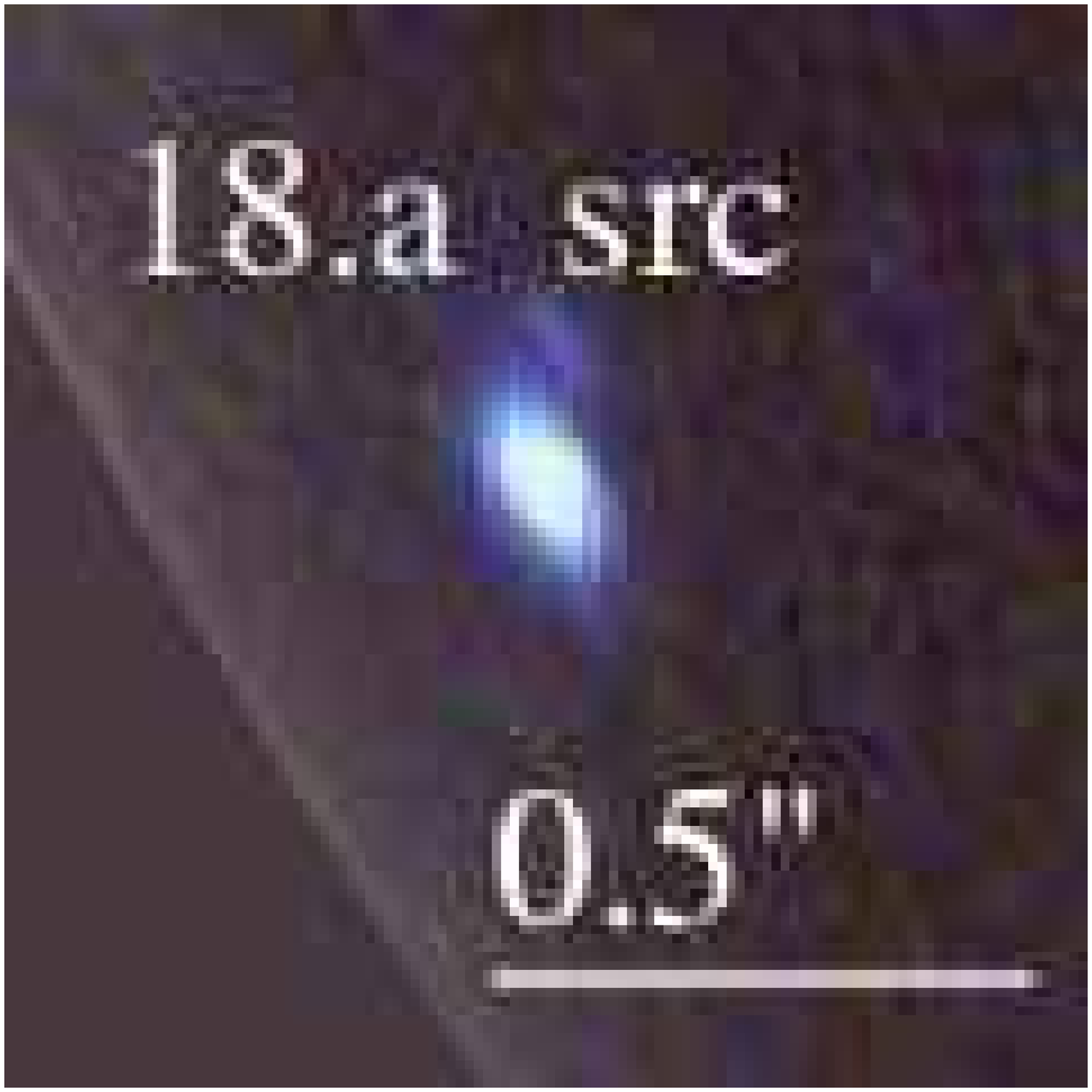}}
    & \multicolumn{1}{m{1.7cm}}{\includegraphics[height=2.00cm,clip]{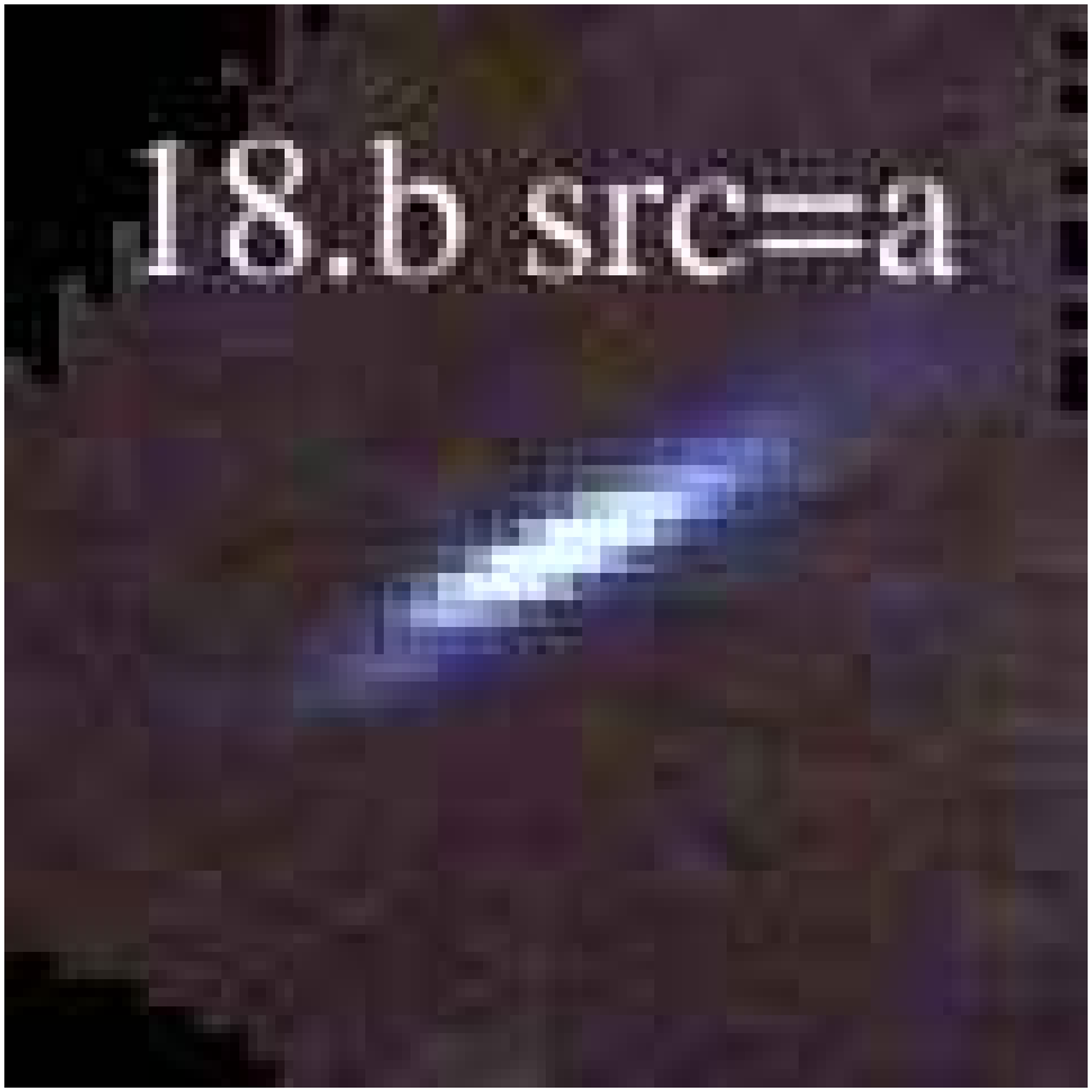}}
    & \multicolumn{1}{m{1.7cm}}{\includegraphics[height=2.00cm,clip]{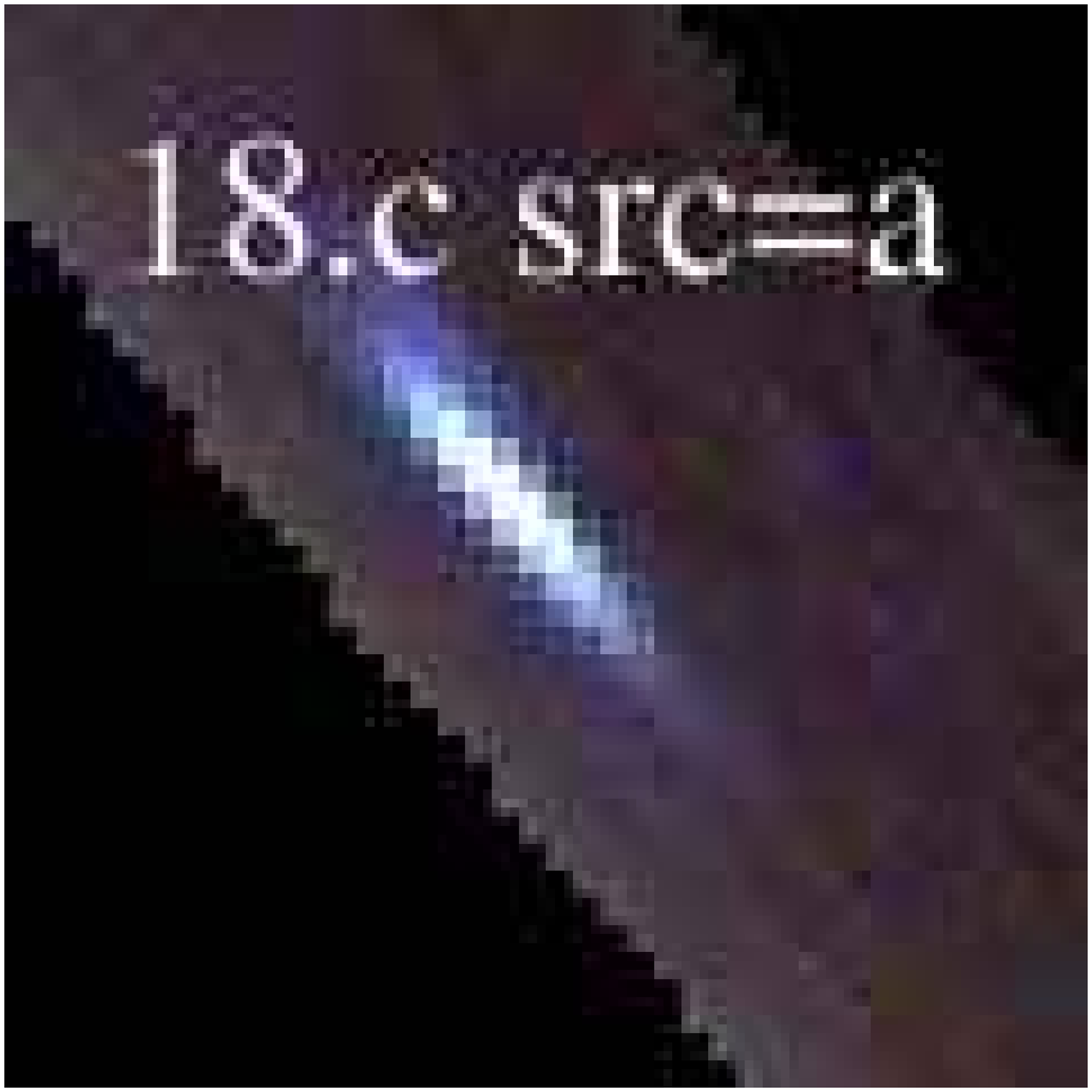}} \\
  \end{tabular}

\end{table*}

\clearpage

\begin{table*}
  \caption{Image system 19:}\vspace{0mm}
  \begin{tabular}{cccc}
    \multicolumn{1}{m{1cm}}{{\Large A1689}}
    & \multicolumn{1}{m{1.7cm}}{\includegraphics[height=2.00cm,clip]{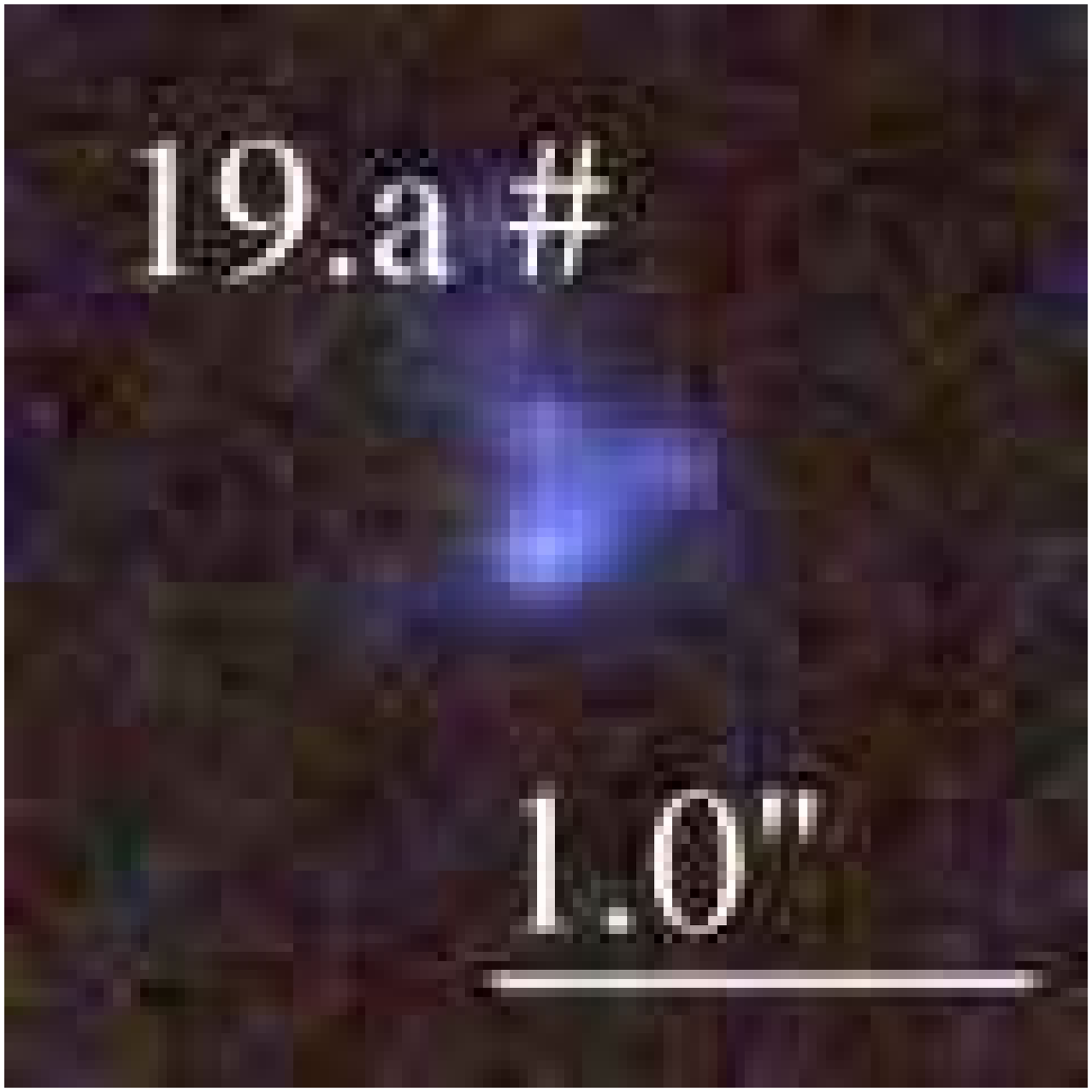}}
    & \multicolumn{1}{m{1.7cm}}{\includegraphics[height=2.00cm,clip]{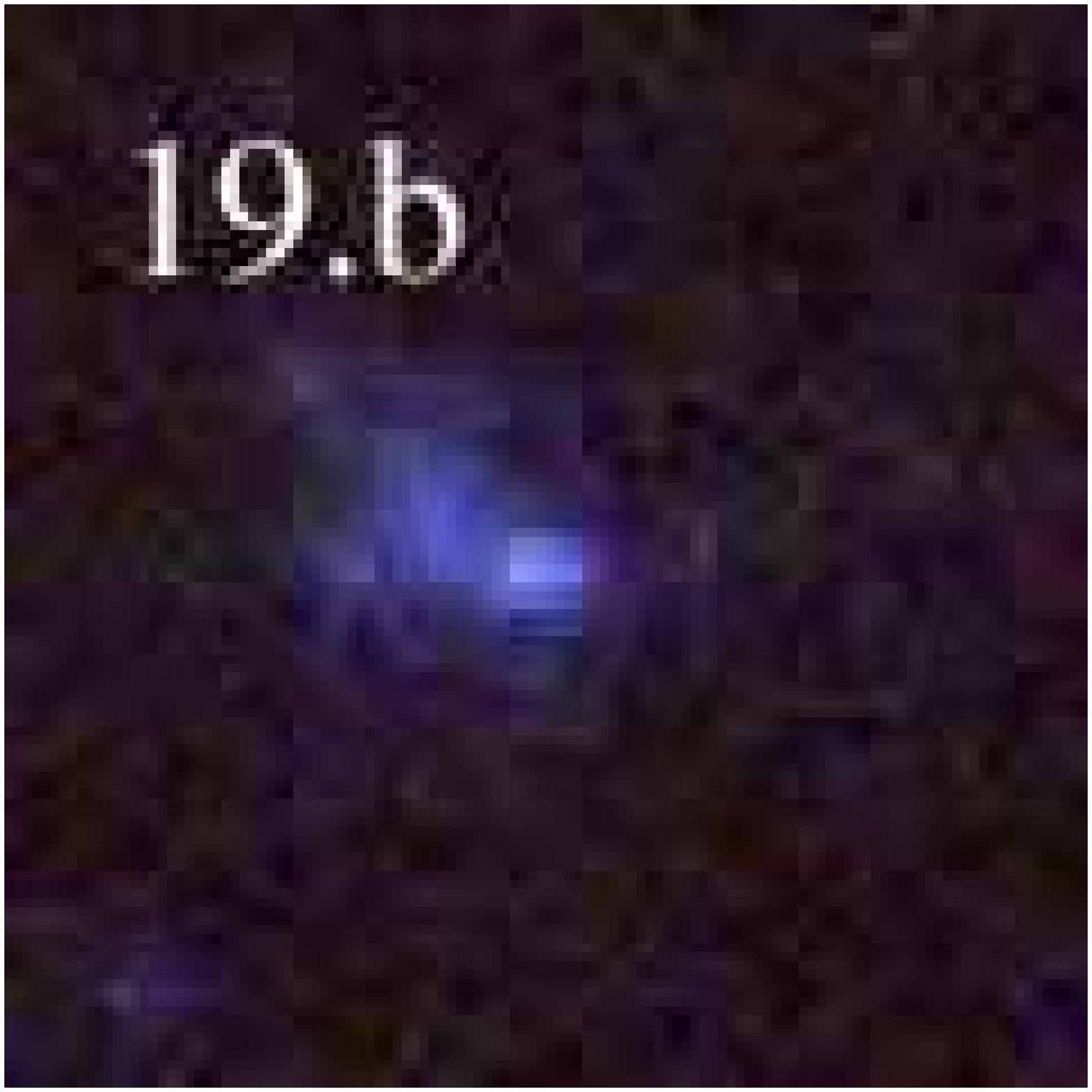}}
    & \multicolumn{1}{m{1.7cm}}{\includegraphics[height=2.00cm,clip]{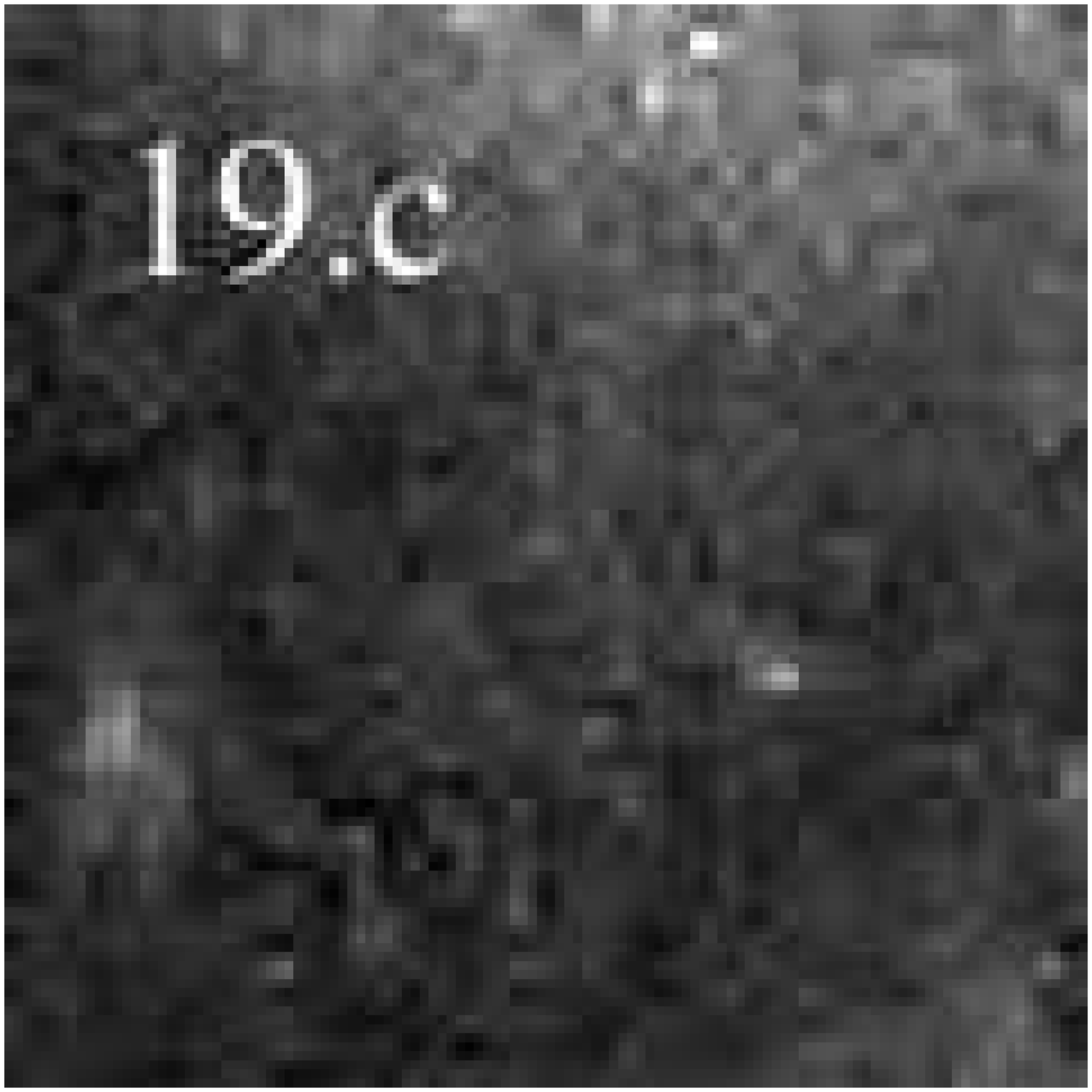}}\\
    \multicolumn{1}{m{1cm}}{{\Large NSIE}}
    & \multicolumn{1}{m{1.7cm}}{\includegraphics[height=2.00cm,clip]{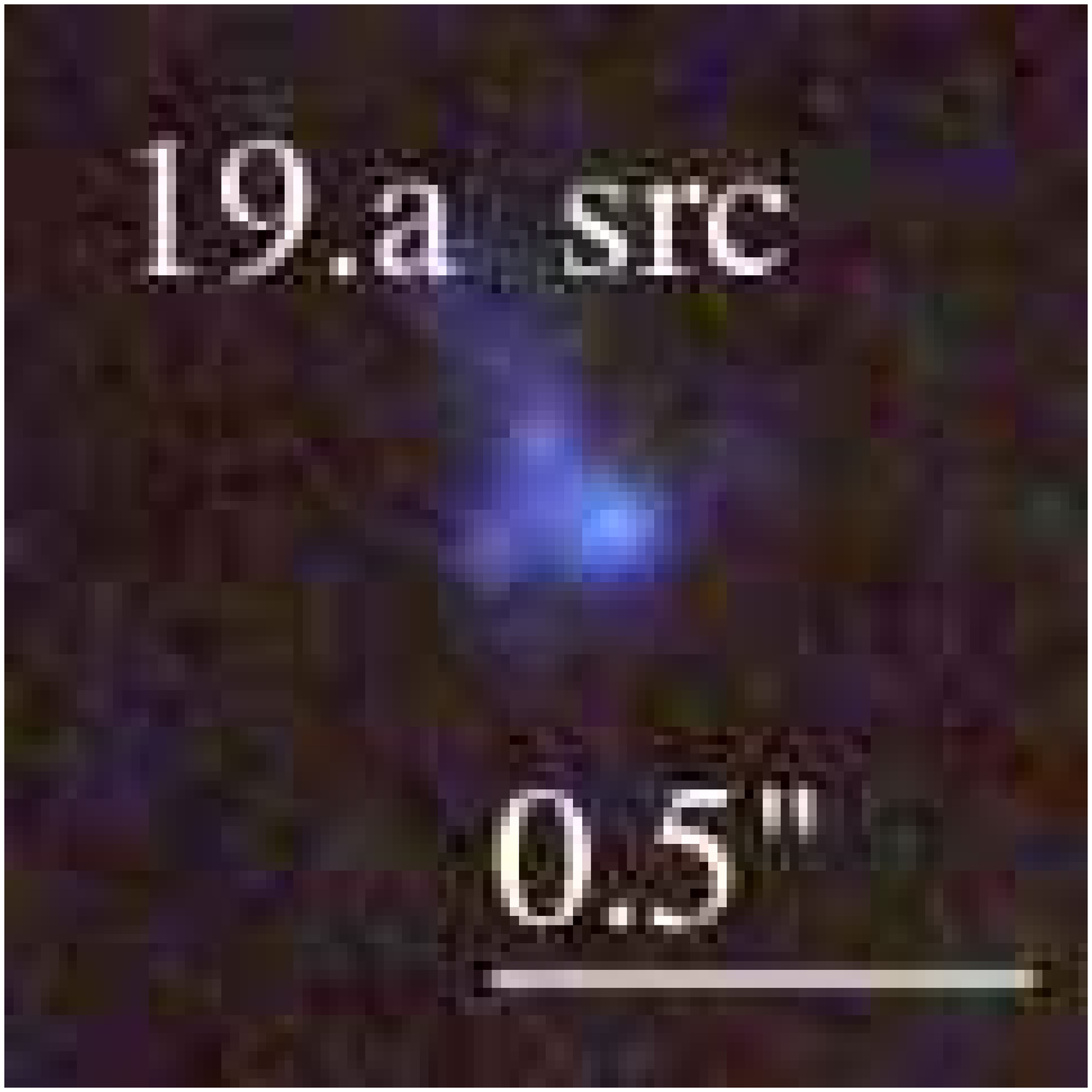}}
    & \multicolumn{1}{m{1.7cm}}{\includegraphics[height=2.00cm,clip]{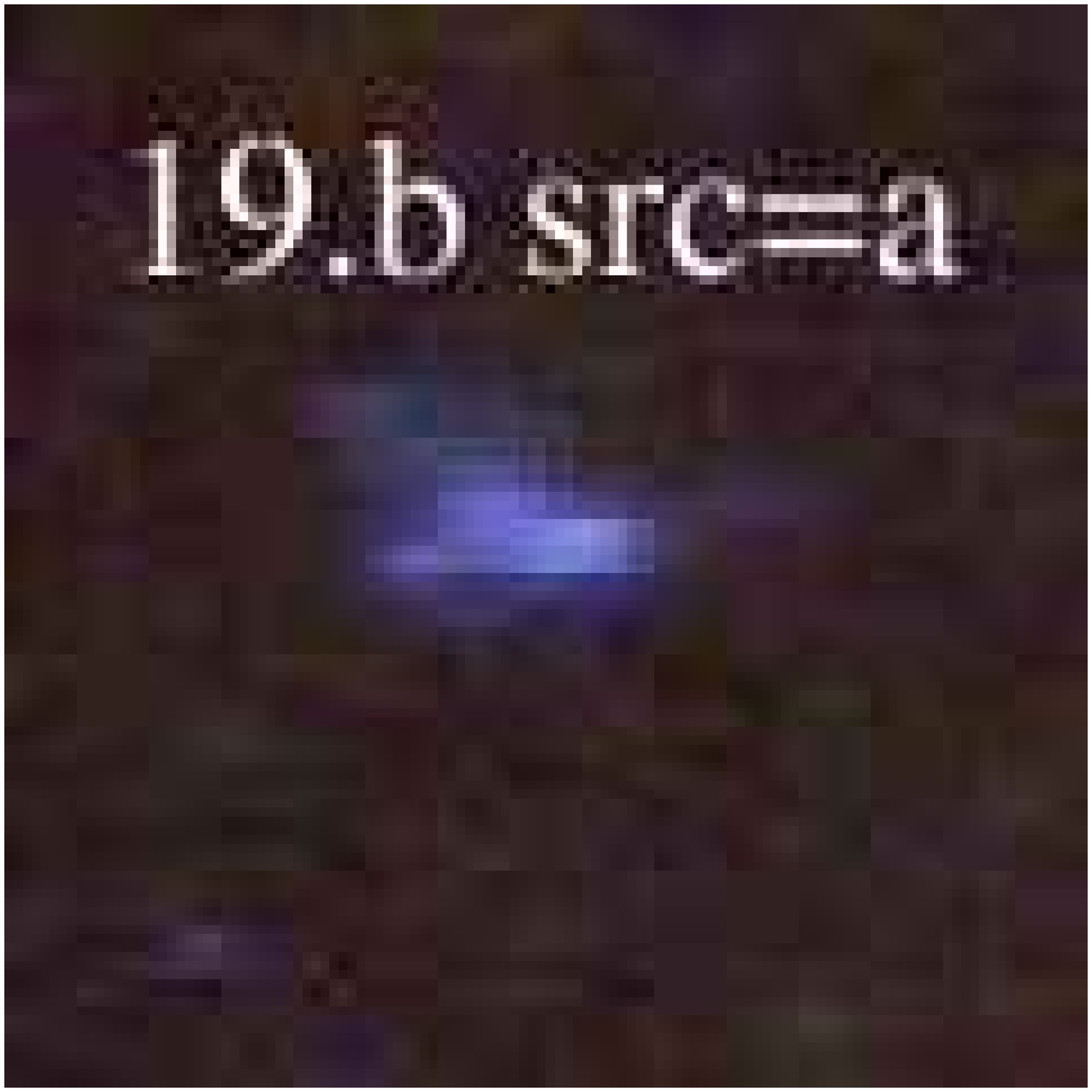}}
    & \multicolumn{1}{m{1.7cm}}{\includegraphics[height=2.00cm,clip]{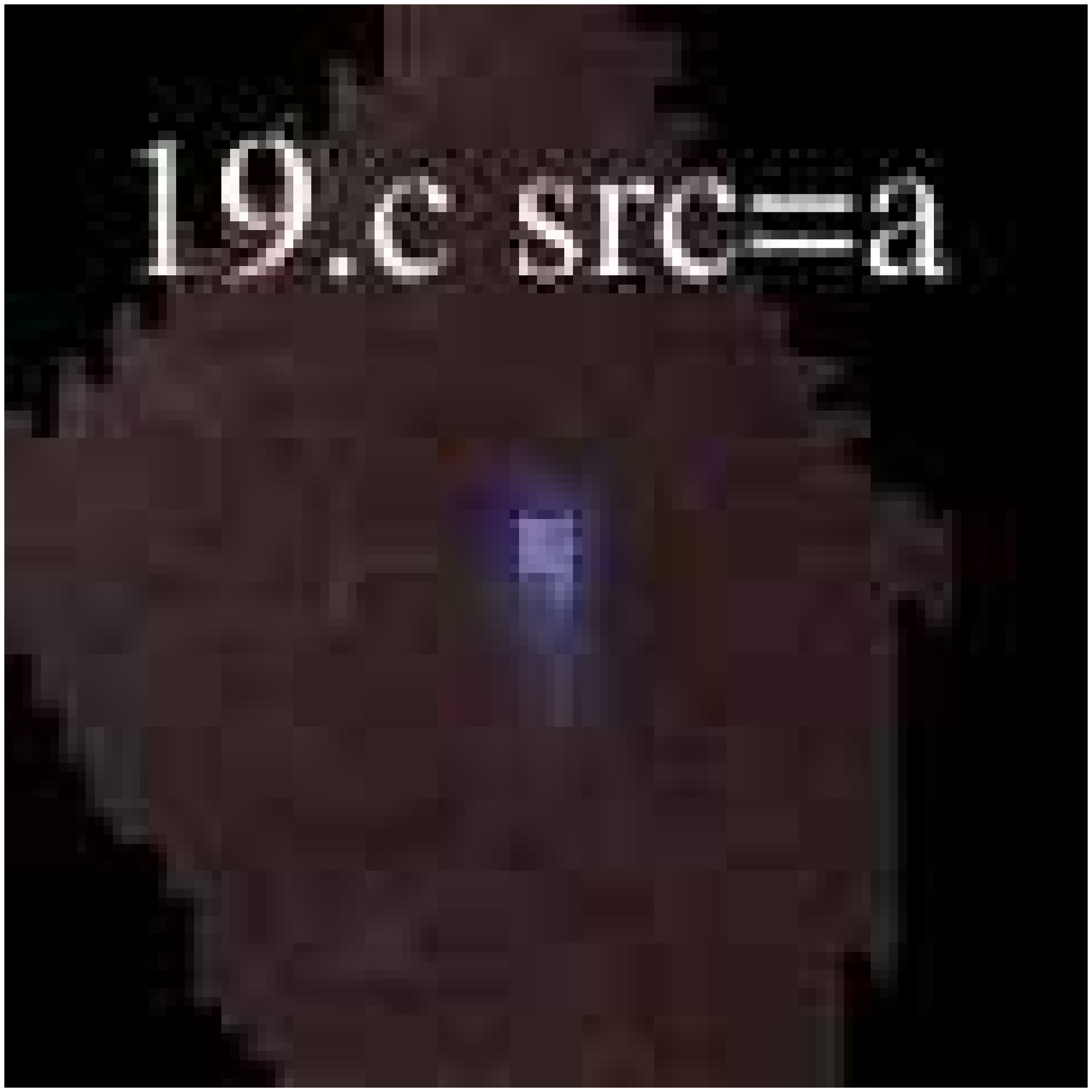}} \\
    \multicolumn{1}{m{1cm}}{{\Large ENFW}}
    & \multicolumn{1}{m{1.7cm}}{\includegraphics[height=2.00cm,clip]{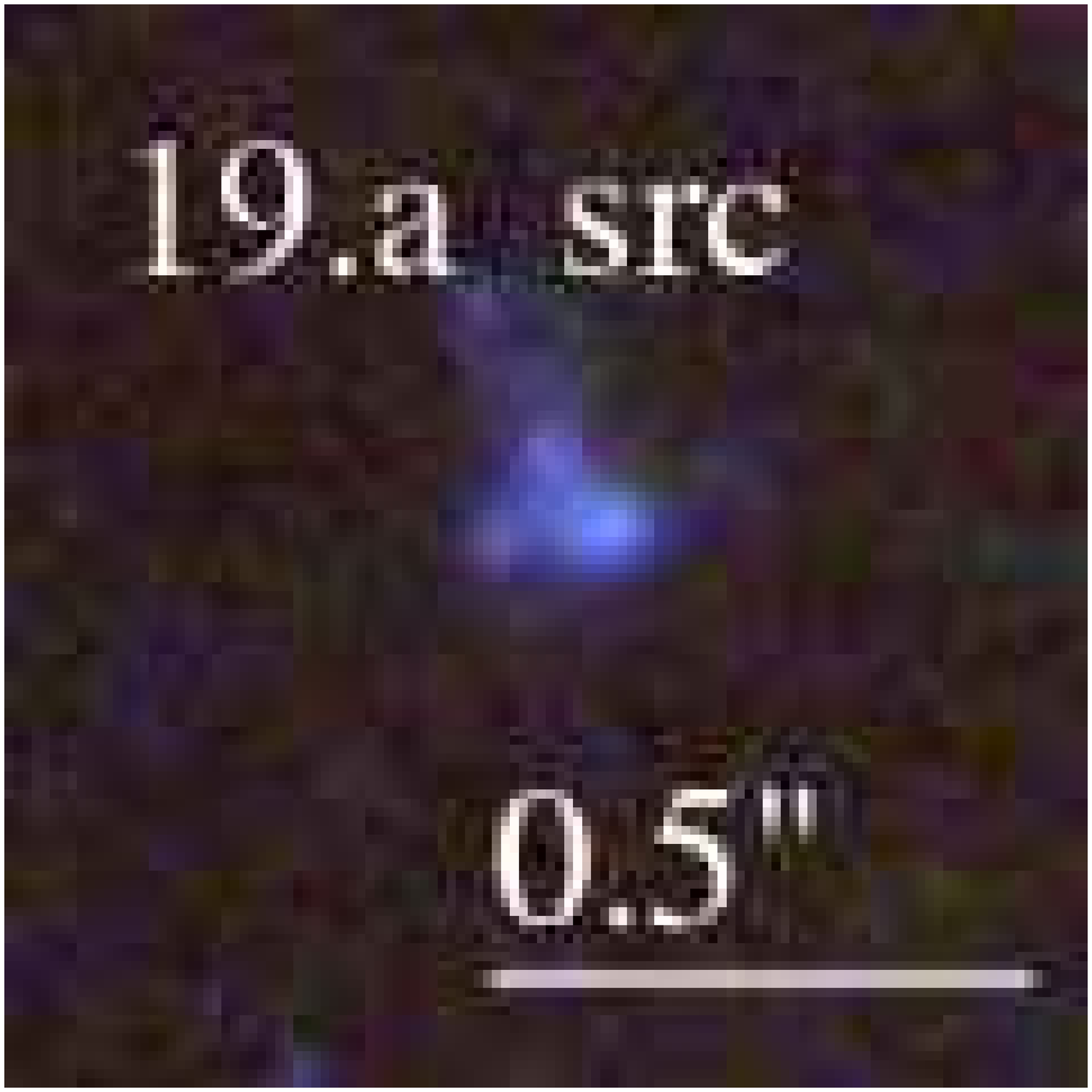}}
    & \multicolumn{1}{m{1.7cm}}{\includegraphics[height=2.00cm,clip]{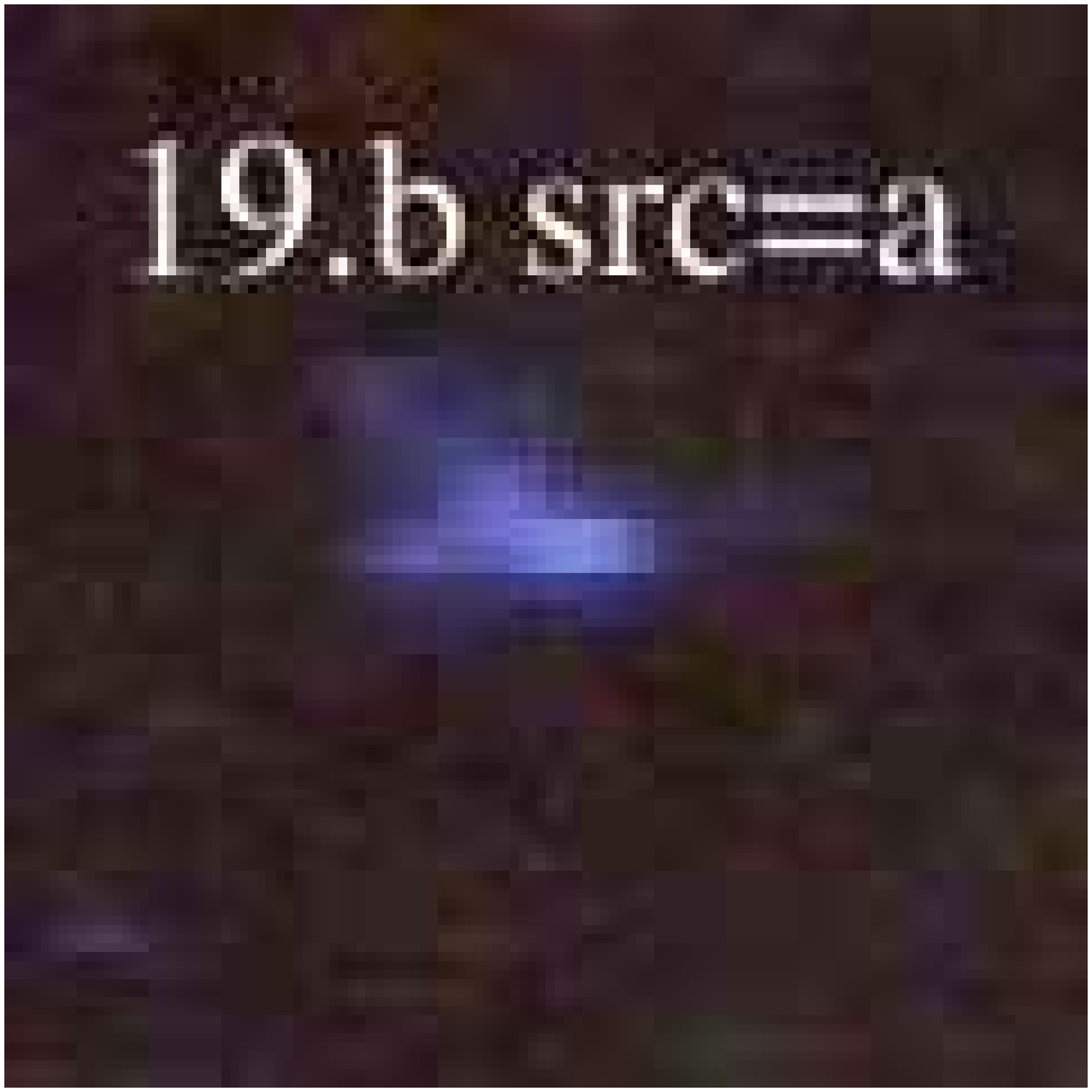}}
    & \multicolumn{1}{m{1.7cm}}{\includegraphics[height=2.00cm,clip]{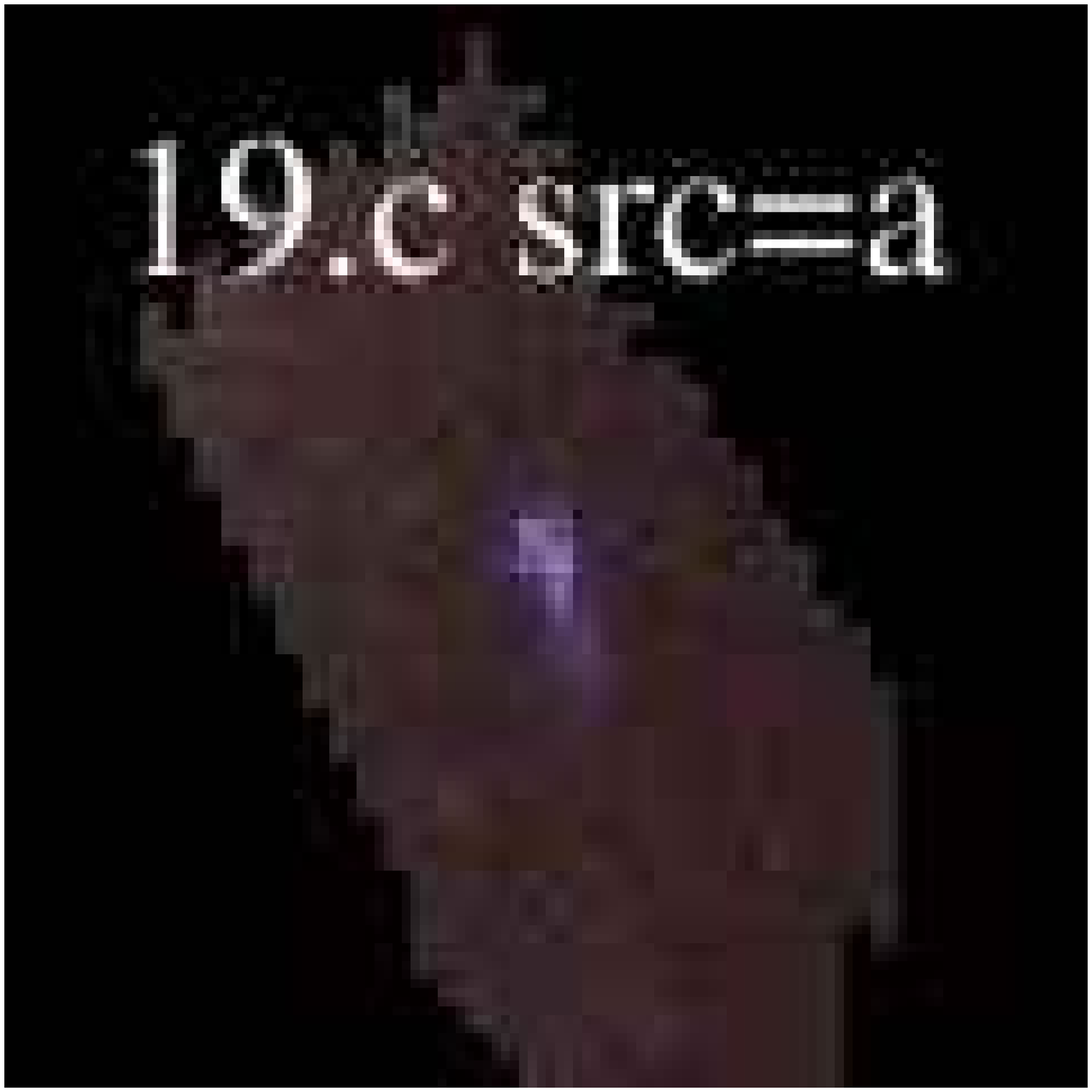}} \\
  \end{tabular}

\end{table*}

\begin{table*}
  \caption{Image system 20:}\vspace{0mm}
  \begin{tabular}{cccccc}
    \multicolumn{1}{m{1cm}}{{\Large A1689}}
    & \multicolumn{1}{m{1.7cm}}{\includegraphics[height=2.00cm,clip]{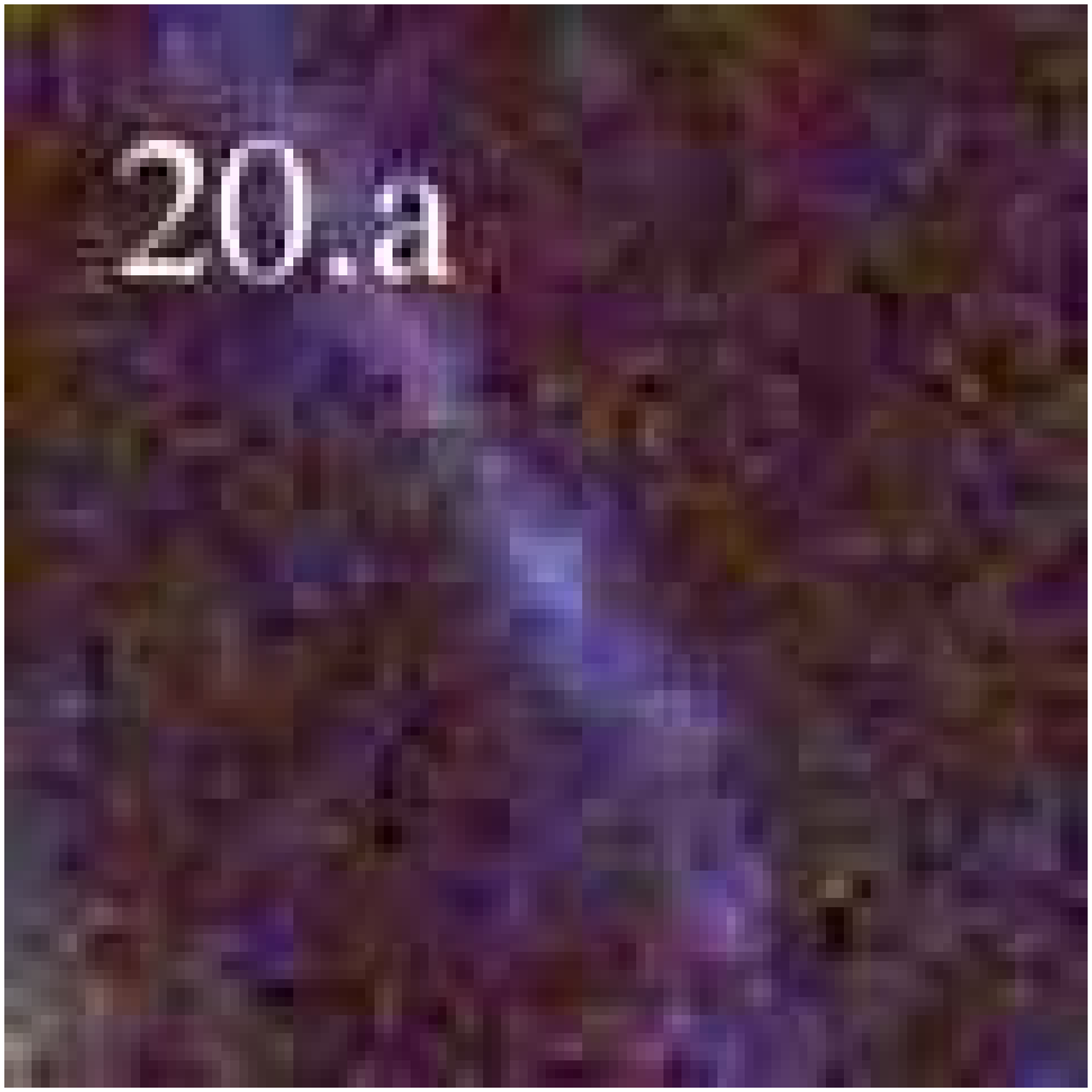}}
    & \multicolumn{1}{m{1.7cm}}{\includegraphics[height=2.00cm,clip]{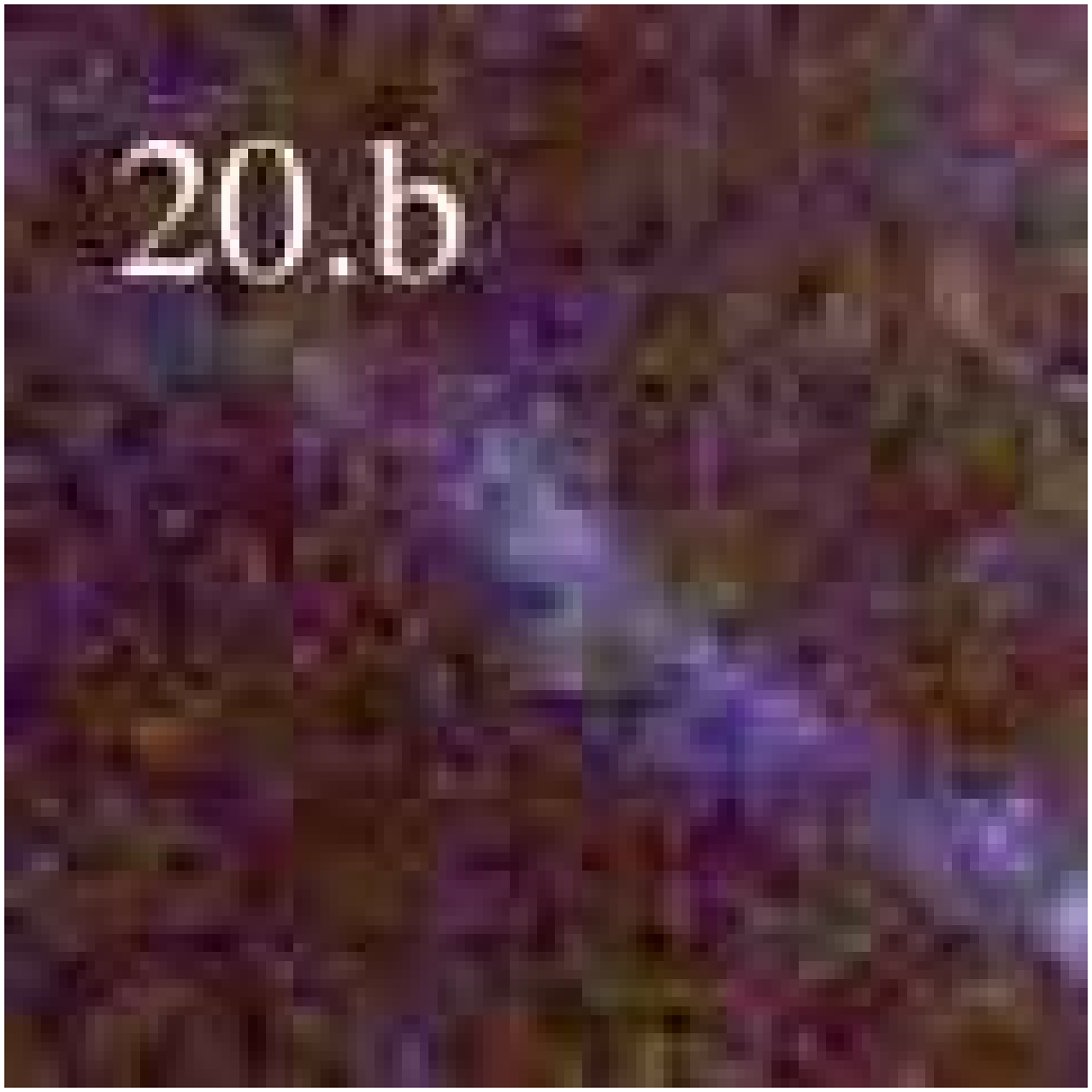}}
    & \multicolumn{1}{m{1.7cm}}{\includegraphics[height=2.00cm,clip]{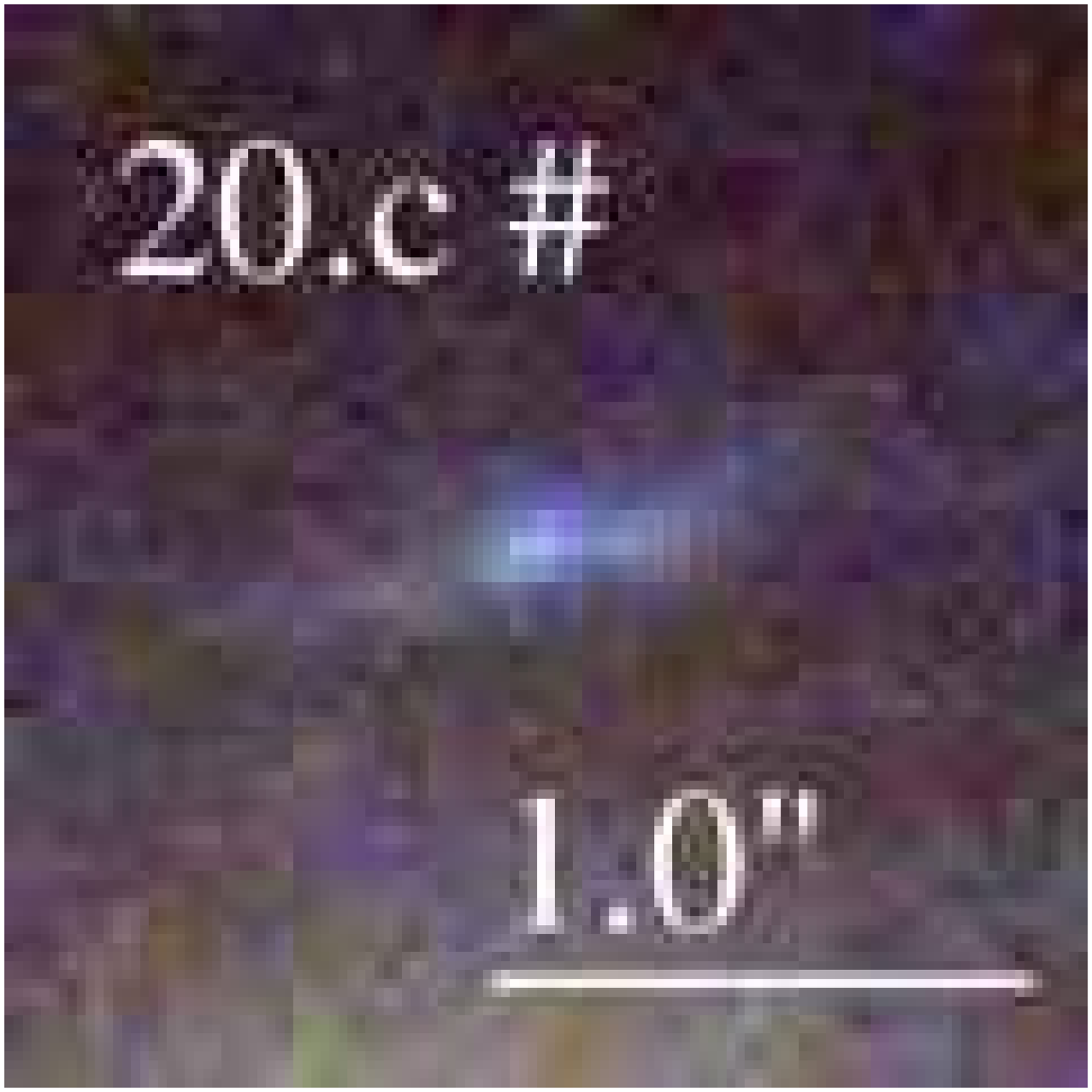}}
    & \multicolumn{1}{m{1.7cm}}{\includegraphics[height=2.00cm,clip]{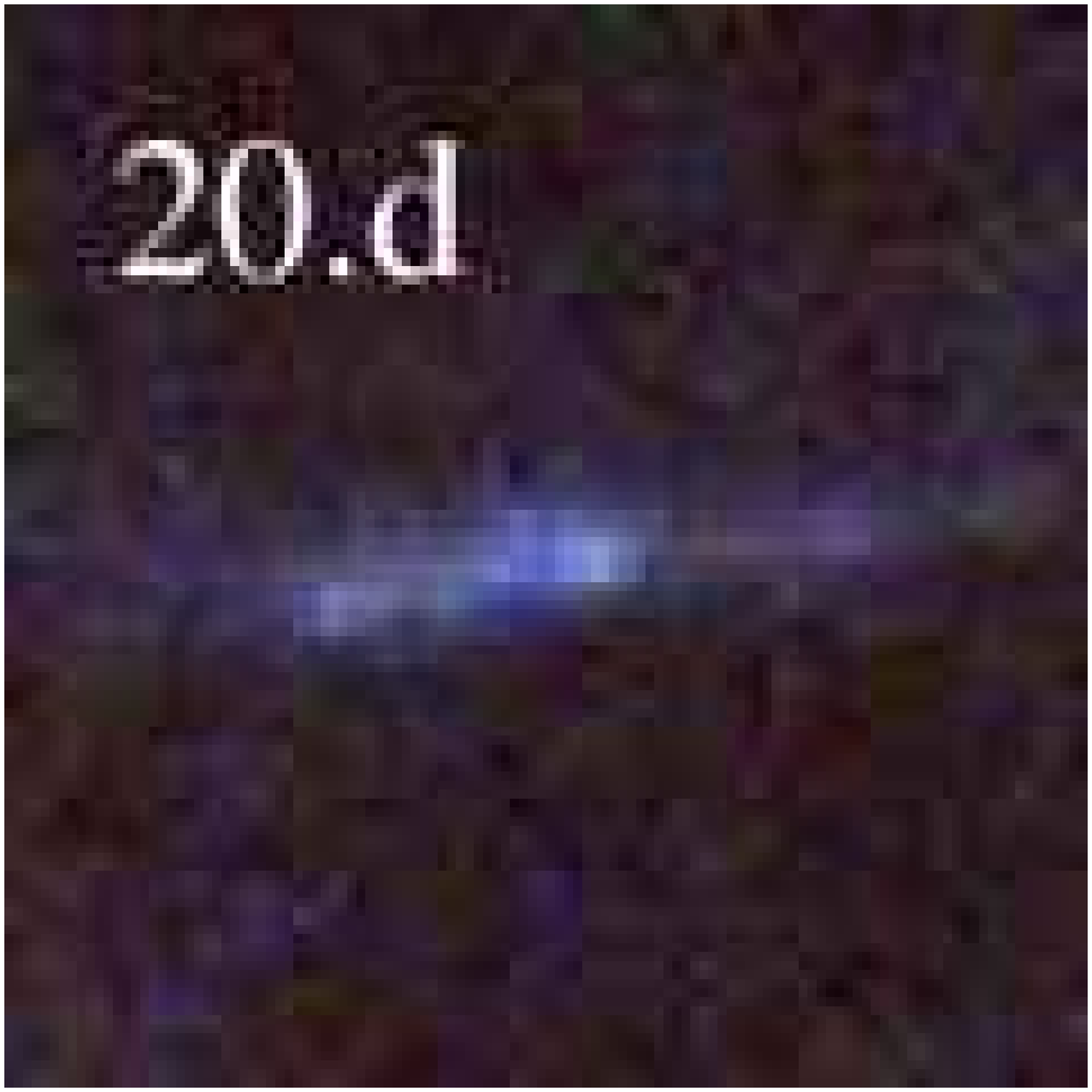}}
    & \multicolumn{1}{m{1.7cm}}{\includegraphics[height=2.00cm,clip]{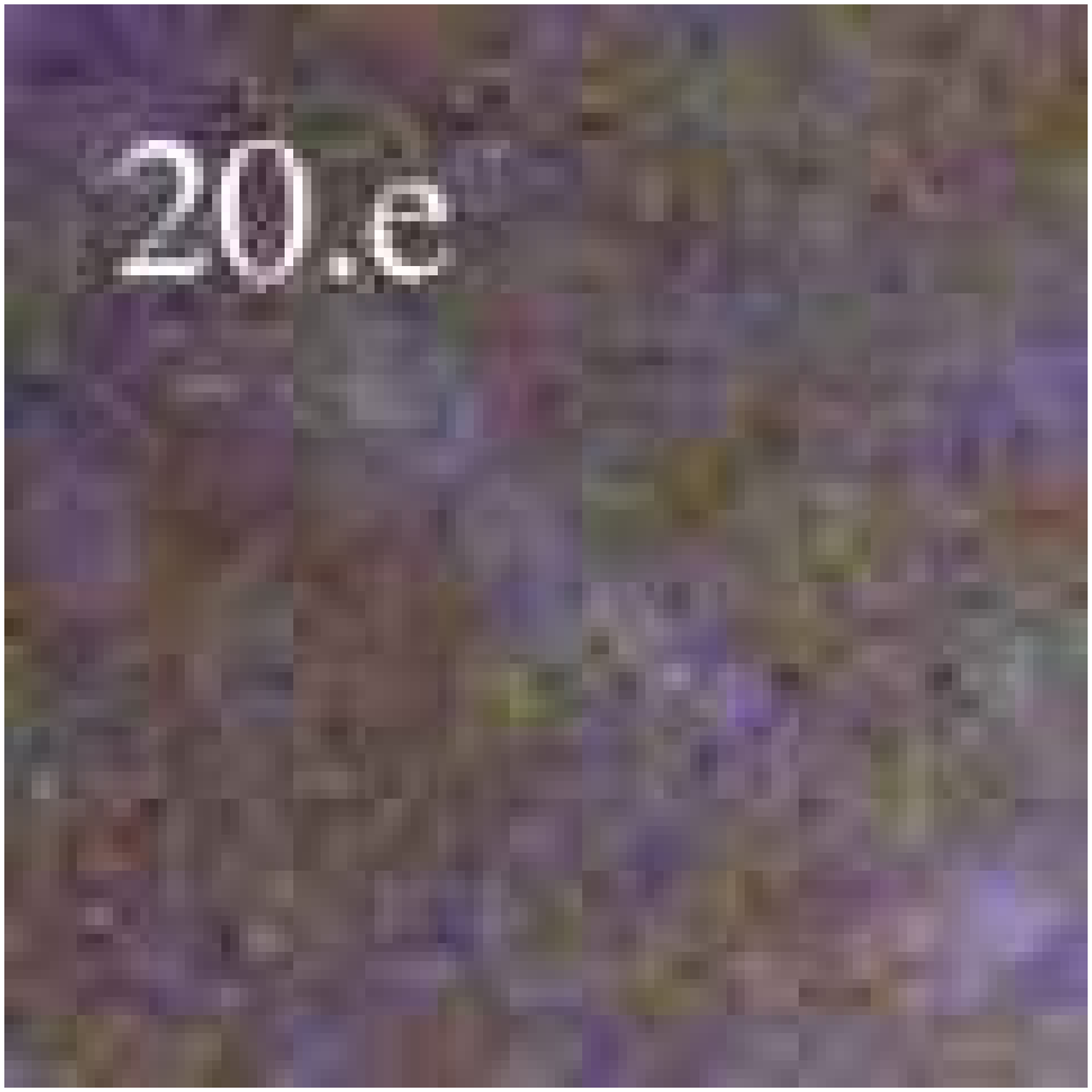}} \\
    \multicolumn{1}{m{1cm}}{{\Large NSIE}}
    & \multicolumn{1}{m{1.7cm}}{\includegraphics[height=2.00cm,clip]{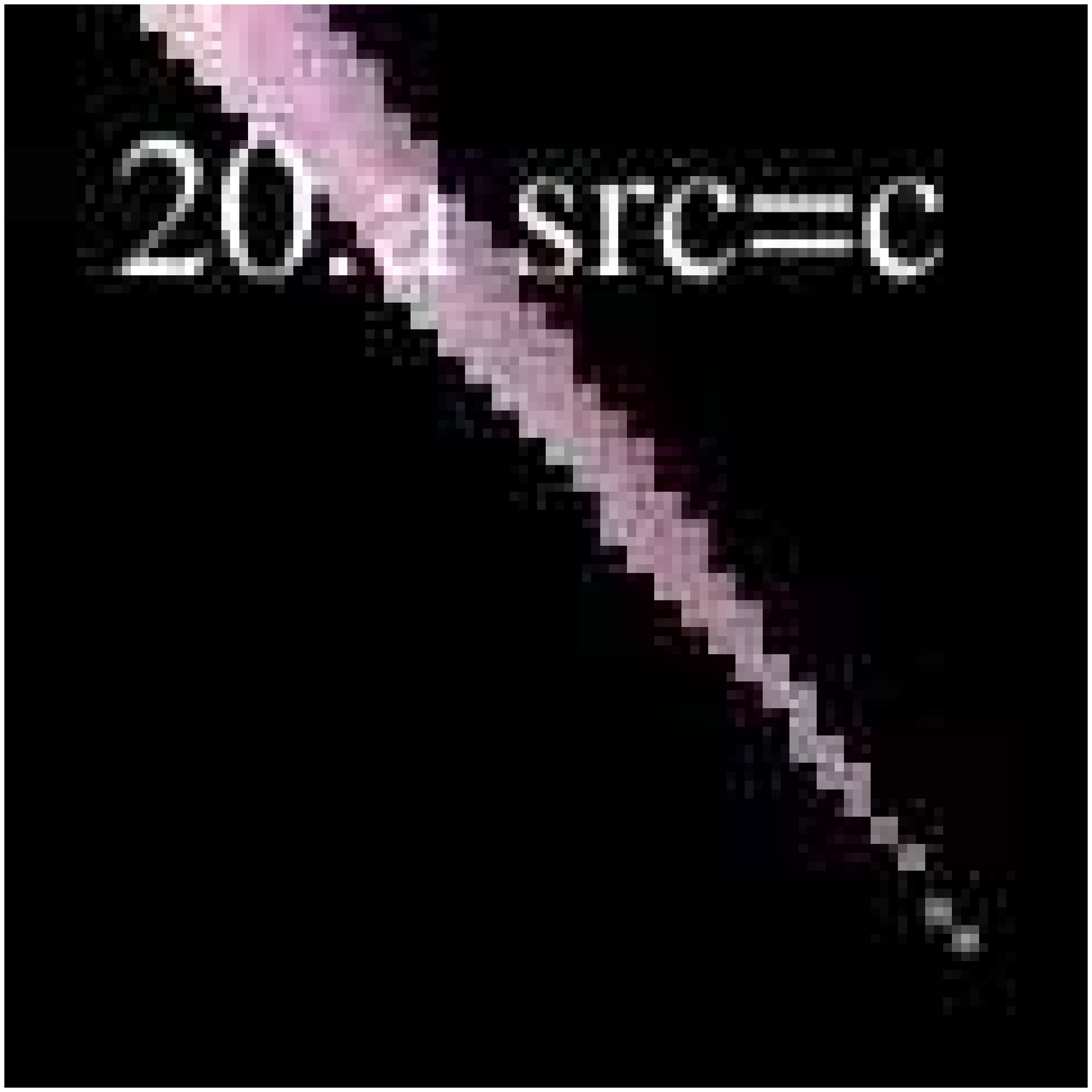}}
    & \multicolumn{1}{m{1.7cm}}{\includegraphics[height=2.00cm,clip]{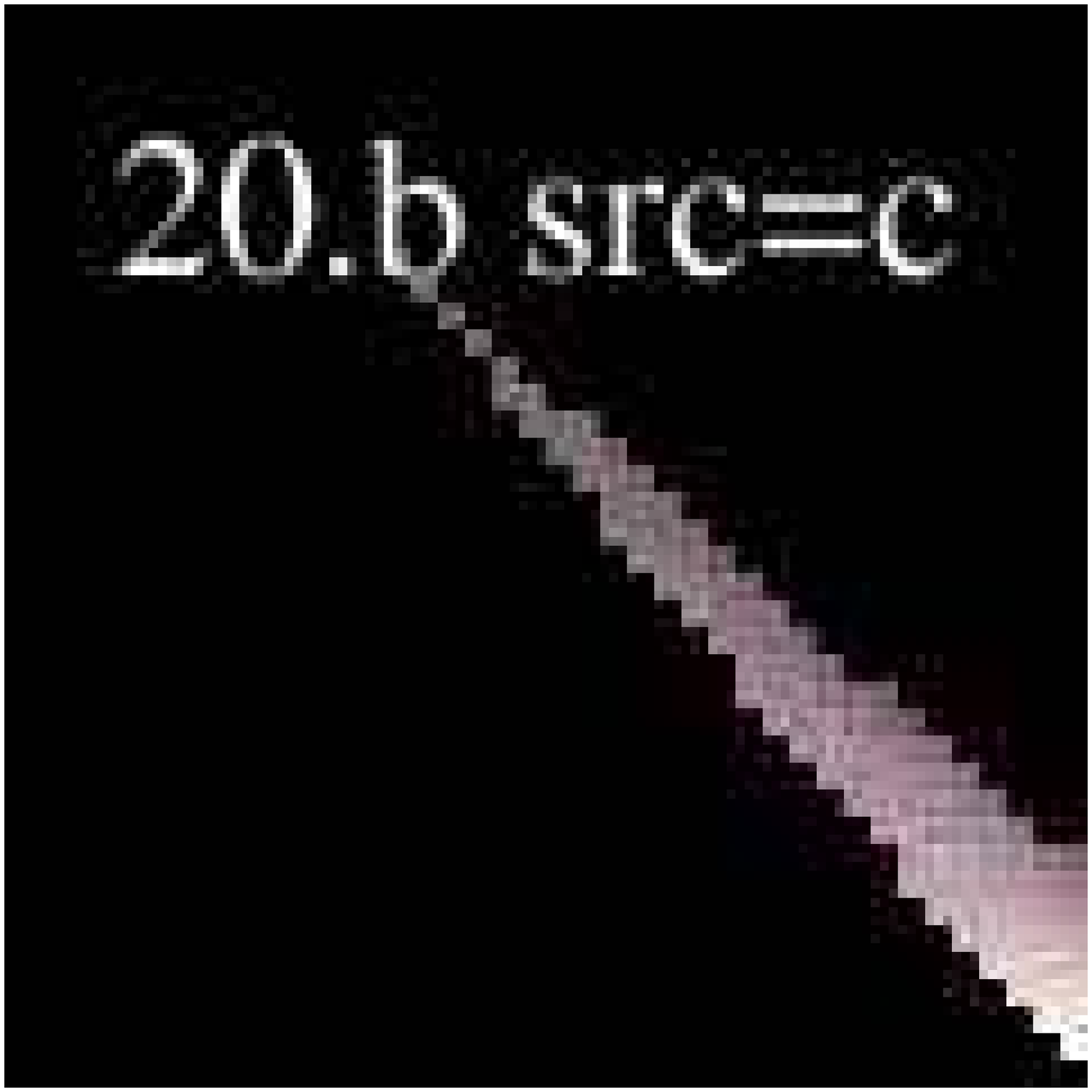}}
    & \multicolumn{1}{m{1.7cm}}{\includegraphics[height=2.00cm,clip]{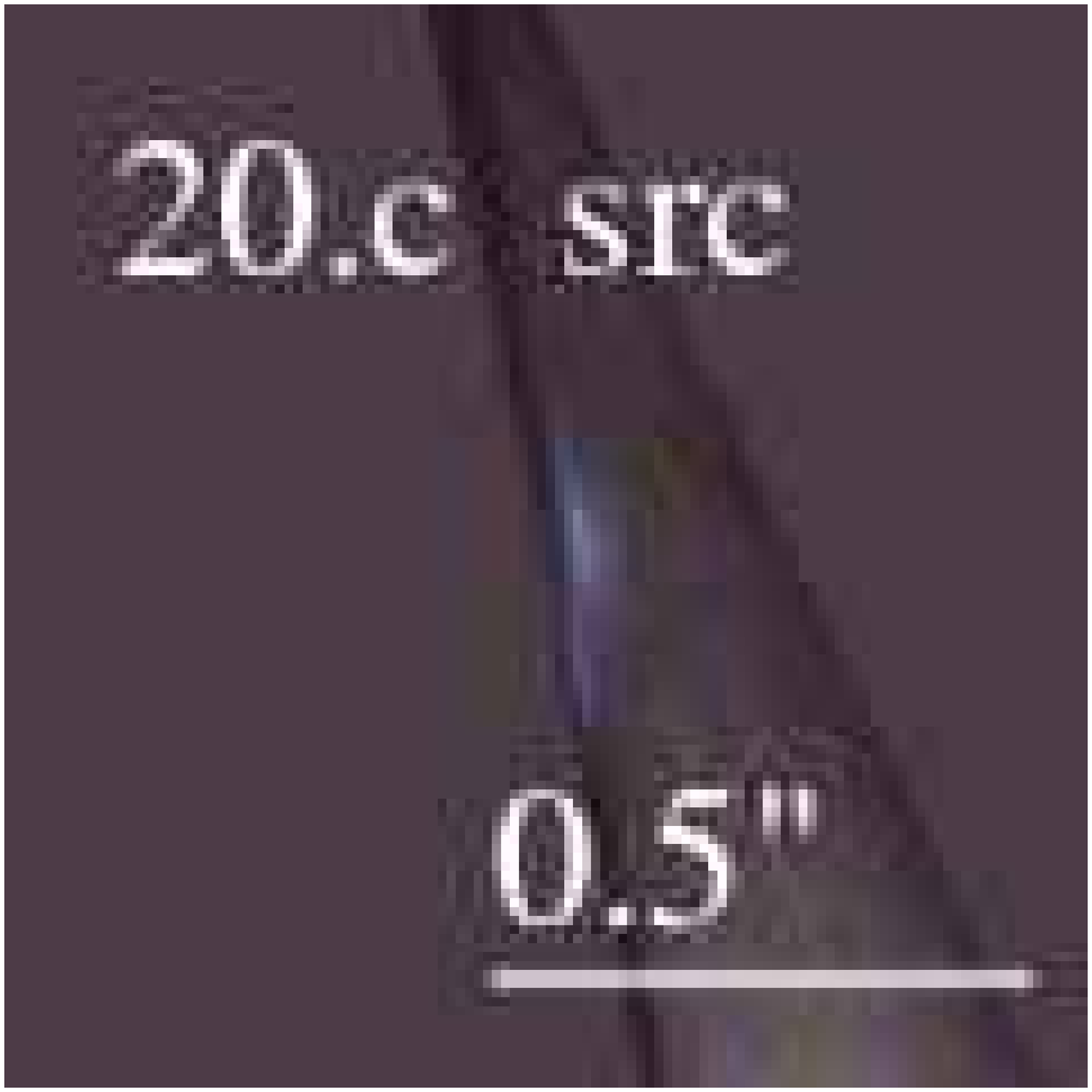}}
    & \multicolumn{1}{m{1.7cm}}{\includegraphics[height=2.00cm,clip]{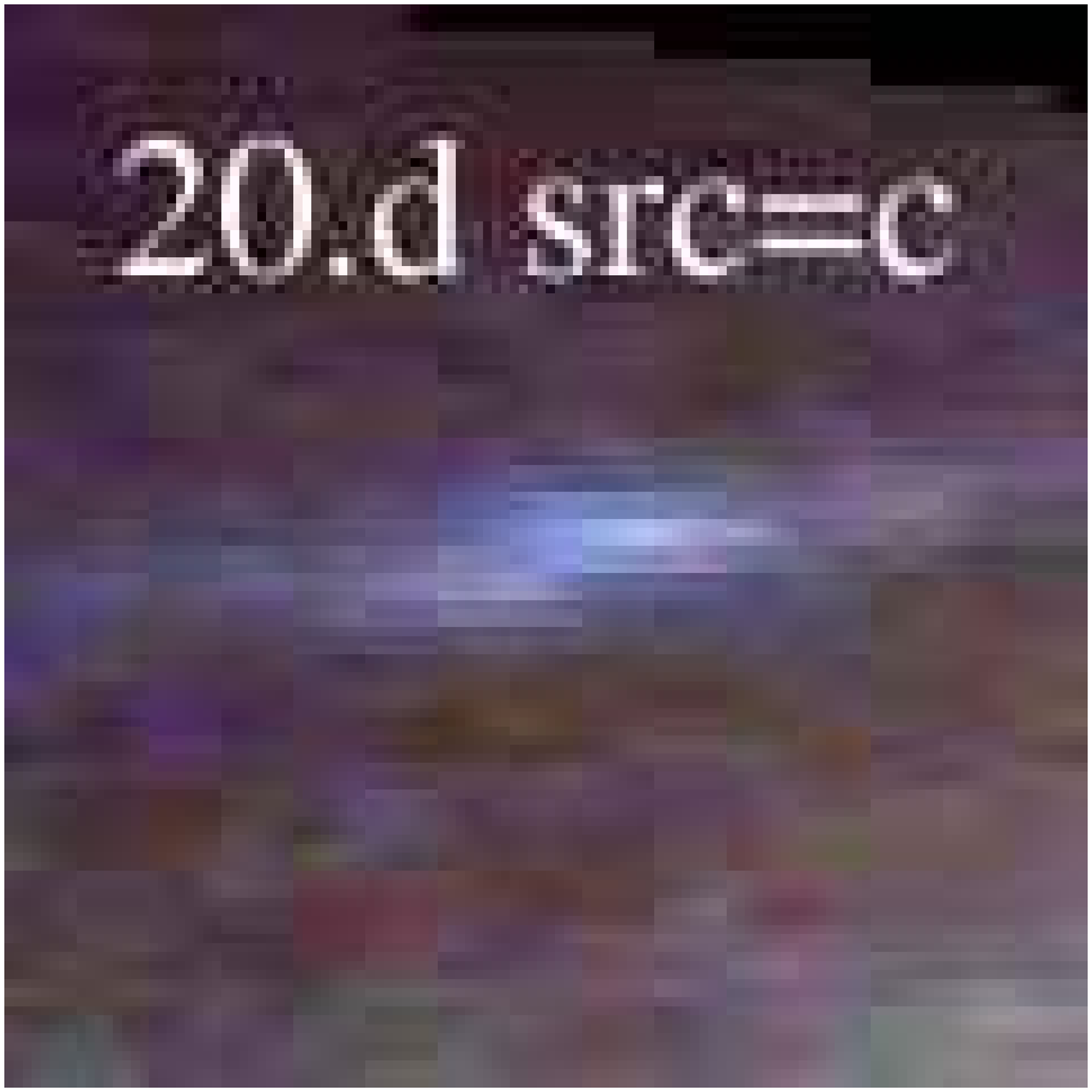}}
    & \multicolumn{1}{m{1.7cm}}{\includegraphics[height=2.00cm,clip]{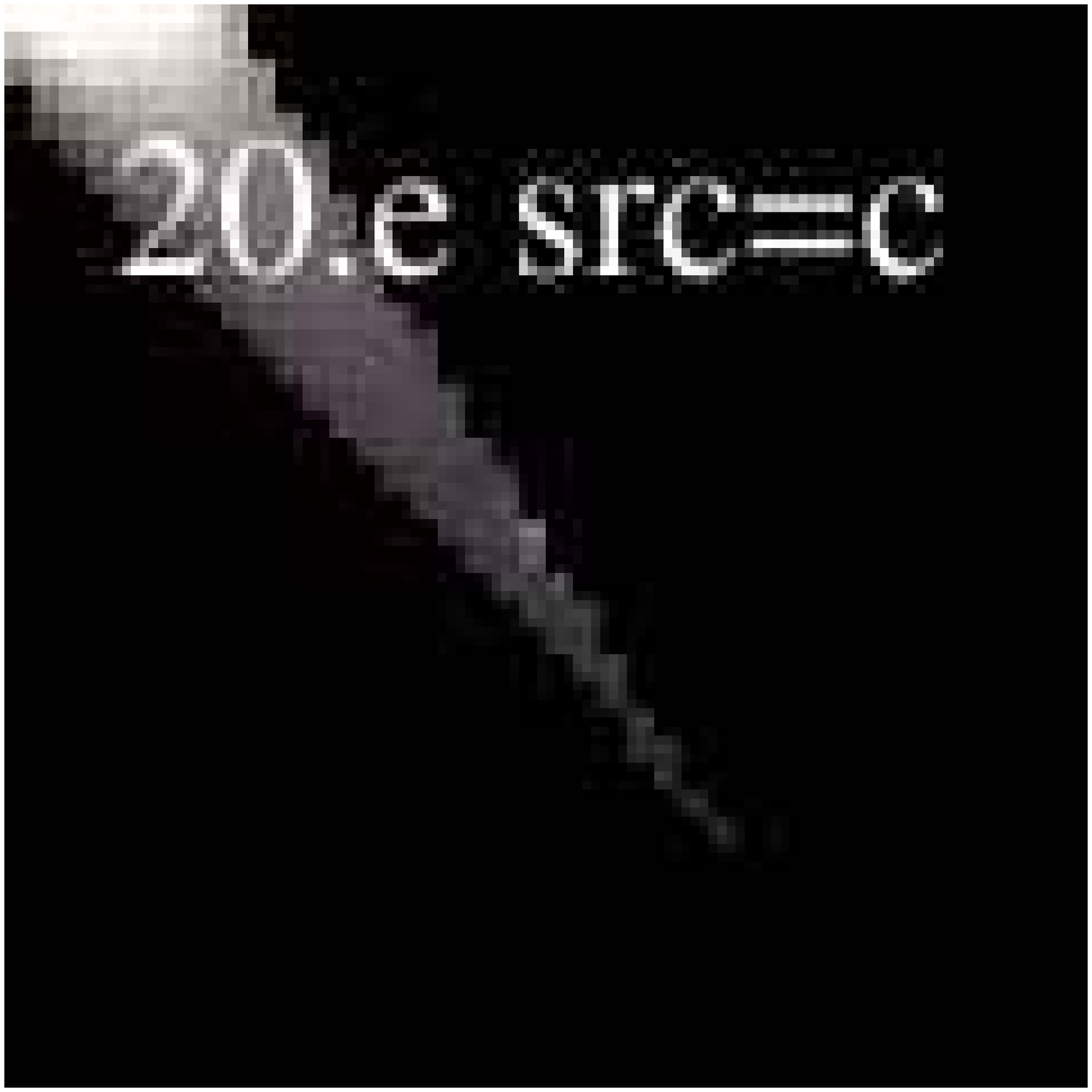}} \\
    \multicolumn{1}{m{1cm}}{{\Large ENFW}}
    & \multicolumn{1}{m{1.7cm}}{\includegraphics[height=2.00cm,clip]{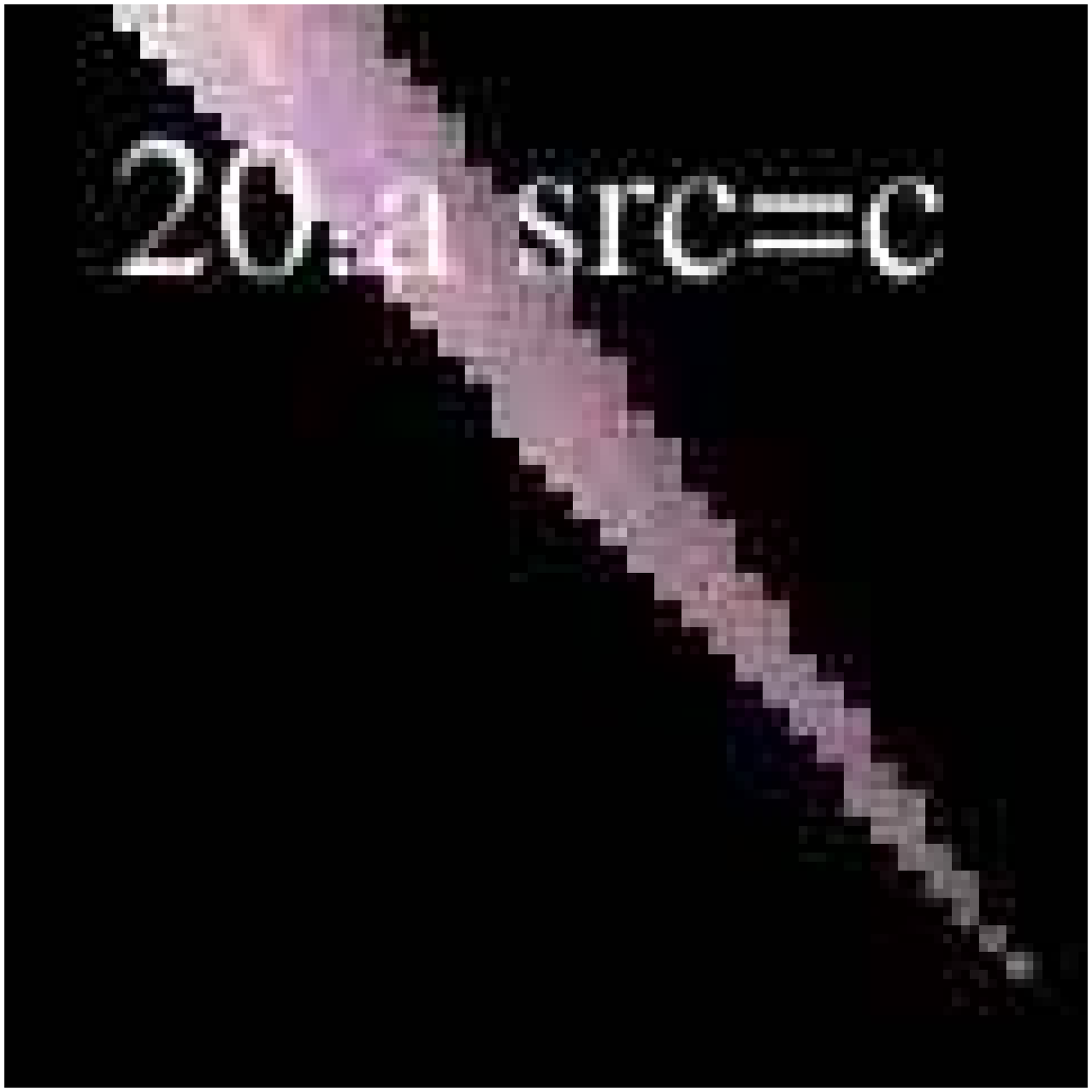}}
    & \multicolumn{1}{m{1.7cm}}{\includegraphics[height=2.00cm,clip]{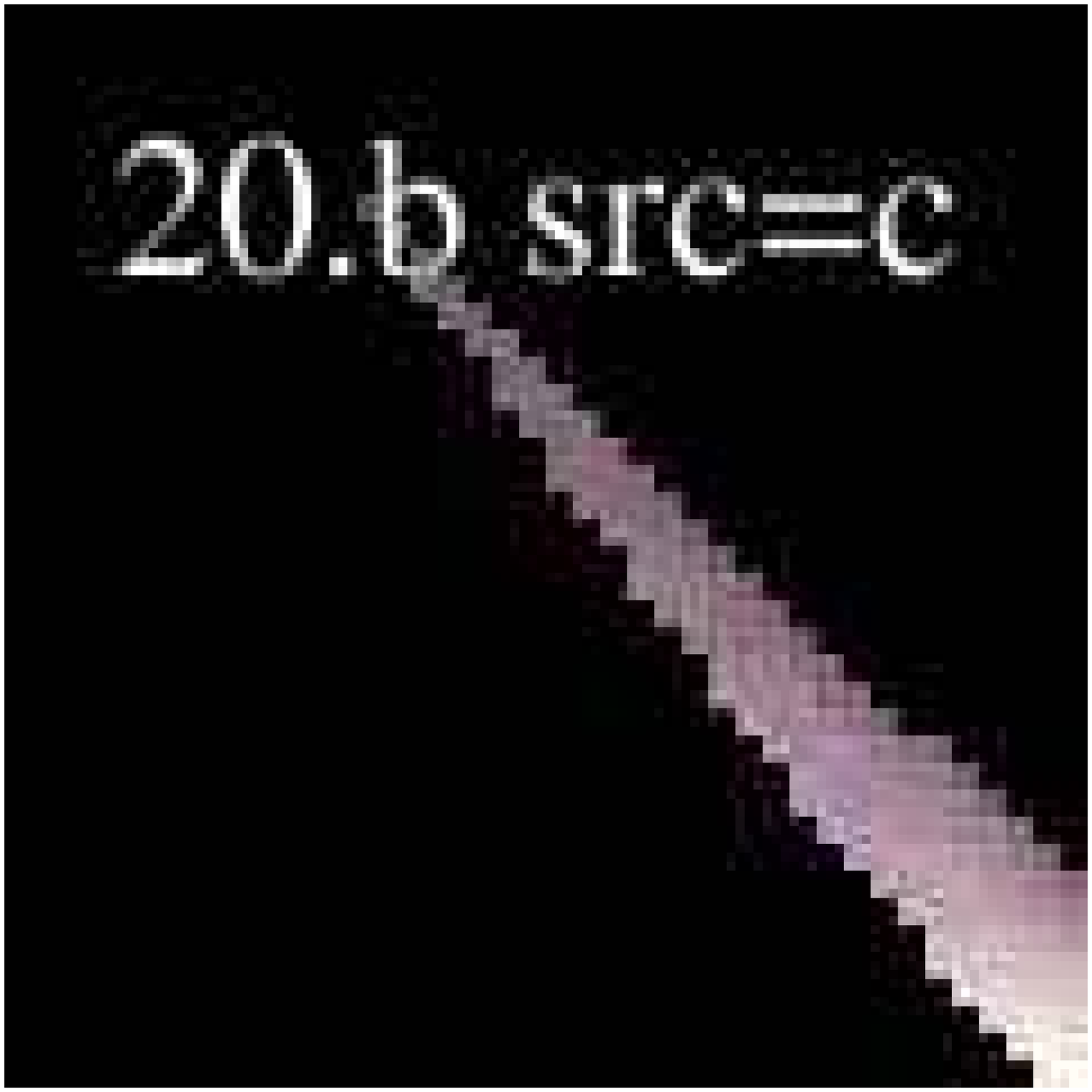}}
    & \multicolumn{1}{m{1.7cm}}{\includegraphics[height=2.00cm,clip]{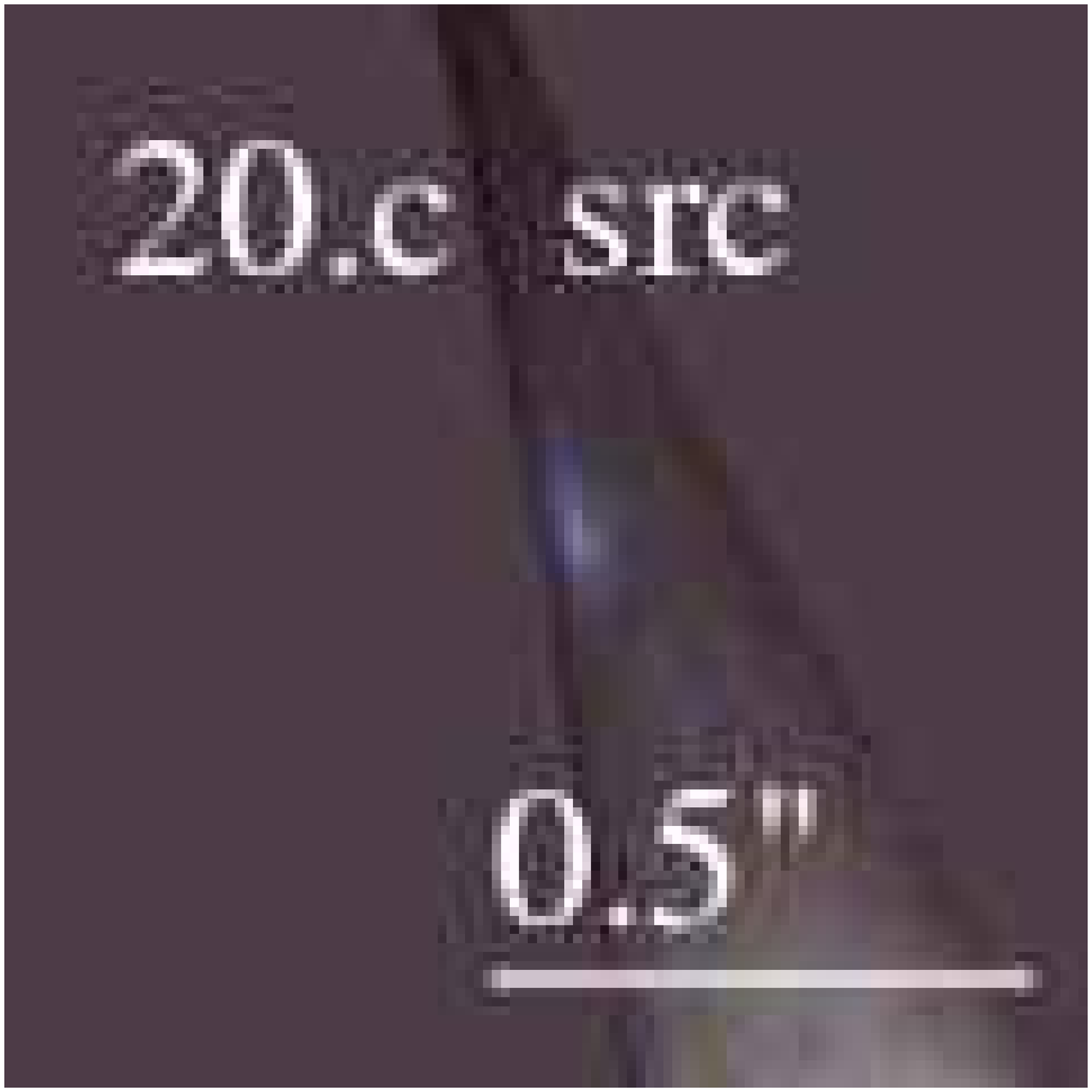}}
    & \multicolumn{1}{m{1.7cm}}{\includegraphics[height=2.00cm,clip]{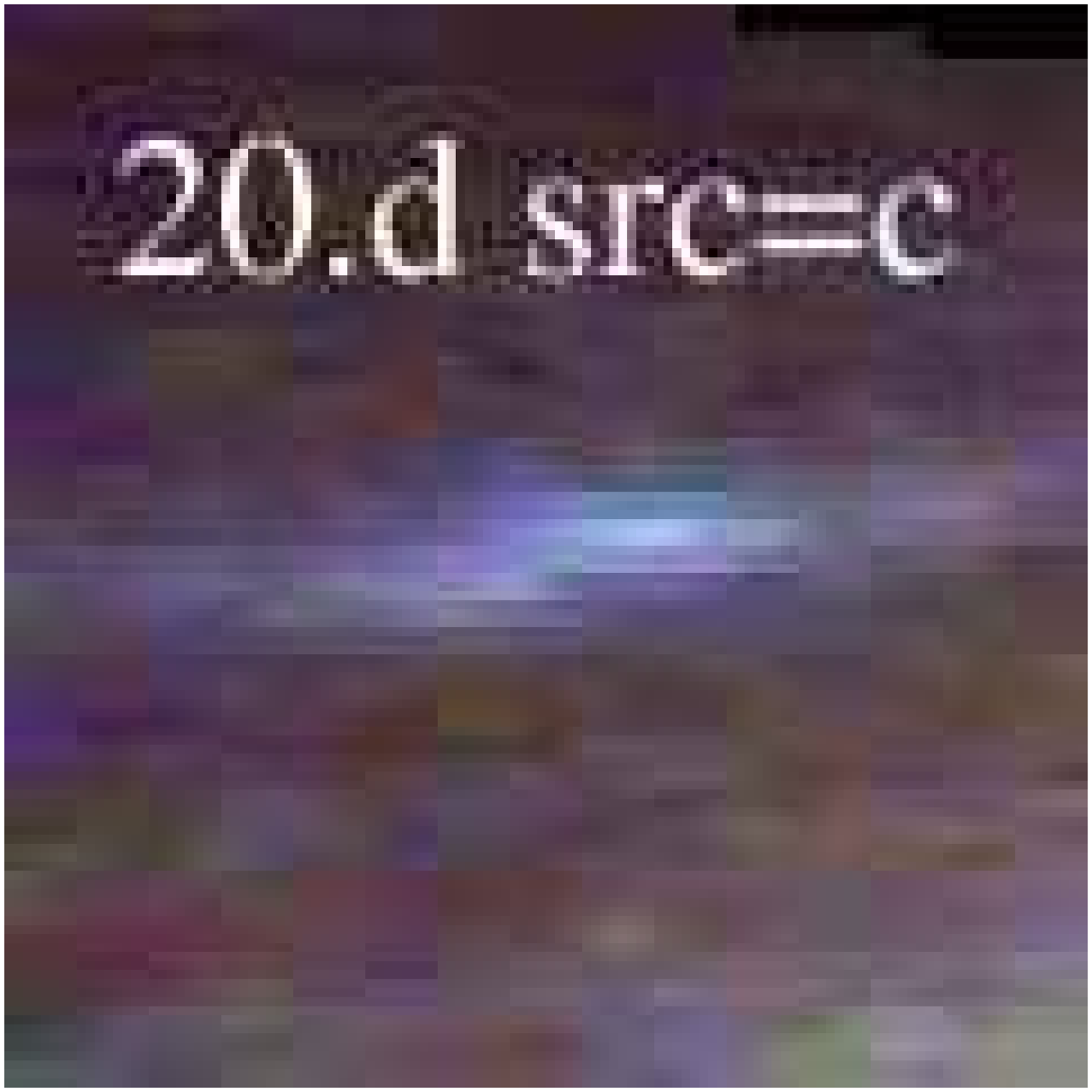}}
    & \multicolumn{1}{m{1.7cm}}{\includegraphics[height=2.00cm,clip]{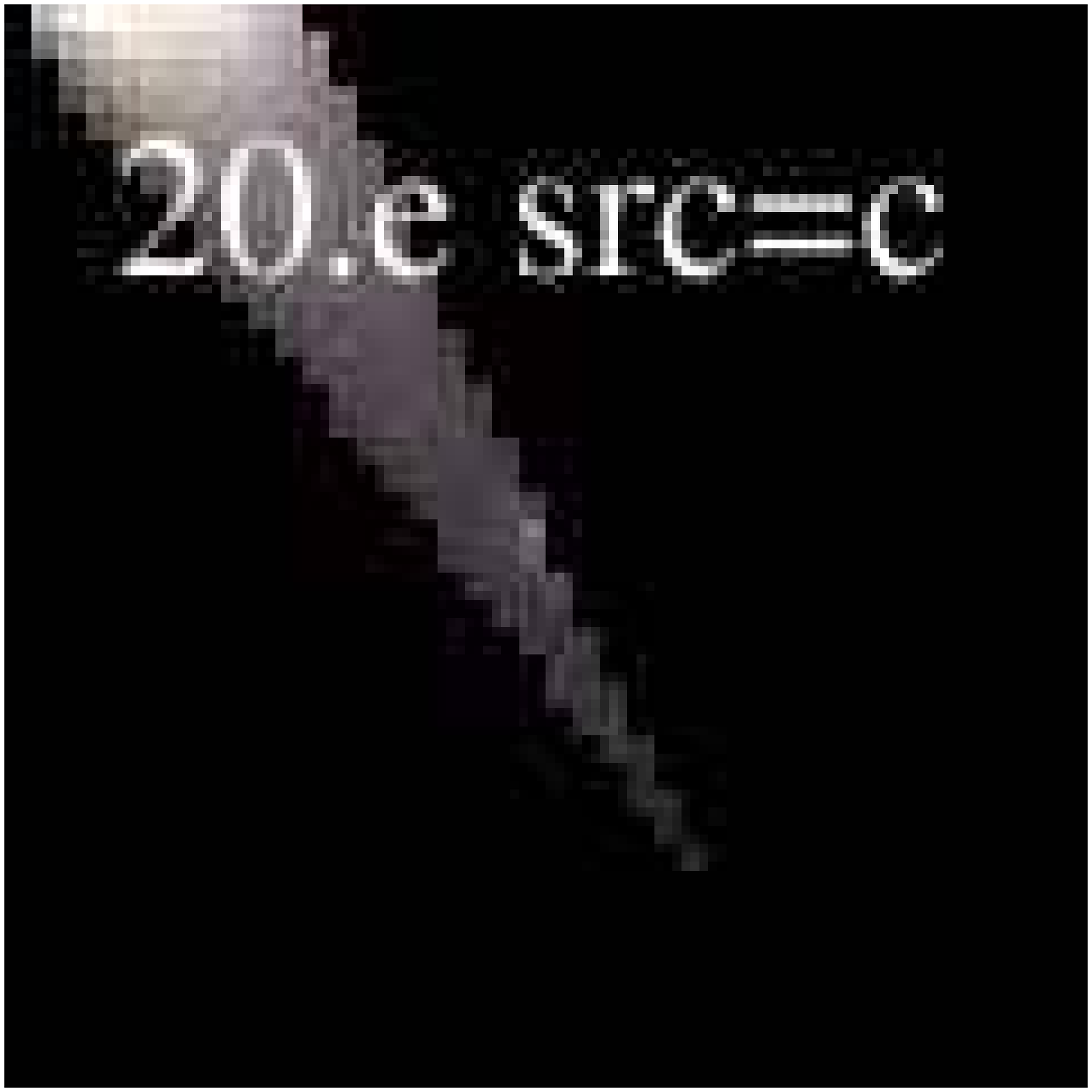}} \\
  \end{tabular}

\end{table*}

\begin{table*}
  \caption{Image system 21:}\vspace{0mm}
  \begin{tabular}{cccc}
    \multicolumn{1}{m{1cm}}{{\Large A1689}}
    & \multicolumn{1}{m{1.7cm}}{\includegraphics[height=2.00cm,clip]{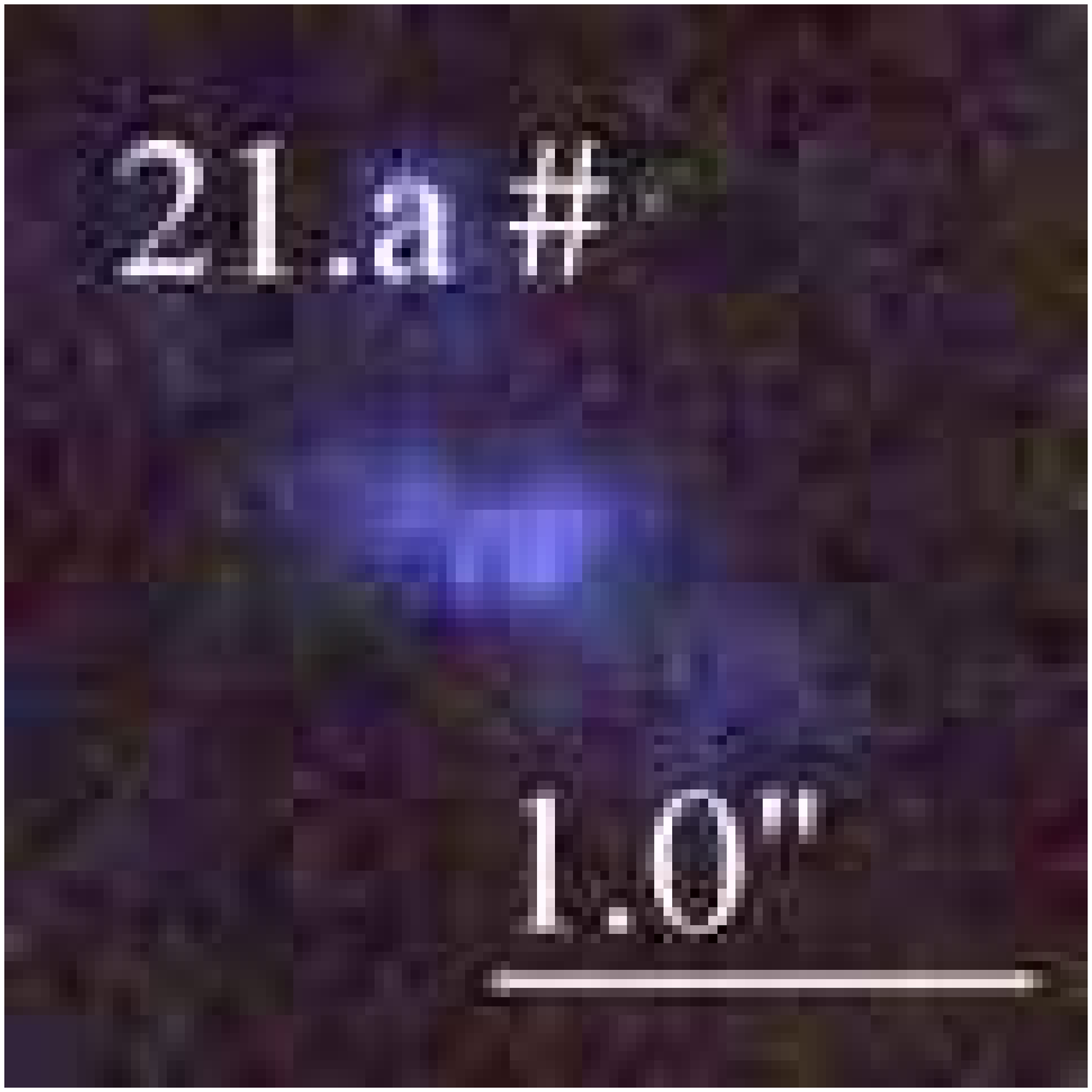}}
    & \multicolumn{1}{m{1.7cm}}{\includegraphics[height=2.00cm,clip]{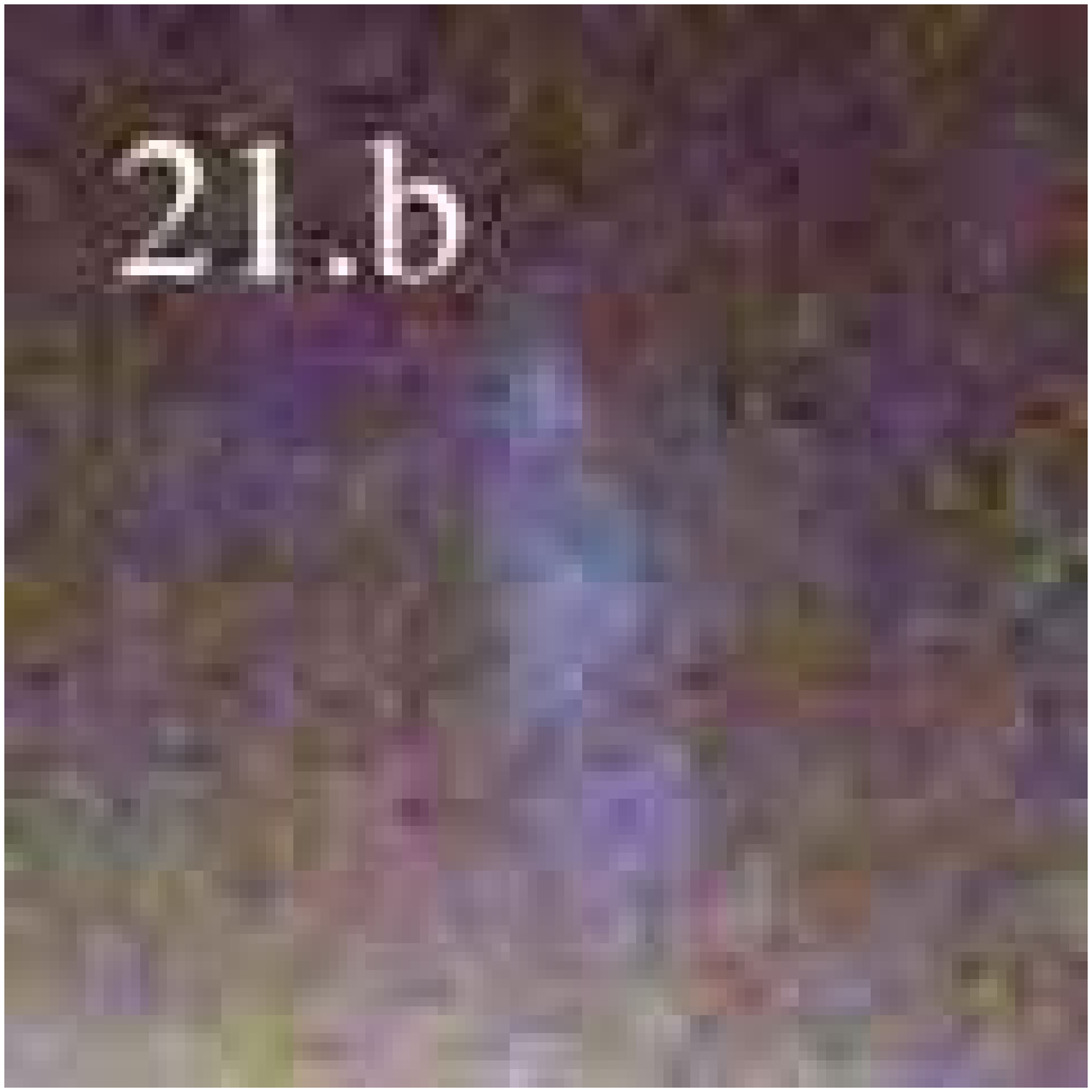}}
    & \multicolumn{1}{m{1.7cm}}{\includegraphics[height=2.00cm,clip]{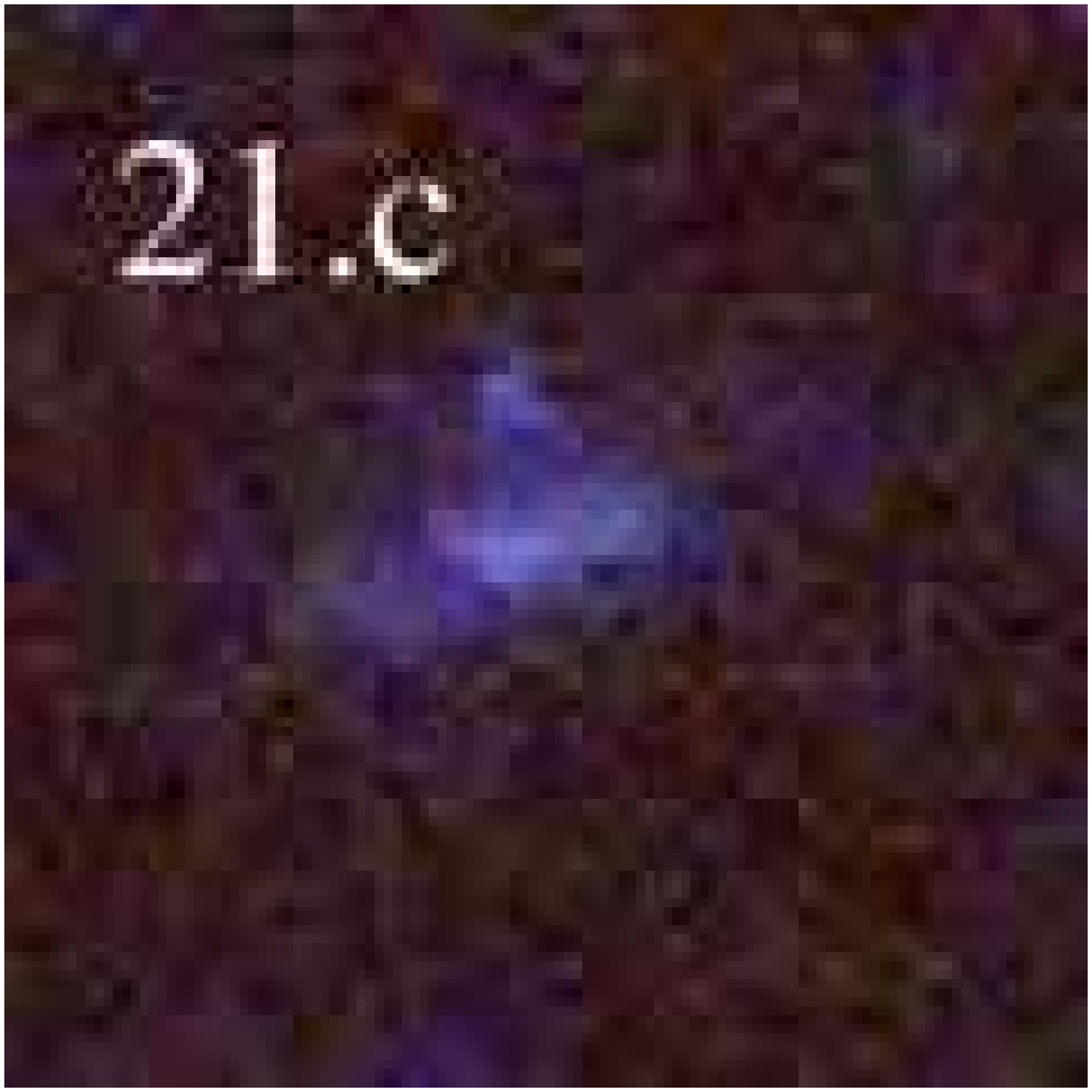}} \\
    \multicolumn{1}{m{1cm}}{{\Large NSIE}}
    & \multicolumn{1}{m{1.7cm}}{\includegraphics[height=2.00cm,clip]{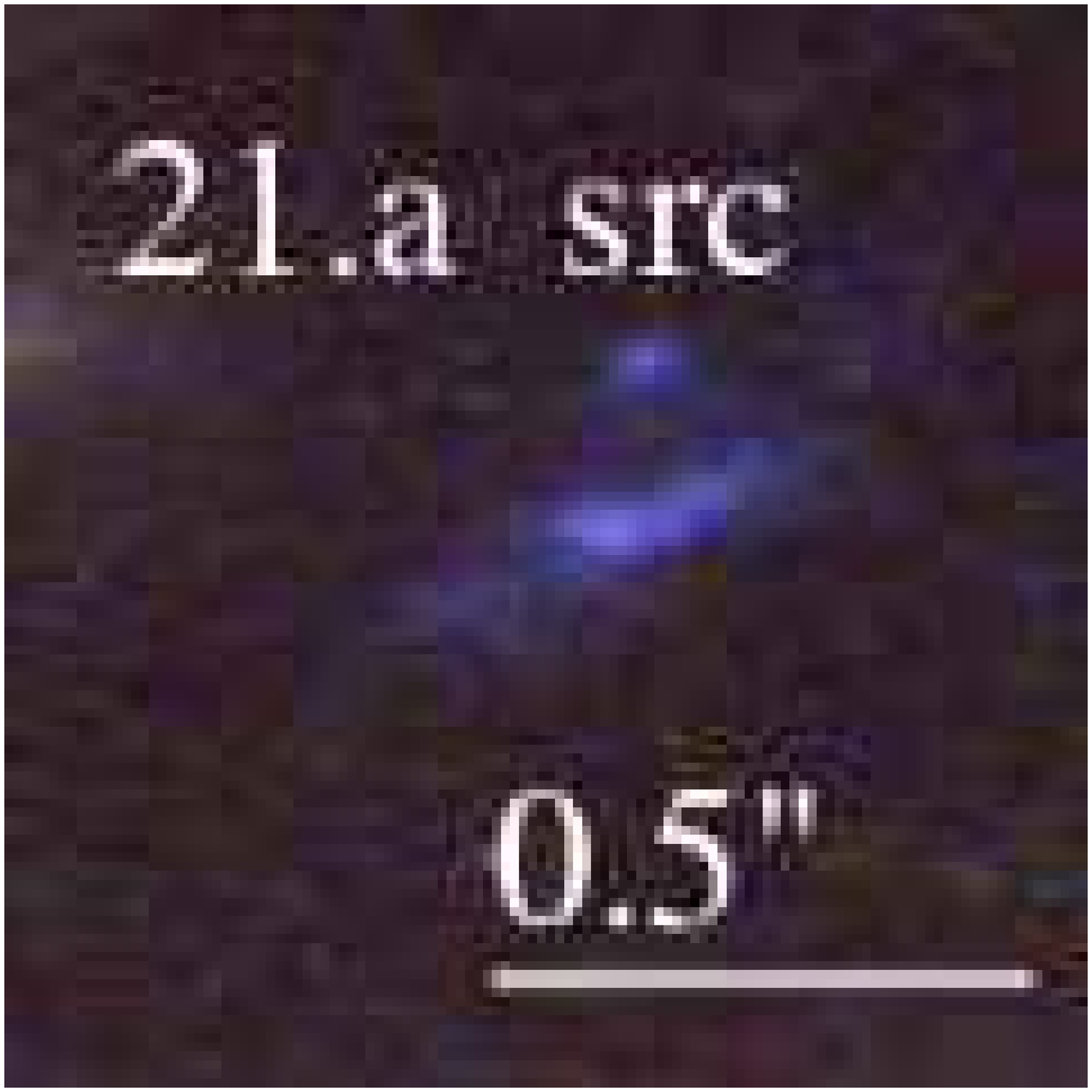}}
    & \multicolumn{1}{m{1.7cm}}{\includegraphics[height=2.00cm,clip]{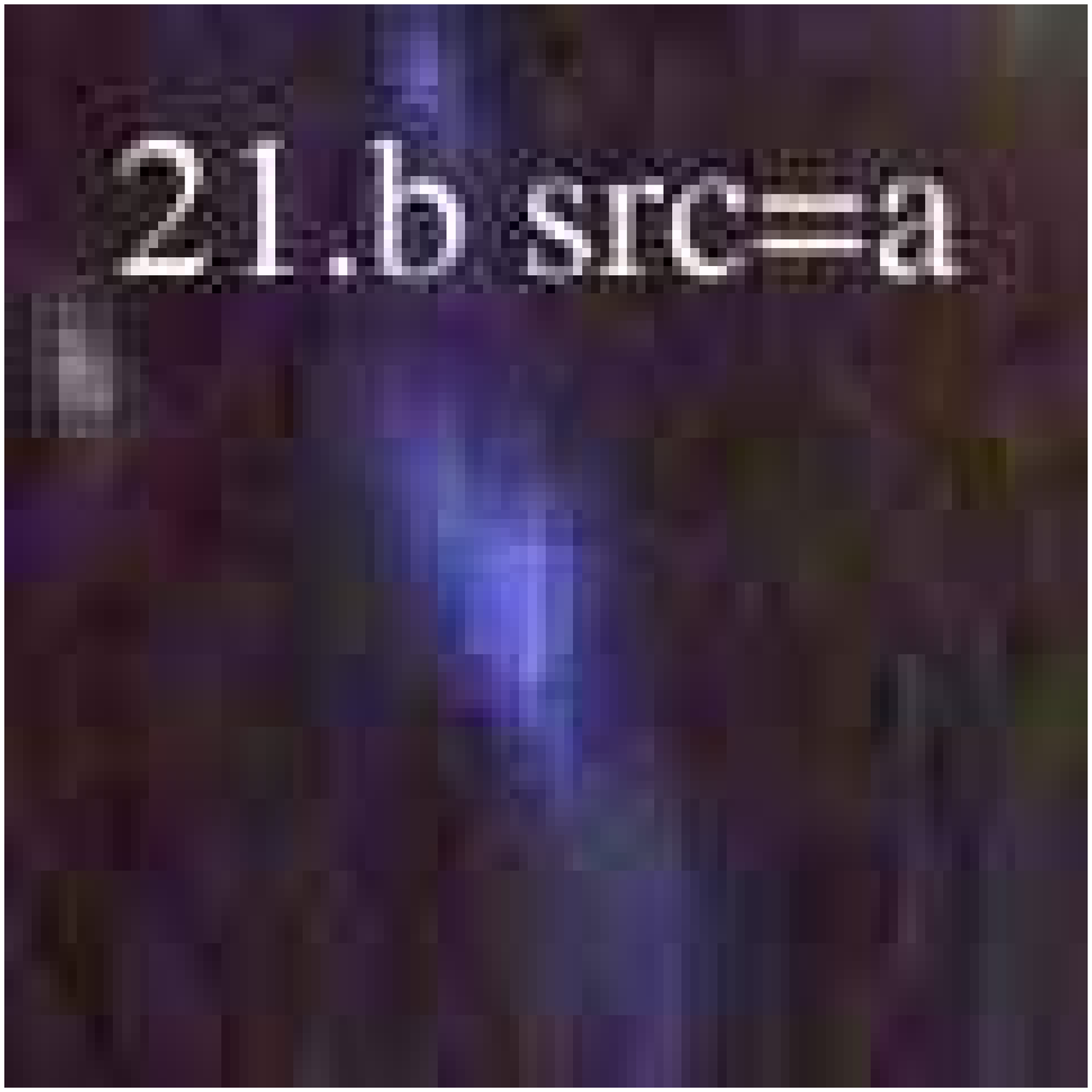}}
    & \multicolumn{1}{m{1.7cm}}{\includegraphics[height=2.00cm,clip]{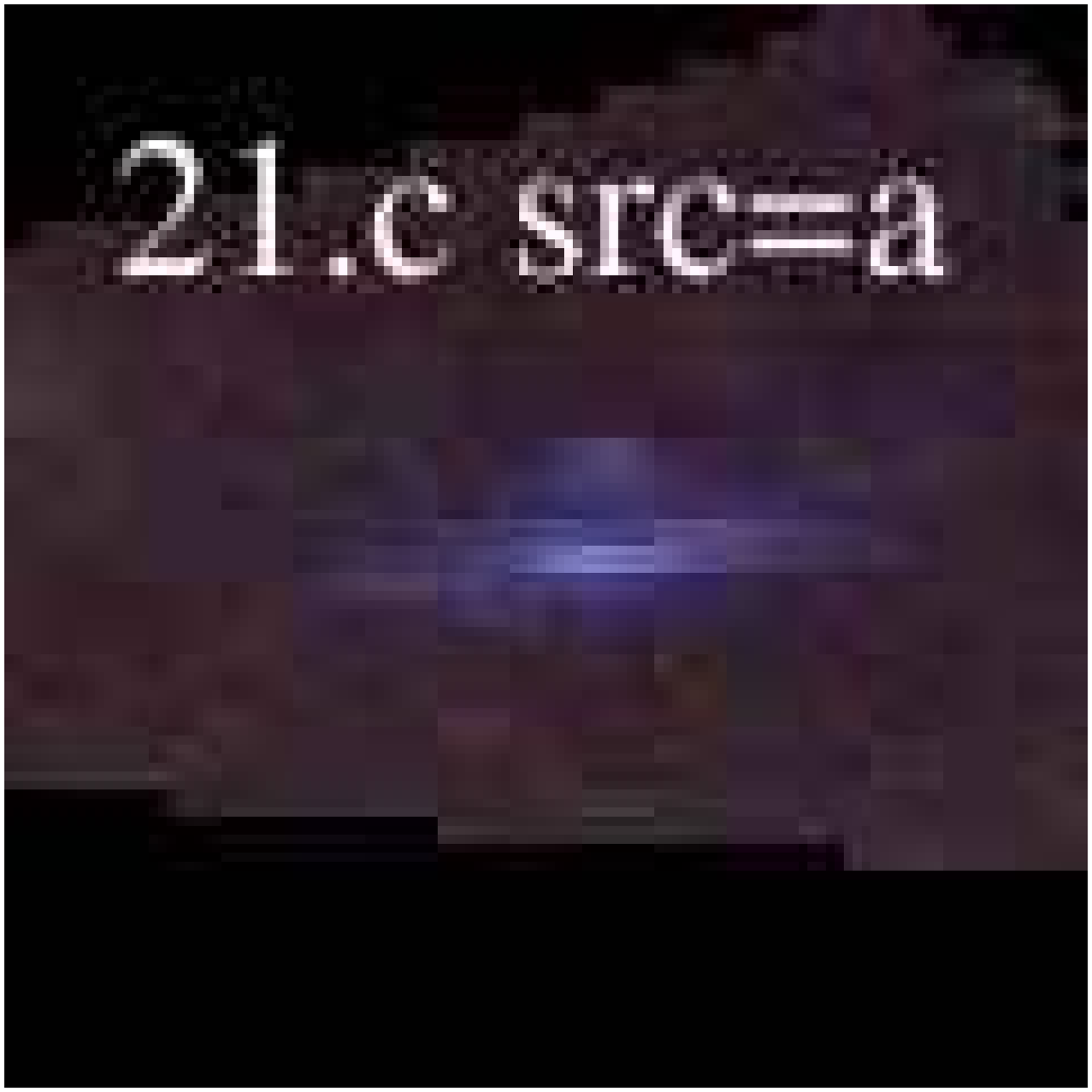}} \\
    \multicolumn{1}{m{1cm}}{{\Large ENFW}}
    & \multicolumn{1}{m{1.7cm}}{\includegraphics[height=2.00cm,clip]{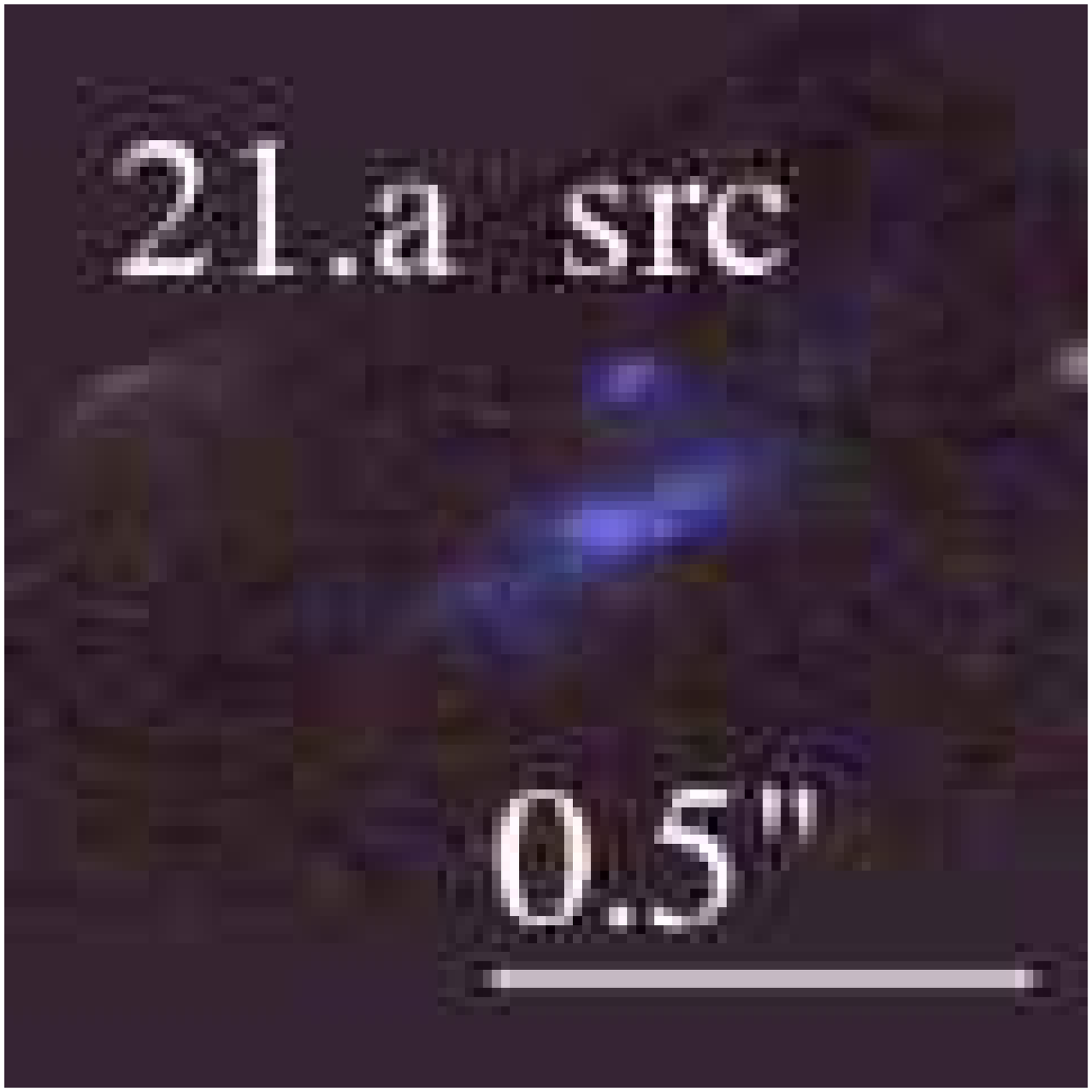}}
    & \multicolumn{1}{m{1.7cm}}{\includegraphics[height=2.00cm,clip]{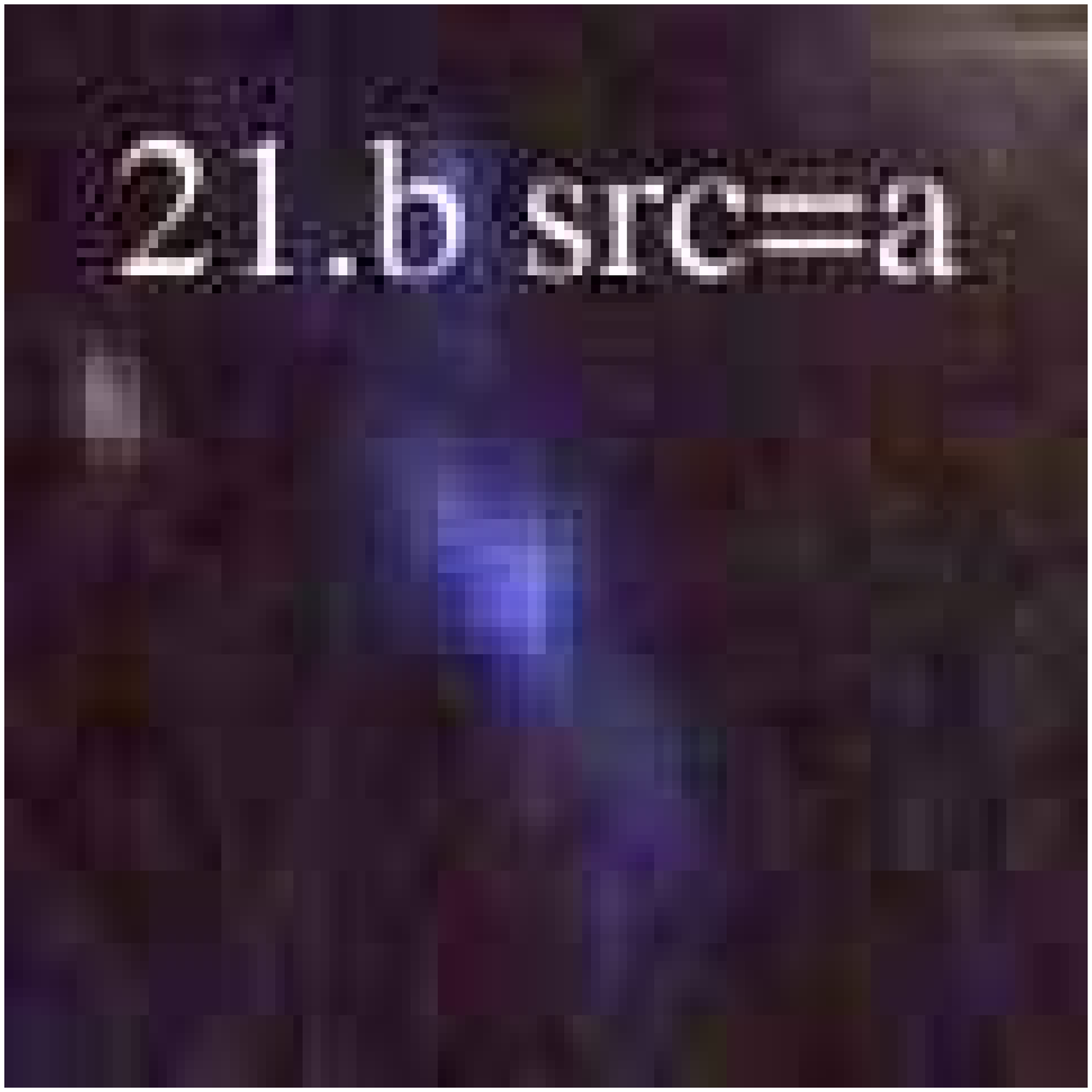}}
    & \multicolumn{1}{m{1.7cm}}{\includegraphics[height=2.00cm,clip]{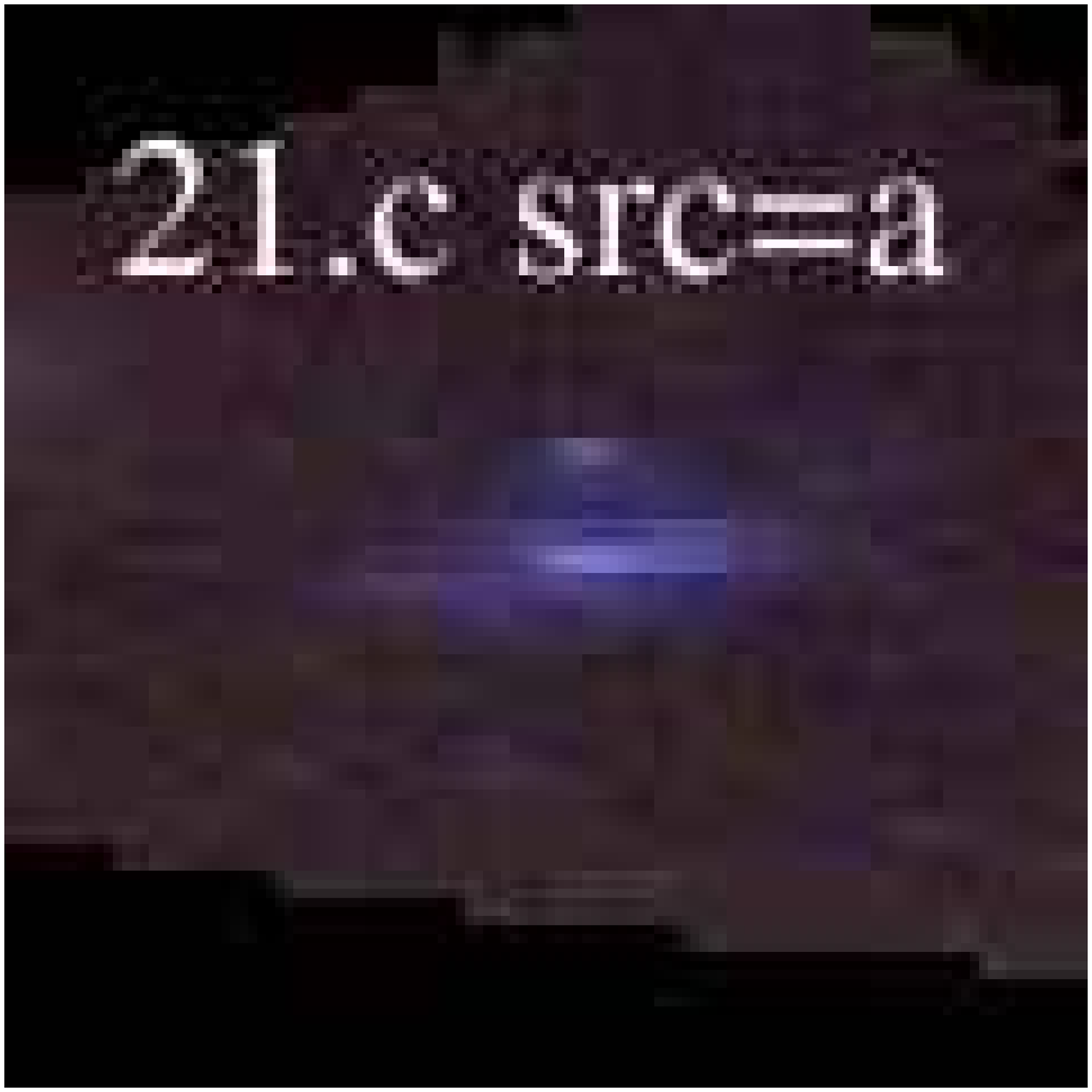}} \\
  \end{tabular}

\end{table*}

\clearpage

\begin{table*}
  \caption{Image system 22:}\vspace{0mm}
  \begin{tabular}{cccc}
    \multicolumn{1}{m{1cm}}{{\Large A1689}}
    & \multicolumn{1}{m{1.7cm}}{\includegraphics[height=2.00cm,clip]{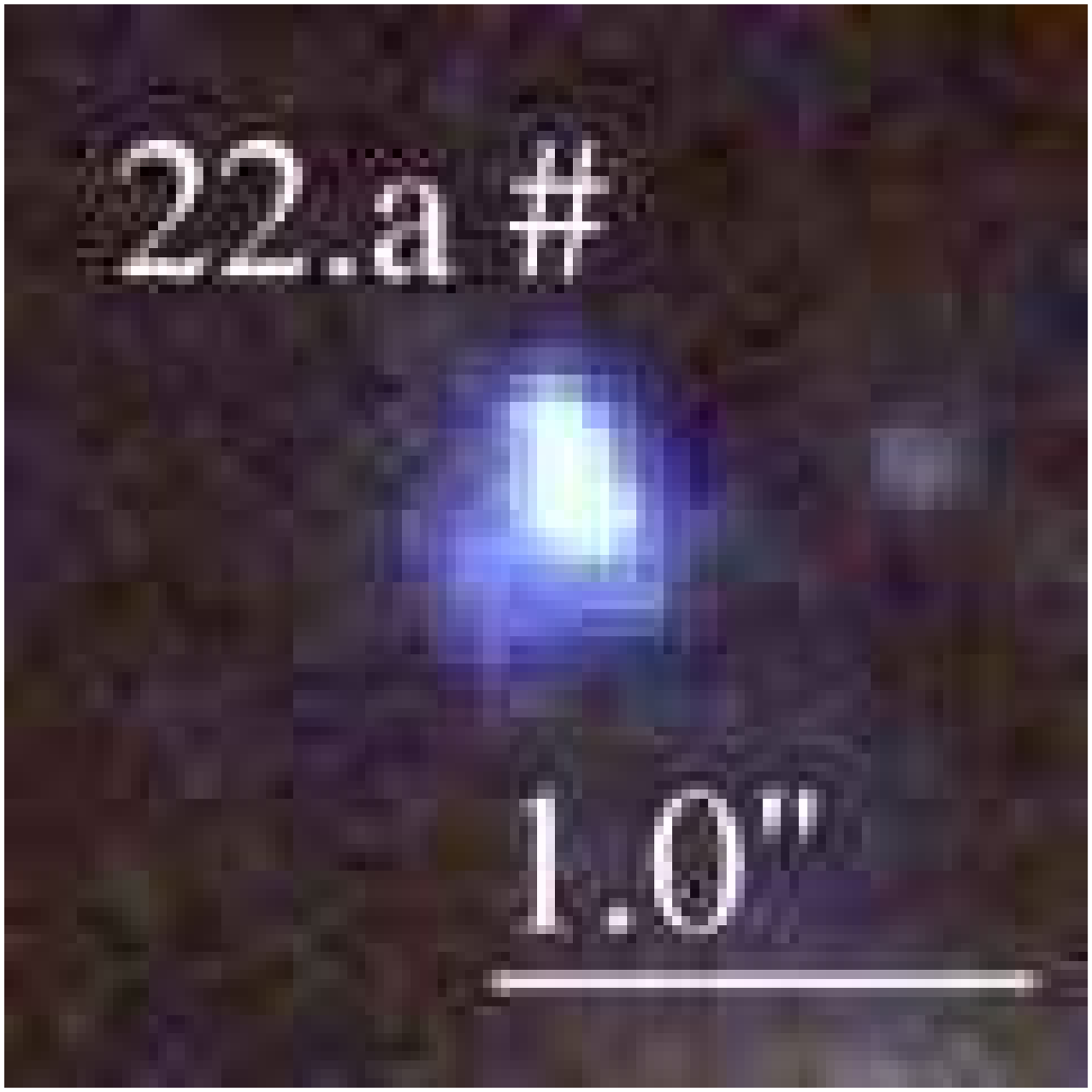}}
    & \multicolumn{1}{m{1.7cm}}{\includegraphics[height=2.00cm,clip]{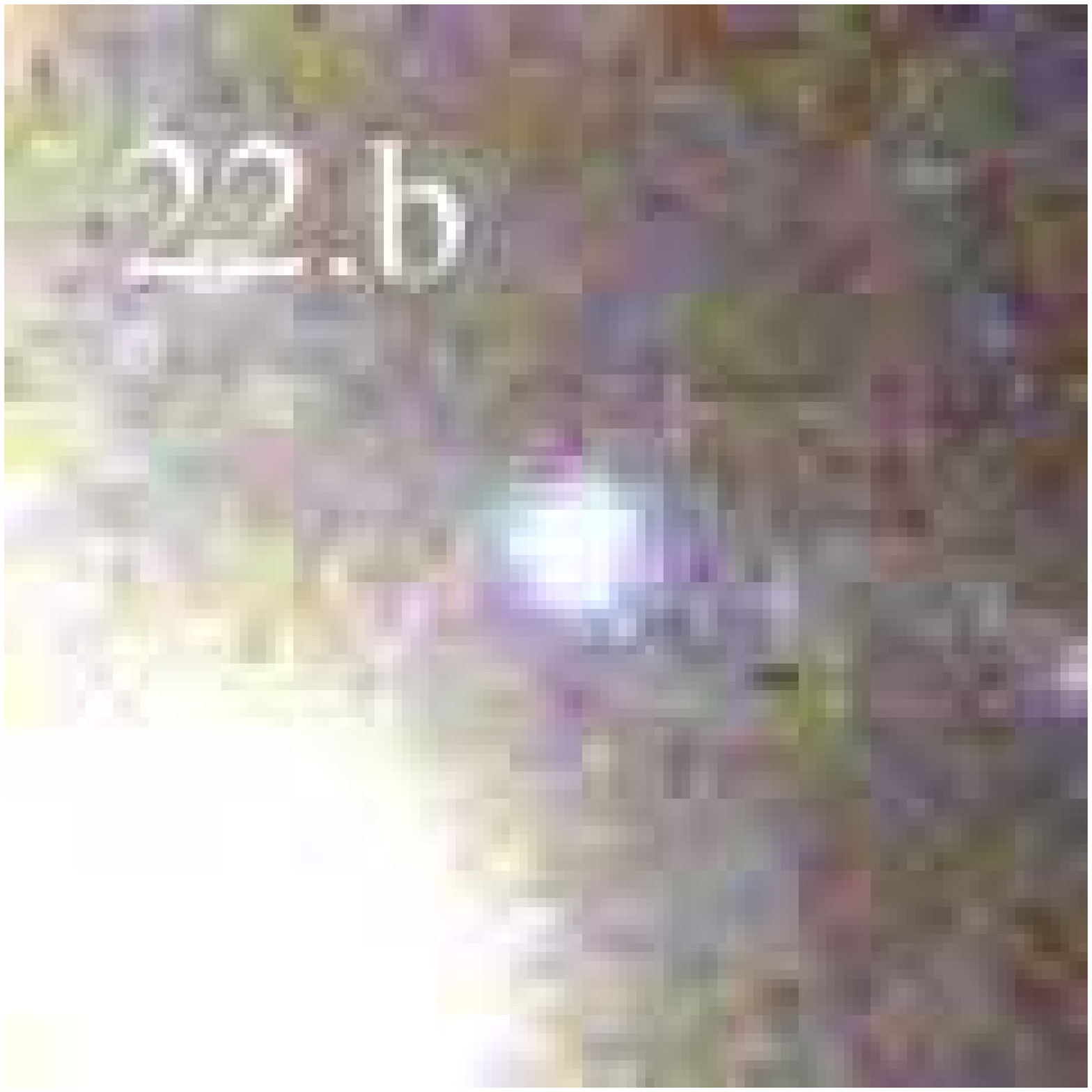}}
    & \multicolumn{1}{m{1.7cm}}{\includegraphics[height=2.00cm,clip]{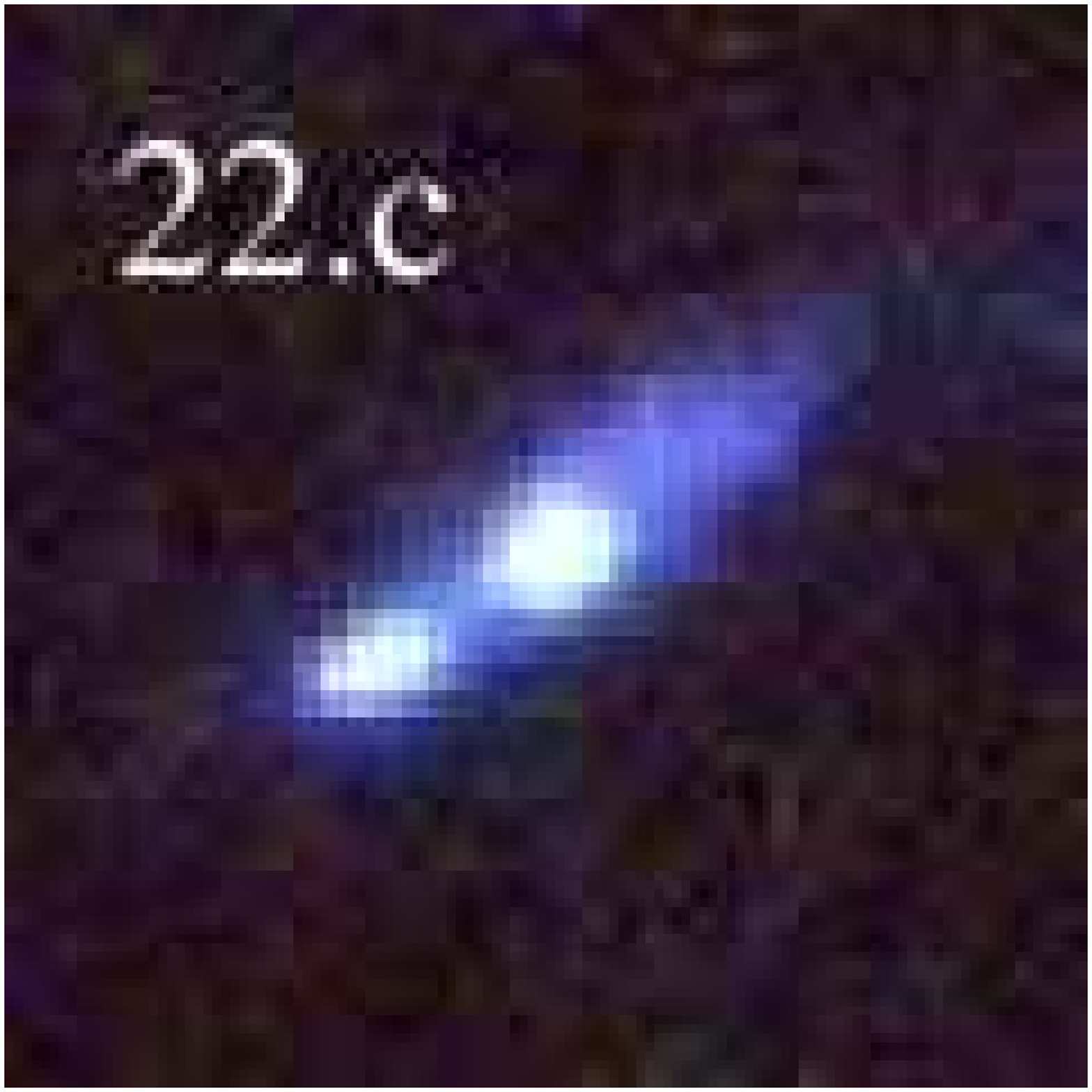}} \\
    \multicolumn{1}{m{1cm}}{{\Large NSIE}}
    & \multicolumn{1}{m{1.7cm}}{\includegraphics[height=2.00cm,clip]{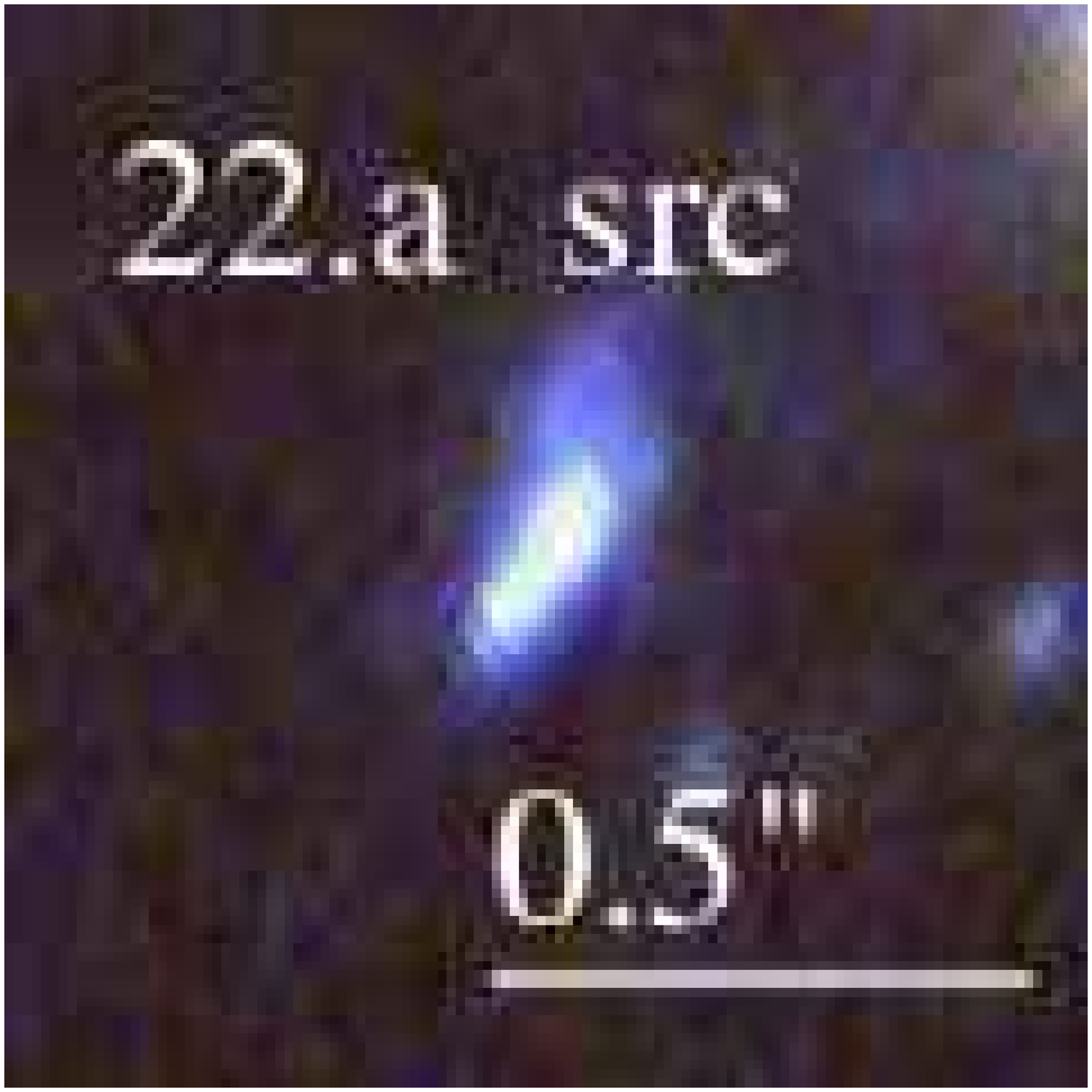}}
    & \multicolumn{1}{m{1.7cm}}{\includegraphics[height=2.00cm,clip]{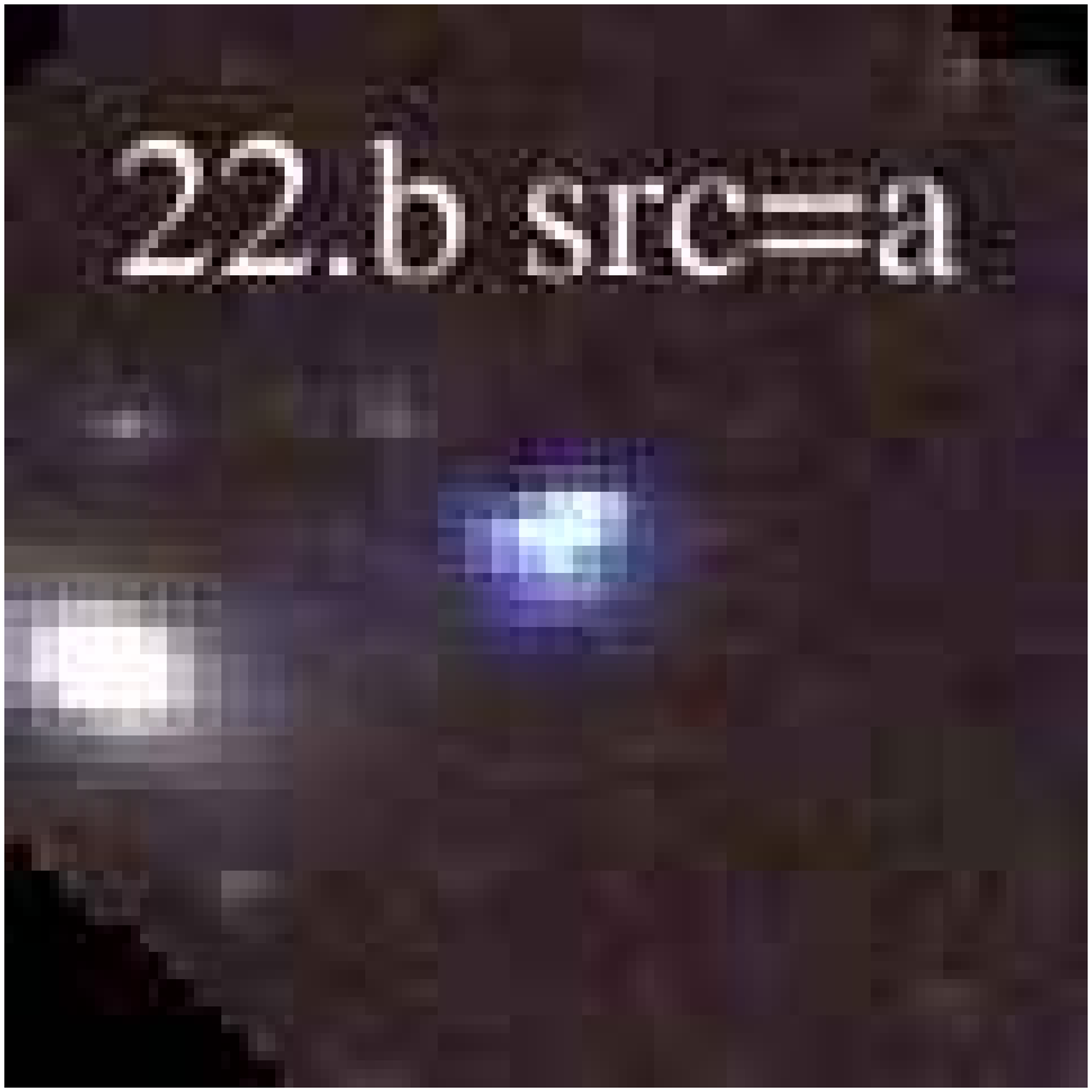}}
    & \multicolumn{1}{m{1.7cm}}{\includegraphics[height=2.00cm,clip]{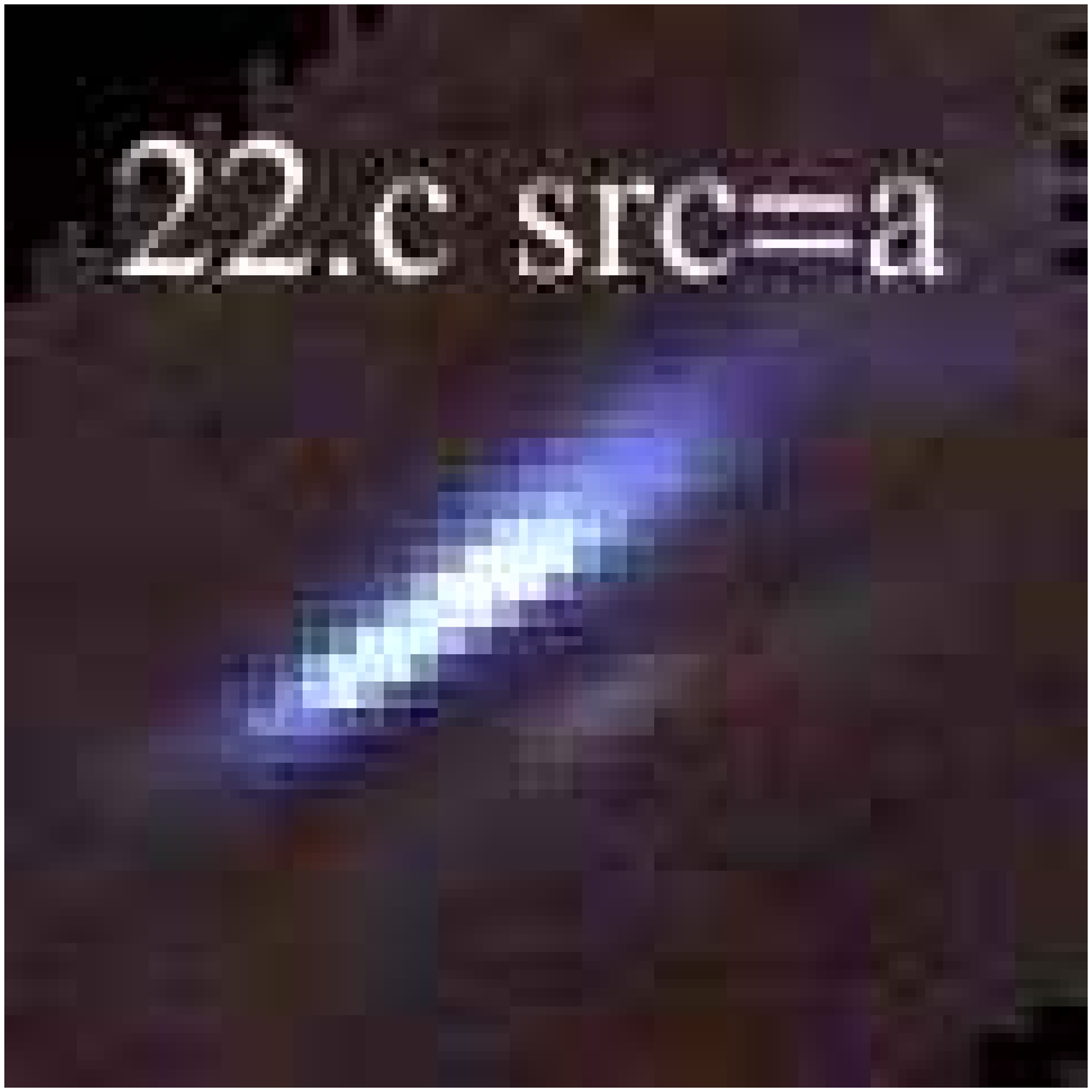}} \\
    \multicolumn{1}{m{1cm}}{{\Large ENFW}}
    & \multicolumn{1}{m{1.7cm}}{\includegraphics[height=2.00cm,clip]{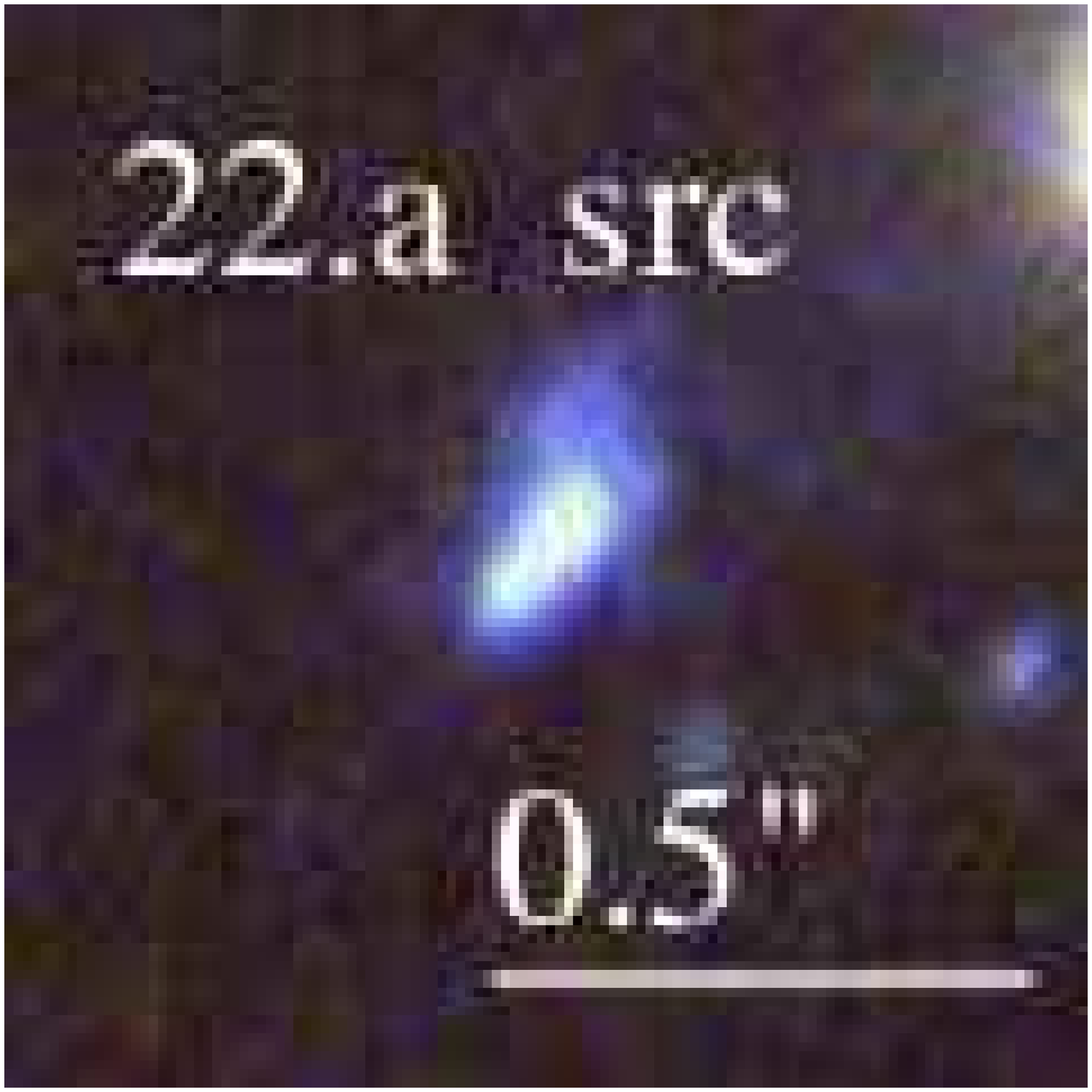}}
    & \multicolumn{1}{m{1.7cm}}{\includegraphics[height=2.00cm,clip]{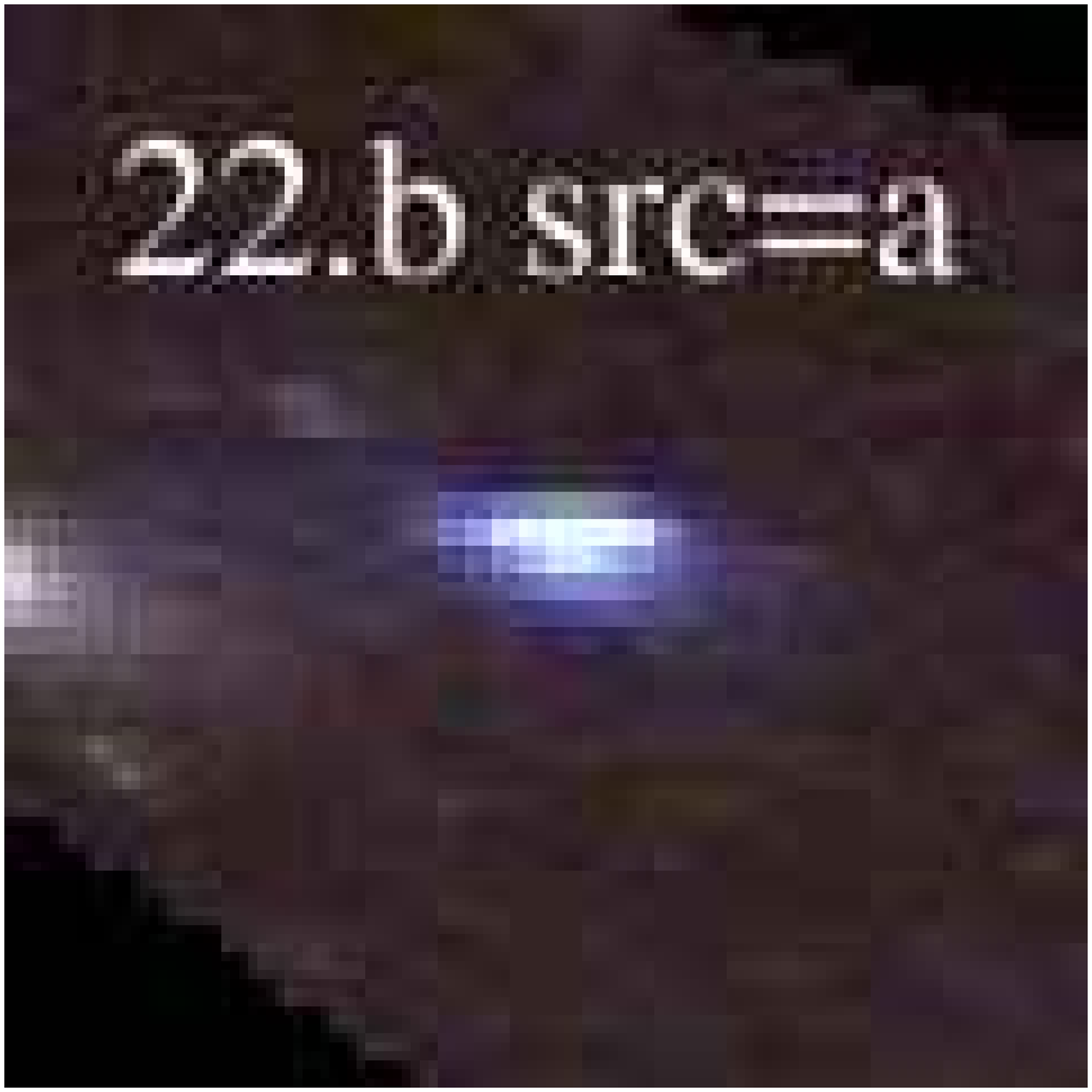}}
    & \multicolumn{1}{m{1.7cm}}{\includegraphics[height=2.00cm,clip]{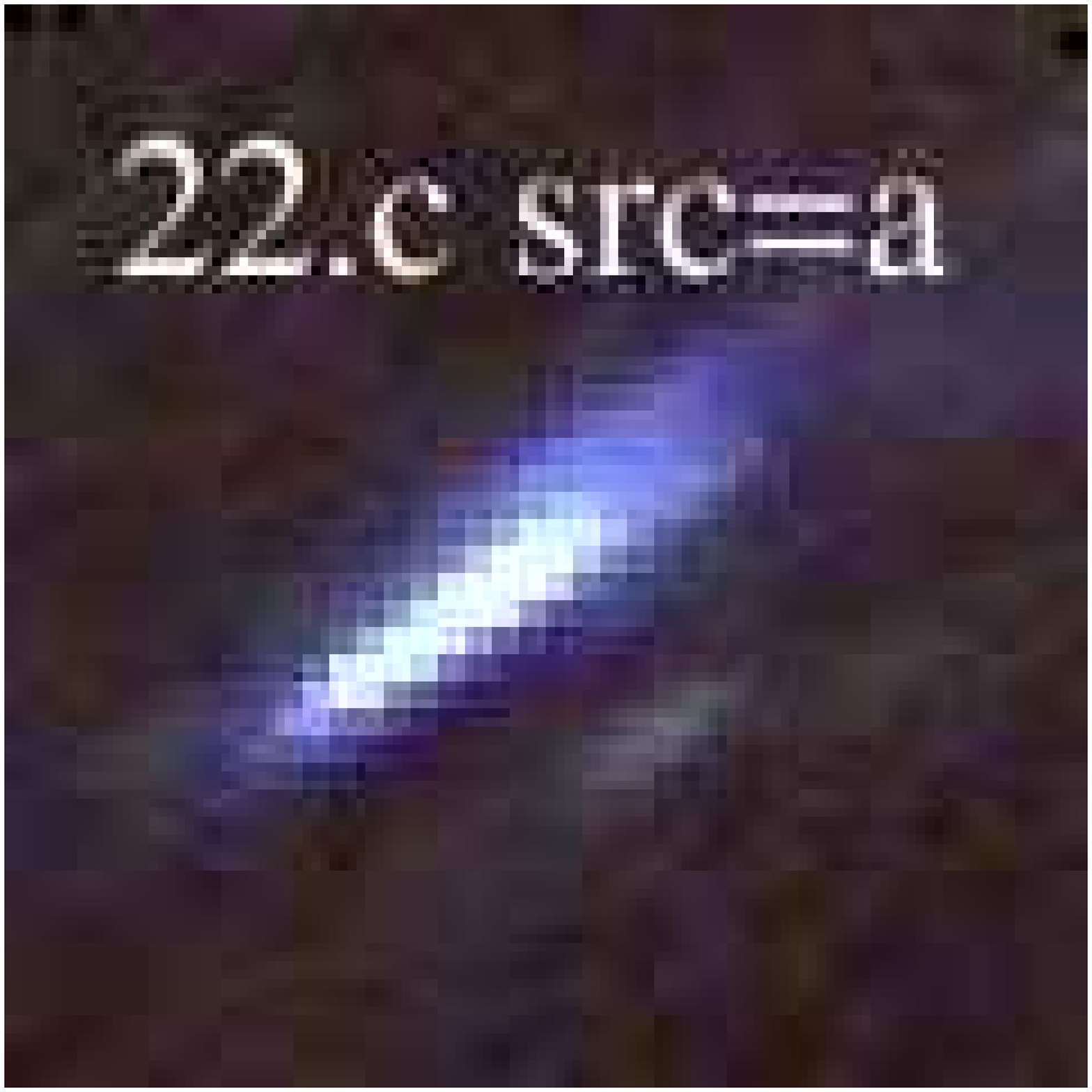}} \\
  \end{tabular}

\end{table*}

\begin{table*}
  \caption{Image system 23:}\vspace{0mm}
  \begin{tabular}{cccc}
    \multicolumn{1}{m{1cm}}{{\Large A1689}}
    & \multicolumn{1}{m{1.7cm}}{\includegraphics[height=2.00cm,clip]{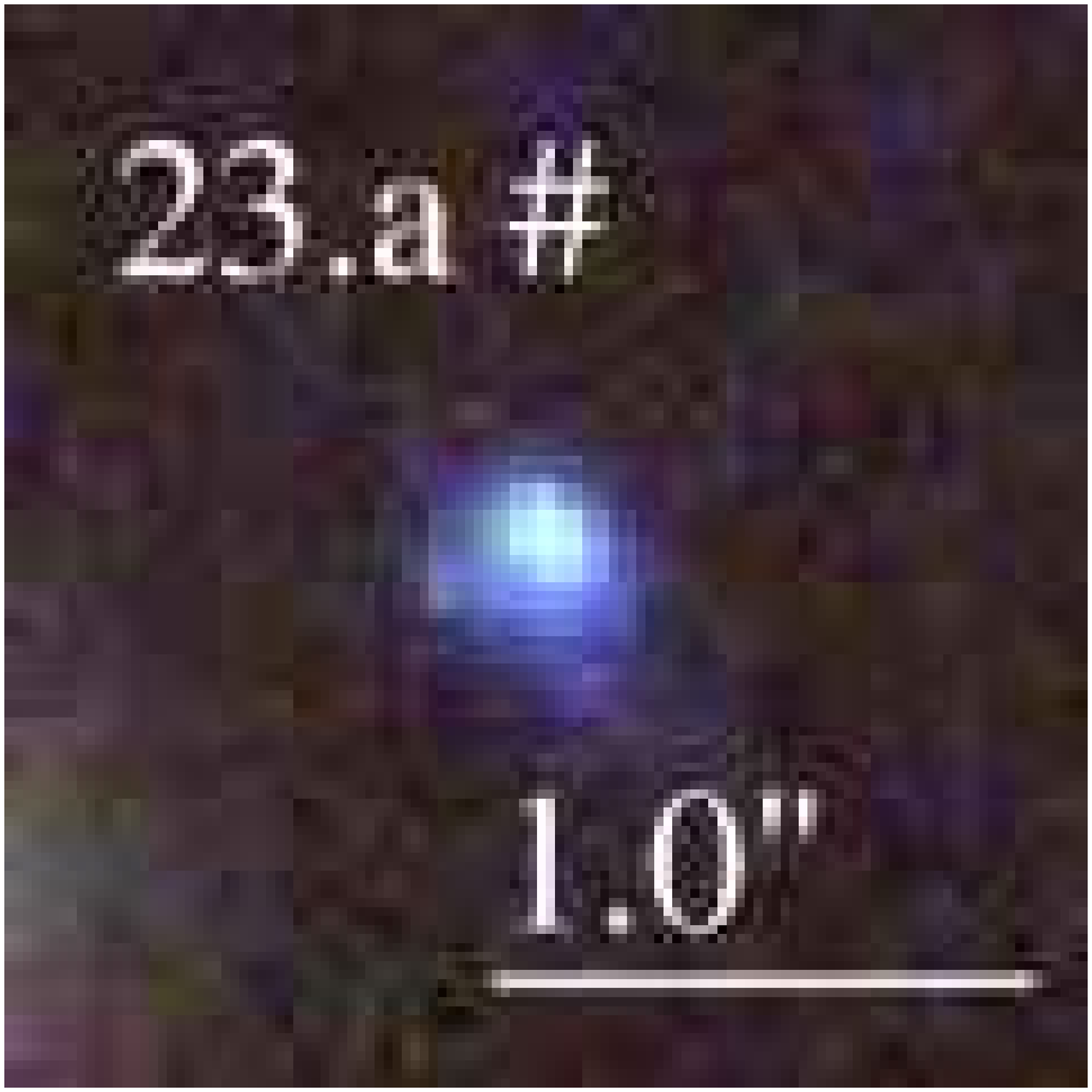}}
    & \multicolumn{1}{m{1.7cm}}{\includegraphics[height=2.00cm,clip]{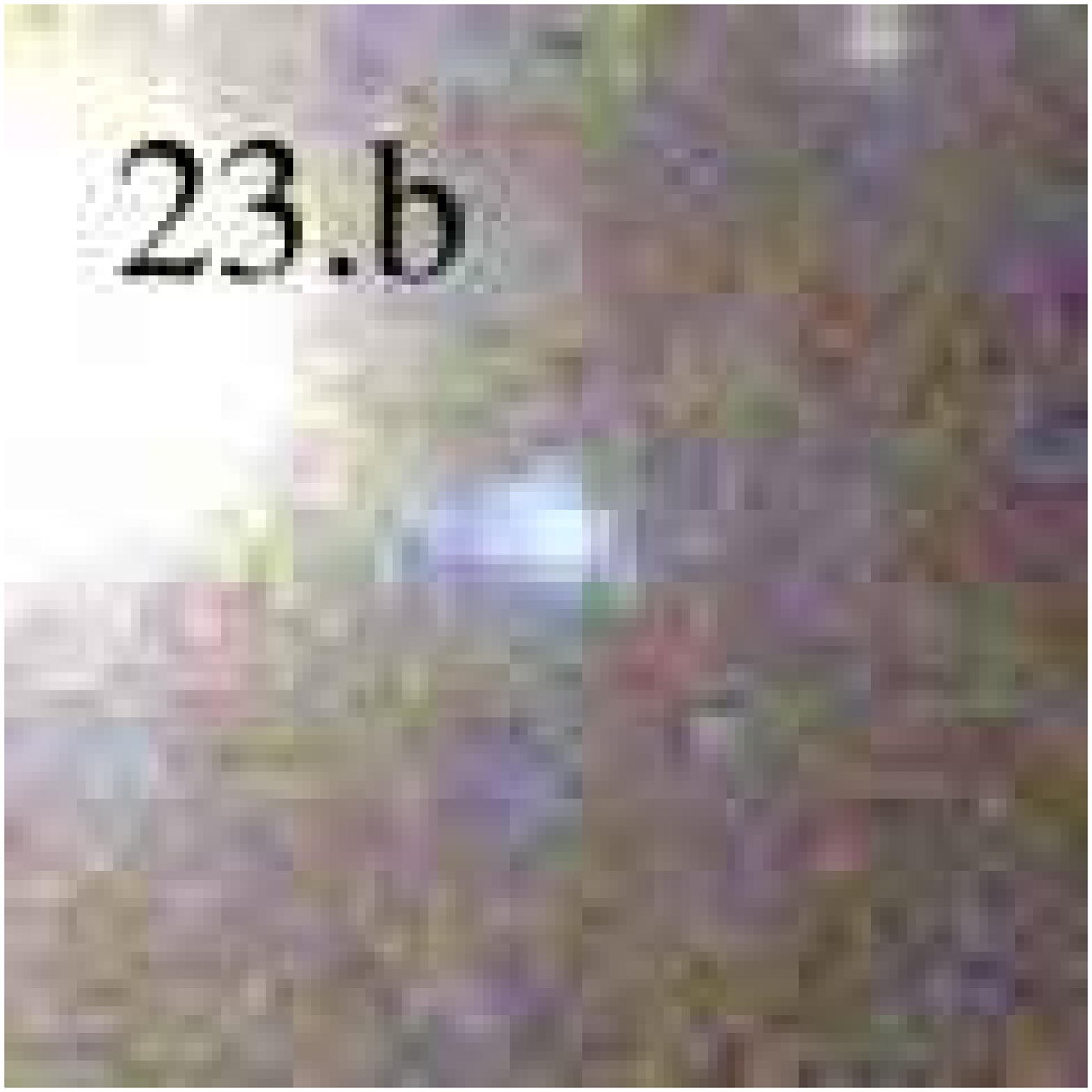}}
    & \multicolumn{1}{m{1.7cm}}{\includegraphics[height=2.00cm,clip]{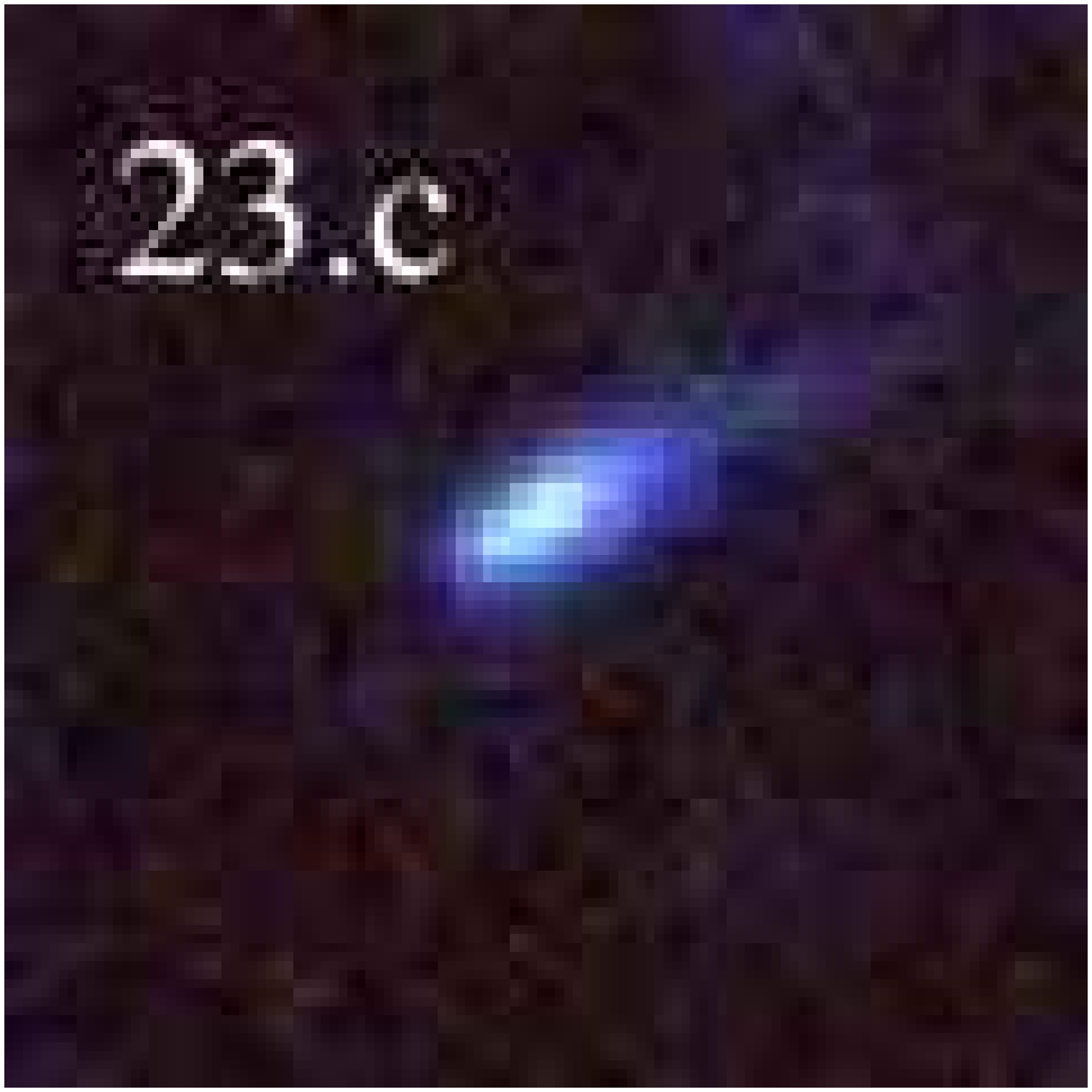}} \\
    \multicolumn{1}{m{1cm}}{{\Large NSIE}}
    & \multicolumn{1}{m{1.7cm}}{\includegraphics[height=2.00cm,clip]{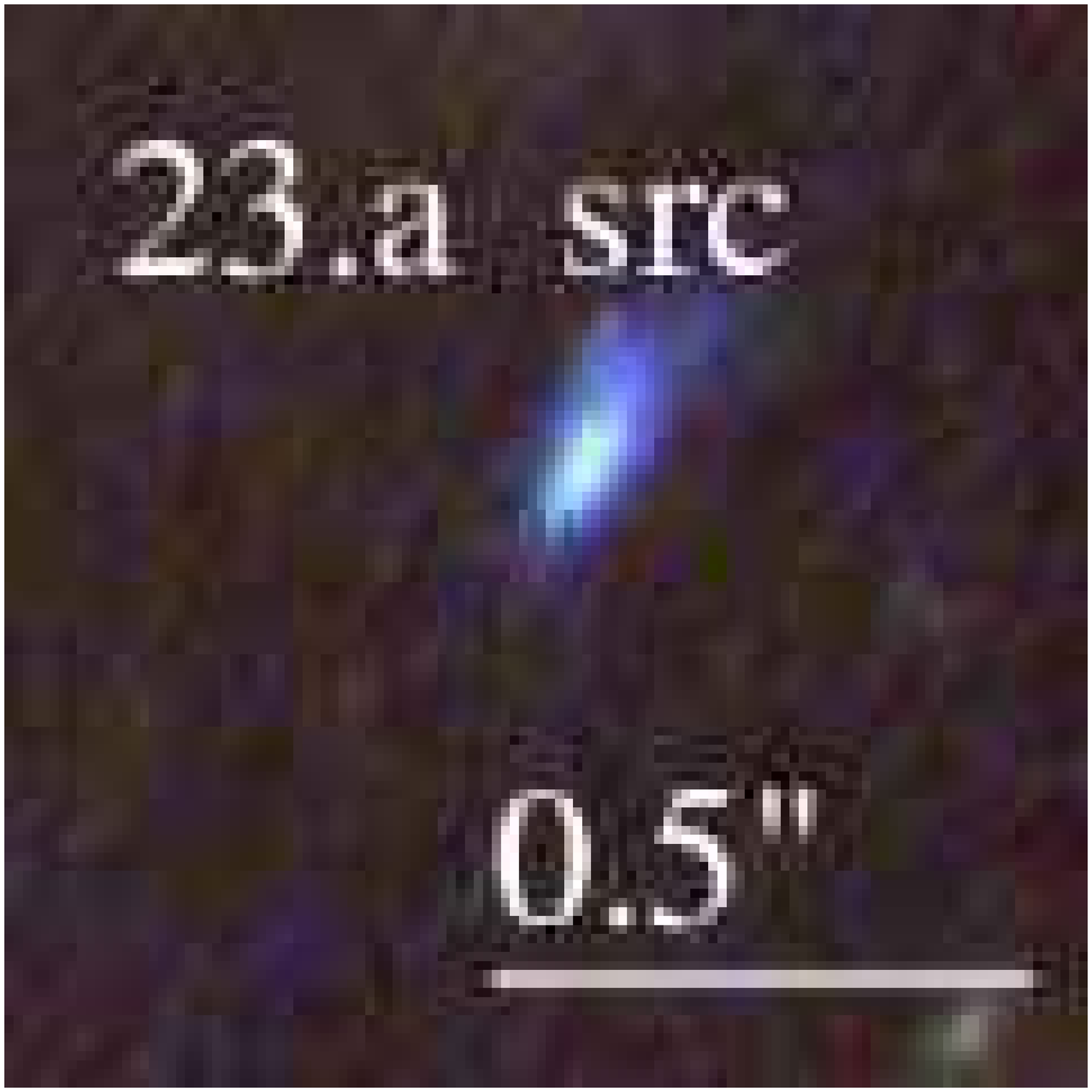}}
    & \multicolumn{1}{m{1.7cm}}{\includegraphics[height=2.00cm,clip]{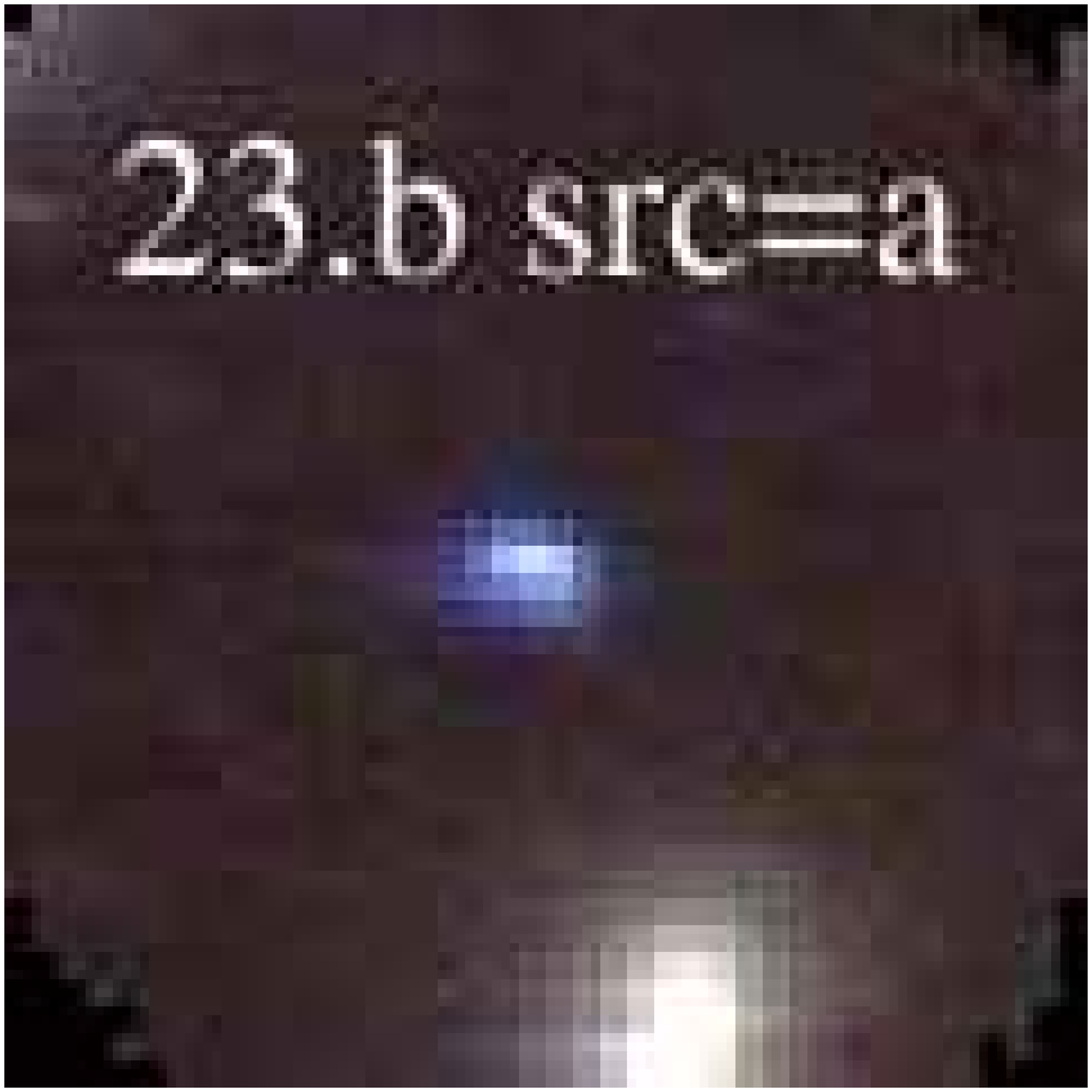}}
    & \multicolumn{1}{m{1.7cm}}{\includegraphics[height=2.00cm,clip]{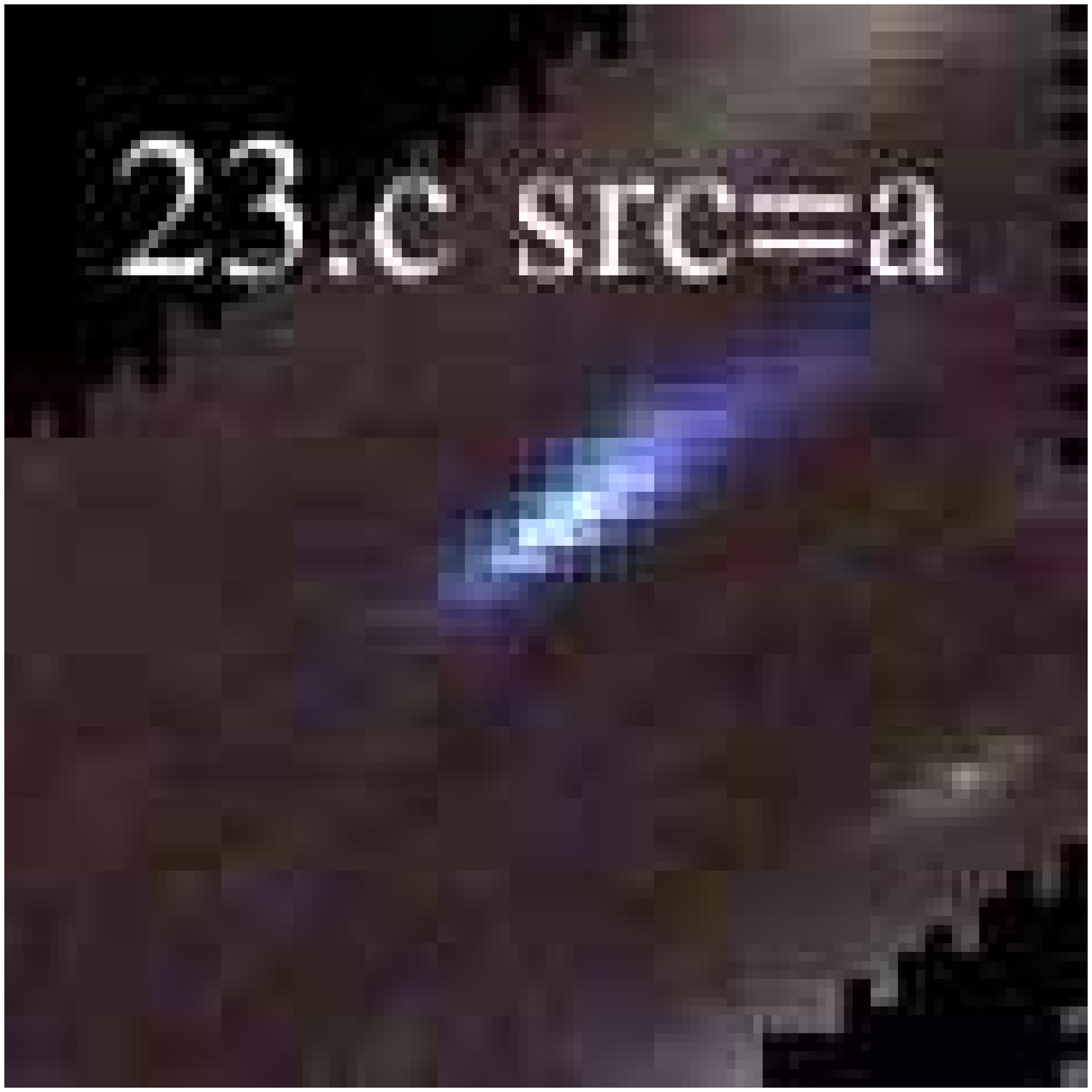}} \\
    \multicolumn{1}{m{1cm}}{{\Large ENFW}}
    & \multicolumn{1}{m{1.7cm}}{\includegraphics[height=2.00cm,clip]{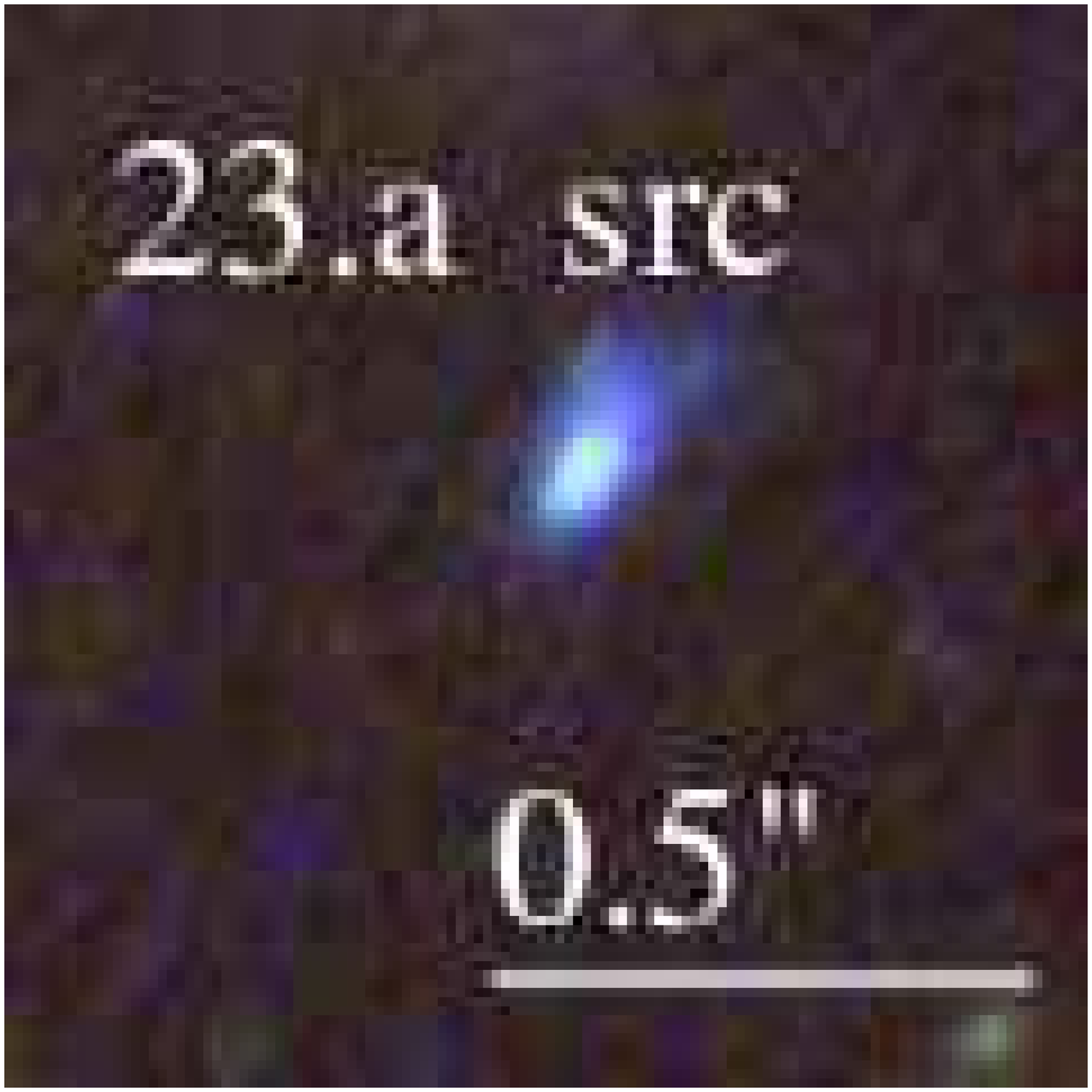}}
    & \multicolumn{1}{m{1.7cm}}{\includegraphics[height=2.00cm,clip]{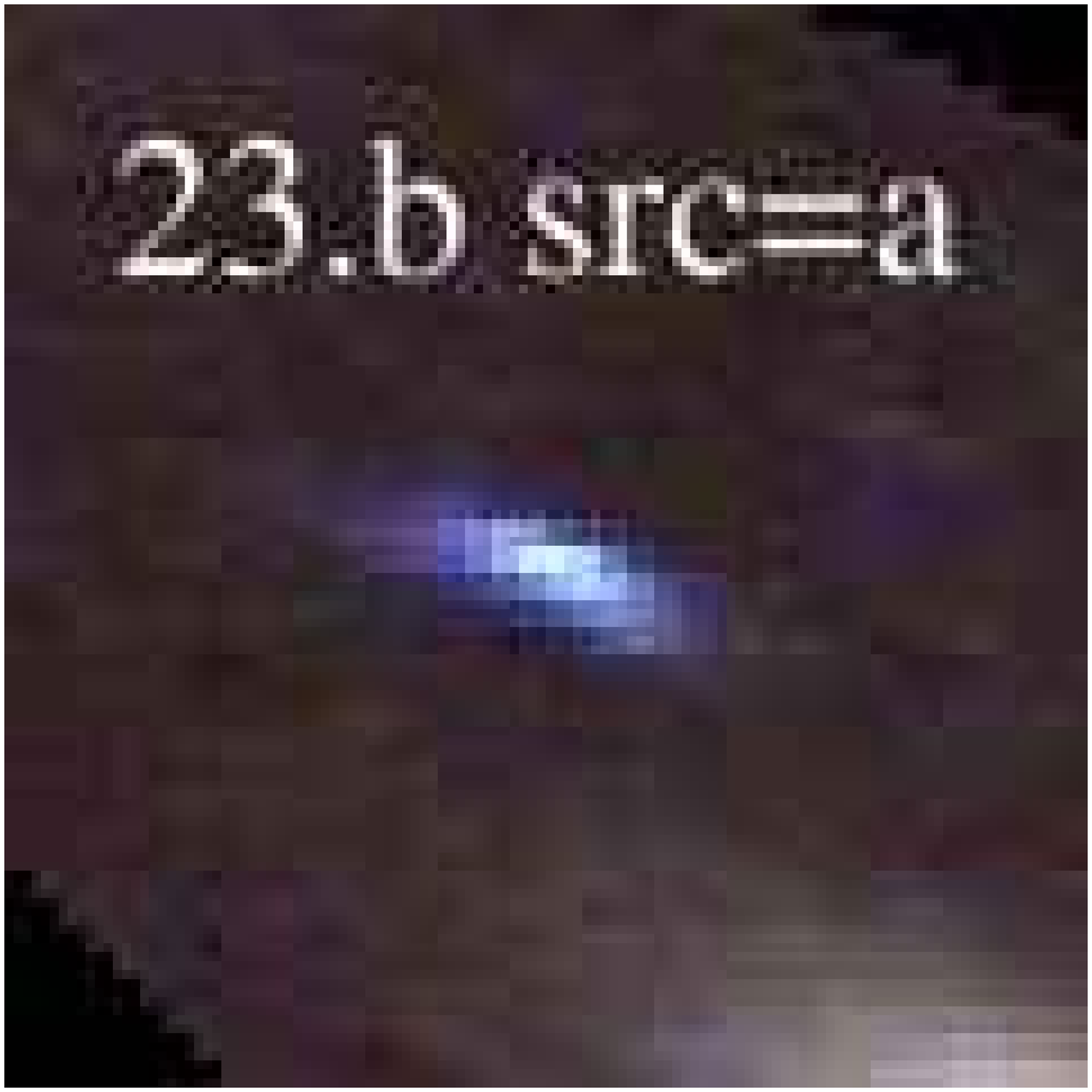}}
    & \multicolumn{1}{m{1.7cm}}{\includegraphics[height=2.00cm,clip]{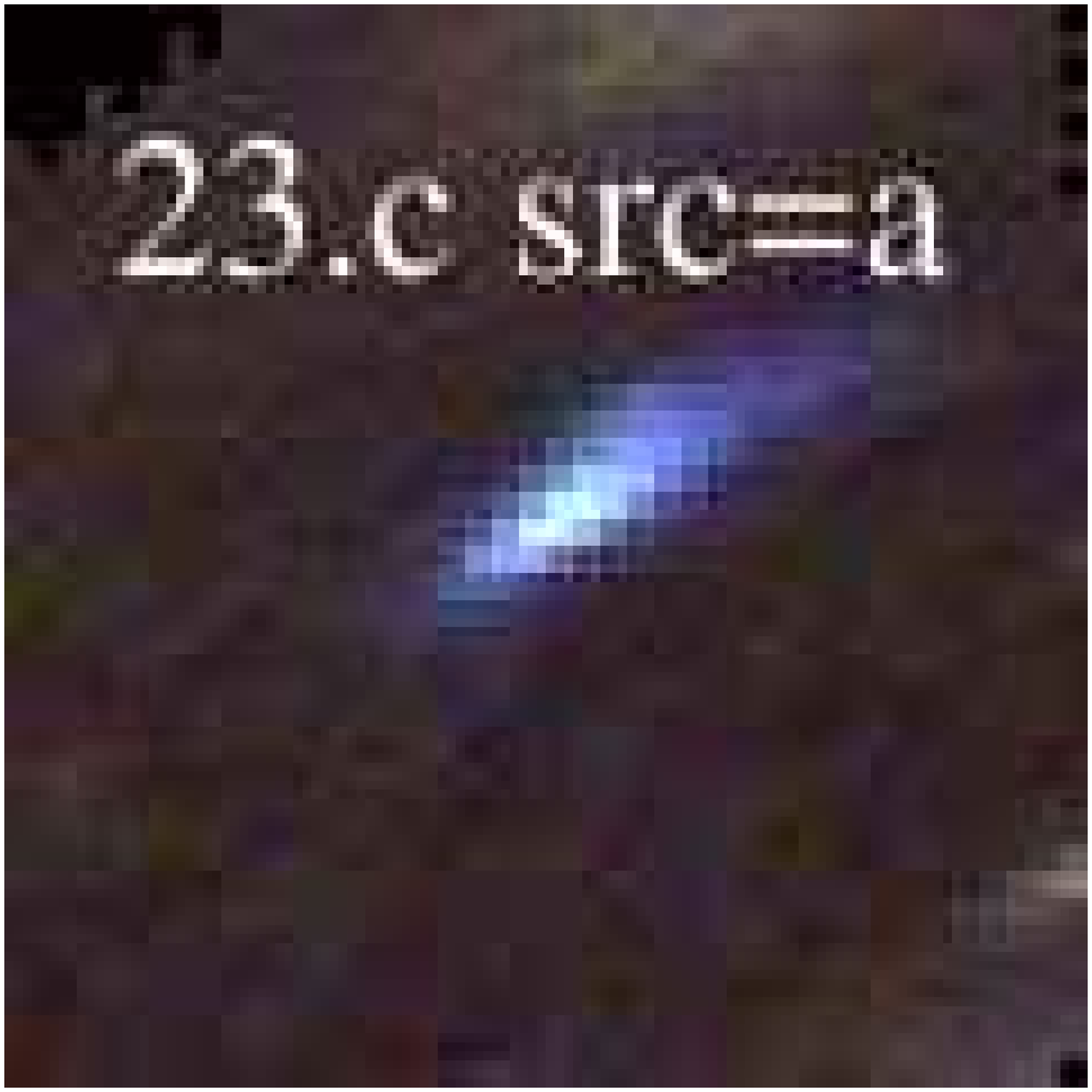}} \\
  \end{tabular}

\end{table*}

\begin{table*}
  \caption{Image system 24:}\vspace{0mm}
  \begin{tabular}{cccccc}
    \multicolumn{1}{m{1cm}}{{\Large A1689}}
    & \multicolumn{1}{m{1.7cm}}{\includegraphics[height=2.00cm,clip]{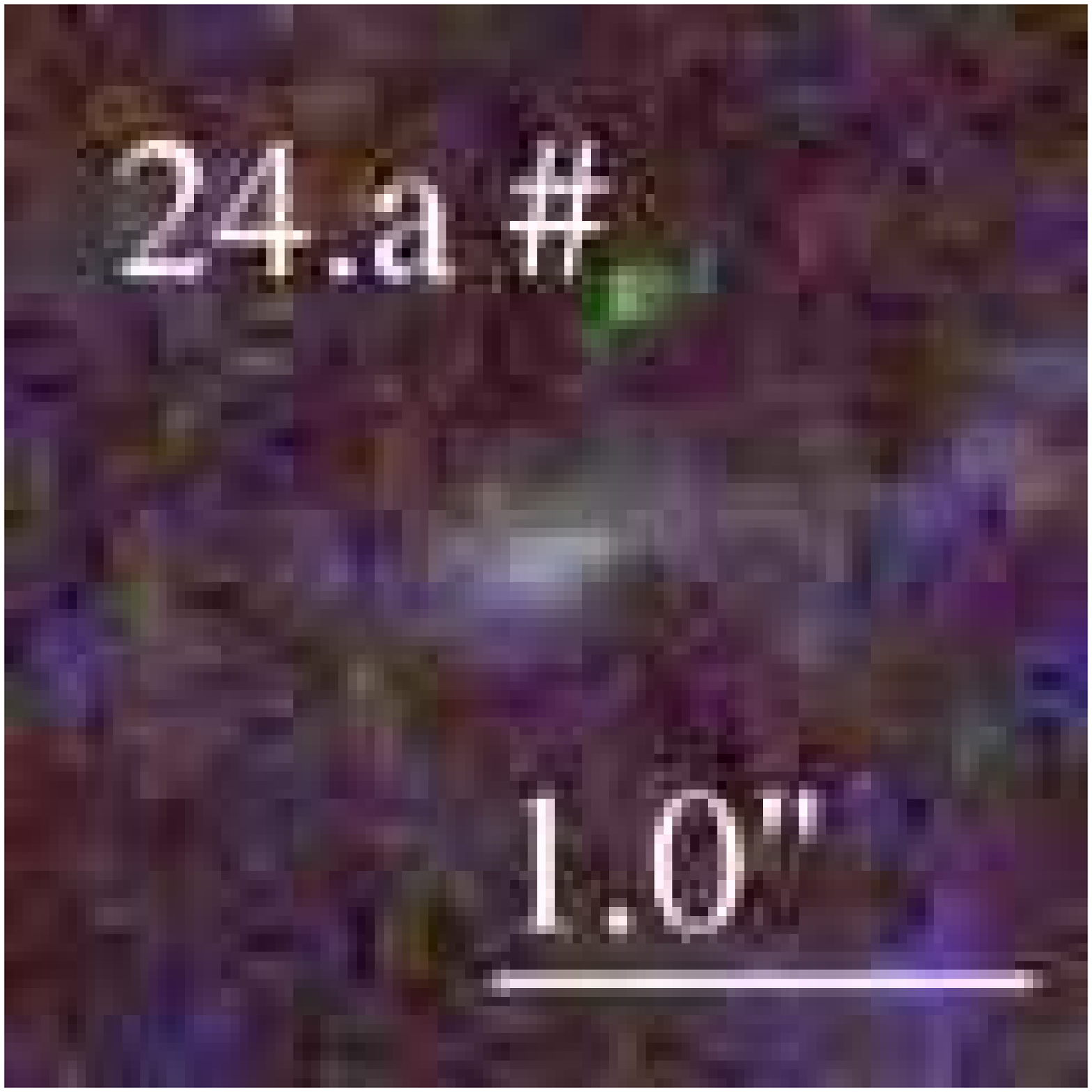}}
    & \multicolumn{1}{m{1.7cm}}{\includegraphics[height=2.00cm,clip]{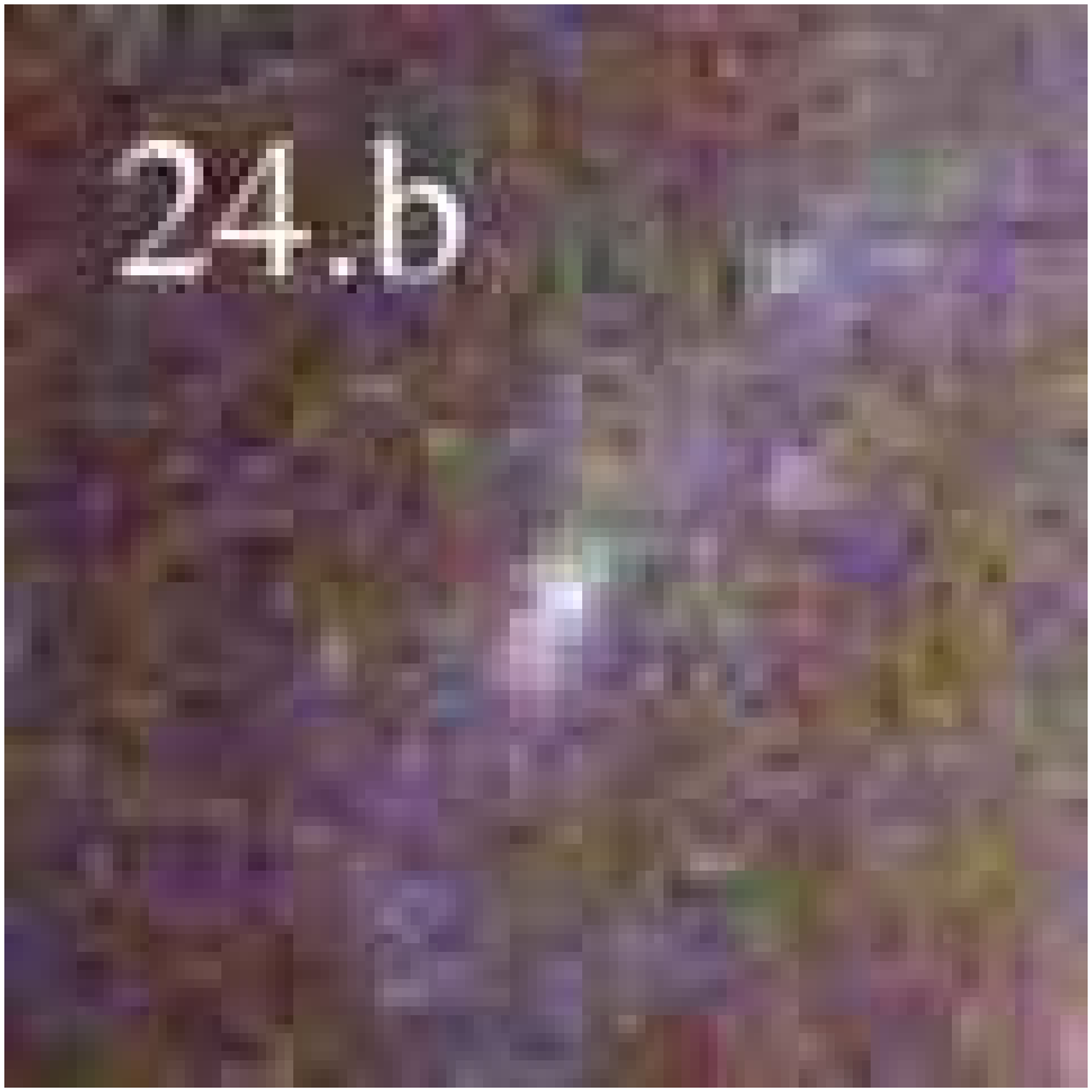}}
    & \multicolumn{1}{m{1.7cm}}{\includegraphics[height=2.00cm,clip]{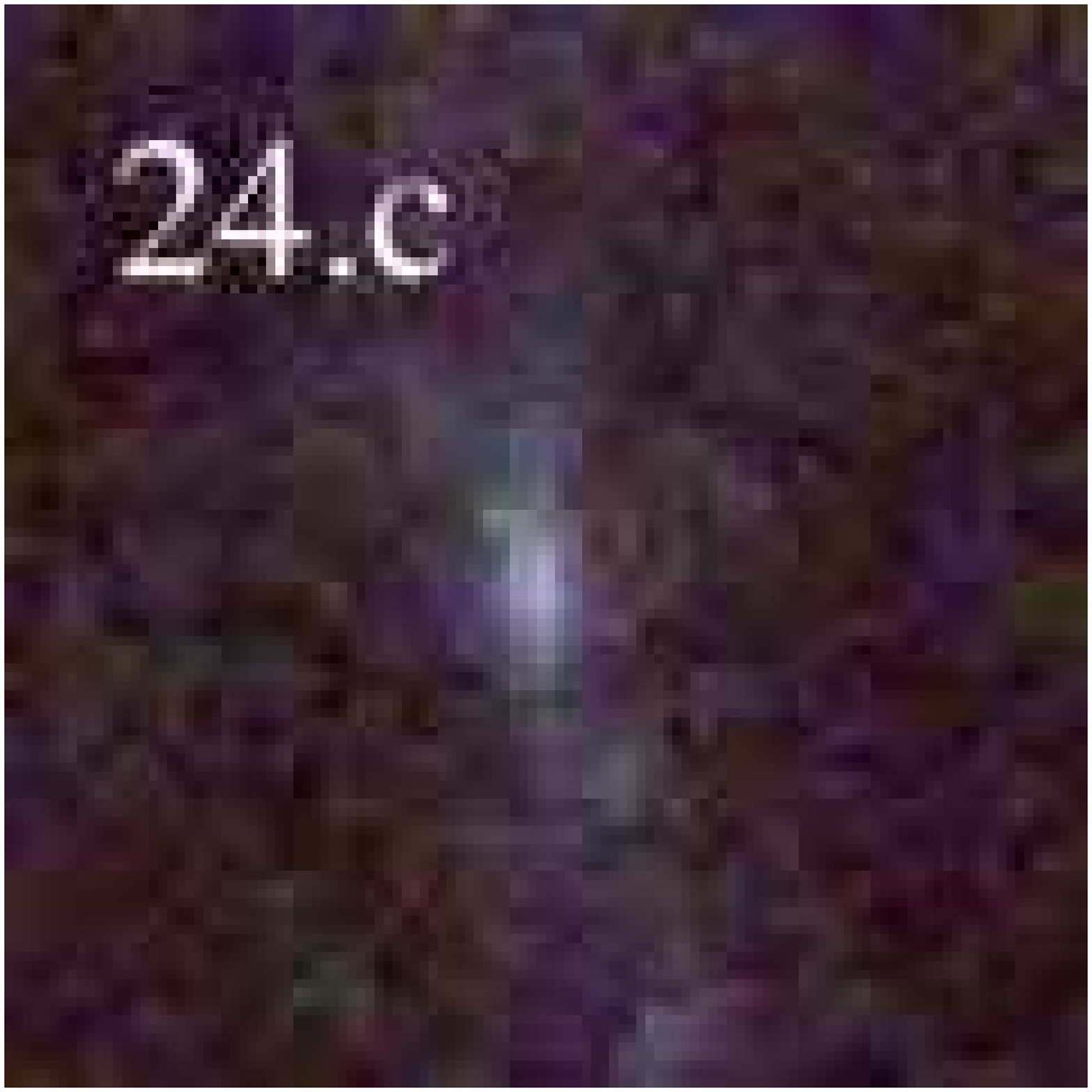}}
    & \multicolumn{1}{m{1.7cm}}{\includegraphics[height=2.00cm,clip]{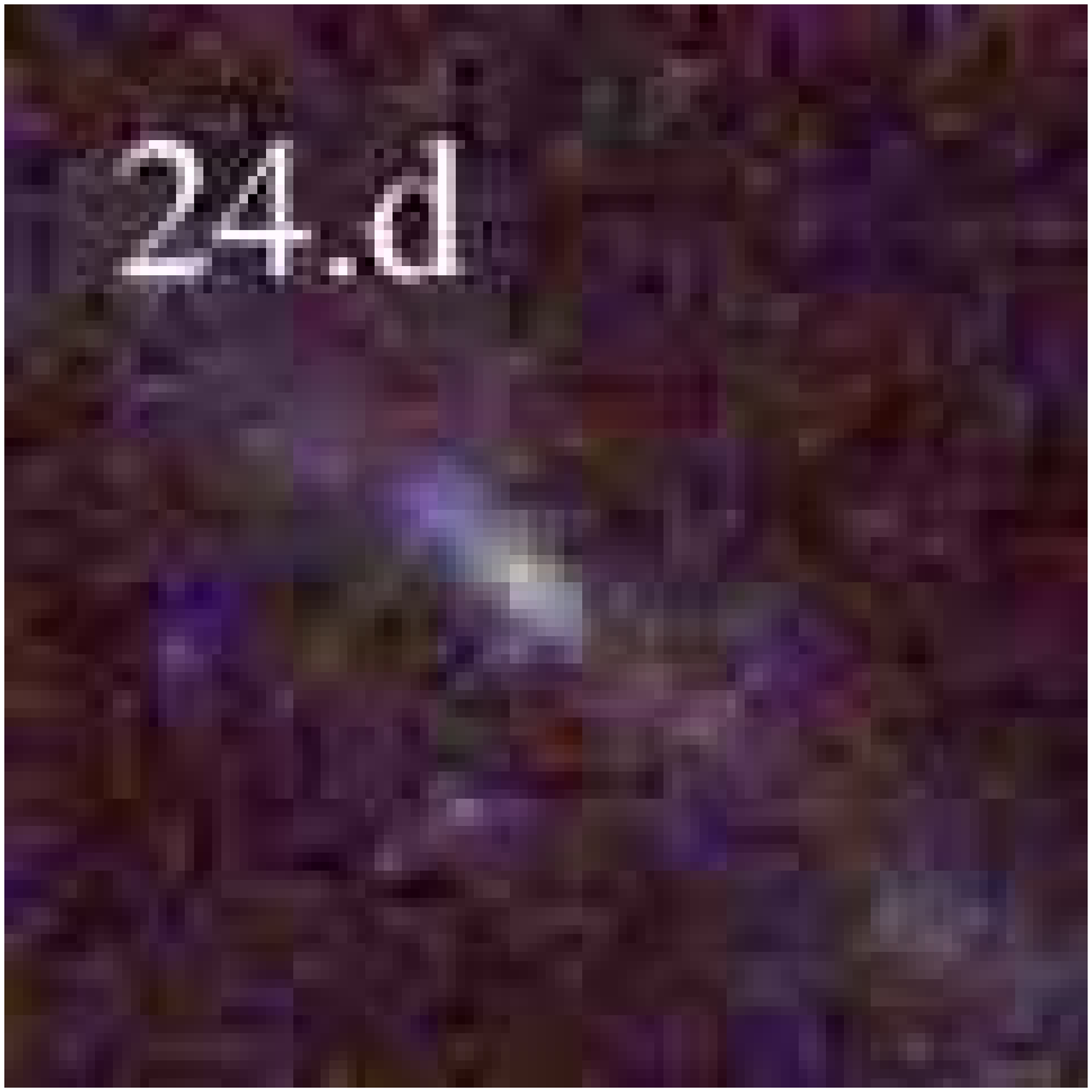}}
    & \multicolumn{1}{m{1.7cm}}{\includegraphics[height=2.00cm,clip]{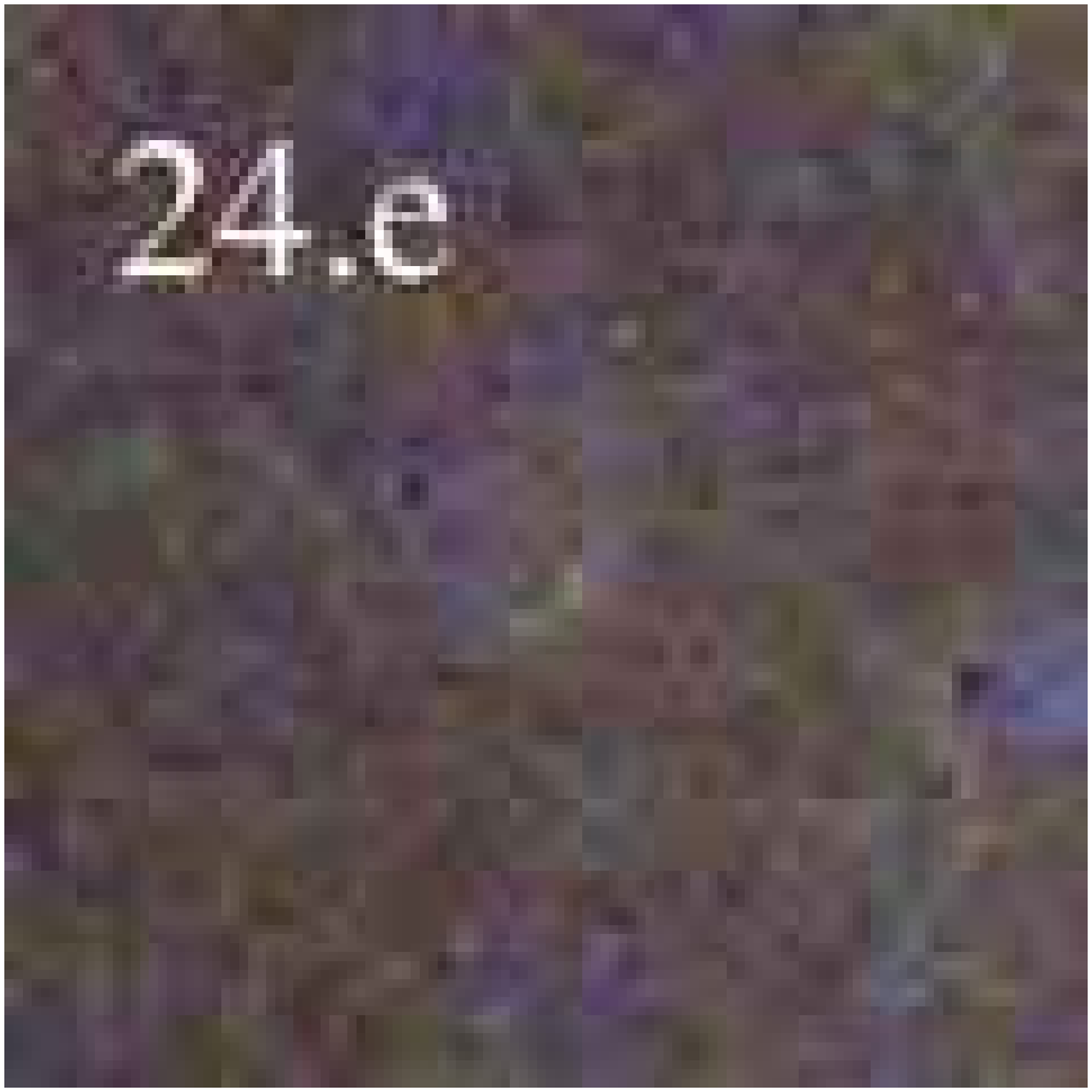}} \\
    \multicolumn{1}{m{1cm}}{{\Large NSIE}}
    & \multicolumn{1}{m{1.7cm}}{\includegraphics[height=2.00cm,clip]{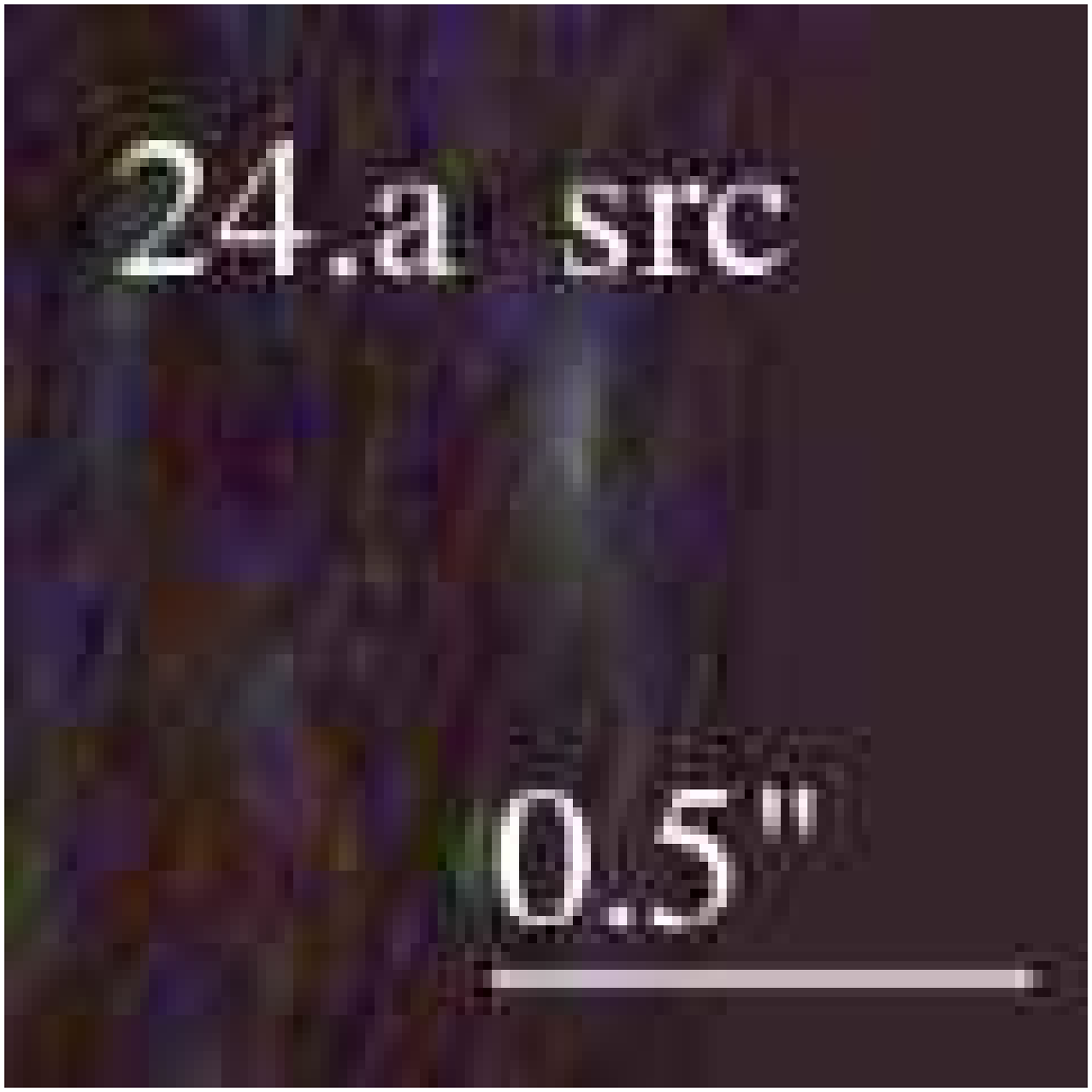}}
    & \multicolumn{1}{m{1.7cm}}{\includegraphics[height=2.00cm,clip]{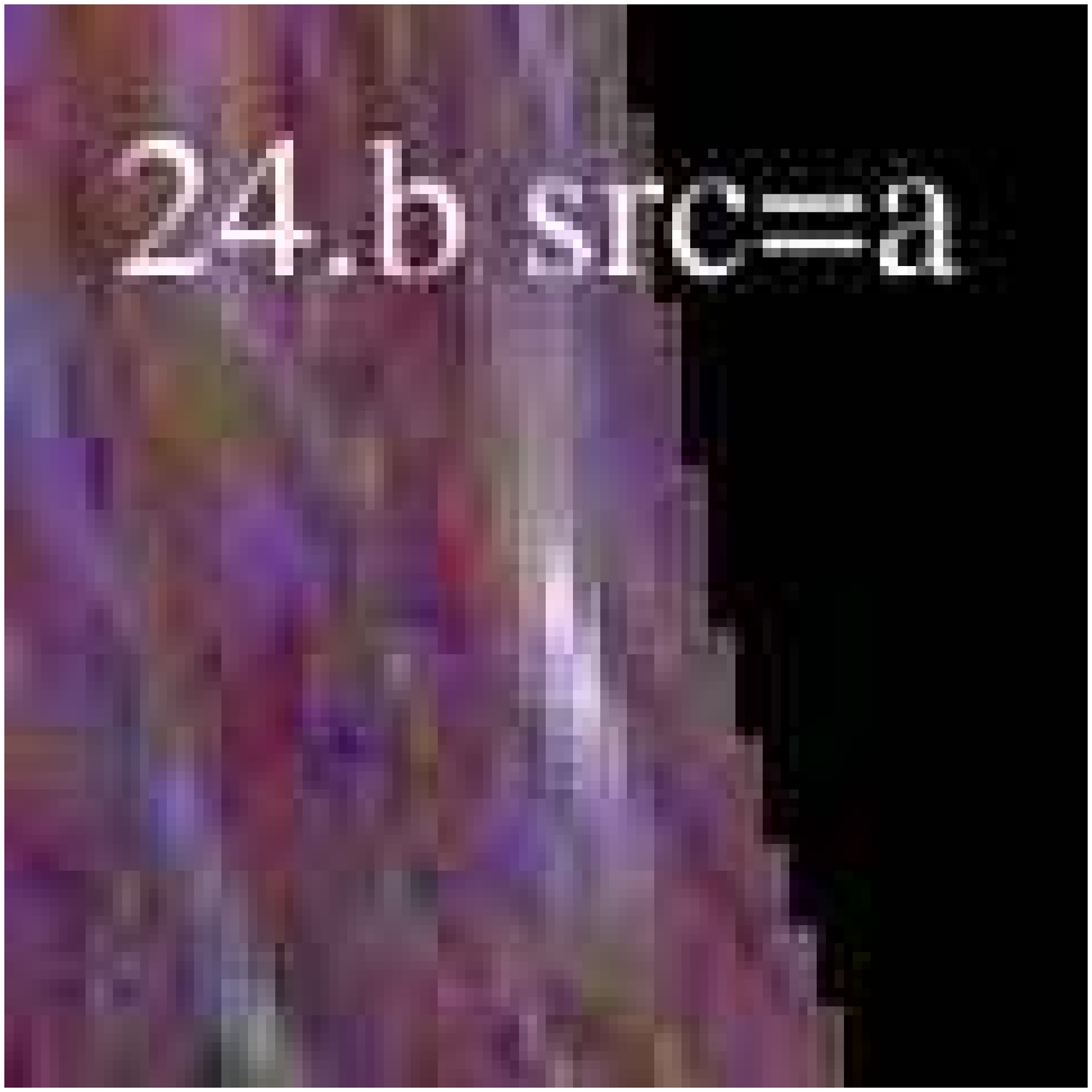}}
    & \multicolumn{1}{m{1.7cm}}{\includegraphics[height=2.00cm,clip]{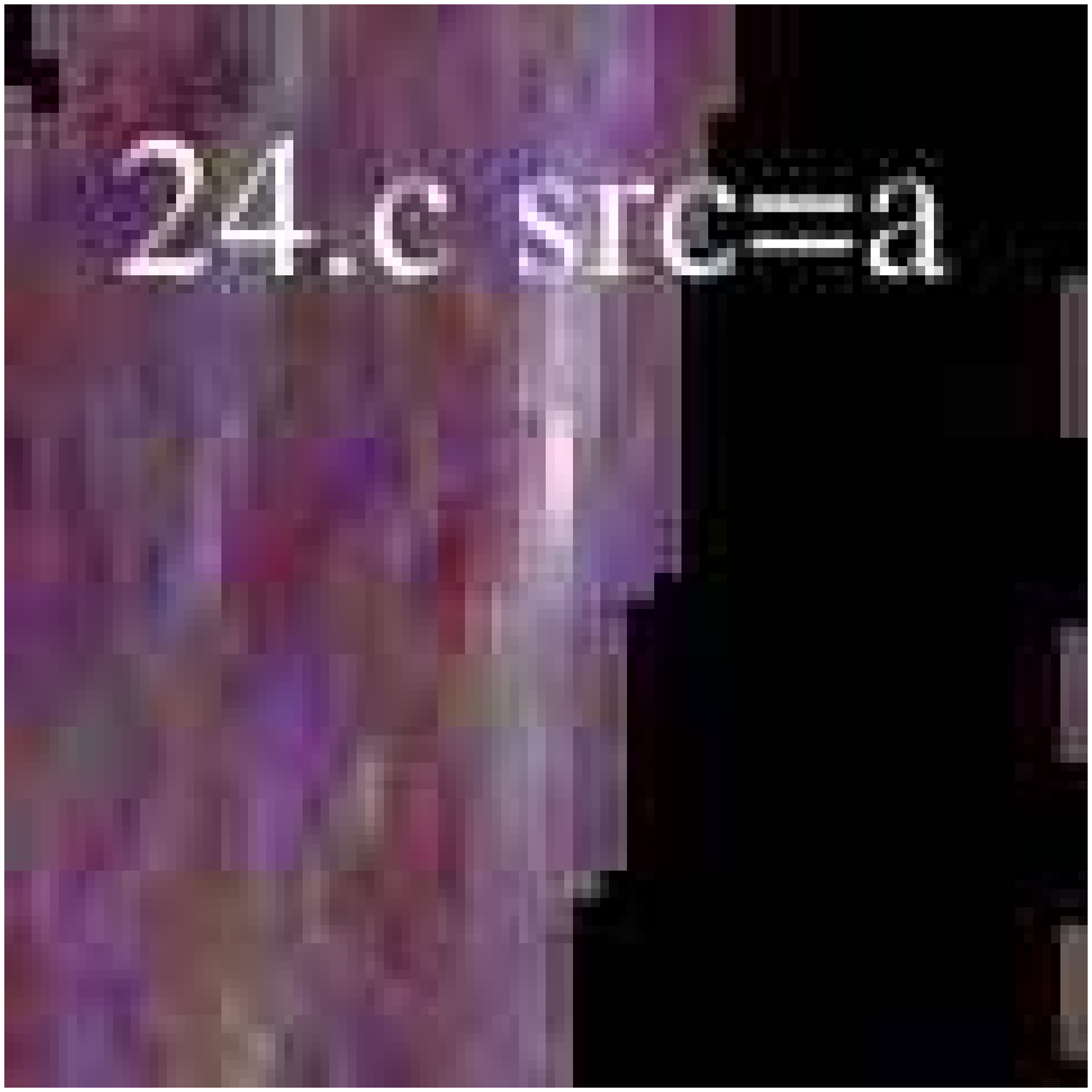}}
    & \multicolumn{1}{m{1.7cm}}{\includegraphics[height=2.00cm,clip]{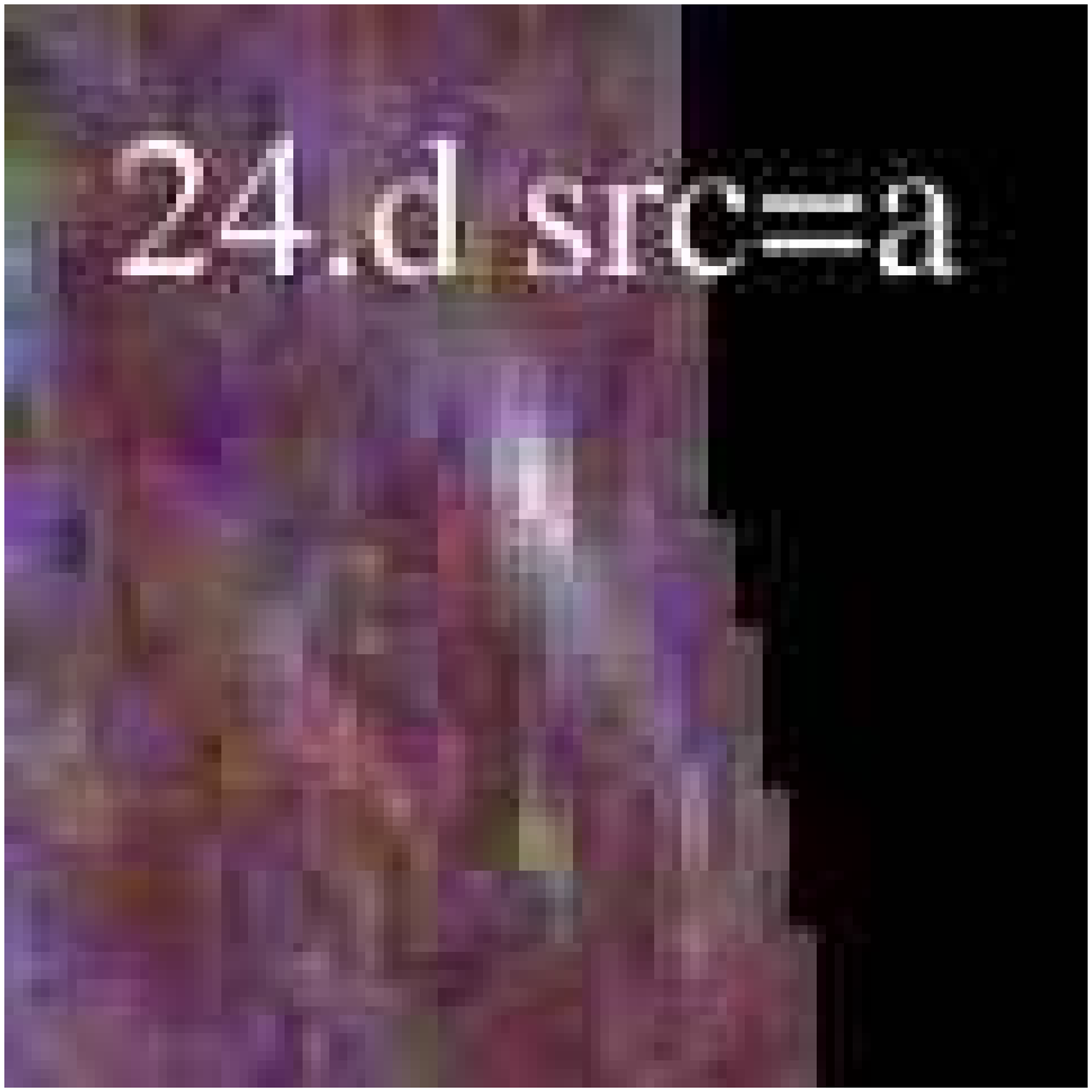}}
    & \multicolumn{1}{m{1.7cm}}{\includegraphics[height=2.00cm,clip]{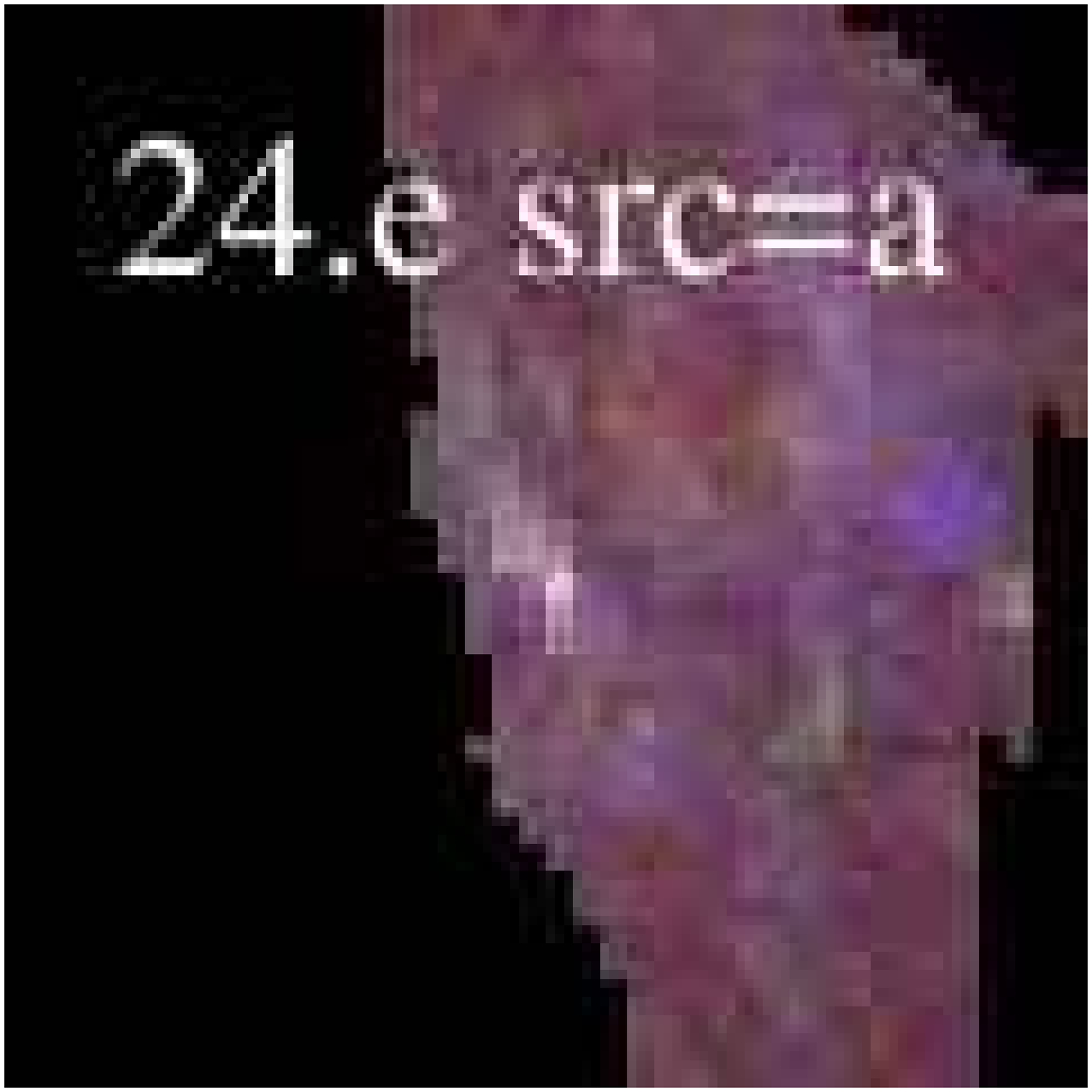}} \\
    \multicolumn{1}{m{1cm}}{{\Large ENFW}}
    & \multicolumn{1}{m{1.7cm}}{\includegraphics[height=2.00cm,clip]{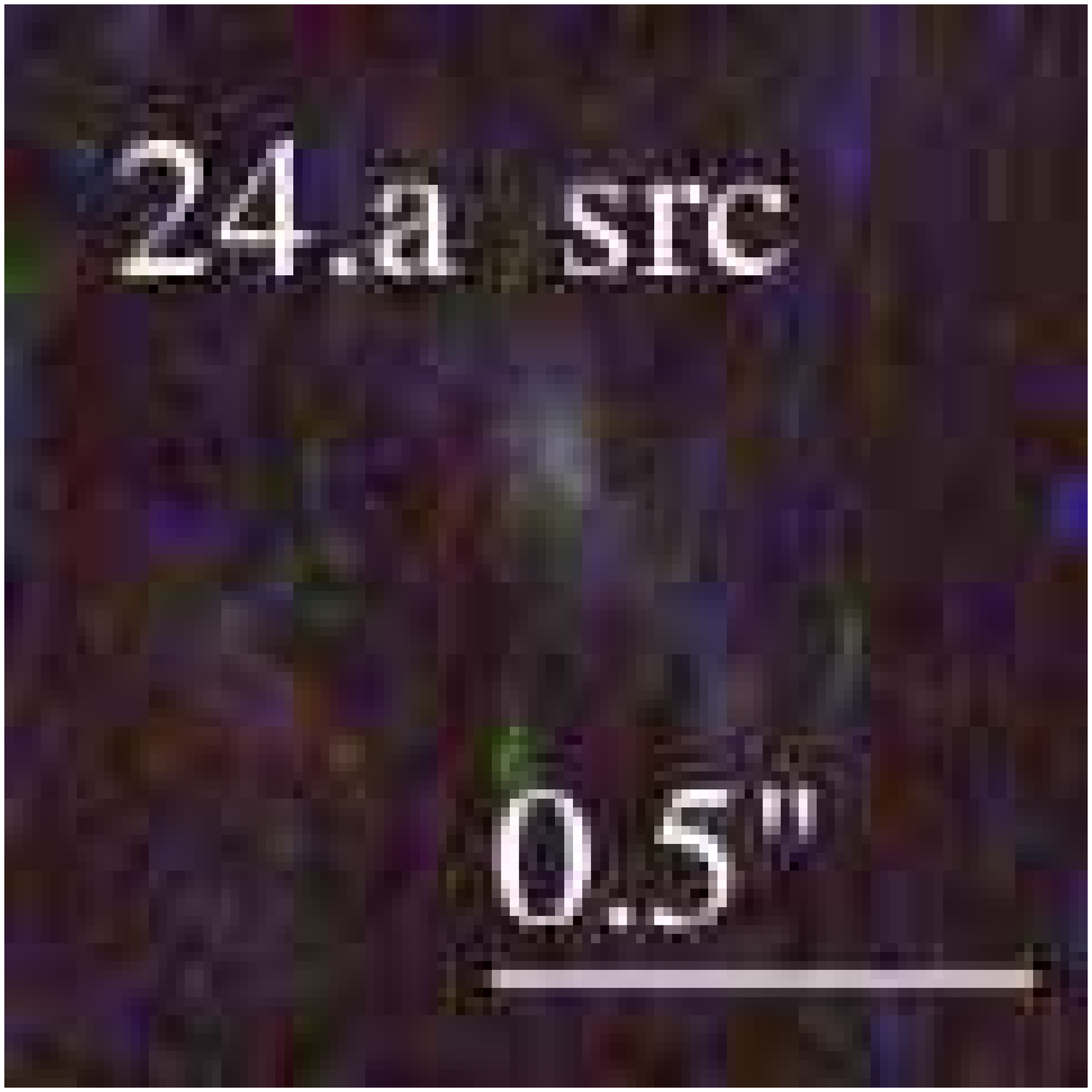}}
    & \multicolumn{1}{m{1.7cm}}{\includegraphics[height=2.00cm,clip]{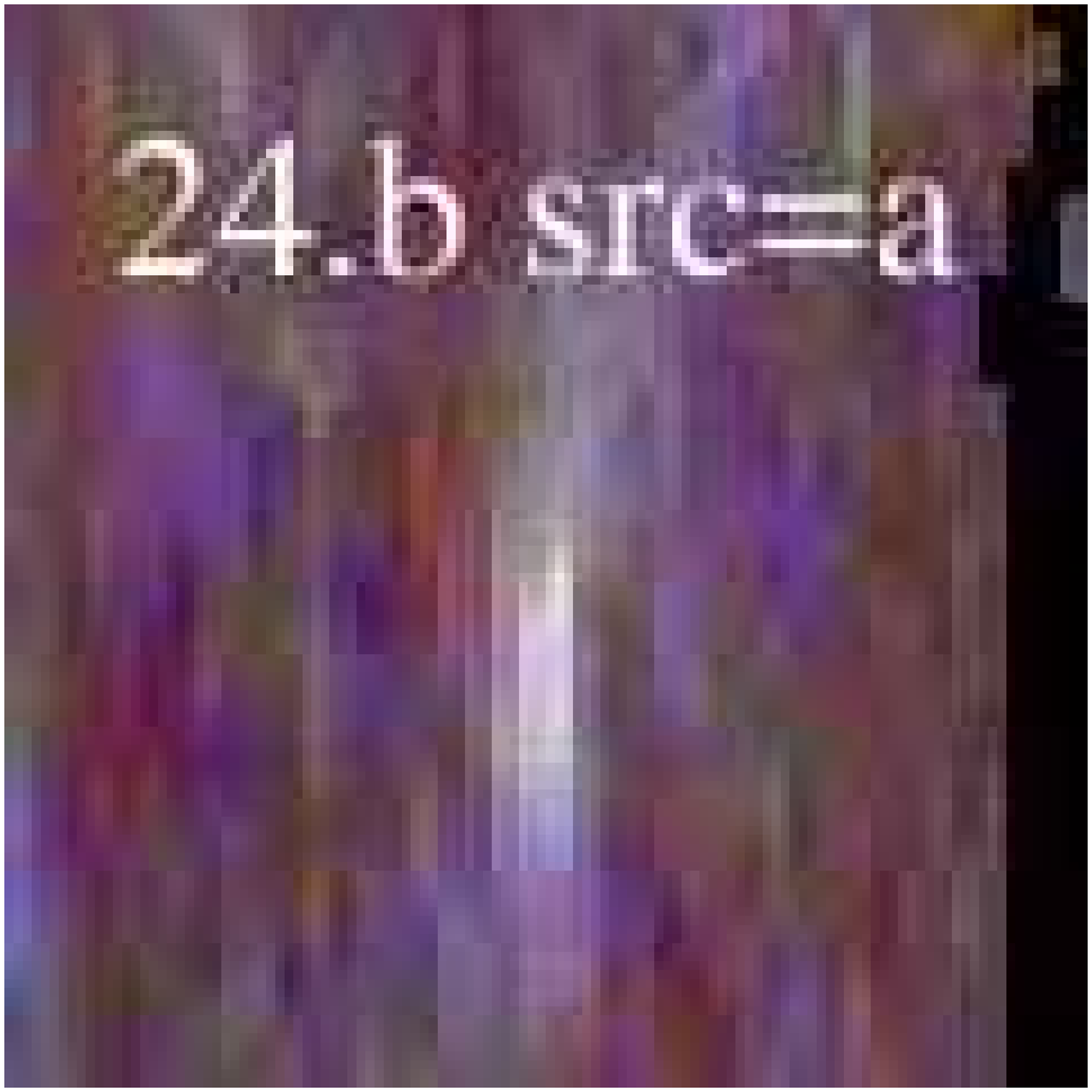}}
    & \multicolumn{1}{m{1.7cm}}{\includegraphics[height=2.00cm,clip]{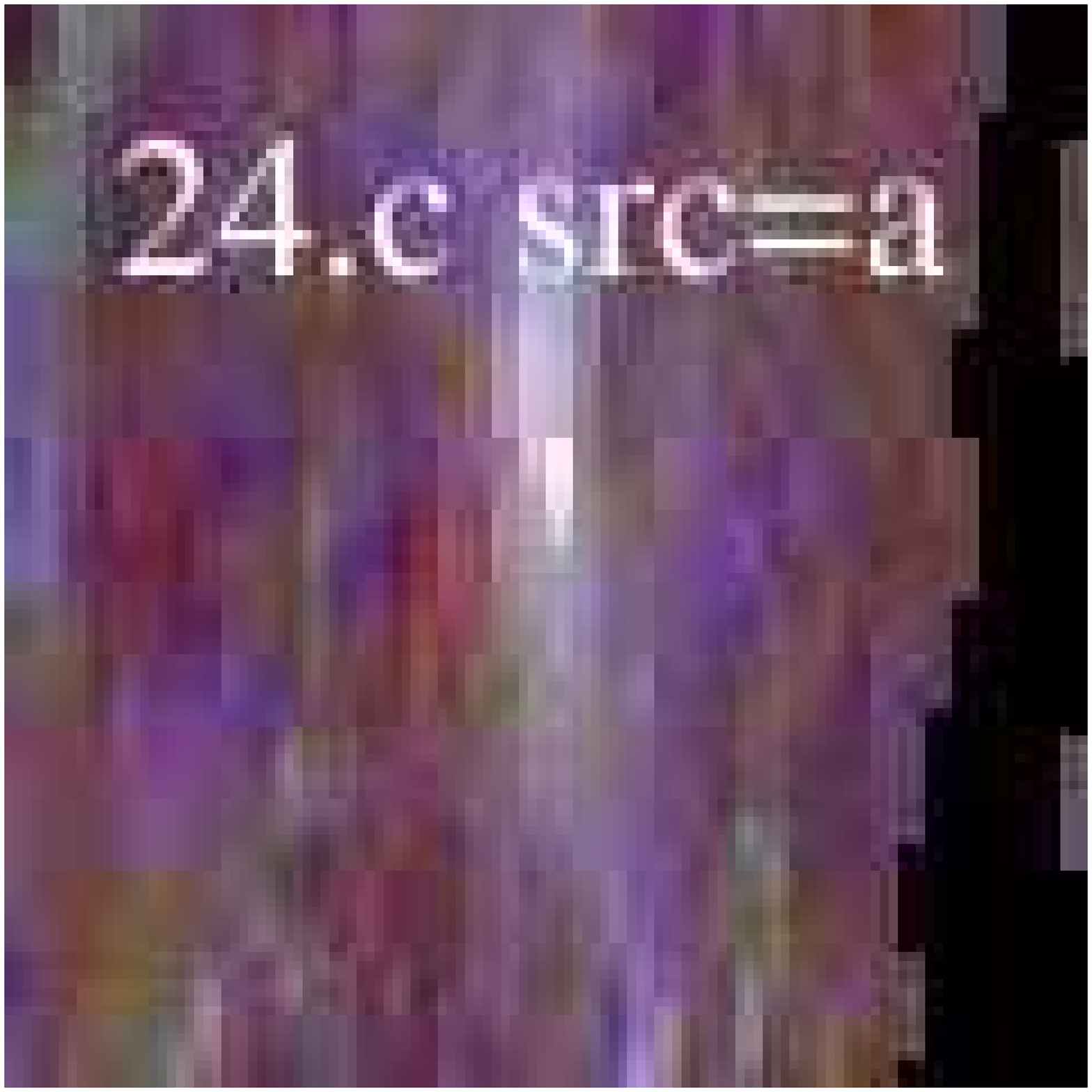}}
    & \multicolumn{1}{m{1.7cm}}{\includegraphics[height=2.00cm,clip]{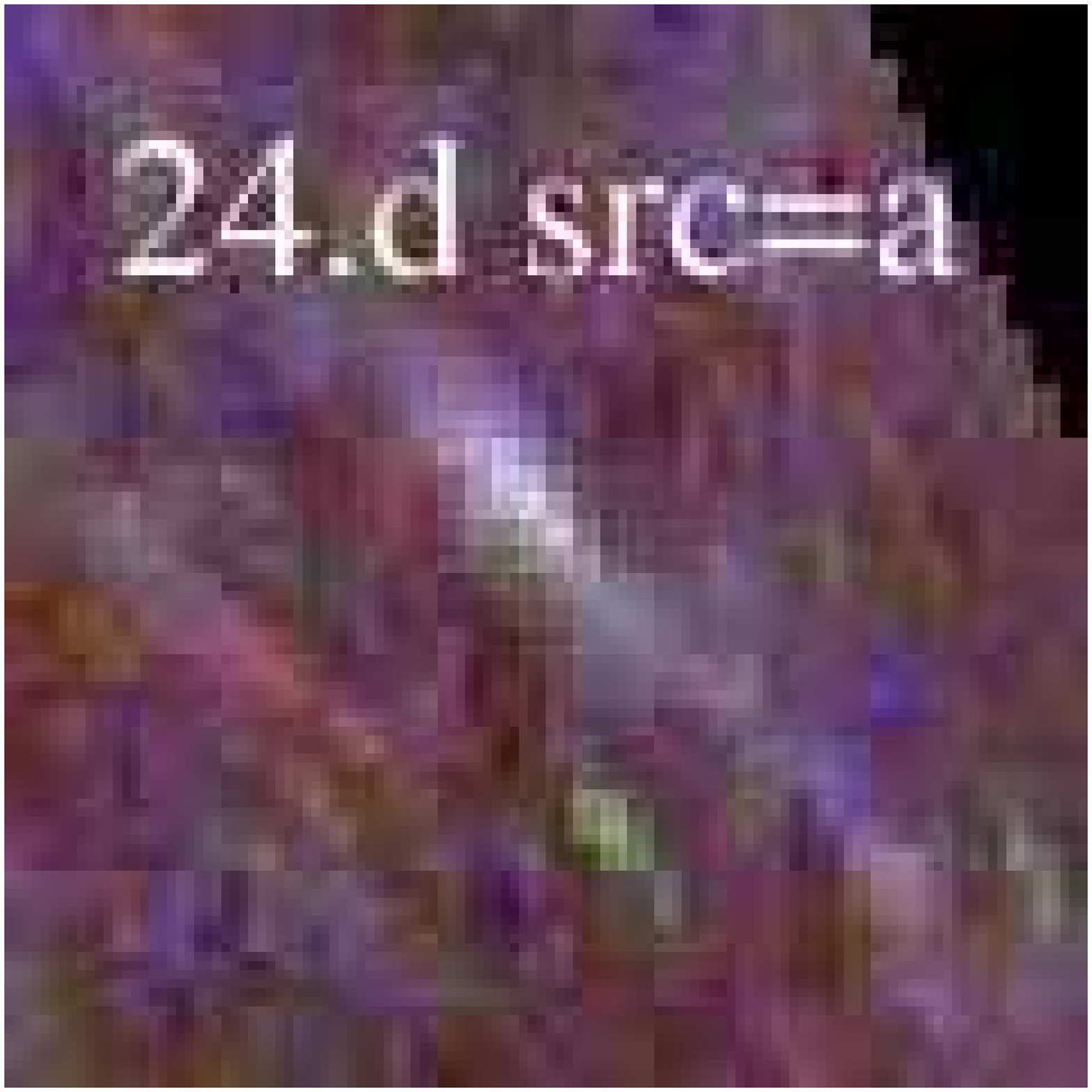}}
    & \multicolumn{1}{m{1.7cm}}{\includegraphics[height=2.00cm,clip]{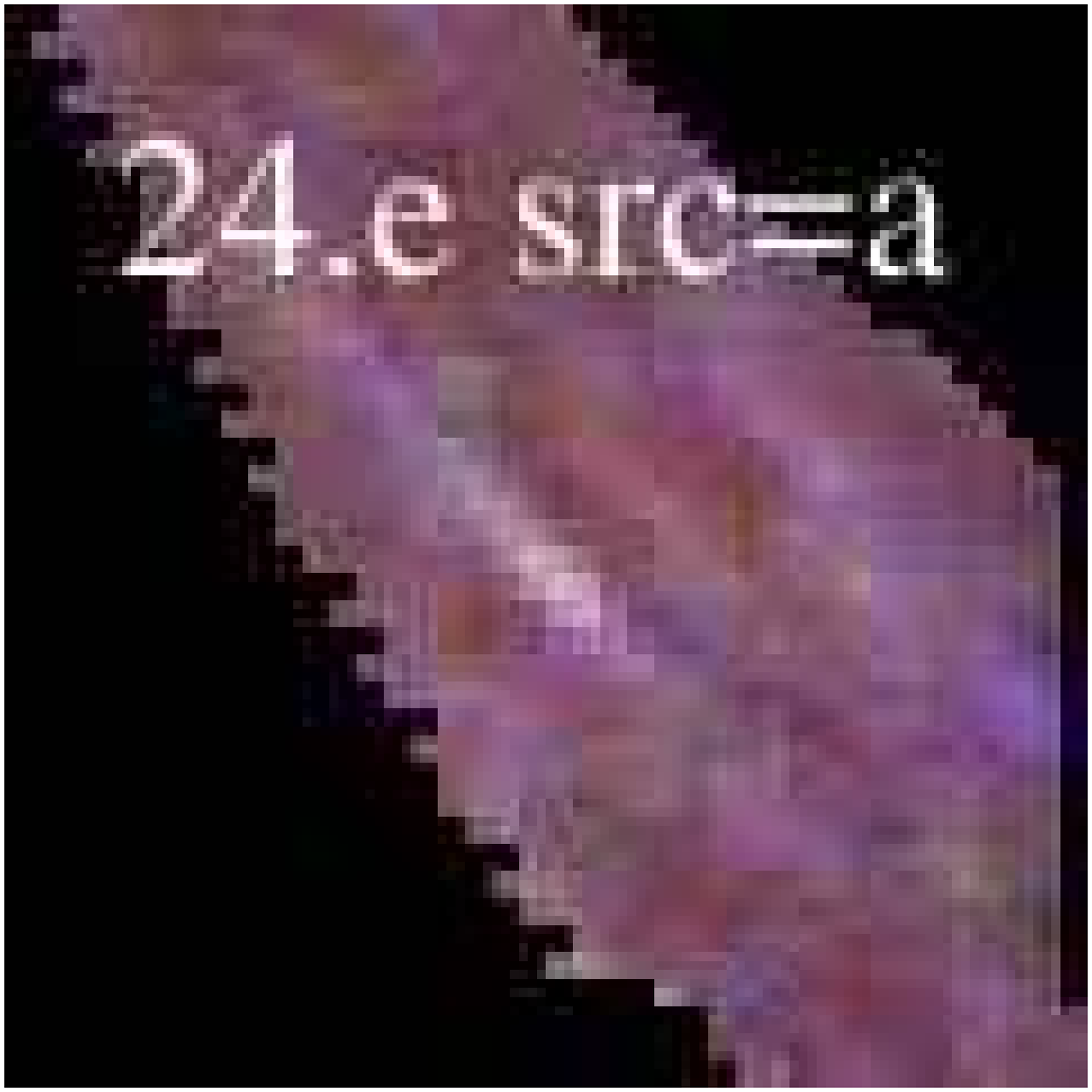}} \\
  \end{tabular}

\end{table*}

\clearpage

\begin{table*}
  \caption{Image system 25:}\vspace{0mm}
  \begin{tabular}{ccc}
    \multicolumn{1}{m{1cm}}{{\Large A1689}}
    & \multicolumn{1}{m{1.7cm}}{\includegraphics[height=2.00cm,clip]{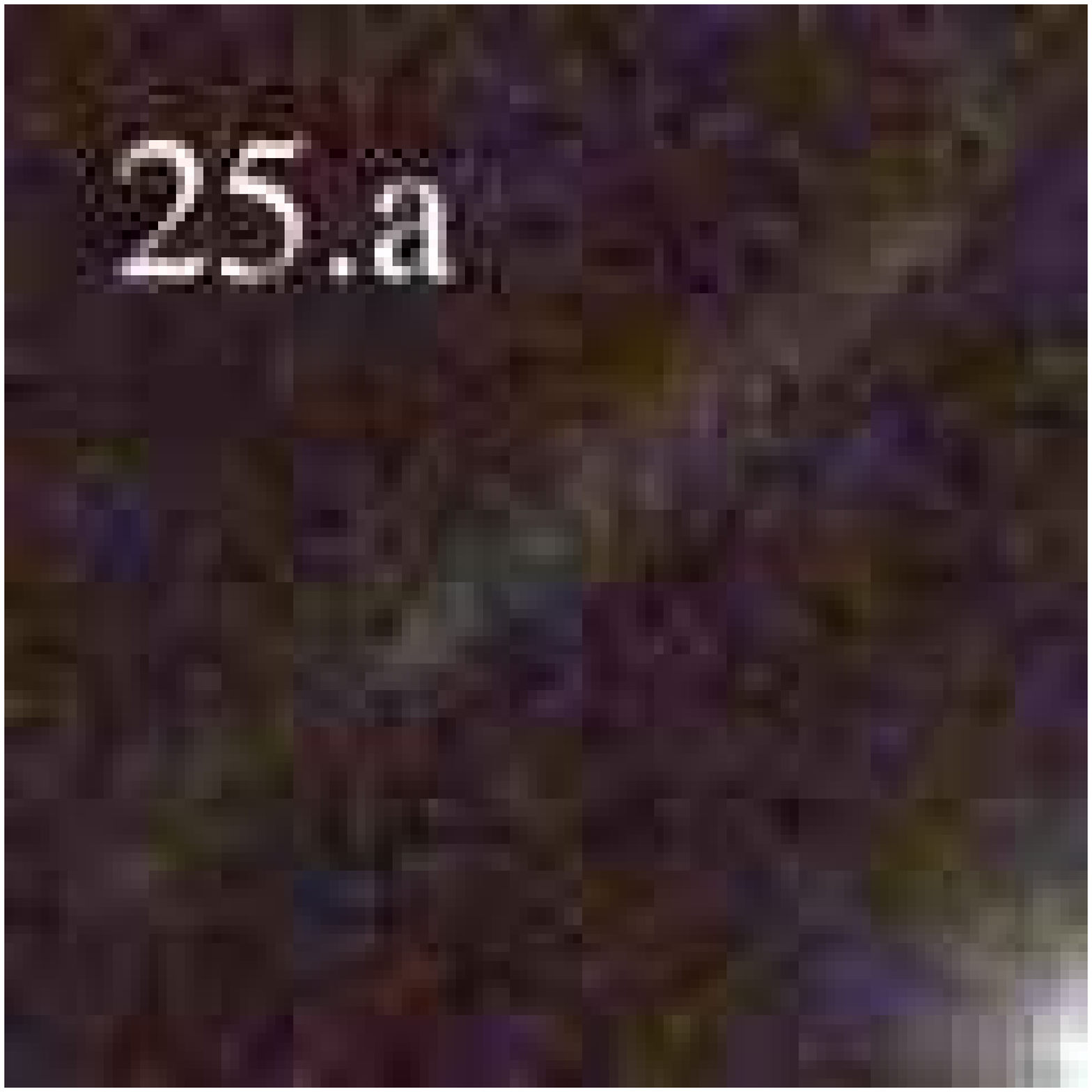}}
    & \multicolumn{1}{m{1.7cm}}{\includegraphics[height=2.00cm,clip]{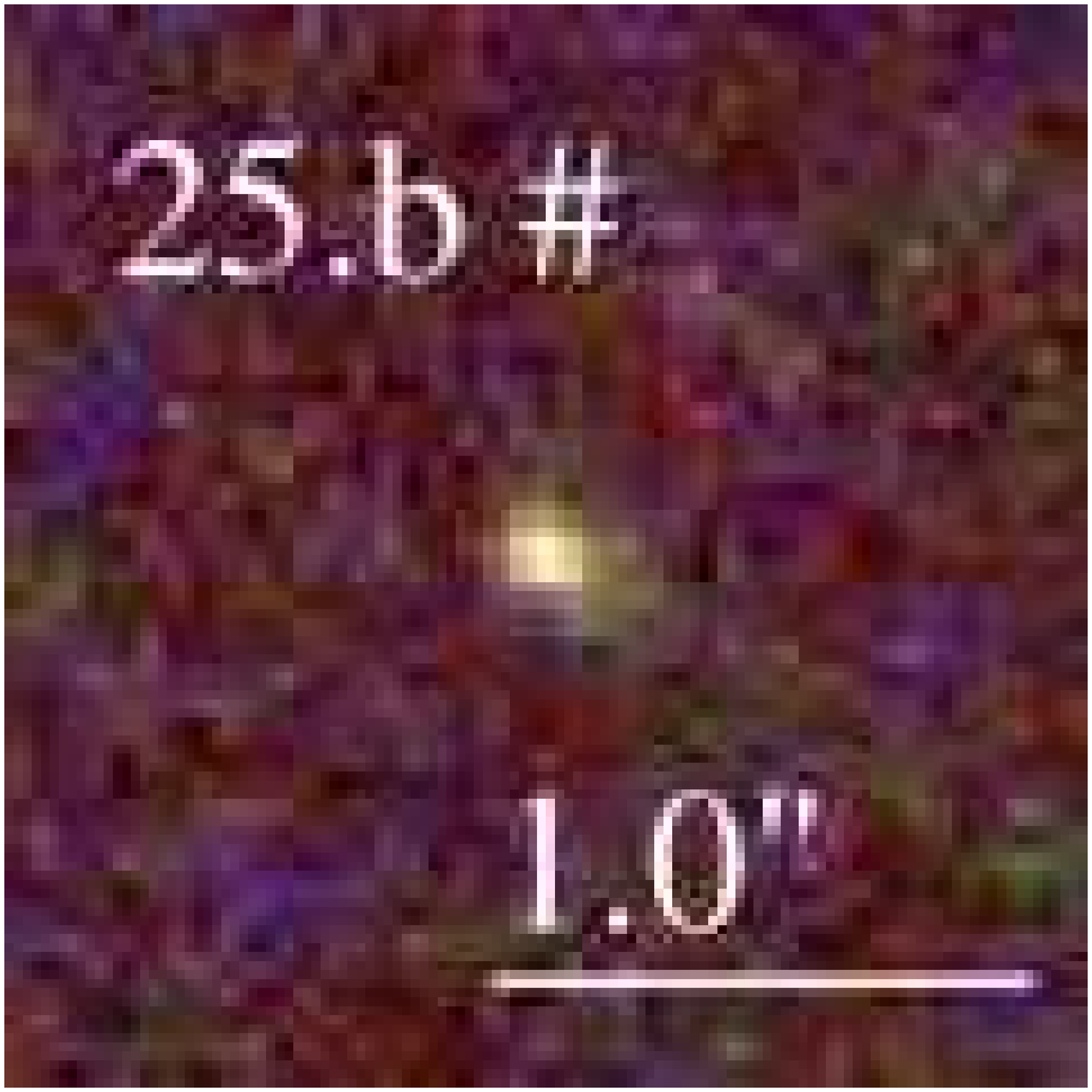}} \\
    \multicolumn{1}{m{1cm}}{{\Large NSIE}}
    & \multicolumn{1}{m{1.7cm}}{\includegraphics[height=2.00cm,clip]{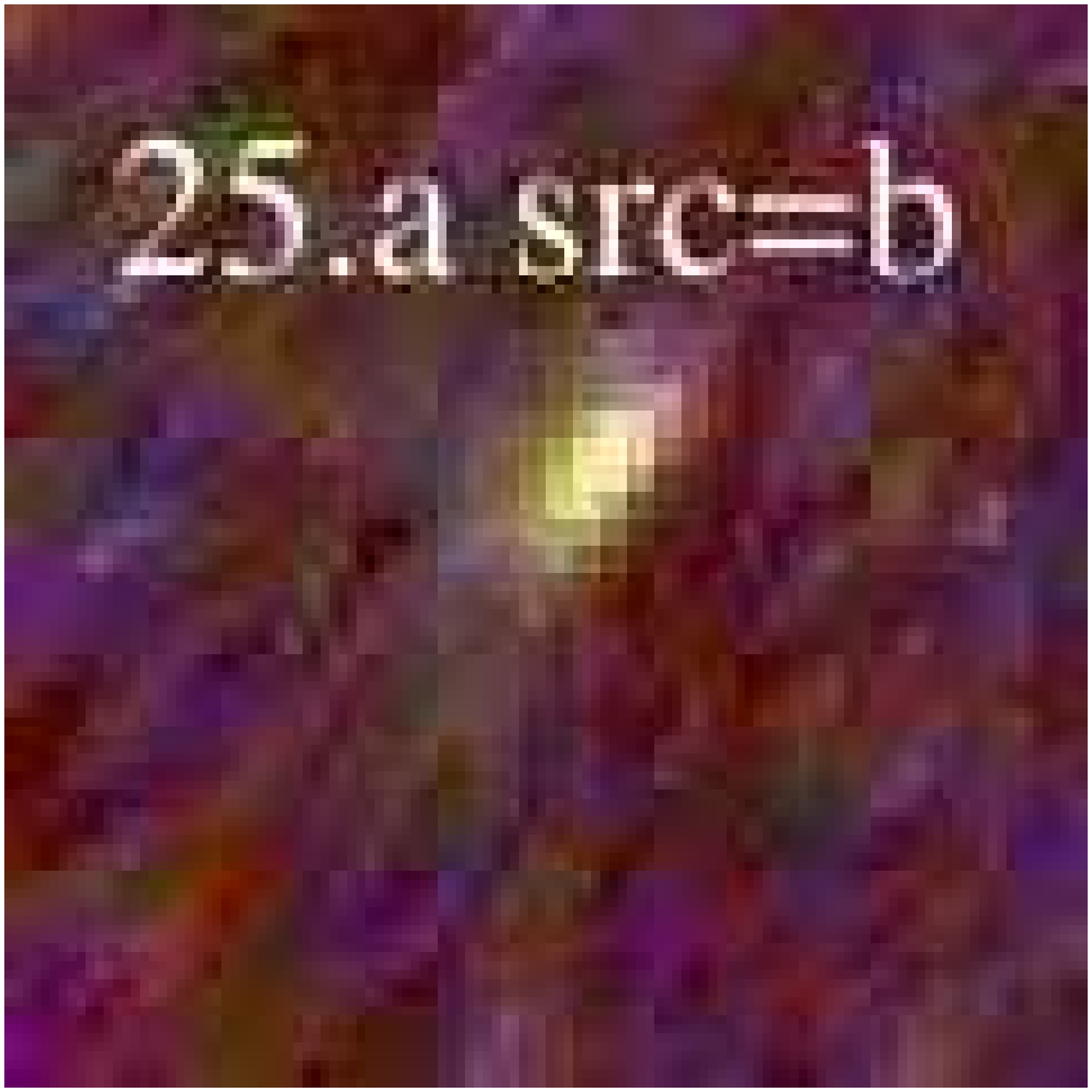}}
    & \multicolumn{1}{m{1.7cm}}{\includegraphics[height=2.00cm,clip]{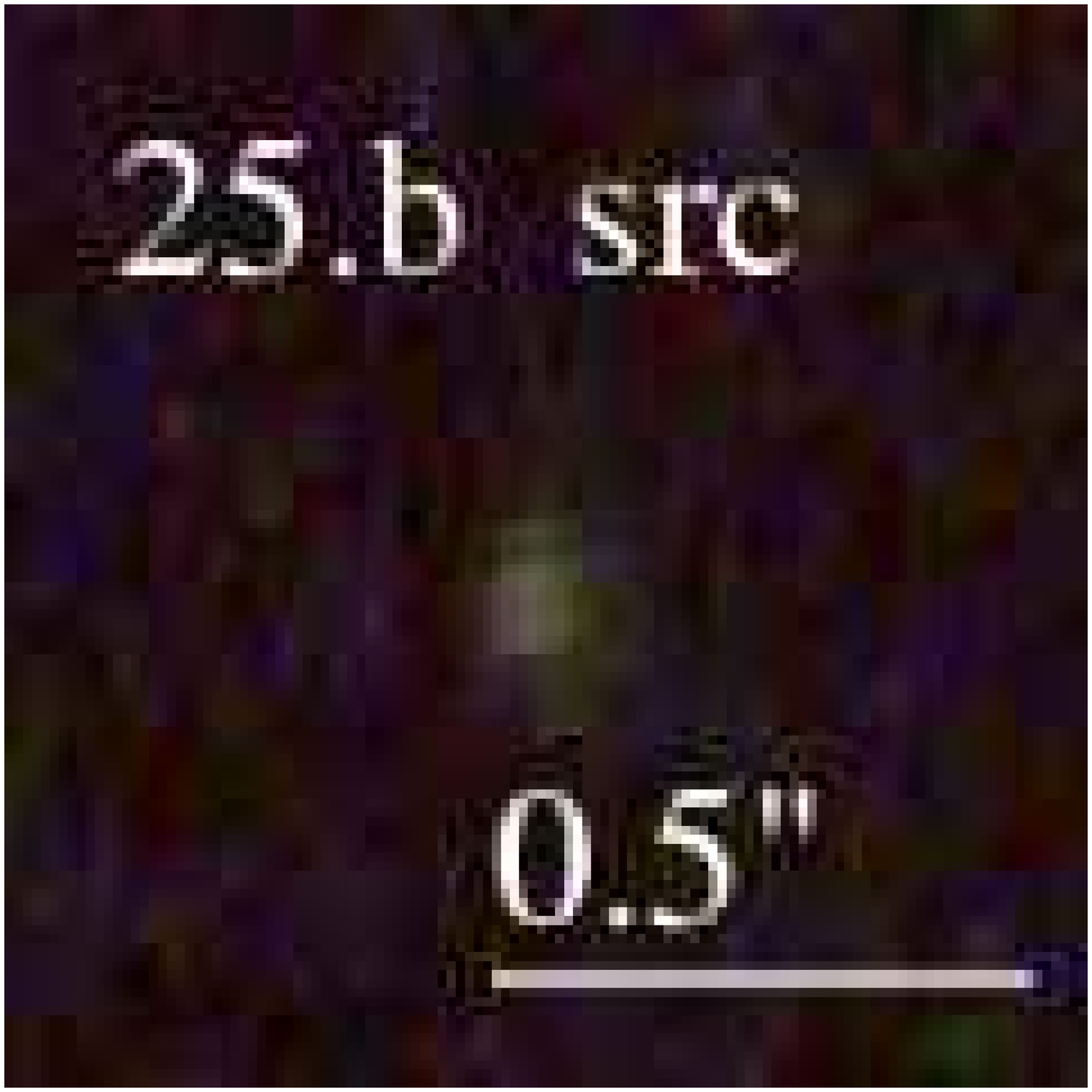}} \\
    \multicolumn{1}{m{1cm}}{{\Large ENFW}}
    & \multicolumn{1}{m{1.7cm}}{\includegraphics[height=2.00cm,clip]{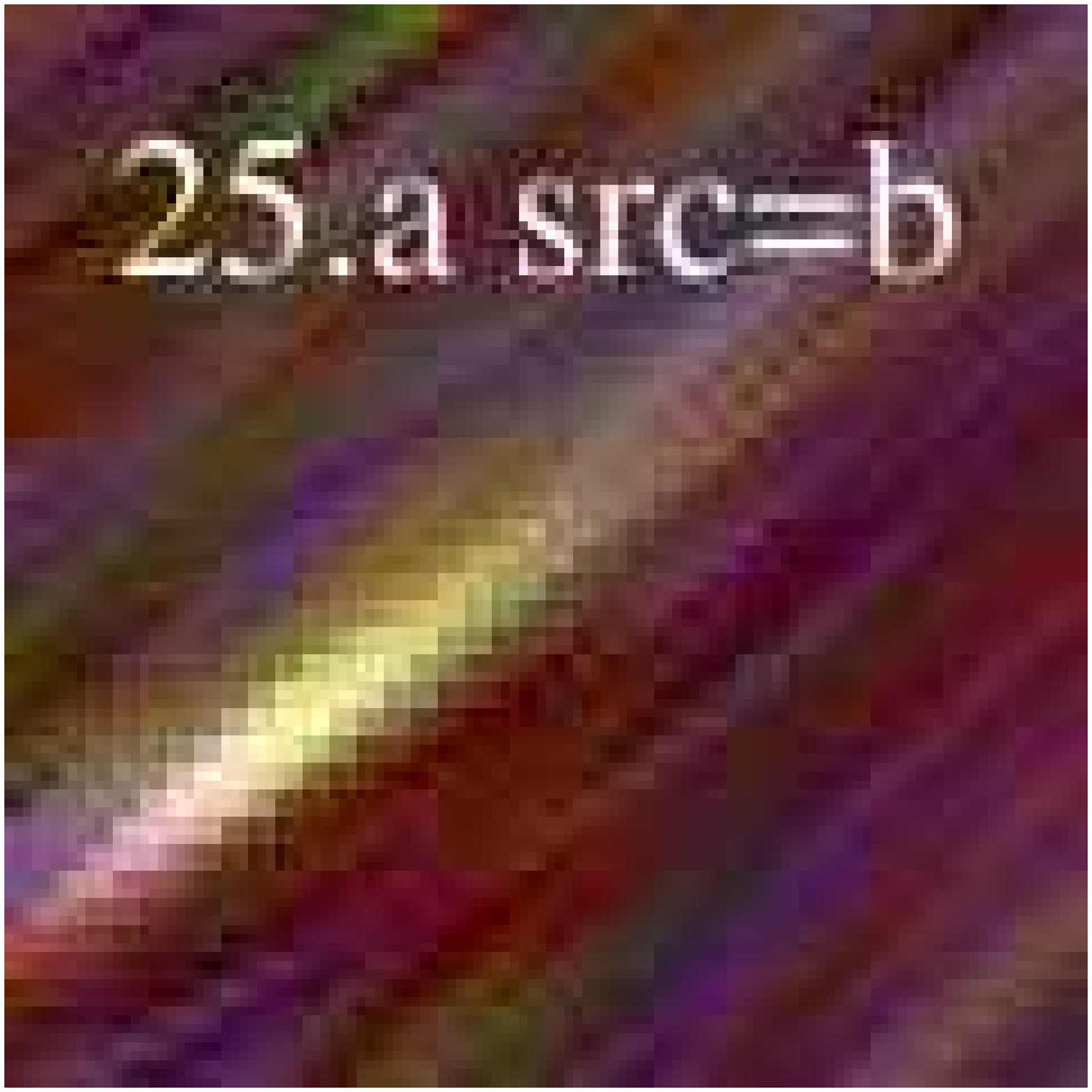}}
    & \multicolumn{1}{m{1.7cm}}{\includegraphics[height=2.00cm,clip]{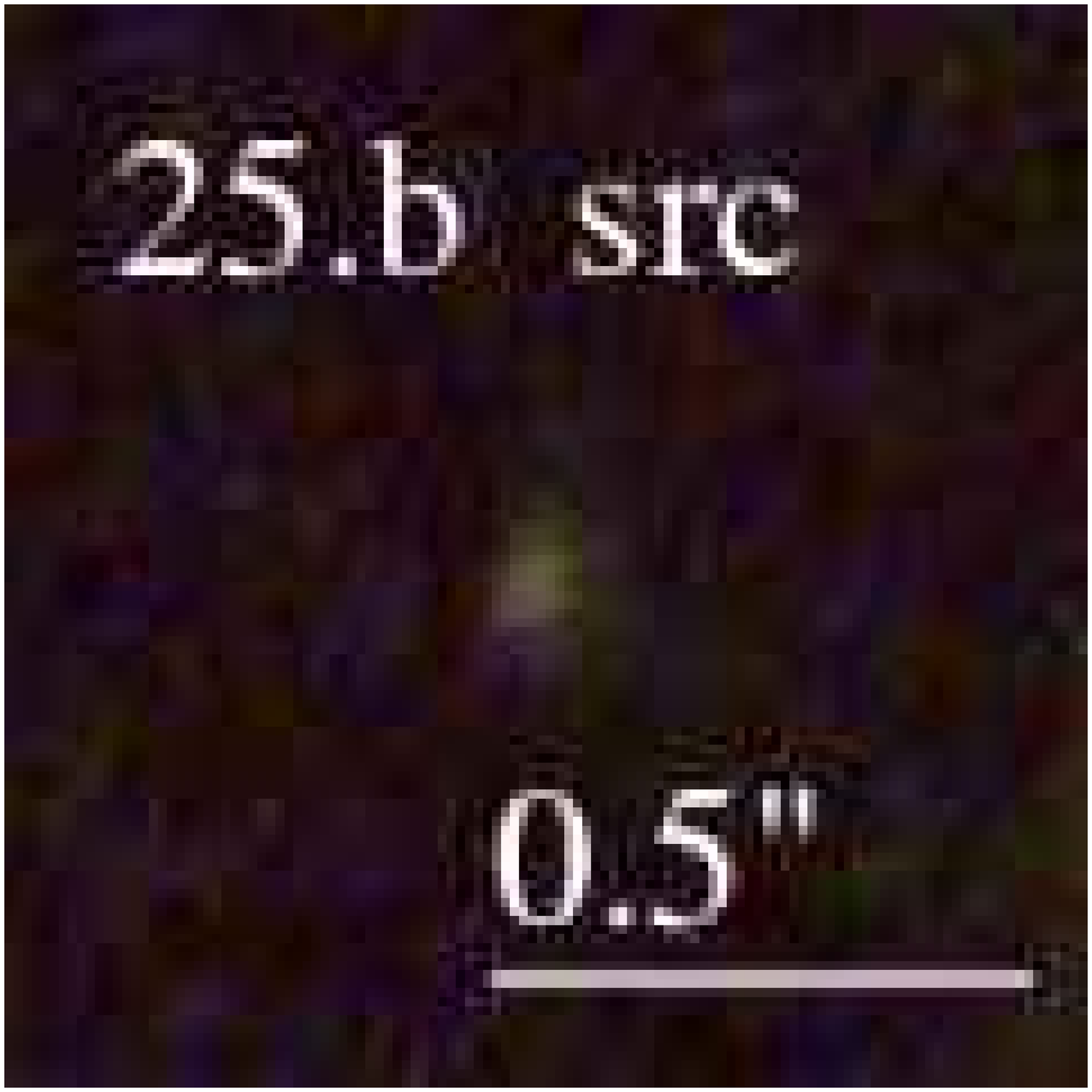}} \\
  \end{tabular}

\end{table*}

\begin{table*}
  \caption{Image system 26:}\vspace{0mm}
  \begin{tabular}{cccc}
    \multicolumn{1}{m{1cm}}{{\Large A1689}}
    & \multicolumn{1}{m{1.7cm}}{\includegraphics[height=2.00cm,clip]{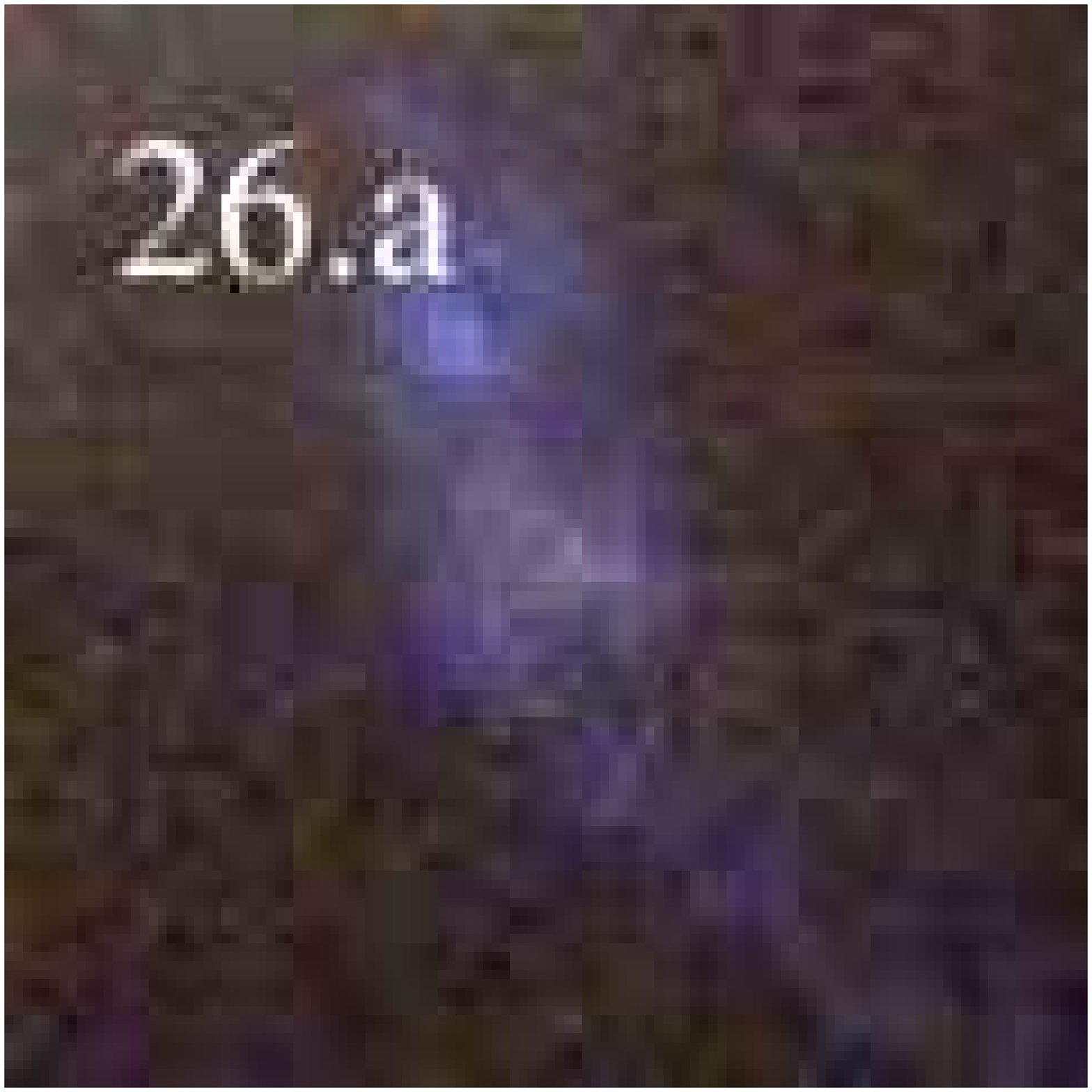}}
    & \multicolumn{1}{m{1.7cm}}{\includegraphics[height=2.00cm,clip]{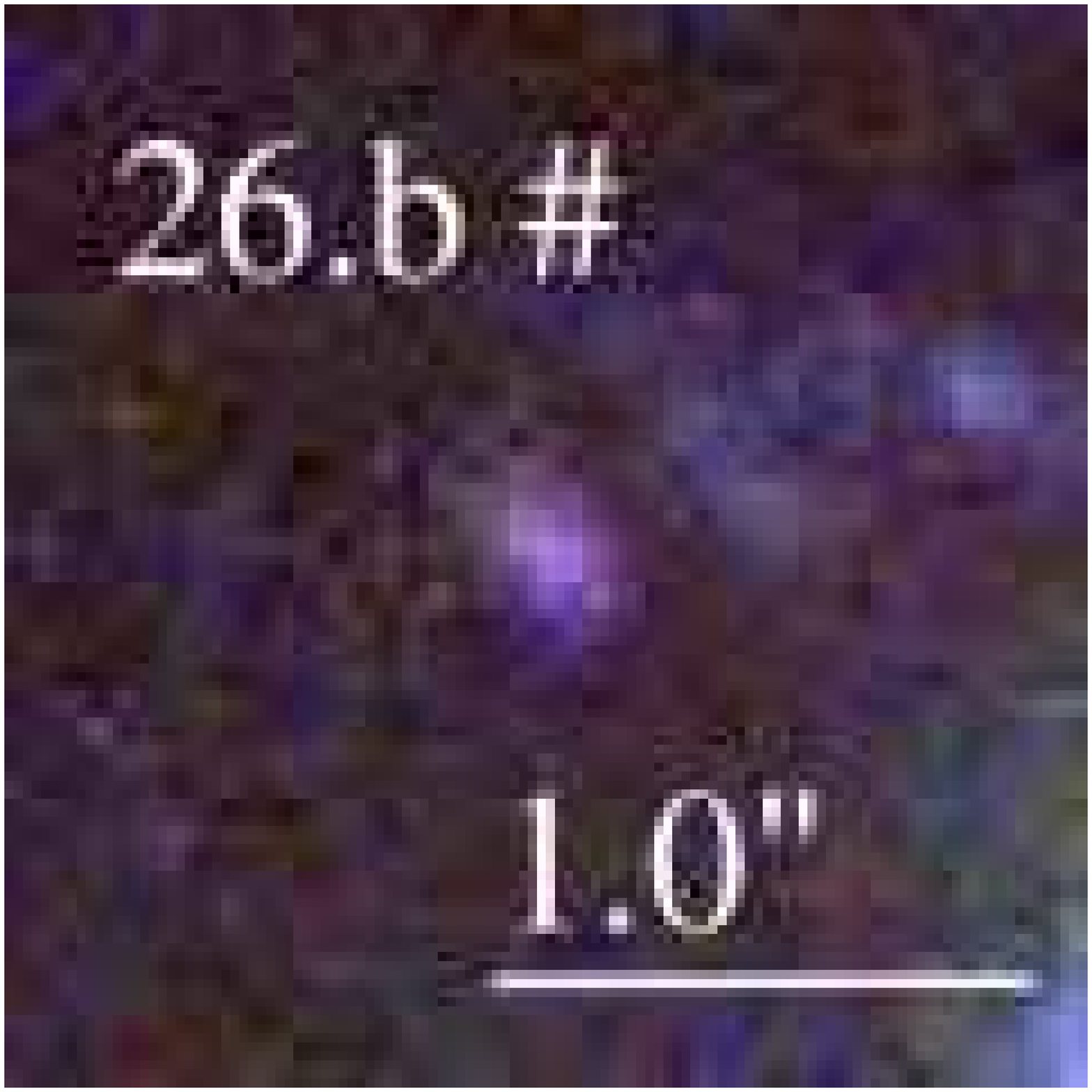}}
    & \multicolumn{1}{m{1.7cm}}{\includegraphics[height=2.00cm,clip]{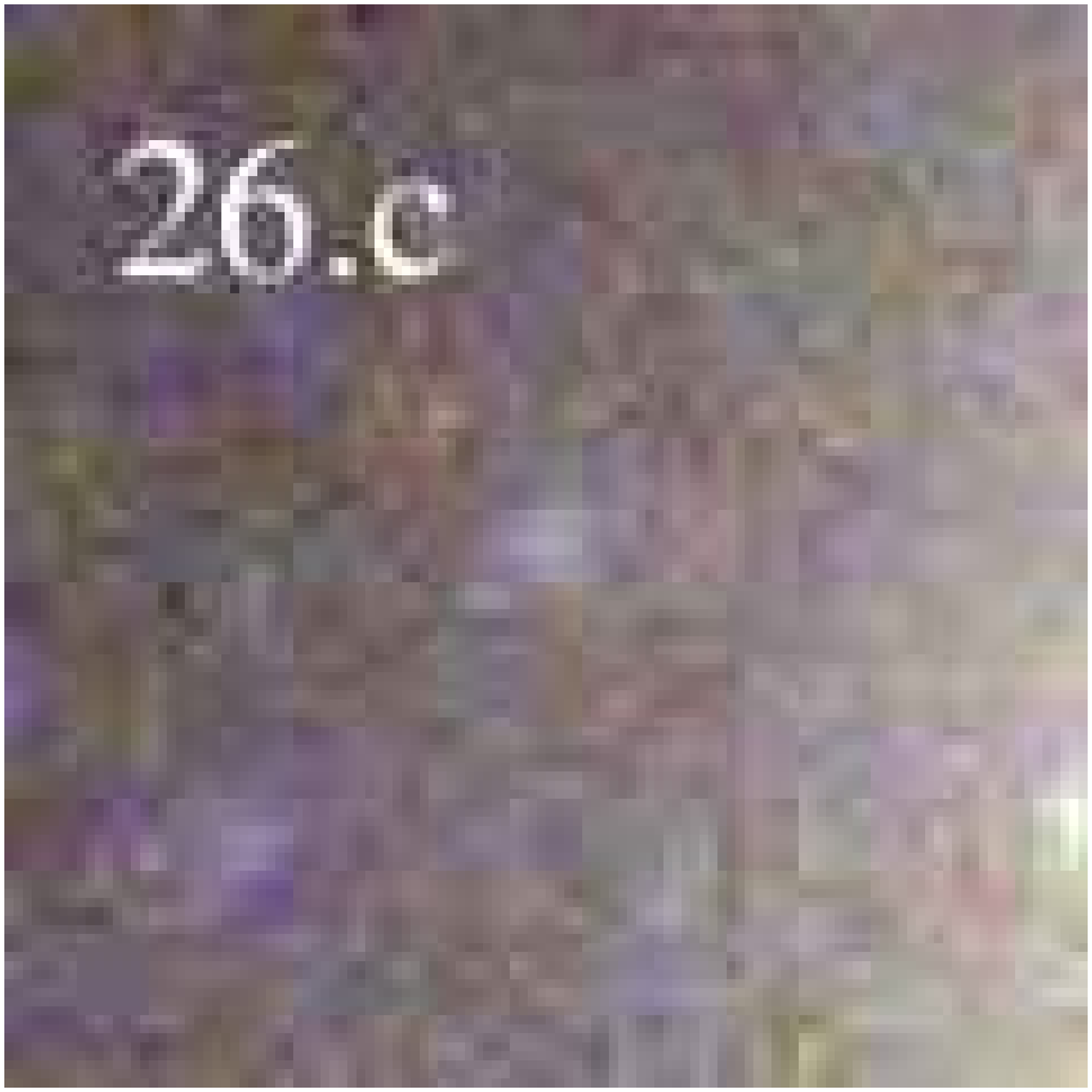}} \\
    \multicolumn{1}{m{1cm}}{{\Large NSIE}}
    & \multicolumn{1}{m{1.7cm}}{\includegraphics[height=2.00cm,clip]{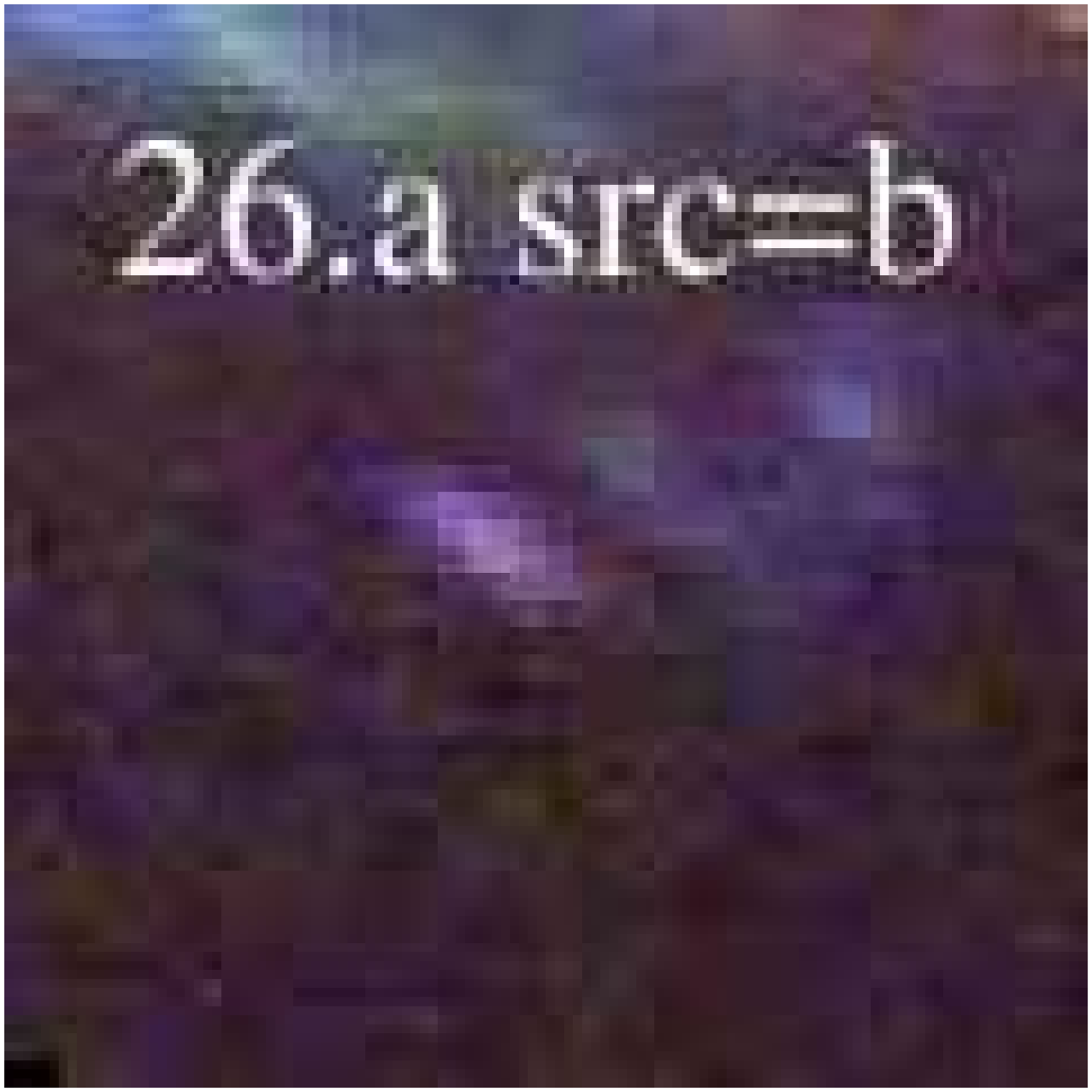}}
    & \multicolumn{1}{m{1.7cm}}{\includegraphics[height=2.00cm,clip]{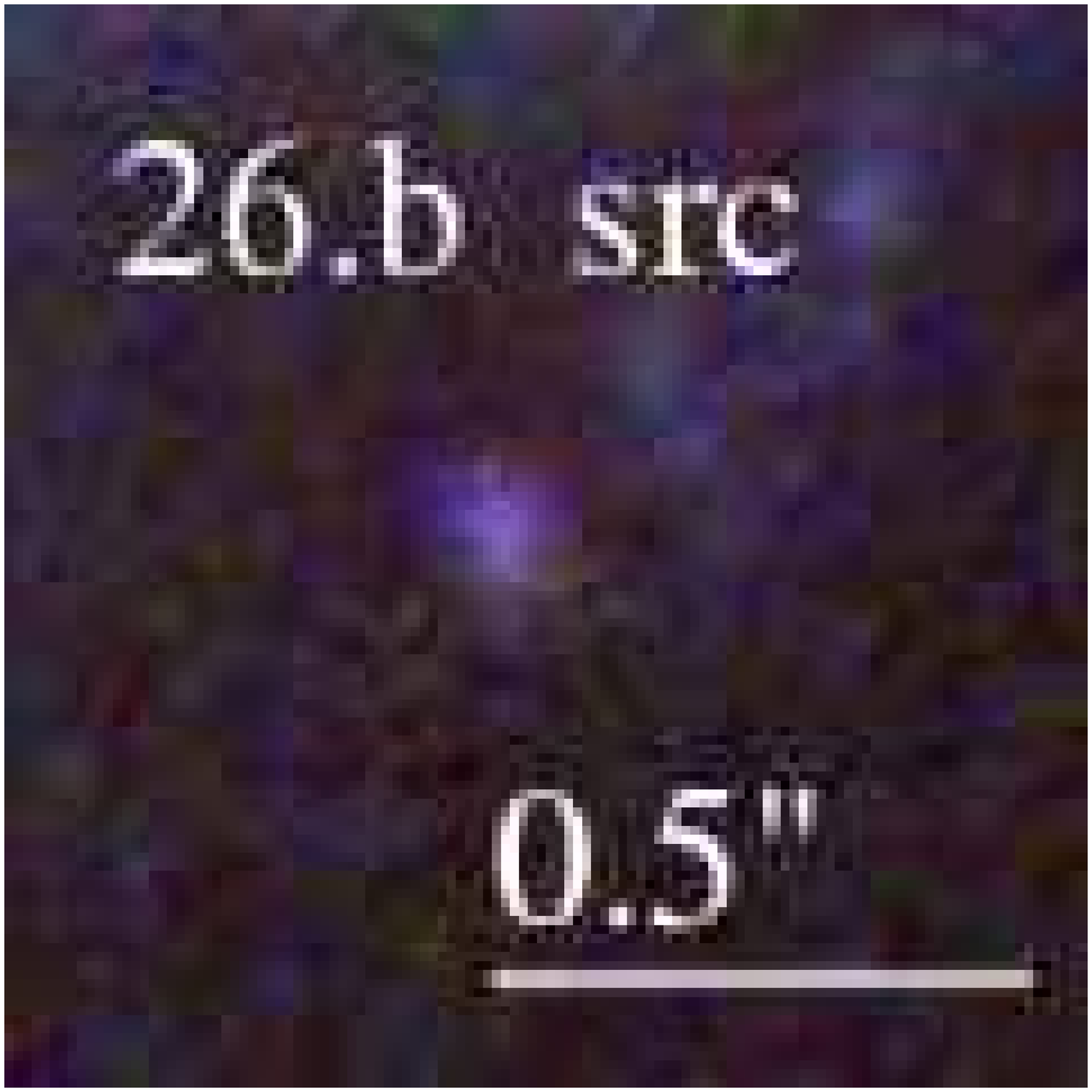}}
    & \multicolumn{1}{m{1.7cm}}{\includegraphics[height=2.00cm,clip]{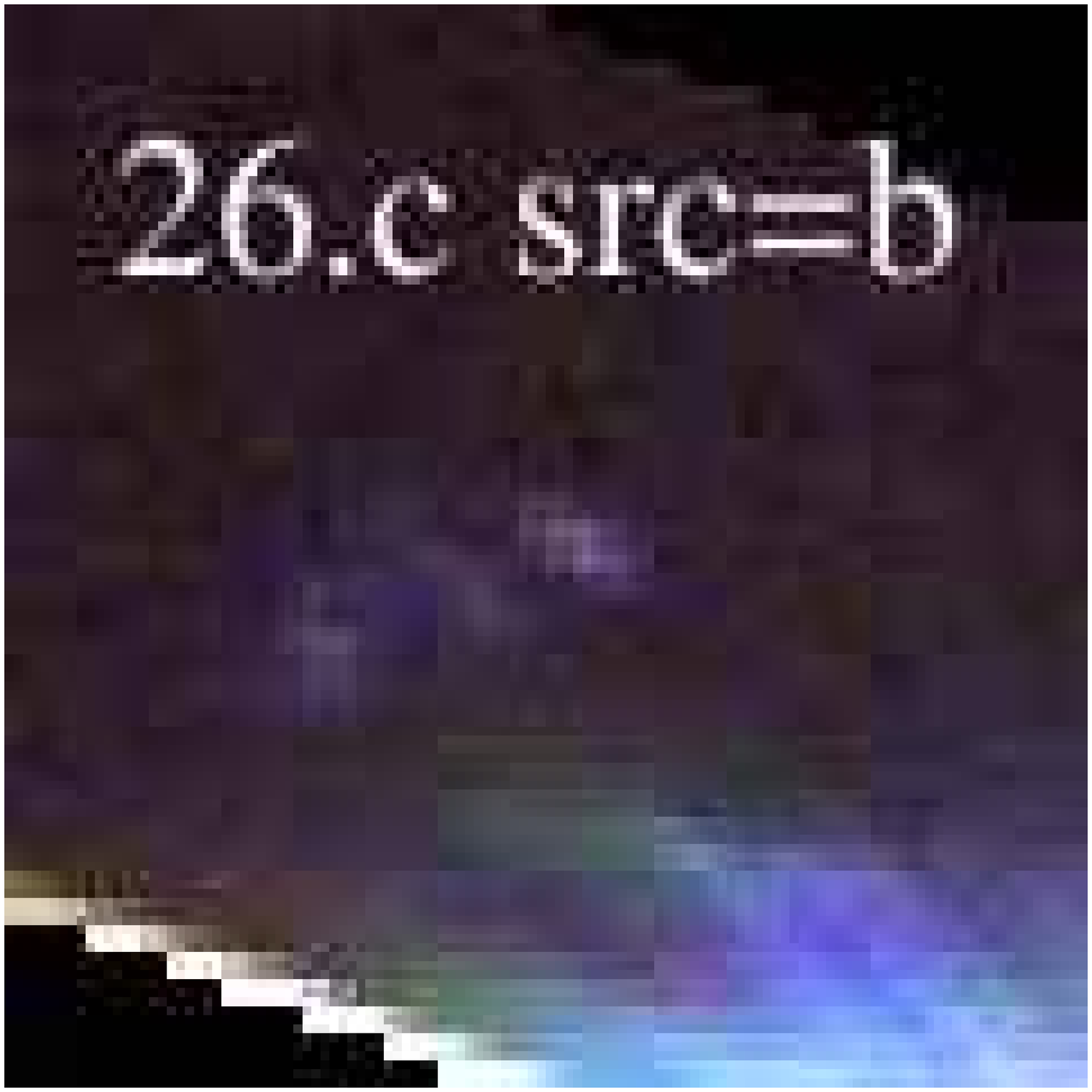}} \\
    \multicolumn{1}{m{1cm}}{{\Large ENFW}}
    & \multicolumn{1}{m{1.7cm}}{\includegraphics[height=2.00cm,clip]{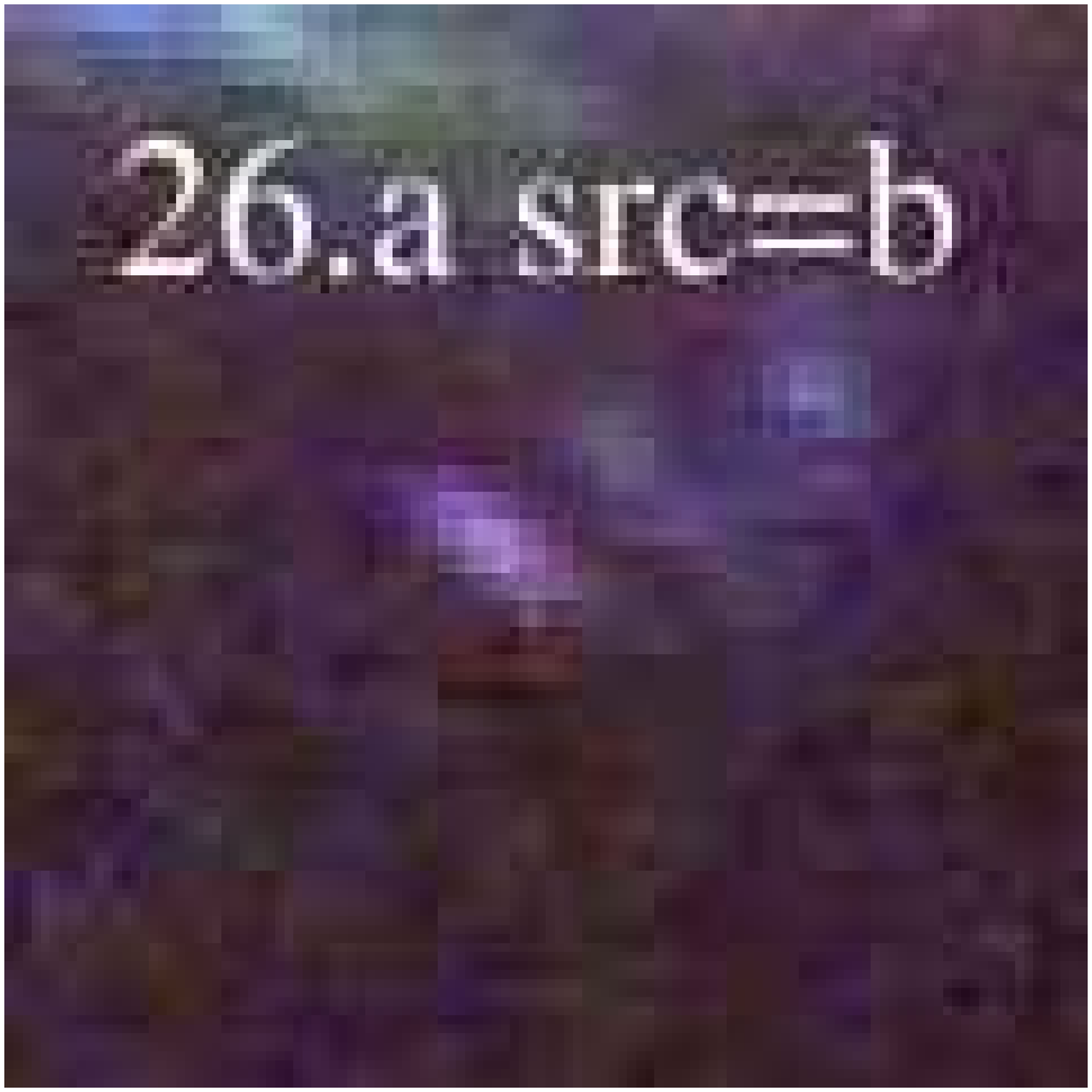}}
    & \multicolumn{1}{m{1.7cm}}{\includegraphics[height=2.00cm,clip]{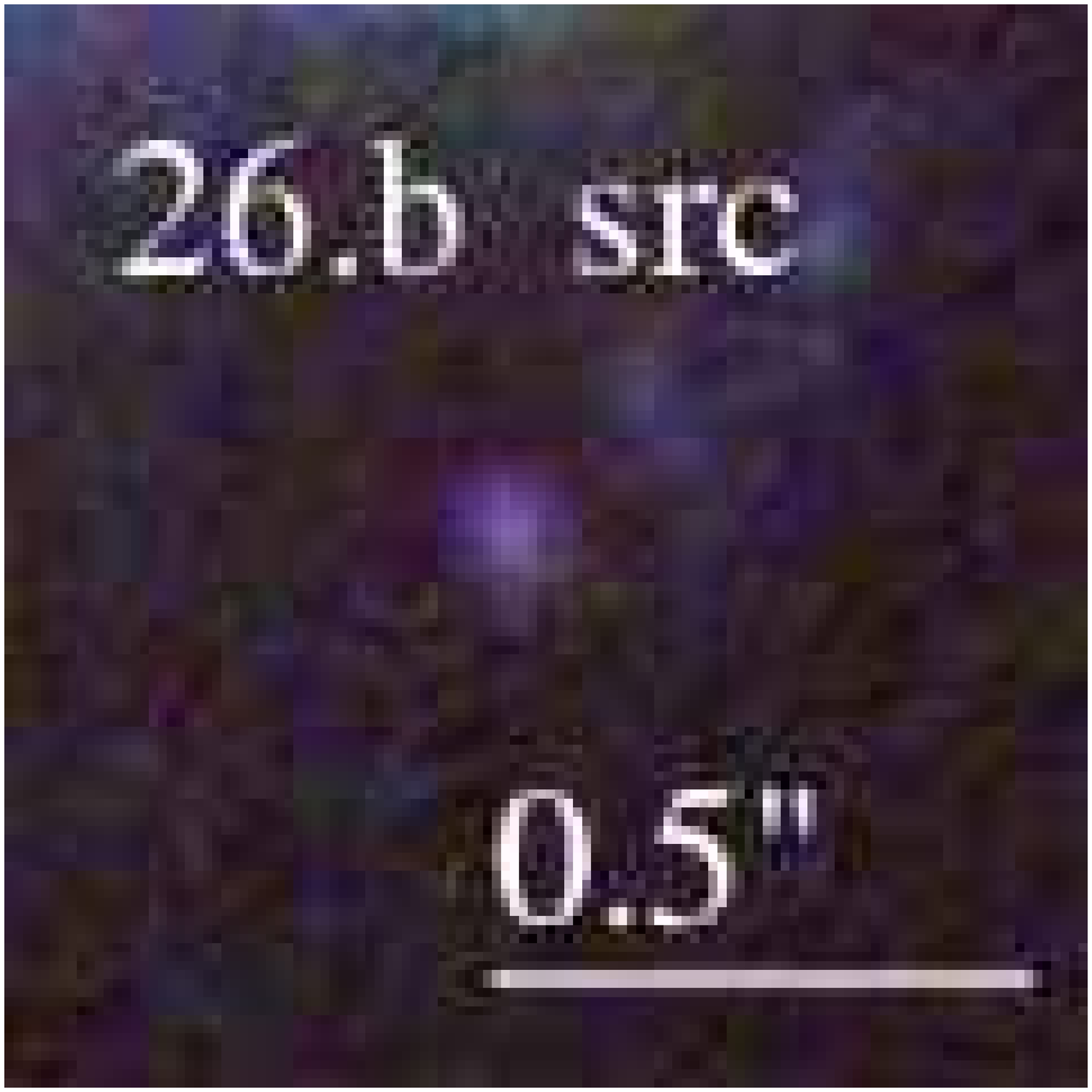}}
    & \multicolumn{1}{m{1.7cm}}{\includegraphics[height=2.00cm,clip]{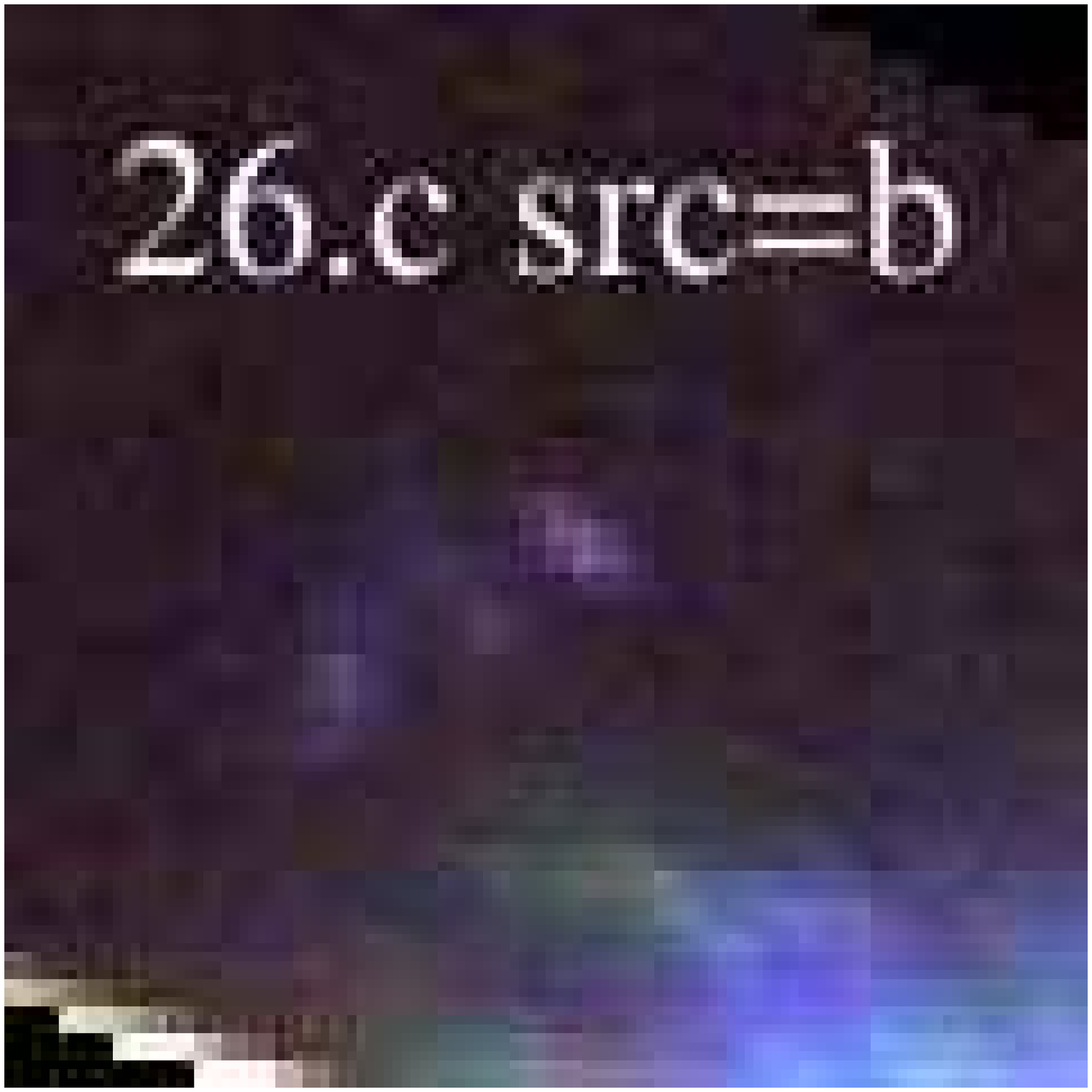}} \\
  \end{tabular}

\end{table*}

\begin{table*}
  \caption{Image system 27:}\vspace{0mm}
  \begin{tabular}{cccc}
    \multicolumn{1}{m{1cm}}{{\Large A1689}}
    & \multicolumn{1}{m{1.7cm}}{\includegraphics[height=2.00cm,clip]{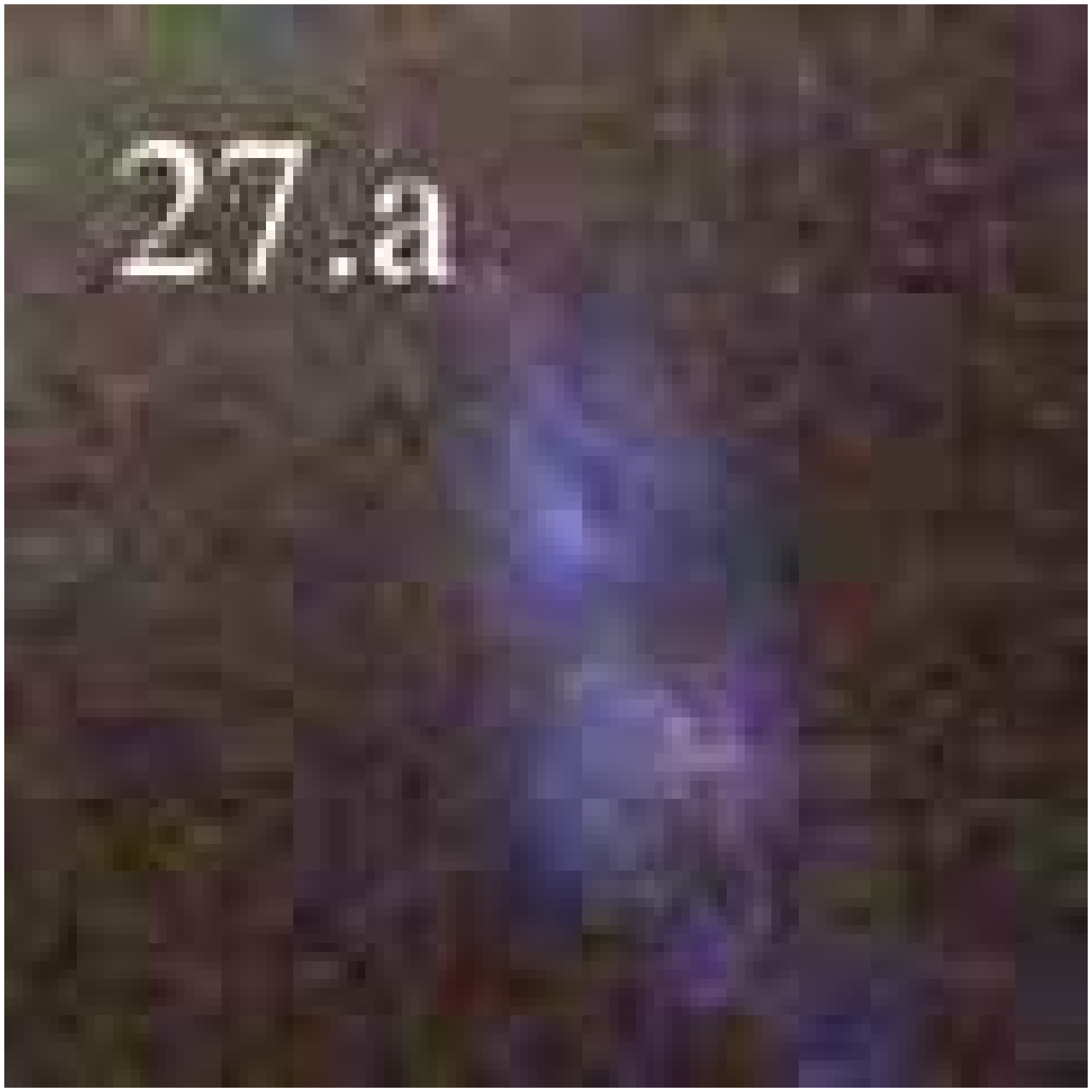}}
    & \multicolumn{1}{m{1.7cm}}{\includegraphics[height=2.00cm,clip]{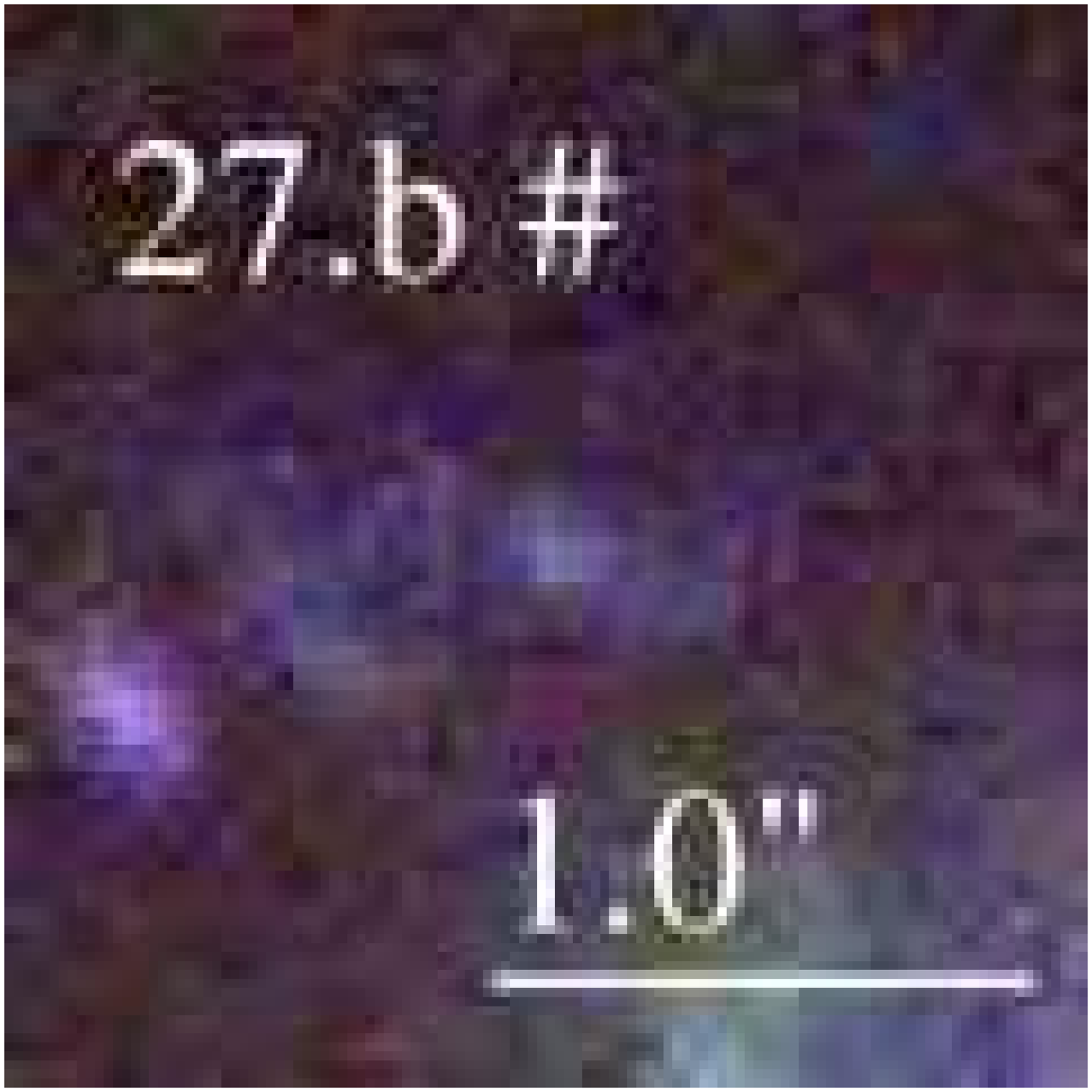}}
    & \multicolumn{1}{m{1.7cm}}{\includegraphics[height=2.00cm,clip]{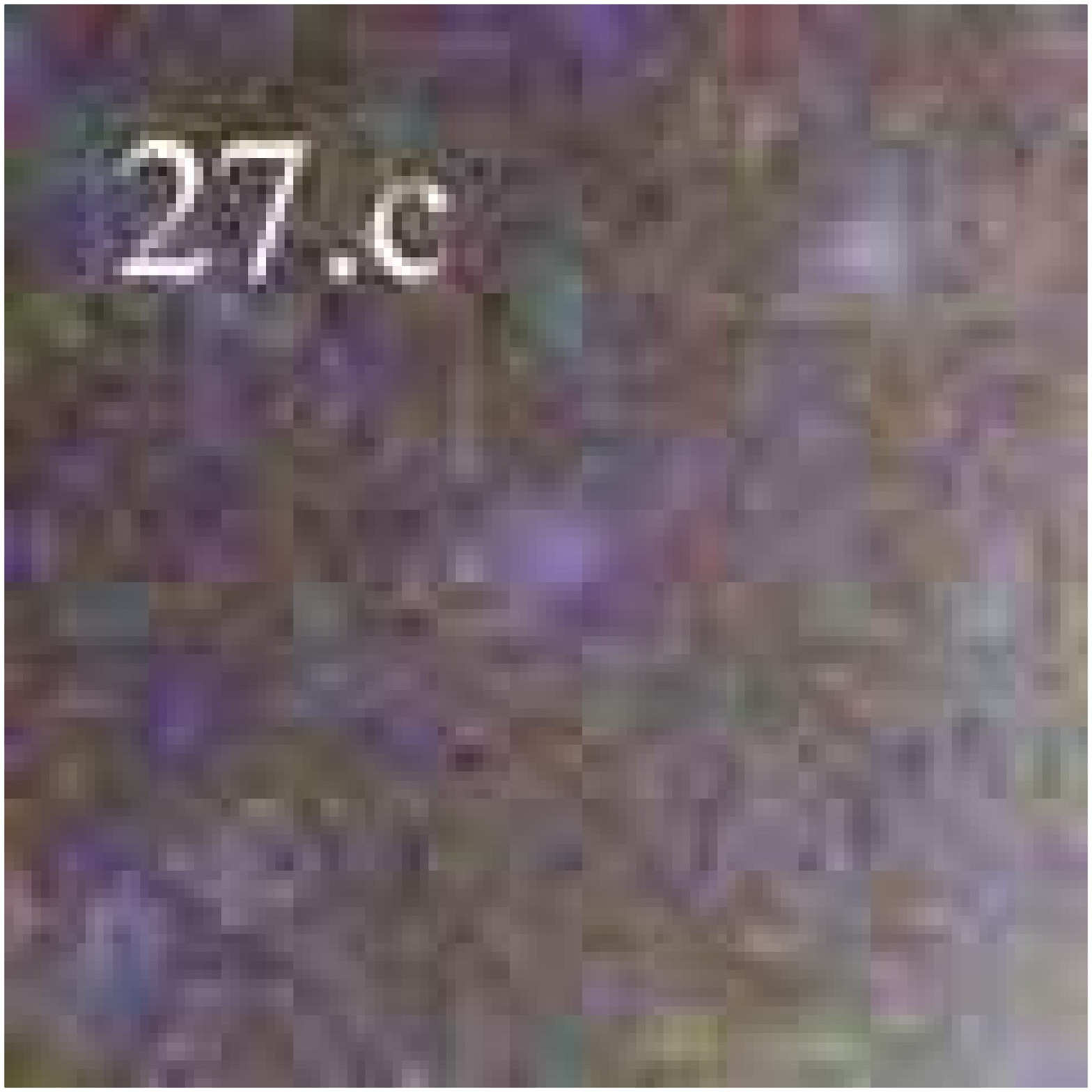}} \\
    \multicolumn{1}{m{1cm}}{{\Large NSIE}}
    & \multicolumn{1}{m{1.7cm}}{\includegraphics[height=2.00cm,clip]{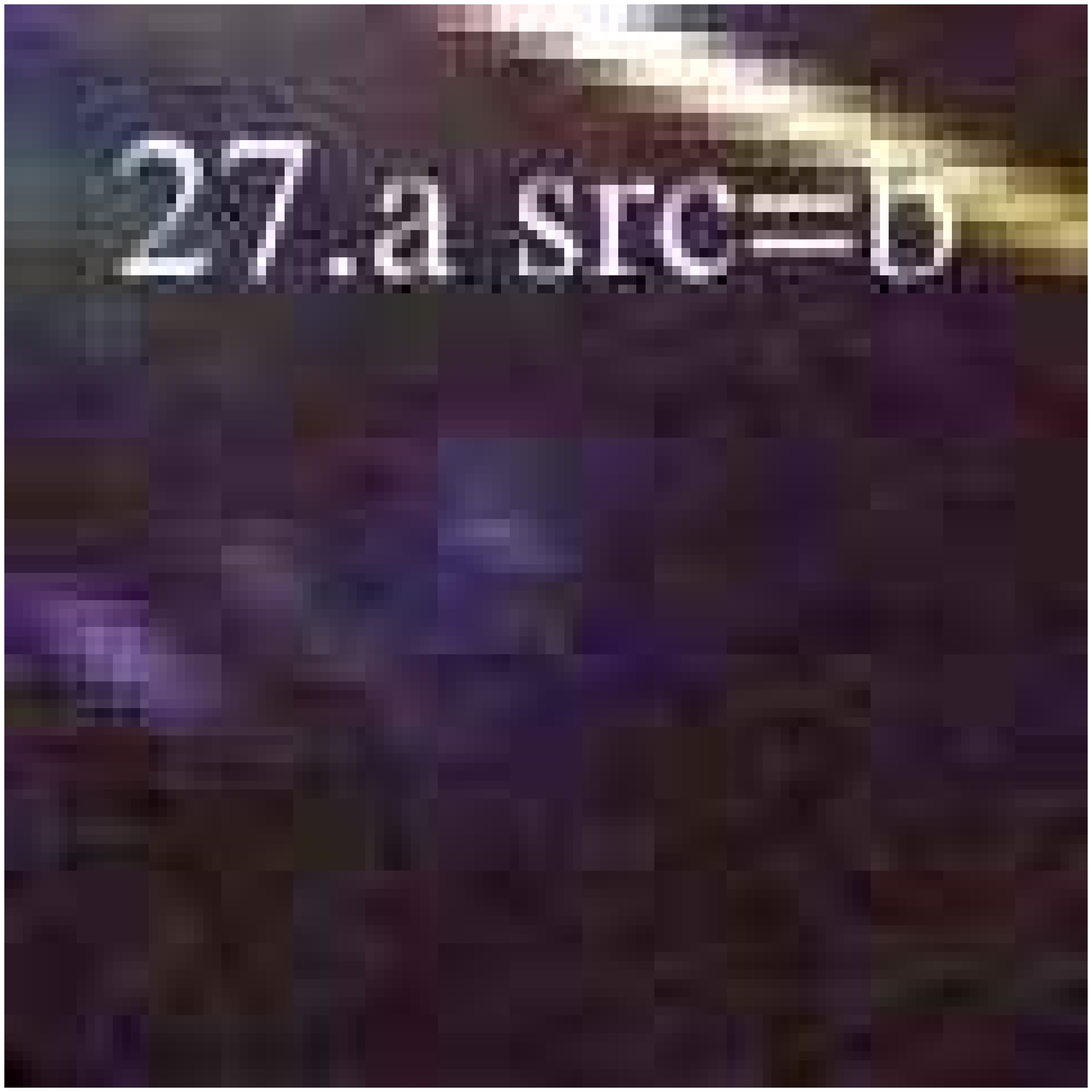}}
    & \multicolumn{1}{m{1.7cm}}{\includegraphics[height=2.00cm,clip]{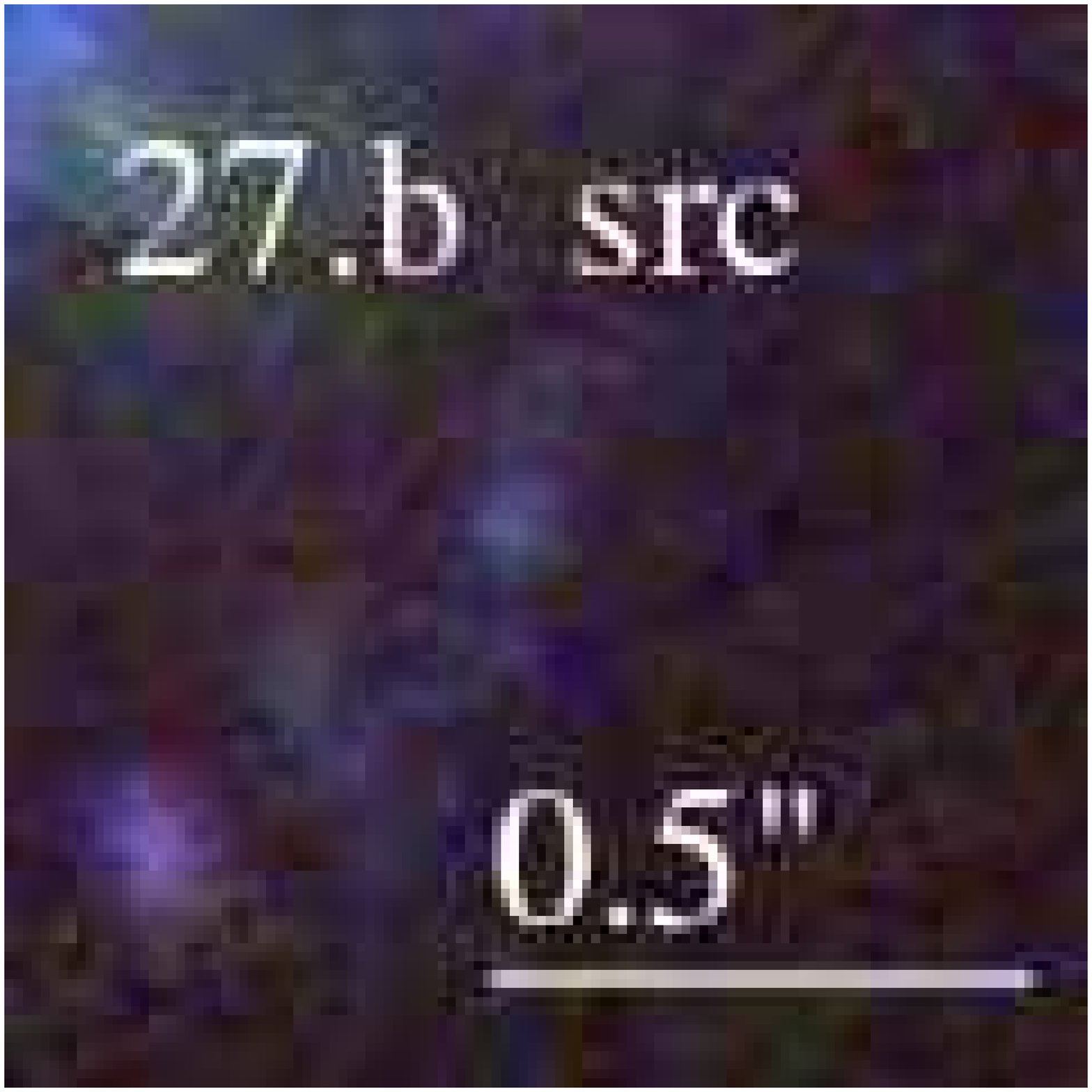}}
    & \multicolumn{1}{m{1.7cm}}{\includegraphics[height=2.00cm,clip]{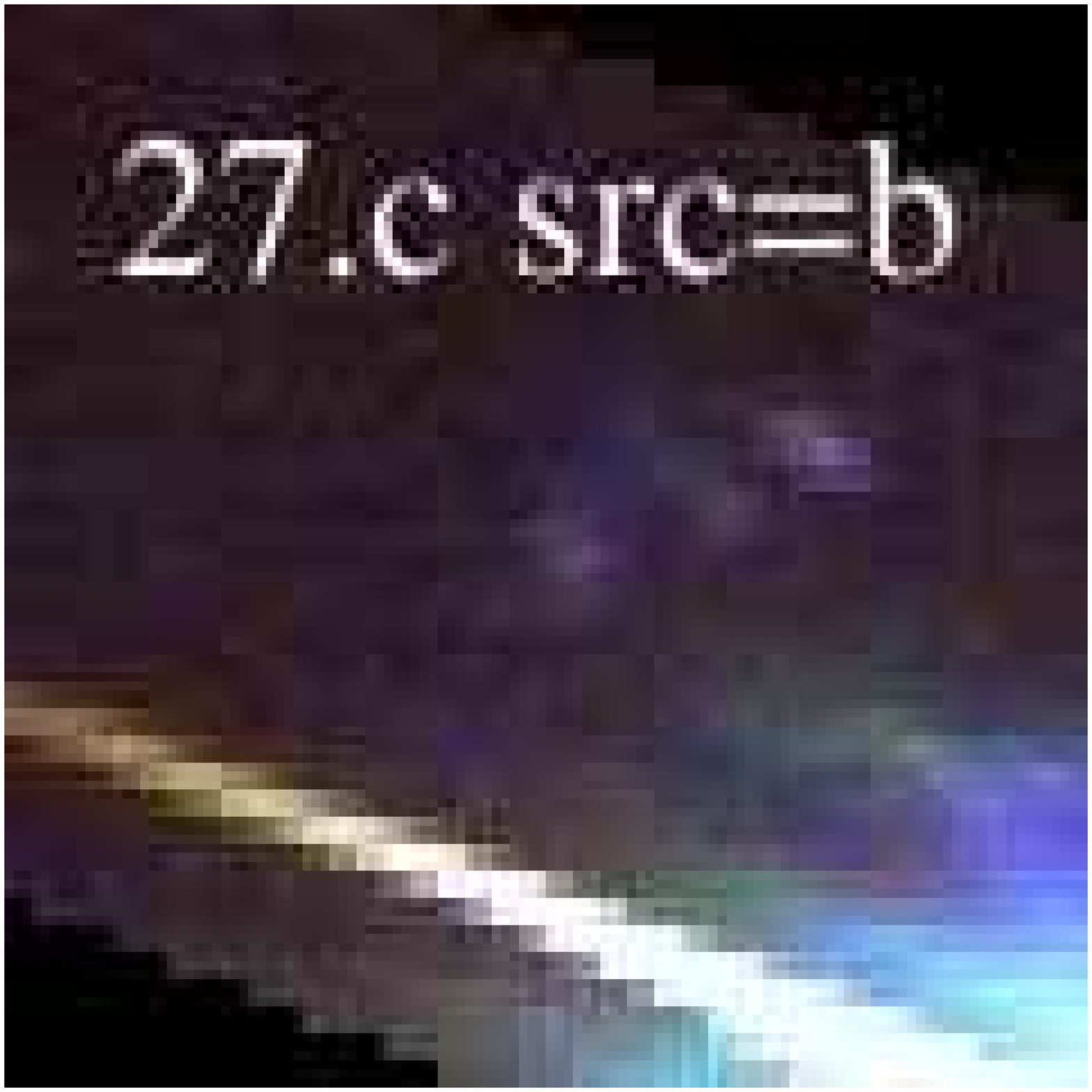}} \\
    \multicolumn{1}{m{1cm}}{{\Large ENFW}}
    & \multicolumn{1}{m{1.7cm}}{\includegraphics[height=2.00cm,clip]{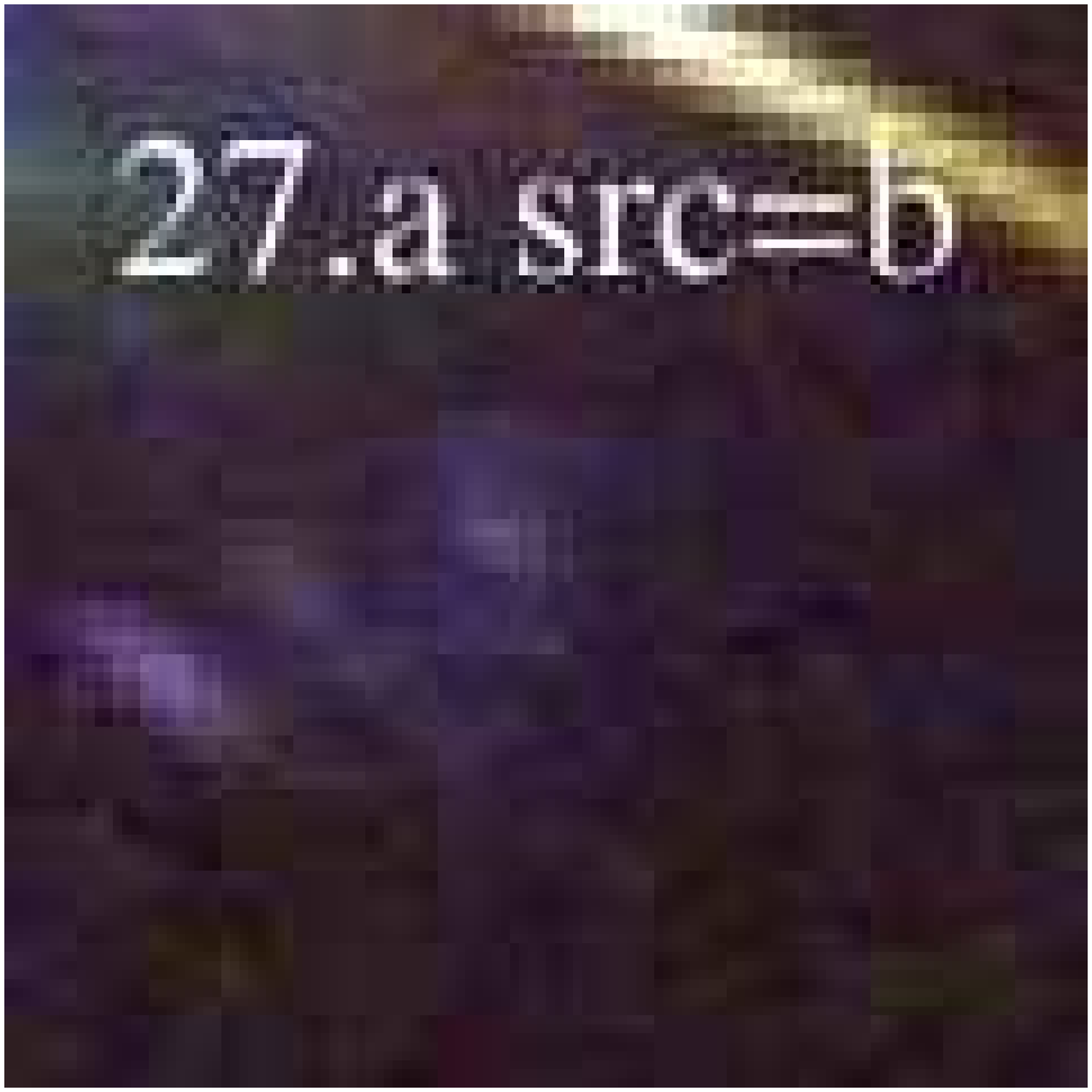}}
    & \multicolumn{1}{m{1.7cm}}{\includegraphics[height=2.00cm,clip]{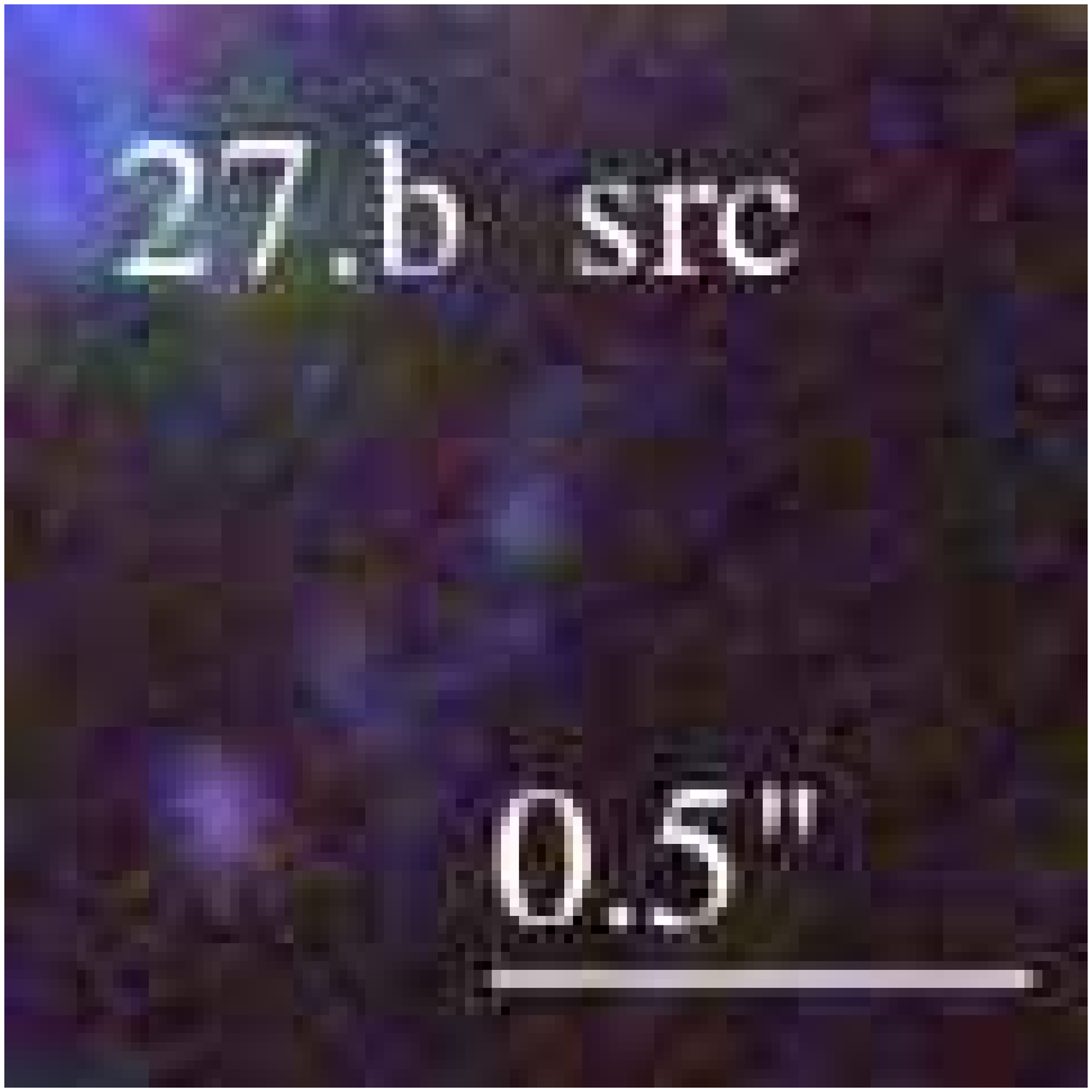}}
    & \multicolumn{1}{m{1.7cm}}{\includegraphics[height=2.00cm,clip]{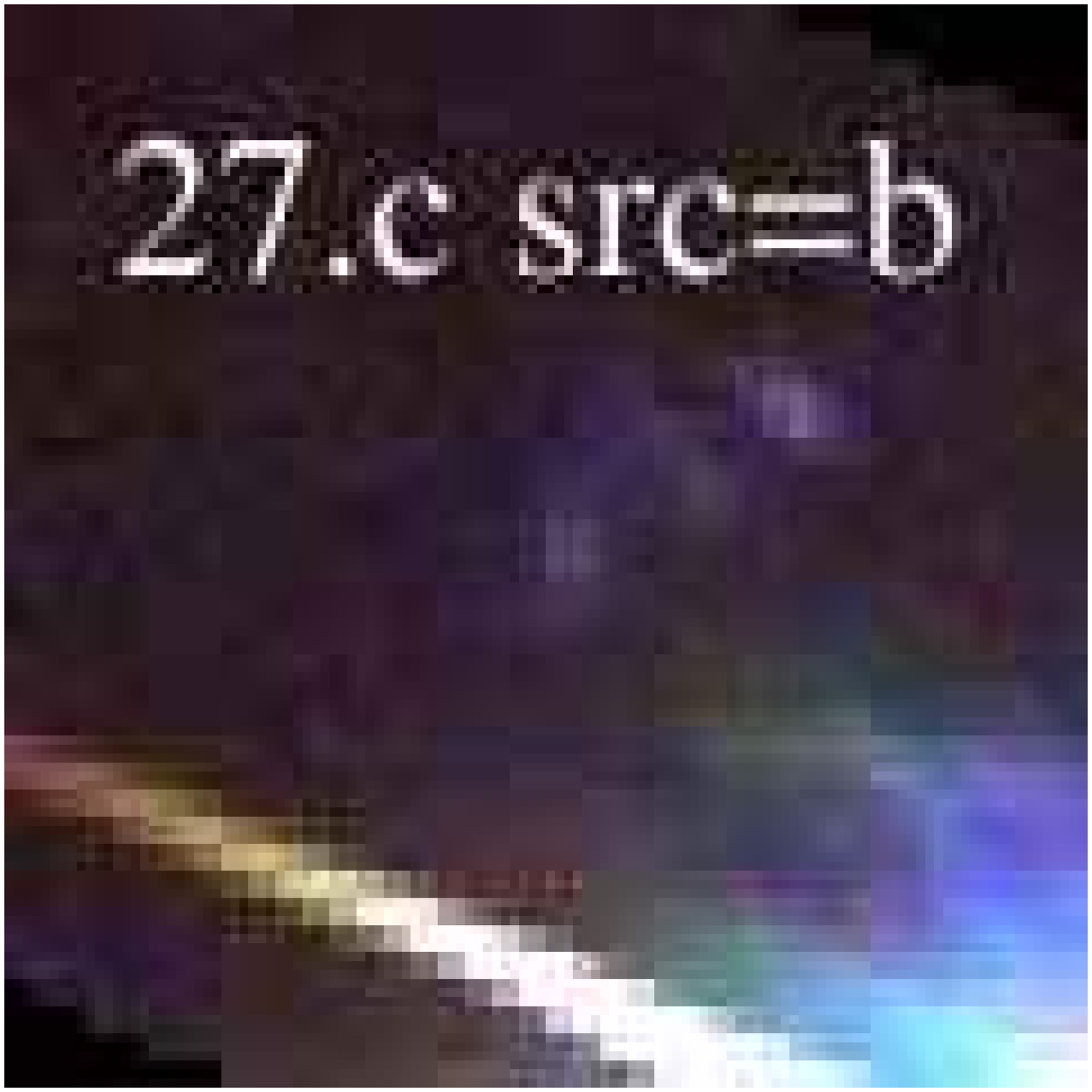}} \\
  \end{tabular}

\end{table*}

\clearpage

\begin{table*}
  \caption{Image system 28:}\vspace{0mm}
  \begin{tabular}{ccc}
    \multicolumn{1}{m{1cm}}{{\Large A1689}}
    & \multicolumn{1}{m{1.7cm}}{\includegraphics[height=2.00cm,clip]{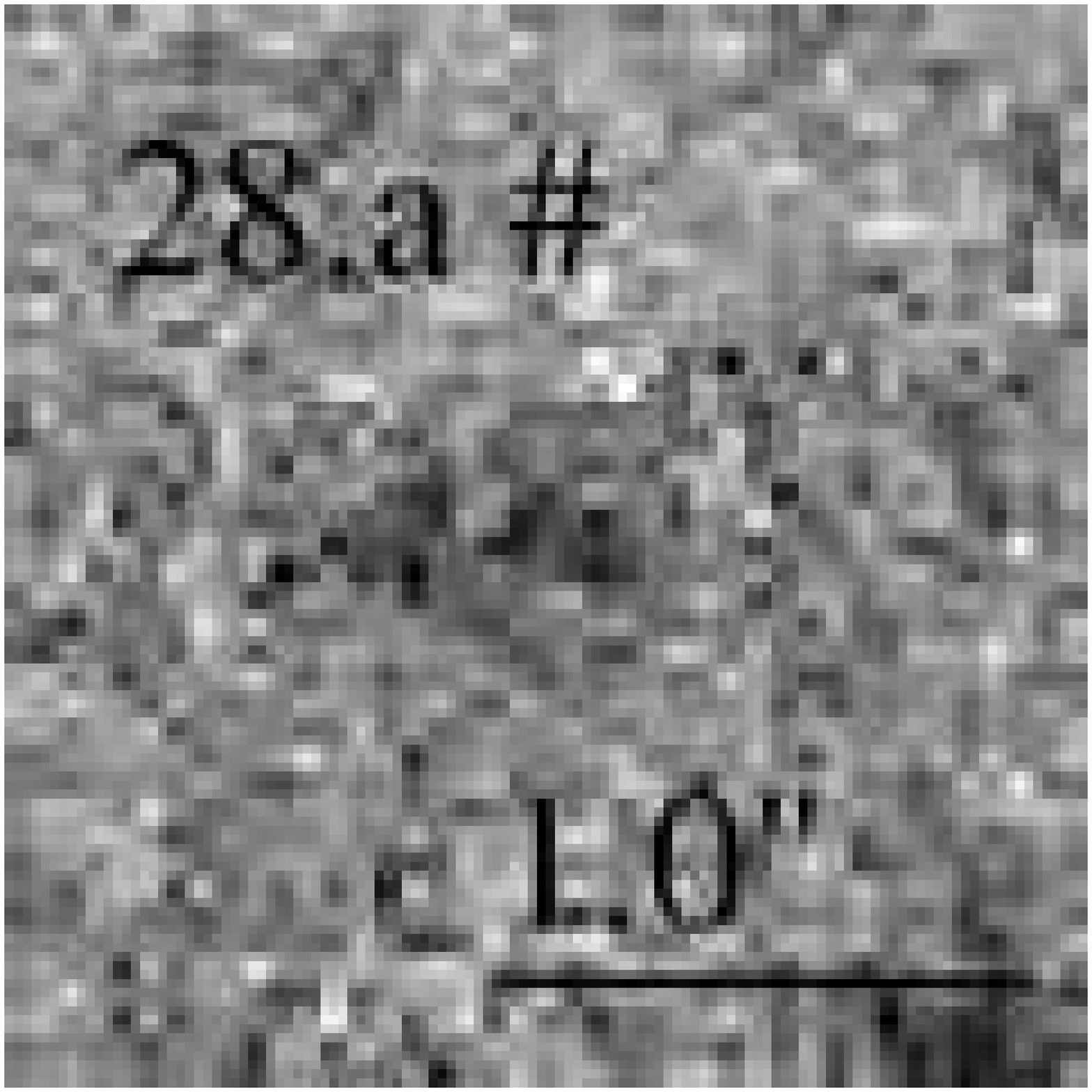}}
    & \multicolumn{1}{m{1.7cm}}{\includegraphics[height=2.00cm,clip]{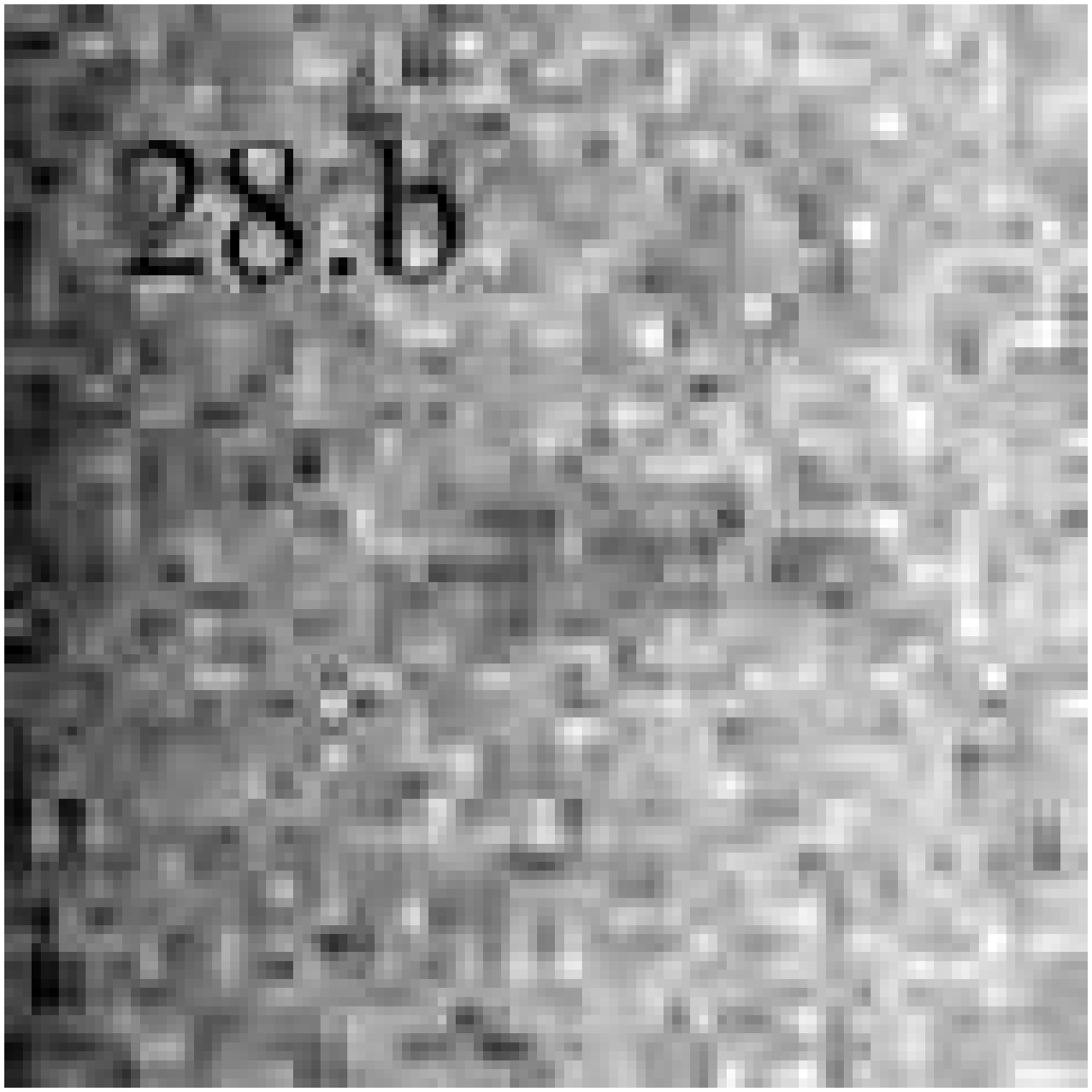}} \\
    \multicolumn{1}{m{1cm}}{{\Large NSIE}}
    & \multicolumn{1}{m{1.7cm}}{\includegraphics[height=2.00cm,clip]{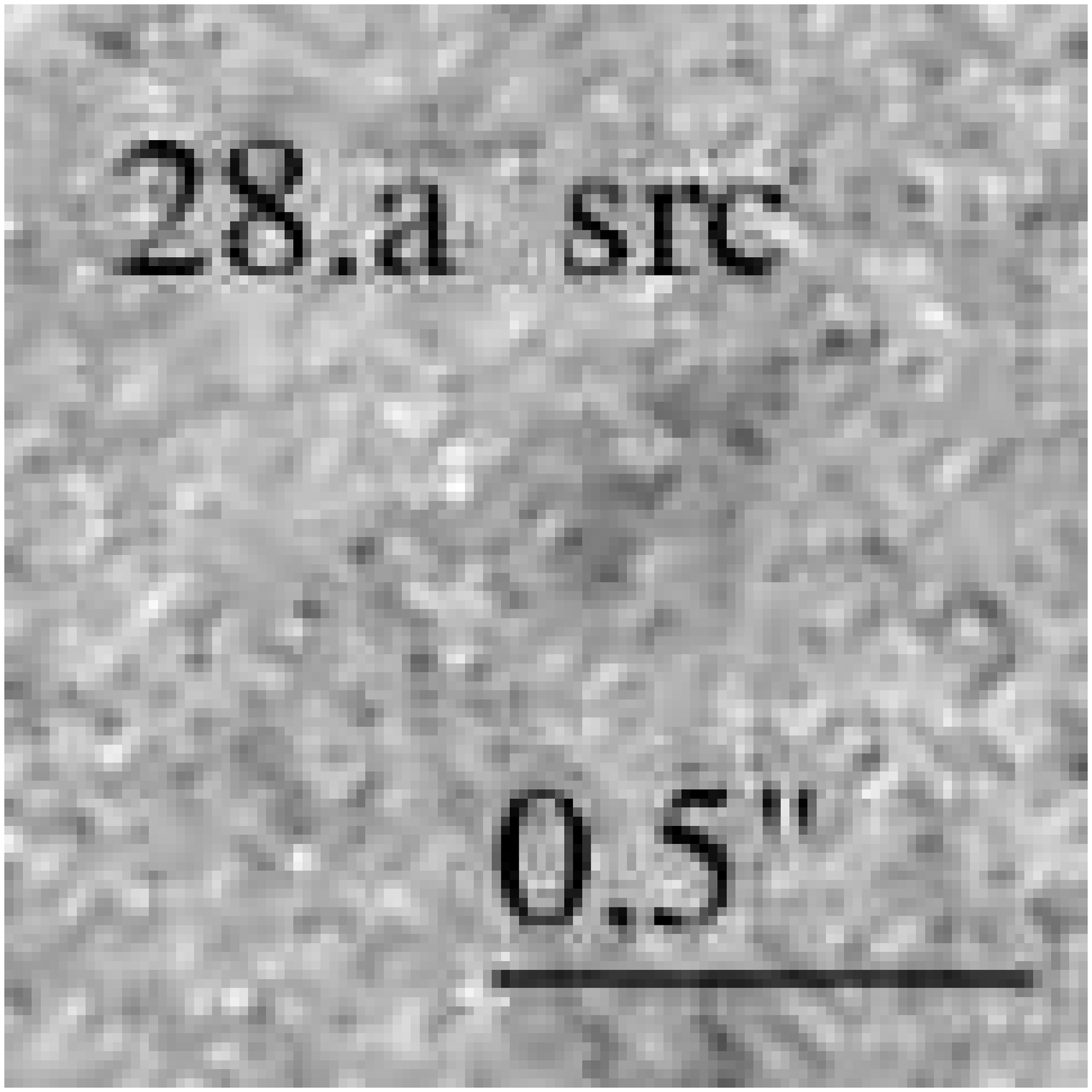}}
    & \multicolumn{1}{m{1.7cm}}{\includegraphics[height=2.00cm,clip]{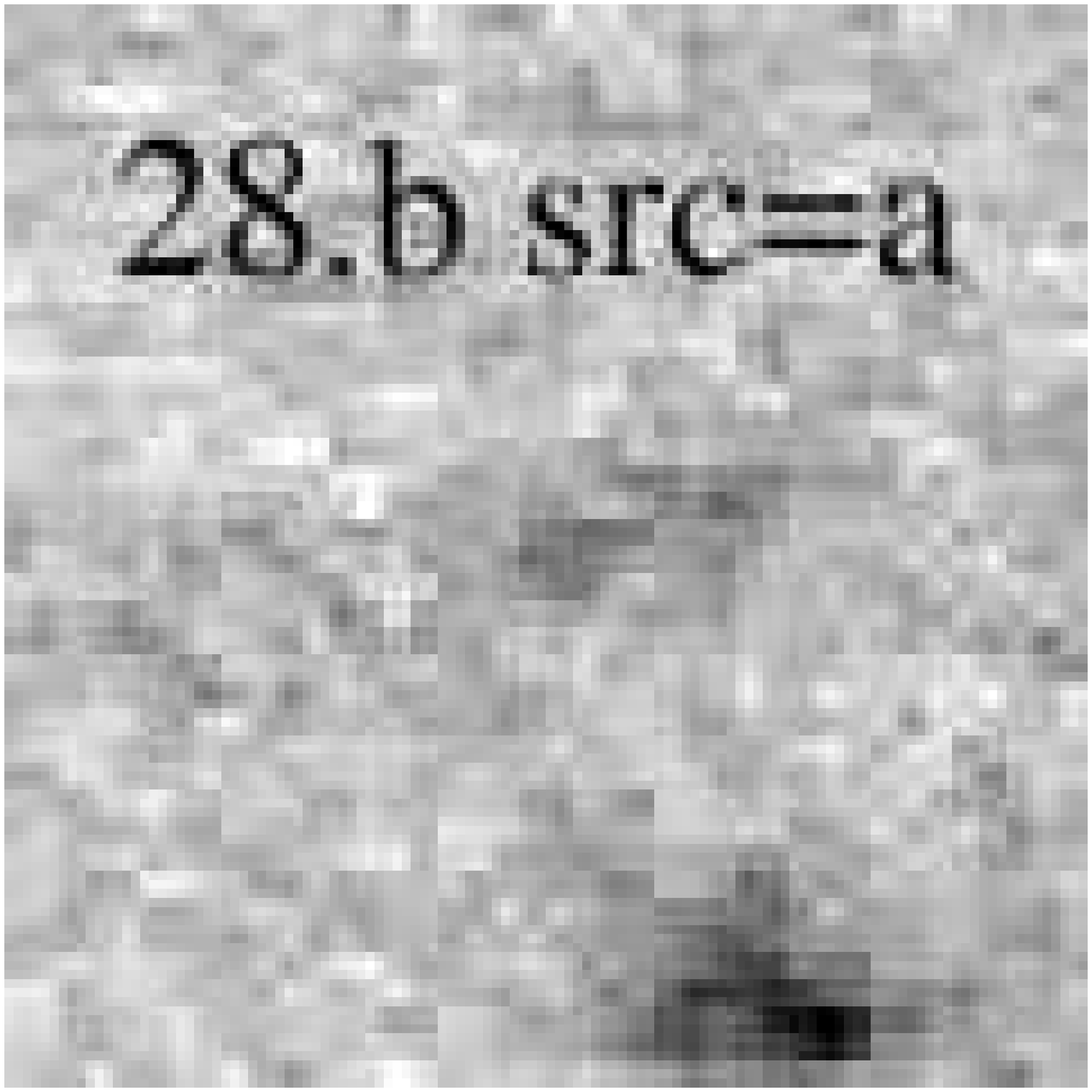}} \\
    \multicolumn{1}{m{1cm}}{{\Large ENFW}}
    & \multicolumn{1}{m{1.7cm}}{\includegraphics[height=2.00cm,clip]{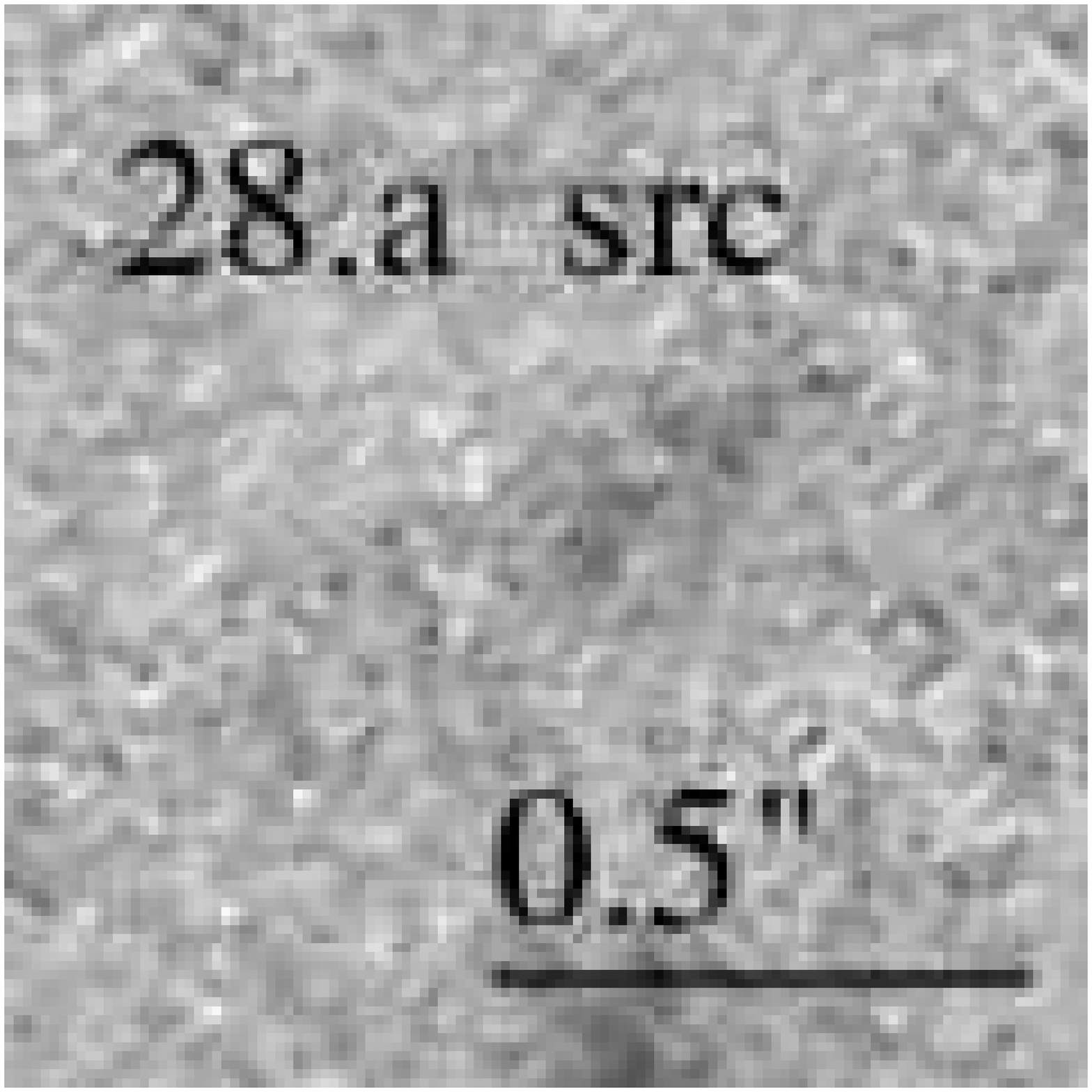}}
    & \multicolumn{1}{m{1.7cm}}{\includegraphics[height=2.00cm,clip]{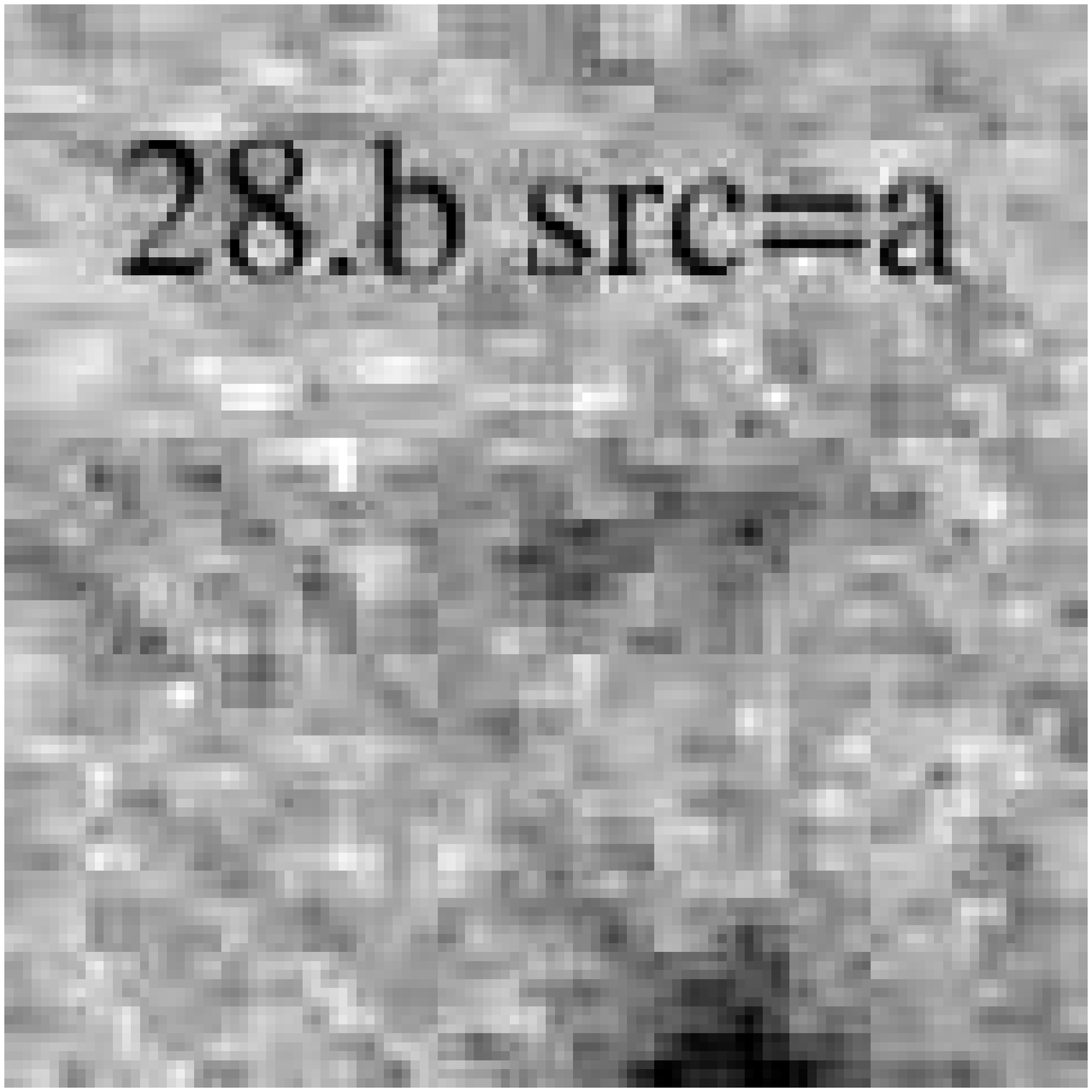}} \\
  \end{tabular}

\end{table*}

\begin{table*}
  \caption{Image system 29:}\vspace{0mm}
  \begin{tabular}{cccccc}
    \multicolumn{1}{m{1cm}}{{\Large A1689}}
    & \multicolumn{1}{m{1.7cm}}{\includegraphics[height=2.00cm,clip]{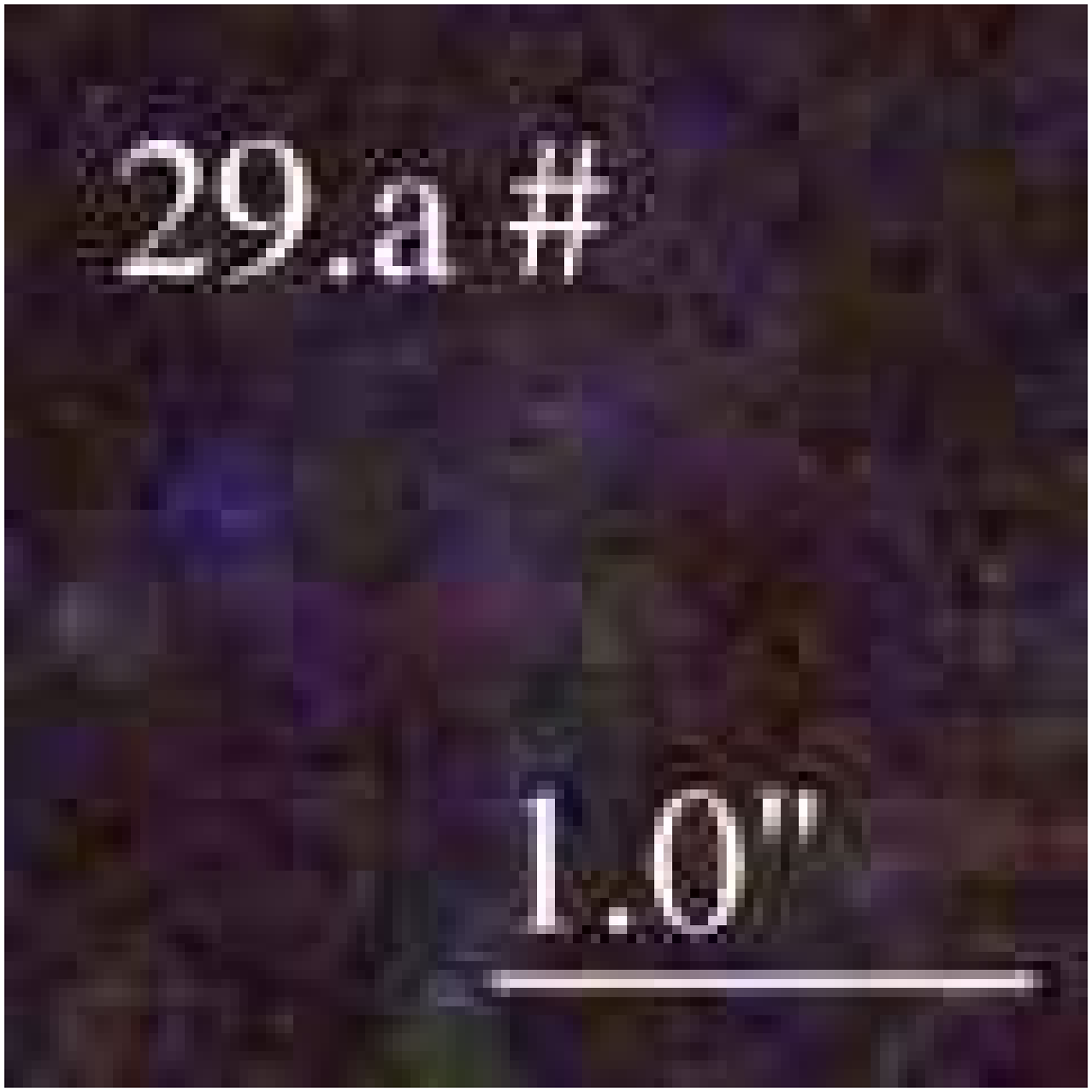}}
    & \multicolumn{1}{m{1.7cm}}{\includegraphics[height=2.00cm,clip]{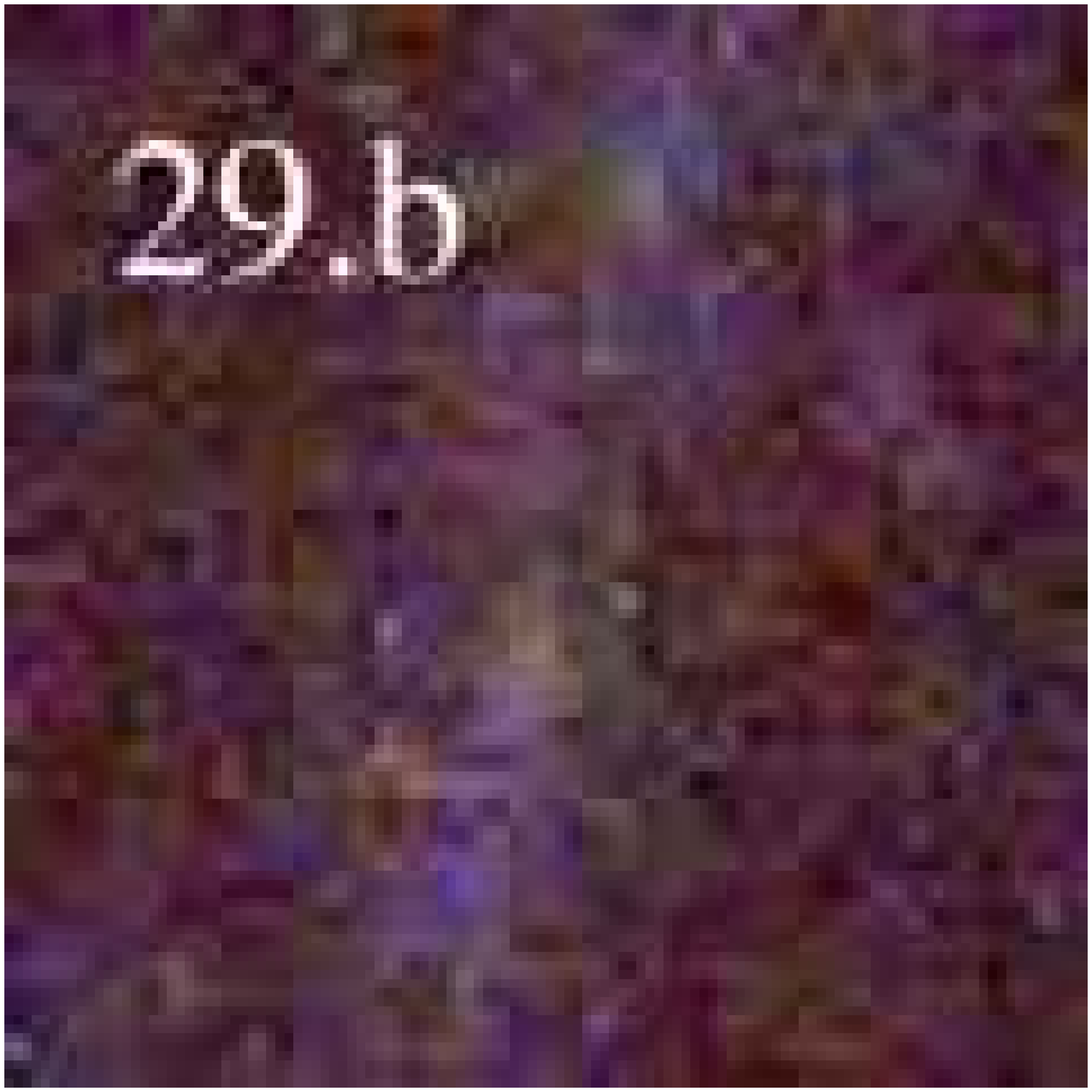}}
    & \multicolumn{1}{m{1.7cm}}{\includegraphics[height=2.00cm,clip]{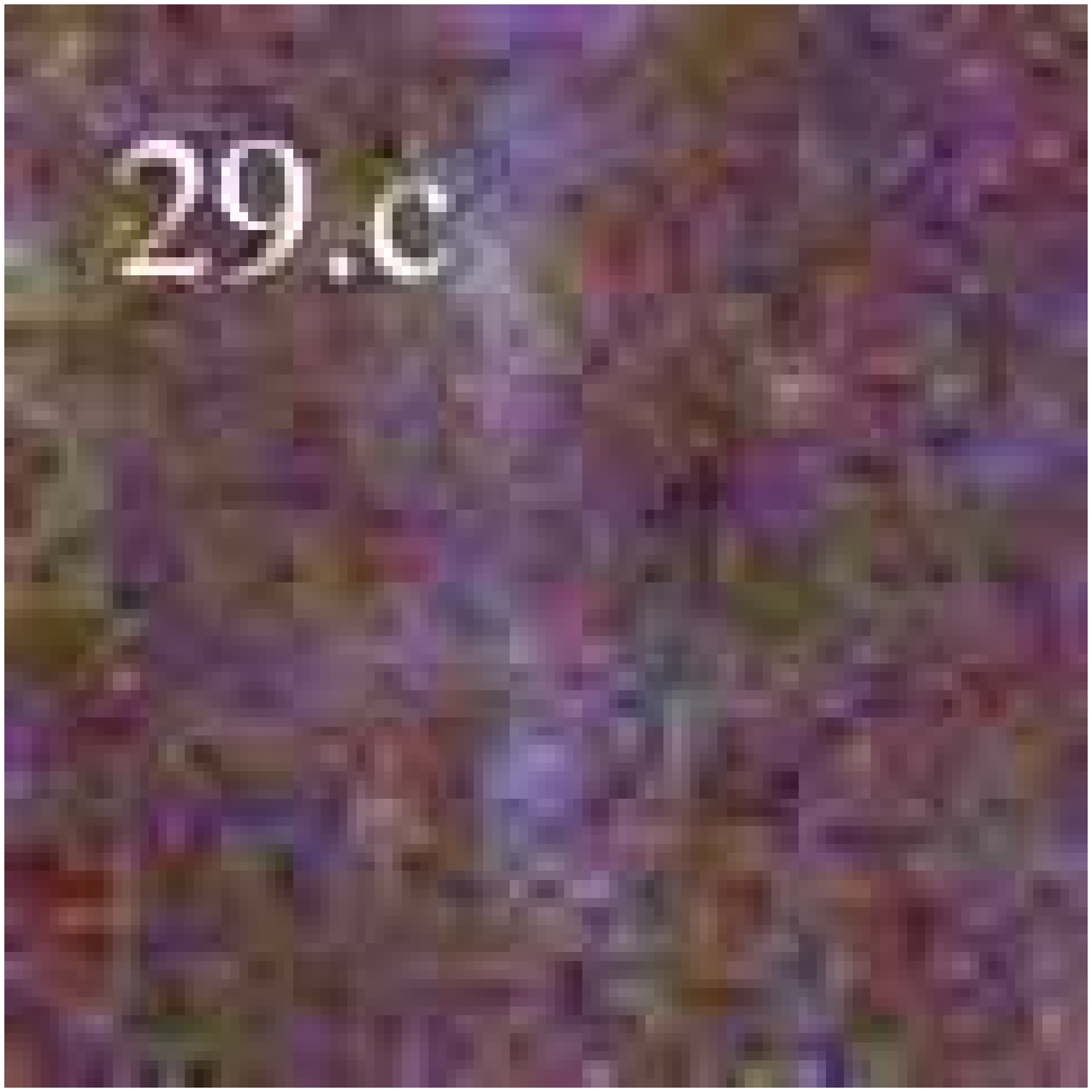}}
    & \multicolumn{1}{m{1.7cm}}{\includegraphics[height=2.00cm,clip]{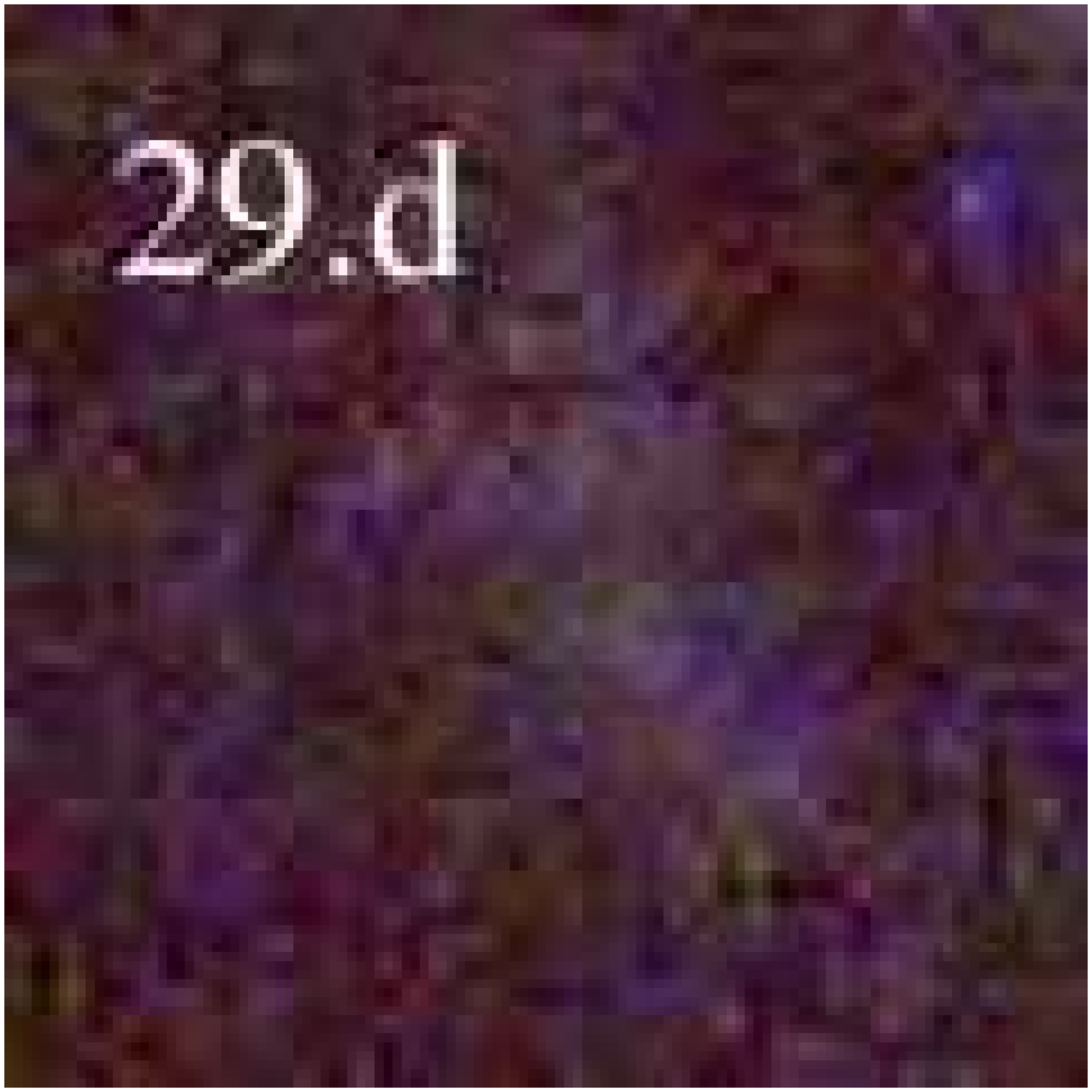}}
    & \multicolumn{1}{m{1.7cm}}{\includegraphics[height=2.00cm,clip]{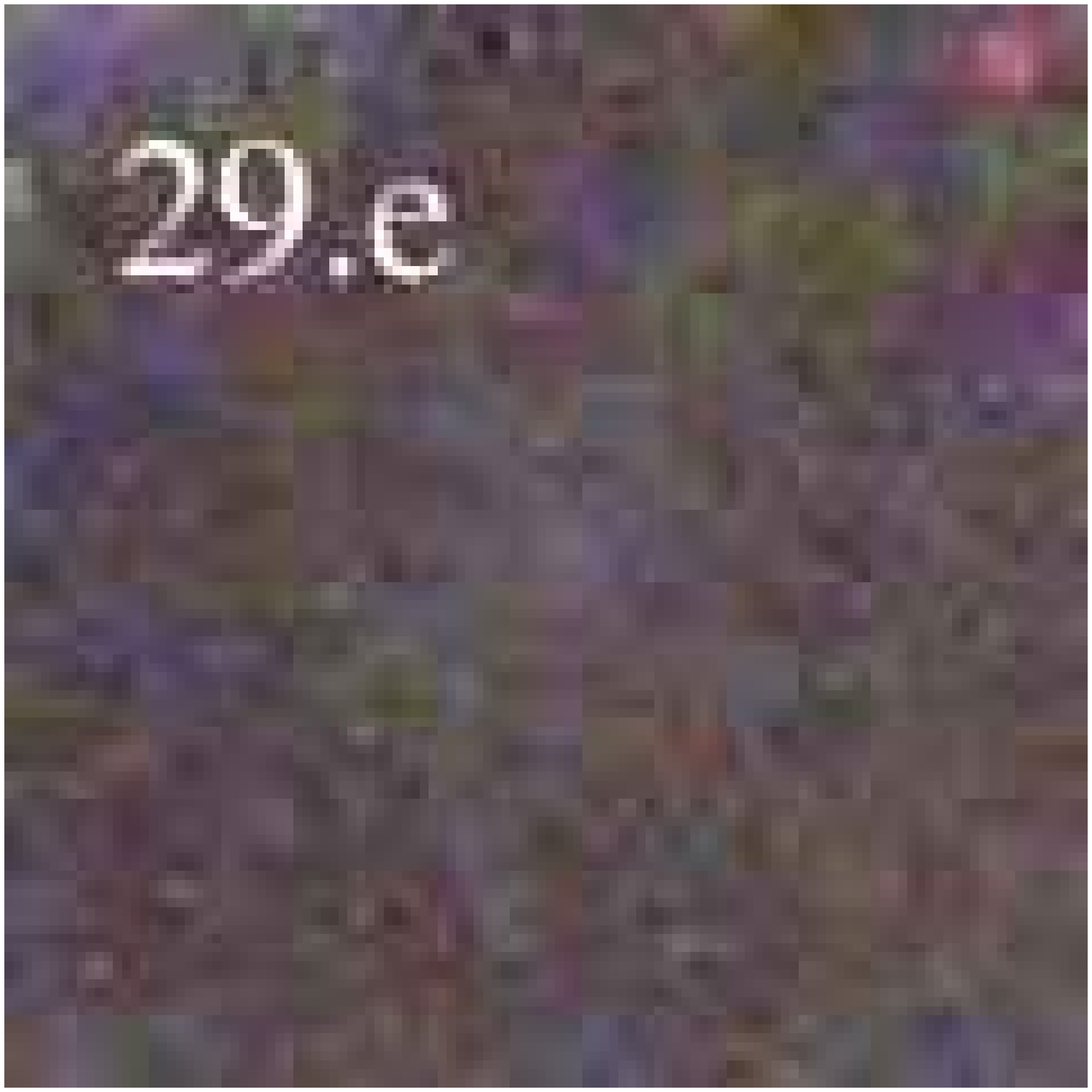}} \\
    \multicolumn{1}{m{1cm}}{{\Large NSIE}}
    & \multicolumn{1}{m{1.7cm}}{\includegraphics[height=2.00cm,clip]{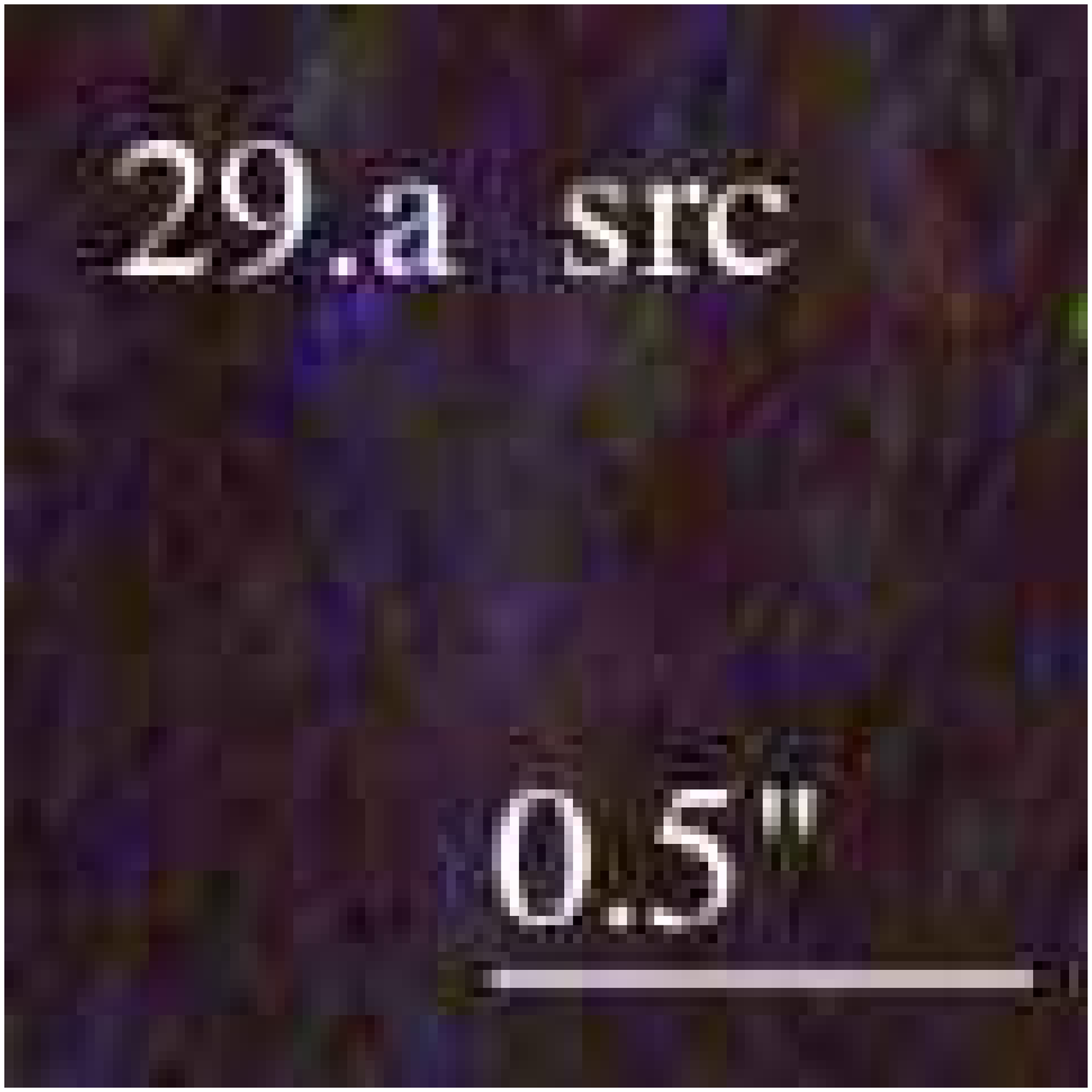}}
    & \multicolumn{1}{m{1.7cm}}{\includegraphics[height=2.00cm,clip]{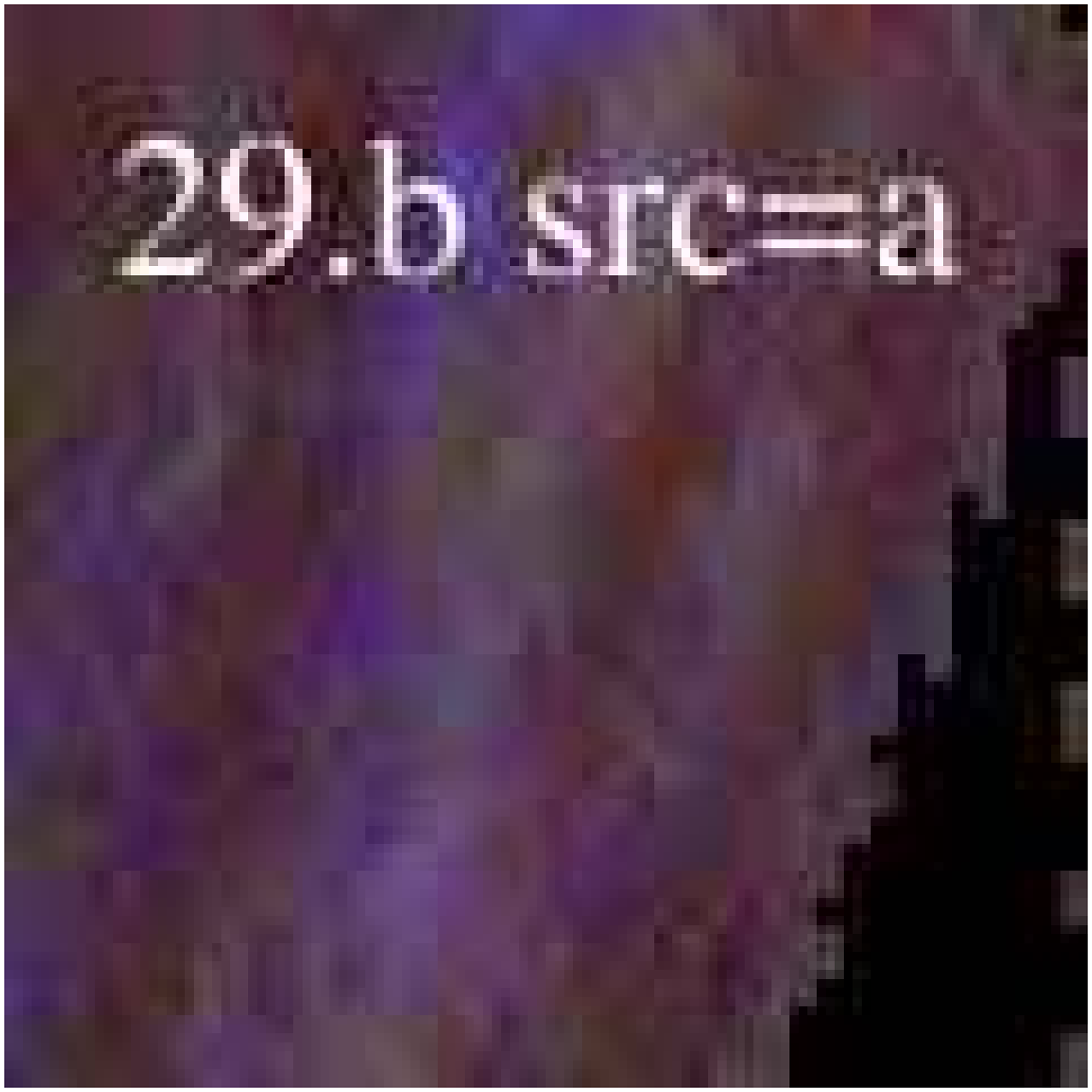}}
    & \multicolumn{1}{m{1.7cm}}{\includegraphics[height=2.00cm,clip]{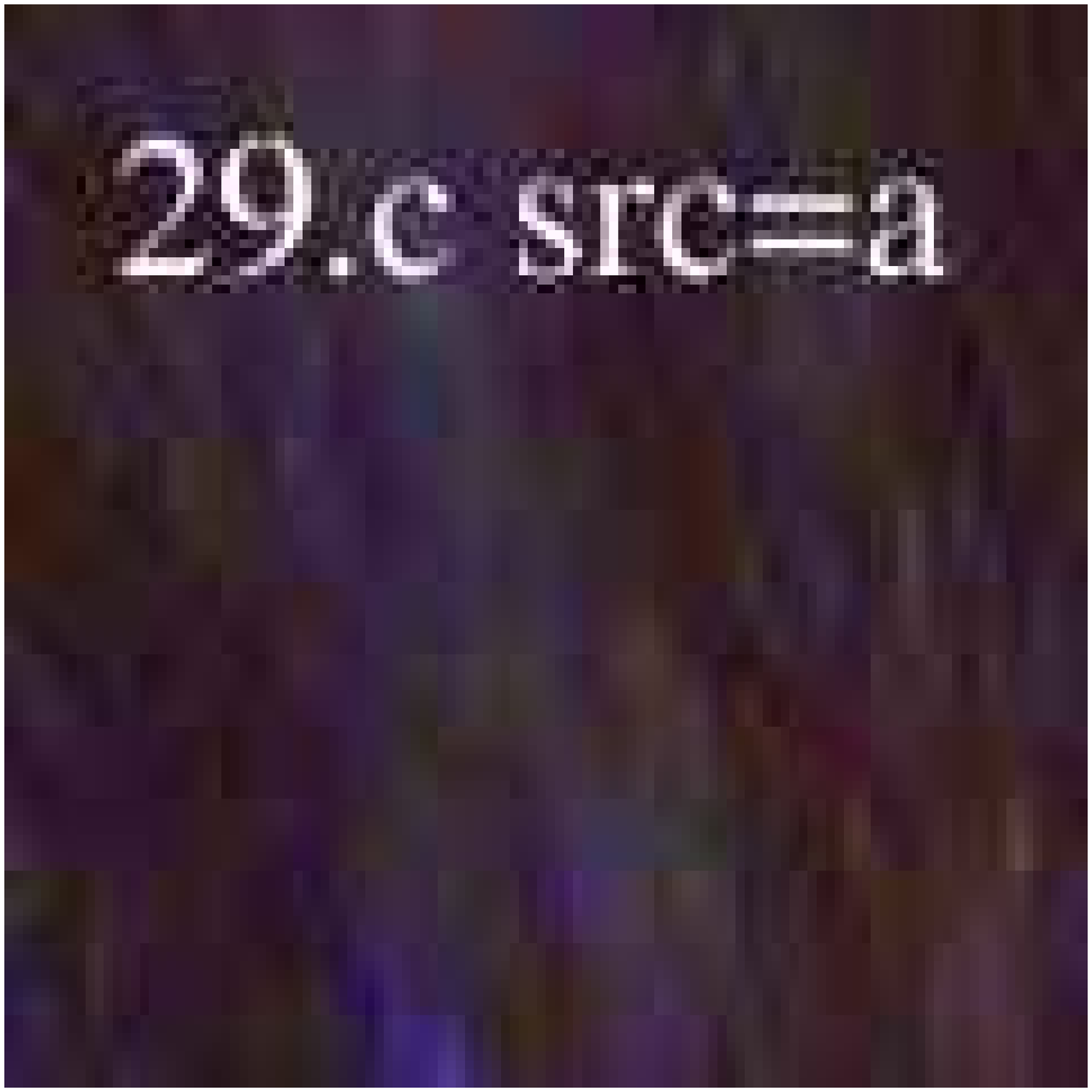}}
    & \multicolumn{1}{m{1.7cm}}{\includegraphics[height=2.00cm,clip]{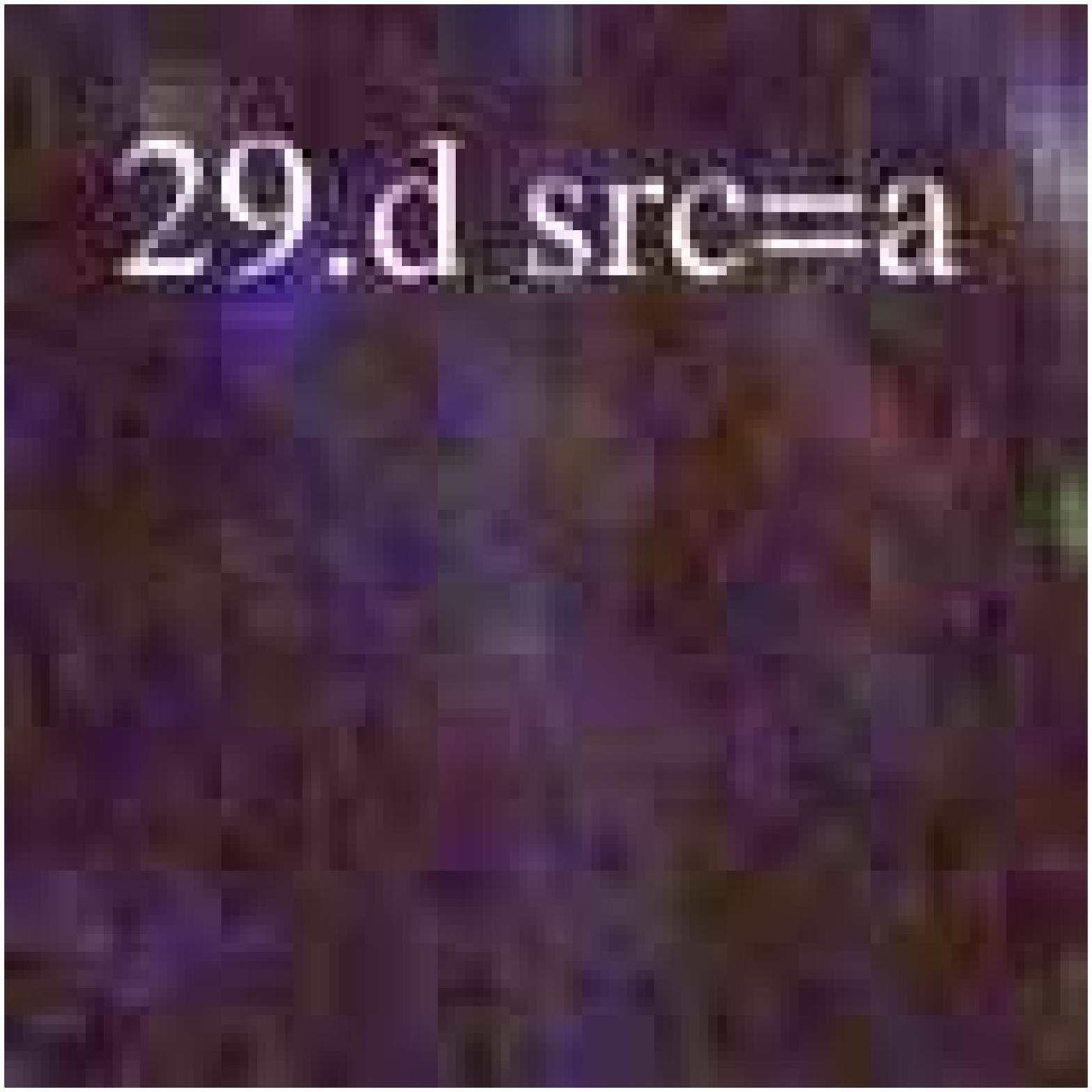}}
    & \multicolumn{1}{m{1.7cm}}{\includegraphics[height=2.00cm,clip]{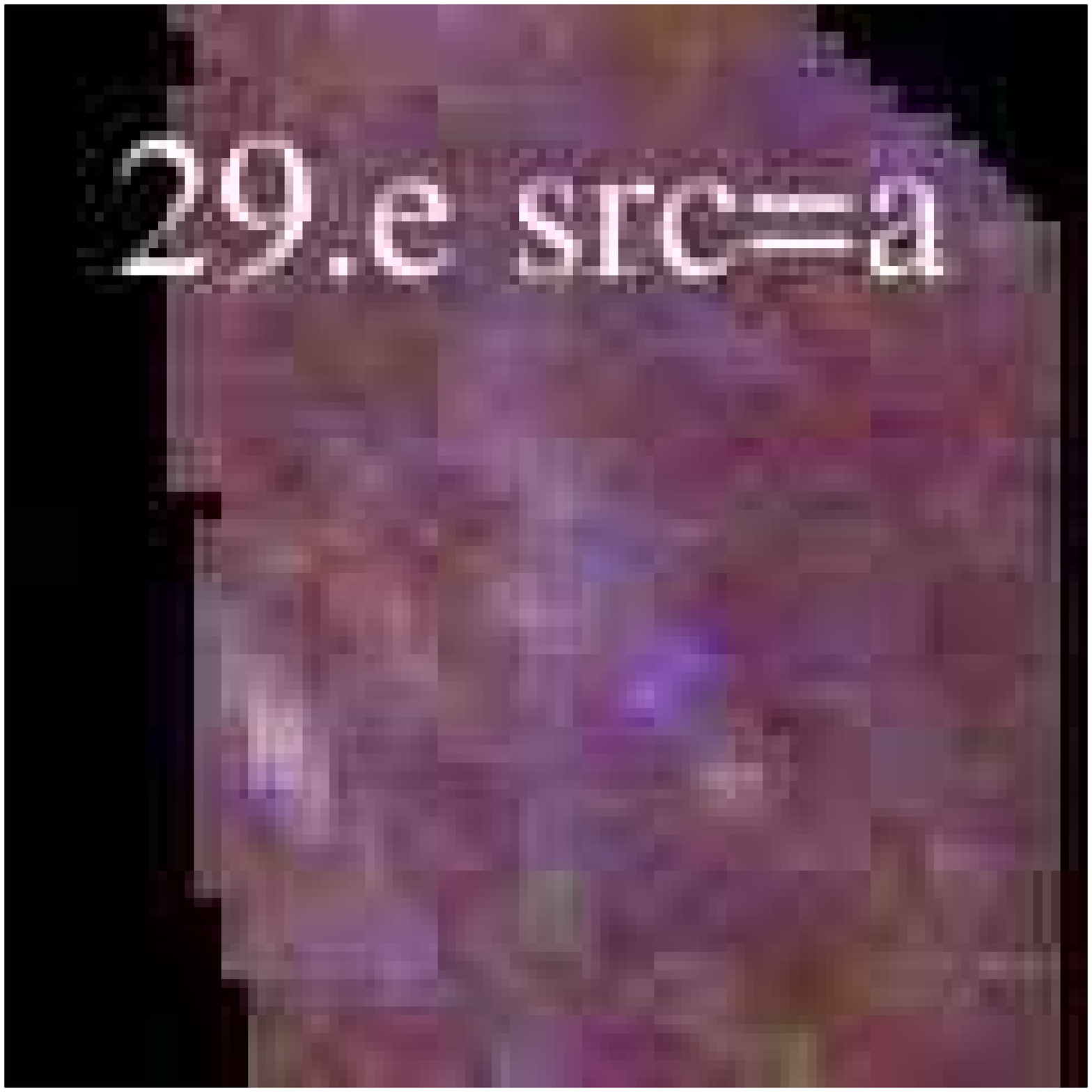}} \\
    \multicolumn{1}{m{1cm}}{{\Large ENFW}}
    & \multicolumn{1}{m{1.7cm}}{\includegraphics[height=2.00cm,clip]{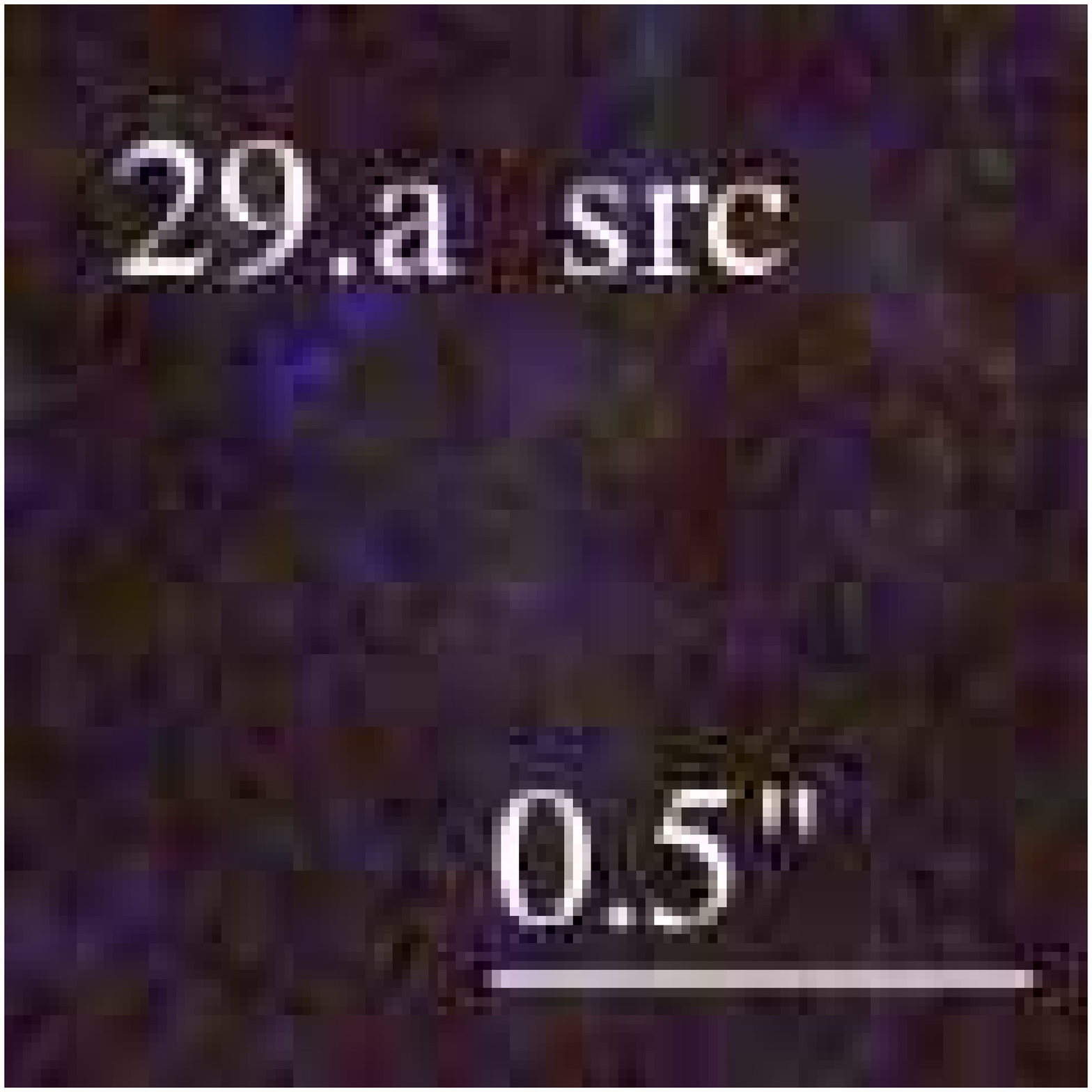}}
    & \multicolumn{1}{m{1.7cm}}{\includegraphics[height=2.00cm,clip]{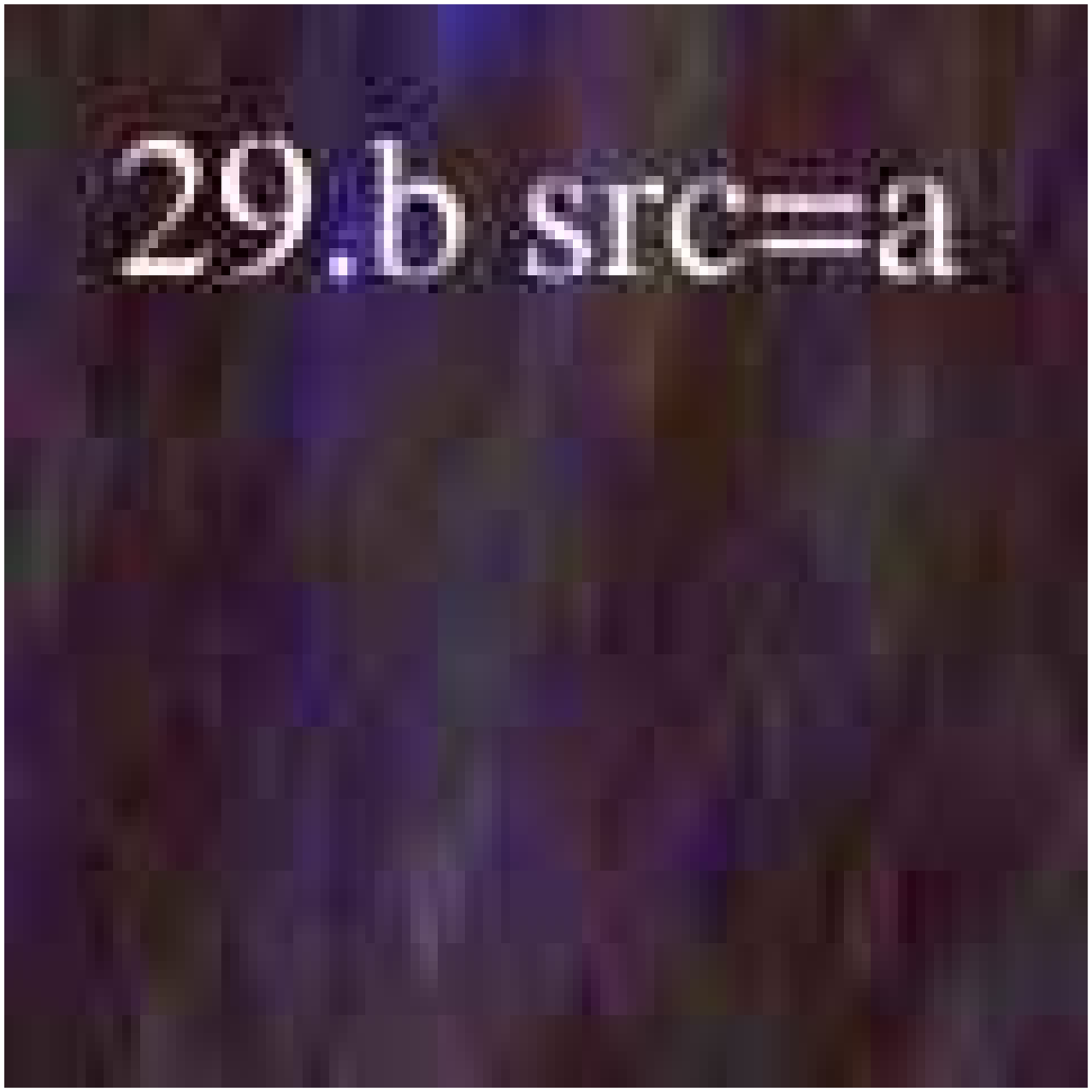}}
    & \multicolumn{1}{m{1.7cm}}{\includegraphics[height=2.00cm,clip]{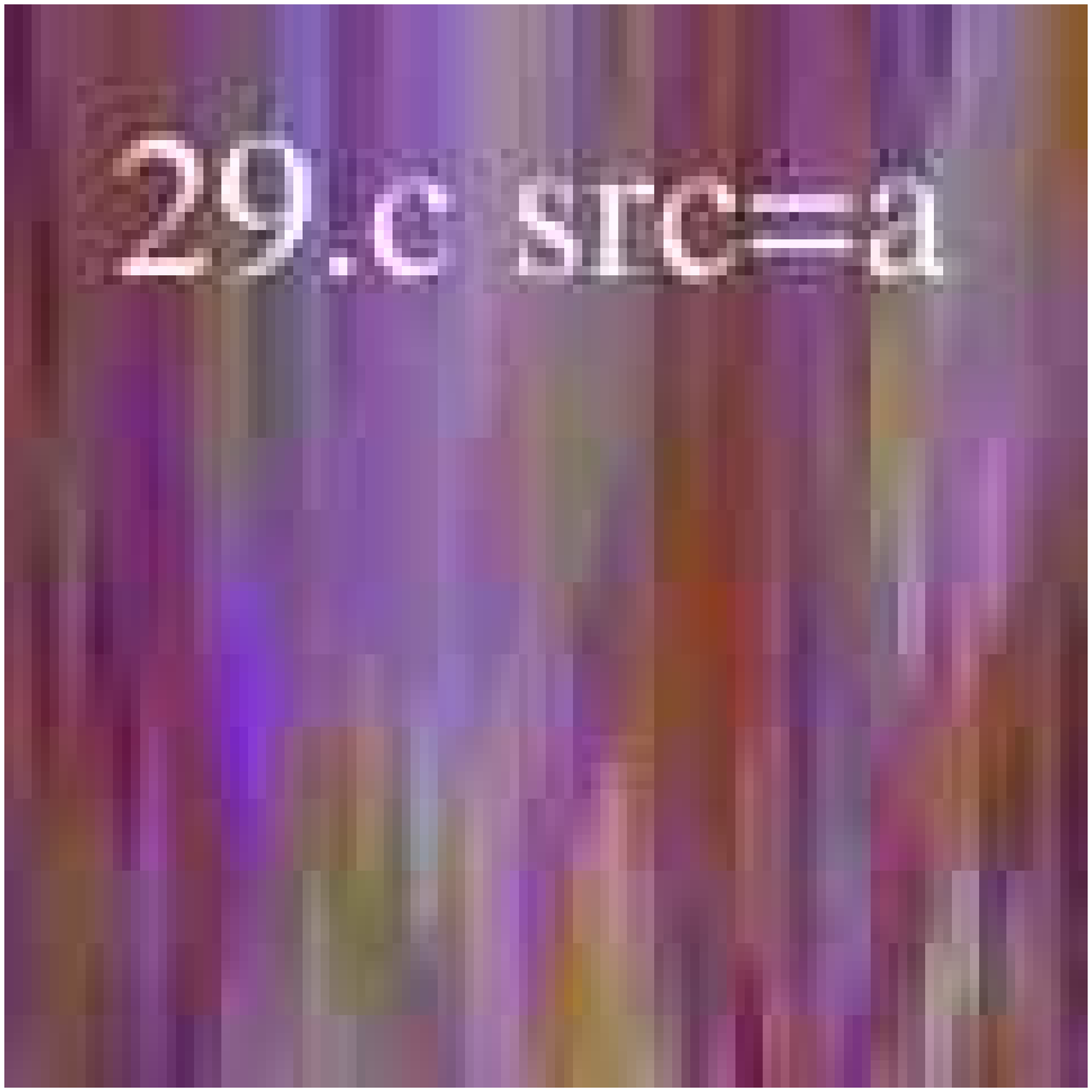}}
    & \multicolumn{1}{m{1.7cm}}{\includegraphics[height=2.00cm,clip]{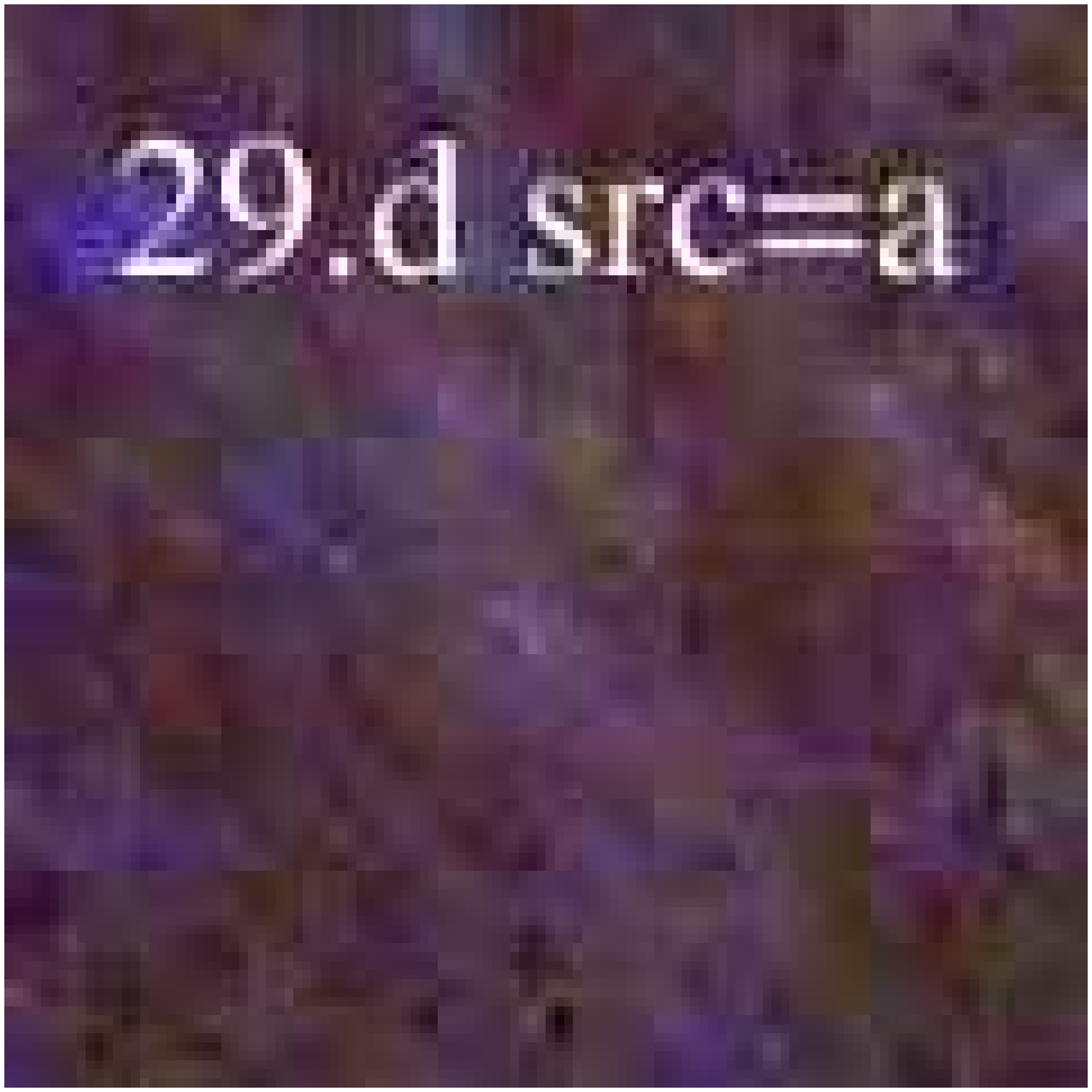}}
    & \multicolumn{1}{m{1.7cm}}{\includegraphics[height=2.00cm,clip]{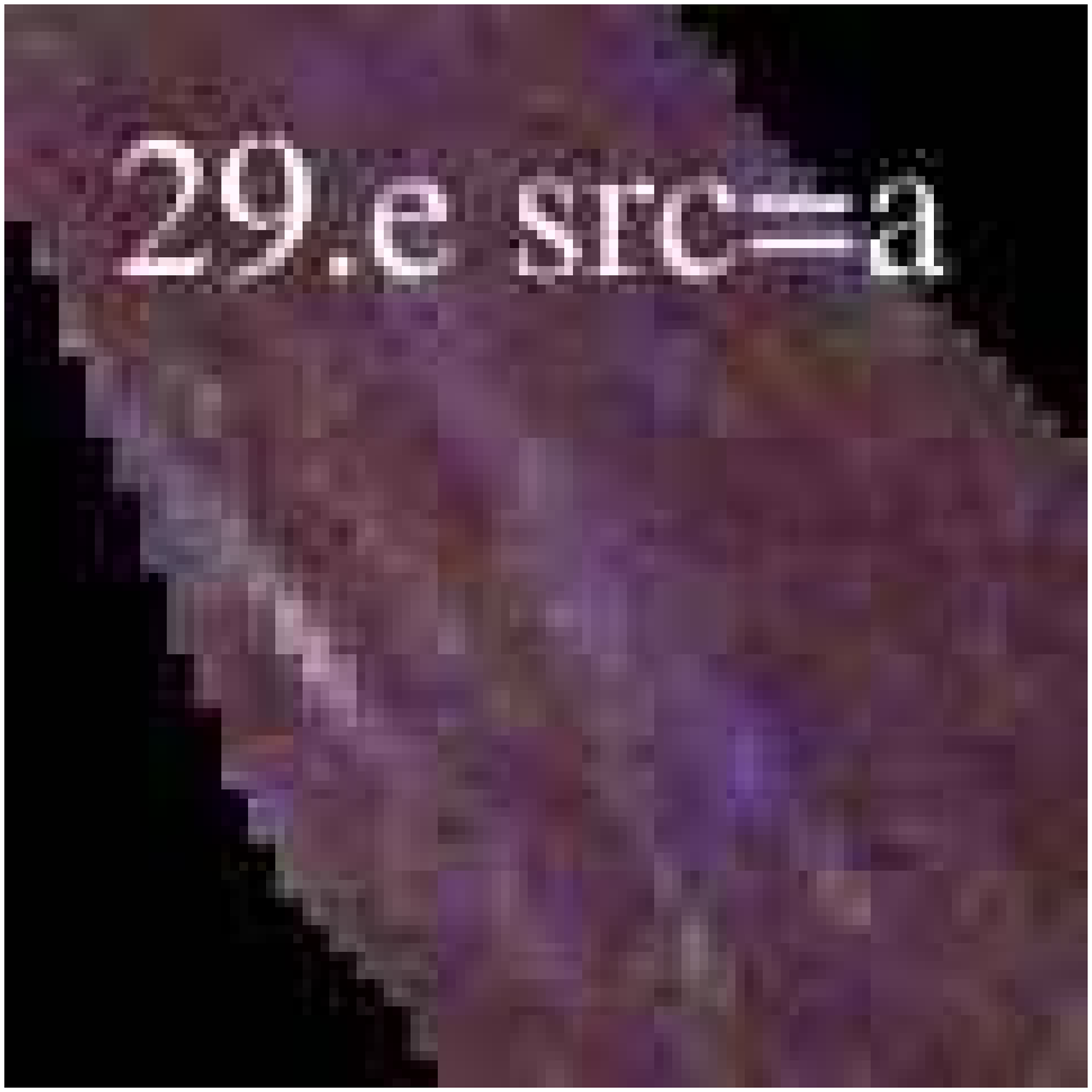}} \\
  \end{tabular}

\end{table*}

\begin{table*}
  \caption{Image system 30:}\vspace{0mm}
  \begin{tabular}{cccc}
    \multicolumn{1}{m{1cm}}{{\Large A1689}}
    & \multicolumn{1}{m{1.7cm}}{\includegraphics[height=2.00cm,clip]{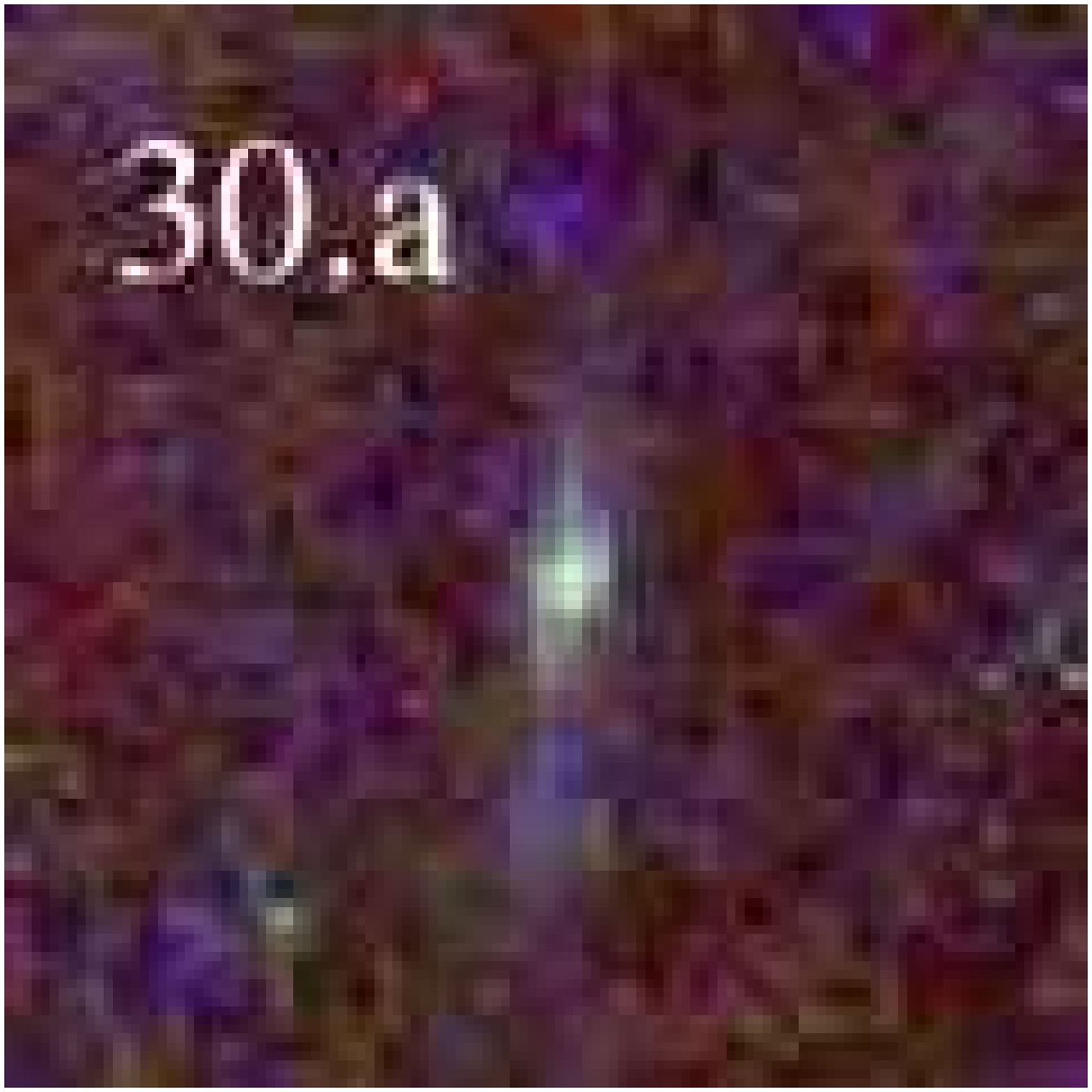}}
    & \multicolumn{1}{m{1.7cm}}{\includegraphics[height=2.00cm,clip]{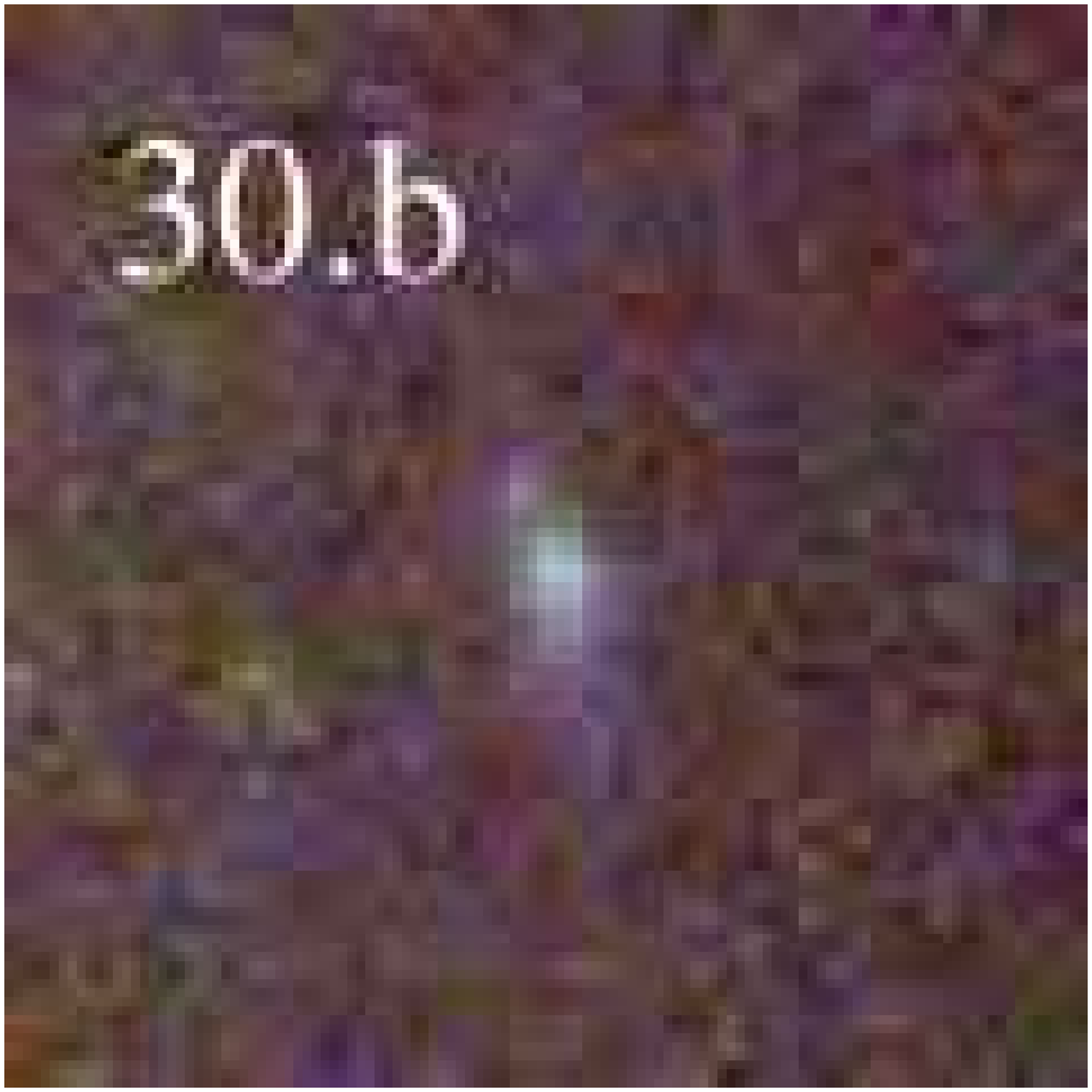}}
    & \multicolumn{1}{m{1.7cm}}{\includegraphics[height=2.00cm,clip]{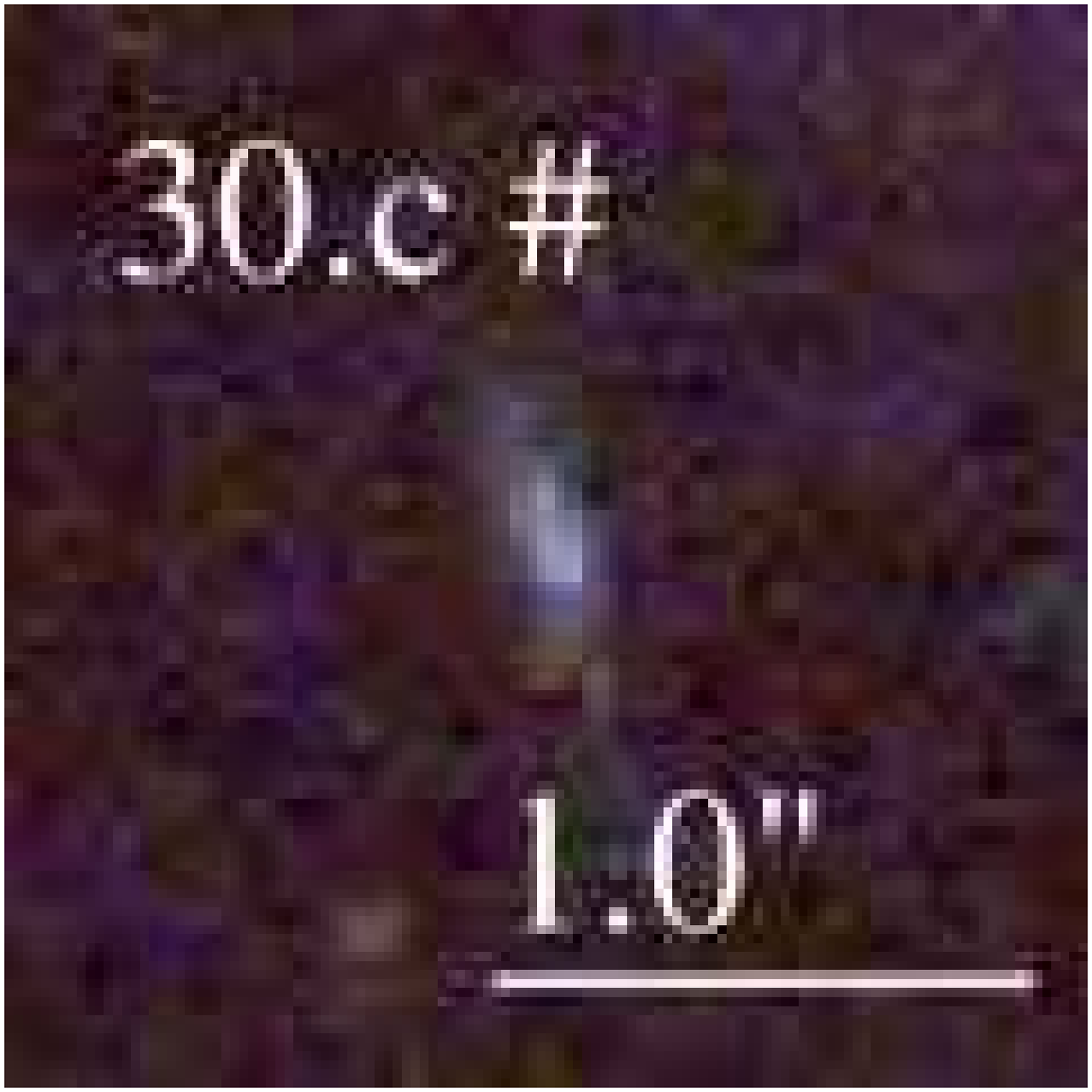}} \\
    \multicolumn{1}{m{1cm}}{{\Large NSIE}}
    & \multicolumn{1}{m{1.7cm}}{\includegraphics[height=2.00cm,clip]{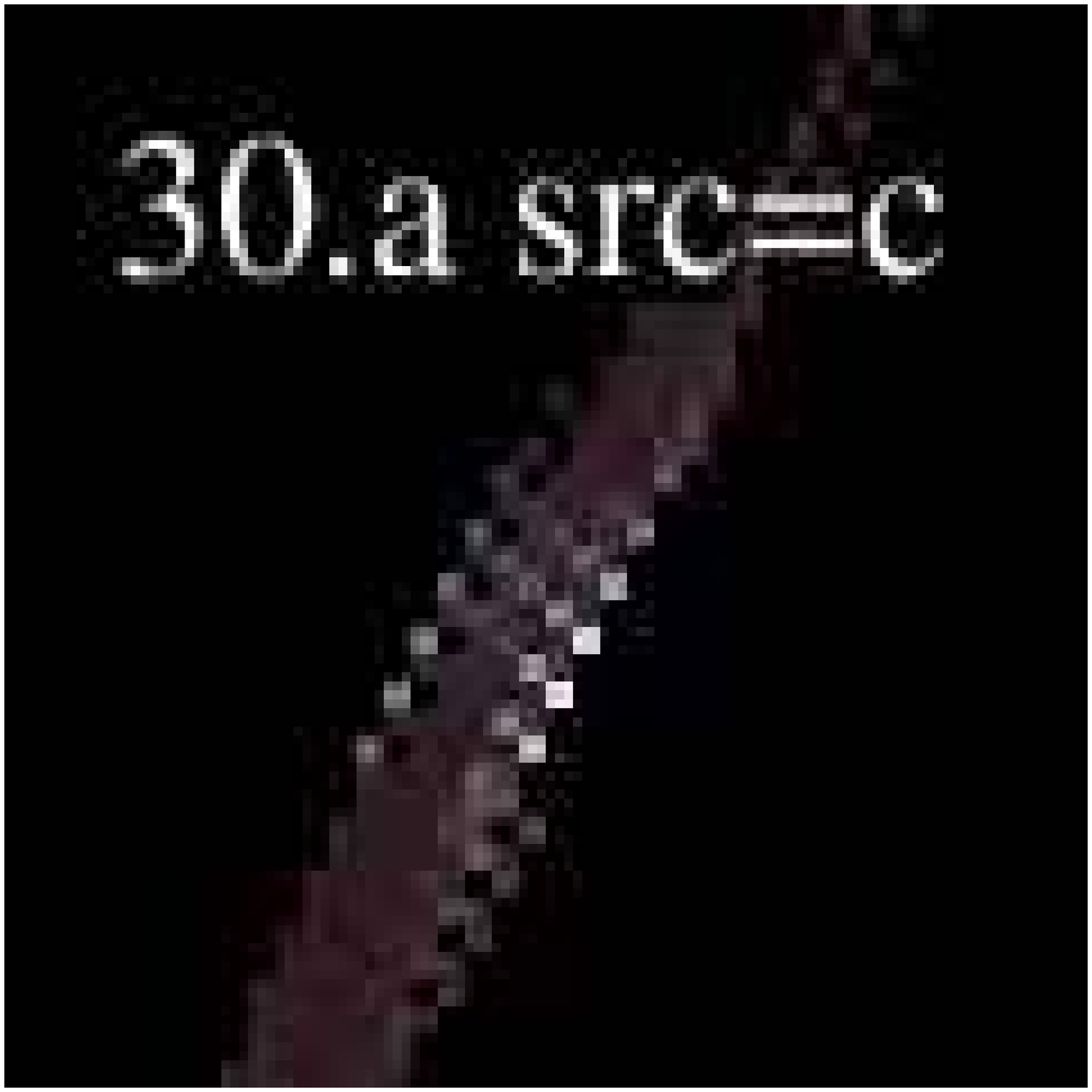}}
    & \multicolumn{1}{m{1.7cm}}{\includegraphics[height=2.00cm,clip]{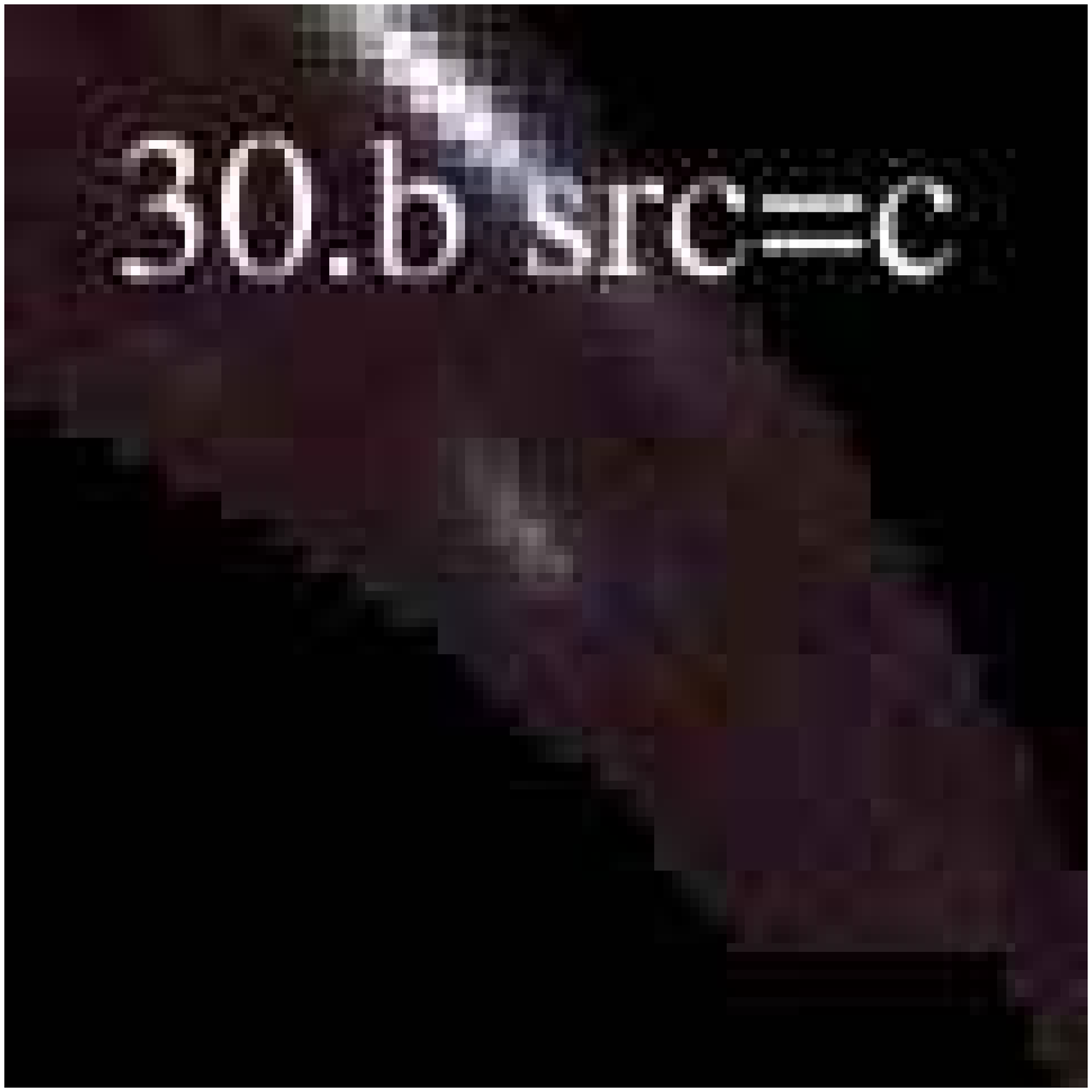}}
    & \multicolumn{1}{m{1.7cm}}{\includegraphics[height=2.00cm,clip]{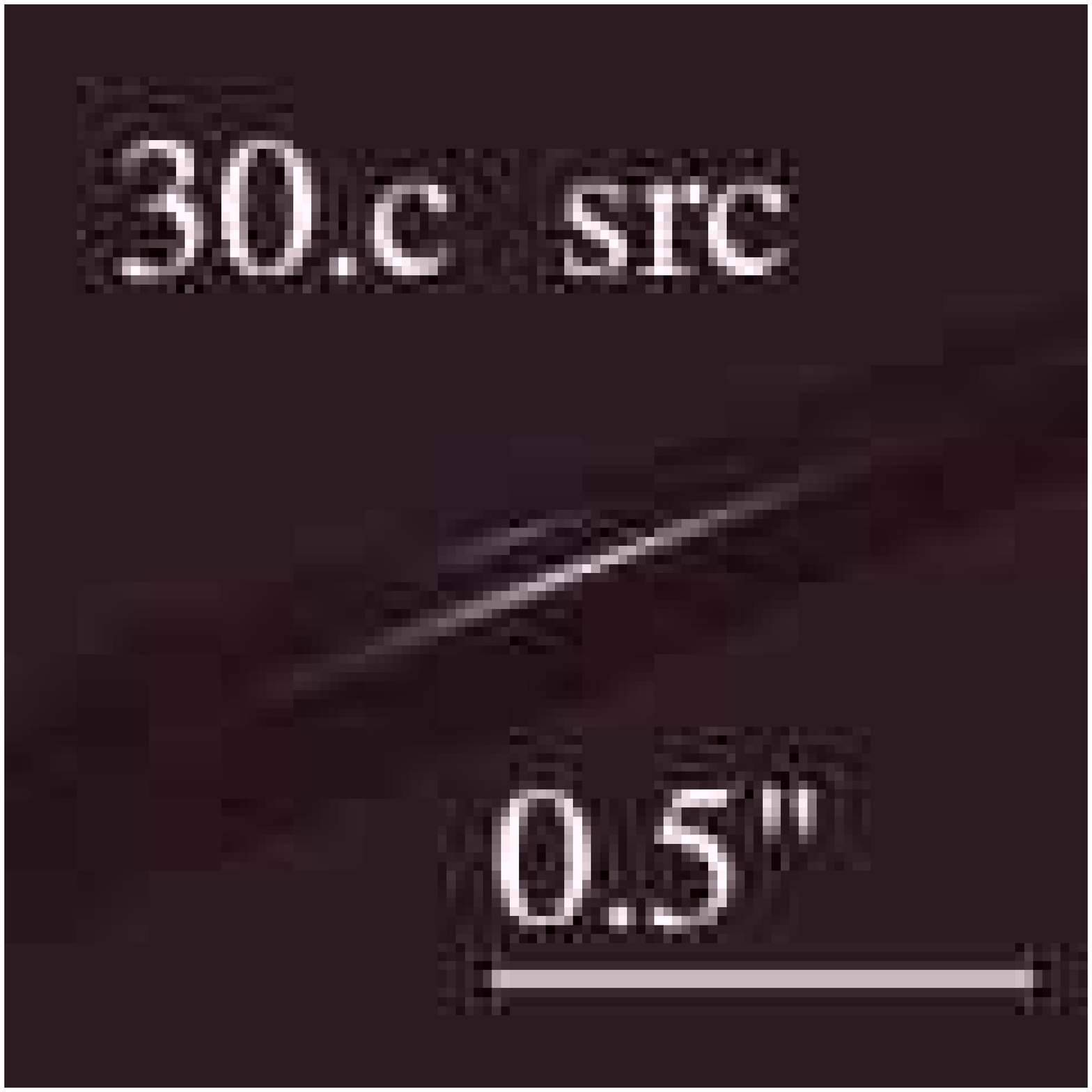}} \\
    \multicolumn{1}{m{1cm}}{{\Large ENFW}}
    & \multicolumn{1}{m{1.7cm}}{\includegraphics[height=2.00cm,clip]{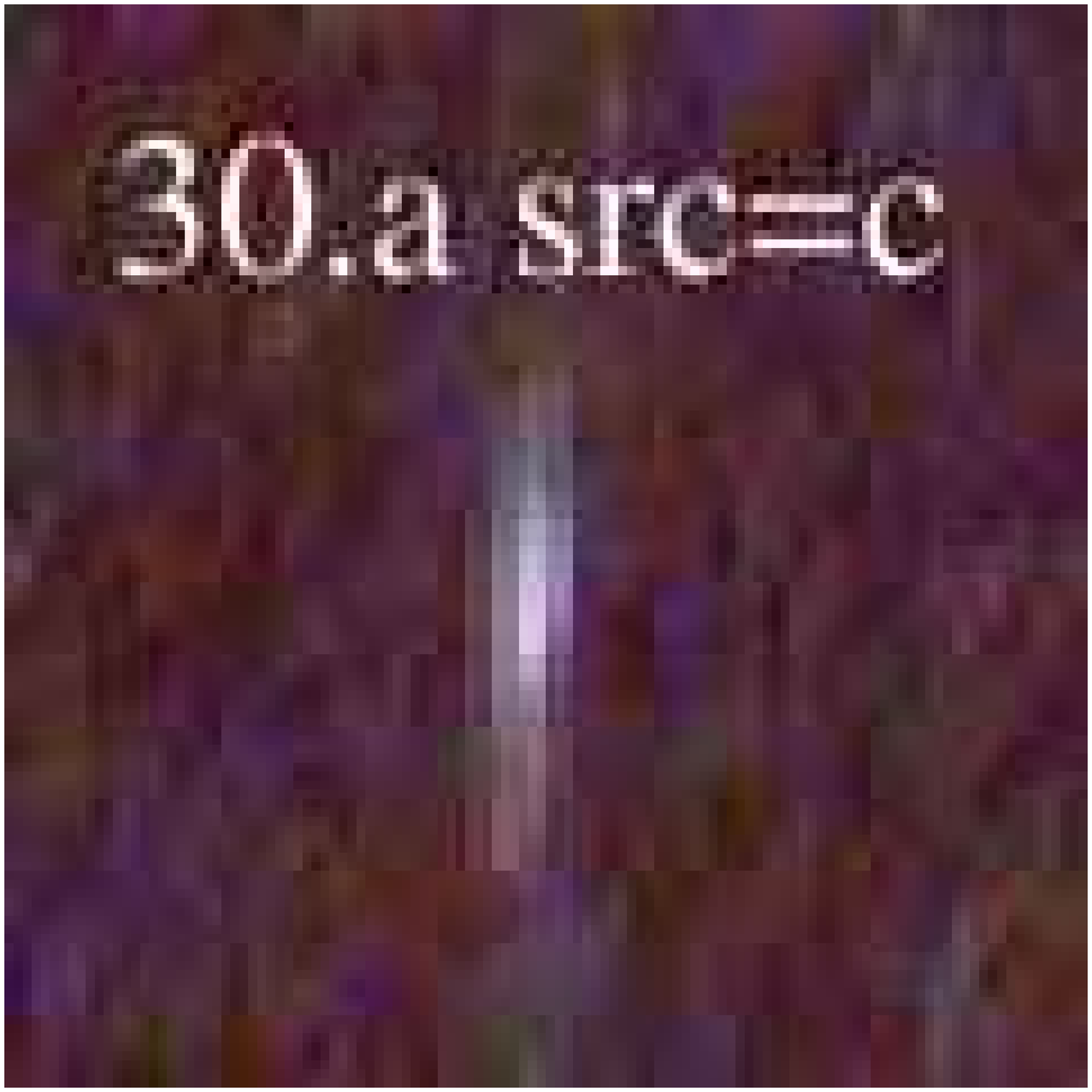}}
    & \multicolumn{1}{m{1.7cm}}{\includegraphics[height=2.00cm,clip]{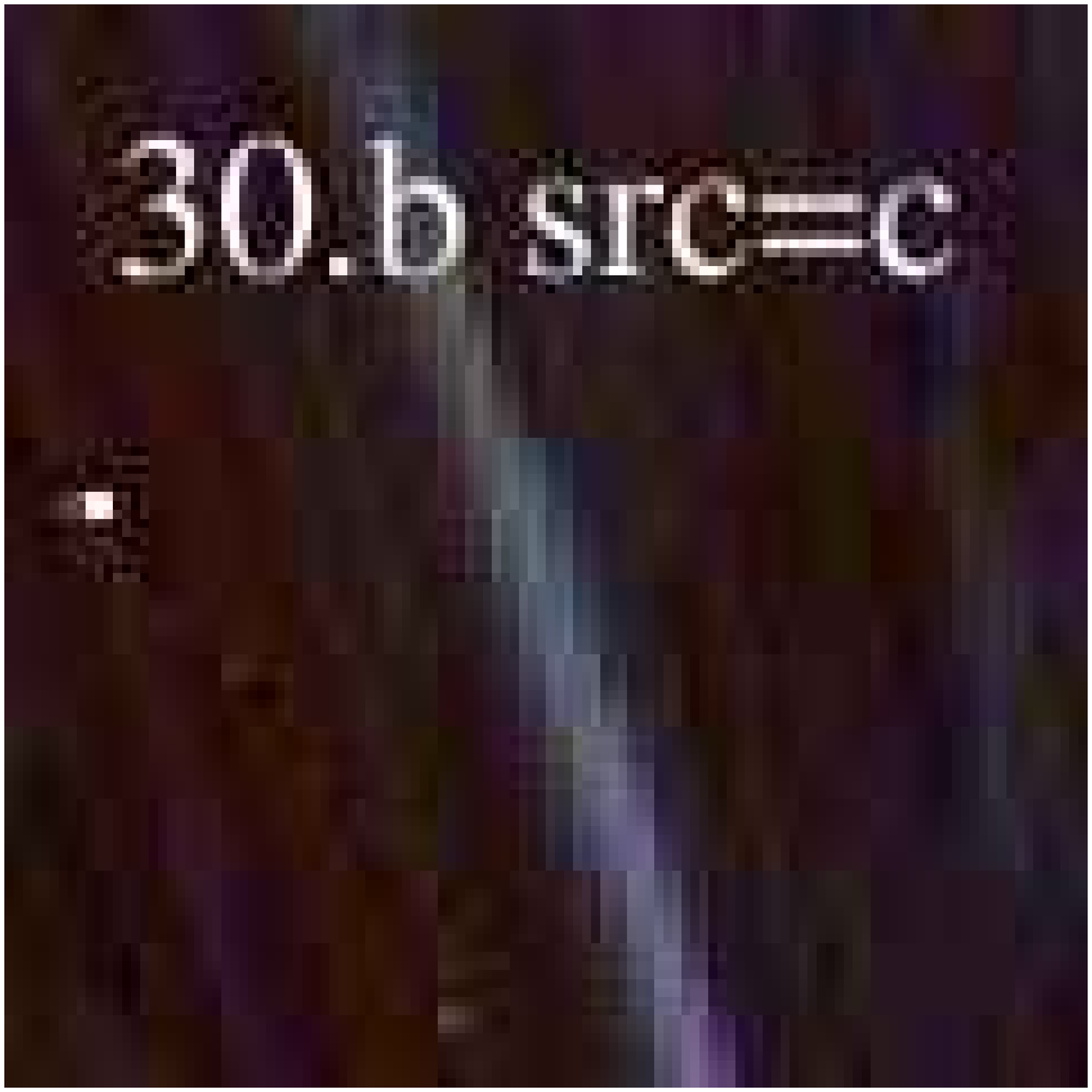}}
    & \multicolumn{1}{m{1.7cm}}{\includegraphics[height=2.00cm,clip]{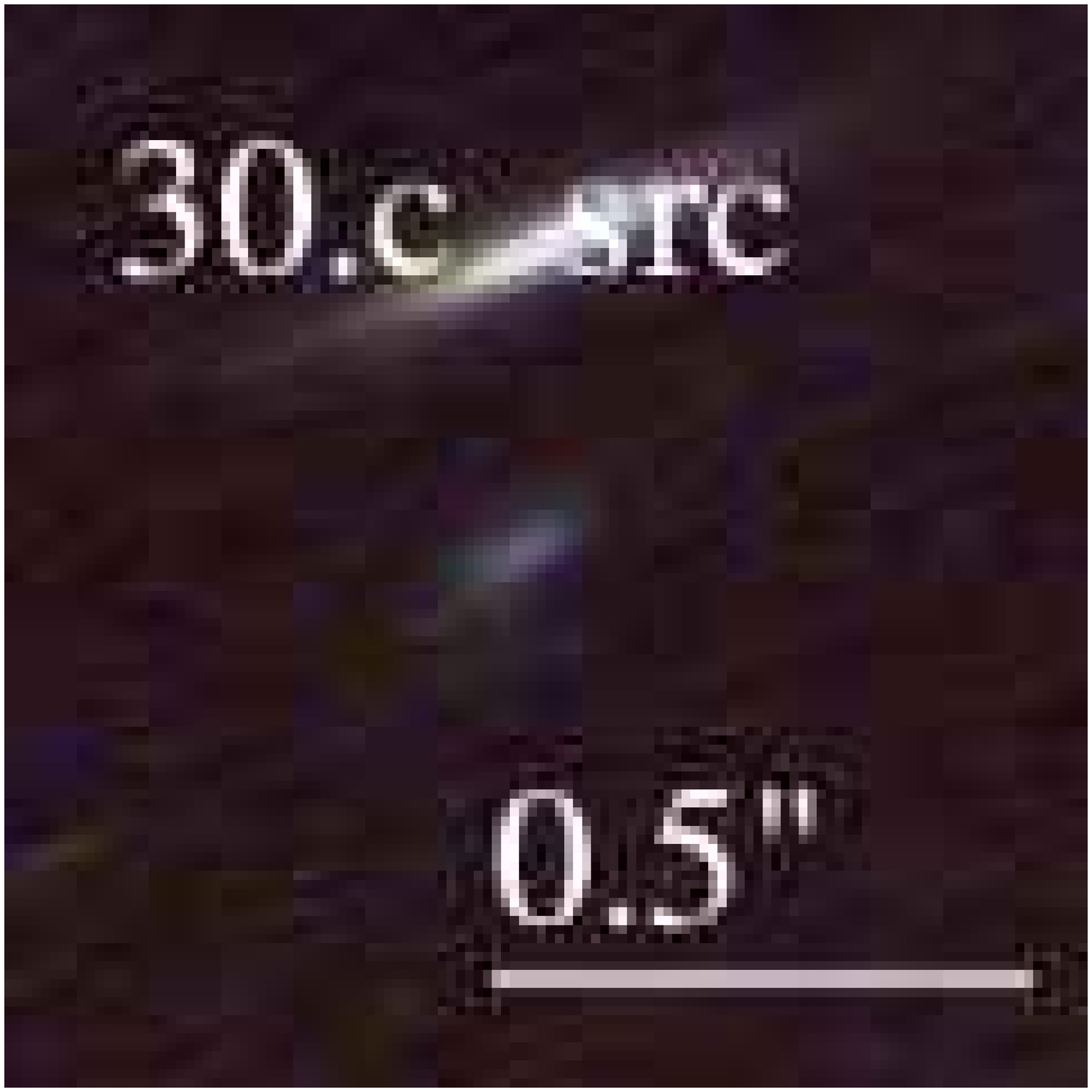}} \\
  \end{tabular}

\end{table*}

\clearpage

\begin{table*}
  \caption{Image system 31:}\vspace{0mm}
  \begin{tabular}{ccc}
    \multicolumn{1}{m{1cm}}{{\Large A1689}}
    & \multicolumn{1}{m{1.7cm}}{\includegraphics[height=2.00cm,clip]{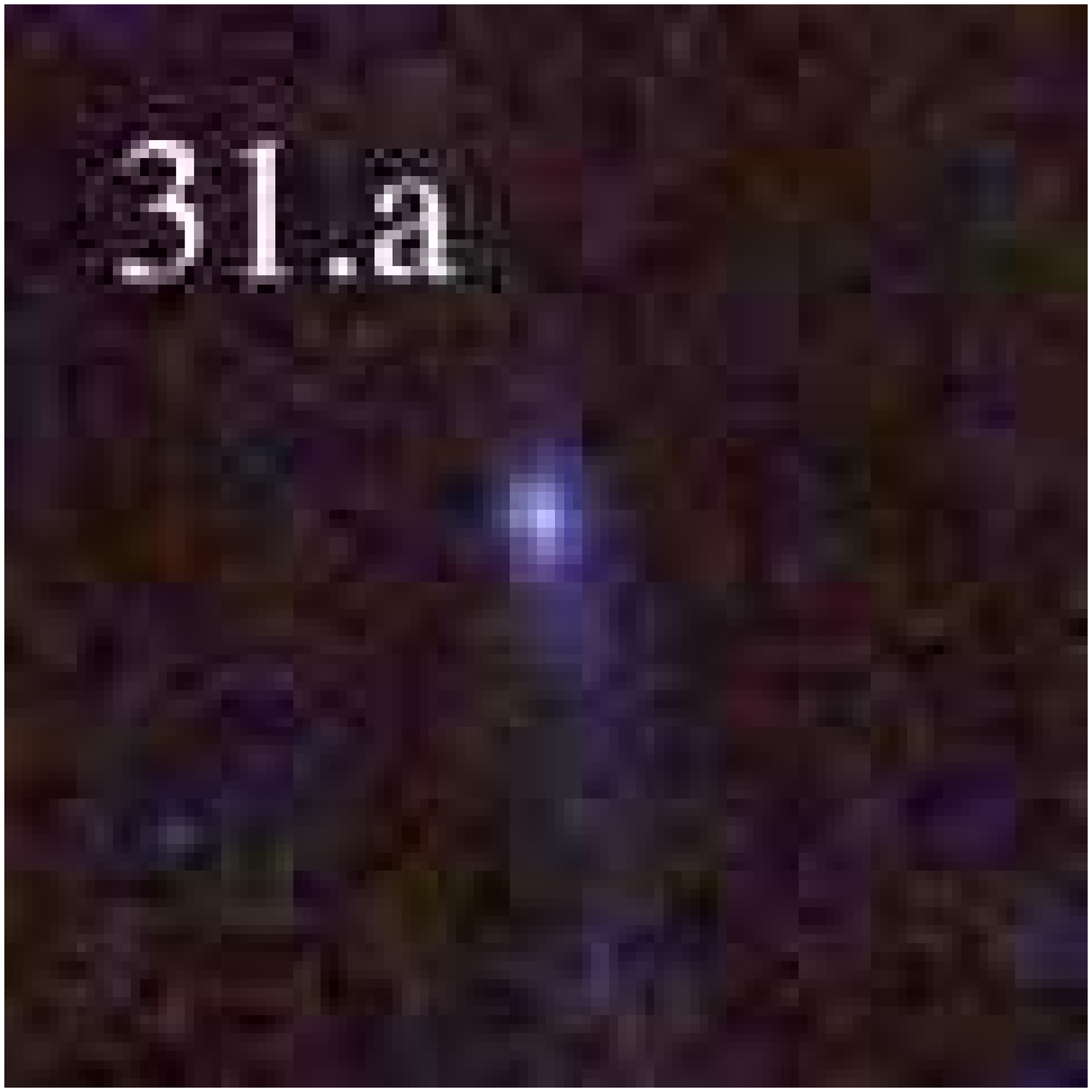}}
    & \multicolumn{1}{m{1.7cm}}{\includegraphics[height=2.00cm,clip]{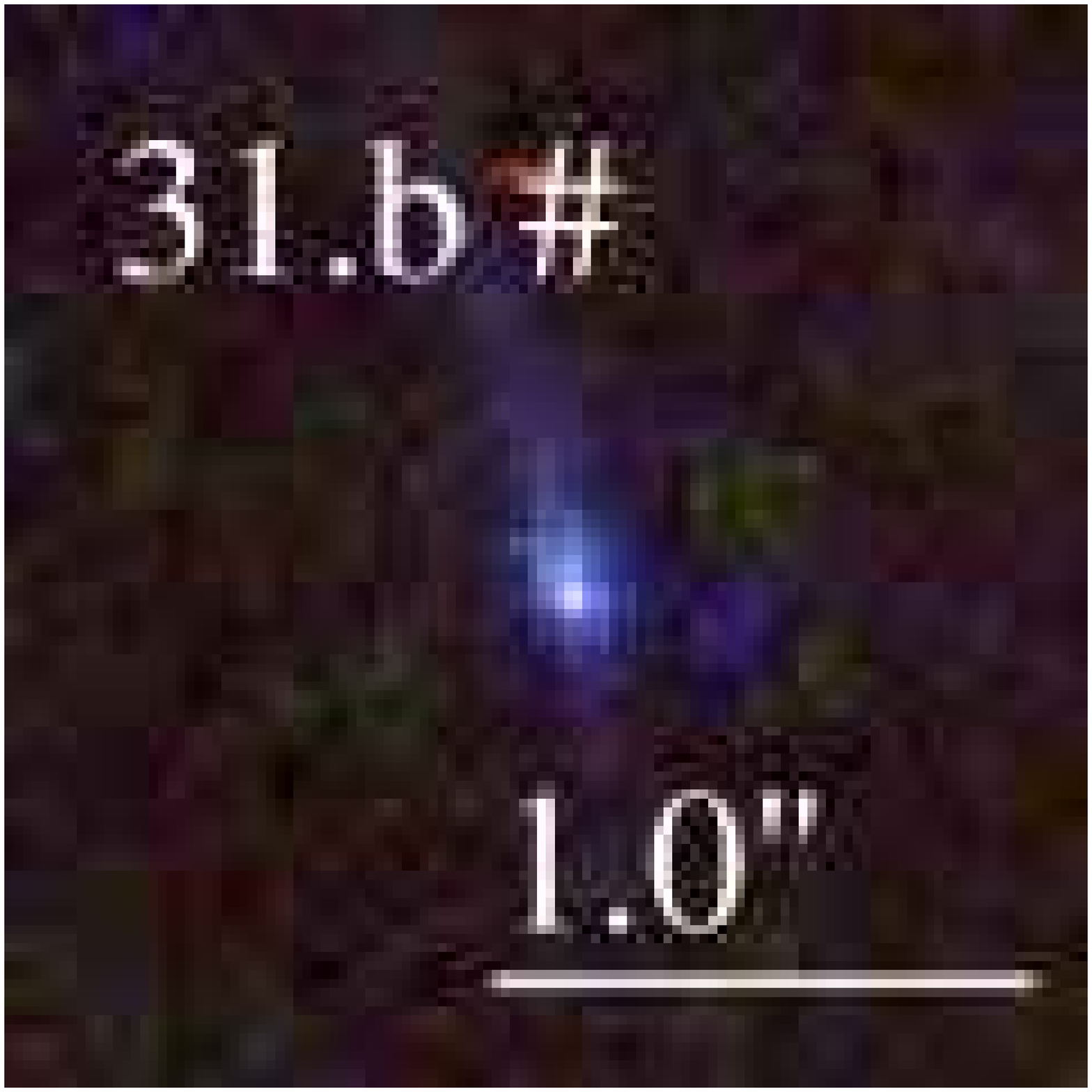}} \\
    \multicolumn{1}{m{1cm}}{{\Large NSIE}}
    & \multicolumn{1}{m{1.7cm}}{\includegraphics[height=2.00cm,clip]{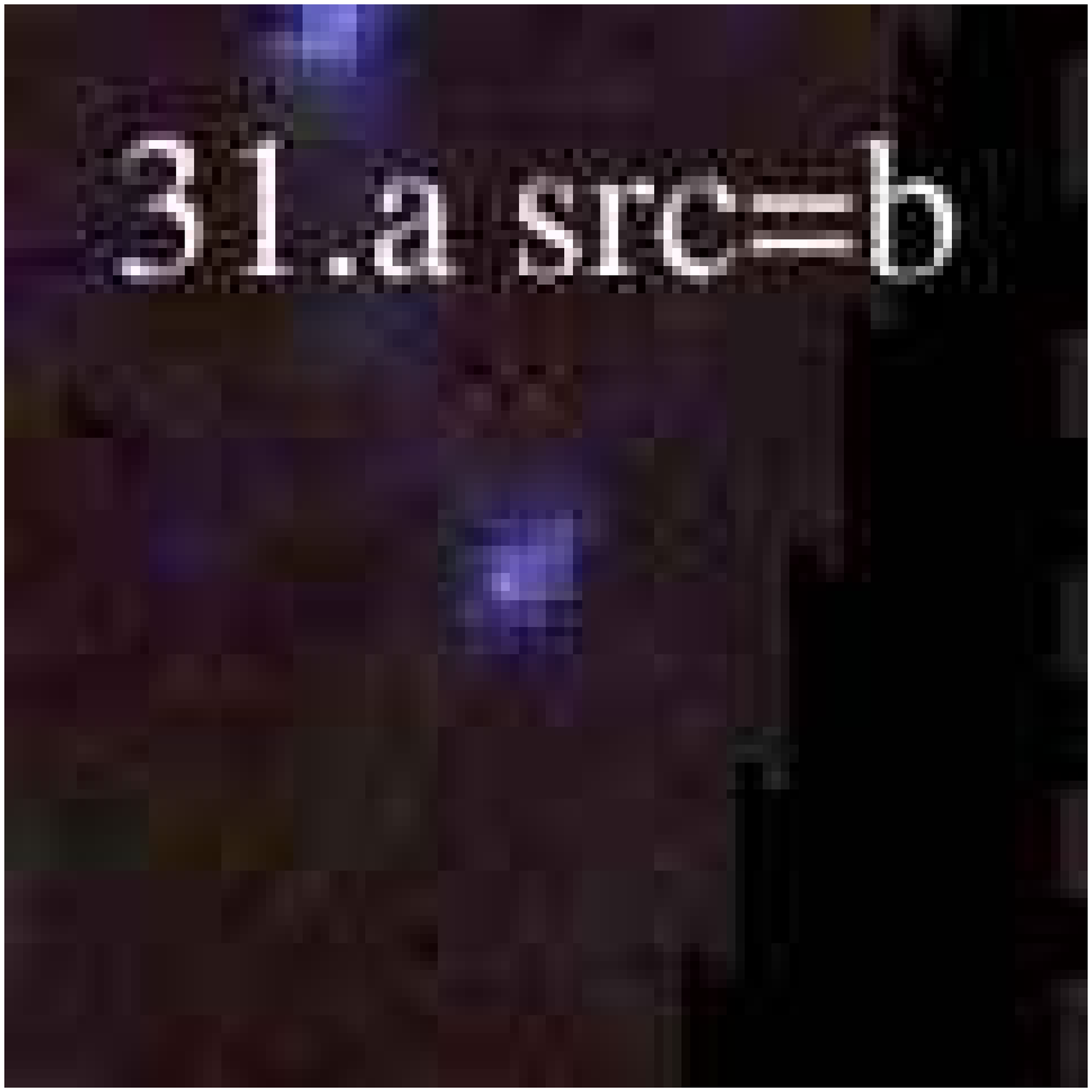}}
    & \multicolumn{1}{m{1.7cm}}{\includegraphics[height=2.00cm,clip]{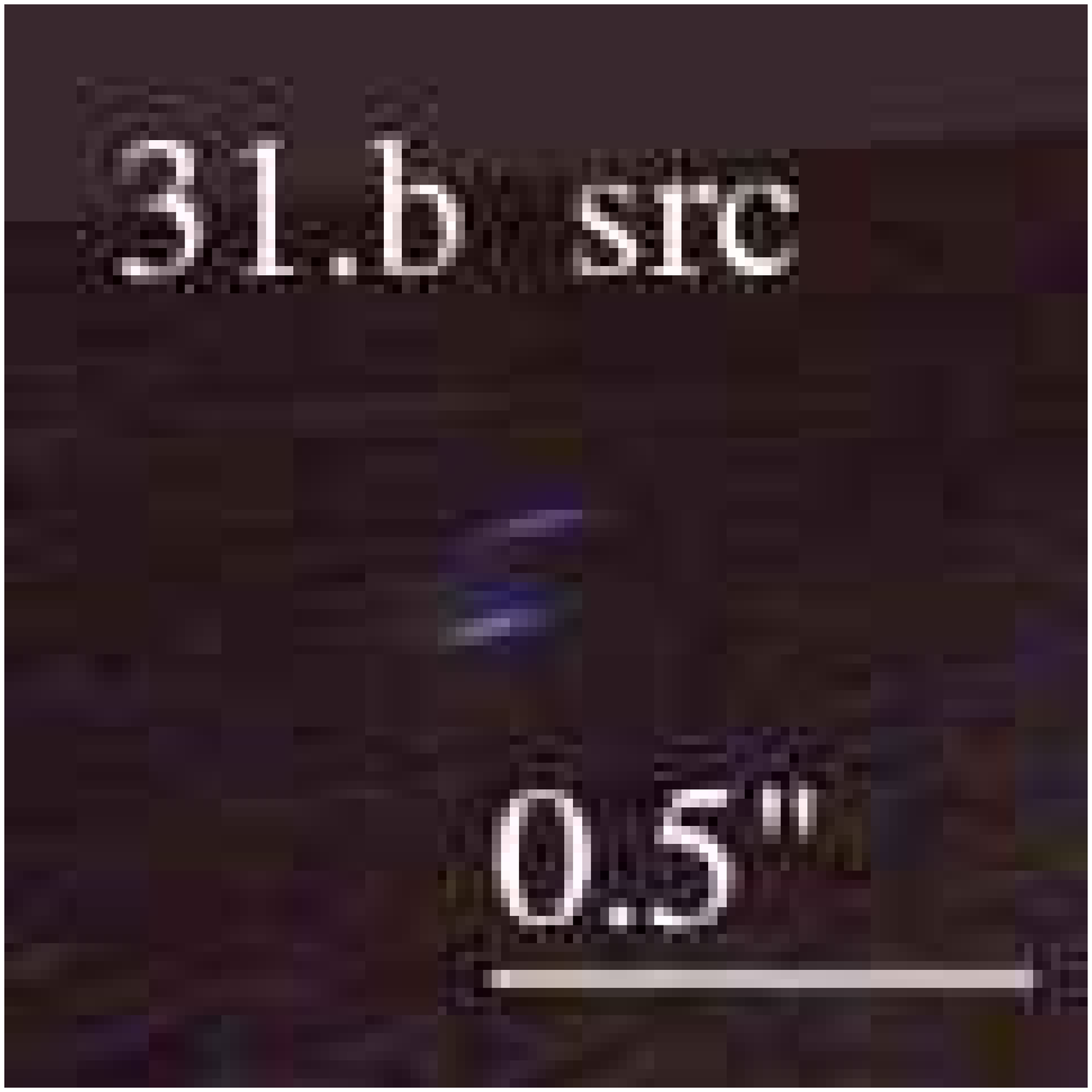}} \\
    \multicolumn{1}{m{1cm}}{{\Large ENFW}}
    & \multicolumn{1}{m{1.7cm}}{\includegraphics[height=2.00cm,clip]{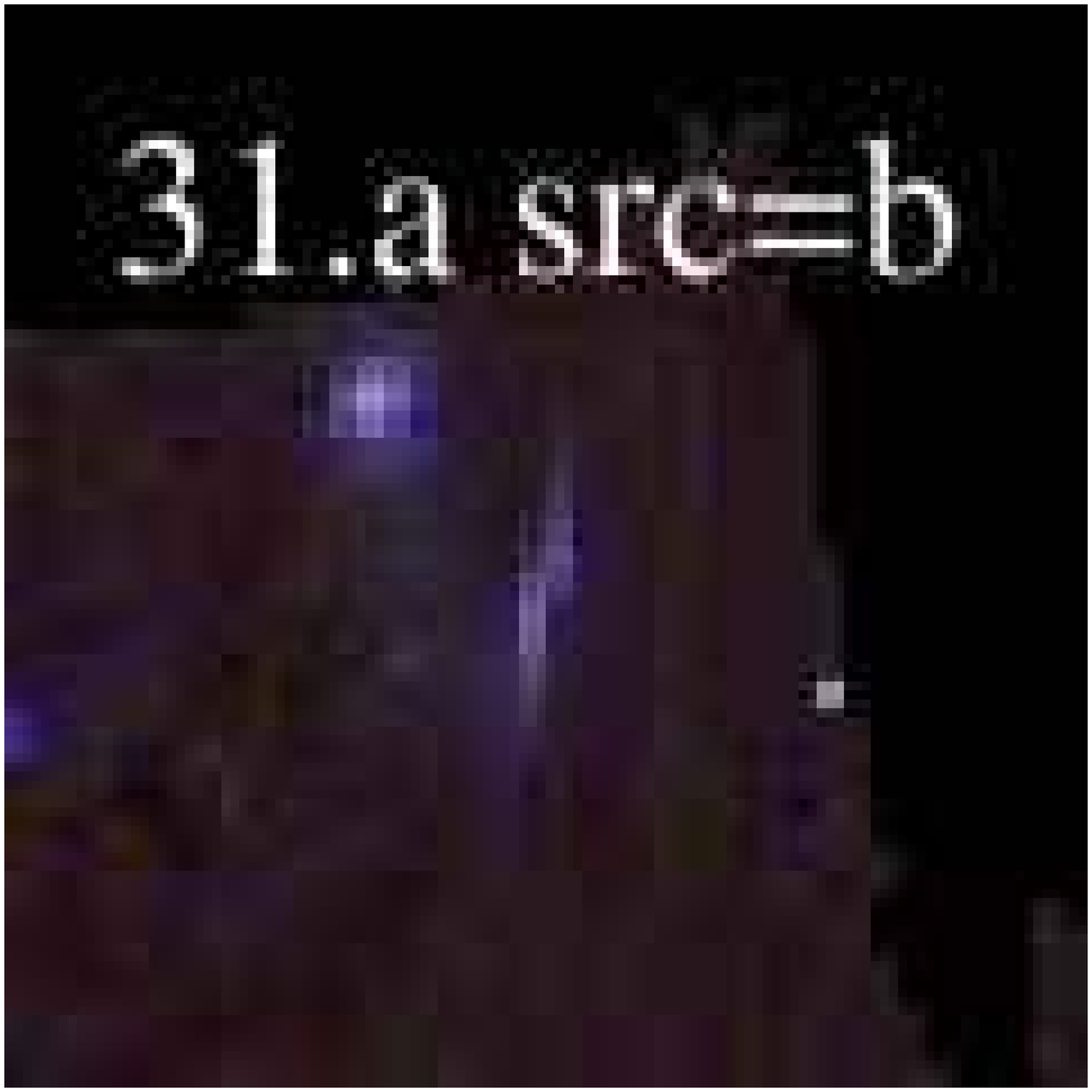}}
    & \multicolumn{1}{m{1.7cm}}{\includegraphics[height=2.00cm,clip]{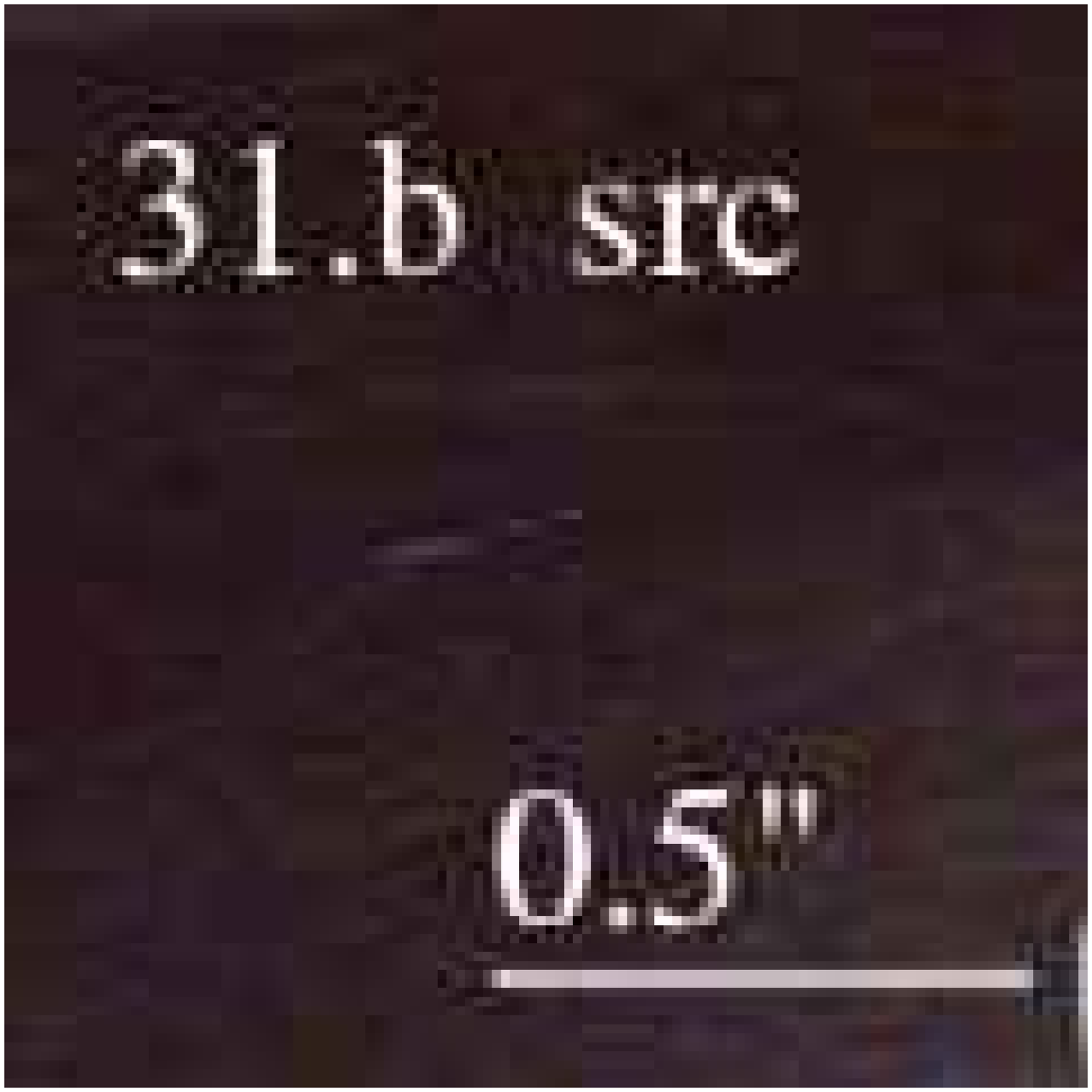}} \\
  \end{tabular}

\end{table*}

\clearpage

\bsp

\label{lastpage}

\end{document}